\documentclass[a4paper,11pt]{book}
\usepackage[english]{babel}
\usepackage[applemac]{inputenc}
\usepackage{dsfont}
\usepackage{amsfonts}
\usepackage{stmaryrd}

\def\be{\begin{equation}}
\def\ee{\end{equation}}
\def\bes{\begin{equation*}}
\def\ees{\end{equation*}}
\def\ba{\begin{eqnarray}}
\def\ea{\end{eqnarray}}

\def\eps{\varepsilon}
\def\lp{l_\text{Pl}}

\def\tr{\text{tr}}

\def\de{\mathrm{d}}

\newcommand{\f}{\frac}
\newcommand{\lb}{\big\lbrace}
\newcommand{\rb}{\big\rbrace}
\newcommand{\SU}{\text{SU}}
\newcommand{\DSU}{\text{DSU}}
\newcommand{\ISU}{\text{ISU}}

\newcommand{\SL}{\text{SL}}
\newcommand{\isu}{\mathfrak{isu}}
\newcommand{\su}{\mathfrak{su}}
\renewcommand{\sl}{\mathfrak{sl}}

\renewcommand\time{\,{\scriptstyle{\times}}\,}

\def\be{\begin{equation}}
\def\ee{\end{equation}}   
\def\ba{\begin{eqnarray}}
\def\ea{\end{eqnarray}}
\def\bas{\begin{subequations}\begin{eqnarray}}
\def\eas{\end{eqnarray}\end{subequations}}

\def\eps{\varepsilon}
\def\lp{l_\text{Pl}}

\def\tr{\text{tr}}

\def\de{\mathrm{d}}

\def\f{\frac}
\def\lb{\big\lbrace}
\def\rb{\big\rbrace}
\def\SU{\text{SU}}

\def\SL{\text{SL}}
\def\su{\mathfrak{su}}

\def\time{\,{\scriptstyle{\times}}\,}

\def\DSU{\text{DSU}}
\def\nn{\nonumber}

\usepackage{subfiles}

\usepackage{fancyhdr}
\fancypagestyle{plain}{
  \fancyhead{}
  
  \fancyfoot{}

}

\usepackage{bm}
\usepackage{color}
\usepackage{amsmath}
\usepackage{amssymb}
\usepackage{caption}
\usepackage{minitoc}   
\usepackage{appendix}
\usepackage{colortbl}  
\usepackage{graphicx}
\usepackage{hyperref}  
\usepackage{multirow}
\usepackage{enumerate}
\usepackage{mathtools} 
\usepackage[euler]{textgreek}

\makeatletter
\def\@makechapterhead#1{%
  \reset@font
  \vspace*{50\p@}%
  \hspace{-1.875cm}%
  \hbox{%
    \vbox{%
      \hsize=1.5cm%
      \begin{tabular}{c}%
        \scshape \strut \@chapapp{} \\
        \colorbox{black}{\vbox{\hbox{\vbox to 1mm{}}\hbox{\color{white} \LARGE \bfseries \hspace{1mm}\thechapter\hspace{1mm}}\hbox{\vbox to 2cm{}}}}%
      \end{tabular}%
      }%
    \vbox{%
      \hspace{20pt}
      \parbox{360pt}{
        \begin{flushleft} \Huge \bfseries #1 \end{flushleft}}%
      }%
    }%
  \vskip 20\p@ 
}
\makeatother

\usepackage{xspace}
\newcommand{\bra}[1]{\ensuremath{\langle #1|}\xspace}          
\newcommand{\ket}[1]{\ensuremath{|#1\rangle}\xspace}           

\newcommand{\clearemptydoublepage}{\newpage{\pagestyle{empty}\cleardoublepage}}
\protect\clearemptydoublepage
\protect\clearemptydoublepage

\newcommand{\cleartoleftpage}{%
  \clearemptydoublepage
  \ifodd\value{page}\thispagestyle{empty}\null\thispagestyle{empty}\newpage{}\thispagestyle{empty}\clearpage\thispagestyle{empty}\fi
}

\graphicspath{{figures/},{../figures/}}

\begin{document}

\begin{titlepage}
\centering
\textsc{\Large Université Paris Diderot (Paris 7) Sorbonne Paris Cité}\\
\large{ED~560 - STEP'UP - ``Sciences de la Terre de l'Environnement\\ et Physique de l'Univers de Paris''}\\
\vspace{0.125cm}
\vspace{1.125cm}
\textsc{\Large \bfseries Thèse de Doctorat }\\
\large{Physique theorique}\\
\vspace{1.125cm}

\hrulefill

\vspace{0.5cm}
\textbf{\huge Towards self dual Loop Quantum Gravity}\\
\vspace{0.25cm}

\hrulefill

\vspace{1.125cm}
présentée par\\
\vspace{0.25cm}
{\Large \bfseries Jibril} \textsc{\Large \bfseries Ben Achour}\\
\vspace{0.75cm}
\large{pour l'obtention des titres de}\\
\vspace{0.25cm}
\textsc{\large Docteur de l'Université Paris Diderot (Paris 7) Sorbonne Paris Cité}\\
\vspace{1.125cm}
Thèse dirigée par Dr Eric \textsc{Huguet} / Dr Karim \textsc{Noui} \\
Laboratoire AstroParticule et Cosmologie \\
\vspace{1.125cm}
soutenue publiquement le 30 septembre 2015 devant le jury composé de~:
{
\renewcommand{\arraystretch}{1.125}
\begin{table}[h!]
  \centering
  \begin{tabular}{llll}
     Dr & Eric & \textsc{Huguet} & Directeur de thèse\\ 
     Dr & Karim & \textsc{Noui} & Co - Directeur de thèse\\
    Dr & Etera & \textsc{Livine} & Rapporteur\\
    Pr & Hanno & \textsc{Sahlmann} & Rapporteur\\
    Dr & Renaud & \textsc{Parentani} & Examinateur\\
    Pr & Alejandro & \textsc{Perez} & Examinateur\\
    Pr &  Pierre & \textsc{Binetruy} & Président\\
  \end{tabular}
\end{table}
}

\end{titlepage}

\clearemptydoublepage

`` Ce qui est plus certain, c'est que nos notions habituelles d'espace et de temps, même assez profondément remaniées par la théorie de la relativité, ne sont pas exactement appropriées à la description des phénomènes atomiques. [...] En vérité, les notions d'espace et de temps tirées de notre expérience quotidienne ne sont valables que pour les phénomènes à grandes échelles. Il faudrait y substituer, comme notions fondamentales valables en micro-physique, d'autres conceptions qui conduiraient à retrouver asymptotiquement, quand on repasse des phénomènes élémentaires aux phénomènes observables à notre échelle, les notions habituelles d'espace et de temps. Est-il besoin de dire que c'est là une tâche bien difficile ?\\
 On peut même se demander si elle est possible, si nous pourrons jamais arriver à éliminer à ce point ce qui constitue le cadre même de notre vie courante. Mais l'histoire de la science montre l'extrême fertilité de l'esprit humain et il ne faut pas désespérer. Cependant, tant que nous ne serons pas parvenus à élargir nos concepts dans le sens indiqué à l'instant, nous devrons nous évertuer à faire entrer, plus ou moins gauchement, les phénomènes microscopiques dans le cadre de l'espace et du temps et nous aurons le sentiment pénible de vouloir enfermer un joyau dans un écrin qui n'est pas fait pour lui.'' \\\\
 
\qquad \qquad \qquad \qquad \qquad \qquad  Louis De Broglie - La Physique Nouvelle et les Quantas (1936)

\clearemptydoublepage

\chapter*{Acknowledgements}

Mes premiers remerciements vont tout d'abord à Eric Huguet et Jacques Renaud, pour m'avoir fait confiance en stage, puis offert la possibilité d'entreprendre cette thèse au sein du laboratoire APC, m'ouvrant ainsi les portes aux joies de la physique théorique, et plus globalement de la recherche scientifique. Je garderai un très bon souvenir de leur bonne humeur, leur disponibilité et des après midi passées devant le tableau à leurs cotés.

Je tiens à remercier tout particulièrement Karim Noui, pour m'avoir fait partager sa passion pour la recherche, ses idées et ses questions, et pour m'avoir fait découvrir le champ de recherche de la gravité quantique à boucles. J'ai particulièrement apprécié sa pédagogie, sa patience sans limites et sa rigueur scientifique. Ce fut un privilège et un très grand plaisir d'apprendre à ses côtés.

Je souhaite aussi remercier mes rapporteurs, Hanno Sahlmann et Etera Livine, pour le temps qu'ils ont pris pour la lecture de ce manuscrit et pour leurs retours et suggestions qui m'ont permis de clarifier plusieurs points.

Je remercie Marc Geiller pour ses conseils (hautement recommandés!), son hospitalité et sa bonne humeur. Ce fut un très grand plaisir de faire mes premiers pas en physique théorique dans le bureau 424A en compagnie de Florian, Marc, Alexis ... et de poursuivre aux côtés de Maxime et Alexis. Un grand merci à vous pour toutes ces discussions, de physique ou autre, et pour les doux craquages de fin de journee. Une pensée va évidemment à tous les compagnons de route, Julien, Ileyk, Georges, Pierre, Vivien, Julian, Ben, Maica, Romain, Lea et à toute l'équipe QUPUC plus généralement ! Une pensée toute particulière va aux tortues de feu, sans qui rien n'aurait pu arriver.

Je tiens à remercier Julien Grain, Antonin Coutant, Daniele Pranzetti et Boris Bolliet pour les bons moments passés en conference, et pour la découverte de Berlin ... et tous les amis, connaissances et rencontres inattendues de Paris qui ont fait de ces trois années une expérience des plus riche !

Finalement, une pensée toute particulière va à ma famille, 
pour leur indéfectible soutien tout au long de cette aventure et depuis si longtemps, un grand merci pour tout.

\clearemptydoublepage

\chapter*{Abstract}

In this PhD thesis, we introduced a new strategy to investigate the kinematical and physical predictions of self dual Loop Quantum Gravity (LQG) and by-passed the old problem of implementing quantum mechanically the so called reality conditions inherent to the self dual phase space.

Since our first motivations come from black holes thermodynamics, we first review, in the third chapter, the loop quantization of spherically isolated horizon based on the $SU(2)$ Ashtekar-Barbero variables leading to the micro-canonical entropy. This approach is based on the effective model of a Chern-Simons connection coupled to point-like particles living on the horizon. Then we present the so called ``gas of punctures'' model for the isolated horizon, which provides the framework to go beyond the micro-canonical ensemble and compute the canonical and grand-canonical entropy. In the context of this model, we investigate how the assumption of a Bose-Einstein statistic for the punctures impacts the semi-classical result and underline a relation between the condensation phenomenon and the presence of purely logarithmic quantum corrections to the entropy. The presentation of this model enable us to point the different limits and drawbacks of the $SU(2)$ loop quantization of spherically isolated horizon. First, one needs to proceed to an unnatural fine tuning on the Immirzi parameter to obtain the right semi-classical limit in the micro-canonical ensemble. Moreover, this quantization does not predict a holographic behavior for the degeneracy of the hole, as one could expect. Therefore, if one wants to avoid to precedent fine tuning on $\gamma$, one is led to assume the so called holographic hypothesis to obtain the right semi classical limit in the context of the gas of puncture model.

The fourth chapter is devoted to studying to what extend the loop quantization based on the self dual variables could cure those problems. Obviously, since no one knows how to quantize the self dual Ashtekar phase space of General Relativity, because of the so called reality conditions, we are led to introduce a new strategy, based on an analytic continuation of the degeneracy from $\gamma \in \mathbb{R}$ to $\gamma = \pm i$. We review in details the construction of the procedure, and present the results. The self dual degeneracy turns out to be naturally holographic, supplemented with some power law corrections which conspired, at the semi classical limit, to provide the expected logarithmic quantum corrections to the entropy. At the leading term, we recover the Bekenstein-Hawking area law. The discrete real area spectrum is turned into a continuous real area spectrum, even if we are now working with $\gamma = \pm i$. Finally, we recognize that our procedure is a well known map which send the Casimir and the character of the $SU(2)$ compact group into the Casimir and character of the continuous representations of the $SU(1,1)$ non compact group. A detailed discussion on the status of our procedure is provided at the end of the chapter.

The fifth chapter is devoted to understanding more precisely the interplay between the disappearance of $\gamma$ in the physical predictions of the quantum theory, the appearance of the $SU(1,1)$ group and the relation with the self dual variables. To to so, we introduce a new toy model of three dimensional gravity admitting a Barbero-immirzi like parameter and $SL(2,\mathbb{C})$ as its symmetry group. The canonical analysis of the toy model in two different gauges, one selecting the compact group $SU(2)$, the second one selecting the non compact $SU(1,1)$ group, allows us to conclude that at least in three dimensional gravity, the presence of $\gamma$ in the $SU(2)$ phase space is a pure gauge artifact. Finally, the loop quantization of the two phases spaces results in two different kinematical area spectrums, which turn out to be related through our analytic continuation procedure. At the end, we show that it is possible to reformulated the $SU(2)$ phase space in terms of new one, based on a complex $SL(2,\mathbb{C})$ connection, supplemented with some reality constraints. Once solved, the reality conditions reduces the complex $SL(2,\mathbb{C})$ phase space to the $SU(1,1)$ phase space as expected. This chapter allows us to exhibit an interesting mechanism concerning the disappearance of the Immirzi parameter in the predictions of three dimensional loop quantum gravity. 

Finally, the sixth chapter is devoted to applying our procedure to the simplest Loop Quantum Cosmology model. By first constructing the LQC dynamics for any arbitrary spin $j$ and then implementing our analytic continuation, we show that our procedure preserves the key features of the LQC models, i.e. we obtain a bouncing universe which admits the right semi classical limit after the bounce.  

\newpage

Dans cette thèse, nous introduisons une nouvelle stratégie afin d'étudier les prédictions cinématiques et physiques de la Gravité Quantique à Boucles (LQG) écrite à partir des variables complexes d'Ashtekar, le but étant d'obtenir une procédure pour contourner la résolution des contraintes de réalité.

Comme nos motivations proviennent avant tout de la thermodynamique des trous noirs (prédite par la LQG), nous présentons dans le troisième chapitre la quantification à boucles des horizons isolées sphériques, basée sur la connexion d'Ashtekar-Barbero $SU(2)$, donnant l'entropie micro-canonique de l'horizon. Cette approche est basée sur la théorie effective d'une connection de Chern-Simons couplée à des particules ponctuelles (défauts topologiques appellés ``piqures'') vivant sur l'horizon. Dans un deuxième temps, nous présentons le modèle dit ``gaz de punctures'' pour l'horizon isolée, qui permet de d'étudier le trou noir non plus seulement dans l'ensemble micro-canonique, mais aussi dans l'ensemble canonique et grand canonique. Dans ce contexte, nous présentons une étude détaillée de l'influence de la statistique de Bose-Einstein sur la limite semi classique de l'horizon. Cette étude permet de mettre à jour une relation entre le phénomène de condensation des piqures et la présence de corrections quantiques purement logarithmiques associée à l'entropie. Plus généralement, la présentation du modèle dit ``gaz de piqures'' nous permet de souligner les faiblesses et limitations de la quantification à boucles des horizons isolées basée sur la connexion réelle d'Ashtekar-Barbero $SU(2)$. Premièrement, il est nécessaire de fixer le paramètre réel d'Immirzi à une certaine valeur afin de reproduire la bonne limite semi-classique (ie l'entropie de Bekenstein Hawking) dans l'ensemble micro-canonique. Qui plus est, cette quantification ne prédit pas une comportement holographique concernant la dégénérescence de l'horizon, comme cela est généralement attendu. Par conséquent, afin d'éviter la fixation du paramètre d'Immirzi pour obtenir la bonne limite semi classique, il est nécessaire d'avoir recours de manière ad hoc, à l'hypothèse holographique dans le cadre du modèle du ``gaz de piqures'', ce qui n'est pas satisfaisant. 

Dans le quatrième chapitre, nous étudions dans quelle mesure la quantification à boucles basée sur les variables complexes d'Ashtekar permettrait une résolution de ces différents problèmes. Comme la quantification directe de la gravité complexe d'Ashtekar n'est pas connue, en raison de la difficulté d'imposer au niveau quantique les contraintes de réalité, nous sommes contraints d'introduire une nouvelle stratégie, basée sur une prolongement analytique de la dégénérescence. Ce prolongement analytique envoie le paramètre d'Immirzi réel à la valeur purement imaginaire. Nous présentons en détails la construction de ce prolongement analytique et les résultats qui en découlent. De manière surprenante, la dégénérescence ainsi calculée s'avère être naturellement holographique. De plus, les lois de puissances présentes dans l'expression de la dégénérescence conspirent pour donner, à la limite semi classique, les corrections logarithmiques attendues. Finalement, nous retrouvons bien en terme dominant l'entropie de Bekenstein-Hawking. Le spectre discret et réel, usuel en LQG, est modifié par notre procédure en un spectre continu et réel, bien que nous travaillons maintenant avec $\gamma = \pm i$. Enfin, nous reconnaissons dans notre procédure un ``mapping'' bien connu qui transforme le Casimir et le caractère du groupe compact $SU(2)$ en le Casimir et le caractère associés aux représentations continues du groupe non compact $SU(1,1)$.

Le cinquième chapitre est dédié à l'étude des liens existants entre différents mécanismes mis en lumière par notre prolongement analytique, à savoir: la disparition du paramètre d'Immirzi $\gamma$ des prédictions de la théorie, l'apparition du groupe $SU(1,1)$ et la relation avec les variables complexes d'Ashtekar. Pour ce faire, nous introduisons un modèle ``jouet'' de gravité trois dimensionnelle admettant un paramètre similaire au paramètre d'Immirzi $\gamma$ et le groupe $SL(2,\mathbb{C})$ comme groupe de symmetry. L'analyse canonique de ce modèle dans deux jauges différentes, l'une sélectionnant le sous groupe compact $SU(2)$ et la seconde sélectionnant le sous groupe non compact $SU(1,1)$, nous permet to montrer que la présence du paramètre d'Immirzi $\gamma$ dans l'espace de phase $SU(2)$ de notre modèle est un pure artefact de jauge. Finallement, la quantification à la boucles de ces deux espaces des phases aboutit à deux spectres d'aire quantiques différents (au niveau cinématique), qui se révèlent être reliés par la continuation analytique étudier dans le contexte des trous noirs. Enfin, nous montrons que l'espace de phase $SU(2)$ peut être réexprimé en un espace des phases, décrit par une connexion complexe $SL(2,\mathbb{C})$ accompagnée de contraintes de réalité. Une fois résolues, ces contraintes réduisent ce nouvel espace des phases complexe $SL(2,\mathbb{C})$ à l'espace des phases réel $SU(1,1)$, comme attendue pour la gravité $2+1$. Ce modèle ``jouet'' permet de mettre en lumière un mécanisme intéressant de disparition du paramètre d'Immirzi dans les prédictions de la Gravité Quantique à Boucles $2+1$. 

Finalement, le chapitre six est dédié à l'application de notre prolongement analytique au modèle le plus simple de Cosmologie Quantique à Boucles (LQC). Généralizant tout d'abord la dynamique de la LQC (pour $\Lambda = 0$ et pour un univers plat $k=0$) à un spin quelconque $j$ (et non plus $j=1/2$), nous appliquons ensuite notre procédure. Nous montrons ainsi que notre continuation analytique, de manière surprenante, préserve les résultats fondamentaux de la LQC, à savoir la résolution de la singularité initiale de l'univers quantique, tout en respectant la bonne limite semi-classique après le rebond.

\clearemptydoublepage

\chapter*{List of publications}

This thesis has given rise to the following articles.

\begin{itemize}
\item \textit{Testing the role of the Barbero-Immirzi parameter and the choice of connection in Loop Quantum Gravity}  (2013) \\ J. Ben Achour , M. Geiller, K. Noui, C. Yu, Published in Phys.Rev. D91 104016, arXiv:1306.3241 [gr-qc]

\item \textit{Spectra of geometric operators in three dimensional LQG: from discrete to continous} (2013) \\ J. Ben Achour , M. Geiller, K. Noui, C. Yu, Published in Phys.Rev. D89 064064, arXiv:1306.3246 [gr-qc]

\item \textit{Analytic continuation of black hole entropy in Loop Quantum Gravity} (2014) \\ J. Ben Achour , A. Mouchet, K. Noui,
Published in JHEP 1506 145, arXiv:1406.6021 [gr-qc] 

\item \textit{Loop Quantum Cosmology with complex Ashtekar variables} (2014) \\ J. Ben Achour, J. Grain, K. Noui,
Published in Class.Quant.Grav. 32  025011, arXiv:1407.3768 [gr-qc] 

\item \textit{Black hole as a gas of punctures : Bose Einstein statistic and logarithmic corrections to the entropy} (2014) \\ O. Azin, J. Ben Achour, M. Geiller, K. Noui, A. Perez,
Published in Phys.Rev. D91 084005, arXiv:1412.5851 [gr-qc]

\item \textit{Analytic continuation of real Loop Quantum Gravity : lessons from black holes thermodynamics} (2015) \\ J. Ben Achour, K. Noui,
Proceeding Conference: C14-07-15.1, arXiv:1501.05523 [gr-qc] 

\end{itemize}

and 

\begin{itemize}
\item \textit{Conformally invariant wave equation for a symmetric second rank tensor (``spin-2'') in a d-dimensional curved background }  (2013) \\ J. Ben Achour, E. Huguet and J. Renaud,
Published Phys.Rev. D89 064041, arXiv:1311.3124 [gr-qc]


\end{itemize}

\pagestyle{fancy}
\frontmatter

\dominitoc \tableofcontents 

\clearemptydoublepage
\mainmatter

\chapter*{Introduction \markboth{Introduction}{Introduction}}
\addstarredchapter{Introduction}
\label{Introduction}
\minitoc

During the three years along which I learned Loop Quantum Gravity and I investigated a particular aspect of the theory, i.e. its self dual version, I had the chance to be the witness of two great advances in our quest of understanding our physical world.
Those two major results are well representative of the fundamental research both in theoretical and experimental physics.

In 2013, during the second year of this PhD, the Large Hadron Collider (LHC) announced the discovery of the Higgs boson \cite{I-HiggsD}. This breakthrough confirmed the Higgs-Brout-Englert model of mass generation in the Standard Model of Particle Physics, predicted almost forty years ago \cite{I-BE, I-H}. This scalar particle, the first to be discovered in the large landscape of observed particles, was introduced in the sixties to explain the mass of the bosons $Z$ and $W$ responsible of the weak interaction. The mechanism relies on a spontaneous symmetry breaking which takes place at very high energy. The mass of the discovered scalar boson is around $125$ Gev/$c^{-2}$, which represents the heaviest observed particle, and therefore a new high energy domain. The same year, the Nobel prize was awarded to Peter Higgs and Francois Englert.

The year before, the satellite Planck concluded its three years of observations of the Cosmological Microwave Background (CMB). The first analysis of the observational data of the Planck mission was announced in 2013 \cite{I-Planck}.
The temperature distribution of the young Universe was established at a incredible level of accuracy, providing the tools to obtained a highly accurate measurement of the parameters of the Cosmological Standard Model. 
The evaluation of the dark components of our universe, i.e. dark energy and dark matter were refined while the possible inflation mechanisms were highly constrained \cite{I-PlanckInf}, excluding a very large class of theoretical models in the literature.

While the Higgs boson discovery concerned the Standard Model of Particles Physics, describing the observed quantum fields and their interactions at the very small length scales, the results of the Planck mission concerned our Standard Model of Cosmology, which deals with the dynamics and content of our universe, and therefore with the very large cosmological scales. Those two models represent our common understanding of the physical world. It is striking that these two models , incredibly successful in predicting the outcomes of experiments and observations,  are based on two conceptually different theories, i.e. Quantum Mechanics and General Relativity. Each one of those theories are built on a set of assumptions which contradict the other one. In a sense, admitting those two set of physical laws acting at different scales, lead us eventually to a schizophrenic picture of the physical world.

Quantum mechanics was developed at the beginning of the century, and the complete mathematical formulation of the theory was formulated in 1927, the seminal papers being \cite{I-QM1, I-QM2, I-QM3, I-QM4}. The theory was rapidly merged with special relativity in order to obtain a Lorentz invariant quantum theory, leading to Quantum Field Theory (QFT).
This theory deals with the physics at atomics and subatomics scales. Its construction required a conceptual revolution in order to understand the law of physics at those very small length scales.
This revolution transformed dramatically our usual notion of motion, measurement and determinism widely accepted in pre-relativistic and classical physics.
The predictions  of the theory are fundamentally probabilistic and no more deterministic. For a given experiment, the theory predicts a set of outcomes each one associated with a probability determined by the fundamental quantum laws. This failure of the determinism at those scales has nothing to do with a lack a precision in the experimental apparatus, but with an intrinsic incertitude in the law of quantum physics.
Moreover, any dynamical field turns out to be quantized, in the sense that it manifests itself in terms of discrete quantas, i.e. the particles. However, just as in pre-relativistic and classical physics, the laws of quantum mechanics are formulated in term of an external time variable $t$ (appearing explicitly in the Schrodinger equation) and describe the quantum motion of the quantum fields on a fixed background space-time. The experimental predictive success of QFT is impressive. It led to particle physics (the so called Standard Model of Particles Physics), atomics physics, nuclear physics, condensed matter physics, semi conductors, computers, lasers, quantum optics, and even to a better understanding of some astrophysical phenomenons, such as Neutron stars. Among all, the mathematical apparatus of the theory allowed to compute the most accurately verified prediction in the history of physics, i.e.  the anomalous magnetic moment of the electron $a_{e}$ \cite{I-QFT}.
The theoretical and experimental values of this quantity are given by:
\begin{align*}
a_{\text{theo}} = 1 159 652 182.79 (7.71)\times 10^{-12} \;\;\;\;\; \text{and} \;\;\;\;\; a_{\text{exp}} = 159 652 180.73 (0.28) \times 10^{-12}
\end{align*}
Again, the predictive success of QFT is incredible and we can safely assume that this theory describes in a fairly convincing manner the physical world at the sub-atomic scales. Therefore, one has no other choice to accept the revolutionary concepts of Quantum Mechanics in order to understand Nature at those scales.

General Relativity was work out at the same epoch, and presented by Einstein \cite{I-Al1} in its final covariant version in 1915. This theory was developed to merge the old Newtonian theory of gravity 
(1666) with the new Special Relativity theory (1905), in order to obtain a general relativistic theory of gravity describing the interaction of the gravitational field with relativistic bodies. Once again, the challenge of merging the two theories required a conceptual revolution in our understanding of what is time, space and causality.
In GR, the gravitational field is encoded in the components of the metric tensor field, which characterized the geometry and the causality of space-time. This field follows some dynamical equations called the Einstein field equations. Consequently, the geometry of space-time becomes a dynamical object, just as the electromagnetic or muon fields. It turns out that the Einstein's field equations are formulated in a covariant fashion and no preferred time variables is picked up to describe the dynamic of the gravitational and matter fields. Instead, an observer is free to choose any time variables and encodes the evolution of physical quantities with it. More, the usual notion of localization is spoiled in the sense that a given point of space-time has no physical meaning in the theory. The localization is understood as a relational procedure, which has to be performed with respect to another dynamical field. In this very elegant theory, the absolute and fixed space-time of pre-relativistic and classical physics is turned into a dynamical field, identified with the gravitational field. There is no more a fixed background on top of which the fields evolve and interact with each other. There is just a set of fundamental fields, among which the gravitational field, interacting with each other without reference to any fixed background. This background independence is the true fundamental novelty of General Relativity \cite{I-BKI, I-Ed1, I-Rov1}.
This theory is a classical theory and is therefore deterministic, i.e. for some given initial conditions (and boundary conditions), one obtains a single physical prediction from the theory.
It led to new areas of physics, such as relativistic astrophysics, cosmology, GPS technology and gravitational radiation, opening maybe the road to gravitational wave astronomy. In this framework, new objects such as horizons and black holes appeared, giving rise to a whole new field of physics.
Among all the postdictions of the theory, it was possible to accurately explain the advance of the perihelia of the Mercury planet, but also to predict the frame dragging effect and the redshift of photons escaping from gravitational objects and the bending of light by astrophysical objects. This theory also predicted the expansion of the Universe, but Einstein was not comfortable with this idea of a dynamical universe, and missed this incredible prediction by modifying the theory, a mistake that he called the biggest mistake of his life. The predictive success of GR is as impressive as the one of QFT and nowadays, the theory has been validated from the millimeter to the parsec. 

However, as any classical theory with a given domain of applicability, it turns out that GR predicted some monstruous objects, i.e. the singularities.
They are ``points'' of space-time where the curvature and the matter density diverge, and corresponds to a prediction of GR at very small scales, even smaller than the usual length scales of QFT.
Those singularities are the signature that the predictive power of GR is lost at those scales and that one can not safely assumed that GR describes the gravitational field in this domain. 
Typically, the singularities occur in cosmology, at the very beginning of the Universe, and at the heart of the black holes. 
If GR fails to describe the gravitational field in this regime of very high energy, what is the right description ?
Since QFT is up to now our best tool to describe the high energy physics and the outcomes of experiments at very small scales, a merging of General Relativity and Quantum Field Theory is mandatory.
However, as we have explained precedently, the two very successful theories are based on very contradicting assumptions with respect to the notion of space, time, causality, motion, and what a theory should predict.

The quest of finding the quantum description of the gravitational field has begun almost one hundred year ago and represents on of the most challenging scientific question of our epoch. The (not yet found) resulting theory is called the Quantum Theory of Gravity. The first role of such theory is to remove the singularities presented in classical GR.

The idea of merging General Relativity and Quantum Field Theory follows the philosophical idea of unification, which has proven to be very successful in the history of sciences physics.
Indeed, the theory of electromagnetism was developed by Maxwell in order to merge the electric and magnetic fields discovered by Faraday into a single object, the electromagnetic field, leading to the discovery of the fundamental nature of light.
Special Relativity was introduced by Einstein in 1905 in order to have a coherent picture of the old Galilean mechanics and the freshly developed electromagnetism.
Ten years later, Einstein unified the old Newtonian theory of gravity and its Special Relativity into an elegant and revolutionary theory of space-time geometry, i.e. General Realtivity.
Finally, incorporating the fundamental principles of Special Relativity in the Quantum Theory led to the very successful Quantum Field Theory. It is therefore natural to apply the same strategy to develop the quantum theory of gravity, by trying to merge General Relativity and Quantum Field Theory. 

The research towards a coherent picture of the quantum gravitational field has taken different directions which can be regroup into three main lines: the covariant approach which used the technics of usual Quantum Field theory, that we can denote the ``flat space quantizations'', the sum over histories which can be thought as a ``Feynman quantization'' and the canonical approach due to Bergman and Dirac, which is based on a ``phase space quantization''. 

 The first attempt to quantize the gravitational field belongs to the first category. Shortly after the construction of Quantum Mechanics (1927) and the quantization of the electromagnetic field (1930), Rosenfeld and then Fierz and Pauli (1952) quantized the linearized version of General Relativity. The graviton was born. The idea was to quantize the metric perturbations propagating on a fixed Minkowskian metric. This program was developed further in the following years and the Feynman rules for linearized General Relativity were soon established by De Witt and Feynman (1957-1967). It was then realized that the transitions amplitudes for the graviton are ill defined at the two loop level (1964). The hint for evidence of the non (perturbative) renormalizability of the graviton quantum field theory was pursued by t'Hooft and confirmed by Deser and Van Nieuwenhuizen (1973) and finally by Gorof and Sagnotti (1986). It was then commonly accepted that the graviton quantum field theory has uncontrollable divergences up to one loop corrections. This is due to the fact the the graviton is radically different from the other ``particles'' since its coupling constant is dimensionful, leading to a quantum field theory that is not perturbatively renormalizable. 
The program of flat quantization of linearized General Relativity in four dimensions finally turned out to be a dead end. Yet, following the example of the Fermi theory for the weak interaction, one can still argue that General Relativity is only the low energy limit of a more fundamental quantum field theory. The hope is that additional higher energy terms in the Lagrangian (either from the gravitational part or from the matter part) will cure the intrinsic divergencies of General Relativity. This is in this spirit that supergravity and supersymmetric string theory emerges (1976). The idea behind string theory is very elegant. The whole set of particles, matter and interaction mediators including the graviton, are understood as different excitations modes of a single extended object, the string. It is a powerful unifying idea and the field grew rapidly to become nowadays the major candidate to the theory of quantum gravity. Very interesting results were obtained within this theory, such as the derivation of the Bekenstein Hawking entropy for extremal and near extremal black holes (1996). However, the price to pay to recover a Lorentz invariant dynamics for the string is to propagate it on a higher dimensional fixed background, the number of dimensions depending on the model (generally $d > 4$). In order to recover a four dimensional space-time at large scales, one has then to compactify the additional dimensions, the choice of the compactification being highly non trivial and leading to a large landscape of string models. Moreover, in this context, the spectrum of the string which provides the content in particles is much larger than what is experimentally known and involve unobserved supersymetric parters for all known particles. Finally, up to now, there is no proof that string theory in its different versions cures the whole divergencies of the four dimensional quantum linearized version of General Relativity. In the mid nineties, important efforts to go beyond the perturbative approach leads to the first non perturbative results in string theory which point towards a single fundamental theory behind the different existing versions, i.e. the so called M theory (1995). Shortly after, the AdS/CFT correspondence conjecture was introduced (1998) and turned out to be a powerful tool to study the M theory. Numerous interesting links between non perturbative string theory and other approachs to quantum gravity appear in the literature, such as relations between strings and non commutative geometry or as the polymer non perturbative attempt to quantize the bosonic string (2004) \cite{I-STh, I-SKa}. Yet, a full background independent formulation of the theory is still elusive.

 The failure of linearized General Relativity to be pertubatively renormalizable can be well understood (the argument extends to any quantum theory of gravity with a fixed background).
 When quantizing the graviton spin-$2$ field, one fixes the background (to be either Minkowskian or another fixed geometry) and then quantizes the perturbations on top of it.
 However, this procedure is ill defined when dealing with the gravitational field. First it assumes that the differentiable and smooth property of the space-time manifold are preserved down to the Planck scale.
 This is a very strong hypothesis and intuitively, we do not expect to recover a smooth manifold at those scales but rather a ``space-time foam'' to quote James Wheeler (1963). Indeed, the gravitation field, i.e. the metric, describes the causality and the geometry of space-time itself. From the general relativistic perspective, there is no fixed background on top of which fields propagate, but only fields (including the gravitational field itself) interacting which each other without any background. Once quantized, the fully quantum metric field will experience quantum fluctuations and quantum superpostions states of the quantum states of the metric will occur, just as for any quantum field. This lead to a fuzzy notion of the causality and the geometry of space-time, i.e. to a \textit{quantum space-time}. In order to have a full quantum theory of the gravitational field , one need to understand precisely this \textit{quantum geometry}.
 From this perspective, fixing a classical background and studying perturbations of the metric on top of it seems to be the wrong strategy to obtain a full quantum theory of the gravitational field. Because of background independence, the quantum theory of gravity is fundamentally non perturbative.

In this spirit, one is pushed to look for non perturbative technics to tackle the quantization of the gravitational field. From the renormalization point of view, General Relatvity, while not perturbatively renormalizable, could well be renormalizable nonpertubatively. This idea led Weinberg to introduce the asymptotic safety program (1976), where one looks for an UV fixed point for General Relativity. Interesting result have already been obtained in this direction \cite{I-Reuter}.

From the ``covariant Feynman quantization'' point of view, the first proposal was introduced by Minser (1957). While formal, the conceptual ideas behind his construction had an important impact on the field.
Twenty years later, Hawking introduced the ``euclidean path integral for gravity'', which has to be understood as a formal Feynman path integral over the riemannian 4-dimensional metric fields (1978). However, defining a suitable measure on the space of metric turns out to be a difficult task and this formalism remains hard to exploit. Moreover, the concept of Wick rotation is generally ill defined and going from the euclidean sector to the Lorentzian one is not as harmless as in the usual path integral quantization of quantum field theory. Yet, it is in this formalism that the notion of the ``wave function'' of the universe was introduced, generating an intense activity in this field. Ten years later, Hartle introduced a sum-over-histories for General Relativity (1983). Finally, the ``Feynman quantization'' for General Relativity will re born through the introduction of the spin foam models inspired from the Loop quantization approach.

The ``phase space quantization program'' was initiated by Bergman and Dirac. The formalism of constrained systems was worked out and applied to General Relativity shortly after. The canonical structure of the theory was rapidly derived by Dirac (1959) and then simplified by Arnold, Deser and Misner \cite{I-ADM}, i.e. in the so called ADM approach (1961). The Hamilton Jacobi equation for General Relativity was then introduced by Perez (1962) and led Wheeler and De Witt to write down the first quantum version of the hamiltonian constraint for quantum General Relativity (1967). This contribution was the starting point of the canonical quantization program. However, for reasons explained in the two first chapters, no one has ever succeeded to build a Hilbert space of the full theory. This goal was reached only in some reduced dimensional models. However, the Wheeler-De Witt equation turns out to be a prolific source of inspiration for the community working in quantum gravity. Yet, fog settles down on the canonical approach for almost twenty years, until Abhay Ashtekar introduced the so called ``complex new variables'' for General Relativity (supplemented with their associated reality conditions) \cite{I-Ash1, I-Ash2}, simplifying drastically the canonical structure of the theory (1986). General Relativity was formulated as a background independent connection gauge theory. This was the starting point of the loop quantization approach. Only two years after, Jacobson and Smolin found interesting solutions of the hamiltonian constraint in the Ashtekar formalism (1988) \cite{I-JacSmo1, I-JacSmo2}. Those solutions consist in closed Wilson loop of the Ashtekar connection. Two years later, the \textit{loop representation} or \textit{polymer representation}  for General Relativity was introduced by Rovelli and Smolin \cite{I-RovSmo1}. This non standard representation is based on earlier work on the loop quantization of gauge theory introduced by Gambini, Trias and other (1980-1983) \cite{I-Gam1}. The polymer quantization is the essential tool to build a background independent quantum field theory of the gravitational field (and of any other field). This quantization relies on the hypothesis, sometimes denoted the \textit{polymer hypothesis}, that the quantum states of the gravitational fields are extended sting-like objects. This polymer quantization was applied to the scalar, Maxwell and graviton field shortly after \cite{I-PolSca, I-PolMax, I-PolGraviton}. However, before applying this quantization procedure to full General Relativity, one needed another step. Indeed, due the difficulty to implement, at the quantum level, the reality conditions inherent to the complex Ashtekar's formalism, a new formulation of General Relativity was introduced by Barbero \cite{I-Bar1}, where the complex Ashtekar $SL(2, \mathbb{C})$ connection is turns into a real $SU(2)$ one, labelled by a real parameter $\gamma$, nowadays called the Barbero-Immirzi parameter (1994). This real $SU(2)$ Ashtekar-Barbero phase space is the starting point of Loop Quantum Gravity. During the ten next years, a rigorous mathematical effort was made by Ashtekar, Lewandowski, Rovelli, Isham, Baez and others, to clarify the quantization procedure based on the Ashtekar-Barbero phase space \cite{I-As1, I-As2, I-As3, I-As4, I-As5, I-As6}. An important progress was realized by Rovelli and Smolin who introduced the spin network basis for the gauge invariant quantum states (1995) \cite{I-RovSmo2}, recovering an independent construction for quantum geometry due to Penrose (1964). The same year, the spectrum of the quantum area and the quantum volume operator were derived \cite{I-RovSmo3}. Those results provided for the first time a concrete quantization of space from a (still incomplete) non perturbative quantum theory of gravity. Finally, a proposal was introduced by Thiemann in order to deal with the reality conditions of the self dual phase space \cite{I-ThRC} in term of a Wick rotation (1995). Then the subject of solving the reality conditions fell in the shadow, the interest being now to exploit the real quantum theory inherited from the real $SU(2)$ Ashtekar-Barbero phase space. The year after, different results appeared (1996). The black hole entropy was computed for the first time by Rovelli \cite{I-RovBH} who recovered the Bekensetin Hawking area law up to a fine tuning on the Immirzi parameter. Moreover, a proposal for a regularization of the complicated hamiltonian constraint was given by Thiemann \cite{I-ThHC1, I-ThHC2, I-ThHC3}. In the mid nineties, a complete canonical quantum theory was then available, but the concrete implementation of the dynamics and the control of the semi classical limit remained obscure. 

Facing the difficulty to implement the dynamics on the canonical side, different authors turned towards a new kind of models, inspired from Ooguri's work on topological quantum field theory \cite{I-Ooguri}. Those spin foam models were first investigated by Rovelli and Reisenberger (1992-1997) \cite{I-RovReis1, I-RovReis2, I-RovReis3} and successfully defined for the first time by Barett and Crane (1997-1999) \cite{I-BarCrane1, I-BarCrane2}. They represent a concrete implementation of the path integral formation for quantum gravity, the sum being over a discrete structure colored by group data, contrary to summing over smooth four dimensional manifold as in the euclidean path integral framework.

At the same epoch, the loop quantization of symmetry reduced models begun. The idea was to applied loop quantization to the two physically relevant situation for quantum gravity, black hole singularity and cosmological singularity. Thiemann, Bojowald, Ashtekar and other authors studied the loop quantization of the spherically symmetric space-time (1994-2005), in the real as well as in the self dual version of the theory \cite{I-SSThKas1, I-SSThKas2, I-SSBoAs}. The conclusion concerning the resolution of the interior singularity depends on the approach. At the same epoch, Bojowald presented the first scenario for the resolution of the Big Bang singularity form quantum geometry effects (1999-2001), leading to a new area called Loop Quantum Cosmology \cite{I-LQCBo1, I-LQCBo2}.  The relation between the polymer quantization and the usual Fock quantization was studied and clarify by Varadarajan, Ashtekar, Lewandowski, Sahlmann and collaborators (2000-2001) \cite{I-Var, I-AsLew1}. In the same spirit, the polymer quantum mechanic of the free particle and the harmonic oscillator were developed and studied by Ashtekar and collaborators \cite{I-PolFP} and by Agullo and collaborators \cite{I-PolCorichi1, I-PolCorichi2}, providing a very enlightening platform to understand this new quantization procedure, the problem of semi-classicality and the link with the usual Schrodinger quantum mechanics appearing at lower energy scale. One important step towards the understanding of the loop quantization procedure was provided by the LOST theorem (2006) due to Lewandowski, Okolow, Salhmann and Thiemann \cite{I-LOST}. This theorem states that the polymer representation used in Loop Quantum Gravity, provided that it carries a unitary action of the diffeomophism group, is unique ! Three years later, Fleishchack demonstrated a similar theorem for the full exponentiated algebra \cite{I-Flei}. This powerful uniqueness theorem represents the corner stone of the loop quantization of General Relativity and put on a solid basis the resulting quantum theory of geometry.

During the ten next years, the main advances were divided between the applications of the theory to concrete situations, such as quantum cosmology and quantum black hole, and the developments of models and technics in order to tackle the difficult questions of the dynamics and the semi classical limit. Since it would be out of reach to mention all the important contributions to theses fields, we only mention some of them in a purely subjective way. The spin foam models on one hand, and the black hole and quantum cosmology fields on the other hand, grew rapidly, with some important improvements and interconnections. For instance, semi classical black hole will obviously turn out to be a very good laboratory to study the semi classical limit of the theory.

From the spinfoam side, the link between the canonical quantum theory and the spin foam approach was established in three dimension (for euclidean gravity without cosmological constant) by Noui and Perez (2004) \cite{I-NouiSF}. In order to go beyond the difficulties present in the Barett Crane model, new technics were introduced in order to deal with the semi classical limit. Coherent states were defined independently by Thiemann \cite{I-ThCS1, I-ThCS2}, and Freidel, Livine, Speziale and Bianchi (2006-2010) \cite{I-FreiCS, I-LivCS, I-BianchiCS}. Those technics allowed the introduction of four dimensional spin foams models solving some difficulties of the Barett Crane model, i.e. the FK model (2008) \cite{I-FK} and the EPRL model (2009) \cite{I-EPRL}. Then Barett and other collaborators succeeded to computed the semi classical approximation of the spin foam amplitude \cite{I-Barrett1, I-Barrett2}, which turns out to match the Regge theory (2010), giving confidence in this approach. A major step in the understanding of those ``path integral quantization'' models appeared with the Group Field Theory reformulation, generalizing the spin-foams construction. A spin foam turns out to be understood as a scalar field theory over a group manifold, i.e., the field taking its arguments in a Lie group and no more in a conventional space-time manifold. Within this reformulation, powerful technics of quantum field theory could be applied, and an important effort was devoted to generalizing the renormalization procedure to those new background independent quantum field theories. Those models are closely related to the so called tensorial quantum field theories, where the renormalization procedure have been intensively studied. For pedagogical reviews on the subject, one can refer to very nice review of Oriti \cite{I-OritiGFT} and to the following thesis manuscript \cite{I-CarrozzaPhD}. In the same spirit, news technics of corse-graining in the very framework of spin-foam model were also introduced and investigated by Dittrich and collaborators, in order to obtain a better control of the semi classical limit \cite{I-BianchaCoarse1, I-BianchaCoarse2}. The idea behind those works is that recovering the smooth classical geometry of General Relativity from a quantum theory of gravity, such as the spin foams models, cannot be addressed simply by taking a naive ``large spin limit''. The smooth geometry is believed to be the result of a collective behavior of the quantum excitations of the gravitational field, and should show up only in some given phases. Therefore, in order to recover the expected differentiable manifold properties in the semi classical limit, it is mandatory to develop the renormalization procedure and corse-graining technics and investigate the possible phase transitions which could occur for a very large number of quantas of the gravitational field. 

Much efforts were also devoted by Bozom, Livine and collaborators to understanding the link between the dynamics provided by the canonical and the spinfoam approaches in the four dimensional case \cite{I-Bon1, I-Bon2, I-Bon3}. In particular, it was shown that the quantum Wheeler Dewitt equation generates a difference equation on the spin network states, which can be related to some recurrence formulas inherent to the transition amplitudes of the spin foams models, i.e. recurrence formulas on the $9j$ or $15j$ symbols \cite{I-Bon4}. More recently, a ``flux formulation'' of LQG was introduced by Dittrich and Geiller (2015) \cite{I-BchGei1, I-BchGei2}. It provides a canonical loop quantization of the Ashtekar-Barbero phase space, based on a new vacuum different from the usual Ashtekar-Lewandowski one, which seems to be more closely related to the spinfoams quantization and is therefore easier to compare.
Finally, the geometrical information contained in the spin network states was unravel by the introduction of the twisted geometry \cite{I-TGFrei1, I-TGFrei2} and spinning geometry \cite{I-SGFrei} (2010-2013). Among other results, those works allowed a reformulation of spin foam models in term of spinors and twistors, which led to a better understanding of the geometry encoded in the spin foam models. Those works culminated with the twistorial description of spin foam transition amplitudes due to Speziale and Wieland \cite{I-Twistor1}, were the transitions amplitudes of the EPRL model were recovered using path integral in twistor space (2012). Finally, a new spinorial action was introduced by Wieland (2014). The resulting equations of motion derived from this action turns out to be solved by the so called twisted geometry \cite{I-TwistorAction}.  

From the black hole point of view, a very intense activity in the field led to a rigorous computation of the micro canonical black hole entropy based on the horizon quantum geometry, as well as the introduction of new quasi local definition of the black hole, i.e. the isolated horizon. This object was defined and quantized by Ashtekar, Lewandowski, Krasnov, Corichi and other collaborators \cite{I-AshBH1, I-AshBH2, I-AshBH3, I-AshBH4}, which underlined the role played by the $U(1)$ Chern Simons quantum field theory on a punctured $2$-sphere to describe the quantum excitations of the horizon (2000). Then, the quantization procedure was generalized by Perez, Engle, Pranzetti and Noui \cite{I-AlejBH1, I-AlejBH2}, who showed that one can reformulate the quantization in term of the very well known quantization of a $SU(2)$ Chern Simons quantum field theory coupled to point like particles (2010-2011). This quantization is based on the notion of quantum group, leading eventually to a general formula for the degeneracy of the quantum black hole given by the so called Verlinde formula \cite{I-AlejBH3}. The study of its asymptotics leads to the micro canonical entropy of the horizon. This procedure is reviewed and summarized in chapter 3. An important ingredient was introduced by Perez, Gosh and Frodden, who derived a new notion of local energy for the classical isolated horizon \cite{I-Fro1}, allowing to go beyond the micro canonical computation, and to define the canonical and grand canonical computation for the entropy (2011). Based on this new way to do local physics at the horizon, a statistical model for the quantum black hole was recently introduced \cite{I-Amit1, I-Amit2, I-Amit3}, the so called ``gas of puncture'' picture (2013). The quantization of spherically symmetry reduced model was recently refined by Gambini and Pullin and the interior singularity was shown to be resolved (2013). The computation of the entropy from spin foam models was undertaken by Bianchi and revisited in a different ways since then. More recently, the computation of the entropy of the isolated horizon from the group field theoretical framework was undertaken by Oriti and collaborators \cite{I-BHOriti} (2015). The entropy of the three dimensional BTZ black hole was obtained by Geiller, Noui, Frodden and Perez \cite{I-Fro2}. Finally, the same authors computed the entropy of the four dimensional black hole with the self dual variables by mean of an analytic continuation \cite{I-Fro3} (2012). This argument was justified and the construction of the analytic continuation prescription was rigorously defined in \cite{I-BA1} (2014), opening a new road to deal with the reality conditions and the self dual version of LQG. Those two computations led to the exact Bekensetin Hawking entropy supplemented with its logarithmic quantum corrections, curing the fine tuning on the Immirzi parameter present in the real computation.
Those recent contributions, and many others \cite{I-Muxin, I-Pranz1, I-Pranz2, I-Pranz3, I-Neiman1, I-Neiman2, I-BA2, I-BA3, I-BA4}, seem to point towards the peculiar status of the self dual variables with respect to the dynamics and to the semi classical limit of LQG.
This analytic continuation prescription is the subject of the thesis and it will be presented in chapter 4.

Finally, from the LQC perspective, the seminal work of Bojowald \cite{I-LQCBo1, I-LQCBo2} (1999-2001) was intensively developed by Ashtekar and collaborators, see \cite{I-AsLQC1, I-IvLQC1} for pedagogical reviews. LQC is nowadays an independent area of research and probably the most interesting candidate to obtain physical predictions that could be actually tested. One of the most important contributions was the introduction of the ``improved dynamics'' \cite{I-IDLQC}, which allows to recover the usual Friedman cosmology on large scale.
The LQC technics were applied to many different cosmological settings, i.e. Bianchi and Gowdy cosmologies \cite{I-BianchiLQC1, I-BianchiLQC2, I-BianchiLQC3, I-GowdyLQC1}, flat or closed universe \cite{I-LQCUC, I-LQCOU}, filled up with a scalar field, radiation or dust \cite{I-LQCRad, I-LQCDust}. From those works, the robustness of the singularity resolution for the resulting quantum universe (while admitting the right semi classical limit) was demonstrated \cite{I-LQCRob}. More refined technics were introduced, such as the the group theoritical quantization of isotropic loop cosmology proposed in \cite{I-EteraLQC}, based on the isomorphism between some phase space observables describing the isotropic universe coupled to a scalar field and the $su(1,1)$ Lie algebra, avoiding quantization anomalies and factor ordering ambiguities.
From the inflation point of view, the link between the pre-inflation and inflation periods was also investigated, as well as a general strategy to compute the cosmological perturbations on a quantum space-time.
Two major approaches were developed, i.e. the dressed metric approach due to Ashetkar, Agullo and Nelson \cite{I-Dressedmetric}, and the so called deformed algebra approach due to Bojowald, Barrau, Cailletaux and Grain \cite{I-LQCDA1, I-LQCDA2}. Those frameworks allowed to compute the power spectrum of the scalars and tensorial perturbations, modified by the existence of the quantum bounce. The probality for the quantum universe to experience a slow role inflation compatible with the observational datas has also been investigated by Ashtekar and Sloan, and showed to be close to one \cite{I-LQCProbInf1, I-LQCProbInf2}.
Finally, some efforts were realized to generalized the quantization strategy used in LQC, such as defining a hamiltonian for higher spin than $j=1/2$ \cite{I-DSLQCKev, I-DSLQCNoui}. Those investigations underline the highly regularization-dependent character of the dynamics in LQC models. Recently, this generalization was used to defined a self dual LQC model based on the very same analytic continuation which was defined in the context of black hole physics \cite{I-DSLQCNoui}. Another model of self dual LQC appeared recently in the literature \cite{I-SDLQC}. Both models recover the bouncing scenario and the right semi classical limit (2014-2015). Finally, the way to recast LQC within the spinfoam framework was studied by Ashtekar, Campiglia and Anderson \cite{I-SFLQCAs} and also by Bianchi, Rovelli and Vidotto \cite{I-SFLQCRov}. Important efforts were also undertaken in the context of group field theory in order to reproduce the homogenous and isotropic geometries of cosmology form the full theory \cite{I-LQCGF1, I-LQCGF2, I-LQCGF3, I-LQCGF4}. LQC represents nowadays the best window to test the loop quantization of General Relativity \cite{I-ObsLQC1, I-ObsLQC2}, which explain the very large literature existing in the field. For a more detailed discussion, see chapter 6.

This concludes our chronological overview of the researches directions in the field of Loop Quantum Gravity.

Before closing this chronological review, we mention the very transversal line of research which is the non commutative geometry due to Connes \cite{I-Conne}. This program, led to important results both for the quantum gravitational field and the standard model of particles. Among all, it provides an very elegant mathematical construction of the whole standard model, including the potential of the Higgs boson and the Einstein-Hilbert action, but also a reformulation of the renormalization procedure from a rigorous mathematical point of view. The general picture which emerges from this theory is that the complicated classification of the particles of the standard model, with their mixing matrix and masses, is just a consequence of the fundamental quantum nature of space-time. This theory has become a serious candidate both for the theory of quantum gravity and for a unified theory of the four interactions. Many other approaches exist, such as the dynamical triangulation, the twistor theory, causal sets and other ... \\

Having sketch the chronological path to the theory and the general context, let us discuss what is the subject of this thesis. With the precedent chronological review, we have seen that the theory admits only one free parameter, i.e. the Barbero-Immirzi parameter $\gamma$.
This parameter was introduced by Barbero when trading the self dual variables for the real ones \cite{I-Bar1}. It turns out that this parameter plays a very peculiar role in the quantum theory, entering in the kinematical quantum predictions such as the area and volume spectrums, the entropy of the black hole, the maximal energy density of the universe and finally in the construction of spin foam models ... This omnipresence of $\gamma$ in the quantum theory have risen an important debate about its status and its interpretation. A quite common interpretation is that $\gamma$ should be understood as a fundamental length scale specifying the area gap of LQG which should be fixed by some future experiments. It has been also argued that the ambiguity related to $\gamma$ is similar to the $\theta$ ambiguity present in the quantization of Yang-Mills gauge theory \cite{I-Gam}. This is due to the fact that the canonical transformation which turns the complex Ashtekar variables into the real Ashtekar-Barbero variables cannot be implemented as a unitary transformation in the quantum theory \cite{I-RovTh}. In the same line, it has been proposed that $\gamma$ could play the role of the free angular parameter of the large gauge transformations and that it should be understood as a field and no more as a constant \cite{I-Merc1, I-Merc2}. In the same spirit, the interpretation of $\gamma$ as a field was investigated in \cite{I-Merc3, I-Merc4, I-Krasnov}. Another possible interpretation was introduced in \cite{I-Rand1, I-Rand2}, where it was shown that the presence of $\gamma$ is related to the PT symmetry violation of the so called generalized Kodama state. Moreover, it has been shown that while totally irrelevant from the predictions of vacuum GR, the Immirzi parameter can be understood as the coupling constant of the four-fermions interaction when the Holst action is supplemented with some lagrangian containing fermions fields \cite{I-AlejRov}. See also \cite{I-Merc5, I-FreiImm} for other investigations on the role of the Immirzi parameter in presence of fermions and some of its effects on the cosmological scenario. Finally, it was recently argued that more than a simple coupling constant, the Immirzi parameter could be understand as a cut off for quantum gravity \cite{I-Liv}.

This PhD is devoted to studying another interpretation, which does not represent the majoritary point of view, but has known important encouraging results in the past three years.
According to this point of view, the Immirzi parameter is interpreted as a regulator in the quantum theory, which keeps trace of the non compactness of the initial gauge group we started with, i.e. the Lorentz group.
The presence of $\gamma$ is due to the particular time gauge selected in the canonical analysis, which gives a prominent role to the compact $SU(2)$ group \cite{I-Alex1, I-Alex2, I-AlexLiv}. It allows to perform the loop quantization until the construction of the diffeo-invariant Hilbert space. However, according to the interpretation adopted in this thesis, which is based on a series of works all pointed in the same direction \cite{I-Fro3, I-BA1, I-Muxin, I-Pranz1, I-Pranz2, I-Pranz3, I-Neiman1, I-Neiman2, I-BA2, I-BA3, I-BA4}, the self dual quantum theory seems to represents a better candidate for the physical quantum theory for the gravitational field than the one based on the real $SU(2)$ Ashtekar-Barbero variables.
In particular, the quantum theory based on the self dual variables  seems to reproduce the expected semi classical limit for  black hole in a much more satisfying way than the quantum theory based on the real Ashtekar-Barbero variables \cite{I-Fro3, I-BA1, I-BA2}. From this point of view, the Immirzi parameter should be send back at some point of the quantization program  to the purely imaginary value, in order to obtain the self dual quantum theory. Building such analytic continuation from the real to the self dual quantum theory is the subject of this PhD.
Whether one has to performed the Wick rotation before or after imposing the dynamics is a totally open question. However, this analytic continuation has to provide somehow a way to solve the reality conditions inherent to the self dual theory. The task unravel in the PhD was to derive such concrete procedure and test it on different settings, such as the entropy of quantum spherically isolated horizon, three dimensional gravity and Loop Quantum Cosmology. This work is presented in chapters $4$, $5$, $6$. Chapter $3$ introduce the so called ``gas of punctures`` model, on which we have also worked during this PhD, and which will be used in chapter $4$.

\clearemptydoublepage

\chapter{Hamitonian formulation of General Relativity : from the ADM to the Ashtekar phase space}
\label{ch:RG}
\minitoc

In this chapter, we present the hamiltonian formulation of General Relativity, first in the ADM (or second order) formalism then in the first order formalism.
This exercise is of first importance to understand the symmetries of the classical theory and therefore, to build the quantum theory which respects those classical symmetries.
In the case of General Relativity, the canonical analysis underlines the role play by the diffeomorphism symmetry of the theory, i.e. the background independence of General Relaltivity.
The analysis is performed following the Dirac formalism for constrained systems.  

Le us first present the metric formulation of General Relativity before proceeding to the canonical analysis.

\section{The ADM hamiltonian formulation of General Relativity}

In General Relativity, the space-time is modeled by a four dimensional differentiable manifold $\mathcal{V}$, equipped with a metric $g$ with  a Lorentzian signature $(-1,+1,+1,+1)$ and a rule of parallel transport $\nabla$. We denote it by $(\mathcal{V}, g, \nabla)$. 
Following the seminal work by Einstein \cite{ch1-Al1}, the rule of parallel transport $\nabla$, or equivalently the space-time connection $\Gamma$ is taken to be the unique torsionless and metric compatible connection, i.e. the so called Levi Civita connection. Under this assumption, the gravitational field is described by the Einstein-Hilbert action:
\begin{align*}
S_{4D} = \frac{1}{2\kappa} \int_{\mathcal{V}} \; dx^{4} \sqrt{-g} R
\end{align*}
where $\kappa = 8 \pi G$. The only dynamical field is the four dimensional metric tensor $g_{\mu\nu}$. The Ricci scalar, build from contracting the Riemann tensor, is the only scalar leading to field equations involving at most second order derivative of the metric tensor. Working with the Levi Civita connection $\Gamma(g_{\mu\nu})$, the Ricci scalar and the Riemann tensor are both defined as:
\begin{align*}
R = g^{\rho\nu} g^{\sigma\mu}R_{\sigma\rho\mu\nu}  \;\;\;\;\;\;\;\; \text{and} \;\;\;\;\;\;\;  [\nabla_{\mu}, \nabla_{\nu}] v^{\sigma} = R^{\sigma}{}_{\rho\mu\nu} v^{\rho} 
\end{align*}
for any vector field $v = v^{\sigma} \partial_{\sigma}$ in $\mathcal{V}$. The Riemann tensor encodes the local curvature of the manifold. It satisfies some geometrical identities which do not constrain the dynamics of the metric. They arise because of the invariance under coordinate transformation of the theory, which is a non dynamical symmetry. Those identities are called Bianchi identities and working with the Levi Civita connection, they reduce to:
\begin{align*}
R_{\sigma\rho\mu\nu} + R_{\sigma\mu\nu\rho}+ R_{\sigma\nu\rho\mu} = 0 \qquad  \nabla_{\alpha}R_{\sigma\rho\mu\nu} + \nabla_{\mu}R_{\sigma\rho\nu\alpha}+ \nabla_{\nu}R_{\sigma\rho\alpha\mu} = 0 
\end{align*}
In term of the Levi Civita connection $\Gamma$, the Riemann tensor is given by:
\begin{align*}
R^{\sigma}{}_{\rho\mu\nu} = \partial_{\mu} \Gamma^{\sigma}{}_{\rho\nu} - \partial_{\nu} \Gamma^{\sigma}{}_{\rho\mu} + \Gamma^{\sigma}{}_{\lambda\mu} \Gamma^{\lambda}{}_{\rho\nu}  - \Gamma^{\sigma}{}_{\lambda\nu} \Gamma^{\lambda}{}_{\rho\mu} 
\end{align*}
where the Levi Civita connection is totally determined by the (first derivative of the) metric tensor:
\begin{align*}
\Gamma^{\sigma}{}_{\mu\nu} = \frac{1}{2} g^{\sigma\rho} ( \partial_{\mu} g_{\rho\nu} + \partial_{\nu} g_{\rho\mu} - \partial_{\rho} g_{\mu\nu} )
\end{align*}
It is symmetric in its two last indices which is the result of asking that torsion is vanishing.
From the Einstein-Hilbert action supplemented with some matter lagrangian $\mathcal{L}_{m}$ , i.e.:
\begin{align*}
S_{4D} = \int_{\mathcal{V}} \; dx^{4} \sqrt{-g} \; ( \frac{1}{2\kappa}R + \mathcal{L}_{m})
\end{align*}
one obtains the famous Einstein's field equations which describe the dynamic of the gravitational field coupled to matter:
\begin{align*}
G_{\mu\nu} = R_{\mu\nu} - \frac{1}{2} g_{\mu\nu} R = 8 \pi G T_{\mu\nu}
\end{align*}\\
where $G_{\mu\nu}$ is the Einstein tensor and $T_{\mu\nu}$ is the source of the gravitational field, i.e. the energy momentum tensor, defined as:
\begin{align*}
T_{\mu\nu} =  \frac{1}{\sqrt{-g}} \frac{\delta \mathcal{L}_{m}}{\delta g^{\mu\nu}}
\end{align*}

The contracted second Bianchi identity leads to the ``conservation law'' for the Einstein tensor $\nabla_{\mu} G^{\mu\nu} = 0$. As explained, the Bianchi identity is a purely geometrical constraint and does not refer to the dynamic, i.e. never we used the Einstein field equations to obtain it. The symmetry which is behind is a non dynamical one, i.e. the invariance of the theory under coordinates transformations. Consequently, the covariant ``conservation'' of the Einstein tensor cannot be regarded as a true conservation law.
 
Once this conservation law has been derived, the Einstein's field equations automatically imply the covariant conservation law for the energy-momentum tensor $\nabla_{\mu} T^{\mu\nu} = 0$. 
Therefore, contrary to the conservation of the Einstein tensor $G_{\mu\nu}$, the conservation of the energy momentum tensor $T_{\mu\nu}$ is obtained only when the field equations, i.e. the dynamic is taken into account. Consequently, there is a dynamical symmetry behind this conservation law which is nothing else than the invariance of the theory under diffeomorphism \cite{ch1-Ed1}.   
Studying how the conservation laws in General Relativity are obtained teach us that since any theory can be formulated in a coordinate free fashion, the invariance under coordinates transformations is dynamically empty and generates only geometrical identities which remain true whatever the dynamic of the field is. However, the invariance under diffeomophisms is a true dynamical symmetry which constrains the dynamic and generates the true conservation law of the energy momentum tensor. 

The fact that the theory is invariant under diffeomorphisms is usually called the background independence of General Relativity.
The metric field which dictates the geometry and the causality of the space-time has now the very same status as any field in physics.
Within this new framework, the classical fields do not propagate in space through time, i.e. in a given space-time but they simply interact with another dynamical field, i.e. the metric tensor.
Due to the gauge invariance of the theory under diffeomorphisms, one cannot speak about the value of a field at the point A in space-time since this field can always be pushed forward to another point B and relabeled by a diffeomorphism which does not modify the physical content of the theory (see the hole argument originally proposed by Einstein and discussed in \cite{ch1-Rov1}). From this observation, one is led to a very unusual picture of reality, where space-time as a fixed arena on top of which the other field live disappear to let only the dynamical gravitational field interacts with the other fundamental fields. There is no more a fixed background on which one can do physics. The physical reality is truly background independent and localization in presence of gravity is purely relational.

This status of the diffeomorphims in General Relativity becomes crystal clear in the hamiltonian formulation of the theory.
We present now the first canonical analysis of General Relativity which was worked out in the sixties by Arnold, Deser and Misner \cite{ch1-ADM}.
It unravels the true dynamical variables of the theory and the gauge symmetry under which the theory is invariant. \\

\textit{The ADM hamiltonian analysis} \\

The idea is to perform a $3+1$ decomposition, by selecting a foliation of the four dimensional space-time $(\mathcal{V}, g)$ into a family of space-like Cauchy hyper surfaces $(\Sigma, \bar{g}, K)$ (a world line intersects once and only once a given Cauchy surface, $\bar{g}$ is the induced three dimensional metric on $\Sigma$ while $K$ is the second fundamental form and will be introduced later). Since they are space-like, at each point $p$ of any hypersurface $\Sigma$, there is a unit vector $n \in T_{p}\mathcal{V}$, normal to $\Sigma$ which is time-like. This means that the vector $n$ links two events in $\mathcal{V}$ for which the time interval is greater than the space interval, i.e. in our convention, $ds^{2}= g_{\mu\nu}n^{\mu}n^{\nu} = -1 < 0$.  Those kind of space-times are called the Cauchy's development of $(\Sigma, \bar{g}, K)$ and correspond to globally hyperbolic space-times, i.e. they are all diffeomorphe to a product $\Sigma \times \mathbb{R}$.

With this foliation, the space-time $\mathcal{V}$ and its tangent space at the point $p$ can be written as follow
\begin{align*}
\mathcal{V} = \Sigma \times \mathbb{R} \;\;\;\;\;\; \text{and} \;\;\;\;\;\;\;  T_{p}\mathcal{V} = T_{p}\Sigma \oplus   \mathbb{R} n 
\end{align*}
Given some local coordinates $\{x^{\mu}\} = \{t,x^{a}\}$ on $\mathcal{V}$, we obtain a basis $(\partial_{t}, \partial_{a}) \in T_{p}\mathcal{V}$ at any point $p \in \mathcal{V}$. The time evolution vector field between two Cauchy hyper surfaces $\Sigma(t)$  and $\Sigma(t+dt)$ can be decomposed into its normal and its tangential part w.r.t. $\Sigma$. Its components read
\begin{align*}
\tau^{\mu}= \frac{\partial x^{\mu}}{\partial t} =  N \; n^{\mu} + X^{\mu}
\end{align*}
Requiring that $\tau^{\mu}$ be timelike and future directed, it implies that $N \in \mathbb{R}^{+ *}$. The scalar $N$ is called the lapse and $X\in \Sigma$ is a three dimensional space-like vector called the shift. A priori, $N$ and $X$ can depend both on space and time. 

 \begin{figure}
\begin{center}
	{\includegraphics[width = 0.5\textwidth]{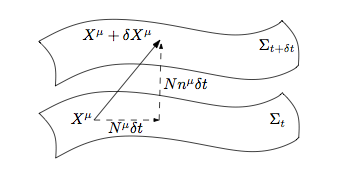}}
	\caption{3+1 decomposition of the space-time into spacelike hypersrufaces $\Sigma_{t}$ evolving in ``time''. Taken from \cite{ch1-Simone1}.}
	\label{fig:complex}
\end{center}
\end{figure}

In term of those quantities, we obtain the ADM metric:
\begin{align*}
ds^{2} &= - N^{2} dt^{2} + q_{ab} ( dx^{a} + X^{a}dt)( dx^{b} + X^{b} dt) \\
& =( - N^{2} + X^{2})dt^{2} +  2 X_{a}  dx^{a}dt + h_{ab} dx^{a} dx^{b}
\end{align*} 
where $h_{\mu\nu} = g_{\mu\nu} + n_{\mu}n_{\nu}$ is the intrisinc metric (projector) on $\Sigma$. Therefore, the four dimensional metric $g_{\mu\nu}$ is replaced by the triple $(q_{ab}, N, X^{a})$. With their associated velocities $(\dot{h}_{ab}, \dot{N}, \dot{X}^{a})$, they are the phase space coordinates  in the ADM formalism.

In order to write the decomposition of the action under this foliation, we need to decompose the Riemann tensor and therefore the covariant derivative $\nabla$. One can show that for two vectors $(X,Y) \in \Sigma$, the covariant derivative can be decomposed as:
\begin{align*}
\nabla_{u} v = D_{v} v + K(u,v) n
\end{align*}
where $D$ denotes the covariant derivative on $\Sigma$ and $K(\;,\;)$ is a bilinear form called the extrinsic curvature given by:
\begin{align*}
K(u,v) = g ( D_{u}n, v) = -  u^{\rho}v^{\sigma} \nabla_{\rho} n_{\sigma}= -  u^{\nu}v^{\mu}h_{\mu}{}^{\rho}h_{\nu}{}^{\sigma} \nabla_{\rho}n_{\sigma} = u^{\nu}v^{\mu} K_{\mu\nu}
\end{align*}
The tensor $K_{\mu\nu}$ can be shown to be symmetric ($n$ is an exact one-form) and can then be written as the Lie derivative of the intrinsic metric w.r.t. the normal vector $n$:
\begin{align*}
\; K_{\mu\nu} = -  \; h_{\mu}{}^{\rho}h_{\nu}{}^{\sigma} \nabla_{(\rho}n_{\sigma)} = - \frac{1}{2} (\; \mathcal{L}_{n} h \;)_{\mu\nu} 
\end{align*}
Note that the indices of the extrinsic curvature can be raise and lower using only the intrinsic metric $q_{\mu\nu}$.
With the precedent decomposition of the covariant derivative and due to the symmetries of the Riemann tensor, one can split the Riemann tensor in three pieces. They are respectively given by the Gauss, the Codazzi and the Mainardi equations and explicitly derived in \cite{ch1-Gour}:
\begin{align*}
h^{\mu}{}_{\alpha} h^{\nu}{}_{\beta} h^{\sigma}{}_{\gamma} h^{\sigma}{}_{\delta} R_{\mu\rho\nu\sigma} & = R^{(3)}_{\alpha\gamma\beta\sigma} + K_{\alpha\beta} K_{\gamma\delta} - K_{\alpha\delta} K_{\beta\gamma} \\
h^{\mu}{}_{\alpha} h^{\nu}{}_{\beta} h^{\sigma}{}_{\gamma}R_{\mu\rho\nu\sigma}  n^{\rho}  & =  D_{\beta} K_{\alpha\gamma} - D_{\gamma} K_{\alpha\beta} \\
h^{\mu}{}_{\alpha} h^{\nu}{}_{\beta} R_{\mu\rho\nu\sigma}  n^{\rho} n^{\sigma} & = \mathcal{L}_{n} K_{\alpha\beta} + K_{\alpha\mu} K^{\mu}{}_{\beta} +  \frac{1}{N} D_{\alpha} D_{\beta} N \\
\end{align*}

From those three equations, one obtains the full knowledge of the components of the Riemann tensor in term of the three dimensional covariant derivative $D$ and the extrinsic curvature $K_{\mu\nu}$.
In particular, under this decomposition, the Ricci scalar becomes:
\begin{align*}
R = R^{(3)} + K_{\mu\nu}K^{\mu\nu} - K^{2} - 2 \nabla_{\mu}[n^{\nu} \nabla_{\nu}n^{\mu} - n^{\mu} \nabla_{\nu}n^{\nu}]
\end{align*}
where $K = K^{\mu}{}_{\mu}$ is the trace of the extrinsic curvature and $R|_{\Sigma}$ is the Ricci scalar associated to $\Sigma$. Notice that the time derivative of the three dimensional metric $q_{ab}$ appears only in the terms involving the extrinsic curvature. 

Plugging this decomposition into the Einstein Hilbert action, the last term will behave as a total derivative and therefore as a boundary term that we disregard.
Thus, the ADM decomposition of the Einstein Hilbert action reads:
\begin{align*}
S_{EH}(q_{ab}, \dot{q}_{ab}, N, \dot{N}, X_{a}, \dot{X}_{a}) & = \frac{1}{\kappa} \int_{\mathcal{V}} dx^{4} \sqrt{g}R \\
& =  \frac{1}{\kappa} \int _{\mathbb{R}}dt \int_{\Sigma} dx^{3}  \; N\sqrt{q} \; [ \; R^{(3)} + K_{\mu\nu}K^{\mu\nu} - (K^{\mu}{}_{\mu})^{2} \; ] \\
& =  \frac{1}{\kappa} \int _{\mathbb{R}}dt \int_{\Sigma} dx^{3}  \; N\sqrt{q} \; [ \; R^{(3)} + K_{ab}K^{ab} - (K^{a}{}_{a})^{2} \; ]
\end{align*}
where we keep only the spatial part since $K^{0a} = q^{0b} q^{ac}K_{bc} = 0$.
The conjugate momentums to the three dimensional metric $q_{ab}$, the lapse $N$ and the shift $N_{a}$ are respectively given by:
\begin{align*}
P_{ab} & = \frac{\partial \mathcal{L}}{\partial \dot{q}^{ab}} = \frac{\partial \mathcal{L}}{\partial K^{ab}}\frac{\partial K^{ab}}{\partial \dot{q}^{ab}} + \frac{\partial \mathcal{L}}{\partial K}\frac{\partial K}{\partial \dot{q}^{ab}} = \frac{1}{\kappa} \sqrt{q} ( \; K_{ab} - q_{ab}K) \\
P & = \frac{\partial \mathcal{L}}{\partial \dot{N}} = 0 \\
P_{a} & =  \frac{\partial \mathcal{L}}{\partial \dot{X^{a}}} = 0 \\
\end{align*}
We conclude that the lapse $N$ and shift $X^{a}$ are non dynamical fields. In the terminology of Dirac, $\dot{N}$ and $\dot{X}^{a}$ are the primary inexpressible velocities which make the lagrangian of General Relativity singular. We will see that the diffeomorphism invariance of the theory generates this singular behaviour.  

The only true dynamical field is the intrinsic metric $q_{ab}$ of the Cauchy hypersurface $\Sigma$. 
Using that:
\begin{align*} 
& P_{ab}P^{ab} = \sqrt{q} \; ( K_{ab} K^{ab} + K^{2}) \;\;\;\;\;\;\;\;\;\;\;\;\;  P^{2} = 4\sqrt{q}  K^{2} \;\;\;\;\;\; \\
& \frac{1}{\kappa}N\sqrt{q} \; [ \; R|_{\Sigma} + K_{ab}K^{ab} - (K^{a}{}_{a})^{2} \; ] = \frac{1}{\kappa}N\sqrt{q} \;  R|_{\Sigma}+  \frac{1}{2} [ \; P^{ab} \dot{q}_{ab} - (\mathcal{L}_{X}q)_{ab} ] 
\end{align*}

we obtain the following action:
\begin{align*}
S_{EH}(q_{ab}, \dot{q}_{ab}, N, \dot{N}, X_{a}, \dot{X}_{a}) &  =  \frac{1}{\kappa} \int _{\mathbb{R}}dt \int_{\Sigma} dx^{3}  \; N\sqrt{q} \; \{ \; R|_{\Sigma} + K_{ab}K^{ab} - (K^{a}{}_{a})^{2} \; \} \\
& = \int _{\mathbb{R}}dt \int_{\Sigma} dx^{3} \; \{ \;  P^{ab}\dot{q}_{ab} - (\mathcal{L}_{X}q)_{ab}  - \frac{N}{\kappa \sqrt{q}} (\; P^{ab}P_{ab} - \frac{1}{2}P^{2} + \;  R|_{\Sigma} ) \; \}
\end{align*}

Therefore, we see that the lapse $N$ plays the role of a Lagrange multiplier.
Finally, we know that the conjugate momentums of the laps $N$ and the shift $X^{a}$ vanish. We need to introduce them, enforcing in the same time their vanishing behaviour through two additional constraints generated by the two news Lagrange multipliers $\lambda$ and $\lambda_{a}$:
\begin{align*}
S_{EH}& = \int _{\mathbb{R}}dt \int_{\Sigma} dx^{3} \{ \;  P^{ab}\dot{q}_{ab} + P^{a} \dot{X}_{a} + P \dot{N}- [ \; X^{a}H_{a}  + N \; H + \lambda C + \lambda_{a} C^{a} ] \; \}
\end{align*}
The Poisson bracket between the canonical conjugated variables $(q_{ab}, P^{ab})$ reads:
\begin{align*}
\{ \; P^{ab}(x), q_{cd}(x')\; \} = \kappa \delta^{a}_{(c}\delta^{b}_{d)} \delta^{3}(x,x')
\end{align*}
and the constraints are respectively given by:
\begin{align*}
& H_{a}  =  - (\mathcal{L}_{X}P)_{ab} \;\;\;\;\;\;\;\;\;\;\;\;\;\;\;  \\
& H = \frac{N}{\kappa \sqrt{q}} (\; P^{ab}P_{ab} - \frac{1}{2}P^{2} + \;  R|_{\Sigma} ) \\
& C  = P  \;\;\;\;\;\;\;\;\;\;\;\;\;  \\
& C_{a} = P_{a} \\
\end{align*}
$H_{a}$ and $H$ are called respectively the vectorial and the Hamiltonian constraints. We can now give the explicit form of the ADM Hamiltonian of General Relativity, which reads:
\begin{align*}
H_{ADM} & = P^{ab}\dot{q}_{ab} + P^{a} \dot{X}_{a} + P \dot{N} - \mathcal{L}_{ADM} \\
& =  X^{a}H_{a}  + N \; H + \lambda C + \lambda_{a} C^{a}
\end{align*}

In the Dirac terminology of the constrained systems, those four precedent constraints are the primary constraints of General Relativity.
In order to have a consistent hamiltonian formulation, the constraints $C \simeq 0$ and $C_{a} \simeq 0$ have to be preserved under the hamiltonian evolution.
Therefore, we compute the Poisson bracket between respectively $C$ and $C_{a}$ and the Hamiltonian $H_{ADM}$. Smearing the two constraints $C$ and $C_{a}$ respectively with a smearing function $f$ and a smearing vector field $f^{a}$, we obtain the secondary constraints:
\begin{align*}
\dot{C}(f) = \{ \; C(f), H_{ADM}\; \} = H(f)  \;\;\;\;\;\;\;\;\;\;\;\;\;\;\; \dot{C}_{a}(f^{a}) = \{ \; C_{a}(f^{a}), H_{ADM}\; \} = H_{a}(f^{a})
\end{align*}
where $H(f)$ and $H_{a}(f^{a})$ are respectively the smeared hamiltonain constraint and the  smeared vectorial constraint. Asing that $\dot{C}(f) \simeq 0$ and $\dot{C}_{a}(f^{a}) \simeq 0$, which is required for consistency, we see that $H(f)$ and $H_{a}(f^{a})$ have to vanish. Therefore, the hamiltonian of General Relaitivty $H_{ADM}$ is nul. This is characteristic of system invariant under reparametrization of time, such as the free particle \cite{ch1-Ed1}.
Finally, studying the evolution of the secondary constraints, one can show that it do not generate tertiary constraint. The hamiltonian analysis of the theory is complete.

The algebra of the constraints (see \cite{ch1-Th1} for further details and explicit computations) reads:
\begin{align*}
&\{ \; \vec{H}(\vec{f}) , \vec{H}(\vec{g})\; \} = - \kappa \vec{H}( \mathcal{L}_{\vec{f}}\; \vec{g}) \\
&\{ \; \vec{H}(\vec{f}) , H(g)\; \} = - \kappa H( \mathcal{L}_{\vec{f}}\; g) \\
&\{ \; H(f) , H(g)\; \} = - \kappa \vec{H}( \vec{X}(f,g,q)) 
\end{align*}
where $X^{a}(f,g,q) = q^{ab}(f \partial_{b}g - g \partial_{b}f)$. It is called the hypersurface deformation algebra. We observe that the two primary constraints $H$ and $\vec{H}$ are first class constraints, i.e. they form a close algebra. We expect therefore that $H$ and $\vec{H}$ are the generators of some fundamental symmetries of the theory. However, contrary to usual gauge field theory based on a Lie group, the algebra of constraints does not form a Lie algebra since the Poisson brackets between first class constraints lead to structure functions and not structure constants. 

Finally, one can compute for instance the Poisson bracket between the hamiltonian constraint $H(N)$ and the momentum $P_{ab}$. This will clarify the role play by the diffeomorphisms group in General Relativity. After a quite long computation \cite{ch1-Th1}, a brave student obtains:
\begin{align*}
& \{ \; H(N) , P_{\mu\nu}\; \} = q_{\mu\nu} \frac{NH}{2} - N \sqrt{q}[q^{\mu\rho}q^{\nu\sigma} - q^{\mu\sigma}q^{\nu\rho}] R_{\rho\sigma} + \mathcal{L}_{Nn} P_{\mu\nu}
\end{align*}
Since the hamiltonian constraint is a first class constraint, we expect that it generates a symmetry on the constrained phase space.
Indeed the last term in the precedent expression is the infinitesimal change of $P_{\mu\nu}$ under a diffeomorphism along the unit vector $n$ normal to the three dimensional surface $\Sigma$.
However, one can safely identify the hamiltonian constraint as the infinitesimal generator of the ``time'' diffeomorphisms only if the vacuum Einstein equations are satisfied $R_{\rho\sigma} = 0$ and if $H=0$, i.e. in a weak sense.
This symmetry refers only to the physical trajectories of the phase space. Therefore, the diffeomophism invariance of the theory is a dynamical symmetry, i.e. it constraints the equations of motion.
This is in contrast with the invariance under coordinate transformation, which is a non dynamical symmetry, it do not refers to the equations of motion of the theory.

The vectorial constraint $H_{a}$ and the hamiltonian constraint $H$ are therefore the generators of the infinitesimal diffeomophisms, respectively along a spatial direction in $\Sigma$ and along the normal vector $n$ to $\Sigma$.
This is obvious form the following Poisson brackets, evaluated on shell, i.e. for $R_{\mu\nu} = H_{a} = H = 0$.
\begin{align*}
& \{ \; H(N) , P_{ab}\; \} \simeq \mathcal{L}_{Nn} P_{ab} \;\;\;\;\;\;\;\;   \{ \; H(N) , q_{ab}\; \} \simeq \mathcal{L}_{Nn} q_{ab} \\
& \{ \; H_{c}(X^{c}) , P_{ab}\; \} \simeq \mathcal{L}_{X^{c}} P_{ab} \;\;\;\;\;\;\;\;  \{ \; H_{c}(X) , q_{ab}\; \} \simeq \mathcal{L}_{X} q_{ab} 
\end{align*}
This conclude our presentation of the ADM hamiltonian formulation of General Relativity.
 
 Having a proper hamiltonian formulation of General Relativity, one can try to build the quantum theory from this classical phase space.
 Unfortunately, in this formulation, the hamiltonian constraint which generates the dynamics (i.e. which is a gauge transformation in this case) is highly non trivial.
 Following the Dirac quantization program, one first builds a representation of the quantum algebra generated by the canonical variables transformed into operators $(\hat{P}^{ab}, \hat{q}_{ab})$.
 Having the Hilbert space structure, one then imposes the first class constraints as operators on the quantum states. This last step extracts the physical quantum states from the whole set of quantum states present in the initial Hilbert space. Those physical states are the ones which respect the symmetries of the theory.
 
 However, because of the highly non trivial character of the scalar constraint, no one has ever succeeded to build a quantum theory from this second order phase space. There is obviously many ways to build a quantum operators corresponding to the scalar constraint and this leads to important ordering ambiguity when defining the hamiltonian quantum operator.
 There are some interesting solutions in symmetry reduced models, where the scalar constraint simplifies drastically, but a full quantum theory from the ADM phase space has so far eluded us.
 
 It is therefore natural to look for another formulation of General Relativity in order to have a proper classical theory from which we can launch the quantization program.
 One way to obtain such formulation is to use the so called first order formalism, which provide the mathematical tools to reformulate General Relativity in terms of connection (and vielbein) just as in Yang Mills theories.

\section{From second order to first order formalism}

We present now this first order formalism on which the rest of the manuscript is based. See \cite{ch1-Zan1} for a pedagogical introduction. The general idea is to trade the structure of a differentiable manifold $\mathcal{V}$ in order to describe the space-time, to a new structure which is a given $G$-principal bundle used in Yang Mills gauge theories, where $G$ is the Lorentz group $SO(3,1)$. The fundamental metric tensor field $g_{\mu\nu}$ is traded for the couple of fields $(e^{I}, \omega^{IJ})$ which are respectively the tetrad one form and the spin connection one form.
The field equations which were second order in term of the metric tensor become first order equations. The shift form the second order to the first order formalism makes closer the treatment of General Relativity to the one of Yang Mills gauge theories. 

Let us introduce this structure in a more physical way.
As noted by Einstein, in a free falling laboratory which is small enough and for experiments which last short enough, the experimental results will be indistinguishable from the one performed in absence of gravity.
This is the essence of the equivalence principle. Therefore, in this small neighborhood represented by the laboratory, the laws of physics are the one valid in Minkowski space, i.e. Lorentz invariance.
We conclude that locally, space-time is Lorentz invariant.
Space-time can then be understood as a smooth manifold $\mathcal{V}$. At each point $p \in\mathcal{V}$, there is a tangent space denoted $T_{p}$ equipped with the Minkowski metric $\eta_{IJ}$.
This tangent space represents an approximation of $\mathcal{V}$ in a small enough neighborhood of $p$.
We can therefore transform vectors, forms and tensors in $T_{p}$ into their counterparts in $\mathcal{V}$. To do so, we introduce the vielbein (or soldering form) $e^{I}$. This is a linear map which transforms locally an orthonormal coordinates system in Minkowski, denoted $z^{I}$, into a system of local coordinate $x^{\mu}$ in an open neighborhood of $p \in\mathcal{V}$:
\begin{align*}
e^{I}_{\mu} = \frac{\partial z^{I}}{\partial x^{\mu}} \;\;\;\;\;\;\;\; \;\;\;\;\;\;\;\; \;\;\;\;\;\;\;\; dz^{I} = e^{I}_{\mu} dx^{\mu}
\end{align*}
The new object transforms as a covariant vector (one form) under space-time diffeomorphism on $\mathcal{V}$, and as a contravariant vector under local Lorentz transformations of $T_{p}$.
The one-one correspondence between vectors from $T_{p}$ to $\mathcal{V}$ can be extended to tensors. In particular, one can write the infinitesimal space-time length $ds^{2}$ as:
\begin{align*}
ds^{2} = \eta_{IJ} dz^{I}dz^{J} = \eta_{IJ} e^{I}_{\mu} e^{J}_{\nu} dx^{\mu}dx^{\nu} = g_{\mu\nu} dx^{\mu}dx^{\nu} \;\;\;\;\;\; \text{whence} \;\;\;\;\;\; g_{\mu\nu} = e^{I}_{\mu}e^{J}_{\nu} \eta_{IJ}
\end{align*}
Therefore, the metric tensor is no more a fundamental object and can now be seen as a composite object. The fundamental ingredient is the vielbien $e^{I}_{\mu}$.
Since a local Lorentz transformation does not affect the Minkowski metric, it also does not affect the metric $g_{\mu\nu}$ describing the geometry in the neighborhood of $p$ in $\mathcal{V}$.
\begin{align*}
ds^{2} =\eta_{IJ} dz^{I}dz^{J} = \eta_{IJ} \Lambda^{I}{}_{M}(x) \Lambda^{J}{}_{N}(x) e^{M}_{\mu} e^{N}_{\nu} dx^{\mu}dx^{\nu} =  \eta_{MN}  e^{M}_{\mu} e^{N}_{\nu} dx^{\mu}dx^{\nu} = g_{\mu\nu} dx^{\mu}dx^{\nu}
\end{align*}
Since $g_{\mu\nu}$ is not affected by a local Lorentz transformation contrary to the vielbein $e^{I}_{\mu}$, we conclude that the vielbein has more independent components than the metric.
Indeed, while the metric, which is a symmetric tensor of rank $2$, has $10$ components (in four dimensions), the vielbein has $16$ components. The 6 additional components refer to the 6 possible Lorentz transformations, i.e. the $3$ rotations and the $3$ boosts of the Lorentz group. They underline the infinite possible reference frames in $T_{p}$ that one can choose. Therefore, for one given metric tensor, there is infinitely many vielbeins which reproduce this metric tensor.

Since we have now an independent action of the Lorentz group at each space-time point $p \in \mathcal{V}$ on tensors in $T_{p}$, it would be natural to ask for a covariant derivative under those local Lorentz transformation, just as in any gauge theories. As usual, the construction of this covariant derivative implies the introduction of a new gauge field, i.e. the Lorentz connection $\omega^{IJ}$. It is usually called the ``spin connection'' since it arises naturally when dealing with spinors. Its action on a vector $v^{I} = v_{\mu}^{I} dx^{\mu} \in T_{p}$ reads:
\begin{align*}
Dv^{I} = d v^{I} + \omega^{IJ} \wedge v_{J} = ( \partial_{[\mu} v_{\nu]}^{I} + \omega^{IJ}_{[\mu} v_{\nu ] \; J} ) \tau_{I} \; dx^{\mu} \otimes dx^{\nu}
\end{align*}
In order to have a Lorentz covariant derivative, the connection has to transformed as follow under local Lorentz transformations:
\begin{align*}
\omega^{I}{}_{J \; \mu}(x) \rightarrow \omega'^{I}{}_{J\; \mu}(x) = \Lambda^{I}{}_{J}(x) \Lambda_{b}{}^{d}(x) \omega^{c}{}_{d\; \mu}(x) + \Lambda^{I}{}_{J}(x) \partial_{\mu} \Lambda_{b}{}^{c}(x)
\end{align*}
This is the usual transformation rule for a connection. The spin connection defines therefore the parallel transport between Lorentz tensors in the tangent spaces, i.e. between neighboring points $x$ and $x+ dx$.
For the moment this connection is totally independent form the vielbien field.
When the metric compatibility condition is required, the two fields are now more independent :
\begin{align*}
\mathcal{D} e^{I} = 0 \;\;\;\;\;\; \text{whence the symmetric part gives} \;\;\;\;\;\; \partial_{\mu} e^{I}_{\nu} + \omega^{I}{}_{\mu\; J} e^{J}_{\nu} - \Gamma^{\sigma}_{\mu\nu} e^{I}_{\sigma} = 0
\end{align*}
where $\mathcal{D}$ is the covariant derivative with respect to the Lorentz indices and the space-time indices, involving both the spin connection and the space-time connection.
Therefore, the two connection are not independent.

Now that we have the two fundamental building blocks of the first order formalism, let us introduce some related objects.
By taking the covariant derivative of the vielbein and the spin connection, one obtain the so called torsion and curvature two forms:
\begin{align*}
T^{I} = De^{I} = de^{I} + \omega^{IJ} \wedge e_{J} \;\;\;\;\;\;\; \text{and} \;\;\;\;\;\;\; F^{IJ} = D \omega^{IJ} = d \omega^{IJ} + \omega^{IM} \wedge \omega_{M}{}^{J}
\end{align*}

Those two objects satisfy the two Cartan equations which are purely geometric identities:
\begin{align*}
DT^{I} = F^{IJ} \wedge e_{J} \;\;\;\;\;\;\;\;\; \text{and} \;\;\;\;\;\;\; DF^{IJ} = 0
\end{align*}
The second equation is called the Bianchi identity. It does not restrict the class of connections but implies that taking successive derivative of the curvature $F^{IJ}$ does not generate new independent tensors. 
When constructing a general action for gravity (geometry), one has to respect the Lorentz invariance required to implement the equivalence principle. Therefore, the lagragian four form has to be a Lorentz scalar w.r.t its internal indices.
In order to construct such scalar from the vielbein and the spin connection, one can only use the Minkowski metric  $\eta_{IJ}$ and the totally antisymmetric tensor $\epsilon_{IJKL}$ in order to contract and raise or lower the internal indices. 
Therefore the building blocks that one can use to build a general Lorentz invariant four form are:
\begin{align*}
e^I \qquad \omega^{IJ} \qquad T^{I} \qquad F^{IJ} \qquad  \eta_{IJ} \qquad \epsilon_{IJKL}
\end{align*}
In four dimension, one can build the following action:
\begin{align*}
S_{4D} = \int_{\mathcal{V}} \epsilon_{IJKL} (\; e^{I} \wedge e^{J} \wedge F^{KL} + \Lambda \; e^{I} \wedge e^{J} \wedge e^{K} \wedge e^{L} \; )
\end{align*}
This is precisely the first order version of the Einstein-Hilbert action supplemented with a non vanishing cosmological constant term. It is usually called the Einstein-Palatini action.
Up to topological term, this is the only action for gravity in four dimensions. To this action, one can add four different terms which do not change the field equations, i.e. the Euler term, the Pontryagin term, the Nieh Yan term and finally the Holst term \cite{ch1-Perez1}. There are respectively given by:
\begin{align*}
\epsilon_{IJKL}F^{IJ} \wedge F^{KL}  \qquad F^{IJ}\wedge F_{IJ} \qquad T^{I}\wedge T_{I} -  e^{I} \wedge e^{J} \wedge F_{IJ} \qquad e^{I}\wedge e^{J} \wedge F_{IJ}
\end{align*}

The field equations of General Relativity in the first order formalism can be derived from the Einstein-Palatini action \cite{ch1-Peldan1}. Let us assume for simplicity that the cosmological constant is vanishing, i.e. $\Lambda = 0$.
The variations of this action w.r.t. the vielbein and the spin connection give:
\begin{align*}
\delta S_{4D} & = \int_{\mathcal{V}}  \epsilon_{IJKL}\; ( \; 2 \; \delta e^{I} \wedge e^{J} \wedge F^{KL} - D (e^{I} \wedge e^{J}) \wedge \delta \omega^{KL} \; ) \\
& = - 2  \int_{\mathcal{V}}  \epsilon_{IJKL}\; ( \; \;  e^{J} \wedge F^{KL} \wedge \delta e^{I} + e^{I} \wedge T^{J} \wedge \delta \omega^{KL} \; )
\end{align*}
The field equations are therefore given by:
\begin{align*}
\epsilon_{IJKL}\;   e^{J} \wedge F^{KL} = 0 \qquad \epsilon_{IJKL}\;   e^{K} \wedge T^{L} = 0
\end{align*}
The last equation implies that the torsion field is vanishing, selecting therefore the unique torsionless connection, i.e. the Levi Civita connection $\omega^{LC}$.
Once this equation has been solved, the first equation reduces to the usual vacuum Einstein equation $R_{\mu\nu} = 0$.
Therefore, this formulation of gravity in terms of  vielbein and spin connection leads to General Relativity. However, contrary to what happen in the second order formalism, General Relativity is obtained only on shell, i.e. when the torsionless equation has been solved and assuming an invertible vielbein. 

Performing the canonical analysis of the Einstein-Palatini action is rather involved since it involves second class constraints. those second class constraints relate some components of the spin connection to some components of the vielbein. Once solved, one ends up with the $ADM$ phase space written in term of the vielbein. Therefore, nothing new appears from this formulation except that now, a new first class constraint shows up, which encodes the Lorentz gauge invariance of the theory, i.e. the Gauss constraint. 

In order to simplify the first class constraints of General Relativity, in view of the quantum theory, one has to go further.
We present now the classical Ashtekar's formulation of General Relativity, which provides such simplification. This is the starting point for the quantization program of Loop Quantum Gravity.

\section{The Ashtekar's hamiltonian formulation of General Relativity}

Following the work by Sen \cite{ch1-Sen1}, Ashtekar realized in 1986 that one can formulate General Relativity in terms of the self dual part of the spin connection \cite{ch1-Ash1, ch1-Ash2}, i.e. a complex $SL(2,\mathbb{C})$ connection, and its self dual conjugated momenta.
The point of performing this reformulation of General Relativity in terms of complex variables was to remove the second class constraints. When proceeding to the canonical analysis of the self dual action, one obtains immediately the first class constraints of self dual gravity, which turns out to be much simpler than the one of the ADM approach.
The Gauss constraint, which enforces now the $SL(2,\mathbb{C})$ internal gauge symmetry, and the vectorial constraint $H_{a}$, remain the same but the scalar constraint becomes now polynomial in the canonical variables.
Within those complex variables, General Relativity was formulated in a very closed form of an $SL(2,\mathbb{C})$ Yang Mills theory. This reformulation opened the hope of applying the Dirac quantization program to General Relativity. We briefly present this self dual phase space at the end of this chapter.

The new canonical variables were obtained by a canonical transformation on the ADM phase space. Those variables being complex valued, one had to impose reality conditions in order to recover real General Relativity (i.e. a real metric). Unfortunately, the complexity which disappear from the scalar constraint reappeared in those reality conditions which are highly non trivial. Up to now, no one knows how to deal with them at the quantum level.

In order to circumvent those reality conditions and begin the quantization program, one is led to look for another class of canonical transformations which does not generate reality conditions.
Such real canonical transformations were proposed by Barbero \cite{ch1-Bar1} and studied by Immirzi  \cite{ch1-Imm1}. Those canonical transformations, labeled by a real parameter $\gamma$ (the so called Barbero Immirzi parameter) lead also to a similar phase space but without the need of imposing any reality conditions.
The drawback of this second approach was that the simplification of the hamiltonian constraint obtained by the initial Ashtekar canonical transformation was lost. However, one could start the quantization program of Dirac and obtain the kinematical Hilbert space using the techniques of LQG discussed in the next chapter. This Ashtekar-Barbero family of phase spaces, labelled by $\gamma$ are therefore crucial to begin the quantization of General Relativity \textit{a la loop}.
The Ashtekar-Barbero canonical transformation consists in trading the canonical variables from the ADM phase space $(q_{ab}, P^{ab})$ into first order canonical variables:
\begin{align*}
(\; K^{i}_{a} , \; E^{a}_{i}\; ) \;\;\;\;\;\; \rightarrow \;\;\;\;\;\;\;  ( \; \gamma K^{i}_{a} , \; E^{a}_{i}\; ) \;\;\;\;\;\; \rightarrow \;\;\;\;\;\;\;  ( \; A^{i}_{a} , \; E^{a}_{i}\; ) \\
\end{align*}
where the indices $a$ and $i$ run over $\{1,2,3\}$. The internal indice $i$ is an $su(2)$-Lie algebra indice. $A^{i}_{a}$ is the $SU(2)$ Ashtekar Barbero connection and $E^{a}_{i}$ is its canonically conjugated momenta, called the electric field. 

Before presenting the Holst action and its hamiltonian analysis, let us explicit this canonical transformation, its infinitesimal generator and mention some of its properties.
Starting from the extrinsic curvature $K^{i}_{b} = e^{i a}K_{ab}$, the first canonical transformation reads:
\begin{align*}
\gamma K = \sum^{\infty}_{n=0} \frac{ (\log{\gamma})^{n}}{n!} \{ K, C\}_{n}
\end{align*}
where $C= \int K^{i}_{a} E^{a}_{i}$ and $\{K, C\}_{0}= K$ and $\{ K, C\}_{n+1} = \{\{ K, C \}_{n} , C \}$. Then a translation of $\Gamma$, generated by the phase space function $F = \frac{1}{\gamma}\int_{\Sigma} \Gamma^{i}_{a} E^{a}_{i}$, leads to the Ashtekar-Barbero connection:
\begin{align*}
A = \Gamma + \gamma K
\end{align*}

As explained earlier, the initial connection proposed by Ashtekar is complex. Starting from the real Ashtekar-Barbero connection with $\gamma = 1$, one can obtain the initial complex connection by using the same generating functional $C= \int K^{i}_{a} E^{a}_{i}$ in order to write:
\begin{align*}
^{\mathbb{C}}A = \sum^{\infty}_{n=0} \frac{ - (i\pi / 2 )^{n}}{n!} \{ A, C\}_{n} = \Gamma - i K
\end{align*}
Since this infinitesimal generator $C$ was originally introduced by Thiemann \cite{ch1-RCTH} in order to build a Wick rotation from the quantum theory based on the real variables to the one based on the complex variables, it has been called the complexifier. It appears also in some coherent states technics under the name of the ``coherent state transform''.

The canonical transformation induced by the infinitesimal generator $C$ is given by:
\begin{align*}
U_{\gamma} : (A, E) \rightarrow (A_{\gamma}, E_{\gamma}) \qquad \qquad \text{where} \qquad U_{\gamma} (A) = \gamma A + (1- \gamma) \Gamma \qquad U(E) = \frac{1}{\gamma} E
\end{align*}
This is this map which was studied by Barbero. For $\gamma = \pm i$, this transformation reduces to the one originally proposed by Ashtekar when he introduced the self dual variables.
This map is a canonical transformation and therefore, it is expected to be implemented unitarily in the quantum theory. However, this expectation turns out to be false, as shown in  \cite{ch1-ThieRov}. This implies that for different values of $\gamma$, the quantization of the real Ashtekar-Barbero phase space leads to inequivalent quantum theories, with different predictions. 

Having mention this important point, let us pursue and present the Holst action and the resulting Ashtekar-Barbero phase space which represents the starting point of the loop quantization.

A bit after the introduction of the complex Ashtekar's variables, Samuel \cite{ch1-Sam1}, Smolin and Jacobson \cite{ch1-Lee1, ch1-Lee2} discovered independently that one can recover this complex formulation of gravity by using the self dual part of the first order Einstein-Palatini action. Finally, Holst realized that one can derived the real Ashtekar-Barbero phase space from the so called Holst action \cite{ch1-Holst1}. 
 This action is equivalent to General Relativity at the classical level and differs only by a topological term called the Holst term.
The Holst action reads:
\begin{align*}
S_{H}(e,\omega) & = \frac{1}{4} \int_{\mathcal{V}} \frac{1}{2}\epsilon_{IJKL} \; e^{I} \wedge e^{J} \wedge F^{KL} (\omega)+ \frac{1}{\gamma} e^{I}\wedge e^{J} \wedge F_{IJ} (\omega)\\
& =  \frac{1}{4} \int_{\mathcal{V}} ( \; \frac{1}{2}\epsilon_{IJKL} +  \frac{1}{\gamma} \eta_{IK} \eta_{JL}\; ) \; e^{I} \wedge e^{J} \wedge F^{KL} (\omega)\\
\end{align*}

where $\gamma$ is the Barbero-Immirzi parameter, and $F(\omega) = D\omega = d\omega + \omega \wedge \omega$ is the curvature of the spin-connection. The internal indices $I \in \{0,1,2,3\}$ are raised and lower with the Minkowski metric $\eta_{IJ}$.
The second term of the first line is the Holst term which reduces to:
\begin{align*}
e^{I}\wedge e^{J} \wedge F_{IJ} (\omega) & = e^{I}_{\mu} e^{J}_{\nu} F_{IJ \; \rho\sigma}  \text{vol} \\
& = e^{I}_{\mu} e^{J}_{\nu} F_{IJ \; \rho\sigma} \; e \; \epsilon^{\mu\nu\rho\sigma}  \;  dx^{0}\wedge dx^{1} \wedge dx^{2} \wedge dx^{3} \\
& = \epsilon^{\mu\nu\rho\sigma} R_{\mu\nu\rho\sigma}\;  \sqrt{g} \; dx^{0}\wedge dx^{1} \wedge dx^{2} \wedge dx^{3} \\
\end{align*}
where $e = \sqrt{g}$ are respectively the determinant of the tetrad and the square root of the determinant of the metric tensor. $\text{vol} = \sqrt{g} \; dx^{\mu}\wedge dx^{\nu} \wedge dx^{\rho} \wedge dx^{\sigma}$ is the four volume form on which we integrate. 

The equation of motion for the fields $e$ and $\omega$ are given by:
\begin{align*}
\delta S_{H}(e, \delta \omega) &  = \int_{\mathcal{V}} \; ( \; \frac{1}{2}\epsilon_{IJKL} +  \frac{1}{\gamma} \delta_{[I|J|K]L}\; ) \; e^{I} \wedge e^{J} \wedge D\; (\delta \omega^{KL})  \\ 
& = -  \int_{\mathcal{V}} \; ( \; \frac{1}{2}\epsilon_{IJKL} +  \frac{1}{\gamma} \delta_{[I|J|K]L}\; ) \; D (e^{I} \wedge e^{J}) \wedge \delta \omega^{KL} \\
& =  - 2 \int_{\mathcal{V}} \; ( \; \frac{1}{2}\epsilon_{IJKL} +  \frac{1}{\gamma} \delta_{[I|J|K]L}\; ) \; T^{I} \wedge e^{J} \wedge \delta \omega^{KL} 
\end{align*}
Therefore the equation of motion derived from the variation according to $\omega^{IJ}$ is:
\begin{align*}
( \; \frac{1}{2}\epsilon_{IJKL} +  \frac{1}{\gamma} \delta_{[I|J|K]L}\; ) \; T^{I} \wedge e^{J} =0 \;\;\;\;\; \text{whence} \;\;\;\;\; T^{I} \wedge e^{J} = T^{I}_{\mu\nu} e^{J}_{\rho} \; dx^{\mu} \wedge dx^{\nu} \wedge dx^{\rho}=0
\end{align*}
since the term in parenthesis never vanishes. After contracting with $e^{\sigma}_{J}$, this reduces to:
\begin{align*}
T^{I}_{\mu\nu} = 0 \;\;\;\;\;\;\;  \text{whence} \;\;\;\;\;\;\;  \omega = \omega^{LC}
\end{align*}
Therefore, the first equation implies that torsion vanishes and the spin connection is the unique metric compatible torsionless  connection, i.e. the Levi Civita connection: $\omega^{LC}$. Plugging this solution into the Holst action, the second term vanishes (on shell) by virtue of the first Bianchi identity that one can express as: $\epsilon^{\mu\nu\rho\sigma} R_{\mu\nu\rho\sigma} ( \omega^{LC}) =0$.
Finally, varying the equation of motion according to the tetrad $e^{I}$, one has:
\begin{align}
 \; \epsilon_{IJKL} \; e^{J} \wedge F^{KL} =0 
\end{align}
Therefore, we obtain the usual first order Einstein field equations from the Holst action.
Notice that the Immirzi parameter simply drops out from the classical theory and therefore, won't have any impact on the classical predictions of the theory. \\

\textit{Hamiltonian analysis in the time gauge} \\

Let us proceed to the hamiltonian analysis of the Holst action. (A detailed computation is presented in \cite{ch1-Noui1} in the euclidean case).
The first step is to proceed to a $3+1$ decomposition of the space-time indices and of the internal indices. The $3+1$ decomposition of the space-time indices corresponds to a foliation of the four dimensional space-time $\mathcal{V}$ into a family of spacelike hypersurfaces $\Sigma$ which evolve in time. The split of the internal indices is crucial in order to impose the gauge fixing which will simplify the canonical analysis. The real $SU(2)$ Ashtekar Barbero phase space is obtained by imposing that the pull back of the tetrad components $e^{0}$ to the spacelike hypersurface $\Sigma$ is zero, i.e. $\underset{\leftarrow}{e^{0}}= 0$ or $e^{0}_{a} = 0$. This gauge, called the time gauge, reduces the non vanishing pieces of the $4 \times 4$ matrix of the tetrad $e^{I}_{\mu}$, which transforms under $SO(3,1)$, to the $3\times3$ matrix $e^{i}_{a}$ which transforms under $SU(2)$, plus the lapse $e^{0}_{0}$ and the shift $e^{i}_{0}$. The indices $i$ and $a$ run over $\{1,2,3\}$. This gauge choice selects therefore the compact subgroup $SU(2)$ from the initial non compact Lorentz group $SO(3,1)$.
Under this gauge fixing, the different components of the tetrad one form field read:
\begin{align*}
& e^{0}_{\mu} dx^{\mu} = e^{0}_{0} dx^{0} + e^{0}_{a} dx^{a} = N dx^{0} \;\;\;\;\;\;\;\;\;\;\; \text{and} \;\;\;\;\;\;\;\;\;  e^{i}_{\mu}dx^{\mu} = e^{i}_{0} dx^{0} + e^{i}_{a} dx^{a} = N^{i}dx^{0} + e^{i}_{a} dx^{a}
\end{align*}
where we have use the notation: $N = e^{0}_{0}$ and $N^{i} = N^{a} e^{i}_{a}$. The scalar $N$ is the lapse function and the vector $N^{a}$ is the shift vector.
Let us perform the $3+1$ splitting of the action:
\begin{align*}
S_{H} & = \frac{1}{4} \int_{\mathcal{V}}( \;  \frac{1}{2} \epsilon^{\mu\nu\rho\sigma} \epsilon_{IJKL} \; e^{I}_{\mu} e^{J}_{\nu} F^{KL}_{\rho\sigma}+ \frac{1}{\gamma} \epsilon^{\mu\nu\rho\sigma} e^{I}_{\mu}e^{J}_{\nu} F_{IJ \; \rho\sigma} \; ) \;dx^{0}\wedge dx^{1} \wedge dx^{2} \wedge dx^{3} \\
& = \frac{1}{4} \int_{\mathcal{V}} \epsilon^{\mu\nu\rho\sigma} \; ( \; \epsilon_{ijk} e^{0}_{\mu} e^{i}_{\nu} F^{jk}_{\rho\sigma} + \epsilon_{ijk} e^{i}_{\mu}e^{j}_{\nu} F^{0k}_{\rho\sigma} + \frac{2}{\gamma} e^{0}_{\mu}e^{i}_{\nu} F_{0i \; \rho\sigma} + \frac{1}{\gamma} e^{i}_{\mu}e^{j}_{\nu} F_{ij \; \rho\sigma} \; )  \;dx^{4} \\
& = \frac{1}{4} \int_{} \epsilon^{abc}  \{ \;2  e^{i}_{a} e^{j}_{b} ( \epsilon_{ijk}  F^{0k}_{0c} + \frac{1}{\gamma} F_{ij \; 0c} )+  2 N^{i} e^{j}_{a} ( \epsilon_{ijk} \; F^{0k}_{bc} + \frac{1}{\gamma}  F_{ij \; bc} \; ) +  N e^{i}_{a}( \epsilon_{ijk} F^{jk}_{bc} + \frac{2}{\gamma} F_{0i \; bc} \; ) \}  dx^{4} \\
& =  \frac{1}{2} \int_{\mathcal{V}} \epsilon^{abc}  \{ \;   \epsilon_{ijk}  e^{i}_{a} e^{j}_{b} ( F^{0k}_{0c} + \frac{1}{\gamma} F^{k}_{0c} )+  N^{i} e^{j}_{a} \epsilon_{ijk} \; ( \; F^{0k}_{bc} + \frac{1}{\gamma}  F^{k}_{bc} \; ) +  N e^{i}_{a}(F_{i \; bc} - \frac{1}{\gamma} F^{0}{}_{i \; bc} \; ) \}  dx^{4} \\
& = \int_{\mathcal{V}}  \{ \; L_{C}+  L_{V} + L_{S} \; \}  dx^{4} \\
\end{align*}
where we have used the convention $\epsilon_{0ijk} =  1$ and $\epsilon^{0abc} = 1$.
In the last line, we have used the condensed notation for the curvature: $F^{k}_{\mu\nu} = \frac{1}{2} \epsilon^{k}{}_{ij}F^{ij}_{\mu\nu}$.
We see already that the time derivatives only appear in the first term, where we have $F_{0c}$.

Now, we introduce the condensed form of the spin connection:
\begin{align*}
& \omega^{ij}_{\mu} = \epsilon^{ij}{}_{k} \omega^{k}_{\mu} \\
& \omega^{k}_{\mu} = \delta^{k}_{k'} \omega^{k'}_{\mu} = \frac{1}{2} \epsilon^{k}{}_{ij} \epsilon_{k'}{}^{ij} \omega^{k}_{\mu} = \frac{1}{2} \epsilon^{k}{}_{ij} \epsilon_{k'}{}^{ij} \omega^{k'}_{\mu} = \frac{1}{2} \epsilon^{k}{}_{ij} \omega^{ij}_{\mu} 
\end{align*}

The curvature of the spin connection $F^{IJ}$ is given by:
\begin{align*}
& F^{IJ}_{\mu\nu} = \partial_{\mu} \omega^{IJ}_{\nu} - \partial_{\nu} \omega^{IJ}_{\mu} + \omega^{IM}_{\mu} \omega_{\nu \; M}{}^{J} - \omega^{IM}_{\nu} \omega_{\mu \; M}{}^{J}
\end{align*}
and can be decomposed as:
\begin{align*}
 F^{0i}_{\mu\nu} & = \partial_{\mu} \omega^{0i}_{\nu} - \partial_{\nu} \omega^{0i}_{\mu} + \omega^{0j}_{\mu} \omega_{\nu \; j}{}^{i} -  \omega^{0j}_{\nu} \omega_{\mu \; j}{}^{i} \\
& =  \partial_{\mu} \omega^{0i}_{\nu} - \partial_{\nu} \omega^{0i}_{\mu} + \omega^{0j}_{\mu} \; \epsilon_{kj}{}^{i} \omega_{\nu \; j}{}^{k} -  \omega^{0j}_{\nu}\;  \epsilon_{kj}{}^{i}\omega_{\mu}^{k} \\
& = \partial_{\mu} \omega^{0i}_{\nu} - \partial_{\nu} \omega^{0i}_{\mu} + \epsilon^{i}{}_{kj} \; \omega^{0j}_{\mu} \; \omega_{\nu \; j}{}^{k} -  \epsilon^{i}{}_{kj}\omega^{0j}_{\nu}\;\omega_{\mu}^{k} \\
& = \partial_{\mu} \omega^{0i}_{\nu} - \partial_{\nu} \omega^{0i}_{\mu} + ( \omega_{\nu} \times \omega^{(0)}_{\mu})^{i} -  (\omega_{\mu} \times \omega^{(0)}_{\nu})^{i} \\
\end{align*}
\begin{align*}
 F^{ij}_{\mu\nu} & = \partial_{\mu} \omega^{ij}_{\nu} - \partial_{\nu} \omega^{ij}_{\mu} + \omega^{im}_{\mu} \omega_{\nu \; m}{}^{j} -  \omega^{im}_{\nu} \omega_{\mu \; m}{}^{j} + \omega^{i0}_{\mu} \omega_{\nu \; 0}{}^{j} -  \omega^{i0}_{\nu} \omega_{\mu \; 0}{}^{j} \\
& =  \epsilon^{ij}{}_{k}\; ( \partial_{\mu} \omega^{k}_{\nu} -  \partial_{\nu} \omega^{k}_{\mu}) +  \epsilon^{im}{}_{k}  \epsilon_{k'm}{}^{j} ( \omega^{k}_{\mu} \omega_{\nu}^{k'} -  \omega^{k}_{\nu} \omega_{\mu}^{k'}) + \eta_{00}\; \omega^{i0}_{\mu} \omega_{\nu}^{0j} -  \eta_{00} \; \omega^{i0}_{\nu} \omega_{\mu}^{0j}  \\
& =  \epsilon^{ij}{}_{k}\; ( \partial_{\mu} \omega^{k}_{\nu} -  \partial_{\nu} \omega^{k}_{\mu}) +  \eta_{pk} \eta^{qj}\epsilon^{mip}  \epsilon_{mk'q} ( \omega^{k}_{\mu} \omega_{\nu}^{k'} -  \omega^{k}_{\nu} \omega_{\mu}^{k'}) + \omega^{0i}_{\mu} \omega_{\nu}^{0j} -  \omega^{0i}_{\nu} \omega_{\mu}^{0j}  \\
& =  \epsilon^{ij}{}_{k}\; ( \partial_{\mu} \omega^{k}_{\nu} -  \partial_{\nu} \omega^{k}_{\mu}) + \omega^{j}_{\mu} \omega_{\nu}^{i} -  \omega^{j}_{\nu} \omega_{\mu}^{i} + \omega^{0i}_{\mu} \omega_{\nu}^{0j} -  \omega^{0i}_{\nu} \omega_{\mu}^{0j}  \\
& =  \epsilon^{ij}{}_{k}\; ( \partial_{\mu} \omega^{k}_{\nu} -  \partial_{\nu} \omega^{k}_{\mu}) - \epsilon^{ij}{}_{k}(\omega_{\mu} \times \omega_{\nu})^{k} + \epsilon^{ij}{}_{k}(\omega^{(0)}_{\mu}\times \omega_{\nu}^{(0)})^{k}   \\
\end{align*}

Finally, we have that:
\begin{align*}
F^{k}_{\mu\nu} = \frac{1}{2} \epsilon^{k}{}_{ij}F^{ij}_{\mu\nu}& = \partial_{\mu} \omega^{k}_{\nu} -  \partial_{\nu} \omega^{k}_{\mu} - (\omega_{\mu} \times \omega_{\nu})^{k}  +(\omega^{(0)}_{\mu}\times \omega_{\nu}^{(0)})^{k}\\ 
\end{align*}
We have therefore the different components of the curvature which enter in the $3+1$ decomposition of the action.
Using those results, the first term in the decomposition of the Holst action reads:
\begin{align*}
L_{C} & = \frac{1}{2} \epsilon^{abc} \;  \epsilon_{ijk} \; e^{i}_{a} e^{j}_{b}  \{ \; \partial_{0}( \omega^{(0)}_{c} +\frac{1}{\gamma} \omega_{c}) \; - \partial_{c} \;  \omega^{(0)}_{0} + \omega_{c} \times \omega^{(0)}_{0} - \omega_{0} \times \omega^{(0)}_{c} \\
& \;\;\;\;\;\;\;\;\;\;\;\;\;\;\;\;\;\;\;\;\;\;\;\;\;\;\;- \frac{1}{\gamma} \partial_{c} \;  \omega_{0}  \; \frac{1}{\gamma} \omega_{0} \times \omega_{c} + \frac{1}{\gamma} \omega^{(0)}_{0} \times \omega^{(0)}_{c} \; \}^{k} \; \\
& = E^{c}. \; \partial_{0}\tilde{A}_{c} + \frac{1}{\gamma} \omega_{0}. [ \;  \partial_{c}E^{c}  - \gamma( \omega^{(0)}_{c} + \frac{1}{\gamma} \omega_{c} ) \times E^{c} \; ] + \omega^{(0)}_{0} . ( \partial_{c} E^{c} - \omega_{c} \times E^{c} ) \\
& \;\;\;\;  + \frac{1}{\gamma} \omega^{(0)}_{0} . ( \omega^{(0)}_{c} \times E^{c}) \\
& =  E^{c} . \; \partial_{0}\tilde{A}_{c} + \frac{1}{\gamma} \; \omega_{0}. \; G +  \omega^{(0)}_{0} . \phi - \frac{1}{\gamma^{2}} \omega^{(0)}_{0} . ( G- \phi) \\
& =  E^{c} . \; \partial_{0}\tilde{A}_{c} + \frac{1}{\gamma} \; \alpha . \; G + (1 + \gamma^{-2}) \omega^{(0)}_{0}.\; \phi \\
\end{align*}
Therefore, the dynamical variables of the theory are given by the $su(2)$-connection $ \tilde{A}^{i}_{a}$ and the so called electric field $E^{a}_{i}$. They are the canonical conjugated variables of the theory and read:
\begin{align*}
 \tilde{A}^{i}_{a} = \omega^{(0)i}_{a} + \frac{1}{\gamma}\omega^{i}_{a} \;\;\;\;\;  E^{a}_{i} = \frac{1}{2} \epsilon^{abc} \; \epsilon_{ijk}  \;  e^{j}_{b} e^{k}_{c} 
 \end{align*}
The components $\omega^{(0)}_{0}$ and $\omega_{0}$ of the spin connection, (or equivalently the combination $\alpha$), turn out to be Lagrange multipliers which enforce the primary constraints $G$ and $\phi$. 
The expression of those constraints and of the Lagrange multiplier $\alpha$ are given by:
 \begin{align*}
 & G =  \partial_{c} E^{c} -\gamma \tilde{A}_{c} \times E^{c} \\
&\phi = \partial_{c} E^{c} - \omega_{c} \times E^{c}  \\
& \alpha = \frac{1}{\gamma} ( \omega_{0} - \frac{1}{\gamma} \omega^{(0)}_{0}) \\ 
\end{align*}

The very last term of the third line is obtained by noticing that:
\begin{align*}
\omega^{(0)}_{c} \times E^{c} = -  \frac{1}{\gamma} \; ( G - \phi)
\end{align*} 
The second term can be computed as follow :
\begin{align*}
2 L_{V} & =\; \epsilon_{ijk}  N^{d} e^{i}_{d} e^{j}_{a} \epsilon^{abc} \{ \; \partial_{b} ( \omega^{(0)}_{c} + \frac{1}{\gamma} \omega_{c}) - \partial_{c} ( \omega^{(0)}_{b} + \frac{1}{\gamma} \omega_{b} ) + \omega_{c} \times ( \omega^{(0)}_{b} + \frac{1}{\gamma} \omega_{b} ) \\
&  \;\;\;\;\;\;\;\;\;\;\;\;\;\;\;\;\;\;\;\;\;\;\;\;\;\;\; \;\; - ( \omega_{b} - \frac{1}{\gamma} \omega^{(0)}_{b} ) \times \omega^{(0)}_{c} \; \}\\
 & =  \epsilon_{ijk}  N^{d} e^{i}_{d} e^{j}_{a} \epsilon^{abc}  . \{ \; \partial_{b} \tilde{A}_{c} - \partial_{c} \tilde{A}_{b} + \gamma \tilde{A}_{c} \times \tilde{A}_{b}  - \gamma ( \omega^{(0)}_{c} + \frac{1}{\gamma} \omega_{c}) \times \tilde{A}_{b} \\
 & \;\;\;\;\;\;\;\;\;\;\;\;\;\;\;\;\;\;\;\;\;\;\;\;\;\;\;\;\;\; + \omega_{c} \times \tilde{A}_{b}- ( \omega_{b} - \frac{1}{\gamma} \omega^{(0)}_{b} ) \times \omega^{(0)}_{c}  \; \} \\
& =  \epsilon_{ijk}  N^{d} e^{i}_{d} e^{j}_{a} \epsilon^{abc}  . \{ \; \partial_{b} \tilde{A}_{c} - \partial_{c} \tilde{A}_{b} + \gamma \tilde{A}_{c} \times \tilde{A}_{b}  - \omega^{(0)}_{c} \times ( \omega_{b} - \frac{1}{\gamma} \omega^{(0)}_{b} - \gamma  \tilde{A}_{b}) \; \} \\
& =   2 \; N^{b} E^{c}  . \{ \; \partial_{b} \tilde{A}_{c} - \partial_{c} \tilde{A}_{b} + \gamma \tilde{A}_{c} \times \tilde{A}_{b}  - (\frac{1}{\gamma} - \gamma)\omega^{(0)}_{c} \times \omega^{(0)}_{b}  \; \} \\
 & =  2 \; N^{b} \{ \;  E^{c}. \; ( \partial_{b} \tilde{A}_{c} - \partial_{c} \tilde{A}_{b} + \gamma \tilde{A}_{c} \times \tilde{A}_{b} ) +  (\frac{1}{\gamma} - \gamma)\omega^{(0)}_{b} \; . \; ( \omega^{(0)}_{c} \times E^{c})  \; \} \\
 & =   2 \;N^{b} \{ \;  E^{c}. \; ( \partial_{b} \tilde{A}_{c} - \partial_{c} \tilde{A}_{b} + \gamma \tilde{A}_{c} \times \tilde{A}_{b} ) + (1 - \frac{1}{\gamma^{2}})\omega^{(0)}_{b} \; . \; ( G -\phi \;) \} \\
 & =   2 \; N^{b} \; H_{b} \\ 
\end{align*}
Therefore, as expected, the vector shift $N^{b}$ turns out to be also a Lagrange multiplier enforcing the constraint $H_{b}$.
The last term in the action can be computed as follow:
 \begin{align*}
2 L_{S} & = N \epsilon^{abc} e_{a} . \{ \;  \partial_{b} \omega_{c} - \partial_{c} \omega_{b} - \omega_{b} \times \omega_{c} + \omega^{(0)}_{b} \times \omega^{(0)}_{c}  \\
 & \;\;\;\;\;\;\;\;\;\;\;\;\;\;\;\;\;  - \frac{1}{\gamma} ( \; \partial_{b} \omega^{(0)}_{c} - \partial_{c} \omega^{(0)}_{b} + \omega_{c} \times \omega^{(0)}_{b} - \omega_{b} \times \omega^{(0)}_{c} \; ) \} \\
 & = N \epsilon^{abc}e_{a} . \{ \; R_{bc}(\omega) + \omega^{(0)}_{b} \times \omega^{(0)}_{c} - \frac{1}{\gamma} [ \; \partial_{b} ( \omega^{(0)}_{c} + \frac{1}{\gamma} \omega_{c} ) - \partial_{c} ( \omega^{(0)}_{b} + \frac{1}{\gamma} \omega_{b} ) \\
 & \;\;\;\;\;\;\;\;\;\;\;\;\;\;\;\;\;\; - \gamma ( \; \omega^{(0)}_{b} + \frac{1}{\gamma} \omega_{b} ) \times ( \omega^{(0)}_{c} + \frac{1}{\gamma} \omega_{c} ) \; ] + \frac{1}{\gamma^{2}} (  \partial_{b} \omega_{c} - \partial_{c} \omega_{b} ) \\
 & \;\;\;\;\;\;\;\;\;\;\;\;\;\;\;\;\;\; - ( \; \omega^{(0)}_{b} + \frac{1}{\gamma} \omega_{b} ) \times ( \omega^{(0)}_{c} + \frac{1}{\gamma} \omega_{c} ) - \frac{1}{\gamma} \omega_{c} \times \omega^{(0)}_{b} + \frac{1}{\gamma} \omega_{b} \times \omega^{(0)}_{c} \} \\
 & = N \frac{E^{b} \times E^{c}}{\sqrt{\text{det}E}} . \{ \; - \frac{1}{\gamma} ( \partial_{b} \tilde{A}_{c} - \partial_{c} \tilde{A}_{b} - \gamma \tilde{A}_{b} \times \tilde{A}_{c} ) + ( 1 + \frac{1}{\gamma^{2}}) R_{bc}(\omega) \; \} \\
 \end{align*}
 In the precedent expressions, we have used different formulas relating the tetrad $e^{i}_{a}$ and the electric field $E^{a}_{i}$.
 The very first step to derive those formulas is to note that for an $n \times n$ matrix $B$, its determinant can be written as:
\begin{align*}
\text{det} (B) = \frac{1}{n!} \epsilon_{a.....b} \; \epsilon^{c......d} \; B^{a}_{c} ......... B^{b}_{d}
\end{align*}
Therefore, using the determinant of the $3\times 3$ tetrad matrix and the expression of the electric field resulting from the Holst action decomposition, we have:
\begin{align*}
\text{det} (e) = \frac{1}{3!} \epsilon_{ijk} \; \epsilon^{abc} \; e^{i}_{a} e^{j}_{b} e^{k}_{c} \;\;\;\; E^{c}_{k} =  \frac{1}{2}\epsilon_{ijk} \; \epsilon^{abc} \; e^{i}_{a} e^{j}_{b}  \;\;\;\;\;\; \text{whence} \;\;\;\;\;\;\; E^{c}_{k} = \frac{1}{2}\text{det} (e) \; e^{c}_{k} = \frac{1}{2} \sqrt{\text{det} (q)} \; e^{c}_{k}
\end{align*}
where $\sqrt{\text{det} (q)}$ is the determinant of the three dimensional induced metric on the space-like hyper surfaces of the foliation.
From those expressions, it is straitforward to show that:
\begin{align*}
e^{i}_{a} = \frac{ \epsilon_{abc} \epsilon^{ijk}  E^{b}_{j} E^{c}_{k}}{2 \sqrt{\text{det} (E)}} \;\;\;\;\;  \epsilon^{abc} e^{i}_{a} = \frac{\epsilon^{ijk}  E^{b}_{j} E^{c}_{k}}{\sqrt{\text{det} (E)}} \;\;\;\;\;  \epsilon_{ijk} e^{i}_{a}e^{j}_{b} = \text{det}(e) \epsilon_{abc} e^{c}_{k} 
\end{align*}
 We have now all the pieces the write the 3+1 decomposition of the Holst action:
 \begin{align*}
 S_{H} & =  \int_{\mathcal{V}} \; \{ \; E^{c} . \; \partial_{0}\tilde{A}_{c} + \frac{1}{\gamma} \; \alpha . \; G +  (1 + \gamma^{-2}) \omega^{(0)}_{0}.\; \phi +  N^{a} \; H_{a} + N \; H \; \}
  \end{align*}
  
  where the constraints are given by:
\begin{align*}
& \phi = \partial_{c} E^{c} - \omega_{c} \times E^{c}  \\
& G =  \partial_{c} E^{c} -\gamma \tilde{A}_{c} \times E^{c} \\
& H_{a} = E^{c}. \; ( \partial_{b} \tilde{A}_{c} - \partial_{c} \tilde{A}_{b} + \gamma \tilde{A}_{c} \times \tilde{A}_{b} ) + (1 - \frac{1}{\gamma^{2}})\omega^{(0)}_{b} \; . \; ( G -\phi \;) \\
& H =  \frac{E^{b} \times E^{c}}{2\sqrt{\text{det}E}} . \{ \; - \frac{1}{\gamma} ( \partial_{b} \tilde{A}_{c} - \partial_{c} \tilde{A}_{b} - \gamma \tilde{A}_{b} \times \tilde{A}_{c} ) + ( 1 + \frac{1}{\gamma^{2}}) R_{bc}(\omega) \; \} 
\end{align*}

All those constraints are primary constraints. In order to have a proper Gauss constraint $G$ as in Yang Mills theories, we have to trade the connection $\tilde{A}$ for the connection $A = - \gamma \tilde{A}$.
Under this change of variable, the precedent constraints become:
\begin{align*}
& \phi = \partial_{c} E^{c} - \omega_{c} \times E^{c}  \\
& G =  \partial_{c} E^{c} +A_{c} \times E^{c} \\
& H_{b} = \frac{1}{\gamma} \; E^{c}. \; ( - \partial_{b} A_{c} + \partial_{c} A_{b} + A_{c} \times A_{b} ) + (1 - \frac{1}{\gamma^{2}})\omega^{(0)}_{b} \; . \; ( G -\phi \;) \\
& H =  \frac{E^{b} \times E^{c}}{2\sqrt{\text{det}E}} . \{ \; - \frac{1}{\gamma^{2}} ( - \partial_{b} A_{c} + \partial_{c} A_{b} +  A_{c} \times A_{b} ) + ( 1 + \frac{1}{\gamma^{2}}) R_{bc}(\omega) \; \} 
\end{align*}

Finally, one can rewrite this set of constraints:
\begin{align*}
& \phi = \partial_{c} E^{c} - \omega_{c} \times E^{c}  \simeq 0 \\
& G =  \partial_{c} E^{c} +A_{c} \times E^{c} \simeq 0\\
& H_{b} \simeq \frac{1}{\gamma} \; E^{c}. \; \mathcal{F}_{cb}(A) \simeq 0 \\
& H =   \frac{1}{2\gamma^{2}} \frac{E^{b} \times E^{c}}{\sqrt{\text{det}E}} . \{ \; \mathcal{F}_{bc}(A) + ( 1 + \gamma^{2}) R_{bc}(\omega) \; \} \simeq 0
\end{align*}

where $\mathcal{F}_{ab}(A)$ is the curvature of the final connection $A$.
Therefore, the true canonical conjugated variables are given by:
\begin{align*}
 A^{i}_{a} = - ( \gamma \omega^{(0)i}_{a} + \omega^{i}_{a} ) \;\;\;\;\;  E^{a}_{i} = \frac{1}{2} \epsilon^{abc} \; \epsilon_{ijk}  \;  e^{j}_{b} e^{k}_{c} 
\end{align*}
The real $su(2)$ connection $A^{i}_{a}$ is called the Ashtekar-Barbero connection.
The theory can be formulated with the dynamical variables $A_{a}$ and $E^{a}$, and the Lagrange multipliers $N$, $N_{a}$, $\omega^{(0)i}_{0}$ and $\omega_{0}$.
However, from the point of view of the canonical analysis, $\omega_{a}$ has to be regarded as a dynamical variable too, since it appears as quadratically in the constraints. We need to introduce its conjugated momenta $\pi^{a}$, enforcing in the same time its vanishing through a new constraint associated to the Lagrange multiplier $\lambda_{a}$. The hamiltonian reads therefore:
\begin{align*}
H_{tot} = -  \int_{\Sigma} \; \{ \; \frac{1}{\gamma} \; \alpha . \; G +  (1 + \gamma^{-2}) \omega^{(0)}_{0}.\; \phi +  N^{a} \; H_{a} + N \; H + \lambda_{a} \pi^{a}\; \}
\end{align*}

For the moment, we have two couple of conjugated variables, each one having 18 components. There is therefore 36 components.
Those components are related through the Gauss constraint $G^{i}$ (3 equations), the second class constraint $\phi^{i}$ (3 equations), the vectorial constraint $H_{a}$ (3 equations), the hamiltonian constraint $H$ (1 equation) and finally by the constraint $\pi^{a}_{i} \simeq 0$ (9 equations). Contrary to the second class constraints $\phi$ and $\pi^{a} \simeq 0$, the first class constraints $G$, $H_{a}$ and $H$ generate symmetries. We need therefore to count them twice in order to take into account all the constraints. Thus, the total number of constraints is: $n_{C} = 2n_{F} + n_{S} = 2\times 3 + 2\times 3 + 2 \times 1 + 3 + 9 = 26$ constraints. At this step, only 10 components among the initial 36 remain independent.
One can show that that the set of constraints $G$, $\phi$, $H_{a}$ and $H$ do not generate secondary constraints.
The last step is to impose the stability of the constraint $\pi^{a} \simeq 0$ during the hamiltonian evolution. This will generate 6 secondary constraints.
Those constraints, which are second class, are obtained by computing the hamiltonian evolution of $\pi^{a}$:
\begin{align*}
\dot{\pi^{a}} = \{\pi^{a} , H_{tot}\} & = - \int_{\Sigma} \{ \pi^{a} , ( 1+ \gamma^{-2} ) \omega^{(0)}_{0} . \; \phi + N H \} \\
& = - \int_{\Sigma} \{ \pi^{a} , ( 1+ \gamma^{-2} ) \omega_{b} . ( \omega^{(0)}_{0} \times E^{b}) + \frac{1}{2}( 1+ \gamma^{-2} ) N \epsilon^{dbc} e_{d}. \; ( \partial_{b} \omega_{c} - \partial_{c} \omega_{b} - \omega_{b} \times \omega_{c} ) \} \\
& = - ( 1+ \gamma^{-2} ) \int_{\Sigma} \; ( \; \omega^{(0)}_{0} \times E^{a} + N  \epsilon^{dac}  ( \; \partial_{c} e_{d} - \omega_{c} \times e_{d} \; ) + \epsilon^{dac} e_{d}  \;  \partial_{c} N )
\end{align*}

Therefore, we obtain the following constraint:
\begin{align*}
 \omega^{(0)}_{0} \times E^{a} + N  \epsilon^{acd}  ( \; \partial_{c} e_{d} - \omega_{c} \times e_{d} \; ) + \epsilon^{acd} e_{d} \;  \partial_{c} N \simeq 0
\end{align*}
From the $9$ equations obtained, $3$ will fix the Lagrange multipliers, while $6$ will lead to secondary constraints.
To remove the three equations related to the fixation of the Lagrange multipliers, we first contract this constraint with $E^{b}$ and then, we antisymmetrize between $a$ and $b$. we cancel the first and second terms.
We obtain:
\begin{align*}
\Psi^{ab} = N \epsilon^{(acd} E^{b)}. \;  ( \; \partial_{c} e_{d} - \omega_{c} \times e_{d} \; ) \simeq 0
\end{align*}
One can show that the second class constraint $\Psi^{ab}$ does not generate tertiary constraints and the Dirac algorithm ends here.

As expected, we obtain 6 new secondary constraints. Therefore, we conclude that from the 10 independent dynamical components remaining in our phase space,  only $10-6= 4$ are truly independent variables.
The four dynamical components remaining are the 2 degrees of freedom of General Relativity supplemented with their two conjugated momentas.
The six precedent constraints together with the 3 constraints $\phi^{i}$ can be regrouped providing 9 second class constraints. Once solved, those constraints imply that the 9 components of the rotational part of the spin connection $\omega^{i}_{a}$ reduce to the components of the Levi Civita connection, i.e. the unique torsionless (metric compatible) connection. They imply that $\omega^{i}_{a}$ takes the following form in term of the electric field
\begin{align*}
\omega^{i}_{a} (E)= - \frac{1}{2} ( \partial_{b} E_{a} - \partial_{a} E_{b} + E^{c} ( E_{a} . \partial_{b}E_{c})) \times E^{b} - \frac{1}{4} ( 2 E_{a} \frac{\partial_{b} \text{det}E}{\text{det}E} - E_{b} \frac{\partial_{a} \text{det}E}{\text{det}E} ) \times E^{b}
\end{align*}

This concludes the canonical analysis of the four dimensional Lorentzian Holst action in the time gauge.

We have obtained a phase space which is coordinated by the commutative real $su(2)$ Ashtekar-Barbero connection $A$ and its conjugated momenta $E$.
Their Poisson brackets reads:
\begin{align*}
\{E^{a}_{i}, A^{j}_{b}\} = \gamma \delta^{j}_{i} \delta^{a}_{b} \;\;\;\;\;\;\;\;\;\; \text{with} \;\;\;\;\;\;\;\;\;\; A^{i}_{a} = \Gamma^{i}_{a} - \gamma \omega^{(0)i}_{a} \;\;\;\;\; \text{and} \;\;\;\;\; E^{a}_{i} = \frac{1}{2} \epsilon^{abc} \; \epsilon_{ijk}  \;  e^{j}_{b} e^{k}_{c} 
\end{align*}
where $\Gamma^{i}_{a} = - \omega^{i}_{a}$ is the Levi-Civita connection, $\omega^{(0)i}_{a}$ is the extrinsic curvature and $\gamma$ the Barbero-Immirizi parameter.
Those canonical variables are constrained by the set of first class constraints:
\begin{align*}
& G =  \partial_{c} E^{c} +A_{c} \times E^{c} \simeq 0 \\
& H_{b} \simeq  \; E^{c}. \; \mathcal{F}_{cb}(A) \simeq 0 \\
&  H =   \frac{1}{2} \frac{E^{b} \times E^{c}}{\sqrt{\text{det}E}} . \{ \; \mathcal{F}_{bc}(A) + ( 1 + \gamma^{2}) R_{bc}(\omega (E)) \; \} \simeq 0
\end{align*} 

Note that the Barbero-Immirizi parameter enters both in the Poisson bracket and in the hamiltonian constraint, i.e. at the dynamical level. The two first class constraints $G$ and $H_{a}$ are polynomial in the canonical variables which is an important advantage in order to quantize the theory (definition of quantum operator, ordering ambiguity). However, the hamiltonain constraint fails to have this nice property because of the very last term which depends on $\omega_{a}(E)$. This term takes a very complicated form when expressed in term of the electric field $E$. This is precisely this term which renders the implementation of the dynamics in canonical Loop Quantum Gravity difficult.
We note that this bad term disappears when $\gamma = \pm i$.

Finally, we compute the algebra of the constraints. To to so, we introduce the following smeared constraints:
\begin{align*}
G(u) = \int_{\Sigma} dx^{3} u^{i} G_{i}  \qquad H(N) = \int_{\Sigma} dx^{3} N C \qquad H_{a}(N^{a}) = \int_{\Sigma} dx^{3} N^{a} ( V^{a} - A^{i}_{a} G_{i})
\end{align*}
where $u$ is a vector belonging to the $su(2)$ Lie algebra, N is the lapse scalar and $N^{a}$ is the shift vector. Note that we have used the expression of the vectorial constraint $H_{a}$ where the second class constraint have been solved. 
The full algebra reads:
\begin{align*}
& \{G(u), G(v)\} = G( [u,v]) \\
& \{G(u), H_{a}(N^{a})\} = - G( \mathcal{L}_{N^{a}} u) \\
&  \{G(u), H(N)\} = 0 \\
& \{ H_{a}(N^{a}) , H_{b}(M^{b}) \} = H_{b}( \mathcal{L}_{N^{a}} M^{b}) \\
& \{ H_{a}(N^{a}) , H(M) \} = - H( \mathcal{L}_{N^{a}} M) \\
& \{ H(N) , H(M) \} = H_{a}( S^{ab}( N \partial_{b} M - M \partial_{b} N)) \\
\end{align*}
where $S^{ab} = E^{i}_{a} E_{b \; i} / \sqrt{\text{det}q}$.
The algebra being closed, it confirms that $G$, $H_{a}$ and $H$ form a set of first class constraints.
The gauge transformations of the canonical variables under the $SU(2)$-gauge transformations and the spatial diffeomorphism are obtained by applying respectively the Gauss constraint and the vectorial constraint.
We use the notation $P^{a}_{i} = \gamma^{-1} E^{a}_{i}$ for which the $\gamma$-dependency of the Poisson bracket drops out.
\begin{align*}
\{P^{a}_{i}, A^{j}_{b}\} = \delta^{j}_{i} \delta^{a}_{b}
\end{align*}
The $SU(2)$ gauge transformations read:
\begin{align*}
\{P^{a}, G(u)\} = u \times P^{a} \qquad \{A_{a}, G(u)\} = - D_{a}u
\end{align*}
while the spatial diffeomorphims are given by:
\begin{align*}
\{P^{a}, H_{b}(N^{b})\} = \mathcal{L}_{N^{b}} P^{a} \qquad \{A_{a}, H_{b}(N^{b})\} = \mathcal{L}_{N^{b}} A_{a} 
\end{align*}
The Poisson bracket between the hamiltonian constraint and the canonical variables is more complicated. The important point is that under time diffeomorphism, the real Ashetkar Barbero connection does not transform properly as a connection as pointed by Samuel \cite{ch1-Sam2}. One recovers the right transformation only for $\gamma = \pm i$.

We have presented the real Ashtekar-Barbero phase space. It is the starting point of the quantization program of Loop Quantum gravity that I will present in the next chapter.
However, it is important to stress that the resulting phase space does not admit a unique formulation, since one can either choose to proceed to the canonical analysis without fixing any gauge, i.e. in a full $SL(2,\mathbb{C})$ fashion, or by making a new gauge choice which selects instead of $SU(2)$, the group $SU(1,1)$ which is the non compact subgroup of $SL(2,\mathbb{C})$. The $SU(1,1)$ gauge fixed hamiltonian analysis was, to the knowledge of the author, never computed and remains to be done. The canonical analysis without fixing any gauge was undertaken by Alexandrov in \cite{ch1-Alex1}. The resulting phase space is quite complicated and the connection turns out to be non commutative, spoiling therefore the used of the loop variables. Yet, we observe in this phase space that $\gamma$ disappears, and the quantum theory based on this phase space should therefore be free of $\gamma$. Moreover, it was argued that the area spectrum computed from this quantum theory turns out to be continuous and $\gamma$-independent \cite{ch1-Alex2, ch1-Alex3, ch1-Alex4, ch1-Alex6}. Finally, it was shown in \cite{ch1-Alex5} that this $SL(2,\mathbb{C})$ phase space is equivalent to the self dual Ashetkar's phase space supplemented with its reality conditions. For the sake of completeness, we present now this self dual phase space, originally introduced by Ashtekar in 1986.
\\

\textit{The self dual Ashtekar phase space}\\

It is a well known fact that for gauge theories, the algebraic form of the hamiltonian simplifies drastically when written in term of certain complex variables. The self dual Ashtekar's variables for gravity belongs to this category.
The new variables that Ashtekar introduced in \cite{ch1-Ash1} are given by:
\begin{align*}
^{\mathbb{C}}A^{i}_{a} = \Gamma^{i}_{a} - i \omega^{(0)i}_{a} \;\;\;\;\;\; \text{and} \;\;\;\;\;\; ^{\mathbb{C}}E^{a}_{i} = \frac{1}{2} \epsilon^{abc} \; \epsilon_{ijk}  \;  e^{j}_{b} e^{k}_{c}
\end{align*}
where $\Gamma^{i}_{a}$ is the Levi Civita connection.
The variable $^{\mathbb{C}}A$ is the self dual connection, i.e. an $SL(2,\mathbb{C})$ connection. The self dual phase space, coordinatized by $(\; ^{\mathbb{C}}A, \; ^{\mathbb{C}}E)$, can be obtained by the canonical analysis of the self dual Einstein Palatini action.
The subscript $\mathbb{C}$ indicates that $^{\mathbb{C}}A$ and $^{\mathbb{C}}E$ have complex valued components.
Their Poisson bracket reads:
\begin{align*}
\{ ^{\mathbb{C}}E^{a}_{i},  ^{\mathbb{C}}A^{j}_{b}\} = i \delta^{j}_{i} \delta^{a}_{b} 
\end{align*}
In term of those variables, the self dual Ashtekar phase space is given by:
\begin{align*}
& G =  \partial_{c} \; ^{\mathbb{C}}E^{c} + \; ^{\mathbb{C}}A_{c} \times \; ^{\mathbb{C}}E^{c} \simeq 0 \\
& H_{b} \simeq \; ^{\mathbb{C}}E^{c}. \; \mathcal{F}_{cb}(\; ^{\mathbb{C}}A) \simeq 0 \\
&  H =  -  \frac{ \; ^{\mathbb{C}}E^{b} \times \; ^{\mathbb{C}}E^{c}}{2\sqrt{\text{det}\; ^{\mathbb{C}}E}} . \; \mathcal{F}_{bc}(\; ^{\mathbb{C}}A)  \simeq 0
\end{align*} 
The first striking observation is the simplicity of the constraints compared to the $ADM$ phase space. The complicated $ADM$ hamiltonian constraint is traded for a simple polynomial hamiltonian constraint.
With such simple constraints for gravity, this opened the hope of applying the Dirac quantization program to General Relativity. However, since the variables $^{\mathbb{C}}A$ and $^{\mathbb{C}}E$ are complex valued, one has to impose some condition to recover real General Relativity.
Since we want a real valued metric at the end of the day, we have to impose:
\begin{align*}
\sqrt{\text{det}q}\;  q_{ab} = \; ^{\mathbb{C}}E^{i}_{a} \; ^{\mathbb{C}}E^{j}_{b} \eta_{ij} = \text{Tr} (\; ^{\mathbb{C}}E_{a}\; ^{\mathbb{C}}E_{b})    \in \mathbb{R}
\end{align*}
Moreover, the self dual connection $^{\mathbb{C}}A$ and its complex conjugate $^{\mathbb{C}}\bar{A}$, considered as independent variables, have to satisfy:
\begin{align*}
^{\mathbb{C}}A^{i}_{a} + \; ^{\mathbb{C}}\bar{A}^{i}_{a} = 2 \Gamma^{i}_{a} ( \; ^{\mathbb{C}}E)
\end{align*}
Those two equations ensures that the self dual theory reproduces correctly real General Relativity. One has also to take into account the stability of those constraints which generates secondary constraints. They are called the reality conditions. The main difficulty when building the quantum theory starting from this phase space, is to impose quantum mechanically those constraints which are highly non trivial.

In the mid nineties, different strategy were proposed to deal with those reality conditions. In 1995, Thiemann proposed to deal with the reality conditions at the quantum level through a Wick rotation of the real $SU(2)$ quantum theory \cite{ch1-RCTH, ch1-WT2}. This work was extended in \cite{ch1-Ash3} for gravity coupled to matter. However, the outcomes remained formal. At the classical level, it was shown that the reality conditions can be implemented as second class constraints \textit{a la Dirac} \cite{ch1-R1, ch1-R2}. Finally, an extension of the phase space, taking into account the complex conjugate of the canonical variables $(E,A)$, was introduced by Alexandrov. In this phase space, the reality conditions were treated using the Dirac bracket machinery and the Lorentzian formulation of the phase space, introduced in \cite{ch1-Alex5}, was shown to be equivalent to the self dual Ashtekar's gravity supplemented with the reality conditions. See \cite{ch1-R3,ch1-R4, ch1-R5} for other investigations on the reality conditions.
Although those constructions and results are very interesting, up to now, no one has succeeded to build a full quantum theory of gravity starting from the self dual phase space.

Yet, in some simplified set up, such as spherically symmetry reduced models, the reality conditions can be solved at the quantum level and the self dual Ashetkar spherical gravity can be quantized.
See \cite{ch1-Th3, ch1-Th4} for further details.

In this context, it would be useful to develop a general strategy which provides a way to circumvent the explicit resolution of the reality conditions.

This thesis is precisely devoted to the development of such strategy. This procedure consists in an analytic continuation prescription \cite{ch1-BA1} which has been tested in different context such as spherically isolated horizon \cite{ch1-BA2}, isotropic and homogenous space-time \cite{ch1-BA3} and three dimensional gravity \cite{ch1-BA4, ch1-BA5}. We stress already that such procedure is a fairly common strategy when one has to deal with lorentzian quantum three dimensional gravity \cite{ch1-Witten1}.
We will see that while a direct quantization of the self dual phase space is still elusive, we can try to elucidate different aspect of the self dual quantum theory by mean of an analytic continuation of the quantum theory based on the kinematical Ashtekar-Barbero phase space form $\gamma \in \mathbb{R}$ to $\gamma = \pm i$. \\

\textit{Lessons from the canonical analysis of General Realtivity in the first order formalism} \\

We have detailed in the precedent section the canonical analysis of General Relativity, written in the first order formalism.
Working with the Holst action, which is equivalent to the Einstein-Palatini action at the classical level, we have derived the real $SU(2)$ Ashtekar-Barbero phase space.
This phase space makes General Relativity looks like a $SU(2)$ Yang Mills theory. Yet, the deep difference between gravity and usual Yang Mills gauge theories is the presence of the vectorial and hamiltonian constraint.
Those two constraints are first class and generate the gauge diffeomorphism symmetry, respectively in the three dimensional space and along ``time''. Those constrains underline the background independence of General Relativity or more generally, of any theory of gravity. This is the real novelty of the Einstein theory.

It is one of the most important lesson of the canonical analysis. 

It implies that Quantum General Relativity is fundamentally different from usual Quantum Mechanic (QM) and usual Quantum Field Theory (QFT) which are defined on a given background, usually taken to be the Minskowski geometry. Since background independence is at the heart of General Relativity, we expect this notion to play a crucial role in the quantum version of the theory. Therefore, (from the LQG point of view) the first task in order to build a quantum theory of gravity is to develop the mathematical tools to write down a background independent quantum field theory and then apply those tools to the gravitational field.

This task was unravel during the nineties and the resulting rigorous mathematical background is nowadays at the basis of the quantum theory of geometry also called (real) Loop Quantum Gravity \cite{ch1-Th1}.
This theory in its canonical version and describe its structure and predictions is presented in the next chapter.


\clearemptydoublepage

\chapter{Real Loop Quantum Gravity}
\label{ch:LQG}
\minitoc

This chapter is devoted to giving a general picture of the theory of quantum geometry that has been developed during the last thirty years, i.e. Loop Quantum Gravity.
This theory is not complete but it has reached a maturity which provides us with an very interesting candidate for the theory of quantum gravity.

 Loop Quantum Gravity is a rather conservative approach w.r.t. its initial assumptions since no other ingredients than four dimensional vacuum General Relativity and Quantum Mechanics are required, but in the same time, very new mathematical tools are introduced in order to merge the two orthogonal theories. Indeed, as explained in the precedent chapter, the novelty of General Relativity is the concept of background independence, which is manifest through the general covariance of the theory. The first challenge is therefore to build a background independent quantum field theory formalism in order to be able to describe the quantum gravitational field. Obviously, this requires new mathematical tools that are quite different from the conventional Fock quantization used in standard Minkowskian Quantum Field Theory. The main hypothesis of the loop quantization is the \textit{polymer-like} nature of the quantum states of a generally covariant quantum field. The resulting formalism is a mathematical generalization of usual Quantum Field Theory which incorporates general covariance. LQG is the application of this formalism to the gravitational field.
 
We first give a justification for the introduction of the loop variables as we understand it, and we present then the quantization strategy and the construction of the Hilbert space. The usual textbook where one can find a very detailed description, proofs and results about the loop quantization of General Relativity is \cite{ch2-Th1}. Another presentation of the theory with a particular emphasis on the concept of background independence both at the classical and at the quantum level can be found in \cite{ch2-Rove1}. An very pedagogical introductory book focusing on the spinfoam theory recently appears \cite{ch2-Rove2}. Finally, there are many good reviews focusing either on the canonical quantization or on the spin foams models such as \cite{ch2-Alejandro1, ch2-Hanno1, ch2-Jerzy1}.

\section{Polymer quantization: introducing the loop variables}

The strategy adopted in Loop Quantum Gravity is to implement the generalized Dirac program for quantizing background independent constrained systems \cite{ch2-Mat1}.
The starting point is a given infinite dimensional phase space  $\Omega$ and its associated Poisson bracket between the canonical conjugated variables and the first class constraints which generate the fundamental symmetries. 
The idea is to first quantize the unconstrained phase space, obtaining therefore the unconstrained Hilbert space $\mathcal{H}$. Then one imposes the first class constraints at the quantum level in order to select the physical quantum states.
This procedure is explained and carried out in details in \cite{ch2-Th1}. Since this mathematical construction can be found in many textbooks, we rather present the main lines and justifications to make a closer contact with physical motivations.

Our starting point is the phase space of General Relativity expressed in term of the $SU(2)$ real Ashtekar-Barbero connection.
This phase space reads:
\begin{align*}
\{E^{a}_{i}, A^{j}_{b}\} = \gamma \delta^{j}_{i} \delta^{a}_{b} \;\;\;\;\;\;\;\;\;\; \text{and} \;\;\;\;\;\;\;\;\;\; \{E^{a}_{i}, E^{j}_{b}\} = 0 = \{A^{a}_{i}, A^{j}_{b}\} 
\end{align*}
Those $18$ canonical conjugated variables are constrained by the $7$ first class constraints which generate the $SU(2)$ gauge transformations, the spatial diffeomorphisms and the temporal diffeomorphisms:
\begin{align*}
& G =  \partial_{c} E^{c} +A_{c} \times E^{c} \simeq 0 \\
& H_{b} \simeq \frac{1}{\gamma} \; E^{c}. \; \mathcal{F}_{cb}(A) \simeq 0 \\
&  H =   \frac{1}{2\gamma^{2}} \frac{E^{b} \times E^{c}}{\sqrt{\text{det}E}} . \{ \; \mathcal{F}_{bc}(A) + ( 1 + \gamma^{2}) R_{bc}(\omega (E)) \; \} \simeq 0
\end{align*} 
Those constraint are obtained when fixing the gauge $e^{0}_{a} = 0$ called the time gauge. Once imposed, the first class constraints reduce the dynamical 18 dynamical components to 4 degrees of freedom per phase space point,
i.e. corresponding to the two degrees of freedom of General Relativity and their conjugated momentas.

The goal is to build the unconstrained Hilbert space $\mathcal{H}$ from this phase space. Usually, one starts by choosing a polarization such that the quantum states are either functional of the connection $\Psi(A)$, either functional of the electric field $\Psi(E)$. In Minkwoskian gauge field theory, one generally chooses the polarization where the quantum states are of the form $\Psi(A)$. In this case the conjugated variables turned into quantum operators have the following action on the quantum states:
\begin{align}
\hat{A} \triangleright \Psi(A) = A  \Psi(A)  \;\;\;\;\;\;\;\; \text{and} \;\;\;\;\;\;\;\; \hat{E} \triangleright \Psi (A) = - i \hbar \frac{\delta \Psi (A)}{\delta A}
\end{align}

However, while in the case of canonical quantum field theory, this approach is very powerful, one encounters important difficulties when dealing with a background independent gauge theory, such as General Relativity. The main problem is to defined the measure on the state space which respects the symmetries of the classical theory. Let us write formally the scalar product between two quantum states:
\begin{align*}
\int_{\Sigma} \bar{\Psi}(A) \Psi(A) d\mu(A) 
\end{align*}
To have a well defined quantum theory of general relativity where the background independence is fully implemented, one need to find a measure $d\mu(A)$ which is diffeomorphism invariant.
This task of finding a suitable background independent measure is highly non trivial and working with the usual Schrodinger representation (1.1) turns out to be a dead end.
Therefore, in the context of background independent quantum field theory, it can be more interesting to adopt another representation. Let us review the construction of the so called loop representation.

One can define the quantum theory from the classical algebra  $\mathcal{B}$ of observables without referring to an Hilbert space. This Poisson algebra will be a subalgebra of the observables $\mathcal{C}^{\infty} (\Omega)$ on the phase space $\Omega$. 
Then one can turn this classical Poisson algebra  $\mathcal{B}$ into a quantum $\ast$-algebra \footnote{ An $\ast$-algebra $\mathcal{B}$ is a vector space equipped with a (generally non commutative) product : $ \mathcal{B} \times \mathcal{B} \rightarrow \mathcal{B}$ and with an involution $\ast : \mathcal{B} \rightarrow \mathcal{B}$ which plays the role of the complex conjugate}   $\bar{\mathcal{B}}$ by trading the Poisson bracket structure for a quantum commutator:
\begin{align*}
\{E^{a}_{i}, A^{j}_{b}\} = \gamma \delta^{j}_{i} \delta^{a}_{b} \;\;\;\;\; \rightarrow \;\;\;\;\;\; [\hat{E}^{a}_{i} , \hat{A}^{j}_{b} ] = i \; \hbar \gamma \delta^{j}_{i} \delta^{a}_{b}
\end{align*}

  Once equipped with the quantum $\ast$-algebra $\bar{\mathcal{B}}$ , one can apply the powerful GNS construction \cite{ch2-GNS}. This construction allow us to find a representation $(\mathcal{H}, \pi)$ of the quantum $\ast$-algebra $\bar{\mathcal{B}}$ , which implement unitarily its symmetries. This representation 
  corresponds to an Hilbert space $\mathcal{H} = ( V\; , < \; , \; >)$, i.e. a complex vector space $V$ with its inner product,  and a linear  map $\pi : \bar{\mathcal{B}} \rightarrow \text{End}(\mathcal{H})$ from the $\ast$-algebra $\bar{\mathcal{B}}$  to the space of endomorphisms of $V$, which preserves the $\ast$-algebra structure. A quantum state is given by a positive linear functional $\omega : \bar{\mathcal{B}} \rightarrow \mathbb{C}$ called the expectation. If one decides to work with quantum states of the form $\Psi(A)$, i.e. functional of the connection, then the space of those functional is called the quantum configuration space and denoted $\mathcal{A}$. The GNS construction is purely algebraic and its power lies in its generality. One can apply it to quantum mechanics with finite degrees of freedom, quantum field theory but also for background independent quantum field theory. 
There is obviously some mathematical subtleties concerning the nature of the operators but we refer the interested reader to \cite{ch2-GNS} for more details.

Therefore, the choice of the initial Poisson alegbra $\mathcal{B}$ is of first importance since it is the starting point of the quantization. It has to be selected both by physical motivations and mathematical simplicity. The crucial point is that, for physical and mathematical reasons, the connection $A$ and its conjugated momentum $E$ are not well suited to describe the gravitational field at the quantum level. This fact can be traced back to the background independence of General Relativity, which drastically complicate the task of finding a suitable measure respecting the symmetry of GR.

We present now the different arguments in order to select the Poisson alegbra used in Loop Quantum Gravity, i.e. the holonomy-flux algebra.

\subsection{The holonomy-flux algebra}

The introduction of the holonomy-flux algebra relies on difference arguments that can be divided into two pieces: they are \textit{a priori} arguments motivated by physical considerations, and \textit{a posteriori} arguments which are motivated by mathematical simplicity and uniquness. Let us first present the physical motivations.

The oldest one dates back to the thirties. At this epoch, a russian physicist, Matvei Bronstein, realized that the behaviour at short distance of a quantum field in presence of a dynamical background is radically different than the one in a fixed background \cite{ch2-Br1, ch2-Br2}.
When the geometry is fixed, one usually assumes that the quanta of a field can probe any arbitrary small distances. Indeed, when computing the loop corrections to a Feynman diagram, one integrates (in principle) over the whole range of energy (i.e. probing therefore the smallest distances). This is not problematic if one does not consider the dynamic of the gravitational field.

However, when gravity become dynamical, it is assumed that above a given energy, the quanta of a quantum field will create a black hole and therefore an horizon. This horizon will act as a screen, which forbids to probe distances smaller than its radius. This will happen when the Compton length of the quanta will have the same order of magnitude than its Schwarschild radius : $r_{c} \simeq r_{s} $.  It turns out that this scale corresponds to the Planck length, where the classical description of the gravitational field provided by General relativity breaks down. It is the quantum gravity regim.
From this simple semi classical argument, we conclude that \textit{a quantum field can not probe any arbitrary small regions when quantum gravity is taken into account}. Quantum gravity acts as a natural cut off. Therefore we know that we can not use local variables (i.e. fields) to describe the gravitational field at the quantum level. Instead, one has to introduce \textit{non local variables}, or integrated variables.
This is the first lesson for the construction of our Poisson algebra.

Moreover, as explained in the introduction and derived in the precedent chapter, General Relativity is a background independent theory. See \cite{ch2-BCKI, ch2-Rove3} for an interesting discussion of the meaning of background independence. The diffeomorphism group generates a dynamical symmetry which constrains the equations of motion, contrary to the invariance under coordinates transformations, which is a non dynamical symmetry \cite{ch2-Ed1}. Therefore, in order to respect this fundamental symmetry, the Poisson algebra has to be build from objects which do not refer to any background. It is the second lesson for the construction of our Poisson algebra.

Note that the two physical arguments of non locality and background independence (of the fundamental variables we want to quantize) are not independent. 

More generally, the idea that the fundamental quantum states of Yang Mills theories are sting-like objects, i.e. loops, is quite old and date back to the eighties and even before. 
Polyakof, Mandelstam but also Wilson, Gambini and Trias argued in this direction for a long time. In spirit, those ideas lay in the continuity of the old Faraday's intuition that the electric and magnetic fields manifest truly through their lines of flux. Following this idea, the loop representation used in LQG relies on the hypothesis that the fundamental excitations of the quantum gravitational field are one dimensional excitations, i.e. loops. Adopting this point of view, the quantum space-time is built up from loops excitations.

From the precedent discussion, we have obtained two physical requirements for building our new variables, and we can now introduce them. They are called the holonomy-flux variables.
Once presented, we will explained the \textit{a posteriori} mathematical arguments.

A very natural non local and background independent variable built from the connection is provided by the Wilson loop. This integrated functional of the connection is commonly used in gauge field theory, for instance when one deals with lattice QCD. 
The holonomy of the connection can be understood as the order product of the exponential of the integral of the gauge connection along a path $e$ embedded in the three dimensional manifold $\Sigma$.
\begin{align*}
h_{e} ( A) =   \overrightarrow{\text{exp} } \int_{e} A  =  \overrightarrow{\text{exp} } \int_{e} A^{i}_{a} \tau_{i} \dot{e}^{a} =  \overrightarrow{\text{exp} } \int^{t(e)}_{s(e)} A^{i}_{a} \tau_{i} \frac{dx^{a}}{d\lambda} d\lambda
\end{align*}
The arrow denoted the order product, $\dot{e}^{a}$ is the tangent vector to the path $e$, $t(e)$ and $s(e)$ are respectively the target and the source points of the path $e$, $\tau_{i} = - \sigma_{i} /2$ are the $su(2)$ algebra generators and $\sigma_{i}$ are the Pauli matrices.

This integration is very natural because the connection $A= A^{i}_{a} \tau_{i} dx^{a}$ being a one form, it is natural to integrate it along a path. The passage to the exponential ensures that the object will have interesting group properties and therefore a good behaviour under internal gauge transformations.
Indeed, the connection $A = A^{i} \tau_{i}$ lives in the $su(2)$ Lie algebra and therefore, the exponential maps it into a group element of $G =SU(2)$. The Wilson loop is obtained by taking the trace of finite dimensional representation of the holonomy, which lead to a gauge invariant quantity with respect to the $SU(2)$ gauge transformations (i.e. the character). 

Let us now study how the holonomy is transformed under $SU(2)$ gauge transformation. The usual gauge transformation of the connection, ensuring the covariance under $SU(2)$ gauge transformations of the covariant derivative, is given by:
\begin{align*}
g \triangleright   A = g A g^{-1} + g d g^{-1}
\end{align*}
The holonomy being a group element, it transforms under the gauge transformations as:
\begin{align*}
g \triangleright  h_{e} = g_{s(e)} h_{e} g^{-1}_{t(e)} 
\end{align*}
Therefore, the introduction of the holonomy turns the complicated gauge transformations of the connection into a simple local and discrete gauge transformation at the source and target points of the path $e$.
From its definition, under a change of the orientation of the path $e$ or a composition of two paths $e_{1}$ and $e_{2}$ into $e = e_{1} \circ e_{2}$, the holonomy satisfies:
\begin{align}
h_{e^{-1}} = h^{-1}_{e}  \;\;\;\;\;\;\;\;\;\; \text{and} \;\;\;\;\;\;\;\;\; h_{e} = h_{e_{1}} h_{e_{2}}
\end{align}
Finally, the action of a diffeomorphism $\phi$ on the connection shifts this connection from one point $C \in \Sigma$ to another point $D \in \Sigma$ and then relabeled the coordinates at the final point $D$ with the same coordinates than the one we started with at the point $C$ []. It is a combination of a pushforward, i.e. the shift from $C$ to $D$, and a coordinates transformation, i.e. the relabeling of the coordinates at $D$.
Applying it to the holonomy of the connection, we observe that only the support of the integration will be affected, i.e. the diffeomorphism simply shifts the path $e$. This action of the diffeomorphism on the holonomy reads:
\begin{align*}
h_{e}(\phi^{\ast} A) = h_{\phi(e)} (A)
\end{align*}
Finally, we note that the Poisson bracket between holonomies defined on the same path or on different paths is given by:
\begin{align*}
\{ \; h_{e} , h_{e'}\; \} = 0
\end{align*}
Having presented the holonomy variable and its behaviour under $SU(2)$ gauge transformations and diffeomorphisms transformations, we introduce now the flux variables. Since we have at our disposal a vector field at value in the dual $su(2)$ Lie algebra, i.e. the electric field $E^{a}_{i}$ conjugated to the Ashtekar-Barbero connection, we need to define a suitable smearing in order to obtain a non local variable conjugated to the holonomy. To have a natural integration of the vector electric field, we first need to turn it into a form, which are the natural objects to integrate. From the vector $E^{a}$, one can define the two form $E_{ab} = \epsilon_{abc} E^{c}$ which is naturally integrated over a two surface, providing us with the flux variable. This flux variables is defined as:
\begin{align*}
(X_{i} )_{S} (E) & = \frac{1}{2} \int_{S} (E_{i })_{ab} dx^{a} \wedge dx^{b} = \frac{1}{2}  \int_{S} \epsilon_{abc} E^{c}_{i} dx^{a} \wedge dx^{b} =  \int_{S} \epsilon_{abc} E^{c}_{i} dx^{a} dx^{b} \\
& =  \int_{S} \epsilon_{abc} E^{c}_{i} \frac{\partial x^{a}}{\partial y^{1}}  \frac{\partial x^{b}}{\partial y^{2}} dy^{1} dy^{2} =  \int_{S} [dy]^{2}  \; E^{c}_{i} \; n_{c}   
\end{align*}
where the one form $n_{c}$, normal to the surface $S$ locally coordinatized by $(y^{1}, y^{2})$, is given by:
\begin{align*}
n_{c} =  \epsilon_{abc} \frac{\partial x^{a}}{\partial y^{1}}  \frac{\partial x^{b}}{\partial y^{2}}  =  (v_{1} \times v_{2})_{c}  
\end{align*}
where $v_{1}$ and $v_{2}$ are the two co-planar vectors living on the surface $S$.

The variable $(X^{i})_{S}(E)$ represents therefore the flux of the electric field $E^{a}_{i}$ across a given surface $S$.
It is a vector field living in the Lie algebra $g = su(2)$, i.e. in the tangent space of the Lie group $G = SU(2)$ at the identity, i.e. $T_{0} G$. It plays the role of the momentum $p$ in usual quantum mechanics.
The Poisson bracket between two flux variables is given by:
\begin{align}
\{ \; X_{S_{e}} , X_{S_{e'}}\; \} = 0
\end{align}

However, for different reasons, it would be interesting to obtain a flux variable which actually satisfies the $g= su(2)$-Lie algebra. This requirement is crucial for instance to build the $SU(2)$ quantum tetrahedron, but also when one develops the so called twisted geometry. In order to obtain such a flux variable, one can modified to above definition. Under a gauge transformations, the initial flux variable transforms as:
\begin{align*}
g \triangleright E(y) = g(y) E(y) g^{-1}(y)  \;\;\;\;\; \text{whence} \;\;\;\;\; X^{i} (g(y) E(y) g^{-1}(y) )
\end{align*}
There, the flux variable does not transform as a vector.
The definition of this flux can be modified as follow \cite{ch2-Frei1}. Consider a oriented path $e$ embedded in $\Sigma$ and a point $z \in e$. Choose a surface $S_{e}$ which is pierced by the path $e$ transversally such that $z = e \cap S_{e}$.
Choose then a set of paths $\pi_{e} : S_{e} \times [0,1] \rightarrow \Sigma$ which are in one to one correspondence with the points $y \in S_{e}$. Those paths go from the source point $s(e)$ of the path $e$ to their associated point $y \in S_{e}$, ie $\pi_{e}(y,0) = s(e)$ and $\pi_{e} (y,1) = y$. Therefore, with this construction, one can defined the modified flux variable as:
\begin{align*}
(X_{i} )_{S_{e}, \pi_{e}} (E, A) & = \int_{S} [dy]^{2}  \; h_{\pi_{e}}\; E^{c}_{i} \; h^{-1}_{\pi_{e}} \; n_{c} 
\end{align*}
where $h_{\pi_{e}}$ is the holonomy of the connection along the path $\pi_{e}$, the integral in $h_{\pi_{e}}$ running from $s(\pi_{e})$ to $y$. The flux variable depends now both on the electric field $E^{a}_{i}$ and on the connection $A^{i}_{a}$.

The gauge transformation of the electric field $E(y)$ and of the holonomy $h_{\pi_{e}}(y)$ are given by:
\begin{align*}
g \triangleright E(y) = g(y) E(y) g^{-1} (y) \;\;\;\;\;\;\;\; \text{and} \;\;\;\;\;\;\;\; g \triangleright h_{\pi_{e}}(y) = g_{s(\pi_{e})} h_{\pi_{e}}(y) g^{-1} (y)
\end{align*}
Those transformations imply that the modified flux variable transforms as:
\begin{align*}
(X_{i}) _{S_{e}, \pi_{e}} ( g \triangleright A , g \triangleright E ) = g_{s(\pi_{e})} \;  [ \; (X_{i}) _{S_{e}, \pi_{e}} (A,E) \; ] \; g^{-1}_{s(\pi_{e})}
\end{align*}
Once, again, with this new definition of the flux variable, the $SU(2)$ gauge transformations of the electric field are turned into discrete gauge transformations.
Moreover, the change of the orientation of the path $e$ implies that $\pi_{e^{-1}} = e^{-1} \circ \pi_{e}$ and for the surface $S_{e^{-1}} = - S_{e}$ which leads to:
\begin{align*}
h_{\pi_{e^{-1}}}(y) = h^{-1}_{e} h_{\pi_{e}}(y) \;\;\;\;\;\; \text{whence} \;\;\;\;\;\; X_{S_{e^{-1}}, \pi_{e^{-1}}} = - h^{-1}_{e} X_{S_{e}, \pi_{e}} h_{e}
\end{align*}

The Poisson bracket for the modified flux variable does not commute anymore and becomes:
\begin{align}
\{ \; X^{i}_{S_{e}, \pi_{e}} , X^{j}_{S_{e'}, \pi_{e'}}\; \} = \epsilon^{ij}{}_{k} \; \delta_{ee'}\; X^{k}_{S_{e}, \pi_{e}}
\end{align}
This is the usual $su(2)$ Lie algebra. Finally, the poisson bracket between the modified flux and the holonomy is given by:
\begin{align*}
\{ \; X^{i}_{S_{e}, \pi_{e}} , h_{e'}\; \} = - \delta_{ee'} \tau^{i} h_{e} + \delta_{ee'^{-1}} h_{e} \tau^{i}
\end{align*}
Those Poisson brackets which define the holonomy-flux algebra are presented in \cite{ch2-Frei1} for instance.
At this stage, we have obtained our classical Poisson algebra $\mathcal{B}$, i.e. the holonomy-flux algebra. This algebra satisfies the two physical requirements to be build with non local and background independent variables, i.e. the holonomy of the connection along a given path and the flux of the electric field across a given surface. While the holonomy of the Ashtekar-Barbero connection is an element of the Lie group $G =SU(2)$, its conjugated momentum $X^{i}$ is a vector belonging to the Lie algebra $g= su(2)$. Therefore, this algebra has the structure of the cotangent bundle $T^{\ast} G = G \times g$ with respect to the group $G= SU(2)$.

At the beginning of this section, we mentionned that in addition of the physical motivations for choosing this Poisson algebra, there are also some mathematical simplicity motivations arising a posteriori. As we have seen, the $SU(2)$ gauge transformations which are non trivial when considering the connection become much simpler when working with the holonomy. This was historically the first mathematical motivations. The point is that the simplicity of the gauge transformations of the holonomy allows one to define easily gauge invariant quantum states under $SU(2)$ gauge transformations. Those states will be introduced in the following sections and are called spin networks states.

The second motivation for working with this Poisson algebra is the uniqueness theorem firstly derived by Lewandowski, Okolow, Salhmann and Thiemann in 2006 \cite{ch2-LOST}, and derived in a slightly different context by Fleichack  in 2009 \cite{ch2-Fl1}.
This argument came a posteriori since this algebra was introduced in the mid nineties but it is nowadays understood as the corner stone of the loop quantization.
Basically, when the GNS construction is applied to the quantum holonomy-flux algebra, the requirement that the representation carries a unitary action of the diffeomorphism group selects uniquely the possible representation (up to unitary equivalence). It is the diffeomorphism counterpart of the uniqueness theorem of Von Ston Nemann in quantum mechanics, which ensures the uniqueness of the usual Schrodinger representation for finite dimensional systems.
The LOST theorem underlines the powerful role played by the diffeomorphism group of General Relativity at the quantum level. Being background independent is a highly restrictive property. One of the key feature of the construction is that in order to implement unitarily the diffeomorphism group on the Hilbert space, one has to relax the continuity property of the scalar product of the representations, i.e. the matrix elements of a quantum operator evaluated on two quantum states is no more continuous. From this observation, we note that the polymer representation resulting from quantizing the holonomy-flux algebra will be non equivalent to the usual Fock quantization. This difference is precisely the key ingredient to quantize the gravitational field, and emerges due to the requirement of background independence. A very pedagogical presentation of this crucial point can be found in \cite{ch2-Ashtekar}.

Let us mention that, as any theorem, the LOST theorem is based on some hypothesis which can be relaxed. Therefore, the unicity property of the Ashtekar-Isham-Lewandowski representation, i.e. the loop representation, could well be by passed by modifying slightly the different hypothesis of the LOST theorem. This is precisely what was done by Koslowski, Sahlmann and Varadarajan \cite{ch2-KosRep, ch2-HannoRep, ch2-VarRep}, and much more recently in \cite{ch2-Geiller1, ch2-Geiller2, ch2-Geiller3}, leading to the so called Flux representation of Loop Quantum Gravity.
This new representation leads to a new notion of quantum geometry different from the one inherited from the loop representation.

Having presented the holonomy-flux algebra and the motivations to work with it in order to describe the quantum gravitational field, we can now discuss its representations, i.e. the quantum configuration space $\mathcal{A}$ of Loop Quantum Gravity, and the general strategy used to defined interesting mathematical structures on it, such as a topology, a measure and a notion of differential geometry.

\subsection{From the quantum configuration space to the unconstrained Hilbert space of Loop Quantum Gravity}

The task of studying the representations of the holonomy-flux algebra was undertaken in the mid nineties by different authors, among who Ashtekar, Isham, Lewandowski and Baez \cite{ch2-Ashte1, ch2-Ashte2, ch2-Ashte3, ch2-Ashte4, ch2-Ashte5, ch2-Ashte6}.
Those works provided a rigorous mathematical construction of the quantum configuration space $\mathcal{A}$ on which the unconstrained background independent Hilbert space $\mathcal{H}$  of Loop Quantum Gravity can be build.
An important point to notice is that the following construction is valid for any background independent quantum gauge field theory (with some certain assumptions on the nature of the gauge group G), i.e. it is a very general strategy which is not restricted to quantum gravity. This is precisely the formalism mentioned in the introduction. This formalism was applied to the Maxwell field \cite{ch2-Ashte7}, to the graviton \cite{ch2-Ashte8} but also to the scalar field \cite{ch2-Ashte9}.

The quantum configuration space $\mathcal{A}$ is the space of smooth connections over the three dimensional manifold $\Sigma$. We denote by $\mathcal{A} / \mathcal{G}$ the space of gauge equivalent connections over $\Sigma$. The idea is to study the space of continuous functional of the connection $f(A) \in C^{0} (\mathcal{A} / \mathcal{G})$. 

Here comes the crucial point. As we have seen in the precedent section, the well suited variables to describe the gravitational field at the quantum level are not the local connection field but its holonomy, which is a non local  and background independent quantity. Therefore, we will work only with functionals of the holonomy of the connection, i.e.  $f(h_{e}(A))$, which are a restricted class of functionals of the connection $f(A)$. 

Such functional depends therefore on the connection $A$ but also on a path $e \in \Sigma$. Those functionals can be generalized in the following way. Consider a collection of paths $\Gamma = (e_{1}, ....., e_{n}) \in \Sigma$ which satisfies the following properties:
\begin{itemize}
\item every path $e_{i} \in \Gamma$ is diffeormorphic to the closed interval $[0,1]$
\item if $e_{1} , e_{2} \in \Gamma$ with $e_{1} \neq e_{2}$, the intersection $e_{1} \cap e_{2}$ is contained in the set of vertices of $e_{1}$ and $e_{2}$, (i.e. the vertices of a given path $e_{i}$ are the couple of its source point $s(e)$ and its target point $t(e)$)
\item every $e_{i} \in \Gamma$ is at both sides connected with another element of $\Gamma$
\end{itemize}

$\Gamma$ is called a graph. To make contact with the notation of LQG, a path will be called an edge for the rest of the chapter. Therefore, the graph $\Gamma$ contains $n$ edges.
To each edges $e_{i} \in \Gamma$, one can associate the holonomy of the connection $A$ along $e_{i}$, i.e. $h_{e_{i}(A)} \in G= SU(2)$. Then, one can consider the following continuous function:
\begin{align*}
f_{\Gamma} : & \;\;\; SU(2)^{n} \rightarrow \mathbb{C} \\
& ( h_{e_{1}}, ......, h_{e_{n}})  \rightarrow f_{\Gamma} ( h_{e_{1}}, ......, h_{e_{n}})  
\end{align*}
 Those functions are called cylindrical functions and depend only on the connection $A$ through the holonomies of this connections along the edges of the associated graph $\Gamma$.
 Therefore, those functions can be viewed as having support on a graph colored by group datas, i.e. the holonomies.
We observe that the whole machinery of graph and group enters at this level into the mathematical construction of LQG.

Thus, starting from the general space of continuous functionals of the smooth gauge equivalent connections  $f(A) \in C^{0} (\mathcal{A} / \mathcal{G})$, we have selected a very particular class of functionals called the cylindrical functions $f_{\Gamma}$ in order to respect the concept of background independence. The space of cylindrical functions over a graph $\Gamma$ will be denoted:
\begin{align}
\text{Cyl}^{0}(\Gamma)
\end{align}
Ovisously, one has to considerate $\text{Cyl}^{0}(\Gamma)$ for all possible graphs $\Gamma$ in order to recover the continuous manifold $\Sigma$.

In order to understand this construction and the possible mathematical structures that one can associate on $\text{Cyl}^{0}(\Gamma)$, it is interesting to introduce the notion of a \textit{projective family of graphs}.
This notion has been of first importance to study the topology and the possible measures on the space of cylindrical functions, but also to develop a differential calculus on this unconventional configuration space.
The different definitions can be found in \cite{ch2-Ashte5} and we only give the main ideas and notions.

The construction of the projective family of graphs works as follow. We first give the general abstract definitions in order to build this object and then we interprete it in our context.
Let us introduce a partially ordered, direct set $L$ of labels. This set is equipped with a relation ``$\geqslant$'' such that for all labels $\gamma, \gamma', \gamma'' \in L$, we have:
\begin{itemize} 
\item  $\gamma \geqslant \gamma $
\item  $ \gamma \geqslant \gamma'$ and $\gamma' \geqslant \gamma$ then $\gamma = \gamma'$ 
\item $ \gamma \geqslant \gamma' $ and $\gamma' \geqslant \gamma'' $ then $ \gamma \geqslant \gamma'' $
\end{itemize}

A projective family $(\mathcal{X}_{\gamma}, P_{\gamma\gamma'})$ with $\gamma, \gamma' \in L$ consists of sets $\mathcal{X}_{\gamma}$ indexed by elements of $L$, together with a family of surjective projections which satisfy:
\begin{itemize}
\item $P_{\gamma\gamma'} : \mathcal{X}_{\gamma'} \rightarrow \mathcal{X}_{\gamma}$ where $\gamma' \geqslant \gamma$
\item $P_{\gamma\gamma'} \circ P_{\gamma'\gamma''} = P_{\gamma\gamma''} $ where  $\gamma' \geqslant \gamma$
\end{itemize}

Applying this construction to our case, a set $\mathcal{X}_{\gamma}$ corresponds to the a given graph $\Gamma$ and the projection $P_{\gamma\gamma'}$ projects a given graph $\Gamma'$ to a smaller graph $\Gamma$. There are therefore an partially ordered infinite tower of graphs, each stage corresponding to a given graph $\Gamma$. To recover the continuous limit from a given graph $\Gamma$, one need to refine it by adding more and more edges, until one obtains the continuous three dimensional manifold $\Sigma$.

This limit is obtained by defining the projective limit $\bar{\mathcal{X}}_{\gamma}$  of a projective family $(\mathcal{X}_{\gamma}, P_{\gamma\gamma'})$.
It is the cartesian product of the whole sets $\mathcal{X}_{\gamma}$ and is defined as:
\begin{align*}
\bar{\mathcal{X}}_{\gamma} = \{ \; ( x_{\gamma}) \in \times_{\gamma \in L} \mathcal{X}_{\gamma}  \; :  \; \gamma'  \geqslant \gamma  \;\;\; \text{then} \;\;\; P_{\gamma\gamma'} x_{\gamma'} = x_{\gamma} \; \}
\end{align*}

Having identified each space $\mathcal{X}_{\gamma}$ with a given graph $\Gamma$, we can now introduce the space of continuous cylindrical functions over a given graph : $\text{Cyl}^{0} (\mathcal{X}_{\gamma} )=\text{Cyl}^{0}(\Gamma)$, i.e. the space of the continuous functions which depend only on the holonomies of the connection along the edges of the graph $\Gamma$. It is the space introduce in (4) which is now define in a more general context.
One can then take the union of all the continuous cylindrical function over the whole set of possible graphs:
\begin{align*}
\bigcup_{\gamma \in L} \text{Cyl}^{0} (\mathcal{X}_{\gamma}) = \bigcup_{\Gamma \in \Sigma} \text{Cyl}^{0} (\Gamma) 
\end{align*}

Finally, for three graphs labelled by  $\gamma_{1}, \gamma_{2}, \gamma_{3}$ such that $\gamma_{3} \geqslant \gamma_{1}$ and $\gamma_{3} \geqslant \gamma_{2}$, one needs to identify through an equivalence relation the cylindrical functions $f_{\gamma_{i}} \in \text{Cyl}^{0} \mathcal{X}_{\gamma_{i}} = \text{Cyl}^{0}( \Gamma_{i})$ such that:
\begin{align*}
f_{\gamma_{1}} \sim f_{\gamma_{2}}  \;\;\;\;\; \text{if} \;\;\;\;\; P^{\ast}_{\gamma_{1} \gamma_{3}} f_{\gamma_{1}} = P^{\ast}_{\gamma_{2} \gamma_{3}} f_{\gamma_{2}} 
\end{align*}
where $P^{\ast}_{\gamma_{i}\gamma_{j}}$ denotes the pull back from the graph labelled by $\gamma_{i}$ to the space labelled by $\gamma_{j}$, i.e. the map which sends the function from the graph $\Gamma_{i}$ to the bigger graph $\Gamma_{j}$. Finally, with this equivalence class, we can defined the space:
\begin{align*}
\text{Cyl}^{0} ( \bar{\mathcal{X}}) = ( \; \bigcup_{\gamma \in L} \text{Cyl}^{0} (\mathcal{X}_{\gamma}) \; ) / \sim
\end{align*}
This is the general quantum configuration space of LQG, i.e. the space of cylindrical functions defined on all the possibles graphs embedded in $\Sigma$, up to equivalence relation. 
Written in term of graphs, one has simply:
\begin{align*}
\text{Cyl}^{0} ( \bar{\Gamma}) = ( \; \bigcup_{\Gamma \in \Sigma } \text{Cyl}^{0} (\Gamma) \; ) / \sim
\end{align*}
where $\bar{\Gamma} =  \bar{\mathcal{X}}$ is the cartesian product of all the possible graphs. 

Now that we have identified the quantum configuration space, we need to equipped it with a measure in order to obtain a well defined unconstrained Hilbert space $\mathcal{H}$. To find this measure, and therefore the scalar product, let us consider two cylindrical functions defined on $\Gamma$, i.e. let say $f_{\Gamma}, g_{\Gamma} \in \text{Cyl}^{0}(\Gamma)$. The graph $\Gamma$ is assumed to have $n$ edges and $v$ vertices.
We know that the arguments of $f_{\Gamma}, g_{\Gamma}$ are group elements $h$ of $SU(2)$, i.e. the holonomies along each edge of $\Gamma$. It turns out that the natural invariant measure on a Lie group is given by the Haar measure $dh$ \footnote{In our case, since we are working with $G= SU(2)$ which is isomorphic to the three sphere, each element $h \in SU(2)$ can be characterized by three angles called the Euler angles. The Haar measure turns out to be simple the spherical measure on the three sphere}. Since $f_{\Gamma}$ and $g_{\Gamma}$ depends on $n$ copies of the group $G= SU(2)$, we need a measure corresponding to $n$ times the $SU(2)$ Haar measure in order to integrate all the arguments of the functions. 
The scalar product of the two cylindrical functions reads therefore:
\begin{align*}
< \; f_{\Gamma}, g_{\Gamma} \; > \; &= \; \int_{SU(2)^{n}} dh^{n} \; f_{\Gamma} (h_{e_{1}} , .... , h_{e_{n}}) \overline{g_{\Gamma} (h'_{e_{1}} , .... , h''_{e_{n}})} \\
& \; = \; \int _{SU(2)^{n}}  (d\mu_{\Gamma})_{AL} \; f_{\Gamma} (h_{e_{1}} , .... , h_{e_{n}}) \overline{g_{\Gamma} (h'_{e_{1}} , .... , h''_{e_{n}})} \\
\end{align*}
where $(d\mu_{\Gamma})_{AL}$ is called the Ashtekar-Lewandowski measure \cite{ch2-Ashte10}.  It is simply a ``cylindrical'' gauge invariant measure build from $n$ copies of the invariant Haar measure on the Lie group $SU(2)$. 
This provides us with a $SU(2)$-invariant measure which can be extended to the quantum configuration space, leading therefore to a well defined Hilbert space structure. Moreover, this measure turns out to be diffeomorphism invariant, which implies that the unconstrained Hilbert space $\mathcal{H}$ carries a natural action of the diffeomorphism group. 
Finally, it is natural to ask that the functions $f_{\Gamma}$ and $g_{\Gamma}$ be square integrable with respect to the Ashtekar-Lewandowski measure in order to obtain finite transition amplitudes.
To a given graph $\Gamma$ having $n$ edges, one can therefore associate the Hilbert space $\mathcal{H}_{\Gamma}$ defined as:
\begin{align*}
\mathcal{H}_{\Gamma} = ( \text{Cyl}^{0}(\Gamma), d\mu_{\Gamma}) = L^{2} [SU(2), dh]^{n}
\end{align*}
where $f \in L^{2} [SU(2), dh]^{n}$ is a square integrable function over the tensor product of $n$ copies of the group $SU(2)$.
By taking the Cauchy completion \footnote{The Cauchy completion of a space consists basically in filling up the ``holes'' of the considered space. For instance, the Cauchy completion of the rational numbers $\mathbb{Q}$ is the real numbers $\mathbb{R}$. The missing points that one had by the Cauchy completion correspond to the limits of the Cauchy sequences of rational numbers which generate the real numbers.} of the space $\text{Cyl}^{0} ( \bar{\Gamma})$ with respect to the measure $d\mu_{\Gamma}$, one obtains the unconstrained Hilbert space of Loop Quantum Gravity:
\begin{align*}
\mathcal{H} = \bigoplus_{\Gamma \in \Sigma} \mathcal{H}_{\Gamma} 
\end{align*}

This Hilbert space is in fact isomorphic to the space of square integrable functionals $L^{2}(\overline{\mathcal{A}}, d\mu_{AL})$ over the space of generalized connections $\overline{\mathcal{A}}$ where the integration is realized with respect to the Ashtekar-Lewandowski measure $d\mu_{AL}$. This result was obtained by Ashtekar and Lewandowski and is explained in \cite{ch2-Jerzy1}. The space of generalized connections \footnote{The denomination connection and holonomy is commonly interchanged in mathematics. Here, the term ``connections'' corresponds to their associated holonomies, i.e. the functional over $\overline{\mathcal{A}}$ correspond to cylindrical function over $\bar{\Gamma}$.} $\overline{\mathcal{A}}$ corresponds to the $C^{k}$ connections over $\Sigma$, with $k = 0$ to $k = \infty$, but also to the connections which are distributional. 
A distributional connection is a map which assign to and edge $e\in \Sigma$ and element $h\in SU(2)$ which satisfies the two properties (1).
Therefore, the unconstrained Hilbert space of LQG can be written as:
\begin{align*}
\mathcal{H} \sim L^{2}[\overline{\mathcal{A}}, d\mu_{AL}]
\end{align*}

Now that we have characterized our unconstrained Hilbert space, we would like to obtain an interesting basis on it, which will allow explicit computations. 
Once again, we can benefit from the group structure of the support of the cylindrical functions. Since the cylindrical functions depend on $SU(2)$ groups elements, a natural idea to decompose those functions is to use the harmonic analysis on the group $SU(2)$. This can be done using the Peter-Weyl theorem which states that square integrable functions over $SU(2)$ with respect to the Haar measure can be decomposed as a sum over irreducible representations of the group \footnote{It is the very same procedure that the usual Fourier transform, which is nothing else than the harmonic analysis on the group $U(1)$, i.e. $f(x) = \sum_{n} a_{n} e^{in x}$ where $e^{inx}$ is the scalar representation of the group $U(1)$ labelled by the integer $n$.}. An irreducible representation of the goup $SU(2)$ is labelled by a half-integer $j$ called the spin, and by another half-integer $m \in \{ -j , ...., j\}$ called the magnetic number.
For a square integrable function $f \in L^{2}(SU(2), dh)$, the harmonic decomposition reads:
\begin{align*}
\Psi( h) = \sum_{j_{i}, m_{i}, n_{i}}  \; (\tilde{f^{j}})^{n}{}_{m} (D^{j}(h))^{m}{}_{n} \;\;\;\;\;\; \text{where} \;\;\;\;\;\; (\tilde{f^{j}})^{n}{}_{m} = \int_{SU(2)} dh \; (D^{j}(h))^{n}{}_{m} \; f(h)
\end{align*}
The elements of matrices $(D^{j}(h_{e}))^{m}{}_{n}$ of the irreducible representation are called the Wigner matrices. 
They satisfy some orthogonality relations that one can found in \cite{ch2-Edmond1}. 

For a cylindrical function $\Psi_{\Gamma} \in \text{Cyl}^{0}(\Gamma)$ on a graph $\Gamma$ containing $k$ edges, the Peter Weyl decomposition generalizes to:
\begin{align*}
\Psi_{\Gamma}( h_{e_{1}}, ......., h_{e_{n}}) = \sum_{j_{i}, m_{i}, n_{i}}  \; (\tilde{f}^{j_{1} ......j_{k}})^{n_{1} ....... n_{k}}{}_{m_{1} ...... m_{k}} (D^{j_{1}}(h_{e_{1}}))^{m_{1}}{}_{n_{1}} ..... (D^{j_{k}}(h_{e_{k}}))^{m_{k}}{}_{n_{k}} 
\end{align*}
The cylindrical function $\Psi_{\Gamma}$ is therefore given by a product of $k$ Wigner matrices, each one corresponding to an edge of the graph $\Gamma$. The component $f^{n_{1} ....... n_{k}}{}_{m_{1} ...... m_{k}}$ is a tensor living in the tensor product $( T (SU(2)) \otimes T^{\ast}(SU(2)))^{k}$ where $T (SU(2))$ is the tangent space of $SU(2)$.

This decomposition provides us with a basis which will turn out to be very interesting when dealing with the $SU(2)$ gauge transformations.
We have now an Hilbert space and a basis on it which characterized the cylindrical functions, i.e. the unconstrained quantum states.
The next step in the Dirac algorithm is to apply the constraints of General Relativity, which will select among all the unconstrained quantum states the ones which actually describes the quantum gravitational field. 

\section{Building the Hilbert space}

We will now construct the quantum states which are invariant under the fundamental symmetries of General Relativity. To select those quantum states, one has first to express the different classical constraints in term of the holonomies and the fluxes which are our fundamental variables to describe the quantum gravitational field. Then, by trading the classical variables into quantum operators, we obtain the quantum constraints.
This step involves some regularization procedures in order to have a well defined quantum operators acting on the unconstrained Hilbert space. This task is explained in great details in \cite{ch2-Th1}.

In LQG, the Gauss and the spatial diffeomorphism constraints have been successfully implemented but the implementation of the scalar constraint is still problematic. While there is some propositions, such as the Master constraint program due to Thiemann, they remain formal and there is still no control on the physical quantum states. 

Let us first describe the $SU(2)$ gauge invariant quantum states.

\subsection{The gauge invariant Hilbert space : spin network quantum states}

We have seen that the phase space of General Relativity contains a constraint imposing the gauge $SU(2)$ symmetry, i.e. the Gauss constraint.
This constraint reads:
\begin{align*}
G_{i} = \partial_{a} E^{a}_{i} + \epsilon_{ij}{}^{k}A^{j}_{a} E^{a}_{k} = D_{a} E^{a}_{i} \simeq 0
\end{align*}
It states that the $SU(2)$ covariant divergence of the electric field $E^{a}_{i}$ vanishes.
The action of the Gauss constraint on the connection $A^{i}_{a}$ and on the electric field $E^{a}_{i}$ reads:
\begin{align*}
\{G(u), A^{i}_{a} \} = - ( \partial_{a} u^{i} +  \epsilon^{i}{}_{jk}A_{a}^{j} u^{k} ) = - D_{a} u^{i}  \;\;\;\;\; \text{and} \;\;\;\;\; \{ G(u), E^{a}_{i} \} = \epsilon_{i}{}^{jk} u_{j} E^{a}_{k} = (u\times E^{a})_{i}
\end{align*}
where $u =u^{i} \tau_{i}$ is a vector living in the $su(2)$ Lie algebra.
Therefore, the Gauss constraint is the generator of the $SU(2)$ gauge transformations. 

In order to implement this constraint at the quantum level, two strategies can be followed.
The first one would be to define the Gauss quantum operator through some regularization, then apply it to the unconstrained Hilbert space and look for the quantum states which are annihilated.
This strategy and the explicit form of the Gauss quantum operator are given in [] (chapter 9, page 265).
However, using the gauge transformations properties of the holonomy, one can simply build directly the general gauge invariant cylindrical functions over a graph $\Gamma$. Those cylindrical functions will corresponds to the solutions of the Gauss quantum operator and provide us with the gauge invariant quantum states.

As we have seen in the beginning of this chapter, the holonomy of the connection along an edge $e$ transforms under $SU(2)$ gauge transformations as:
\begin{align*}
g \triangleright h_{e} =  g_{s(e)} \; h_{e} \; g^{-1}_{t(e)}
\end{align*}
where $s(e)$ and $t(e)$ are respectively the source and the target points of the edge $e$.
Therefore, under a gauge transformation, a cylindrical function over a graph $\Gamma$ with $n$ edges transforms as:
\begin{align*}
g \triangleright \Psi_{\Gamma} ( h_{e_{1}}, ......, h_{e_{n}} ) = \Psi_{\Gamma} ( \; g_{s(e_{1})} \; h_{e_{1}} \; g^{-1}_{s(t_{1})} \; , ......, \; g_{s(e_{n})} h_{e_{n}}g^{-1}_{t(e_{n})} \; ) 
\end{align*}
Asking for gauge invariant cylindrical functions over the graph $\Gamma$ implies that:
\begin{align*}
\Psi_{\Gamma} ( h_{e_{1}}, ......, h_{e_{n}} ) = \Psi_{\Gamma} ( \; g_{s(e_{1})} \; h_{e_{1}} \; g^{-1}_{s(t_{1})} \; , ......, \; g_{s(e_{n})} h_{e_{n}}g^{-1}_{t(e_{n})} \; ) 
\end{align*}

A very interesting solution to the precedent equation is the Wilson loop, which corresponds to the trace of the holonomy of the connection along a closed edge, i.e. for which the source and the target points are identified: $s(e)=t(e)$.
For such a cylindrical function, the gauge transformation reads:
\begin{align*}
g \triangleright \Psi_{\Gamma} (h_{e}) =  \text{Tr} ( \; g_{s(e)}  h_{e} g^{-1}_{s(e)} \;  )=   \text{Tr} (   h_{e}   ) = \Psi_{\Gamma} (h_{e}) 
\end{align*}
This was the first basis of functionals introduced in the mid nineties by Jacobson, Smolin and Rovelli.
However, a more general basis was introduced in 1995 called the spin networks basis. There are of first importance in LQG and encode the gauge invariant quantum geometry resulting from the loop quantization.

We will introduce this basis by using an example, i.e. the cylindrical function over the theta graph $\Gamma_{\theta}$. This graph contains three edges $(e_{1}, e_{2}, e_{3})$ which recouple at two vertices $(v_{1}, v_{2})$.
We note $s_{e_{1}} = s_{e_{2}}= s_{e_{3}} = v_{1}$ and $t_{e_{1}} = t_{e_{2}}= t_{e_{3}} = v_{2}$. Using the Peter-Weyl theorem, a cylindrical function $\Psi_{\Gamma_{\theta}} \in \text{Cyl}^{0}(\Gamma_{\theta})$ can be expressed as:
\begin{align*}
\Psi_{\Gamma}( h_{e_{1}}, h_{e_{2}}, h_{e_{n}}) = \sum_{j_{i}, m_{i}, n_{i}}  \; (\tilde{f}^{j_{1}j_{2}j_{3}})^{n_{1} n_{2} n_{3}}{}_{m_{1} m_{2} m_{3}} (D^{j_{1}}(h_{e_{1}}))^{m_{1}}{}_{n_{1}}(D^{j_{2}}(h_{e_{2}}))^{m_{2}}{}_{n_{2}} (D^{j_{3}}(h_{e_{3}}))^{m_{3}}{}_{n_{3}} 
\end{align*}
Let us apply a gauge transformation on this cylindrical function and see how we have to modify its structure to obtain the gauge invariance. It reads:
\begin{align*}
& g \triangleright \Psi_{\Gamma}( h_{e_{1}}, h_{e_{2}}, h_{e_{3}})  =  \Psi_{\Gamma}( g_{v_{1}}h_{e_{1}} g^{-1}_{v_{2}}, g_{v_{1}} h_{e_{2}} g^{-1}_{v_{2}} , g_{v_{1}} h_{e_{3}} g^{-1}_{v_{2}} )  \\
& = \sum_{j_{i}, m_{i}, n_{i}}  \; (\tilde{f}^{j_{1}j_{2}j_{3}})^{n_{1} n_{2} n_{3}}{}_{m_{1} m_{2} m_{3}} (D^{j_{1}}(g_{v_{1}}h_{e_{1}} g^{-1}_{v_{2}}))^{m_{1}}{}_{n_{1}}(D^{j_{2}}(g_{v_{1}} h_{e_{2}} g^{-1}_{v_{2}}))^{m_{2}}{}_{n_{2}} (D^{j_{3}}(g_{v_{1}} h_{e_{3}} g^{-1}_{v_{2}}))^{m_{3}}{}_{n_{3}} \\
& = \sum_{j_{i}, m_{i}, n_{i}}  \; (\tilde{f}^{j_{1}j_{2}j_{3}})^{n_{1} n_{2} n_{3}}{}_{m_{1} m_{2} m_{3}} (D^{j_{1}}(g_{v_{1}}))^{m_{1}}{}_{\alpha_{1}}(D^{j_{2}}(g_{v_{1}}))^{m_{2}}{}_{\alpha_{2}} (D^{j_{3}}(g_{v_{1}}))^{m_{3}}{}_{\alpha_{3}} \\[-4.5mm]
& \;\;\;\;\;\;\;\;\;\;\;\;\;\;\;\;\;\;\;\;\;\;\;\;\;\;\;\;\;\;\;\;\;\;\;\;\;\;\;\;\;\;\;\;\;\;\;\;\; (D^{j_{1}}(g^{-1}_{v_{2}}))^{\beta_{1}}{}_{n_{1}}(D^{j_{2}}(g^{-1}_{v_{2}}))^{\beta_{2}}{}_{n_{2}} (D^{j_{3}}(g^{-1}_{v_{2}}))^{\beta_{3}}{}_{n_{3}} \\
& \;\;\;\;\;\;\;\;\;\;\;\;\;\;\;\;\;\;\;\;\;\;\;\;\;\;\;\;\;\;\;\;\;\;\;\;\;\;\;\;\;\;\;\;\;\;\;\;\;  (D^{j_{1}}(h_{e_{2}}))^{\alpha_{1}}{}_{\beta_{1}}(D^{j_{2}}(h_{e_{2}}))^{\alpha_{2}}{}_{\beta_{2}} (D^{j_{3}}(h_{e_{3}}))^{\alpha_{3}}{}_{\beta_{3}} \\
\end{align*}
This object is obviously not invariant under $SU(2)$ gauge transformation. One way to obtain a gauge invariant object which does not depend on the gauge transformation at the vertices, i.e. on $g_{v_{1}}$ and $g^{-1}_{v_{2}}$, is to integrate with the invariant Haar measure over $g_{v_{1}}$ and $g^{-1}_{v_{2}}$. Introducing those integrals, we obtain a new object which takes the form:

\begin{align*}
 & \Psi^{inv}_{\Gamma_{\theta}}( h_{e_{1}}, h_{e_{2}}, h_{e_{3}})  =  \int_{SU(2)^{2}} dg_{v_{1}} dg^{-1}_{v_{2}} \Psi_{\Gamma}( g_{v_{1}}h_{e_{1}} g^{-1}_{v_{2}}, g_{v_{1}} h_{e_{2}} g^{-1}_{v_{2}} , g_{v_{1}} h_{e_{3}} g^{-1}_{v_{2}} )  \\
& = \sum_{j_{i}, m_{i}, n_{i}}  \; (\tilde{f}^{j_{1}j_{2}j_{3}})^{n_{1} n_{2} n_{3}}{}_{m_{1} m_{2} m_{3}} \int_{SU(2)} dg_{v_{1}} (D^{j_{1}}(g_{v_{1}}))^{m_{1}}{}_{\alpha_{1}}(D^{j_{2}}(g_{v_{1}}))^{m_{2}}{}_{\alpha_{2}} (D^{j_{3}}(g_{v_{1}}))^{m_{3}}{}_{\alpha_{3}} \\[-4.5mm]
& \;\;\;\;\;\;\;\;\;\;\;\;\;\;\;\;\;\;\;\;\;\;\;\;\;\;\;\;\;\;\;\;\;\;\;\;\;\;\;\;\;\;\;\;\;\;\;\;\; \int_{SU(2)} dg^{-1}_{v_{2}}  (D^{j_{1}}(g^{-1}_{v_{2}}))^{\beta_{1}}{}_{n_{1}}(D^{j_{2}}(g^{-1}_{v_{2}}))^{\beta_{2}}{}_{n_{2}} (D^{j_{3}}(g^{-1}_{v_{2}}))^{\beta_{3}}{}_{n_{3}} \\
& \;\;\;\;\;\;\;\;\;\;\;\;\;\;\;\;\;\;\;\;\;\;\;\;\;\;\;\;\;\;\;\;\;\;\;\;\;\;\;\;\;\;\;\;\;\;\;\;\;  (D^{j_{1}}(h_{e_{1}}))^{\alpha_{1}}{}_{\beta_{1}}(D^{j_{2}}(h_{e_{2}}))^{\alpha_{2}}{}_{\beta_{2}} (D^{j_{3}}(h_{e_{3}}))^{\alpha_{3}}{}_{\beta_{3}} \\
& =  \sum_{j_{i}, m_{i}, n_{i}}  \; (\tilde{f}^{j_{1}j_{2}j_{3}})^{n_{1} n_{2} n_{3}}{}_{m_{1} m_{2} m_{3}} \; P^{m_{1}m_{2}m_{3}}{}_{\alpha_{1}\alpha_{2} \alpha_{3}} \; P^{\beta_{1}\beta_{2}\beta_{3}}{}_{n_{1}n_{2}n_{3}}  (D^{j_{1}}(h_{e_{1}}))^{\alpha_{1}}{}_{\beta_{1}}(D^{j_{2}}(h_{e_{2}}))^{\alpha_{2}}{}_{\beta_{2}} (D^{j_{3}}(h_{e_{3}}))^{\alpha_{3}}{}_{\beta_{3}} \\
\end{align*}

where we have introduced the tensorial projector over the invariant tensor product of $SU(2)$- irreducible representations:
\begin{align*}
P^{m_{1}m_{2}m_{3}}{}_{\alpha_{1}\alpha_{2} \alpha_{3}} & = \int_{SU(2)} dg_{v_{1}} (D^{j_{1}}(g_{v_{1}}))^{m_{1}}{}_{\alpha_{1}}(D^{j_{2}}(g_{v_{1}}))^{m_{2}}{}_{\alpha_{2}} (D^{j_{3}}(g_{v_{1}}))^{m_{3}}{}_{\alpha_{3}} \\
& = \iota^{m_{1}m_{2}m_{3}} \iota_{\alpha_{1}\alpha_{2} \alpha_{3}} \\
& = \begin{pmatrix}
 m_{1} & m_{2} & m_{3} \\
 j_{1} & j_{2} & j_{3}
\end{pmatrix} 
\begin{pmatrix}
 j_{1} & j_{2} & j_{3}  \\
  \alpha_{1} & \alpha_{2} & \alpha_{3}
 \end{pmatrix} 
\end{align*}

Here, we have introduced the $SU(2)$ invariant tensor $\iota$. Its components are given by the Clebsh-Gordan coefficient also called the $3j$ symbol.
Therefore, the projector $P$ is given by the successive action of two intertwiners $\iota$. Considering two representations and their associated vector spaces $V^{j_{1}}$ and $V^{j_{2}}$, the intertwiners maps the tensorial product of  $V^{j_{1}} \otimes V^{j_{2}}$ into the vector space $V^{j}$, i.e. they provides the usual recoupling of two representations into a another one. It reads:
\begin{align*}
\iota : V^{j_{1}} \otimes V^{j_{2}} \rightarrow V^{j} \;\;\;\;\;\;\;\;\;\;\;\;\;\; \text{and} \;\;\;\;\;\;\;\;\;\;\;\;\;\; \iota = \begin{pmatrix}
 j_{1} & j_{2} & j  \\
 m_{1} & m_{2} & m_{3} 
 \end{pmatrix} \bra{j_{1}, m_{1}} \otimes \bra{j_{2}, m_{2}} \otimes \ket{j,m}
\end{align*}
The intertwiner is an element of the invariant tensor product:
\begin{align*}
[ V^{j_{1}} \otimes V^{j_{2}} \otimes V^{j_{3}} ]_{inv}
\end{align*}
From the recoupling theory of angular momentums, we know that tensor product of the two kets $\ket{j_{1}, m_{1}} \in V^{j_{1}}$ and $\ket{j_{2}, m_{2}} \in V^{j_{2}}$ can be decompose as:
\begin{align*}
\bra{j_{1}, j_{2}, m_{1}, m_{2}} \ket{j,m} =  \begin{pmatrix}
 j_{1} & j_{2} & j  \\
 m_{1} & m_{2} & m_{3}
 \end{pmatrix} 
\end{align*}
One recovers the usual notation for the Clebsh-Gordan coefficient in term of the $3j$ symbol.

Now let us come back to our new definition for the gauge invariant cylindrical function $ \Psi^{inv}_{\Gamma}( h_{e_{1}}, h_{e_{2}}, h_{e_{3}}) $.
In order to have a gauge invariant functional, we have introduced a new object, i.e. the projector $P$, which is built up from two intertwiners $\iota$.
Those intertwiners are associated to the vertices of the graph $\Gamma_{\theta}$ and ensure the gauge invariance of the cylindrical function.
Indeed, applying a gauge transformation to the modified cylindrical function do not change its arguments since they group elements of the gauge transformations can always be reabsorbed in the two projectors P, while the Haar measure stays invariant under the right and left group multiplication.

We can now rewrite this functional over the graph $\Gamma_{\theta}$ as:
\begin{align*}
  \Psi^{inv}_{\Gamma_{\theta}}( h_{e_{1}}, h_{e_{2}}, h_{e_{3}})  
& =  \sum_{j_{e}} \sum_{m_{i}, n_{i}}  \; (\tilde{f}^{j_{1}j_{2}j_{3}})^{n_{1} n_{2} n_{3}}_{m_{1} m_{2} m_{3}} \; \iota^{m_{1}m_{2}m_{3}} \iota_{\alpha_{1}\alpha_{2} \alpha_{3}} \; \iota^{\beta_{1}\beta_{2}\beta_{3}}\iota_{n_{1}n_{2}n_{3}}  \prod^{3}_{i= 1}(D^{j_{i}}(h_{e_{i}}))^{\alpha_{i}}{}_{\beta_{i}}\\
& = \sum_{j_{e}} \sum_{m_{i}, n_{i}}  \; (\tilde{f}^{j_{1}j_{2}j_{3}})^{n_{1} n_{2} n_{3}}_{m_{1} m_{2} m_{3}} \; \iota^{m_{1}m_{2}m_{3}} \iota_{n_{1}n_{2}n_{3}}  \prod_{v_{i}} \iota_{i} \; \cdot  \prod_{e_{k}} D^{j_{k}}(h_{e_{k}}) \;  \\
\end{align*}
where the $\cdot$ denotes the contraction of the indices between the intertwiners at the vertices and the representation associated to the edges.
From this expression, a basis for the gauge invariant cylindrical functionals emerges. To each edges, one can associate a given representation of the holonomy $h_{e}$, and to obtain a gauge invariant quantity, one associates to each vertex an intertwiner which couples the ingoing representations to the outgoing ones.
This construction generalizes to any graph $\Gamma$ with a finite number $k$ of edges $e$ and $i$ of vertices $v$ and reads:
\begin{align*}
S (\Gamma, \vec{j}, \iota ) (h_{e_{k}})=  \prod_{v_{i}} \iota_{i} \; \cdot  \prod_{e_{k}} D^{j_{k}}(h_{e_{k}}) 
\end{align*}
which can also be written more generally:
\begin{align*}
S (\Gamma, \vec{j}, \iota ) [h_{e}] =  \bigotimes_{v \in \Gamma} \iota_{i} \; \cdot  \bigotimes_{e \in \Gamma} D^{j_{k}}(h_{e}) 
\end{align*}
This object is called a spin network and corresponds to the basis of $SU(2)$ gauge invariant cylindrical functionals over a graph $\Gamma$. They were introduced in \cite{ch2-Smolin2} by Rovelli and Smolin, and even before by Penrose (1964).
Having integrated out the gauge freedom at the $v$ vertices, the space of cylindrical functionals solutions to the quantum Gauss constraint can be equipped with Ashtekar-Lewandowski measure. This is the gauge invariant Hilbert space associated to the graph $\Gamma$, which we can denote:
\begin{align*}
\mathcal{H}^{G}_{\Gamma} = L^{2} [ SU(2)^{e} / SU(2)^{v} , dh] = \bigoplus_{j_{e}}\bigotimes_{v \in \Gamma} [\mathcal{H}_{v}]_{inv}
\end{align*}
where we have introduce the gauge invariant Hilbert space associated to each vertex which corresponds to:
\begin{align*}
\mathcal{H}_{v} = \big{(} \bigotimes_{e| s(e) = v}\mathcal{H}^{j_{e}}_{SU(2)} \big{)} \otimes \big{(} \bigotimes_{e| t(e) = v}\mathcal{H}^{j_{e}}_{SU(2)}  \big{)}^{\ast} 
\end{align*}
To understand the introduction of this vertex Hilbert space, let us describe more precisely the structure of an edge $e$ and a vertex $v$.
Consider an edge $e$ equipped with a representation given by the Wigner matrice $D^{j_{e}}(h_{e})^{m}{}_{n}$. This Wigner matrice is a linear map from the Hilbert space $\mathcal{H}^{j_{e}}_{SU(2)}$ to its dual $(\mathcal{H}^{j_{e}}_{SU(2)})^{\ast}$, i.e. :
\begin{align*}
D^{j_{e}}(h_{e}) : \mathcal{H}^{j_{e}}_{SU(2)} \rightarrow (\mathcal{H}^{j_{e}}_{SU(2)})^{\ast}
\end{align*}
Therefore, graphically, the source point $s(e)$ of the edge $e$ carries the Hilbert space $\mathcal{H}^{j_{e}}_{SU(2)}$ while the target point $t(e)$ carries its dual $(\mathcal{H}^{j_{e}}_{SU(2)})^{\ast}$.
The edge $e$ with its source and target points represents the linear map between those two Hilbert spaces.
Consider now a $n$ valent vertex $v$ containing $p$ ingoing edges and $q = n -p$  outgoing edges. Each $q$ source points $s(e) = v$ associated to the $p$ outgoing edges carry therefore an Hilbert space $\mathcal{H}^{j_{e}}_{SU(2)}$ while each $p$ target points $t(e) = v$  associated to the $p$ ingoing edges carry a dual Hilbert space $(\mathcal{H}^{j_{e}}_{SU(2)})^{\ast}$. The Hibert space associated to the vertex $\mathcal{H}_{v} $ is therefore given by the tensorial product of all those Hilbert spaces associated to the source points $s(e)$ and dual Hilbert spaces associated to the target points $t(e)$. 

For the moment, we have considered only the Hilbert space $\mathcal{H}^{G}_{\Gamma} $ associated to the graph $\Gamma$. As previously, one obtain the full gauge invariant Hilbert space of LQG by summing over all the graphs over $\Sigma$:
\begin{align*}
\mathcal{H}^{G} = \bigoplus_{\Gamma \in \Sigma} \mathcal{H}^{G}_{\Gamma} 
\end{align*}

The spin networks represents the basis of the cylindrical functions living in this Hilbert space and we will denote a general spin network quantum state \ket{S}. They are build over graphs which are still embedded in the three dimensional manifold $\Sigma$, i.e. a vertex remains located at a given point of the manifold. Therefore, they are not diffeomorphism invariant. Our next task is to apply the spatial diffeomorphism quantum constraint , i.e. the vectorial quantum constraint, in order to obtain our gauge invariant kinematical Hilbert space.

\subsection{The spatial diffeomorphsim Hilbert space : knot as quantum states}

We present now the construction of the spatially diffeomorphism invariant quantum states by following the pedagogical review \cite{ch2-Hanno1}. The implementation of the spatial diffeomorphism symmetry is one of the key step in order to implement the background independence at the quantum level.
The vectorial constraint inducing the diffeomorphisms on the three dimensional spacelike hypersurface $\Sigma$ is given by:
\begin{align*}
H_{a}(N^{a}) = \int_{\Sigma} N^{a} E^{b}_{i} F^{i}_{ab} \simeq 0
\end{align*}
where $N^{a}$ is the shift vector lying in $\Sigma$.
This constraint transforms the connection and the electric field as follow:
\begin{align*}
\{H_{a}(N^{a}) , A_{b} \} = \mathcal{L}_{N^{a}} A_{b} \;\;\;\;\;\;\; \{H_{a}(N^{a}) , E^{b} \} = \mathcal{L}_{N^{a}} E^{b}
\end{align*}
This constraint generates the Lie derivative of the connection one form and of the electric vector field along the direction of the shift vector $N^{a}$.
Once again, the naive idea in order to implement the Dirac quantization program would be to promote this constraint into operator expressed in term of the fundamental holonomy and flux operators.
Then one would look for quantum states $\ket {S} \in \mathcal{H}^{G}$ which satisfy:
\begin{align*}
\hat{H}_{a} \ket {S} = 0
\end{align*}
Therefore, the task is to look for the eigenvectors associated to the eigenvalues zero.
However, this equation is not well defined since the solutions do not span a subspace of  $\mathcal{H}^{G}$. 
This is due to the non compactness of the diffeomorphism group $\text{Diff}(\Sigma)$ and the undoundness of the operator $\hat{H}_{a} $. First, an unbounded operator will be well defined only on a dense subspace of $\mathcal{H}^{G}$. Secondly, for a non compact operator, the spectrum can be both discrete and continuous on different ranges.
If the eigenvalue zero is in the continuous spectrum, one has to extend the Hilbert space in order to include distributional quantum states. The procedure which consists in selecting a dense subspace of an intermediate Hilbert space and then extend the solutions space to include distributional quantum states is known as the raffined algebraic quantization. An interesting review can be found in \cite{ch2-Ga1}.

For a given theory which is quantized \textit{a la Dirac}, the general idea is to introduce an intermediate or auxiliary Hilbert space $H_{aux}$. The first step corresponds to choose the largest dense subspace $\mathcal{D}$ in it. This space will be the home for the unbounded operators.
 The second step is to introduce the algebraic dual of the dense subspace $\mathcal{D}$, which is usually denoted $\mathcal{D}^{\ast}$.
This space contains the linear distributional functionals $f$ which act on $\mathcal{D}$ as :
\begin{align*}
& f : \mathcal{D} \rightarrow \;\;\; \mathbb{C} \\
& \;\;\;\;\;\;  \phi  \rightarrow f(\phi)
\end{align*}
Then one imposes the constraint operator in this enlarged space $\mathcal{D}^{\ast}$ and selects the subspace $H_{phy} \in \mathcal{D}^{\ast}$ corresponding to the solutions of the quantum constraint.
The two dual spaces $\mathcal{D}$  and $\mathcal{D}^{\ast}$ are related by the so called ``ringed'' map $\eta$, i.e.:
\begin{align*}
& \eta : \mathcal{D} \rightarrow \mathcal{D}^{\ast} \\
& \;\;\;\;\; \phi \rightarrow \eta(\phi) = f
\end{align*}
This map provides a natural tool to build an inner product on $\mathcal{D}^{\ast}$ and therefore on $H_{phy}$.
From the topological point of view, one obtains the following inclusion:
\begin{align*}
\mathcal{D} \subset H_{aux} \subset \mathcal{D}^{\ast}
\end{align*}
This structure is usually called the ``Gelfand triple''. 

We follow this strategy to implement the spatial diffeomorphism constraint in LQG. Only the infinitesimal diffeomorphism are taken into account, i.e. the one connected to the identity.

The auxiliary Hilbert space which represents the starting point of the procedure corresponds to the gauge invariant Hilbert space $\mathcal{H}^{G}$.
Its largest dense subspace is given by the space of cylindrical functionals built from finite linear combinations of spin networks that we denote $S = (\text{Cyl}_{G}, d\mu_{AL})$. Following the refined algebraic quantization procedure, the solutions to the vectorial constraint have to be selected in the dual of this space, denoted $S^{\ast}$, which contains gauge invariant linear cylindrical functional not necessarily continuous.

Having identified our spaces of interest, we would like to implement the spatial diffeomorphism symmetry constraint.
 At this point, the fact that the Ashtekar-Lewandowski measure is diffeomorphism invariant is crucial.
This property of the measure ensures us that the Hilbert space $\mathcal{H}^{G}$ carries therefore a unitary action of the diffeomorphism group as well as its largest dense subspace $S$. This measure and therefore this property can be extended to its dual $S^{\ast}$. 
Consequently, we do not need to find a regularized expression for the vectorial constraint and act on  $S^{\ast}$ to select the diff-invariant quantum states. Instead, we can work directly with the self adjoint diffeomorphism operator acting on cylindrical functions $\Psi \in S$ as:
\begin{align}
U_{\phi} \Psi = \Psi'' 
\end{align}
This operator allow us to introduce a map called $P_{Diff} : S \rightarrow S^{\ast}$ which can be build as:
\begin{align*}
(P_{Diff} \Psi )(\Psi') = \sum_{\Psi'' = U_{\phi} \Psi} < \;  \Psi'', \Psi' \; >
\end{align*}
where $\Psi'' = U_{\phi} \Psi$ corresponds to all the distinct states generated by a diffeomorphism $\phi$ for some $\phi \in \text{Diff}(\Sigma)$.
This sum turns out to be well defined. The map $P_{Diff} $ is the projector on the diff-invariant quantum states since a diff-invariant quantum state $\Psi= U_{\phi} \Psi$ is projected onto itself
by this map, i.e:
\begin{align*}
P_{Diff} \Psi  = P_{Diff} (U_{\phi} \Psi)
\end{align*}

Those diff-invariant quantum states spanned therefore the Hilbert space $H_{Diff} \in S^{\ast}$.
The inner product on this diff-invariant Hilbert space is given by:
\begin{align*}
< \; P_{Diff} \Psi, P_{Diff} \Psi'\; > \; = P_{Diff} \Psi (\Psi')  \;\;\;\;\;\;\; \text{where}  \;\;\;\;\;\;\; \Psi, \;\; \Psi' \in S 
\end{align*}

Having define the diff-invariant Hilbert space, let us see how the diffeomorphism acts on a given spin network $S(\Gamma, \vec{j}, \iota)$. 
As we have seen. under a diffeomorphism, the holonomy of the connection along an edge $e$ is affected only through a shift of the edge and the source point $s(e)$ and $t(e)$ can be modified.
However, the coloring of the edge, i.e. the spin $j$ representation does not change under the transformation.
Now, considering a graph, a diffeomorphism can act non trivially on the nodes. For a given spin network having some fixed intertwiners at each nodes, a diffeomorphism can modified this intertwiner structure without changing the valence at each nodes. 

The action of a diffeomorphism $\phi \in \text{Diff}(\Sigma)$ on a spin network $S(\Gamma, \vec{j}, \iota)$ reads:
\begin{align*}
U_{\phi} \; S(\Gamma, \vec{j}, \iota) = S(\Gamma' = \phi \circ \Gamma, \vec{j}, \iota')
\end{align*}

One can show that this operator is not weakly continuous, and that even if we are working with infinitesimal diffeomorphism, i.e. connected to the identity, those diffeomorphism act as finite transformations at the level of the gauge invariant cylindrical functions.
 
 At this point, one can define the equivalence class of graphs $[\Gamma]$ under a diffeomorphism $\phi \in \text{Diff}(\Sigma)$ applied to $\Gamma$. They are all the possible distinct graphs that one can obtained by applying a diffeomorphism $\phi$ for some $\phi \in \text{Diff}(\Sigma)$. Outside of this equivalence class, there are some graphs which can not be obtained by applying a diffeomorphism $\phi$ to $\Gamma$ and which are in another equivalence class of another graphs. Those diffeomorphism equivalence class of graph are called knots. Basically, they corresponds to abstract graph which are no more embedded in a ambient space.

Asking for diffeomorphism invariance of the quantum states, one obtains a spin network defined over a knot. Such a diff-invariant quantum state is called and s-knot.
However, this is not the end of the story. For a spin network defined over a knot, there is still some degeneracy. Indeed, for there are a family of transformtions acting one the space of intertwiner over the nodes  which can be obtain by a diffeomorphims on the graph $\Gamma$. Those transformations modified the coloring of the graph while they do not modify the valence of the nodes. Therefore, a diff-invariant quantum state is labelled by its diffeomorphism equivalent class of graph denoted usually $\mathcal{K}_{d}$ and a coloring of its nodes denoted $c$, leading to the notation: $\ket{s} = \ket{\mathcal{K}_{d}, c}$.

 The inner product in $H_{Diff}$ implies that two s-knots are orthogonal if their knot is different, i.e. if they are defined over two graphs which do not belong to the same diffeomorphism equivalence class.
Finally, an important property of the diff-invariant Hilbert space is that it does not admit a countable basis, i.e. it is not separable. The reason is that the knots are labelled both by a discrete numbder and by a continuous one called the moduli, leading to an uncountable basis. However, the non separability of the Hilbert space $H_{Diff}$ can be cure by slightly modifying the precedent construction to obtain a separable Hilbert space, allowing the cylindrical functions to be not differentiable at a finite number of isolated points \cite{ch2-Rovelli}.

Those diiff-invariant states represent the non dynamical physical quantum states of the quantum gravitational field.
The resulting picture is an abstract graph colored by group data. This graph is no more embedded in an ambient space and therefore, it is not localized in a fixed background. This background has been removed by the spatial diffeomorphism symmetry. The localization can only be relational, i.e. with respect to some other dynamical field. this is the spirit of the background independence of General Relativity. The quantum geometry is fully encoded in the combinatorial structure of the graph, i.e. on the recoupling at each nodes of the group elements associated to the edges of the graph. This graph represents the quantum space which interacts with the matter quantum fields through the hamiltonian constraint. 

Having briefly describe the construction of the diffeomorphim invariant quantum states of the quantum gravitational field, we will now discuss the implementation of the scalar constraint.

\subsection{Implementing the Hamiltonian constraint : towards the physical Hilbert space}

Up to now, the implementation of the hamiltonian constraint in LQG remain an hard task and has generated a lot of work both from the canonical and the covariant approach of LQG. This is not surprising since solving this final constraint would imply to fully solve the quantum Einstein equations.
At the classical level, it would mean to have a general solution of the highly non linear Einstein's field equations, which is obviously not possible.

As we have seen, the reformulation of General Relativity in the complex Ashtekar's formalism has brought a new hope concerning the implementation of the hamiltonian constraint at the quantum level.
The complicated ADM scalar constraint was turned into a harmless polynomial constraint. However, due to the highly non linear reality conditions, this formalism was given up for the real Ashtekar-Barbero one.
With those variables, the polynomial hamiltonian constraint inherited an complicated additional term:
\begin{align*}
H(N) =  \frac{N}{2\gamma^{2}} \frac{E^{b} \times E^{c}}{\sqrt{\text{det}E}} . \{ \; \mathcal{F}_{bc}(A) + ( 1 + \gamma^{2}) R_{bc}(\omega (E)) \; \} \simeq 0
\end{align*}

The first difficulty to implement this constraint at the quantum level is the second term which take a non linear complicated form since it depends on the Levi Civita connection expressed in term of the electric field. Its is therefore an important obstacle to define a quantum operator from the classical expression. (It turns out that this term is not present in the self dual theory, i.e. when $\gamma = \pm i$). Moreover, the factor $1/\sqrt{\text{det}q}$ has to be treated with due care. Therefore, from the canonical point of view, one has to first find a suitable regularization of this constraint and then apply it on the diff-invariant quantum states living in $H_{Diff}$.  The first proposal for such a quantum operator was introduced by Thiemann in 1996 \cite{ch2-HCTh1, ch2-HCTh2, ch2-HCTh3}. While very difficult to extract concrete physical states, it nevertheless brought a precise mechanism of the background independent dynamic generated by the hamiltonain constraint of General Relativity, which is already an impressive progress. 

Basically, the hamiltonian constraint acts as a annihilation or creation operator which generates or suppress quanta of volume.
From the point of view of a given quantum state, represented by a spin network associated to a knot, the application of the hamiltonian constraint enlarge or suppress the initial set of nodes.
This gives rise to a new quantum states, represented by a spin network with a different combinatorial structure. Yet, while very appealing, this picture of the background independent dynamics inherited from LQG is not robust since the full concrete implementation of the hamiltonian constraint is still far from being controlled.

A second difficulty, is that the algebra of the constraints is not a true Lie algebra.
This point motivates the so called Master constraint program \cite{ch2-MCTh1, ch2-MCTh2, ch2-MCTh3}. The strategy was to define a new single constraint which is the sum of the square of each of the three constraints, i.e. the Gauss, vectorial and scalar constraints, i.e:
\begin{align*}
M = \int_{\Sigma} dx^{3} \frac{[H]^{2} + q^{ab} H_{a} H_{b} + \delta^{ij} G_{i}G_{j}}{\sqrt{\text{det}q}}
\end{align*}
This trick allows one to work with a very simplified algebra and thus applied the refined algebraic quantization technics. 
\begin{align*}
& \{ H_{a}(N^{a}) , H_{b}(\tilde{N}^{b}) \} = - \kappa H_{b}( \mathcal{L}_{N^{a}} \tilde{N}^{b}) \\
& \{ H_{a}(N^{a}) , M \} = 0\\
& \{ M , M \} = 0 \\
\end{align*}
An important result obtained from this program is a proof of the \textit{existence} of the physical Hilbert space of LQG \cite{ch2-MCTh4}.
However, this program remains formal up to now and concrete physical predictions of the quantum theory (i.e., with a full implementation of the dynamics) are still out of reach. 

One could mention a third difficulty concerning the scalar constraint. When applied to the real Ashtekar-Barbero connection, the hamiltonian constraint does not generate its Lie derivative along the orthogonal direction to the three dimensional hyper surfaces $\Sigma$. Instead it lead to a complicated transformation for the connection one form, which reduce to the expected Lie derivative only for $\gamma = \pm i$. In this formulation, due to the peculiar choice of the time gauge in the hamiltonian analysis, the one form connection does not transform as expected under ``time'' diffeomorphisms \cite{ch2-Alex1, ch2-Samuel, ch2-Liv}. This is potentially problematic. It implies that the Ashtekar-Barbero is probably not well suited to implement the dynamic in LQG and that one should maybe come back the self dual variable in order to deal with the dynamic.
A general strategy to circumvent this problem and work with the self dual variables will be presented and discuss in chapter $4$ , $5$ and $6$. This is the subject of this PhD.

The difficulty encountered on the canonical side to define a quantum operator for the hamiltonian constraint led to the development of the spin foam models at the end of the nineties.
Those models, inspired from topological quantum field theory popularized by Atyiah, Ooguri and Witten, were refined during the last fifteen years. They gave rise to the Barett-Crane model \cite{ch2-BC} and then culminated with the introduction of  the $FK$ \cite{ch2-FK} and the $EPRL$ \cite{ch2-EPRL} spin foam models. Those models are a concrete realization of the sum over histories models which use the techniques and ideas of LQG. They represent the path integral or covariant version of LQG. However, both of those models suffers form difficulties concerning the imposition of secondary class constraints and they do not represent a final answer to the implementation of the dynamics in LQG. See \cite{ch2-Alejandro1} for a pedagogical review.

Recently, new investigations on the canonical side led to interesting results. A regularization of the second term appearing in the hamiltonian constraint was proposed, and a general regularization of the hamiltonian constraint was introduced by mixing this new curvature operator and the old proposition of Thiemann \cite{ch2-Jerzy3, ch2-Jerzy4}. 

This conclude our presentation of the implementation of the quantum constraint in LQG.
At this point, the $SU(2)$ Gauss constraint and the spatial diffeomorphism constraints have been successfully imposed, leading to a kinematical quantum theory of the gravitational field which implement the requirement of background independence. Even if the dynamics is not well controlled and poorly understood, very powerful proposals exist to deal with this final step of the quantization program.
In this PhD, we will be interested in one of this proposal, which relies on the particular status of the self dual variables. We will discuss this strategy in chapter four.

We will now present the main results of kinematical LQG which is the quantization of are and volume of some region of space.

\section{The geometric operators : area and volume operators}

As in any quantum theory, one is interested in computing the spectrum of some quantum operator representing a physical quantity.
Not surprinsigly, in Loop Quantum Gravity, those operators turns out to corresponds to the geometrical quantities such as the area or the volume of a region, but also to angles.
The quantum spectrum of the area and volume operator was first computed in 1996 by Rovelli and Smolin \cite{ch2-Rove4}. The striking result is that one obtains a discrete spectrum which forbids the value zero, i.e. the area and the volume are quantized and possess an minimal gap \cite{ch2-Ashte11}. This property of the quantum geometry at the Planck scale is at the heart of the different results obtained in black hole thermodynamics and quantum cosmology. The existence of the area gap cures the well known singularities of classical General Relativity. It lead to a the resolution of the singularity in the Big Bang scenario and of the singularity at the heard of the spherically symmetric space-time with some mass at the center (such as the Schwrazschild black hole). Moreover, it explains the large but finite value of the Bekenstein Hawking entropy for the black hole.
This is therefore a crucial result that we will now reproduce.

Let us consider the classical area of a given region of space. While the space is locally coordinated by the coordinates $\{x^{a}\}$ with $a \in \{ 1,2, 3\}$, we denote the coordinates of the two dimensional surface $S$ by $\{y^{\alpha}\}$ with $\alpha \in \{ 1,2\}$. The two dimensional metric is denoted $h_{\alpha\beta}$. The area of the surface is given by:
\begin{align*}
A_{r}(S) = \int_{S} dy^{2} \; \sqrt{\text{det}h} = \int_{S} dy^{2} \sqrt{\text{det} (q_{ab} \frac{\partial x^{a}}{\partial y^{\alpha}}\frac{\partial x^{b}}{\partial y^{\beta}})}
\end{align*}

One can directly compute the determinant which reads:
\begin{align*}
\; \text{det} (q_{ab} \frac{\partial x^{a}}{\partial y^{\alpha}}\frac{\partial x^{b}}{\partial y^{\beta}}) & = q_{ab}  \frac{\partial x^{a}}{\partial y^{1}}\frac{\partial x^{b}}{\partial y^{1}} q_{cd} \frac{\partial x^{c}}{\partial y^{2}}\frac{\partial x^{d}}{\partial y^{2}} - q_{ab}  \frac{\partial x^{a}}{\partial y^{1}}\frac{\partial x^{b}}{\partial y^{2}} q_{cd} \frac{\partial x^{c}}{\partial y^{1}}\frac{\partial x^{d}}{\partial y^{2}} \\
& = 2 q_{ab} q_{cd}  \frac{\partial x^{a}}{\partial y^{1}}\frac{\partial x^{c}}{\partial y^{2}} \frac{\partial x^{[b}}{\partial y^{1}}\frac{\partial x^{d]}}{\partial y^{2}} \\
& = 2 q_{a[b} q_{cd]}  \frac{\partial x^{a}}{\partial y^{1}}\frac{\partial x^{c}}{\partial y^{2}} \frac{\partial x^{b}}{\partial y^{1}}\frac{\partial x^{d}}{\partial y^{2}} \\
\end{align*}
Now using the fact that: 
\begin{align*}
\text{det}(q) = \frac{1}{3!} \epsilon^{abc} \epsilon^{bef} q_{ab}q_{ce} q_{df} \;\;\;\;\; \text{and therefore} \;\;\;\; \text{det}(q) q^{ab} = \frac{1}{2} \epsilon^{acd} \epsilon^{bef} q_{ce} q_{df} =  \frac{1}{2} \epsilon^{acd} \epsilon^{bef} q_{c[e} q_{df]}  
\end{align*}
we obtain:
\begin{align*}
2 q_{m[p} q_{np]} & =  ( \delta^{e}_{p}\delta^{f}_{q} - \delta^{e}_{q} \delta^{f}_{p}) q_{m[e} q_{nf]} = ( \delta^{c}_{m}\delta^{d}_{n} - \delta^{c}_{n} \delta^{d}_{m}) ( \delta^{e}_{p}\delta^{f}_{q} - \delta^{e}_{q} \delta^{f}_{p}) q_{ce} q_{df} \\
& = \epsilon_{amn} \; \epsilon^{acd} \; \epsilon_{bpq} \; \epsilon^{bef} \; q_{ce} q_{df} =\frac{1}{2} \; \epsilon_{amn} \; \epsilon_{bpq} \; \text{det}(q) q^{ab}
\end{align*}

Therefore, the precedent determinant takes the form:
\begin{align*}
\; \text{det} (q_{ab} \frac{\partial x^{a}}{\partial y^{\alpha}}\frac{\partial x^{b}}{\partial y^{\beta}}) & =  2 q_{a[b} q_{cd]}  \frac{\partial x^{a}}{\partial y^{1}}\frac{\partial x^{c}}{\partial y^{2}} \frac{\partial x^{b}}{\partial y^{1}}\frac{\partial x^{d}}{\partial y^{2}}  = \; \epsilon_{fac} \; \epsilon_{ebd} \; \text{det}(q) q^{fe} \frac{\partial x^{a}}{\partial y^{1}}\frac{\partial x^{c}}{\partial y^{2}} \frac{\partial x^{b}}{\partial y^{1}}\frac{\partial x^{d}}{\partial y^{2}} \\
& =  \text{det}(q) q^{fe} n_{f} \; n_{e}
\end{align*}

where we have introduced the normal vector $n_{a}$ to the surface $S$:
\begin{align*}
n_{a} = \epsilon_{abc} \frac{\partial x^{b}}{\partial y^{1}}\frac{\partial x^{c}}{\partial y^{2}} = (v_{1} \times v_{2})_{a} \;\;\;\;\;\;\; \text{where} \;\;\;\;\;\; v^{b}_{1} =  \frac{\partial x^{b}}{\partial y^{1}}
\end{align*}

Finally, we need to express this determinant with respect to the electric field $E^{a}_{i}$ since at the quantum level, we would like to have an expression for the area operator in term of the flux operators.
Therefore, we use the fact that $E^{a}_{i} = \text{det{(e)}} \; e^{a}_{i}$ and we obtain:
\begin{align*}
A_{r}(S) & = \int_{S} dy^{2} \sqrt{ \text{det}(q) q^{ab} n_{a} \; n_{b} } = \int_{S} dy^{2} \sqrt{( \text{det}(e) )^{2} \; e^{a}_{i} e^{b}_{j} \delta^{ij} \; n_{a} \; n_{b}} \\
& = \int_{S} dy^{2} \sqrt{ \; E^{a}_{i} E^{b}_{j} \delta^{ij}  \; n_{a} \; n_{b}} \\
\end{align*}

In order to quantize this classical quantity, we first need to regularize the surface $S$. To do so, we decompose it into $N$ cells $c_{i}$, and we assume that each cell has a surface equal to $\sigma^{2}$.
We can then write the classical area such that:
\begin{align*}
A_{\text{reg}}(S) & = \sum^{N}_{n= 1} \int_{c_{i}} dy^{2} \sqrt{ \; E^{a}_{i} E^{b}_{j} \delta^{ij}  \; n_{a} \; n_{b}} \\
& \simeq  \sum^{N}_{n= 1} \sigma^{2} \sqrt{ \; E^{a}_{i} E^{b}_{j} \delta^{ij}  \; n_{a} \; n_{b}} \\
& = \sum^{N}_{n= 1} \sqrt{ \; (\sigma^{2}  E^{a}_{i} n^{a}) ( \sigma^{2}  E^{b}_{j} n_{b}) \delta^{ij} } \\
\end{align*}

The quantity $\sigma^{2}  E^{a}_{i} n^{a}$ is nothing else than the flux of the electric field $E^{a}_{i}$ across the surface of a cell $c$. As we have seen, this flux is one of our fundamental operator (the other one being the holonomy of the connection). The regularized classical area is therefore given by:
\begin{align*}
A_{\text{reg}}(S) = \sum^{N}_{n= 1} \sqrt{ \;X^{i}_{c} (E) X^{j}_{c}(E) \delta_{ij} } \
\end{align*}

The decomposition is choosen such that in the limit $N \rightarrow \infty$, one recovers the classical area $A_{r}(S)$. 
Since we have a well defined regularization of the area of a given surface, expressed in term of the flux of the electric field, we can now turn it into a quantum operator and compute its spectrum.
Let us consider a gauge invariant quantum state of the gravitational field given by a spin network $S(\Gamma, \vec{j}, \iota)$. In order to simplify the computation, it is interesting to choose the decomposition of the surface such that each cell is pierced by an edge of the graph $\Gamma$ once and only once.

The action of the quantum area operator on a spin network is therefore given by:
\begin{align*}
\hat{A}_{r}(S) \; S(\Gamma, \vec{j}, \iota) & = \lim\limits_{N \to \infty}  \sum^{N}_{n = 1} \sqrt{ \;\hat{X}^{i}_{c} (E) \hat{X}^{j}_{c}(E) \delta_{ij} } \; S(\Gamma, \vec{j}, \iota) \\
& =  \lim\limits_{N \to \infty} 8 \pi l^{2}_{p} \gamma \sum^{N}_{n = 1} \sqrt{C_{j_{e}}} \; S(\Gamma, \vec{j}, \iota) \\
& = 8 \pi l^{2}_{p} \gamma  \sum_{e| e \cap S = \emptyset} \sqrt{j_{e}(j_{e}+ 1)} \; S(\Gamma, \vec{j}, \iota) \\
\end{align*}
 where we have denoted $C_{j_{e}}$ the Casimir of $SU(2)$ in the representation $j_{e}$ associated to the edge $e$. Here, the final sum runs over all the edges which intersect the surface $S$.
 Since the spin $j_{e}$ is an half integer, the quantum area turns out to be quantized. This is one of the most important result of LQG. Note the discrete spectrum of the area operator relies heavily on the presence of the compact group $SU(2)$, which can be traced back to the choice of the time gauge during the canonical analysis.
 If one would have work with a non compact group, it is not clear that the area spectrum would remain discrete.
 Moreover, it is important to stress that this prediction of LQG is derived at the gauge invariant level. The area (and volume) operator do not commute with the spatial diffeomorphism and the hamiltonian constraint. Therefore, we do not know if the discrete spectrum will survive once the two constraints generating background independence will be taken into account. However, it has been proven that the area (and volume) spectrum is not affected by the implementation of the spatial diffeomorphism constraint, provided that the corresponding surface and volume are defined intrinsically either by matter field or by some geometrical conditions \cite{ch2-Th1}. 
One example of such a definition of a surface can be found in the context of black hole physics, where the isolated surface horizon is defined by some geometrical conditions without any reference to a coordinate system.
Therefore, in this case, the spectrum of the area derived above becomes a diffeomorphism invariant prediction. However, there is not a similar proof concerning the scalar constraint.
Therefore, we do not know if this discrete spectrum will remain unaffected by the imposition of the background independent dynamic of LQG. It has been argued by some authors that this final step of the quantization could lead to a continuous spectrum. See \cite{ch2-BT1, ch2-Ro1} for an interesting discussion about this point.

The discussion concerning the physical observability of the kinematical predictions is related to the status of the Dirac observables and the so called partial observables, which play a crucial role in the context of a background independent quantum theory \cite{ch2-Rove1, ch2-Rovelli1, ch2-Tam1}.

Finally, let us present briefly the volume operator in LQG.
Consider a closed three dimensional region $M$ of space coordinatized by the local coordinates $\{x^{a}\}$ where $a \in \{1,2,3\}$. The induced metric on such region is denoted $q_{ab}$.
The volume of this classical region reads :
\be
V(M) =  \int_{M} \sqrt{\text{det}(q)} \; dx^{3}
\ee
This determinant can be expressed using the tetrad field as:
\be
\sqrt{\text{det}(q)} = \text{det}(e) = \frac{1}{3!} \epsilon_{ijk} \; \epsilon^{abc} \; e^{i}_{a} e^{j}_{b} e^{k}_{c}
\ee
\footnote{Here, we have used the definition of the determinant of  an $n \times n$ matrix $A$ given by:
\begin{align*}
\text{det} (A) = \epsilon_{a.....b} \; \epsilon^{c......d} \; A^{a}_{c} ......... A^{b}_{d}
\end{align*}}
Now, just as for the area, we want to express the volume in term of the electric field $E^{i}_{a}$. Its expression is given by:
\be
E^{c}_{k} =  \epsilon_{ijk} \; \epsilon^{abc} \; e^{i}_{a} e^{j}_{b} = e \; e^{c}_{k} = \sqrt{\text{det}q} \; e^{c}_{k} \;\;\;\;\;\;\;  e^{i}_{a} e^{j}_{b} =  \epsilon^{ijk} \;\epsilon_{abc} \;  E^{c}_{k}
\ee
It is clear that we have:
\begin{align*}
\text{det}(e)^{2} & = \frac{1}{3!} \frac{1}{3!} \epsilon_{ijk} \; \epsilon^{abc} \; e^{i}_{a} e^{j}_{b} e^{k}_{c} \; \epsilon_{lmn} \; \epsilon^{rst } \; e^{l}_{r} e^{m}_{s} e^{n}_{t} \\
& = \frac{1}{3!} \frac{1}{3!}  E^{c}_{k} \; e^{k}_{c} \; E^{t}_{n} \; e^{n}_{t} = \epsilon^{knq} \; \epsilon_{ctp} \; E^{p}_{q} \; E^{c}_{k} \; E^{t}_{n} \\
& =\frac{1}{3!} \frac{1}{3!}   \epsilon^{ijk} \; \epsilon_{abc} \; E^{a}_{i} \; E^{b}_{j} \; E^{c}_{k}
\end{align*}

Using this result, the volume of the region $M$ becomes:
\begin{align*}
V (M)  & =  \int_{M} \; dx^{3}\; \sqrt{\text{det}(q)} =   \int_{M} \; dx^{3} \; \text{det}(e) =  \int_{M}  \; dx^{3} \;  \sqrt{ \; \frac{1}{3!} \; \epsilon^{ijk} \; \epsilon_{abc} \; E^{a}_{i} \; E^{b}_{j} \; E^{c}_{k}} \\
& =  \int_{M}  \; dx^{3} \; \sqrt{\;\frac{1}{3!} \; \epsilon_{abc} \; E^{a} .(E^{b} \times E^{c})} 
\end{align*}
 
Once again, we obtain an expression for the volume of a given region $M$ which involved only the electric field $E^{a}_{i}$.
In order to define the corresponding quantum operator, one has to regularize this expression. Two types of volumes operators exist in the literature \cite{ch2-Rove4, ch2-Ashte12}. The properties of the volume spectrum were presented and discussed in details in \cite{ch2-J1, ch2-J2}.
It turns out that the volume operator acts only at the nodes of a spin network, and has a non vanishing action only for nodes at least four valent.
This is easily understood by noticing that the simplest elementary volume that one can build is a tetrahedron, which has four faces, and corresponds therefore to a four valent node from the point of view of its dual graph.
 \begin{figure}
\begin{center}
	{\includegraphics[width = 0.3\textwidth]{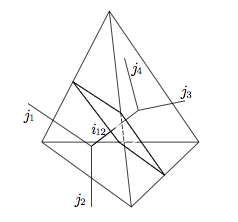}}
	\caption{A quantum tetrahedron and its associated dual graph. The four valent node has been decomposed into two three valent nodes. Taken from \cite{ch2-Bonzom1}.}
	\label{fig:complex}
\end{center}
\end{figure}

The computation of the spectrum of the volume operator involve the evaluation of several sum of complicated $6j$ symbols. An important step towards the computation of the volume spectrum was provided by Thiemann and Brunneman in 2004. They noticed that the use of a $SU(2)$ group identity, called the Elliot - Biedenharn identity, simplify drastically the whole computation, leading to a very compact result. One can refer to \cite{ch2-Th2} for a detailed presentation of this computation. One of the difficulty is that the volume operator fails to be diagonal in the spin network basis. However, the result of this computation shows that, juts as for the area operator, the volume spectrum is discrete, proportional to the Planck volume $l^{3}_{p}$ and admits a volume gap.

Finally, the lenght operator was discussed in \cite{ch2-L1, ch2-L2, ch2-L3}.

The resulting (quantum) geometry encoded in the spin network quantum states is quite elegant \cite{ch2-TW1}. At each $n$ valent node of the network is associated a polyhedre with $n$ faces.
The $n$ faces are encoded in the spin label of each one of the $n$ edges which met at the node. Since there is as many polyhedra as nodes, i.e. for nodes at least four valent, one can picture the resulting geometry as a gluing of polyhedra which do not have necessarily the same number of faces. This gluing is rather subtle, since two faces attached to the same edge and corresponding to two polyhedras have not necessarily the same shape.
This point was unravel in \cite{ch2-TW2,ch2-TW3}, were the authors introduced the notion of twisted geometry. Basically, this non matching between adjacent faces implies the existence of a non vanishing torsion at the quantum level.

Another very interesting construction is the so called spinning geometry, where the notion of spin labeling the graph appears rather naturally \cite{ch2-SP1}.
Therefore, the loop quantization of General Relativity, based on the hypothesis that the quantum states of the gravitational field are extended string-like objects, leads to a very elegant notion of quantum geometry.
Important efforts are now realized to understand the semi classical limit of those quantum geometry, and how one can recover the classical notion of geometry present in General Relativity.

\section{The Immirzi ambiguity}

In this short section, we discuss briefly the so called Immirzi ambiguity, which will be of first importance in the following chapters.
As we have seen in the first chapter, the Immirzi parameter $\gamma$ enters in the Holst action as a coupling constant in front of the topological term.
When the field equations are derived, and the torsion less condition solved, the Immirzi parameter drops out from the classical theory.
Therefore, it won't affect the classical predictions of the vacuum solutions of General Relativity. \footnote{When matter is coupled to gravity, then the Immirzi parameter can have non trivial effects.
Indeed, when one couples gravity to fermions, it has been shown that the Immirzi parameter enters as a coupling constant in the four fermions vertex interaction.}

However, proceeding to the canonical analysis of the Holst action in the time gauge, one ends up with a $\gamma$-dependent phase space.
The Immirzi parameter enters both in the Poisson bracket between the electric field and the Ashtekar-Barbero connection, and in the scalar constraint.
Following the loop quantization procedure, one quantizes the gravitational field and computes the geometric quantum operators such as the area and volume spectrums.
Interestingly, those predictions are $\gamma$-dependent. More precisely, the Immirzi parameter enters explicitly in the expression of the area and volume gap.

This point raises the following questions. Why the Immirzi parameter should play such a crucial at the quantum level while it is totally absent from the quantum theory ?
Is the presence of $\gamma$ in the kinematical spectrum of the area and volume operators a gauge artifact due to the peculiar gauge that we have choosen for the canonical analysis ?
Do the $\gamma$-dependency of the kinematical spectrum will survive once one will take into account the scalar constraint ?

Those questions are still matter of debate. 
It has been argued that the ambiguity related to $\gamma$ is similar to the $\theta$ ambiguity present in the quantization of Yang-Mills gauge theory \cite{ch2-Gamb}. This is due to the fact that the canonical transformation which turns the complex Ashtekar variables into the real Ashtekar-Barbero variables cannot be implemented as a unitary transformation in the quantum theory \cite{ch2-RovTh}. In the same line, it has been proposed that $\gamma$ could play the role of the free angular parameter of the large gauge transformations and that it should be understood as a field and no more as a constant \cite{ch2-Merc1, ch2-Merc2}. In the same spirit, the interpretation of $\gamma$ as a field was investigated in \cite{ch2-Merc3, ch2-Merc4, ch2-Krasnov}. Another possible interpretation was introduced in \cite{ch2-Rand1, ch2-Rand2}, where it was shown that the presence of $\gamma$ is related to the PT symmetry violation of the so called generalized Kodama state. Moreover, it has been shown that while totally irrelevant from the predictions of vacuum GR, the Immirzi parameter can be understood as the coupling constant of the four-fermions interaction when the Holst action is supplemented with some lagrangian containing fermions fields \cite{ch2-AlejRov}. See also \cite{ch2-Merc5, ch2-FreiImm} for other investigations on the role of the Immirzi parameter in presence of fermions and some of its effects on the cosmological scenario. Finally, it was recently argued that more than a simple coupling constant, the Immirzi parameter could be understand as a cut off for quantum gravity \cite{ch2-Liv}.

However, different investigations in three dimensional gravity, in black hole thermodynamics and in spin foams models seem to point towards a rather different interpretation of $\gamma$.
In this interpretation, the self dual variables, i.e. associated to a purely imaginary Immirzi parameter, seems to provides a better candidate in order to deal with the semi classical limit and the dynamics of LQG \cite{ch2-BA1, ch2-BA2, ch2-BA3, ch2-BA4, ch2-Fro3,  ch2-Muxin, ch2-Pranz1, ch2-Pranz2, ch2-Pranz3, ch2-Neiman1, ch2-Neiman2}. 

Indeed, the presence of $\gamma$ in the kinematical quantum theory is related to the choice of working in the time gauge, or equivalently to work with the compact $SU(2)$ group \cite{ch2-Alex1}.
Indeed, without fixing any gauge, i.e. working with the full Lorentz group $SO(3,1)$,  Alexandrov showed that the resulting phase space, albeit complicated, is $\gamma$-independent \cite{ch2-Alex2}. While developing a quantum theory from this phase space is elusive, it was argued that the kinematical geometrical operators admit a $\gamma$-independent spectrum \cite{ch2-Alex3, ch2-Alex4}. From this point of view, the presence of $\gamma$ seems to keep trace of the non compactness of the initial gauge group.

Moreover, the question of the status of the Immirzi parameter is crucial for one more reason. It is well known that the real $su(2)$ Ashtekar-Barbero connection does not transform properly under time diffeomorphism.
The Poisson bracket of the hamiltonian constraint with the real connection does not lead to the expected transformation for a one form connection \cite{ch2-Alex1, ch2-Samuel, ch2-Liv}. Instead, one obtains a very complicated expression which reduce to the expected Lie derivative only for $\gamma= \pm i$. Therefore, only the self dual connection transforms properly under time-diffeomorphism. This point could be potentially problematic since working with the real Ashtekar-Barbero connection when imposing the dynamics could lead to some anomalies in the quantum theory. Those anomalies could well be the presence of the Immiri parameter in the kinematical spectrum of the geometrical quantum operators.
In this perspective, the self dual theory could well be the only physically acceptable quantum theory. 

Going further in our interpretation, the Immirzi parameter could be regarded as a kind of regulator, allowing to Wick rotate the kinematical quantum theory. At the end of the quantization procedure, $\gamma$ should be sent to the purely imaginary value, in order to recover to self dual quantum theory.  This is the point of view adopted in this manuscript.

This interpretation relies on some solid results obtained in very different contexts, such as the semi classical limit of black hole and the loop quantization of three dimensional gravity \cite{ch2-BA1, ch2-BA2, ch2-BA3, ch2-BA4}.
We will present those results in the following chapters and discuss the tests needed to develop this research direction.

\clearemptydoublepage

\chapter{Black hole as a gas of punctures}
\label{ch:BH}
\minitoc

This chapter is devoted to the treatment of black hole in Loop Quantum Gravity. 

We first describe the important shift from the global definition of the black hole to the quasi-local one given by the concept of isolated horizon.
Then, the canonical quantization of this new object is reviewed, and we present the central formula which gives the number of micro-states in the micro-canonical ensemble, i.e. the Verlinde formula.  The Bekenstein Hawking area law derives from this formula.

We present in the next part the statistical model called the ``gase of punctures picture'', which allows to treat the black in the canonical and the grand canonical ensembles.
We review the important inputs used to write the partition function of the black hole: i.e. the local version of the first law of black hole thermodynamics, the holographic degeneracy, the indistinguishability of the punctures and the chemical potential associated to the punctures. 
Having this model at hand, we perform the thermodynamical study of the system in the grand canonical ensemble for the Maxwell Boltzman statistic. In the last section, we explain how the introduction of a quantum statistic for the punctures can modify the semi classical behaviour of the quantum black hole. We perform the thermodynamical study with the Bose Einstein statistic in the grand canonical ensemble and exhibit under which conditions a Bose Einstein condensation occurs. 

Before presenting all this material, we need to mention another investigations in the same area.
In order to describe the quantum black hole using the loop quantization technics, a large amount of work has been developed in the so called spherically symmetry reduced models of gravity.
Those models represent another promising road towards a deeper understanding of the quantum black hole. They have provided an interesting framework to study the fate of the classical singularity in the quantum theory but also to study the evaporation process. In spirit, they follow closely the Loop Quantum Cosmology models. However, completing the quantization of those models remains harder than for the LQC counterpart. The main conclusion  of those models is the resolution of the interior Schwarzschild classical singularity. In the following sections, we won't address those models but the interested readers can refer to \cite{ch3-Boj1, ch3-Gam1, ch3-Gam2, ch3-Gam3}.

\section{Introduction}

Black holes are one of the more emblematic predictions of General Relativity. Paradoxically, they stand among the most simple non trivial solutions to the Einstein equations and yet, their characteristics reveal already the limitations of the Einstein theory. Indeed, the singularity present at the heart of the hole underlines the need of going beyond the classical theory and to develop the quantum theory of gravity. 
Forty years of studying those solutions have led to a rich ensemble of results, paving the way towards a more fundamental understanding of those gravitational systems. 
In light of those results, black holes seems to provide an interface between General Relativity, quantum physics and statistical mechanics, the three pillars of fundamental physics \cite{ch3-Asht1}.
This unexpected interface was unraveled by pioneers work in the seventies by Penrose, Carter, Hawking and Bekenstein and Israel. At this time, it was shown that black holes solutions follow laws with a remarkable similarity with the well known four laws of  thermodynamics.

First, the variation of the mass of the black hole $M$ is naturally related to the variation of the other instrinsic properties of the black hole such as its area $A$, its charge $Q_{e}$ and its angular momentum $J$. The law relating the modifications of the four parameters $(M, A,Q,J)$ mimics the first law of thermodynamics, relating the variation of the internal energy $U$ of a statistical system with its work variation $\delta W$ and its heat variation $\delta Q_{H}$. 

Second, Hawking showed that during any physical process, the variation of the area of the horizon can only grow. This is in strong analogy with the second law of thermodynamic which states that the entropy of a statistical system can only remain constant or grow.

Finally, the third law of thermodynamics states that a physical system cannot reach the zero temperature by any physical process. The analogy is found by defining the so called ``extremal'' black hole, for which the surface gravity vanish. It was shown that an ``extremal'' black hole cannot be reached by a finite number of physical process from a non ``extremal'' black hole. To complete the analogy, it was also shown that just as the temperature of an equilibrium system is constant, the surface gravity of a black hole remain constant on its horizon. 

From those results, the analogy becomes natural if one identified the area of the horizon as the entropy of the system, while the surface gravity $k$ as to be identified with the temperature of the horizon.
The famous result obtain by Hawking in 74' justified this identification. Indeed, using the recently developed machinery of quantum field theory in curved space-time, he showed that a black hole radiates with a black body spectrum at the temperature $T_{H}$ proportional to the surface gravity of the hole and that the entropy of the black hole is proportional to the area of the horizon as follow:
\begin{align*}
T_{H} = \frac{\hbar}{2\pi} k \;\;\;\;\;\;\;\;\;\;\;\;\;\;\;\;\;\;\;\;\;\;\;  S_{H} = \frac{a_{H}}{4 l^{2}_{p}}
\end{align*} 
where $k$ is the surface gravity (to be define below) and $l^{2}_{p}$ is the Planck length.
The Hawking temperature is the temperature measured by an inertial observer at infinity. 

Therefore, a black hole is a thermodynamical system to which we can associate an entropy and a temperature.

In thermodynamics, one describes a physical system using some macroscopic variables. This macroscopic description totally wash away the quantum microscopic details of the system.
For a given value of the macroscopic variables, the physical system can be in different quantum states (micro states). The notion of entropy precisely captures this degeneracy.
It account for the volume of the microscopic phase space which share the same macroscopic variables.
To account for this entropy, one first have to identify the quantum degrees of freedom of the system.
Therefore we are led to the following question : \textit{what are the quantum degrees of freedom responsible for the large entropy of a black hole ?}

One of the major achievement of Loop Quantum Gravity has been to account for this entropy through a rigorous description of the quantum geometry of the horizon \cite{ch3-Asht2}.
In this approach, a black hole is treated as a null boundary in the space-time i.e. an isolated horizon.
We will come back in the next paragraph on the definition of an isolated horizon and the constraints it satisfy. 
In LQG, the quantum states of the gravitational field are spin-networks (at least at the gauge invariant level) living in the bulk. Each edge of the bulk spin network pierce the horizon at a given puncture. They are the fundamental excitations of the black hole horizon.
Each edges and therefore the corresponding puncture, comes with an quanta of area, the area of the macroscopic hole being the sum of all the punctures contributions.
This heuristic picture of the quantum black hole relies heavily on a rigorous mathematical set of results which involve Chern Simons theory and its quantized version.
The first step is to define properly the notion of a isolated black hole. This is done using the notion of an \textit{isolated horizon} \cite{ch3-Asht3}.
It turns out that in the connection formulation of GR, because of the symmetries of the horizon and the conditions required to be isolated, the symplectic current inherit a boundary term on the horizon which is exactly the Chern Simon symplectic current.
Therefore, one can interpret the classical degrees of freedom living on the horizon as being those of a Chern Simons theory coupled to $2+1$ particles, i.e. punctures.
Then the phase space of GR containing an isolated horizon is quantized, treating separately the bulk degrees of freedom and the horizon degrees of freedom. The quantization of the horizon degrees of freedom is done following the well known quantization of Chern Simons theory coupled to point like particles. In this lines, one can derive the dimension of the Hilbert space, its logarithm giving the entropy of the hole in the micro canonical ensemble.
In order to go beyond the micro canonical ensemble, there is up to now only one proposal. This is the model of the gas of punctures picture introduced in \cite{ch3-Alej1}.
This statistical model is build on different ingredients which have not yet derived from first principles. However, it allows to compute the canonical and grand canonical entropy. This model represents a theoretical laboratory where one can study the impact of the quantum statistic for the punctures but also the role played by the chemical potential. This leads to interesting results such as the Bose Einstein condensation of the punctures \cite{ch3-Asin1}. This could be relevant when debating on the status of the large/small spin limit in the semi classical limit. Finally, even if the status of the chemical potential associated to the punctures is still obscure, it could provide an interesting tool in order to describe the evaporation process.

It is important to stress here that all the material presented in the three following sections has been borrowed from various sources, and that our goal is to summarize the long way from the notion of isolated horizon to the computation of its entropy. In the following, we follow \cite{ch3-Ashtekar4, ch3-Nouiii1} for the isolated horizon properties and \cite{ch3-Frodden1} for the symplectic structure derivation, the quantization part and the discussion about the local first law of thermodynamics. The goal is not to give an original point of view but rather to regroup some important results for the following.

Let us first describe the very first step : the notion of an isolated horizon.

\section{A new paradigm : the isolated horizon}

The usual notion of a black hole is given by the concept of an \textit{event horizon}.
The definition reads :

\textit{A black hole is a space-time region that is causally disconnected to the null future infinity $\mathcal{I}^{+}$.}

Therefore, if an observer want to know if he is at a point belonging to a black hole region, he has to know if a light ray sent from its position will reach or not the null future infinity  $\mathcal{I}^{+}$. This is obviously impossible because it would imply to know the whole space-time. This is why the notion of an event horizon is not physical and is not acceptable to describe physical black hole.
Static black hole for example, are defined using space-time equipped with a time-translational killing vector field everywhere. On this manifold, not just the black hole is static, but the whole space time also. Therefore, this description of a static black hole is overly restrictive, since nothing can happen outside the hole, i.e. the whole world is frozen. However, we know that astrophysical black holes live in a highly dynamical environment close and far away from them.

 \begin{figure}
\begin{center}
	\includegraphics[width = 0.4\textwidth]{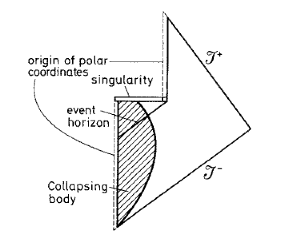}
	\caption[This is the short caption]{Penrose diagram for a collapsing object (bold curved) giving birth to an ``evaporating'' Schwarzschild black hole. The event horizon is a null horizon which separates points connected to the null future infinity $I^{+}$, i.e. the exterior of the hole, from the one causally disconnected form $I^{+}$, i.e. the interior of the hole. This definition is highly non physical. (Taken from \cite{ch3-Hawking1})}
\end{center}
\end{figure}

From this observation, one is led to search for a more physical definition which could encompass the true astrophysical black holes.
For this, the definition should not refer to any properties of the space-time at infinity but only on the intrinsic geometry of the horizon. This is precisely the sense of the shift from the global definition to the quasi local one.

In order to go beyond the notion of an event horizon, different kind of horizons have appeared in the literature. 
In this perspective, one can discriminate between a spacelike, null or timelike horizon. Those three types of horizons provide a way to describe a dynamical black, a isolated black hole or simply a membrane for which matter can simply go in and out. For all those horizons, matter can cross the horizon inwards. However, the causal nature of the horizon tells us to which extent  massive or masseless particles can escape from the hole. For a spacelike horizon, matter will never escape from the hole. This kind of horizon can only grow in time. An example is provided by the dynamical horizon introduced in \cite{ch3-Hay1, ch3-Asht5}. A null horizon corresponds to the limit case. Matter cannot escape from the hole,  but the massless particles emitted at the horizon can finally get out of the hole if the singularity is resolved (Hawking radiation). They follow the diagonal line representing the null horizon on a Penrose diagram and are released in the bulk at the point where the singularity is removed. A non expanding horizon or a weakly isolated horizons corresponds to this category, i.e. they are of a null type. However, those horizons fulfill additional requirement which ensure that they remain isolated, i.e. that there is no flux of radiation through and along the horizon. One can then restrict even more the definition, leading to the notion of the isolated horizon used in the rest of the chapter. Finally, a timelike membrane provides a more realistic object from which matter can either cross the horizon inwards and also escape freely from the hole. They have been used in order to derive a very interesting dictionary between gravity and thermodynamics out of equilibrium \cite{ch3-Frei1, ch3-Frei2}.

We will now describe the way to define an isolated horizon. The starting point is the introduction of a non expanding horizon. One can then constraint the definition in order to be able to derive the first and zero thermodynamical law, introducing the weakly isolated horizon then the isolated horizon definition. The reader can refer to \cite{ch3-Ashtekar4} for all the details.

Let $\mathcal{M}$ be a $4$-dimensional manifold equipped with a metric $g_{ab}$ of signature $(-, +,+,+)$. Let $\triangle$ be a null hyper surface of $(\mathcal{M}, g_{ab})$.
A future directed null normal to $\triangle$ will be denoted by $\chi$. \\

\textit{Definition of a non expanding horizon} \\

A non expanding horizon is a $2+1$-dimensional sub-manifold $\triangle$ which satisfy three constraints:
\begin{itemize}
\item i) $\triangle$ is topologically $S^{2} \times \mathbb{R}$ and null.
\item ii) The expansion $\theta_{l}$ of $\chi$ vanishes on $\triangle$ for any null normal $\chi$.
\item iii) All equations of motion hold at $\triangle$ and the stress-energy tensor $T_{ab}$ of matter fields at $\triangle$ is such that $-T^{a}_{b}\chi^{b}$ is future directed and causal for any future directed null normals $\chi$.
\end{itemize}

With those three requirements, we can already derive important consequences on the null normal $\chi$. 

From the precedent definition, i.e. because $\chi$ is null and expansion free, and using the Raychauduri equation on $\triangle$, it implies that the null normal $\chi$ be also twist and shear free. Moreover, because it is a null vector, we have that:
\begin{align*}
\chi^{a} \nabla_{a} \chi^{b} = \kappa \chi^{b} 
\end{align*} \\
The constant $\kappa$ is the surface gravity of the isolated horizon. However, it is not defined uniquely and depend on the choice of null normal $\chi$. Indeed, under a rescaling of the null normal by a function f, we have:
\begin{align*}
\chi^{'} = f \chi \;\;\;\;\; \text{whence} \;\;\;\;\;  \kappa' = f \kappa + \mathcal{L}_{\chi} f
\end{align*} 
Finally, from the properties of the null normal $\chi$, we can show that a natural connection $1$-form on $\triangle$ exists such that:
 \begin{align*}
\underset{\leftarrow}{\nabla_{a}} \chi^{b}= \omega_{a} \chi^{b}  \;\;\;\;\; \text{whence} \;\;\;\;\;  \omega_{a} \chi^{a} = \kappa  \;\;\;\;\; \text{and} \;\;\;\; \mathcal{L}_{\chi} q_{ab} = 0
\end{align*}
Therefore, the null normal $\chi$ is a Killing field when restricted to the horizon $\triangle$.
Finally, the precedent definition implies some restriction on the Ricci tensor and on the Weyl tensors. All those properties are presented and discussed in great details in \cite{ch3-Ashtekar4, ch3-Nouiii1}. 
We just have listed briefly the ones we need for the following discussion.

The definition of a non expanding horizon, while simple, has important geometrical consequences. However, the definition is not rigid enough to derive thermodynamical law such as the zeroth law. Indeed, the surface gravity defined above is not constant on the horizon because one can always rescaled the null normals $\chi$ by a function, modifying therefore $\kappa$.
To be able to do so, we need to go beyond the precedent definition and introduced the notion of a weakly isolated horizon. This new definition will restrict the choice of null normals $\chi$ and give the tools to derive the zeroth law.
The idea goes as follow. We already know that $\chi$ is a Killing vector of the intrinsic metric. Defining the tensor field $K_{ab} = D_{a} \chi_{b}$ where $D$ is the covariant derivative intrinsic to $\triangle$, we can show that this tensor plays a similar role of the extrinsic curvature.  Asking that $K_{ab}$ be time independent, i.e. $\mathcal{L}_{\chi} K_{ab} = 0$  is equivalent to ask that $\mathcal{L}_{\chi} \omega = 0$. Now, we observe that under a constant rescaling, the connection $\omega$ introduced above do not change. If we define an equivalence class $[\chi]$ of null normal under constant rescaling, then each $\chi$ belonging to this equivalence class will satisfy the law $\mathcal{L}_{\chi} \omega = 0$. Therefore, \\

\textit{Definition of a weakly isolated horizon} \\

A weaky isolated horizon $(\triangle, [\chi])$ consists of a non expanding horizon  $\triangle$, equipped with an equivalence class $[\chi]$ of null normals to it satisfying: 
\begin{align*}
\mathcal{L}_{\chi} \omega \; \hat{=} \; 0 \;\;\;\;\;\;\;   \text{for all  $l\in [\chi]$} 
\end{align*}\\
The equivalence class $[\chi]$ is defined up to constant rescaling:
\begin{align*}
 \chi' \;\;\; \text{and} \;\;\; \chi \;\;\; \in [\chi]  \;\;\;\;  \text{iff} \;\;\;\; \chi' = c \chi \;\;\; \text{with} \;\;\; c \in \mathbb{R} 
\end{align*}\\

With this definition, we see that under a constant rescaling, going from one null normal of the equivalence class to another, the surface gravity changes as:
\begin{align*}
\chi^{'} = c \chi \;\;\;\;\; \text{whence} \;\;\;\;\;  \kappa' = c \kappa 
\end{align*}

Therefore, the surface gravity is also not uniquely defined in this context. With this definition at hand, one can show explicitly that $d \kappa = 0$ on the isolated horizon $\triangle$ for all $\chi \in [\chi]$.
even if there is obviously different possible choices of $[\chi]$.
Yet, we conclude that for a choosen $[\chi]$, a weakly isolated horizon satisfy the zeroth law of black holes thermodynamics.

Futhermore, a (family of) first law of thermodynamics can be shown to hold for weakly isolated horizons. This is the very same law which was derived for usual black hole, except that now, the quantities such as the energy (i.e. the mass), the angular momentum, the electric charge and electric potential, and the surface gravity respectively given by $E_{IH} $, $J_{IH}$, $Q_{IH}$, $\phi_{IH}$ and $\kappa_{IH}$,  are intrinsic to the horizon and are not defined at infinity. Its derivation assume only asymptotic flans at infinity. The case of rotating (weakly) isolated horizons have been tackle in \cite{ch3-Asht6}. The first law reads:
\begin{equation}
\delta E_{IH} = \frac{\kappa_{IH}}{8 \pi} \delta A_{IH} + \Omega_{IH} \delta J_{IH} + \phi_{IH} \delta Q_{IH}
\end{equation}
where $ A_{IH}$ is the area of the isolated horizon. Therefore, the weakly isolated horizon definition provides the tool to reproduce in a quasi local fashion the week known thermodynamics of black hole.
Note that for a weakly isolated horizon, there is a given first law for each possible choices of equivalence class $[\chi]$. 

The introduction of a more restrictive definition can select uniquely the equivalence class $[\chi]$. This is the isolated horizon definition.
This definition reads:\\

\textit{Definition of an isolated horizon} \\

A weaky isolated horizon $(\triangle, [\chi])$ is said to be isolated if:  
\begin{align*}
\big{[}\mathcal{L}_{\chi}, \mathcal{D} \big{]} V \; \hat{=} \; 0 \;\;\;\;\;\;\;   \text{for all vectors fields $V$ tangential to $\triangle$ and all $\chi \in [\chi]$} 
\end{align*}

An isolated horizon definition describes a null horizon which does not grow in time, i.e. its area remains constant. However, contrary to the event horizon, the space-time outside the hole can be dynamical as well as the geometry of the horizon. One can generalized this definition of a spherically isolated horizon to a rotating and a distorted isolated horizon, which are respectively the isolated horizons of type $I$, $II$ and $III$. This object is therefore much more general than an event horizon and provides a way to describe in a quasi local way a black hole.

For the rest of this chapter, we will adopt the new definition of an isolated horizon of type $I$, i.e. a non rotating spherically isolated horizon, when speaking about a black hole. \\

\textit{Consequences for an isolated horizon} \\

Having state the definition of an isolated horizon, we explicitly give the constraints for such an object, constraints which derive directly form the definition above.
One can derive three constraints which are :

\begin{align*}
\underset{\Leftarrow}{F^{i}_{ab}} = - \frac{2\pi}{a_{H}} \underset{\Leftarrow}{\Sigma^{i}_{ab}}
\end{align*}

where $F^{i}_{ab}$ is the curvature of the Ashtekar-Barbero connection $A^{i}_{a}$ and $\Sigma^{i}_{ab}$ is related to the conjugated momentum of $A$. Those quantities have been presented in the first chapter and will be used in the following sections. The area of the horizon is denoted $a_{H}$.
The double arrow refers to the pull back of the object to $H = \triangle \cap \mathcal{M}$, i.e. the restriction of the object to $H =\triangle  \cap \mathcal{M}$.
For concrete calculation, the double arrow means that we only keep the component of the two form on $H$, i.e. of the kind $d\theta \wedge d \phi$ while the single arrow means that we keep only the component of the two form on $\triangle$, i.e. of the kind $d\theta \wedge d \phi$, $d t \wedge d \phi$ and $d t \wedge d \theta$. This condition encodes somehow the ``black hole'' nature of the horizon, i.e. since one need to take into account the vanishing of the expansion to derive it. This equation appears as a crucial ingredient in order to derive the Chern Simons theory boundary term on the horizon. 

The imaginary part of the precedent equation reads:
\begin{align*}
\underset{\Leftarrow}{d_{\Gamma } }K^{i} = 0
\end{align*}

Finally, we have that:
\begin{align*}
\epsilon^{i}{}_{jk} \underset{\Leftarrow}{ K^{j}} \wedge \underset{\Leftarrow}{ K^{k}}  = \frac{2\pi}{a_{H}} \underset{\Leftarrow}{\Sigma^{i}}
\end{align*}

Those properties of an isolated horizon has been derived in \cite{ch3-Asht7} using the Newman Penrose formalism and in \cite{ch3-Asht8} using the spinors formalism. Another way to demonstrate those laws is to work on an explicit Einstein solution.
The precedent laws were derived in \cite{ch3-Nouiii1} in the case of the Reissner Norsdrom space time and in \cite{ch3-Frodden1} for the Schwarschild space time.

Finally, we note that the first equation plays the role of an equation of motion for the connection $A$ which is a dynamical field living on the horizon and where $\Sigma$ is the source generating the curvature of $A$ on the horizon surface. However, up to know, there is now derivation of those equation of motion from a variational principle, i.e. from the Lagrangian of General Relativity constrained to have an isolated horizon boundary condition. This task remains to be done.

Now we describe how the Chern Simons theory appears in this approach. The very first step is to derive the symplectic potential of General Relativity.
This is done by looking at the boundary term generated when differentiating the action, i.e.:
\begin{align*}
\delta S & = \frac{1}{\kappa} \int_{M} \epsilon_{IJKL} \delta (e^{I} \wedge e^{J} \wedge F^{KL}(\omega) ) \\
& = \frac{2}{\kappa} \int_{M} \; ( \epsilon_{IJKL} e^{J} \wedge F^{KL}(\omega)) \; \wedge \delta e^{I} - \epsilon_{IJKL} D (e^{I} \wedge e^{J} ) \; \wedge \delta \omega^{KL} + \frac{1}{\kappa} \int_{\partial M} \epsilon_{IJKL} e^{I} \wedge e^{J} \wedge \delta \omega^{KL} \\
& = \frac{1}{\kappa} \int_{\partial M}  \epsilon_{IJKL} e^{I} \wedge e^{J} \wedge \delta \omega^{KL} 
\end{align*}

The two first terms in the second line vanish because of the equation of motion of General Relativity.
Now one can decide to work in the gauge where $\underset{\leftarrow}{e^{0}}= 0$, where the arrow denote the pullback of the one form $e^{I}$ to the three dimensional boundary $\partial M$.
This lead to:
\begin{align*}
\delta S & = \frac{1}{\kappa} \int_{\partial M}  \epsilon_{IJKL} e^{I} \wedge e^{J} \wedge \delta \omega^{KL}  \\
& = \frac{2}{\kappa} \int_{\partial M}  \epsilon_{0ijk} e^{0} \wedge e^{i} \wedge \delta \omega^{kl} + \epsilon_{ij0k} e^{i} \wedge e^{j} \wedge \delta \omega^{0k} \\
& = \frac{2}{\kappa} \int_{\partial M} \Sigma_{i}\wedge \delta K^{i} 
\end{align*}

where we have used the notation $\Sigma_{i} = \epsilon_{ijk} e^{j} \wedge e^{k}$ for the densitized electric field and $K^{i} = \omega^{0i}$ for the extrinsic curvature.
Therefore, the symplectic potential of General Relativity in the first order formalism used here is given by:
\begin{align*}
\Theta( \Sigma, \delta K) & = \frac{1}{\kappa} \Sigma_{i}\wedge \delta K^{i} 
\end{align*}
This is a 3-form, as expected since we are working in four dimensions. Now the symplectic current $\omega$ is defined as:
\begin{align*}
\omega ( \delta_{1} \Sigma, \delta_{2} \Sigma, \delta_{1} K, \delta_{2} K) & = \delta_{1} \Theta( \Sigma, \delta_{2} K) - \delta_{2} \Theta( \Sigma, \delta_{1} K) \\
& =  \frac{1}{\kappa}  \delta_{[1} \Sigma \wedge \delta_{2]} K
\end{align*}
We denote $\delta$ the vector fields living in the tangent manifold of the phase space. They generates the symmetries of the horizon.
For example, we denote $\delta_{v} \Sigma$ the diffeomorphism of $\Sigma$ tangent to the horizon, i.e. $\delta_{v} \Sigma = \mathcal{L}_{v} \Sigma$ where $\mathcal{L}_{v} \Sigma$ is the Lie derivative of $\Sigma$.
The 3-form $\omega $ can now be integrated over the three dimensional boundary manifold $\partial M$ to give the symplectic two form $\Omega ( . , . )$:
\begin{align*}
\Omega (\delta_{1}, \delta_{2}) & =  \int_{\partial M} \omega = \frac{1}{\kappa}  \int_{\partial M}\delta_{[1} \Sigma \wedge \delta_{2]} K
\end{align*}

The important point is that this integral is independent of the three dimensional Cauchy surface $\partial M$ \cite{ch3-Nouiii1}.
Now that we have the symplectic two form, we would like to see how it is modified when expressed in term of the Ashtekar's variables.
To do so, we will need an intermediate result. This way of presenting the symplectic structure of Isolated horizon is borrowed from \cite{ch3-Frodden1} because it seem to us the most economical way to do it.
We use the notation $\Gamma^{i}= - \frac{1}{2} \epsilon_{ijk} \omega^{jk} $. Using the gauge $\underset{\leftarrow}{e^{0}}= 0$, the Cartan equations projected on $\partial M$ become:
\begin{align*}
\underset{\leftarrow}{de^{I}} + \underset{\leftarrow}{\omega^{IJ}} \wedge \underset{\leftarrow}{e_{J}} = 0  \;\;\;\;\; \rightarrow \;\;\;\;   \left \{ \begin{array}{l}
\underset{\leftarrow}{\omega^{0j}} \wedge \underset{\leftarrow}{e_{j}} = 0  \\
\underset{\leftarrow}{de^{i}} + \underset{\leftarrow}{\omega^{ij}} \wedge \underset{\leftarrow}{e_{j}} = 0
\end{array} \right .
\end{align*}
Differentiating the second equation immediately implies that:
\begin{align*}
\underset{\leftarrow}{ \delta \omega^{ij}} \wedge \underset{\leftarrow}{e_{j}} = - \underset{\leftarrow}{d\delta e^{i}} - \underset{\leftarrow}{\omega^{ij}} \wedge \underset{\leftarrow}{ \delta e_{j}} 
\end{align*}
Ommiting the arrow under the forms, we can now compute the following term, which turns out to be usefull:
\begin{align*}
 \Sigma_{i} \wedge  \delta \Gamma^{i} & = - \frac{1}{2} \epsilon_{ijk}\;  e^{j} \wedge e^{k} \wedge \epsilon^{i}{}_{mn} \; \delta \omega^{mn} \\
& = - \;  e_{j} \wedge e_{k} \wedge \delta \omega^{jk} \\
& = - \;  e_{i} \wedge d \; \delta e^{i} - e_{i} \wedge \omega^{ij} \wedge \delta e_{j} = - \;  e_{i} \wedge \; d \; \delta e^{i} + d e^{i}  \wedge  \delta e_{j} \\
& = - d ( e^{i} \wedge \delta e_{i} )
\end{align*}

Coming back to the sympletic two form of General Realtivity, and considering the boundary as being an isolated horizon, we note: $\partial M = \triangle $ the $2+1$ dimensional horizon and $H = \triangle \cap \mathcal{M}$ its $2$-dimensional spatial section. The Ashtekar's variables reads : $A^{i} = \Gamma^{i} + \gamma K^{i}$ where $\gamma$ is the Immirzi parameter. Under this canonical transformation, the symplectic two form becomes:
\begin{align*}
\gamma \kappa \Omega (\delta_{1}, \delta_{2}) & = \int_{\triangle}\delta_{[1} \Sigma_{i} \wedge \delta_{2]} A^{i} - \int_{\triangle}\delta_{[1} \Sigma_{i} \wedge \delta_{2]} \Gamma^{i} =  \int_{\triangle}\delta_{[1} \Sigma_{i} \wedge \delta_{2]} A^{i} + \int_{\triangle} d (\delta_{[1}e_{i} \wedge \delta_{2]} \delta e^{i} )\\
& = \int_{\triangle}\delta_{[1} \Sigma_{i} \wedge \delta_{2]} A^{i} +  \int_{H}\delta_{[1} e_{i} \wedge \delta_{2]} \delta e^{i}
\end{align*}

We see that, working with the Ashtekar's variables, the symplectic two form inherits a boundary term.
Studying carefully the variation of the fields at the horizon $H$, where the only permissible variations are the tangent diffeomorphisms to the sphere $H$ and the $SU(2)$ gauge transformations of the connection, one can demonstrate the following equality \cite{ch3-Nouiii1, ch3-Frodden1}:
\begin{align*}
\int_{H}\delta_{[1} e_{i} \wedge \delta_{2]} \delta e^{i} = - \frac{a_{H}}{2\pi(1- \gamma^{2})} \int_{H} \delta_{[1} A_{i} \wedge \delta_{2]} \delta A^{i}
\end{align*}

where $a_{H}$ is the area of the sphere $H$. This result is derived using the boundary condition of the isolated horizon definition and is therefore intimately linked with this new object.
Interestingly, the right term is the Chern Simons symplectic two form for the Ashtekar connection restricted to the horizon. Thus, the Chern Simons theory enter into the game at this level.

We can now look at our ``black hole'' as an isolated horizon on which a dynamical connection, i.e. the $2+1$ Ashtekar connection $A$, lives. This connection has the symplectic structure of a Chern Simons connection and satisfy the following equation of motion:
\begin{align*}
F^{i}_{ab} (A)= - \frac{2\pi}{a_{H}} \Sigma^{i}_{ab}
\end{align*}

\section{Quantizing the isolated horizon}

Starting from the precedent equation of motion, we will now explain how to deal with the quantum theory for such a connection. The reader can refer to \cite{ch3-Noui2} for a detailed presentation.
We know from LQG that the kinematical quantum states, at least at the gauge invariant level, are given by spin networks living in the bulk.
A spin network is a graph $\Gamma$, a collection of edges $e$ intersecting at vertices $v$, equipped with group data. The graph is embedded in the bulk manifold and at this level, diffeomorphism invariance has not been imposed. Each edges carries a representation of the $SU(2)$ group and the vertices carry invariant tensors under $SU(2)$ transformations, which couple the ingoing representations to the out going representations at each node of the graph. 

In our context, a spherically symmetric boundary is present in the manifold, i.e. the isolated horizon. Therefore, for a given graph $\Gamma$ embedded in the manifold, some of its edges will pierce the horizon surface at a given point $p \in \triangle$ on the horizon, as represented on Figure $2$.
Within this framework, one can use the area operator defined in LQG to compute the area of the horizon. For a given surface, the contributions to the area come only from edges which pierce the area. For a single edge carrying a $j$-representation of $SU(2)$ and piercing the surface at $x_{p}$, the action of the operator $\hat{\Sigma}$ is:
\begin{align*}
\epsilon^{ab} \hat{\Sigma}^{i}_{ab} (x) |\Gamma, j, m \; > \; & =  2 \kappa \gamma \delta(x,x_{p}) \tau^{i}_{p} \; |\Gamma, j, m \; > \; 
\end{align*} 
where $x$ denotes the coordinate surface and $\Gamma$ is the graph with a single edge piercing the surface at the point $x_{p}$. $\tau^{i}$ is a generator of the $su(2)$ Lie algebra. Extrapoling for an given graph $\Gamma$ which pierce the horizon $n$ times at $( x_{1}, ..., x_{n})$, we obtain:
\begin{align*}
\epsilon^{ab} \hat{\Sigma}^{i}_{ab} (x) |\Gamma, j_{1}, m_{1}, ....., j_{n}, m_{n} \; > \; & =  2 \kappa \gamma \sum^{n}_{p= 1} \delta(x,x_{p}) \tau^{i}_{p} \; |\Gamma, j_{1}, m_{1}, ....., j_{n}, m_{n} \; > \; 
\end{align*} 
With this result for the quantum operator $\hat{\Sigma}$, we can now write the quantum version of the ``equation of motion'' derived earlier.
For a Chern Simons connection living on an isolated horizon $\triangle$, the quantum dynamics is given by:
\begin{align*}
\hat{F}^{i}_{ab} (A) = - \frac{4\pi \kappa \gamma}{a_{H}} \sum^{n}_{p= 1}\delta(x,x_{p}) \tau^{i}_{p}
\end{align*}
At this point, the Hilbert space on which the precedent quantum operator acts has not been defined. In order to do so, we will proceed by identification with a well known theory, i.e. the Chern Simons theory coupled to point like particles \cite{ch3-CSCPP, ch3-PP, ch3-Noui3}.  

 \begin{figure}
\begin{center}
	\includegraphics[width = 0.4\textwidth]{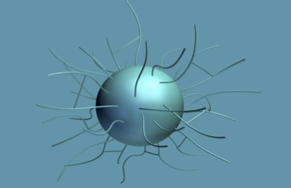}
	\caption{Heuristic picture of the quantum horizon pierced by some edges of the spin networks living in the bulk. (Taken form \cite{ch3-Rovelli1}).}
	\label{fig:complex}
\end{center}
\end{figure}

As mentionned earlier, the ``equation of motion'' for the Chern Simons connection A has not been derived from a variational principle. Is there a three dimensional action leading to this equation of motion or at least to a similar form ?
The answer is in the affirmative and this action is the well known Chern Simons action coupled to point like particles, given by:
\begin{align*}
S =  S_{CS} + S_{p} = \frac{k}{4\pi} \int_{\triangle} A^{i} \wedge dA_{i} + \frac{2}{3} A^{i} \wedge \epsilon_{ijk} A^{j} \wedge A^{k}  + \sum^{n}_{p=1} \lambda_{p} \int_{c_{p}} \tau_{3}  (\Lambda^{i}_{p} )^{-1} (D \Lambda_{p})_{i}
\end{align*}
where the first term is the usual Chern Simons action and the second term is the matter content. The coupling constant $k$ is called the Chern Simons level. In the second term, $\Lambda_{p} \in SU(2)$ represents a $2+1$ particle \cite{ch3-Noui3}. $D$ is the covariant derivative w.r.t the $SU(2)$ Chern Simons connection $A$ which represent the gravitational interaction in $2+1$ dimensions. $\tau_{3}$ belongs to the $su(2)$- Lie algebra, $c_{p}$ is the world line of the p-particle in $\triangle$ and $\lambda_{p}$ is the coupling constant for each particle. Interestingly, when this action is varied w.r.t. the connection $A$, it leads to the following equation of motion:
\begin{align*}
\frac{k}{4\pi} \epsilon^{ab}F^{i}_{ab} (A) = \sum^{n}_{p=1} \delta(x,x_{p}) S^{i}_{p} \;\;\;\;\;\;\;  \text{where} \;\;\;\;\;\;\;  S^{i}_{p} = \lambda_{p} \tau_{3} \Lambda^{j}_{p} \tau^{i} (\Lambda^{-1}_{p})_{j}
\end{align*}

Up to overall factors,  the ``equation of motion'' derived from the Isolated Horizon definition and the LQG framework and the true equation of motion derived from the action of the Chern Simons action coupled to point like particles are the very same. This push us to identify the two systems. This is a conceptual jump and a major hypothesis of the construction. Wether one is loosing some information about the ``black hole nature'' of the object we are quantizing when proceeding to this identification is still not clear. However, from the symplectic structure derived and the equation of motion obtained for the connection $A$ living on the isolated horizon, this identification is natural. Another attempt to quantize the isolated horizon without going through this identification, and therefore without referring to the Chern Simons theory has been introduced in \cite{ch3-Hannno}.

In the following we proceed to the identification Chern Simons theory coupled to point like particles. Therefore, we can work with this action and quantize the theory. The quantization of the Chern Simons theory coupled to $SU(2)$ point like particles is a vast subject and we will obviously not detailed the different steps of this quantization.
It involves the notion of quantum groups. A quantum group is an object build from the deformation of the Lie algebra of a classical Lie group. They have played an important role on the construction of quantum field theory on non combative space-time. In three dimensional gravity, it was argue that a self gravitating scalar euclidean QFT turns out to be described by a QFT on a non commutative space-time without gravity \cite{ch3-Livine1, ch3-Livine2}. The isometry group of such a space-time is given by a deformation of the $SU(2)$ group, i.e. $U_{q}(SU(2))$. Quantum groups are fascinating and complicated objects which provide a tool to study the fate of classical symmetries at the quantum level. Indeed, since most of the approach to quantum gravity predict a minimal length scale, one is led to wonder if the Poincare symmetry is either broken, hidden or deformed at this scale. For an interesting and pedagogical paper on this issue, the reader can refers to \cite{ch3-Noui4}. Finally, quantum groups have been used by Witten (1988) in order to quantize three dimensional gravity \cite{ch3-Witten1, ch3-Witten2}. An extended introduction on quantum groups can be found in \cite{ch3-QG}.

Now, the Hilbert space of the Chern Simons connection living on a punctured two sphere $S^{2}$ is well know to be defined by the $U_{q}(SU(2))$ invariant tensors living in the tensors product $\otimes V_{l}$.
The vector space $V_{l}$ are representation of the quantum group $U_{q}(SU(2))$ with dimension $d_{l} = 2j_{l} +1$. The deformation parameter $q$ is defined to be the root of unity $q = \text{exp}(\pi / (k +2))$ where $k$ is the Chern Simons level. The physical Hilbert space is therefore denoted:
\begin{align*}
\mathcal{H}^{k}_{CS}(j_{1}, ....., j_{p} ) = \text{Inv}_{k} (j_{1} \otimes ..... \otimes j_{p} )
\end{align*}
The dimension of this Hilbert space will provide us the degeneracy of the quantum black hole in the micro canonical ensemble.
This dimension was computed in details in \cite{ch3-Noui5} and reproduced also in \cite{ch3-Frodden1}. It is based on the recoupling theory of the quantum group $U_{q}(SU(2))$. The computation being quite long and heavy, we just give directly the result.
The dimension of the precedent Hilbert space is given by the so called Verlinde formula:
\begin{align*}
g_{k}(p , d_{l}) = \frac{2}{2+k} \sum^{k+1}_{d=1} \sin^{2}(\frac{\pi d}{k+2}) \prod^{p}_{l=1} \frac{\sin(\frac{\pi}{k+2}dd_{l})}{\sin(\frac{\pi }{k+2}d)}
\end{align*}

This is the number of possibilities to recouple the $p$ $U_{q}(SU(2))$-representations $(j_{1}, ...., j_{p})$ associated to the $p$ punctures. There is therefore $g_{k}(p , d_{l})$ quantum states which account for the same macroscopic area of the black hole. Put in differently, the asymptotics of $g_{k}(p , d_{l})$ gives the quantum degeneracy of the horizon and its logarithm the micro canonical entropy for a quantum spherical isolated horizon.

Assuming that all the punctures are carrying the same color, i.e. the same spin, the asymptotic of the Verlinde formula is obtained by taking the limit $k \gg 1$ (large area), $d \gg 1$ (large spins) and finally, $p \gg 1$ (large number of punctures). The result reads:
\begin{align*}
g_{\infty}(p, d) \simeq d{}^{p}
\end{align*}

It turns out that the asymptotic of the degeneracy is not holographic, as usually expected for a black hole (and more generally for a given finite region of space) \cite{ch3-RB}.
Therefore, the $SU(2)$ loop quantization of the isolated horizon, through the identification with the Chern Simons theory coupled to point like particles, does not predict an holographic behavior for the quantum degeneracy of the horizon. The degeneracy computed above account for the ``magnetic degeneratecy'' associated to each puncture. Indeed, for each puncture colored by a spin $j$, the associated magnetic quantum number $m$ can take $d=2j+1$ different values since $-j < m < j$.

Finally, the micro-canonical entropy of the isolated horizon is simply obtained by taking the logarithm of $g_{\infty}(p, d)$. This gives:
\begin{align*}
S \simeq \frac{a_{H}}{4l^{2}_{p}} \frac{\log{d}}{\pi \gamma d} + o(a_{H}) 
\end{align*}
where we have used that $a_{H} = 4\pi l^{2}_{p} \gamma n d$. Therefore, using this quantization procedure for the isolated horizon, one recovers the expected Bekenstein-Hawking entropy only if the Immirzi parameter is fixed to the value:
\begin{align*}
\gamma = \frac{\log{d}}{\pi d}
\end{align*}

\section{The local point of view : the local first law of black holes}

As we have seen, the isolated horizon satisfies a first law of thermodynamics which emerges as a necessary and sufficient condition for the null tangent vector to the horizon $\chi$ to be hamiltonian, i.e.  to generates an hamiltonian flow on the phase space (its dual has to be a closed form on the phase space) \cite{ch3-Ashtekar4}. This first law is quasi local in the sense that the different concepts of energy, charge and angular momentum are intrinsic to the horizon and do not refers to infinity:
\begin{equation}
\delta E_{IH} = \frac{\kappa_{IH}}{8 \pi} \delta A_{IH} + \Omega_{IH} \delta J_{IH} + \phi_{IH} \delta Q_{IH}
\end{equation}

However, the coefficient such as the surface gravity is not defined uniquely for an isolated horizon, and consequently this first law do not single out a preferred notion of energy for the isolated horizon.
Such preferred notion of energy was first work out in \cite{ch3-Frodden2}. It corresponds to the energy that a suitably choosen local observer, close to the horizon, will attribute to the horizon.
This local first law can be derived in different ways. The very first task in order to obtain it is to choose to work with a family of local observers $\mathcal{O}$ which are located close to the horizon, at the distance $L \ll r_{S}$.
Those observers are accelerating to remains stationary a bit above the horizon. Each one defines a world line while the whole set of observers around the horizon defined a worksheet $\mathcal{W}(\mathcal{O})$.
For example, for a Kerr black hole, such observer follow the integral curves of the Killing vector field: 
\begin{equation}
\zeta = \partial_{t} + \Omega \partial_{\phi}
\end{equation}

where $\Omega$ is the horizon angular velocity. Such observer are at rest w.r.t the horizon.
Throwing a charged test particle into the black hole and studying what is the energy variation of the horizon seen by those observers during these process, one obtain the local first law.
\begin{equation}
\delta E_{loc} = \frac{\bar{\kappa}}{8 \pi} \delta A
\end{equation}
where $\bar{\kappa} = \kappa / L $ is the local surface gravity.

However, this derivation only refers to black hole space time admitting Killing field, which is not the case in general.
Since an isolated horizon do not possess any Killing field, we rather show the demonstration directly for this object to avoid any restriction in the conclusion and underline the generality of the local first law introduced in \cite{ch3-Frodden2}.
We refer the interested reader to \cite{ch3-Frodden1} for a pedagogical discussion and the different demonstrations about this local first law.

By definition, an isolated horizon is a null $2+1$ hypersruface $\triangle$ equipped with an equivalence class of null normal vector $\chi$. Two null vectors belonging to this equivalence class are related by a constant rescaling: $\chi'= c \chi$. The acceleration of this vectors defines (not uniquely) a notion of surface gravity for the isolated horizon by: $\chi^{b} \nabla_{b} \chi_{a} = \kappa_{IH} \chi_{a}$. In this class, we can choose a null vector $\chi$ and introduce a affine parameter $v$ such that: $\chi_{a} (d v)^{a} = 1$. This implies that we can write: $\chi_{a} = (\partial_{v})_{a}$. This vector induces a foliation of the isolated horizon of topological two spheres $S_{v}$. Now, we introduce the future directed null normal vector orthogonal to $S_{v}$ , called $n_{a}$. This vector is normalized by asking that: $\chi^{a} n_{a} = -1$. In the same line, we can choose an affine parameter $r$ such that: $- n_{a} (d r)^{a} = 1$. Therefore, we can write $n_{a} = - (\partial_{r})_{a}$. The affine parameter $r$ is choosen such that $r=r_{0}$ on $\triangle$.
Finally, we choose two coordinates $(x_{1}, x_{2})$ which remains constant along the integral curves of $\chi_{a}$ and $n_{a}$, being two coordinates for the two sphere $S_{v}$, i.e. $\chi_{a} (d x_{i})^{a}= n_{a} (dx_{i})^{a}= 0$.
We also ask that the coordinate $v$ remain constant along the integral curve of $n_{a}$, i.e. $n_{a} (dv)^{a}= 0$. We have therefore a coordinates system $(v,r,x_{1},x_{2})$ in the neighborhood of $\triangle$.
It was shown in [] that in this coordinates, the near horizon geometry can be approximated by the following metric:
\begin{align*}
g_{ab} = q_{ab} + 2 dv_{(a}dr_{b)} - 2(r -{r_{0}}) [ \; 2dv_{(a}\omega_{b)} - \kappa_{IH} dv_{a} dv_{b} ] + o[(r-r_{0})^{2}]
\end{align*}
where $q_{ab}$ is the intrinsic metric on the two sphere $S_{v}$. By construction, we have that: $q_{ab}n^{a} = q_{ab} \chi^{a} = 0$. The one form $\omega_{a}$ is defined above as: $\nabla_{a} \chi_{b} = \omega_{a} \chi_{b}$ where the indice $a$ is restricted to $\triangle$. On $\triangle$, since $\chi_{a}$ is a null vector and using the definition of the surface gravity, we have that: $\chi^{a} \omega_{a} = \kappa_{IH}$. Finally, from the isolated horizon definition, we know that the vector $\chi$ is a killing field on $\triangle$, i.e.: $\mathcal{L}_{\chi} g_{ab} |_{\triangle} = 0$.
We have listed all the properties that we need for the derivation. Let us show that they are indeed recovered by using the precedent metric.
We have at our disposal the following basis:
\begin{align*}
\chi_{a} = (\partial_{v})_{a} \;\;\;\;  - n_{a} = (\partial_{r})_{a} \;\;\;\; \zeta_{a} = (\partial_{x_{1}})_{a} \;\;\;\;\; \eta_{a} = (\partial_{x_{2}})_{a}
\end{align*}
Those vectors are build from an extension of the vectors $\chi$ and $n$, intrinsic to $\triangle$, to the neighborhood of $\triangle$.

The norm of the vector $\chi_{a}$ is given by:
\begin{align*}
g_{ab} \chi^{a} \chi^{b} = -2(r -{r_{0}}) [ \; 2 \omega_{b}\chi^{b} - \kappa_{IH} ] = - 2 (r-r_{0}) \kappa_{IH}  \;\;\;\;\;\;\; \| \chi \|^{2} = 2 \kappa_{IH} (r-r_{0})
\end{align*}
We note that this vector is null only on the isolated horizon, for $r=r_{0}$, i.e. $\chi^{a} \chi_{a} |_{\triangle}= 0$.

The normalization of the vector $n$ is given by:
\begin{align*}
g_{ab} \chi^{a} n^{b} = -1 - 2(r -{r_{0}}) \omega_{b}n^{b} = - 1
\end{align*}
This vector is a null vector even outside (but close to) the horizon, i.e. $g_{ab} n^{a} n^{b}= 0$.

We would like now to define our family of stationary observers close to the isolated horizon. Such observers form a world sheet $\mathcal{W}(\mathcal{O})$ which stand at the proper distance $L\ll r_{S}$ from the isolated horizon.
To compute this very small distance, we use the inward normal  $N_{a}$ to the observers worldsheet $\mathcal{W}(\mathcal{O})$. This vector satisfies:
\begin{align*}
N^{a} \chi_{a} |_{\mathcal{W}(\mathcal{O})} = 0 \;\;\;\;\;  \text{and} \;\;\;\;\;\; q_{ab} N^{b} |_{\mathcal{W}(\mathcal{O})} = 0
\end{align*}

Let us compute the first part:
\begin{align*}
g_{ab} N^{a} \chi^{b} & = dr_{a} N^{a} -2(r -{r_{0}}) [ \; dv_{a} N^{a}  \omega_{b}\chi^{b} + \omega_{b} N^{b}  - \kappa_{IH} dv_{a}N^{a} ] \\
& = dr_{a}N^{a} - 2 (r-r_{0}) \omega_{b}N^{b}
\end{align*}

Therefore, we have that:
\begin{align*}
g_{ab} N^{a} \chi^{b} |_{\mathcal{W}(\mathcal{O})}  & = 0 \;\;\;\;\;  \text{if} \;\;\;\;\;  N_{a} = ( \partial_{r})_{a} + \frac{1}{2(r-r_{0}) \kappa_{IH}} (\partial_{v})_{a}
\end{align*}
To compute the proper distance $L$, we need to compute first the norm of the vector $N_{a}$, which is given by:
\begin{align*}
g_{ab} N^{a} N^{b} & = 2 dv_{a} N^{a} dr_{b} N^{b} - 2 (r-r_{0})[ 2 dv_{a} N^{a} \omega_{b}N^{b} - \kappa_{IH} dv_{a} N^{a} dv_{b} N^{b}] = \frac{1}{2 \kappa_{IH} (r-r_{0})}
\end{align*}

Having our ``radial'' vector at hand, we can compute the proper distance $L$ which separates the local observer from the horizon.
\begin{align*}
L = \int^{r}_{r_{0}} \sqrt{g_{ab} N^{a}N^{b}} dr = \int^{r}_{r_{0}} \frac{dr}{\sqrt{2 \kappa_{IH} (r - r_{0})}} = \sqrt{\frac{2(r-r_{0})}{\kappa_{IH}}} = \frac{\| \chi \|}{\kappa_{IH}} = \frac{1}{\bar{\kappa}}
\end{align*}
where we introduce the local surface gravity denoted $\bar{\kappa}$.

Now, we consider a small amount of matter falling through the isolated horizon $\triangle$. This matter is described by a small perturbation of the energy momentum tensor i.e. $\delta T_{ab}$.
Let consider now the vector $J_{a}= \delta T_{ab} \chi^{a}$.
From the Gauss theorem, we know that the flux of the vector $J_{a}$ across the worldsheet observer $\mathcal{W}(\mathcal{O})$ is equal to the flux of $J_{a}$ across the isolated horizon $\triangle$.
\begin{align*}
\int_{\triangle} J_{a}k^{a} = \int_{\mathcal{W}(\mathcal{O})} J_{a} N^{a}
\end{align*}
where $k_{a}$ is a null normal to $\triangle$. We associate to this vector an affine parameter $V$, i.e. $k_{a} = (\partial_{V})_{a}$. We recall that $N^{a}$ is the inward normal to $\mathcal{W}(\mathcal{O})$.
This equality gives explicitly:
\begin{align*}
\int_{\triangle} \; \delta T_{ab} \; \chi^{b} \; k^{a} = \int_{\mathcal{W}(\mathcal{O})} \delta T_{ab} \; \chi^{b} \; N^{a}
\end{align*}
where the integral is done one $dV dS$.
Now, we use that fact that the observer velocity can be expressed as: $u^{a} = \chi^{a}/ \| \chi \|$ and that one can show that the two null normals $\chi_{a}$ and $k_{a}$ to $\triangle$ are proportional, with the explicit scaling given by: $\chi_{a} = \kappa V k_{a}$ \cite{ch3-Frodden1}. This gives:
\begin{align*}
\kappa \int_{\triangle}  \; V \;  \delta T_{ab} \; k^{b} \; k^{a} = \int_{\mathcal{W}(\mathcal{O})} \delta T_{ab} \; \| \chi \| \; u^{b} \; N^{a}
\end{align*}
At this step, one uses two ingredients. The first one is that $\| \chi \|$ depends only on the radial coordinate $r$ and can be taken out from the integral.
The second one is the Raychauduri equation evaluated at the horizon. By definition, the expansion $\theta$, the shear $\sigma_{ab}$, the twist $\omega_{ab}$ of the null normal $k^{a}$ vanish at the horizon. (However, the derivative of the expansion do not vanish). Using the Einstein equation projected on the null normal $k_{a}$, one has: $R_{ab}k^{a}k^{b} = 8 \pi G \delta T_{ab} k^{a}k^{b}$. Consequently, the Raychauduri equation reduces to:
\begin{align*}
\frac{d \theta}{dV}  = - 8 \pi \delta T_{ab} k^{a} k^{b} 
\end{align*}
and we obtain the following equality:
\begin{align*}
- \frac{ \kappa}{8 \pi \| \chi \| } \int_{\triangle}  \; V \; \frac{d \theta}{dV} =   \int_{\mathcal{W}(\mathcal{O})} \delta T_{ab} \; u^{b} \; N^{a}
\end{align*}

The right hand side of this equation is the energy-flux associated to the local observers $\mathcal{W}(\mathcal{O})$ and we will denote it $\delta E$.
Finally, the expansion $\theta = \frac{d}{dS} \frac{d}{dV} dS$ describes the rate of an infinitesimal area change. Here, by integrating over the affine parameter $V$, we are computing the sum of all the infinitesimal area change during the whole evolution.

\begin{align*}
-  \int^{\infty}_{V_{1}} dV  \; V \; \oint_{H} dS \; \frac{d \theta}{dV} =  - \oint_{H} dS\; V_{1} \theta(V_{1}) \; + \oint_{H} \; dS \; \int^{\infty}_{V_{1}} dV \; \theta = \delta A
\end{align*}

where we have used that at infinity, $\theta (\infty) = 0$. The last integral do not depend on $V_{1}$. 
The conservation of the energy-flux across the two membranes leads to the local first law of thermodynamics for the isolated horizon:
\begin{align*}
\delta E = \frac{\kappa}{ 8 \pi \| \chi \| } \delta A = \frac{1}{8 \pi L} \delta A = \frac{\bar{\kappa}}{8 \pi} \delta A
\end{align*}
All the precedent details can be found in \cite{ch3-Frodden1}.
We first note that since any stationary solutions of the Einstein equations satisfy the isolated horizon boundary condition, they also satisfy this local first law.
Therefore, such law is valid for the Schwarschild black hole, the Kerr black hole, the De Sitter cosmological horizon etc ...

Moreover, as we explained earlier, the general first law derived satisfied by an isolated horizon do not single out a preferred notion of energy, since the surface gravity $\kappa$ is not uniquely defined for an IH.
Therefore, there is a family of first law that are associated to a given IH. Indeed, when the null normal $\chi$ get rescaled by a constant rescaling $\chi' = c \; \chi$, the associated surface gravity becomes : $\kappa ' = c \kappa$.

Interestingly, the situation is different when one considers the local first law just derived. Indeed the local surface gravity $\bar{\kappa}$ is uniquely defined for the whole equivalent class of null normal of the IH.
This reads:
\begin{align*}
\chi' = c \chi \;\;\;\;\;\;\;\;  \rightarrow \;\;\;\;\;\;\;\;  \bar{\kappa}' = \frac{\kappa'}{\| \chi' \|} = \frac{ c \;  \kappa}{c \; \| \chi \|} = \frac{\kappa}{\| \chi\|} = \bar{\kappa}
\end{align*}

Therefore, this local first law do single out a preferred notion of energy for the horizon.
By integrating, we obtain the quasi local notion of energy for an isolated horizon, i.e. the Frodden-Gosh-Perez notion of energy:
\begin{align*}
E = \frac{A}{ 8 \pi L} 
\end{align*}

This formula is very appealing when dealing with Loop Quantum Gravity. Indeed, at the quantum level, the area becomes an operator and its spectrum is well known.
One can therefore try to used this local notion of energy to develop a statistical treatment of black hole, and work in the canonical and grand canonical ensemble.
The eigenvalues of the energy of the black hole will be given by the eigenvalues of its quantum area.

Finally, we note that the first law, written in term of $E= TS$ where $T$ is a temperature and $S$ the entropy, implies that : $S= \frac{A}{4l^{2}_{p}}$ for $T = \frac{l^{2}_{p}}{2\pi L}$. We call this temperature the Unruh temperature $T_{U}$ since we work in a local fashion, close to the horizon. This temperature, or its inverse $\beta_{U}$ will be of first importance in the following.

\section{A statistical treatment : the gas of punctures picture}

In this section, we present in details the gas of punctures picture for the quantum black hole. This is for the moment the only proposal to go beyond the micro canonical computation, and study the quantum black hole in the canonical and the grand canonical ensemble. The model was first introduced in \cite{ch3-Alej1} and studied further in \cite{ch3-Alej2, ch3-Alej3}. It is important to stress that the different inputs on which the model is built are not yet derived from first principles. This task remains to be done. However, the results obtained in this context justify a posteriori the different building blocks of the model.
This chapter is devoted to explain the different ingredients used to write down the partition function and successfully derive the right semi classical Hawking entropy from this model.
We proceed to the explicit derivation of the entropy in the grand canonical ensemble first in the Maxwell Botzman statistic, then in the Bose Einstein statistic. The results obtained in the canonical ensemble, presented in \cite{ch3-Alej3}, can be recovered simply by requiring $\mu = 0$ in the grand canonical partition function.

As explain in the precedent section, the new concept used to define a physical black hole is an isolated horizon. When the IH is quantized, the quantum degrees of freedom are topological defects called punctures which live on the horizon. From the LQG point of view, they are the fundamental quantum excitations accounting for the black hole entropy. The punctures play the same role for the quantum black hole than the molecules for a gas. It is therefore natural to look at the quantum black hole as a gas of punctures. This is the heuristic picture adopted in this chapter.

In usual statistical mechanics, one can study a system through different situations: the so called micro canonical, canonical and grand canonical ensembles \cite{ch3-Stat}. 

The thermodynamical results obtained in the Loop literature up to now, only refers to the micro canonical situation, i.e. when the system is defined by a fixed value of its energy and a fixed value of the number of punctures $(E,N)$. In this context, the entropy comes only from the quantum degeneracy, i.e. from the number of micro states accounting for the same value of $(E, N)$. The Boltzman law gives the entropy of the isolated system as the logarithm of the quantum degeneracy $\text{g}_{tot}$:
\begin{equation}
S = \text{log} ( Z_{MC} )  \;\;\;\;\;\;\;   \text{where} \;\;\;\;\;\;\;   Z_{MC} = \text{g}_{tot}
\end{equation}
where $Z$ is the partition function in the micro canonical ensemble and the Boltzman constant $\kappa$ has been set to 1. In this situation, all the configurations refering to the value $(E,N)$ are equiprobable.
To go further and study the thermal fluctuations $\triangle T$ or the variation of the number of punctures $\triangle N$, one has to go beyond the micro canonical ensemble and to define respectively the canonical ensemble and the so called grand canonical ensemble..

In this first situation, the system is in equilibrium with a thermal bath. In our context, a constant incoming flux of radiation at the Hawking temperature $T_{H}$ (as measured at infinity) and a constant outgoing flux of radiation also at the Hawking temperature ensure this equilibrium situation. Therefore, the energy of the system can experienced some thermal fluctuations and the energy is no more fixed. Only the number of puncture $N$ remains fixed.
The canonical entropy reads:
\begin{equation}
S = \beta E + \text{log} (Z_{C} (\beta))  \;\;\;\;\;\;\;   \text{where} \;\;\;\;\;\;\;   Z_{C} (\beta)= \sum \text{g}_{tot} \; e^{-\beta E}
\end{equation}
where $Z_{C} (\beta)$ is the canonical partition function and $\beta$ is the usual inverse temperature.

Finally, the grand canonical ensemble refers to a system (such as a gas) coupled to a thermal bath and to a particles reservoir.
The gas can now exchange particles. This is described by the introduction of a chemical potential $\mu$ which encodes the energy lost when the system loose one particle.
Defining a similar quantity for the quantum isolated horizon turns out to be tricky as explain below. However, the difficulty is purely interpretational and writing down the equations with a non vanishing ``chemical potential'' for the punctures is straitforward. The grand canonical entropy is given by:
\begin{equation}
S = \beta E + \text{log} ( Z_{GC}(\beta, \mu))  - \beta \mu N \;\;\;\;\;\;\;   \text{where} \;\;\;\;\;\;\;   Z_{GC}(\beta, \mu) = \sum g_{tot} \; e^{-\beta (E - \mu N)}
\end{equation}
where $Z_{GC} (\beta, \mu)$ is the grand canonical partition function.

Obviously, the three ensembles share the same thermodynamical limit and the leading term of the entropy calculated in the different ensembles should be the same.
However, each ensembles will disagree on the subleading terms.  It is therefore crucial to compute the
entropy in the various ensembles to discuss the quantum corrections to the entropy, which are generically expected to be logarithmic \cite{ch3-Carlip1, ch3-Das1, ch3-Sen1}.

From the previous part, the statistical isolated horizon can be seen as a set of punctures, each one labelled by a quantum number, its spin $j_{k}$.
Therefore, one can treat this statistical system as a gas of punctures and write down the partition function.
To do so, one has to identify three ingredients :
\begin{itemize}
\item a local notion of energy $E_{loc}$

We will use the notion of energy derived above, i.e. the Frodden Gosh Perez notion of energy
\begin{equation}
E_{loc} = \frac{A}{8 \pi l}
\end{equation}
This is the energy measured by a stationary observer at the fixed proper distance $l\ll r_{S}$ of the horizon \cite{ch3-Frodden2}. As we have seen, its expression is unique for the whole equivalence class of null vectors $\chi$ associated to the isolated horizon and it is therefore a natural candidate.

However, importing this classical notion of energy down to the Planck scale is the key hypothesis of this model.
There is no proof that it is the right notion of energy at the quantum level. Only the results of the model will justify a posteriori this hypothesis.
This classical notion of energy is very appealing for our concerns since it is proportional to the area of the horizon. In LQG, the area is quantized and we can directly use its spectrum in the model to
derive the thermodynamical properties of the hole.
This is the true input from LQG in this model. 
\item the degenerescence experienced by each punctures

Since each puncture carries a quantum spin $j_{k}$, its degeneracy $g_{spin}$ is given by the number of values available for its magnetic number $m$ which is:
\begin{equation}
g_{spin} = 2j_{k} + 1 \;\;\;\;\;\;\;  \text{since} \;\;\;\;\;\;  - j_{k} < m < j_{k}
\end{equation}
However, in order to remain general, we allow other possible sources of degeneracy. 
We denote therefore the total degeneracy:
\begin{equation}
g_{tot} = g_{spin} \; g_{M}  \in \mathbb{N}
\end{equation}
where $g_{M}$ is the additional degeneracy that we do not specify for the moment.
We will see that $g_{tot}$ is constrained to satisfy an inequality called the holographic bound.
\item the quantum statistic of the puncture : Maxwell Boltzman, Bose Einstein, Fermi Dirac

An important lessons of quantum mechanics and quantum field theory is that quantum particles are indistinguishable. Either there are bosons and satisfy the Bose Einstein statistic, either they are fermions and obey the Fermi Dirac statistic. In the case of $2+1$ quantum field theory, another statistic can play a crucial role, the anyonic statistic \cite{ch3-Anyo1}. Even if it could be relevant for the thermodynamics of the punctures, which are intrinsically $2+1$ dimensional objects, we won't study this particular statistic in the following. For the beginning, we will first work out the Maxwell Boltzman statistic by introducing a Gibbs factor, which is the usual procedure in statistical mechanics to implement indistinguishability. The Bose Einstein and the Fermi Dirac statistics for the punctures were first presented in \cite{ch3-Alej3} in the framework of the canonical ensemble. Here, we go further and describe how the introduction of the Bose Einstein statistic modifies the semi classical behaviour of the gas of punctures in the grand canonical framework, i.e. with a non vanishing chemical potential \cite{ch3-Asin1}.
\end{itemize}

From the list above, we can already write down the general expression of the partition function describing the gas of punctures in the canonical ensemble.
For each puncture labelled by a spin $j_{k}$, there is $g_{tot}$ micro states available. For such a configuration, one has to associate the Boltzman weight $e^{- \beta E(j_{k})}$. Finally, we can have an arbitrary set of $n_{k}$ punctures caring the same spin $j_{k}$. Therefore, to take into account all the possibilities, one has to sum over $n_{k}$. 
\begin{equation}
Z_{C}(\beta) = \sum_{n_{k}} \prod_{k} \frac{1}{n_{k} !}\{  \;  g_{tot} e^{- \beta E_{k}} \; \}^{n_{k}} \;\;\;\;\;\;\;  \text{with} \;\;\;\;\;\;\;  E = \sum_{k} n _{k}E_{k} = \frac{a_{H}}{8 \pi l} \;\;\;\;\;  N = \sum_{k} n_{k}
\end{equation}

The term $n_{k} !$ is the Gibbs factor introduced to implement the indistinguishability of the punctures. 
At this stage, we have almost all the material to derive the entropy in the grand canonical ensemble. However, we still need to introduce a chemical potential $\mu$ and discuss its interpretation in this peculiar context.
Then we will show how the form of the total degeneracy $g_{tot}$ is constrained, leading to so called holographic degeneracy which play an important role in the model.

\subsubsection{Introducing a chemical potential for the punctures}

Since we would like to compute the entropy of the gas of punctures in the most general situation, i.e. the grand canonical situation, we need to introduce a chemical potential for the punctures.
However, its interpretation as a true chemical potential is not direct. 
In usual statistical mechanics, the grand canonical situation refers to a system in contact with a thermal bath with which it can exchange energy through heat, but also with a particles reservoir with which it can exchange particles.
Both the system and the particles reservoir are $3+1$ dimensional objects and the exchanged particles keep their identity during the exchange process.
The chemical potential $\mu$ is introduced as the conjugated variable of the number of particles $N$ and encodes the energy lost by the system when it looses one particle.
In this case, the number of particles $N$ becomes a conserved quantity.
The introduction of a chemical potential modifies the first law of thermodynamics as follow:
\begin{equation}
dE = T dS  + \mu dN + \text{work terms}\;\;\;\;\;\;\;  
\end{equation}

However, the situation when dealing with quantum isolated horizon is more tricky. Indeed, the quantum degrees of freedom accounting for the entropy of the hole are the punctures.
Those objects are intrinsically $2+1$ objects ( i.e. topological defects on the horizon) and are not defined in the bulk which is $3+1$ dimensional. Therefore, the role of the chemical potential associated to the punctures do not describe an exchange where the ``particles'' keep their identity. It rather describes a conversion between $2+1$ gravitational quantum degrees of freedom, i.e. punctures, and bulk matter degrees of freedom. 
Therefore, to understand the role of the chemical potential in this model, one has to shift the notion of exchange to the notion of conversion. Its role is drastically different from usual statistical mechanics. We stress that this discussion do not rely on a precise mathematical derivation concerning the conversion process, but it rather interpretational.
To get a precise picture of the conversion phenomenon, one would have to understand the mechanism describing the detachement of punctures, which is a dynamical process. This is still out of reach up to now.
For a interesting account on the role of the chemical potential associated to the punctures, one can refer to \cite{ch3-mu}.

Moreover, the introduction of a non vanishing chemical potential can seem unnatural for one more reason. If we considers that the large spins will dominate the semi classical limit as shown in \cite{ch3-Alej2}, then the spacing between large area eigenvalues becomes negligible. Therefore, the energy needed to create or annihilate a puncture having a large spin vanish. Consequently, the chemical potential should be set to zero from the beginning, and the number of punctures $N$ is not conserved.
However, if we don't adopt this point of view, the presence of a non vanishing chemical potential for the punctures can lead to interesting results that we will present below. For example, if one assume a Bose Einstein statistic for the punctures, we can show that the gas of punctures will experienced a Bose Einstein condensation in some particular regim. In this regim, the small spin will dominate the semi classical regim and the argument presented above against the introduction of a non vanishing chemical potential is bypassed. 
For a non vanishing chemical potential, in turns out that the number of punctures $N$ is now conserved and $N$ becomes an observable. Therefore, the characterization of a quantum black hole by the usual classical charges such as its mass $M$, its charge $Q$ and its angular momentum $J$ have to be completed. It is now characterized by an additional purely quantum charge which is its number of punctures $N$. This new charge can be understood as a new quantum black hole hair. 

For our concern, this problem concerning the role of the chemical potential associated to the punctures is purely interpretational. Concretely, one can write down the partition function with a non vanishing chemical potential and proceed to the derivation without difficulty. We can now give the expression of the grand canonical partition function. Since the number of puncture $N$ is no more fixed, we have now to sum over N:
\begin{equation}
Z_{GC}(\beta, \mu) = \sum_{n} z^{n}\sum_{n_{k}} \prod_{k} \frac{1}{n_{k} !}\{ \; g_{tot} e^{- \beta E_{k}} \; \}^{n_{k}} \;\;\;\;\;\;\;  \text{with} \;\;\;\;\;\;\;  E = \sum_{k} n _{k}E_{k} = \frac{a_{H}}{8 \pi l} 
\end{equation}

We have introduced the quantity $z$, called the fugacity:
\begin{equation}
z = e^{\beta \mu}
\end{equation}

where $\mu$ is the chemical potential.

\subsubsection{The holographic degeneracy}



Having the form of our grand canonical partition function at hand, the last step consists in finding a general explicit expression for the total degeneracy $g_{tot}$ (at each punctures). The degeneracy $g_{spin}$ that we have written in the partition function only refers to all the possible magnetic numbers $m$ that a puncture can admit for a given spin $j_{k}$ : $-j_{k}<m<j_{k}$. 
However, to remain as general as possible, we should authorize each puncture to experience additional degeneracy. Therefore, we denote the total degeneracy at each puncture $g_{tot}$ and we have:
\begin{align}
g_{tot} = g_{spin} g_{M}
\end{align}

The additional degeneracy is denoted $g_{M}$. In \cite{ch3-Alej2}, it was proposed that this additional degeneracy could account for ``matter'' degeneracy present at each punctures.
The argument works as follow. One can show by direct calculation \cite{ch3-Frodden2} that an observer at fixed proper distance $l$ from the horizon experiences a thermal bath at the temperature $T_{U} = l^{2}_{p}/2 \pi l$.
Since $l \ll r_{S}$, the temperature measured by the stationary observer can be considered infinite. Thus, the thermal bath is filled by particles of all energies. Those degrees of freedom can be denoted matter degrees of freedom even if they can account for gravitational radiation also. For the following, we will denote matter degrees of freedom the ones which do not contribute to the hamiltonian, i.e. (that is to say all possible degrees of freedom except the punctures). 
Although this proposal to interpret the additional degeneracy is very appealing, let us keep in mind that $g_{M}$ is introduced only to account for an additional degeneracy than the one related to the magnetic number $m$ of each punctures.

Because of the form of the Hamiltonian $(2.3)$, we know that those degrees of freedom will only affect the partition function through the degeneracy term $g_{tot}$. 

\begin{equation}
 Z_{GC}(\beta, \mu) = \sum_{n} z^{n}\sum_{n_{k}} \prod_{k} \frac{1}{n_{k} !}\{ \; g_{tot} \; e^{- \beta E_{k} } \; \}^{n_{k}}  \;\;\;\;\;  \text{with} \;\;\;\;  E_{k} = \frac{a_{k}}{8 \pi l} = \frac{1}{\beta_{U}} 2 \pi \gamma \sqrt{j_{k}(j_{k} + 1)} 
\end{equation}
where we have :
\begin{equation}
g_{tot} = (2j_{k} +1) \; g_{M}\;\;\;\; \in \;\;\;\;  \mathbb{N}
\end{equation}
The additional degeneracy $g_{M}$ introduced above can be identified with the matter degeneracy. For the following, we write the total degeneracy $g_{tot}$ as an exponential of an unkown function $\alpha$ which depends on the different coupling constants of matter: $g_{tot} = e^{\alpha}$.
At the Unruh temperature, the requirement of convergence of the precedent partition function reads:
\begin{equation}
\alpha < \beta_{U} E_{k} = \beta_{U} \frac{1}{\beta_{U}}  \; 2 \pi \gamma \sqrt{j_{k}(j_{k} + 1)} = \frac{a_{k}}{4 l^{2}_{p}}
\end{equation}

Thus, without knowing the exact expression of $g_{M}$, we conclude that $g_{tot}$ has to satisfy:
\begin{equation}
g_{tot} \; < \; \text{exp} ( \frac{a_{k}}{4 l^{2}_{p}})
\end{equation}

Therefore, the total degeneracy is bounded by the so called \textit{holographic bound}. The precise form of the local energy used in this model, and its quantum version coming from the LQG theory imply the holographic bound.
At least from the point of view of this model, one can therefore understand this bound as a ``prediction'' of LQG. It is interesting to note that similar bounds have been derived in the literature, such as the Bousso entropy bound \cite{ch3-Bousso1}. Moreover, the holographic behavior of the degeneracy has been obtained from very different approachs, among which the one considering the entanglement entropy of matter degrees of freedom across the horizon \cite{ch3-Ryu1}.
A clear relation between all those similar results, if it exists, is still to build.

We arrive therefore at the second hypothesis of the model. In view of the condition (1,19), we assume that the total degeneracy has an exponential form and can be written such that:
\begin{align}
g_{tot} \; = \; \text{exp} ( \lambda \frac{a_{k}}{l^{2}_{p}} ) \;\;\;\;\;\;  \text{with} \;\;\;\;\;\;  \lambda = \frac{1}{4} ( 1- \delta_{h})
\end{align}

The free parameter $\delta_{h}$ measures the failure to saturate the holographic bound, which is obtained for $\delta_{h} = 0$.
In the next part, we will show that the free parameter $\delta_{h}$ has to converge to zero to obtain the right semi classical limit. Consequently, to recover the right semi classical result with this model, the holographic bound has to be saturated.

Working with an exponential form for the degeneracy and not with another form, such as a power law, is one of the strong hypothesis of the model.
This assumption is not justified from first principles yet. Ideally, one would have choosen the degeneracy predicted by the loop counting of the degeneracy of the quantum isolated horizon in the micro canonical ensemble. However, the quantization of the isolated horizon based on the $SU(2)$ Ashtekar-Barbero connection does not lead to an holographic degeneracy. We are therefore led to postulate it. As we will see, this ingredient is crucial to recover to right semi classical limit, i.e. the Bekensetin Hawking area law at the leading term. One could interpret this failure of the loop quantization based on the $SU(2)$ Ashtekar-Barbero variables to reproduce the holographic degeneracy for the black holes as an indication that those real variables are not well suited to deal with the semi classical limit of the theory. We will develop this argument later on.

Concernng the subleading order, it has been shown in the appendixes of \cite{ch3-Alej3} that this exponential degeneracy supplemented with a power law lead to logarithmic corrections as expected. In this case, the degeneracy reads:
\begin{equation}
g_{tot}^{n_{k}} \; = \; \; \big{(} \; \frac{a_{H}}{l^{2}_{p}} \; \big{)}^{p} \; \text{exp} ( \lambda \frac{n_{k} \; a_{k}}{l^{2}_{p}} ) \;\;\;\;\;\;  \text{with} \;\;\;\;\;\;  \lambda = \frac{1}{4} ( 1- \delta_{h}) \;\;\;\; p \in \mathbb{N}
\end{equation}
Once again, this precise form of the degeneracy is postulated and not derived.
However, we will see that interestingly, while the exponential form of the degeneracy, supplemented with a power law has to be postulated in this model where $\gamma \in \mathbb{R}$, it can be derived from first principles, i.e. from the quantum theory, when $\gamma = \pm i$ \cite{ch3-Geiller1, ch3-BA1}. In this case, the spin of the punctures has to continuous and is denoted $s_{k}$ instead of $j_{k}$. This is the subject of the next chapter. In this case, the degeneracy obtain from the asymptotics of the Verlinde formula reads, at the semi classical limit:
\begin{equation}
g_{tot}^{n} \; \simeq  \; \text{exp} ( \lambda \frac{a_{H}}{l^{2}_{p}} ) \;\;\;\;\;\;  \text{where} \;\;\;\;\;\;\;  \gamma = \pm i \;\;\;\;\;\; \text{and} \;\;\;\;\; \lambda = \frac{1}{4}
\end{equation}
where we have omitted the power law corrections to the degeneracy.
It is assumed for simplicity that all the punctures carry the same spin $s$ and the area of the hole is given by: $a_{H}= 4 \pi l^{2}_{p} ns$. The symbol $\simeq$ means that this result hold at the semi classical limit when the number of punctures $n$ and the spin $s$ of the punctures are large.
The case where different spins are allowed lead to the same form for the degeneracy expect that the power law depends on the number of colors. The fact that the quantum theory based on the self dual variables
reproduces correctly the holographic degeneracy for the quantum black hole in LQG is a strong indication that the self dual variables are the right one to work with in order to deal with the semi-classical limit of the quantum theory.
This point of view will be develop and  studied in gret details in the next chapter.

For the moment, let us come back to the model with $\gamma \in \mathbb{R}$. Now that we have choosen an explicit expression for the degeneracy which respect the holographic bound, we can study the thermodynamics of the system.
For the rest of the discussion, we will work only with the exponential degeneracy (1.20).

\subsection{The Maxwell Boltzman statistic}

We will proceed to the computation of the entropy, the mean energy and the mean number of punctures in the grand canonical ensemble starting from the partition function presented above. This section will only refers to the Maxwel Boltzman statistic. This is the common limit of the Bose Einstein and Fermi Dirac statistics at high temperature.
Before diving in the technicalities, we only summarize the result obtained in the canonical ensemble using this model.
The detailed study of this model assuming a vanishing chemical potential $\mu=0$ or equivalently $z=1$ was presented in \cite{ch3-Alej3}. The different thermodynamical quantities of interest (entropy, mean energy, mean number of punctures , mean spin color and heat capacity) were computed at the semi classical limit, for the Maxwell Boltzman, the Bose Einstein and the Fermi Dirac statistics.
For this three choices of statistics, it can be shown that:
\begin{itemize}
\item The requirement that the semi classical limit occurs at the Unruh temperature $T_{U}$ for the local observers, which corresponds to the inverse temperature $\beta_{U} = 1/ T_{U} = 2 \pi l \; / \; l^{2}_{p}$, imposes that the holographic degeneracy be saturated, i.e. that $\lambda - 1/4 \ll 1$ or equivalently, that $\delta_{h} \ll 1$. 
(One could wonder why the semi classical limit is required to occur at the Unruh temperature. It was shown in \cite{ch3-Frodden2} that for classical black holes, i.e.  for $a_{H}\gg l^{2}_{p}$, a local observer close to the horizon measures precisely the Unruh temperature $T_{U} = \; l^{2}_{p} / 2 \pi l $).
\item The large spins dominate at the semi classical limit, in the sense that the mean spin of the punctures becomes large when the area of the hole is large. Moreover, the mean number of punctures became large also. Both scale as:
\begin{equation}
\bar{j} \propto \sqrt{a_{H}} \;\;\;\;\;\;  \bar{n} \propto \sqrt{a_{H}} 
\end{equation}
The precise behaviour depends on the quantum statistics. 
\item The entropy of the gas of punctures reproduces the Bekenstein-Hawking result $S= a_{H}/ 4 l^{2}_{p} + S_{cor}$ with the correct factor $1/4$, while the subleading corrections are of the form $S_{cor} \propto \sqrt{a_{H}}$ (the precise value of the multiplicative factor is not important and depends also on the statistics). Additional logarithmic quantum corrections can be obtained by introducing a power law in the holographic degeneracy.
\end{itemize}

We present now the derivation of the same thermodynamical quantities assuming a non vanishing chemical potential, i.e. in the grand canonical ensemble.
The system is described by two couple of conjugated variables $(\beta, E)$ and $(\mu, N)$. The point of view usually adopted is to regard the system as specified by a fixed value of the temperature and a fixed value of the chemical potential, while the energy and the number of particles are not fixed and adapt themselves when we modify $(\beta, \mu)$.
We introduce two free parameter $(\delta_{\beta}, \delta_{h})$. The first one quantifies the departure from the Unruh temperature $\beta_{U}$, while the second one has already been presented. It account for the failure to saturate the holographic bound.
This two parameters enter in the equations as follow:
\begin{equation}
\beta = \beta_{U} ( 1+ \delta_{\beta}) \;\;\;\;\;\;  \text{and} \;\;\;\;\;\;\;  \lambda = \frac{1}{4}(1 - \delta_{h}) 
\end{equation}

The grand canonical partition function reads :
 
\begin{align*}
 Z_{GC}(\beta, \mu) & = \sum_{n} z^{n}\sum_{n_{k}} \prod_{k} \frac{1}{n_{k} !}\{ \; g_{tot} e^{- \beta E_{k} } \; \}^{n_{k}} \\
 & =  \sum_{n} z^{n}\sum_{n_{k}} \prod_{k} \frac{1}{n_{k} !} \; q_{k}^{n_{k}}  = \sum_{n} \frac{z^{n}}{n!} \sum_{n_{k}}  \frac{n !}{n_{1} ! ..... n_{k} !} q_{1} ...... q_{k} \\
 & = \sum_{n} \frac{z^{n}}{n!} \{ q_{1} + ...... + q_{k}\}^{n} = \sum_{n} \frac{z^{n}}{n!} \{ \sum_{j_{k}}q_{k} \}^{n}
 = \sum_{n} \frac{z^{n}}{n!} \mathcal{Q}(\beta)^{n} \\
& = e^{z \mathcal{Q}(\beta)} 
\end{align*}

From the second line to the third line, we have used the binomial formula of Newton.

We give also the explicit expression of $q_{k}$:
\begin{align*}
q_{k} & =  g_{tot} \; e^{- \beta E_{k}}  = \text{exp} \; (\; \lambda \frac{a_{k}}{4l^{2}_{p}} - \beta E_{k}\; ) \\
& =   \text{exp} \; (\; (1- \delta_{h} - 1 - \delta_{\beta} ) 2 \pi \gamma \sqrt{j_{k} (j_{k} + 1)}\; ) \\
& =  \text{exp} \; (\; - 2 \pi \gamma \delta \sqrt{j_{k} (j_{k} + 1)}\; ) \\
\end{align*}

where we have introduce the quantity $\delta = \delta_{\beta} + \delta_{h}$. 

The expression of  $\mathcal{Q}(\beta)$ (or $\mathcal{Q}(\delta)$) follows directly:
\begin{align*}
\mathcal{Q} (\delta) & =  \sum^{\infty}_{k = 1/2} \text{exp} \; (\; - 2 \pi \gamma \delta \sqrt{j_{k} (j_{k} + 1)}\; ) \simeq  \sum^{\infty}_{k = 1/2} \text{exp} \; (\; - 2 \pi \gamma \delta {j_{k}} \; ) \\
& = \sum^{\infty}_{n = 1} \text{exp} \; (\; - \pi \gamma \delta  \; )^{n}  = \frac{1}{1 - \text{exp}( - \pi \gamma \delta)} - 1 = \frac{1}{\text{exp}( \pi \gamma \delta) - 1}
\end{align*}
The approximation done in the second step turns the non linear area spectrum into a linear one.
It is valid only for large spins. We will see that this approximation is self consistent since the mean color of the puncture turns out to be large at the semiclassical limit.

From the partition function, we obtain the expression of the mean energy $\bar{E}$ and the mean number of punctures $\bar{N}$.
They reads:
\begin{align*}
\bar{E}  & = - \frac{\partial}{\partial \beta} \text{log} \; Z_{GC} (\beta, \mu) = z \sum^{\infty}_{j_{k}= 1/2} E_{j_{k} } \; \text{exp} ( - 2 \pi \gamma \delta \sqrt{j_{k} (j_{k} + 1)})  \\
& =  z \pi \gamma T_{U} \sum^{\infty}_{k= 1} \sqrt{k (k + 2)} \; \text{exp} ( -  \pi \gamma \delta \sqrt{k (k + 2)})\\
\bar{n}  & = z \frac{\partial}{\partial z} \text{log} \; Z_{GC} (\beta, \mu) = z \mathcal{Q}(\delta) = z \sum_{j_{k}}\text{exp} \; (\; - 2 \pi \gamma \delta \sqrt{j_{k} (j_{k} + 1)}\; )  \\
& = z \sum^{\infty}_{k= 1} \; \text{exp} ( -  \pi \gamma \delta \sqrt{k (k + 2)}) 
\end{align*}

We can treat the mean energy without the need of the linear spectrum approximation. Following the result of \cite{ch3-Alej3}, we first assume that the semi classical limit is reached at the Unruh temperature, i.e. for $\delta_{\beta} = 0$.
This assumption implies that $\delta_{h} \neq 0$. Moreover, the semi classical limit is defined as $a_{H} \gg l^{2}_{p}$.
We will now show that at the semi classical limit, the black hole degeneracy saturate the holographic bound, i.e. $\delta_{h} \rightarrow 0$.
To do so, let us first note that : $k \leqslant  \sqrt{k (k + 2)} \leqslant k +1$. Consequently, we can write for the mean energy the following inequalitiy :
\begin{align*}
 z \pi \gamma T_{U} \sum^{\infty}_{k= 1} k  \; \text{exp} ( -  \pi \gamma \delta (k + 1))  \leqslant \bar{E}  \leqslant  z \pi \gamma T_{U} \sum^{\infty}_{k= 1} (k + 1) \; \text{exp} ( -  \pi \gamma \delta k) 
\end{align*}
The two terms of the inequality can be computed and lead to:
\begin{align*}
 z \pi \gamma T_{U} \frac{\text{exp}( - 2 \pi \gamma \delta)}{(1 - \text{exp}( - \pi \gamma \delta))^{2}} \leqslant \bar{E}  \leqslant  z \pi \gamma T_{U} \frac{1}{(1 - \text{exp}( - \pi \gamma \delta) )^{2}}
\end{align*}
At this point, we can already see that for the mean energy to be larger than a given value, $\delta = \delta_{\beta} + \delta_{h}$ has to be small. Finally the asymptotics of the precedent inequality for $\delta \ll 1$ read:
 \begin{align*}
 z_{U} T_{U} \frac{1 + ( \mu\beta_{U} - 2 \pi \gamma) \delta_{\beta} - 2\pi \gamma \delta_{h}}{\pi \gamma \delta^{2}} + \mathcal{O}(1) \leqslant \bar{E}  \leqslant  z_{U} T_{U} \frac{1 + ( \mu\beta_{U} - 2 \pi \gamma) \delta_{\beta} }{\pi \gamma \delta^{2}} + \mathcal{O}(1)
\end{align*}
where $z_{U} = \text{exp}(\mu\beta_{U})$ is the fugacity at the Unruh temperature.
This inequality immediately implies that there is a couple of constant parameters $(x,y)$ such that:
 \begin{align*}
\bar{E} =  z_{U} T_{U} \frac{1 + x \delta_{\beta} + y \delta_{h}}{\pi \gamma \delta^{2}} + \mathcal{O}(1)   \;\;\;\;\;  \text{and} \;\;\;\;\; \frac{\bar{a_{H}}}{4 l^{2}_{p}} = \frac{z_{U}}{\pi \gamma \delta^{2}} ( 1 + x \delta_{\beta} + y \delta_{h}) +\mathcal{O}(1) 
\end{align*}
We obtain therefore that the mean energy $\bar{E}$ and the mean area of the hole $\bar{a}_{H} = 8 \pi l \bar{E}$ scale as $\delta^{-2}$ at the Unruh temperature, when $\beta = \beta_{U}$ (i.e. for $\delta_{\beta} = 0$).
This means that the mean area becomes large, i.e. that the semi classical limit is reached, only if the holographic bound is saturated, i.e. for $\delta_{h} \rightarrow 0$. 

We proceed along the same lines for the mean number of punctures $\bar{n}$.
\begin{align*}
 z  \sum^{\infty}_{k= 1}   \; \text{exp} ( -  \pi \gamma \delta (k + 1))  \leqslant \bar{n}  \leqslant  z \sum^{\infty}_{k= 1} \; \text{exp} ( -  \pi \gamma \delta k) 
\end{align*}

which can be rewritten as:
\begin{align*}
z\frac{e^{-\pi \gamma \delta}}{e^{\pi \gamma \delta} - 1}  \leqslant \bar{n}  \leqslant  z \frac{1}{e^{\pi \gamma \delta} - 1} 
\end{align*}

The asymptotics of this inequality leads to :
\begin{align*}
\frac{z_{U}}{\pi \gamma \delta} + \mathcal{O}_{1}(1)  \leqslant \bar{n}  \leqslant  \frac{z_{U}}{\pi \gamma \delta} + \mathcal{O}_{2}(1) 
\end{align*}
We conclude that the mean number of punctures is given by:
\begin{align*}
\bar{n}  = \text{log} \; Z_{GC} (\delta, \mu)= \frac{z_{U}}{\pi \gamma \delta} + \mathcal{O}(1) 
\end{align*}
We immediately observe that $\bar{n}$ scales as $\delta^{-1}$. Since we have shown that $a_{H}$ scales as  $\delta^{-2}$, we have the following behaviour for the mean number of punctures:
\begin{align*}
\bar{n}  \propto \sqrt{\bar{a}_{H}}
\end{align*}
Therefore, at the semi classical limit, when $a_{H}$ is large, the number of punctures becomes also large. 
We can now compute the entropy of the black hole. In the grand canonical ensemble, it reads:
\begin{align*}
S_{GC} (\beta, \mu) & = \beta \bar{E} - \beta \mu \bar{n} + \text{log} \; Z_{GC} (\beta, \mu) \\
& = \beta_{U} ( 1+ \delta_{\beta}) z_{U} T_{U} \frac{1}{\pi \gamma \delta^{2}} - \beta_{U}(1+ \delta_{\beta}) \mu \frac{z_{U}}{\pi \gamma \delta} + \frac{z_{U}}{\pi \gamma \delta} + \mathcal{O}(1) \\
& = \frac{z_{U}}{\pi \gamma \delta^{2}} + \frac{z_{U}}{\pi \gamma \delta} [ \delta_{\beta} ( \frac{1}{\delta} - \mu \beta_{U})] - \frac{z_{U}}{\pi \gamma \delta} ( 1 - \mu \beta_{U}) + \mathcal{O}(1)
\end{align*}

For $\delta_{\beta} = 0$, i.e. at the Unruh temperature, as we have assumed from the beginning, the entropy reduces to:
\begin{align*}
S_{GC} (\beta_{U}, \mu)& = \frac{a_{H}}{4l^{2}_{p}} - \frac{z_{U}}{\pi \gamma \delta_{h}} ( 1 - \mu \beta_{U}) + \mathcal{O}(1)
\end{align*}

This result is consistent with the one obtain in \cite{ch3-Alej3}, i.e. in the canonical ensemble (for $\mu=0$).
Since the area of the horizon scales as $\delta^{-2}$, we see that the subleading term is proportional to $\sqrt{\bar{a}_{H}}$. This quantum corrections are too large compared to the one expected, i.e. the logarithmic corrections.
However, we note that for a chemical potential fixed at the Unruh temperature: $\mu = T_{U}$, this too large quantum corrections disappear. 
\begin{align*}
S_{GC} (\beta_{U}, T_{U})& =  \frac{a_{H}}{4l^{2}_{p}}  + \mathcal{O}(1)
\end{align*}

This fine tuning of the chemical potential and the possibility to cancel the square root quantum correction was first noticed in \cite{ch3-BA1}. 

Another way to reach the semi classical limit exists. It consists in assuming from the beginning that the holographic bound is saturated, i.e. that $\delta_{h} = 0$.
In this situation, $\delta_{\beta} \neq 0$. The semi classical limit is reached for $\delta_{\beta} \rightarrow 0$ and the entropy reads:
 \begin{align*}
S_{GC} (\beta_{U}, \mu)& = \frac{z_{U}}{\pi \gamma \delta^{2}_{\beta}} - \frac{z_{U}}{\pi \gamma \delta_{\beta}} ( 2 - \mu \beta_{U}) + \mathcal{O}(1) \\
& = \frac{a_{H}}{4l^{2}_{p}} - \frac{z_{U}}{\pi \gamma \delta_{\beta}} ( 2 - \mu \beta_{U}) + \mathcal{O}(1) 
\end{align*}
To cancel the too large quantum correction, the chemical potential has to be fixed to the value: $\mu = 2T_{U}$.
Therefore, in this case, for $\delta_{h} = 0$, we obtain:
\begin{align*}
S_{GC} (\beta_{U}, 2T_{U})& =  \frac{a_{H}}{4l^{2}_{p}}  +\mathcal{O}(1)
\end{align*}

Therefore, working with the Maxwell Boltzman statistic, we have recover the right semi classical limit in the most general case, i.e. in the grand canonical ensemble.
Moreover, we have recovered the same scaling for the quantum corrections to the entropy than the one found in \cite{ch3-Alej3}. However, the grand canonical ensemble provides us a way to cancel those too large quantum corrections through a fine tuning of the chemical potential $\mu$. Fixing $\mu$ to the Unruh temperature $T_{U}$ (or two times the Unruh temperature depending on the way we reach the semi classical limit) leaves us only with the Bekenstein Hawking area law.
This particular value of the chemical potential is very appealing since it represents the energy lost by the black hole when one puncture is removed.
Therefore, it draw a picture of the semi classical black hole as a gas of puncture which can convert its quantum building blocks , i.e. the punctures, into thermal quanta with an energy equal to the Unruh temperature.
When such conversion from a puncture to a therm quanta occurs, the area of the hole shrinks by one quanta of area. This heuristic picture could provide an interesting framework for describing the evaporation process. 
However, this scenario encounters some difficulties regarding its coherence. Indeed, since we are working with semi classical black hole which a large spin domination, the energy required to annihilate a puncture is close to zero.
This is in conflict with the large value $\mu = T_{U}$ used for our fine tuning. This means that even if the evaporation scenario proposed above is very appealing, the Maxwell-Boltzamn statistic is not the right framework to develop it.

This conclude our presentation of the gas of punctures with the Maxwell Boltzamn statistic in the grand canonical ensemble.

\subsection{The Bose Einstein statistic : entropy, logarithmic correction and condensation of the punctures}

 The goal of this section is to reproduce the statistical physics analysis outlined above, but now assuming that the punctures satisfy the Bose--Einstein statistics. We are going to show that, under certain conditions, the presence of a non-vanishing chemical potential leads to the elimination of the too large subleading corrections to the
entropy. The analysis is however more involved than in the case of punctures satisfying the Maxwell--Boltzmann statistics. Furthermore, we will see that at the semi-classical limit the corrections to the entropy are indeed logarithmic only when a condensation phenomenon appears, in the sense that the number $n_{1/2}$ of punctures carrying
a spin $1/2$ becomes much larger to the number of ``excited" (i.e. with spin $j>1/2$) punctures.

First, in order to simplify the analysis, let us assume that the degeneracy saturates the holographic bound. This means that $\delta_\text{h}=0$. The case $\delta_\text{h}\neq0$ will be treated later on. When $\delta_\text{h}=0$, our starting point is the computation of the grand canonical partition function $\mathcal{Z}_\text{B}(\beta,\mu)$ which, following \cite{ch3-Stat}, is defined by the expression
\begin{align*}
\mathcal{Z}_\text{B}(\beta,\mu)=\prod_j\mathcal{Z}_j(\beta,\mu),\qquad\text{with}\qquad\mathcal{Z}_j(\beta,\mu)=\big(1-z\exp(\beta_\text{U}E_j)\exp(-\beta E_j)\big)^{-1}.
\end{align*}
The product runs once again over the set of half-integers. There is an important difference between this black hole partition function and usual bosonic partition functions of quantum physics. The difference lies in the ``degeneracy" term $\exp(\beta_\text{U}E_j)$, which appears here in the denominator whereas ``degeneracies" typically appear in the nominator in standard quantum systems. The presence of this term here can be traced back to the holographic hypothesis for the degeneracy of quantum microstates. One consequence of this fact is that the partition function is defined only when the following condition is satisfied:
\begin{align}
(\beta-\beta_\text{U})E_j-\beta\mu>0,\qquad\forall\,j\in\mathbb{N}/2.
\end{align}
The mean energy and the mean number of punctures are respectively given by
\begin{align*}
\bar{E}=\sum_jE_j\bar{n}_j\qquad\text{and}\qquad
\bar{n}=\sum_j\bar{n}_j,\qquad\text{with}\qquad\bar{n}_j=\big(z^{-1}\exp\big((\beta-\beta_\text{U})E_j\big)-1\big)^{-1},
\end{align*}
and where $\bar{n}_j$ represents the mean number of punctures colored with the spin representation label $j$. The condition $(1.25)$  ensures that $\bar{n}_j$ is always positive and thus well-defined. In what follows, it will be useful to decompose the mean number of punctures as $\bar{n}=\bar{n}_{1/2}+\bar{n}_\text{ex}$, where the number of excited punctures is given by
\be
\bar{n}_\text{ex}=\sum_{j\geq1}\bar{n}_j.
\ee
The same decomposition can be done with the mean energy in the form $\bar{E}=\bar{E}_{1/2}+\bar{E}_\text{ex}$, with
\be
\bar{E}_{1/2}=E_{1/2}\bar{n}_{1/2},\qquad \text{and}\qquad\bar{E}_\text{ex}=\sum_{j\geq1}E_j\bar{n}_j.
\ee

We now want to tackle the study of the thermodynamical properties (in particular the entropy) of the system of punctures at the semi-classical limit. If we require only that the mean energy $\bar{E}$ (or equivalently the mean horizon area $\bar{a}_\text{H}$) become large in Planck units at the semi-classical limit, then this semi-classical limit is ill-defined. The reason for this is that, in the grand canonical ensemble, the system admits two intensive free parameters, which are the (inverse) temperature $\beta$ and the chemical potential $\mu$. Indeed, one cannot achieve the ``large" energy condition by tuning independently only one out of these two parameters. Therefore, we need an extra condition in order to define more precisely the semi-classical limit.

In the first subsection, we are going to impose that the temperature is fixed to $\beta_\text{U}$ at the semi-classical limit. Physically, this condition is easily understood from the point of view of quantum field theory in curved space-time. We will show that for the system to become semi-classical, its chemical potential must approach zero. This ensures that the punctures behave as photons at the semi-classical limit and that their number is not fixed.

In the second subsection, we are then going to assume that $\mu$ is fixed (at least in the semi-classical limit) to a non-vanishing (fundamental) value. In this case, we can show that $\beta\rightarrow\beta_\text{c}$ at the semi-classical limit, where $\beta_\text{c}$ is a priori different from the Unruh temperature. However, when $\mu$ is ``small", then $\beta_\text{c}$ approaches $\beta_\text{U}$.

The outcome of these two detailed computations will be the entropy $S_\text{B}$ (where the subscript $\text{B}$ refers to the Bose--Einstein statistics). We are going to show that the leading order term of $S_\text{B}$ reproduces as expected the Bekenstein--Hawking formula, but that the subleading corrections depend specifically on the choice of semi-classical regime. The corrections turn out to logarithmic only when the black hole exhibits a condensation phenomenon, i.e. when the spin $1/2$ representations are dominating in the sense that $\bar{n}_{1/2}\gg\bar{n}_\text{ex}$.

\subsubsection{First semi-classical regime, $\boldsymbol{\mu\rightarrow0}$ and $\boldsymbol{\beta\rightarrow\beta_{\text{U}}}$}
\label{subsec:mu zero}

\noindent Standard Bose--Einstein condensation (i.e. for usual systems) occurs for a gas of particles when one controls the total number $n$ of particles and the (inverse) temperature $\beta$. There are only two free parameters in the theory. The total energy and the fugacity (or equivalently the chemical potential) ``adapt'' themselves to a situation in which $n$ and $\beta$ are fixed. For instance, the fugacity is fixed by the equation $n=\bar{n}$, where $\bar{n}$ is the mean number of particles. In this section, we assume that the gas of punctures describing the black hole is defined at first sight in a similar way, i.e. that there exists a virtual physical process where one controls the number of punctures and the temperature of the black hole. From this point of view, (3.25) should be interpreted as a condition on the chemical potential of the system, i.e.
\be
\mu<\left(1-\f{\beta_\text{U}}{\beta}\right)E_j,\qquad\forall\,j\in\mathbb{N}/2.
\ee
This condition takes a very different form depending on wether we are above or below the Unruh temperature:
\bas
&&\text{if}\qquad\beta>\beta_\text{U},\qquad\text{then}\qquad\mu<\left(1-\f{\beta_\text{U}}{\beta}\right)E_{1/2},\\
&&\text{if}\qquad\beta<\beta_\text{U},\qquad\text{then}\qquad\mu=-\infty.
\eas
From this point of view, the value $\beta=\beta_\text{U}$ appears to be very peculiar, and marks a certain transition. At high temperature ($\beta < \beta_\text{U}$), the system necessarily has a negative infinite chemical potential, and therefore it could in principle be described by a classical Maxwell--Boltzmann statistics\footnote{However, the ``classical" Maxwell--Boltzmann partition function is only formally defined, and one should regularize properly the condition $\mu=-\infty$ in order to make it mathematically well-defined. There exists a simple and natural way to do so, which consists in first truncating the sum over $j$ in the canonical partition function $\mathcal{Q}(\beta)$ to a maximal value $j_\text{max}$, then fixing the chemical potential to an arbitrary value $\mu<(1-\beta_\text{U}/\beta)E_{j_\text{max}}$, and finally taking the limit where $j_\text{max}\rightarrow\infty$.}. At low temperature, the system is well-described by a quantum statistics. Therefore, a quantum-to-classical transition seems to occur at $\beta=\beta_\text{U}$. In what follows, we are going to focus on the ``quantum'' (or low temperature) regime $\beta>\beta_\text{U}$.

In what follows, we are first going to give a brief characterization of the low temperature regime and of the associated physical properties, and then we are going to derive these results more precisely. \\

\textit{Semi-classical limit and Bose--Einstein condensation} \\

\noindent As we said in the introduction to this section, we are interested in situations in which the mean energy (or equivalently the mean horizon area) becomes macroscopic (in Planck units). Indeed, this is the minimal requirement in order for the system to be semi-classical. For the time being and for the sake of generality, let us impose no specific conditions on the temperature other than $\beta>\beta_\text{U}$. From the expression (3.27), it is immediate to see that the mean energy becomes large at a fixed temperature only when the chemical potential approaches a particular value given by the condition $\epsilon=0$, where the quantity $\epsilon$ (which has the dimension of an energy) is defined by
\be\label{condition on mu}
\epsilon=\left(1-\f{\beta_\text{U}}{\beta}\right)E_{1/2}-\mu.
\ee
More precisely, the straightforward inequality
\be
\bar{E}>\bar{E}_{1/2}=\f{E_{1/2}}{\exp(\beta\epsilon)-1}
\ee
ensures that the horizon area can be arbitrarily large provided that $\epsilon$ is sufficiently small. On the contrary, if $\epsilon$ is finite, one can see that the mean energy is bounded from above, and therefore that the black hole cannot become semi-classical. In fact, when $\beta\neq\beta_\text{U}$, one recovers the usual Bose--Einstein condensation phenomenon which occurs at small temperatures (with respect to the Unruh temperature). Indeed, it is easy to see that the mean number of punctures carrying a spin label $1/2$ is given by
\be
\bar{n}_{1/2}=\f{1}{\exp(\beta\epsilon)-1},
\ee
which is not bounded from above and can take any arbitrarily large value provided that the chemical potential is such that\footnote{When $\beta_\text{U}=0$ (i.e. for usual statistical systems), this condition reduces to the well-known condition that the chemical potential $\mu$ must tend to the energy $E_{1/2}$ of the ground state.} $\epsilon\rightarrow0$. In contrast to this, the mean number $\bar{n}_\text{ex}$ of excited punctures is necessarily bounded according to
\be\label{series Nex}
\bar{n}_\text{ex}\leq\bar{n}_\text{ex}^\text{max},\qquad\text{where}\qquad\bar{n}_\text{ex}^\text{max}=\sum_{j\geq1}\f{1}{\exp\big((\beta-\beta_\text{U})(E_j-E_{1/2})\big)-1}.
\ee
Here we have assumed that we are below the Unruh temperature. Exactly as in Bose--Einstein condensation, when the total number $\bar{n}$ of punctures increases at a given fixed temperature and exceeds the maximal value $\bar{n}_\text{ex}^\text{max}$ for $\bar{n}_\text{ex}$, the punctures ``condensate'' in the spin $1/2$ representation. Therefore, the number of punctures carrying a spin $1/2$ becomes macroscopic.\\

\textit{Low temperature regime, $\beta\gg\beta_\text{U}$}\\

\noindent In order to make the previous observation more precise, it would be interesting to obtain a simple expression for $\bar{n}_\text{ex}^\text{max}$ in terms of $\beta$. Unfortunately, this is not possible in general because the series \eqref{series Nex} cannot be written in a simple closed form. Nevertheless, it becomes possible to simplify the expression for $\bar{n}_\text{ex}$ if we assume for instance that the temperature is way below the Unruh temperature, i.e. that $\beta\gg\beta_\text{U}$. This hypothesis is furthermore consistent with the fact that Bose--Einstein condensation occurs experimentally at ``low" temperatures. In this case, it is easy to show that
\be
\text{when}\qquad\beta\gg\beta_\text{U},\qquad\bar{n}_\text{ex}^\text{max}\simeq\exp\big(-\beta(E_1-E_{1/2})\big),\qquad\text{and}\qquad\bar{n}_{1/2}\simeq\f{1}{\beta(E_{1/2}-\mu)}.
\ee
Therefore, we can conclude that at very low temperatures there are almost no excited punctures, and that almost all of the punctures are colored with representations of spin $1/2$. Concerning the chemical potential, its values approaches the minimal energy $E_{1/2}$ at the semi-classical limit according to
\be
\mu\simeq E_{1/2}-\f{1}{\beta\bar{n}}.
\ee
However, the low temperature regime is not the one we are interested in because we would like the black hole temperature at the semi-classical limit (i.e. when the energy is large) to be given by the Unruh temperature and not by $T=0$.

This is the focus of the next subsection, which is devoted to the study of the properties of the system close to the Unruh temperature.\\

\textit{At Unruh temperature, $\delta_\beta\ll 1$}\\
\label{asymptotics}

\noindent Now, we assume that the temperature is very close to the Unruh temperature and we introduce, as in \cite{ch3-Alej3}, the parameter $\delta_\beta=\beta/\beta_\text{U}-1$ (so we have $\delta_\beta\ll 1$). Furthermore, we impose the condition $\epsilon\ll1$ which, in the case in which $\delta_\beta\ll 1$, is equivalent to $\mu \ll 1$ for the black hole at the semi-classical limit. Therefore, we study the case $\mu\rightarrow0$ and $\beta\rightarrow\beta_\text{U}$. As we said, the limit $\mu\rightarrow0$ is physically interpreted by the fact that systems of punctures (here quantum excitations of the gravitational field) behave in a way analogous to systems of photons when the chemical potential vanishes \cite{ch3-Stat}.

In addition to being physically relevant, the regime in which $\epsilon\rightarrow0$ and $\delta_\beta\rightarrow0$ (which implies $\mu\rightarrow0$) is also technically interesting because it allows for explicit calculations (for asymptotic quantities). In this regime, the mean number of particles in the representation of spin $1/2$ is necessarily large, and its asymptotic expansion is given by
\be\label{meanN12}
\bar{n}_{1/2}=\f{1}{\exp(\beta\epsilon)-1}\simeq\f{1}{\delta_\epsilon},\qquad\text{with}\qquad\delta_\epsilon=\beta_\text{U}\epsilon.
\ee
Contrary to what happens in usual Bose--Einstein condensation, the number of excited punctures is also large when $\delta_\beta$ approaches zero. This can be seen on the expression
\ba
\bar{n}_\text{ex}&=&\sum_{j\geq1}\bar{n}_j\nn\\
&=&\sum_{j\geq1}\f{1}{\exp\big(\beta\epsilon+\beta_\text{U}\delta_\beta(E_j-E_{1/2})\big)-1}\nn\\
&=&\sum_{k=2}^\infty\f{1}{\exp\left(\beta\epsilon+\pi\gamma\delta_\beta\big(\sqrt{k(k+2)}-\sqrt{3}\big)\right)-1}\nn\\
&\simeq&\f{1}{\pi \gamma \delta_\beta}\int_{\delta_\beta\Gamma}^\infty\f{\de x}{\exp(\delta_\epsilon)e^x-1},
\ea
where we have introduced $\Gamma=\beta_\text{U}(E_1-E_{1/2})$. The analysis of the last integral leads immediately to the result
\be\label{meanNex}
\bar{n}_\text{ex}\simeq-\f{\log(\Gamma\delta_\beta+\delta_\epsilon)}{\pi\gamma\delta_\beta},
\ee
which shows that the number of excited punctures increases when $\beta$ approaches the Unruh temperature. Therefore, at first sight, it seems that there is no condensation phenomenon at Unruh temperature. However, the situation is not that simple. Indeed, even if $\bar{n}_\text{ex}$ can be arbitrarily large when $\beta$ approaches $\beta_\text{U}$, it can still be negligible when compared to $\bar{n}_{1/2}$. This is the case when
\be\label{BEC condition}
\f{\bar{n}_\text{ex}}{\bar{n}_{1/2}}\simeq-\f{\delta_\epsilon}{\pi\gamma}\f{\log\delta_\beta}{\delta_\beta}\ll1,\qquad\Rightarrow 
\qquad\delta_\epsilon\ll-\f{{\delta_\beta}}{\log\delta_\beta}.
\ee
Conversely, there are situations in which at the semi-classical limit the number of punctures colored with spin $1/2$ is very small. This happens when the condition $\delta_\epsilon\gg-{{\delta_\beta}}/{\log\delta_\beta}$ is satisfied. Therefore, there are different semi-classical regimes depending on the way in which $\beta$ approaches $\beta_\text{U}$ compared to the way in which $\mu$ approaches $0$.

In order to have a more physical meaningful characterization of these conditions, it is useful to rewrite them as conditions involving the mean number $\bar{n}$ of punctures and the mean energy $\bar{E}$. This is in principle possible because, as we said previously, in the presence of a chemical potential the system admits two independent parameters which can be fixed independently and can be chosen freely. To achieve this we need to compute the mean energy, and a straightforward calculation shows that its asymptotic expansion is
\be\label{meanE}
\bar{E}=\bar{E}_{1/2}+\bar{E}_\text{ex},\qquad\text{with}\qquad\bar{E}_{1/2}\simeq E_{1/2}\f{1}{\delta_\epsilon};\qquad\text{and}\qquad\bar{E}_\text{ex}\simeq\f{\pi}{6\gamma\beta_\text{U}}\f{1}{\delta_\beta^2},
\ee
when $\delta_\epsilon\ll1$ and $\delta_\beta\ll1$. Now, we have all the necessary ingredient to classify the different semi-classical regimes. For this, we need to compare $\bar{n}$ given by \eqref{meanN12} and \eqref{meanNex} to $\bar{E}$ given by (3.40). We will assume here that $\gamma$ is a constant of order 1.
\begin{enumerate}
\item There is a semi-classical regime such that $\bar{n}\ll\beta_\text{U}\bar{E}$. For this to be the case, we must necessarily have $\delta_\epsilon\gg\delta_\beta^2$, and therefore the mean energy reduces to
\be
\bar{E}\simeq\f{\pi}{6\gamma\beta_\text{U}}\f{1}{\delta_\beta^2}.
\ee
Then, to extract the leading order term in the expression for $\bar{n}$, we have to further distinguish between the two following subcases:
\bas
\text{(a)}&&\qquad\big(\delta_\beta^2\ll\big)\delta_\epsilon\ll-\f{\delta_\beta}{\log\delta_\beta},\qquad\Rightarrow\qquad\bar{n}\simeq\bar{n}_{1/2}\simeq\f{1}{\delta_\epsilon},\\
\text{(b)}&&\qquad\delta_\epsilon\gg-\f{\delta_\beta}{\log\delta_\beta}\big(\gg\delta_\beta^2\big),\qquad\Rightarrow\qquad\bar{n}\simeq\bar{n}_\text{ex}\simeq-\f{\log(\Gamma \delta_\beta + \delta_\epsilon)}{\gamma \pi\delta_\beta} \,.
\eas
Note that there is a condensation in the first subcase. 
\item There is a semi-classical regime such that $\log\bar{n}\simeq\log(\beta_\text{U}\bar{E})$. For this condition to be satisfied, necessarily $\delta_\epsilon=\mathcal{O}(\delta_\beta^2)$, which implies immediately that
\be
\log(\beta_\text{U}\bar{E})\simeq-\log{\delta_\epsilon},\qquad\text{and}\qquad\bar{n}\simeq\f{1}{\delta_\epsilon}.
\ee
The first condition means that $\bar{E}$ scales as $\delta_\epsilon^{-1}$. Note that there is again a condensation in this semi-classical regime. To be more explicit and to extract precisely the leading order term in the expression of $\bar{E}$, one distinguishes the two different sub cases:
\bas
\text{(a)}&&\qquad\delta_\epsilon\simeq\alpha\delta_\beta^2\qquad\Rightarrow\qquad\bar{E}\simeq\left(E_{1/2}+\f{\alpha\pi}{6\gamma\beta_\text{U}}\right)\f{1}{\delta_\epsilon}\label{E2a},\\
\text{(b)}&&\qquad\delta_\epsilon\ll\delta_\beta^2\hphantom{\alpha}\!\qquad\Rightarrow\qquad\bar{E}\simeq E_{1/2}\f{1}{\delta_\epsilon}.\label{E2b}
\eas
These two regimes are qualitatively the same. 
\item There is no semi-classical limit such that $\bar{n} \gg \beta_\text{U} \bar{E}$. This is clear given the fact that $\bar{E}_j=E_j\bar{n}_j$ for any $j$, with $\beta_\text{U}E_j=2\pi\gamma\sqrt{j(j+1)}$. This implies that there exists a constant $C$ such that $\beta_\text{U}\bar{E}>C\bar{n}$. Therefore, it is not possible to have $\bar{n}\gg\beta_\text{U}\bar{E}$.
\end{enumerate}

Finally, one can say about the chemical potential that it vanishes at the semi-classical limit according to the following behavior:
\be
\mu\simeq E_{1/2}\delta_\beta-T_\text{U}\delta_\epsilon.
\ee
Note that the case $\mu=0$ has been studied in great details in \cite{ch3-Alej3}. With our notations, this situation is equivalent to setting $\delta_\epsilon=\beta_\text{U}E_{1/2}\delta_\beta$ (when $\beta$ is close to $\beta_\text{U}$), which corresponds to the case 1(b). As expected from quantum physics, there is no condensation with zero chemical potential. \\

\textit{Entropy and corrections to the area law} \\
\label{asymptotic entropy}

\noindent Now that the different (asymptotic) semi-classical regimes have been properly defined and classified, we can go further into the study of thermodynamical properties of the system. The next step is the computation of the semi-classical entropy $S_\text{B}$. The leading order term is easily obtained, and an immediate calculation shows that
\be
S_\text{B}=\f{\bar{a}_\text{H}}{4\lp^2}+S_\text{cor},\qquad\text{with}\qquad S_\text{cor}=\delta_\beta\beta_\text{U}\bar{E}-\beta\mu\bar{n}+\log\mathcal{Z}_\text{B}(\beta,\mu).
\ee
Now, let us compute $\log\mathcal{Z}_\text{B}(\beta,\mu)$ at the semi-classical limit. It is given by
\ba
\log\mathcal{Z}_\text{B}(\beta,\mu)&=&-\sum_j\log\big(1-\exp\big(\beta\mu+(\beta_\text{U}-\beta)E_j\big)\big)\\
&\simeq&-\log\big(1-\exp(-\delta_\epsilon)\big)-\sum_{j\geq1}\log\big(1-\exp\big(\delta_\beta\beta_\text{U}(E_{1/2}-E_j)-\delta_\epsilon\big)\big).
\ea
Here we have distinguished the spin $1/2$ contribution to the (logarithm of the) partition function from the excited contributions, and we have neglected terms proportional to $\delta_\beta\delta_\epsilon$ in the argument of the last exponential because we are interested only in the leading order terms. An immediate calculation now leads to the asymptotic expression
\ba
\log\mathcal{Z}_\text{B}(\beta,\mu)&\simeq&-\log\delta_\epsilon-\sum_{k=2}^\infty\log\left(1-\exp\left(-\pi\gamma\delta_\beta(\sqrt{k(k+2)}-\sqrt{3})-\delta_\epsilon\right)\right)\\
&\simeq&-\log\delta_\epsilon-\f{1}{\pi\gamma\delta_\beta}\int_{\Gamma\delta_\beta+\delta_\epsilon}^\infty\de x\log\big(1-e^{-x}\big).
\ea
Here we have used the fact that the series can be viewed as a Riemann sum and therefore can be approximated by its corresponding Riemann integral when $\delta_\beta\ll1$. The integral obtained in this way converges, and its main contribution, when the parameters $\delta_\beta$ and $\delta_\epsilon$ are small, is simply given by
\be
\int_{\Gamma\delta_\beta+\delta_\epsilon}^\infty\de x\log\big(1-e^{-x}\big)\simeq\int_{0}^\infty\de x\log\big(1-e^{-x}\big)=-\f{\pi^2}{6}.
\ee
Finally, we end up with the following expression for $\log\mathcal{Z}_\text{B}(\beta,\mu)$ in the semi-classical regime:
\be
\log\mathcal{Z}_\text{B}(\beta,\mu)\simeq-\log\delta_\epsilon+\f{\pi}{6\gamma\delta_\beta}.
\ee

We now have all the necessary ingredients to compute the subleading correction to the entropy as a function of the two independent parameters $\bar{E}$ and $\bar{n}$. There is no simple and explicit formula for $S_\text{cor}$ in general, but from the previous study we can say that the leading order term to $S_\text{cor}$ is necessarily given by the leading order term of the following sum:
\be\label{expansion of S}
\f{\pi}{3\gamma\delta_\beta}+\f{1}{\pi\gamma}\left(\beta_\text{U}E_{1/2}-\f{\delta_\epsilon}{\delta_\beta}\right)\log(\Gamma\delta_\beta+\delta_\epsilon)-\log\delta_\epsilon.
\ee
To write down the leading order term as a function of the macroscopic variables $\bar{E}$ and $\bar{n}$, we need to distinguish between the different cases that were studied in the previous subsection.
\begin{enumerate}
\item Case $\delta_\beta^2\ll\delta_\epsilon$. In this case, $\delta_\beta$ is directly related to the mean energy through
\be
\delta_\beta\simeq\sqrt{\f{\pi}{6\gamma\beta_\text{U}\bar{E}}}.
\ee
The subleading correction to the entropy is then dominated by the first term in \eqref{expansion of S}, which can be written as
\be
S_\text{cor}\simeq\sqrt{\f{2\pi}{3\gamma}\beta_\text{U}\bar{E}}.
\ee
For the sake of completeness, let us give formulae which express the remaining physical parameters $\mu$ and $\bar{n}$ in terms of the small parameters $\delta_\epsilon$ and $\delta_\beta$. In order to do so, we further need to distinguish between different subcases.
\begin{enumerate}
\item When $\delta_\beta^2\ll\delta_\epsilon\ll-\delta_\beta/\log\delta_\beta$, we have
\be
\delta_\epsilon\simeq\f{1}{\bar{n}},\qquad\text{and}\qquad\mu\simeq E_{1/2}\delta_\beta.
\ee
\item When $\delta_\epsilon\gg-\delta_\beta/\log\delta_\beta$, we have again to distinguish between different subcases.
\begin{enumerate}
\item If $\delta_\epsilon\gg\delta_\beta$, then
\be
\delta_\epsilon\simeq\exp\left(-\pi\bar{n}\sqrt{\f{\gamma \pi}{6\beta_\text{U}\bar{E}}}\right),\qquad\text{and}\qquad\mu\simeq-T_\text{U}\delta_\epsilon.
\ee
\item If $\delta_\epsilon\simeq\delta_\beta$, there exists a constant $C\in\mathbb{R}$ such that $\delta_\epsilon\simeq C\delta_\beta$, and then
\be
\delta_\epsilon\simeq C\sqrt{\f{2\pi}{3\gamma\beta_\text{U}\bar{E}}},\qquad\text{and}\qquad\mu\simeq(E_{1/2}-CT_\text{U})\delta_\beta.
\ee
\item If $\delta_\epsilon\ll\delta_\beta$, then the expression for $\delta_\epsilon$ in terms of $\bar{E}$ and $\bar{n}$ is rather complicated and not very useful. For this reason, we will not write it. However, the chemical potential is given by $\mu\simeq E_{1/2}\delta_\beta$.
\end{enumerate}
\end{enumerate}
\item Case $\delta_\epsilon=\mathcal{O}(\delta_\beta^2)$. We have seen that in this regime we have $\bar{E}\simeq E_0\bar{n}$, where $E_0=E_{1/2}$ when $\delta_\epsilon \ll\delta_\beta^2$ (see \eqref{E2b}), and $E_0=E_{1/2}+\alpha\pi\beta_\text{U}^2/(6\gamma)$ when $\delta_\epsilon\simeq\alpha\delta_\beta^2$ (see \eqref{E2a}). Therefore, we have that
\be
\delta_\epsilon\simeq\f{1}{\bar{n}},\qquad\text{and}\qquad\delta_\beta^2\simeq\f{\pi}{6\gamma}\f{1}{\beta_\text{U}(\bar{E}-E_{1/2}\bar{n})},
\ee
which implies that the subleading corrections are of the form
\be\label{log corrections}
S_\text{cor}\simeq\log(\beta_\text{U}\bar{E})+\f{\pi}{3\gamma}\f{1}{\delta_\beta}.
\ee
\end{enumerate}

The study of all these different asymptotic cases leads to the conclusion that the corrections to the entropy cannot be logarithmic in $\bar{E}$ (and therefore in $\overline{a}_H$) in case $(1)$, where they indeed turn out to be of the form $\mathcal{O}(\sqrt{\bar{E}})$. However, the correction can be logarithmic in case $(2)$ if, in addition, the condition  
$-\delta_\beta\log\delta_\epsilon\gg1$ (implying that $\delta_\epsilon^{-1}\gg\exp(\delta_\beta^{-1})$) is satisfied. In this case, we see that \eqref{log corrections} leads to
\be
S_\text{B}\simeq\f{\bar{a}_\text{H}}{4\lp^2}+\log\bar{a}_\text{H}.
\ee
The physical meaning of the condition $-\delta_\beta \log\delta_\epsilon\gg1$ is not so clear, but at the mathematical level it says that $\delta_\epsilon$ approaches zero much faster (in fact at least exponentially faster) than $\beta$ approaches $\beta_\text{U}$. This happens for instance when $\delta_\beta$ is small (compared to $\beta_\text{U}$) but constant. In this case, when the black hole horizon becomes larger and larger, a Bose--Einstein condensation occurs (at ``high" temperature). When expressed in terms of means values, the previous condition can be written as
\be
\bar{E}\simeq\bar{E}_{1/2}\gg E_{1/2}\exp\sqrt{\f{6\gamma\beta_\text{U}}{\pi}\bar{E}_\text{ex}} .
\ee
Then, asking that the energy of the punctures carrying a spin $1/2$ be exponentially larger than the square root of the excited energy, one gets logarithmic subleading corrections to the entropy.

\subsubsection{Second semi-classical regime, $\boldsymbol{\mu\neq0}$}
\label{subsec:mu non zero}

\noindent Now, we investigate the case in which the chemical potential is a fundamental constant of the theory (at least at the semi-classical limit), and does therefore not depend on the number of punctures or on the temperature. This is in sharp contrast to what usually happens in Bose--Einstein condensation. There is only one free parameter in the theory. Furthermore, we assume that there exists a physical process (matter collapse for instance) which causes the horizon area (or equivalently the energy of the black hole) to increase and to become macroscopic. The number of punctures and their colors change during this process, and we are going to show how these two quantities behave when the horizon area becomes macroscopic.

As opposed to the case in which $\mu\rightarrow0$, there exists here only one semi-classical regime for which the horizon area becomes macroscopic. We are going to see that this semi-classical regime is also characterized by a condensation of the spin labels to the value $1/2$. However, this condensation occurs at a temperature different from the Unruh temperature. \\

\textit{Asymptotic expansion of $\bar{n}$  and $\bar{E}$ at the semi-classical limit} \\

\noindent In order to characterize the semi-classical limit when $\mu$ is fixed, it will be convenient to use the following expression for the mean number of punctures colored by the spin $j$:
\be\label{Phi and delta}
\bar{n}_j=(\exp\Phi_j-1)^{-1},\qquad\text{with}\qquad\Phi_j=\beta_\text{U}(E_j-\mu)\delta_\mu+\mu\beta_\text{U}\f{E_j-E_{1/2}}{E_{1/2}-\mu},
\ee
where we have introduced
\be
\delta_\mu=\f{\beta-\beta_\text{c}}{\beta_\text{U}},\qquad\text{with}\qquad\beta_\text{c}=\f{E_{1/2}}{E_{1/2}-\mu}\beta_\text{U}.
\ee
Note that $\delta_{\mu=0}=\delta_\beta$ (where $\delta_\beta$ was introduced in sections). The system is the only defined when $\Phi_j>0$, which implies necessarily that
\be
\mu<E_{1/2},\qquad\text{and}\qquad\delta_\mu>0.
\ee
Since $\mu$ is now fixed, the condition on $\beta$ is now changed, and the semi-classical limit occurs a priori at $\beta_\text{c}\neq\beta_\text{U}$. However, we will see later on that it is possible to take the limit $\mu\rightarrow0$ in order to fix the inverse temperature to $\beta_\text{U}$ at the semi-classical limit.

Now, notice that we have
\be\label{Phi behavior}
\Phi_{1/2}=\beta_\text{U}(E_{1/2}-\mu)\delta_\mu,\qquad\text{and}\qquad\Phi_{j\geq1}>\mu\beta_\text{U}\f{E_j-E_{1/2}}{E_{1/2}-\mu}.
\ee
This ensures that $\Phi_{1/2}$ can be as close to zero as we want, whereas the quantities $\Phi_j$ have non-zero minima when $j\geq1$. These minima are reached when $\beta$ approaches $\beta_\text{c}$. These properties are the signature of a condensation phenomenon when $\beta$ approaches $\beta_\text{c}$.
Notice that this is a priori no longer true when $\mu=0$, since in this case every $\Phi_j$ can approach zero with no restriction when $\beta$ tends to $\beta_\text{c}$.

The bound \eqref{Phi behavior} on $\Phi_j$ implies immediately that $\bar{n}_\text{ex}$ and $\bar{E}_\text{ex}$ are bounded from above, and cannot exceed the maximal values $N_\text{ex}^\text{max}$ and $E_\text{ex}^\text{max}$ defined by
\ba
&&E_\text{ex}^\text{max}=\f{\gamma\pi}{\beta_\text{U}}\xi\left(\f{\mu\beta_\text{c}}{\sqrt{3}}\right),\qquad\text{with}\qquad\xi(x)=\sum_{k=2}^\infty\f{\sqrt{k(k+2)}}{\exp\left(x\big(\sqrt{k(k+2)}-\sqrt{3}\big)\right)-1},\\
&&N_\text{ex}^\text{max}=\zeta\left(\f{\mu\beta_\text{c}}{\sqrt{3}}\right),\hphantom{\f{\gamma\pi}{\beta_\text{U}}}\qquad\text{with}\qquad\zeta(x)=\sum_{k=2}^\infty\f{1}{\exp\left(x\big(\sqrt{k(k+2)}-\sqrt{3}\big)\right)-1}.
\ea
Since $\mu$ is fixed (to a non-vanishing value), $N_\text{ex}^\text{max}$ and $E_\text{ex}^\text{max}$ depend only on the fundamental constants of the theory.

One consequence of this fact is that when the mean energy $\bar{E}$ (or equivalently the mean horizon area) exceeds the value $E_\text{ex}^\text{max}$, only spin $1/2$ punctures contribute to the increasing energy (or mean horizon area), and we recover the Bose--Einstein condensation. Furthermore, the energy becomes macroscopic only when the temperature approaches the critical temperature, i.e. when $\beta\rightarrow\beta_\text{c}$. In this case, we have the asymptotic expansion
\be
\bar{E}_{1/2}=\f{E_{1/2}}{\beta_\text{U}(E_{1/2}-\mu)}\f{1}{\delta_\mu}+\mathcal{O}(1),
\ee
where $\bar{E}_{1/2}$ is viewed as a function of $\delta_\mu$. As a consequence, when the temperature approaches the critical temperature such that the mean area is large enough, i.e.
\be
\f{\bar{a}_\text{H}}{4\lp^2}>\gamma\pi\xi\left(\f{\mu\beta_\text{c}}{\sqrt{3}}\right),
\ee
then the number of excited punctures is ``saturated'' and any increase in the area of the horizon is due to the addition of punctures carrying a spin $1/2$. In this case, the mean area is related to the temperature as follows:
\be
\f{\bar{a}_\text{H}}{4\lp^2}=\f{\beta_\text{c}}{\beta_\text{U}}\f{1}{\delta_\mu}+\mathcal{O}(1).
\ee 

Outside of this regime, the equation of state is more complicated to obtain. For the same reasons, the asymptotic behavior of the mean number of punctures when $\delta_\mu\rightarrow0$ is given by
\be
\bar{n}=\bar{n}_{1/2}+\mathcal{O}(1),\qquad\text{with}\qquad\bar{n}_{1/2}=\f{1}{\beta_\text{U}(E_{1/2}-\mu)}\f{1}{\delta_\mu}+\mathcal{O}(1)=\f{1}{\gamma\pi\sqrt{3}}\f{\beta_\text{c}}{\beta_\text{U}}\f{1}{\delta_\mu}+\mathcal{O}(1).
\ee
Therefore, at the semi-classical limit, when the mean horizon area $\bar{a}_\text{H}$ becomes macroscopic, the number of punctures carrying spin $1/2$ increases drastically. There is a condensation of spin $1/2$ representations, exactly as there is a Bose--Einstein condensation of particles in the ground state in usual quantum physics. The condition for the condensation to occur is given by
\be\label{condensation condition}
\bar{n}_{1/2}\gg\zeta\left(\f{\mu\beta_\text{c}}{\sqrt{3}}\right).
\ee
\\

\textit{Entropy and corrections to the area law} \\

\noindent In order to compute the entropy, we need to compute the asymptotic expansion of $\log\mathcal{Z}_\text{B}(\beta,\mu)$ in the limit $\beta\rightarrow\beta_\text{c}$. Using the notations introduced above, the partition function is given by
\be
\log\mathcal{Z}_\text{B}(\beta,\mu)=-\sum_j\log\big(1-\exp(-\Phi_j)\big).
\ee
Again, we study separately the contribution from the punctures with spin $1/2$ and from the excited punctures. The contribution of the punctures carrying spin $1/2$ is given by
\be
-\log\big(1-\exp(-\Phi_{1/2})\big)=-\log\big(1-\exp\big(\beta_\text{U}(\mu-E_{1/2})\delta_\mu\big)\big)=-\log\delta_\mu+\mathcal{O}(1).
\ee
The contribution of the excited punctures is irrelevant for the asymptotic of $\log\mathcal{Z}_\text{B}(\beta,\mu)$ when $\beta$ approaches $\beta_\text{c}$. Indeed, when $\delta_\mu=0$ (i.e. when $\beta=\beta_\text{c}$), contribution from the excited punctures is well-defined (i.e. the series is convergent), as we now show. First, if $\beta=\beta_\text{c}$,
\be
\sum_{j\geq1}\log\big(1-\exp(-\Phi_j)\big)=\sum_{j\geq1}\log\left(1-\exp\left[\mu\beta_\text{c}\left(1-\f{E_j}{E_{1/2}}\right)\right]\right).
\ee
This series is easily proven to be convergent since the summand is equivalent (when $j$ becomes large) to
\be
\log\left(1-\exp\left[\mu\beta_\text{c}\left(1-\f{E_j}{E_{1/2}}\right)\right]\right)\simeq-\exp\left[\mu\beta_\text{c}\left(1-\f{E_j}{E_{1/2}}\right)\right]
\simeq-\exp\left[\mu\beta_\text{c}\left(1-\f{n}{\sqrt{3}}\right)\right],
\ee
whose series is convergent. As a consequence, we obtain the asymptotic expansion
\be
\log\mathcal{Z}_\text{B}(\beta,\mu)=\log\delta_\mu+\mathcal{O}(1).
\ee
The computation of the semi-classical expansion of the entropy $S_\text{B}=\beta(\bar{E}-\mu\bar{n})+\log\mathcal{Z}_\text{B}(\beta,\mu)$ is therefore immediate and leads to
\be
S=\f{\bar{a}_\text{H}}{4\lp^2}+\log\bar{a}_\text{H}+\mathcal{O}(1).
\ee
We recover the Bekenstein--Hawking expression with logarithmic corrections. \\
 
\textit{The limit $\mu\rightarrow0$} \\

\noindent To finish with this case, let us finally assume that we start with a finite value of $\mu$, first consider the limit $\delta_\mu\rightarrow0$ (as we did in this section), and then we assume that $\mu\rightarrow0$ together with the condition \eqref{condensation condition}. This condition ensures that a condensation occurs and also that the corrections to the Bekenstein--Hawking entropy are logarithmic.
 
To show that this is indeed the case, it is useful to establish a relationship between the classical regime describe here ($\mu$ fixed to a non-zero value) and the regimes described in the previous subsections. A straightforward calculation shows that the limit $\beta\rightarrow\beta_\text{c}$ corresponds to
\be
\delta_\beta=\delta_\mu+\f{\mu}{E_{1/2}-\mu}\rightarrow\f{\mu}{E_{1/2}-\mu},\qquad\text{and}\qquad T_\text{U}\delta_\epsilon=-\mu+\f{\beta-\beta_\text{U}}{\beta}E_{1/2}\rightarrow0.
 \ee
Therefore, if we first take the limit $\delta_\epsilon\rightarrow0$ with $\mu$ fixed, and then send $\mu$ to zero, we are clearly in case (2), where $\delta_\epsilon\ll\delta_\beta^2$. In this case, the correction to the entropy is indeed logarithmic.
 
\subsubsection{Holographic bound with a chemical potential}
\label{subsec:holographic mu}

\noindent This last subsection is devoted to the study of the case $\delta_\text{h}\neq0$, i.e. when we do not assume from the beginning that the holographic bound is saturated. The expressions for the grand canonical partition function grand canonical partition and the related thermodynamical quantities mean values are formally the same as above, simply with $\beta_\text{U}$ replaced by $(1-\delta_\text{h})\beta_\text{U}$. Therefore, the system now admits three free parameter: the temperature (or equivalently $\delta_\beta$), the chemical potential $\mu$, and the holographic parameter $\delta_\text{h}$. There is a priori more freedom to reach the semi-classical regime. \\

\textit{Semi-classical regime and holographic bound}\\

\noindent An immediate analysis shows that in order for the black hole to become macroscopic, we must impose that the parameter $\epsilon_\text{h}$ (which satisfies $\epsilon_\text{h}>0$), defined by
\be
\epsilon_\text{h}=\f{\beta_\text{U}}{\beta}E_{1/2}\delta-\mu,\qquad\text{with}\qquad\delta=\delta_\text{h}+\delta_\beta,
\ee
approaches zero, i.e. $\epsilon_\text{h}\ll1$. Note that $\epsilon_{\text{h}=0}=\epsilon$, with $\epsilon$ given by \eqref{condition on mu}.

If in addition to this we add the condition that the (inverse) temperature tend to $\beta_\text{U}$, then we get necessarily that $\delta_\beta\ll1$, and the condition $\epsilon_\text{h}\ll1$ becomes
\be
E_{1/2}\delta_\text{h}-\mu\ll1.
\ee
As shown in \cite{ch3-Alej3}, the requirements that the area be large and the temperature be fixed to $\beta_\text{U}$ necessarily imply that the holographic bound is saturated when there is no chemical potential $\mu=0$. We also recover what we have just shown in the previous section, namely that the two previous semi-classical requirements imply that the chemical potential vanishes in the semi-classical regime if we set $\delta_\text{h}=0$ from the beginning.

Here, we see that neither $\delta_\text{h}$ nor $\mu$ are uniquely fixed by the requirement of semi-classicallity, as opposed to what could have been expected from the introduction of a new free parameter in the model. However, in light of the physical reasons discussed in the previous section, an additional natural requirement is to ask that the chemical potential of the black hole be ``small"  ($\beta_\text{U}\mu\ll1$) in the semi-classical regime. This condition implies at the end of the day that the degeneracy bound must be saturated in the semi-classical regime, which can be summarized as follows:
\be
\text{Semi-classical regime}\quad\left(\beta\rightarrow\beta_\text{U},\qquad\mu\rightarrow0,\qquad\f{\bar{a}_\text{H}}{\lp^2}\rightarrow\infty\right),\qquad
\Rightarrow\qquad\delta_\text{h}\rightarrow0.
\ee
Adding the condition $\mu\rightarrow0$ as a requirement for the semi-classical limit makes the system equivalent (in this semi-classical limit) to the one studied in \cite{ch3-Alej3} where $\mu=0$. Therefore, it is not a surprise that we here recover the saturation of the holographic bound. \\

\textit{Entropy and corrections to the area law}\\

\noindent The study of the asymptotic expansion of the mean numbers of punctures ($\bar{n}_{1/2}$ and $\bar{n}_\text{ex}$) and the mean energies ($\bar{E}_{1/2}$ and $\bar{E}_\text{ex}$) is exactly the same as in subsection \eqref{asymptotics}. The only difference here is that we simply have to replace the small parameter $\delta_\beta$ by the small parameter $\delta=\delta_\beta+\delta_\text{h}$. However, the analysis of the semi-classical entropy and of the subleading corrections could differ.

Following exactly the same analysis as that of the previous subsection, we have
\be\label{S with h}
S_\text{cor}\simeq\beta_\text{U}(\delta_\beta\bar{E}-\mu\bar{n})+\log\mathcal{Z}_\text{B}(\beta,z),
\ee
whose leading order term is necessarily the leading order term of the sum
\be
\left(1+\f{\delta_\beta}{\delta}\right)\f{\pi}{6\gamma \delta}-\beta_\text{U}E_{1/2}\f{\delta_\text{h}}{\delta_\epsilon}+\f{1}{\pi\gamma}\left(\beta_\text{U}E_{1/2}-\f{\delta_\epsilon}{\delta}\right)\log(\Gamma\delta+\delta_\epsilon)-\log\delta_\epsilon .
\ee
Note that, at the difference with the mean energy and the mean number of punctures, the entropy is not symmetric under the exchange $\delta_\beta\leftrightarrow\delta_\text{h}$ because of the first term in \eqref{S with h}. For this reason, the behavior of the entropy at the semi-classical limit is different from what was studied in the previous subsection.

When $\delta_\text{h}=0$, we recover the asymptotic expansion \eqref{expansion of S}. If we assume that $\delta_\text{h}$ is ``small enough" in comparison to $\delta_\beta$ and $\delta_\epsilon$, the analysis of subsection \eqref{asymptotic entropy} still holds as well. The general case, in which $\delta_\text{h}=\mathcal{O}(\delta_\beta)$ (which means that $\delta_\beta$ does not tend to zero faster than $\delta_\text{h}$), is more subtle to study. Since it is not physically relevant for the present study, we will not perform its analysis here.

Here we are more interested in the case $\delta_\beta=0$, which means that we fix the temperature to be the Unruh temperature (as in \cite{ch3-Alej3}). The novelty compared to 
the previous subsection is the presence of a term proportional to $\delta_\text{h}/\delta_\epsilon$ in the asymptotic expansion \eqref{S with h}. This implies that for the subleading corrections to the entropy to be logarithmic it is necessary to satisfy the following requirements:
\begin{enumerate}
\item The term $-\log\delta_\epsilon$ dominate the asymptotic expansion, which implies in particular that 
\be
-\delta_\epsilon\log\delta_\epsilon\gg\delta_\text{h}.
\ee
\item The condition $\delta_\epsilon\ll\delta_\text{h}^2$ must hold in order to have $\delta_\epsilon\propto1/\bar{E}$ (i.e. we are in the case (2) studied above).
\end{enumerate}

These conditions are clearly contradictory. Therefore, the entropy cannot a priori have logarithmic corrections if we impose $\beta=\beta_\text{U}$ and $\delta_\text{h}\neq0$, even if $\delta_\text{h}$ is arbitrary small. \\

\subsubsection{Summary of the results for the Bose Einstein statistics}

The precedent discussion being quite long and technical, we summarize the different results obtain assuming that the punctures satisfy the Bose Einstein quantum statistic.
In this grand canonical situation, the system is characterized by two conjugated variables which are $(E, \beta)$ and $(\mu, n)$.
Since we are interested in semi classical black hole, i.e. with a large area (or equivalently with a large mean energy), we have to define properly our semi classical limit. \\

\textit{The first semi classical limit studied :  $\bar{E} \gg 1$ and $\beta$ fixed ( i.e. $\beta \gg \beta_{U}$ or $\beta \simeq \beta_{U}$)} \\

In this first case, $(\beta, \bar{n})$ are fixed while $(\bar{E}, \mu)$ are free parameters.
First, from the requirement of convergence of the partition function, one obtain the condition (26) on $\mu$.  
Two cases have to be distinguished regarding if the temperature is above or below the Unruh temperature $T_{U}$.
\begin{itemize}
\item The first one corresponds to $\beta < \beta_{U}$, i.e. to high temperature. This is the regim where the gas can be described by Maxwell Boltzman statistic (with $\mu \rightarrow - \infty$).

\item The second one correspond to $\beta > \beta_{U}$, i.e. to low temperature. This is the regim where the gas is well described by quantum statistics. In this regim, one introduces the quantity $\epsilon$. The link between the semi classical limit and the behaviour of $\epsilon$ is as follow:
 \begin{align}
\epsilon = ( 1 - \frac{\beta_{U}}{\beta} ) E_{1/2} - \mu ~ =   \left \{ \begin{array}{l}
\rightarrow 0 \; \;\;  \text{a semi classical limit exists / a condensation can occur }  \\
\text{finite} \;\;\; \text{ no semi classical limit /  no condensation } 
\end{array} \right .
\end{align}
When $\epsilon \rightarrow 0$ and the temperature is different but below the Unruh temperature, i.e. $\beta > \beta_{U}$, we can studied explicitly the two different asymptotic cases.

\begin{itemize}
\item The first one is the low temperature regim, when $\beta \gg \beta_{U}$. In this case, we are looking for a semi classical limit occurring at $T = 0$.
The number of punctures with spin $j= 1/2$ becomes large, i.e. there is a condensation and the chemical potential tends to the minimal energy $E_{1/2}$, i.e.:
\begin{align*}
\bar{n}_{1/2} \gg 1 \;\;\;\;\;\;\; \text{and} \;\;\;\;\;\;\; \mu \rightarrow E_{1/2}
\end{align*}

 \item The second one, more interesting for physical reasons, is when the semi classical limit occurs at the Unruh temperature $\beta = \beta_{U}$ or $\delta_{\beta} \ll 1$.
Since $\epsilon \ll 1$, these two requirements imply that $\mu \rightarrow 0$. Therefore, the punctures behave as photons and the energy required to annihilate a puncture is close to zero.
In this case, the situation is more subtle than before. Both $\bar{n}_{1/2}$ and $\bar{n}_{ex}$ can become large at the semi classical limit. They are given by:
\begin{align*}
\bar{n}_{1/2} \simeq  \frac{1}{\beta_{U} \epsilon} = \frac{1}{\delta_{\epsilon}} \;\;\;\;\;\; \text{and} \;\;\;\;\;\;\; \bar{n}_{ex} \simeq - \frac{\text{log} \delta_{\beta}}{\delta_{\beta}}
\end{align*}
Therefore, we need to compare them to classify the possible semi classical regims. The regim where a condensation occurs satisfy the following condition:
\begin{align*}
\frac{\bar{n}_{ex}}{\bar{n}_{1/2}} = - \frac{\delta_{\epsilon}}{\delta_{\beta}} \text{log} \delta_{\beta} \ll 1
 \end{align*}
Since the system depends on two free parameters $(\bar{E}, \mu)$, we can used the fixed value of $\bar{n}$ to classify the different regime, i.e. $\bar{n} \gg \beta_{U} \bar{E}$ , $\text{log}\bar{n} \simeq \text{log} {\beta_{U} \bar{E}}$ and $\bar{n} \ll \beta_{U} \bar{E}$. From those conditions, we distinguish three regims and five sub regims. The condensation phenomena occurs only in three sub regims and the quantum corrections to the entropy are logarithmic only in two of them, i.e. :
\begin{align*}
& \text{Case 1 : }\bar{n} \gg \beta_{U} \bar{E} \;\;\;\;\;  \delta^{2}_{\beta} \ll \delta_{\epsilon} \ll - \frac{\delta_{\beta}}{\text{log}\delta_{\beta}} \;\;\;\;\; \bar{n} \simeq \bar{n}_{1/2} \simeq 1 / \beta_{U} \epsilon \;\;\;\;\; S_{cor} \propto  \sqrt{a_{H}/l^{2}_{P}} \\
&\text{Case 2(a): }\text{log}\bar{n} \simeq  \text{log} {\beta_{U} \bar{E}} \;\;\;\;\;  \delta_{\epsilon} \simeq \alpha \delta^{2}_{\beta} \;\;\;\;\; \bar{n} \simeq \bar{n}_{1/2} \simeq 1 / \beta_{U} \epsilon \;\;\;\;\; S_{cor} \propto \text{log} (a_{H}/l^{2}_{P}) \\
&\text{Case 2(b): }\text{log}\bar{n} \simeq  \text{log} {\beta_{U} \bar{E}} \;\;\;\;\; \delta_{\epsilon}  \ll \delta^{2}_{\beta} \;\;\;\;\; \bar{n} \simeq \bar{n}_{1/2} \simeq 1 / \beta_{U} \epsilon \;\;\;\;\; S_{cor} \propto \text{log} (a_{H}/l^{2}_{P}) 
\end{align*} 

All the results for $\beta > \beta_{U}$, in the two asymptotic regimes $\beta_{U} \gg \beta_{U}$ and $\beta \simeq \beta_{U}$, are summarized in the table (3.1).\\

\begin{table}[]
\centering
\caption{Summary table}
\label{lolo}
\begin{tabular}{| l | l | l |}
\cline{1-3}
 & $\bar{E} \gg 1$ and $\beta$ fixed ( $\beta \gg \beta_U$)  & $\bar{E} \gg 1$ and $ \beta$ fixed ( $\beta  \simeq \beta_U$)   \\ \cline{1-3}
$\bar{n} \ll \beta_U \bar{E}$  & \begin{tabular}[c]{@{}l@{}}condensation phenomenon \\ log corrections\\ $\mu \rightarrow E_{1/2}$\end{tabular} & \begin{tabular}[c]{@{}l@{}}(2 subregims) condensation only for one \\ no log corrections \\ $\mu \rightarrow 0$\end{tabular} \\ \cline{1-3}
$\log{\bar{n}} \simeq \log{\beta_U \bar{E}}$ & impossible & \begin{tabular}[c]{@{}l@{}}(2 subregims) condensation \\ log corrections for both \\ $\mu \rightarrow 0$ \end{tabular}  \\ \cline{1-3} $ \bar{n} \gg \beta_{U} \bar{E}$ & impossible & impossible  \\ \cline{1-3}
\end{tabular}
\end{table}

\textit{The second semi classical limit studied : $\bar{E} \gg 1$ and  $\mu$ fixed (ie $\mu \neq 0$)}.\\

In this second case, we fixe the chemical potential $\mu$ to a given value.
We introduce the quantities:
\begin{align*}
\delta_{\mu} = \frac{\beta - \beta_{c}}{\beta_{U}} \;\;\;\;\; \text{where} \;\;\;\;\;\; \beta_{c} = \frac{E_{1/2}}{E_{1/2} - \mu} \beta_{U}
\end{align*}
The partition function is well defined only for:
\begin{align*}
\mu < E_{1/2} \;\;\;\;\;\;\;\;  \text{and} \;\;\;\;\; \delta_{\mu} > 0
\end{align*}
Computing $\bar{E}_{ex}$ and $\bar{n}_{ex}$, we note that they are bounded from above. Once $\bar{E} > \bar{E}^{max}_{ex}$, a condensation occurs.
In order to have a well defined semi classical limit, i.e. $\bar{E} \gg 1$, it is required that $\beta \rightarrow \beta_{c}$. In this case, the mean area (energy) and the mean number of punctures and the quantum corrections to the entropy are given by:
\begin{align*}
\frac{\bar{a}_{H} }{4l^{2}_{p}} \simeq \frac{\beta_{c}}{\beta - \beta_{c}} \;\;\;\;\;\;\; \bar{n} \simeq \bar{n}_{1/2} = \frac{1}{\gamma \pi \sqrt{3}} \frac{\beta_{c}}{\beta - \beta_{c}} \;\;\;\;\;\; S_{cor} \propto \text{log} (a_{H}/l^{2}_{P})
\end{align*}

\end{itemize}
\end{itemize}
 
In the two cases ($\mu$ free and $\mu$ fixed) we recover as expected the Bekenstein-Hawking formula for the entropy. Furthermore, we obtain logarithmic
corrections systematically when $\mu$ is fixed and also in the case $(2)$ when $\mu$ is not fixed. Recovering logarithmic corrections is a non trivial result. In those regime, while $\gamma$ is still present in the expressions of the mean number of punctures and mean energy, it drops out form the expression of the entropy.

Finally, when the chemical potential vanishes from the beginning then the corrections are much more larger than logarithmic corrections. 
Therefore, the presence of the chemical potential $\mu$ associated to the punctures which obey the Bose Einstein statistics seems to be highly related to the logarithmic corrections in this description of quantum black holes.

This concludes our discussion of the gas of punctures. 

\section{Limits of the model}

In this chapter, we have presented the quantum black hole as treated in Loop Quantum Gravity. 
Although this approach have successfully reproduced the semi classical results of black holes thermodynamics, it suffers from difficulties.

The very first one occurs in the micro canonical computation. We have seen that the Bekenstein Hawking area law is recovered up to a fine tuning on the Immirzi parameter.
This unnatural value have raised some doubts about the counting procedure or at a deeper level, on the quantization process. Why $\gamma$ should play such a crucial role in the quantum predictions of the theory such as the entropy of the quantum isolated horizon ? Should we understand it as a fundamental new constant, even if it doesn't play any role at the classical level ?
Different interpretations exist, but none are up to now totally convincing.

In order to avoid the fine tuning of $\gamma$ in the context of black hole thermodynamics, we have seen that the introduction of a chemical potential $\mu$ in the model could shift the dependency in $\gamma$ from the leading term to the subleading terms. More, assuming a Bose-Einstein statistic for the punctures, $\gamma$ disappear totally form the entropy expression even if it remains in the expression of other thermodynamical quantities.
Therefore, even if one recovers the right semi classical limit without fine tuning within this approach, i.e. the Bekenstein Hawking area law, the whole result is still $\gamma$ dependent, i.e. for the quantum corrections to the entropy. So from the point of view of the status of $\gamma$, the problem remains unsolved.

One way out is to adopt another interpretation concerning the status of $\gamma$ in the quantum theory. Based on a series of papers focusing on the self dual variables, i.e. on $\gamma= \pm i$, one is led to interpret $\gamma$ as a regulator in the theory allowing to wick rotate the real theory to the self dual one. The main lesson from those series of papers is that the self dual theory seems to represent a much more satisfying quantum theory with respect to different results, among which the semi classical limit of the quantum black holes.
From this point of view, $\gamma$ should be send to its purely imaginary value at some point of the quantization process, and therefore should disappear from the final quantum theory. It seems that the presence of $\gamma$ in this physical prediction of the quantum theory can be regarded as an anomaly due to the initial gauge fixing, i.e. the choice of working in the time gauge. This point will be explained and illustrated in chapter 5.

The second weakness of the ``gas of punctures'' model is the ad hoc ingredients that we have used to write down the partition function in the canonical and the grand canonical ensemble.
Two choices are made in the construction of the partition function. The first one it the choice of working with the Frodden-Ghosh-Perez notion of energy.
Even if it is a strong hypothesis, it is rather natural. Moreover, it led to an interesting entropy bound which should be investigated further and compare with the covariant entropy bound of Bousso.
The second choice is to use an exponential form for the total degeneracy, which we denote the holographic hypothesis. This is crucial in order to recover the right semi classical limit. However, this choice does not rely on any justification and even worse, is not predicted by the loop quantization of the isolated horizon based on the $SU(2)$ real Ashtekar-Barbero connection. Ideally, one would have chosen the degeneracy (the dimension of the Hilbert space) as predicted at the end of the loop quantization. Yet, the rest of this quantization is not sufficient to recover the right semi classical limit and one needs additional inputs. This point could be interpreted as a first evidence of the failure of the loop quantization based on the $SU(2)$ real Ashtekar-Barbero connection to reproduce the right semi classical limit.

We will see in the next chapter that a very elegant strategy can be applied to circumvent those problems. The idea is to define the quantum black hole from the self dual theory, i.e. with $\gamma = \pm i$.
In this case, there is obviously no need to fine tune $\gamma$. Starting from the Verlinde formula, one can build an analytic continuation of this formula which gives the dimension of the Hilbert space for the quantum black hole in the self dual quantum theory. This quantum degeneracy turns out to be naturally holographic and one does not need to postulate it. In this sense, the self dual theory seems to be a more satisfying candidate to reproduce the right semi classical limit. The construction of this analytic continuation is reviewed in details in the next chapter.

\clearemptydoublepage

\chapter{Black hole entropy with $\gamma = \pm i$}
\label{ch:ABH}
\minitoc

In this chapter, we present the construction of the self dual black hole in Loop Quantum Gravity. The starting point is the Verlinde formula presented in the precedent chapter.
This formula provides the quantum degeneracy of the spherically isolated horizon when working with real Loop Quantum Gravity, i.e. where $\gamma \in \mathbb{R}$.
Working with the self dual quantum theory implies to build an analytic continuation of this formula where the Immirizi parameter is sent to $\gamma = \pm i$.
This simple requirement has important consequences due to the definition of the isolated horizon. In the self dual quantum theory, the area of the hole remains positive and real only when the Chern Simons level is itself purely imaginary: $k  \in i \;  \mathbb{R}$. This observation is the starting point of the procedure. The Verlinde formula is rewritten as an integral of an holomorphic function on the complex plane, which enable us to control the analytic continuation from $k \in \mathbb{R}$ to $k \in i \;  \mathbb{R}$.
We present the details of the construction and obtain the quantum degeneracy for the self dual quantum isolated horizon. We perform then the thermodynamical study of this system first in the micro canonical ensemble.
Then we extend this study to the canonical and to the grand canonical situations following the same steps than the one presented in the precedent chapter. More precisely, we use the Frodden-Ghosh-Perez notion of local energy in order to write the partition function of our gas of punctures \cite{ch4-Amit1}. However, contrary to the precedent chapter, we do not need to postulate an holographic degeneracy to obtain the right semi classical limit.
The holographic degeneracy is computed directly form the ``self dual'' Verlinde formula obtained as a result of the analytic continuation. Moreover, the holographic behaviour of the degeneracy is supplemented with some power law corrections which at the effective level turns out to provide the expected logarithmic quantum corrections to the entropy.
More, we recover the fine tuning of the chemical potential in the grand canonical situation in order to remove the too large square root quantum corrections.
Already at the micro canonical level, this analytic continuation prescription provides a way to obtain exactly the right Bekenstein Hawking entropy at the leading term.
At the canonical and the grand canonical level, our ``self dual'' model seems to be more solid than the one presented in the precedent chapter, where $\gamma \in \mathbb{R}$, since the ad hoc hypothesis of the holographic degeneracy can be eliminated. The only assumption which remains is the use at the quantum level of the classical notion of energy provided by Frodden, Ghosh and Perez . 
Finally, this work permit us to derive for the first time a unique prescription which provide a map between a physical prediction of real Loop Quantum Gravity, i.e the entropy of the hole with $\gamma \in \mathbb{R}$ and a physical prediction of the self dual version of Loop Quantum Gravity, i.e. the entropy of the hole with $\gamma = \pm i$.

Before presenting the whole construction, let us review why $\gamma$ is expected to disappear form the quantum predictions of Loop Quantum Gravity.

\section{Getting rid of the Immirzi ambiguity}

In order to understand the peculiar status of the Barbero Immirizi parameter in Loop Quantum Gravity, it is interesting to come back on the reasons for its introduction.
As explained in the first chapter, the initial variables that were introduced by Ashtekar in the late eighties (1986), to simplify the first class constraints of General Relativity are complex valued \cite{ch4-Ash1}.
In term of those variables, the hamiltonain constraint of General relativity becomes polynomial w.r.t. the conjugated variables. This is a general fact that for gauge theories, the algebraic form of the hamiltonian simplifies drastically when written in term of certain complex variables. The self dual Ashtekar's variables for gravity belongs to this category. Having a polynomial hamiltonian constraint in gravity is of first importance in order to avoid ordering ambiguities and definition of quantum operator when implementing the dynamic at the quantum level. This simplicity of the constraints in the self dual Ashtekar's formalism fed the hope of quantizing the theory following the Dirac quantization program for constrained system. 

However, the drawback of this approach is that the conjugated self dual variables being complex valued, one has to impose reality conditions in order to recover a real metric for instance.
Those reality conditions, presented at the end of the first chapter, are highly non trivial since some of them involve the Levi Civita connection which has a complicated expression in term of the electric field.
Moreover, they involve the complex conjugate of the canonical variables, which bring another difficulties. Solving this reality conditions at the quantum level is up to now, an open issue.

Different proposal have appeared during the two last decades and an interested reader can refer to \cite{ch4-Alex1, ch4-R4}.
One of the first proposal was made by Thiemann \cite{ch4-Th1} twenty years ago (1995). We mention it in particular because the general philosophy is close to the strategy developed in this thesis.
In this work, he showed that for a general gauge theory (including gravity) which turns out to be simpler written in term of some complex variables, the real representation of the canonical commutation relations (supplemented with certain conditions) can be mapped to an holomorphic representation of the same commutation relations, this map keeping in the same time the simplification provided by the complex variables. This map turns out to be unique. It is build from a so called complexifier and acts as a Wick rotation on the quantum theory. Although very encouraging, this work remains formal and no physics from the self dual quantum theory could be extracted.

At the same epoch (1994), Barbero introduced the real $SU(2)$ connection nowadays called the Ashtekar-Barbero connection \cite{ch4-Bar1}, which depends on a free real parameter $\gamma$. The striking result was that in this formulation of lorentzian gravity, the interesting features of the complex variables of Ashetkar were preserved without requiring the imposition of some reality conditions. Above all, this new real commutative connection allows to introduce the loop variables and to follow the quantization program of Loop Quantum Gravity (at least at the kinematical level) \cite{ch4-Th2}. However, in order to avoid the presence of the reality conditions, the hamiltonian constraint takes a more complicated form. Two years later, Immirzi investigated the impact of this free parameter $\gamma$ in the kinematical predictions of the quantum theory, nowadays called the Barbero-Immirzi parameter $\gamma$. This parameter labels the family of canonical transformation sending the ADM phase space (written in the first order formalism) to the real Ashtekar-Barbero phase space. Thus, $\gamma$ does not play any role at the classical level. It is therefore expected that those canonical transformation corresponds to unitary transformation at the quantum level which do not alter the predictions of the quantum theory. Unfortunately, this is not what is found.

The Barbero-Immirzi parameter enters explicitly in the kinematical spectrum of the geometric operators, such as the area and volume operator. 
The spectrum of the area operator, derived in the second chapter, can be formally written as follow:
\begin{align*}
\hat{A}_{S}  \triangleright S( \Gamma, \vec{j}, \iota ) = 8 \pi \hbar G \gamma \sum_{j_{l}} \sqrt{j_{l}(j_{l}+1)} \; S( \Gamma, \vec{j}, \iota )
\end{align*}
where $\hat{A}_{S}$ is the area operator of a certain region of space acting on the quantum state $S( \Gamma, \vec{j}, \iota )$, i.e. a spin network.
One could argue that this ``prediction'' of Loop Quantum Gravity can not be trusted because it is purely kinematical and that we need therefore to impose the dynamic to conclude on the presence of $\gamma$ in the physical prediction of the quantum theory. This argument is receivable in the case of an arbitrary area such as the one considered above. It is not diffeomorphism invariant.
However, in the context of black hole, the area of the horizon can be shown to be a physical area, in the sense that it is not defined by some coordinate but by intrinsic geometrical hypothesis which are diffeomorphism invariant.
Black hole, or isolated horizon provide therefore a way to study a true physical prediction of the quantum theory. As we have seen in the precedent chapter, already at the micro canonical level, the loop computation of the entropy results in a $\gamma$-dependent quantity. Only when one work with a particular value of this parameter that we recover the right semi classical limit, i.e.
\begin{align*}
S =  \frac{a_{H}}{ 4 l^{2}_{p}}  \qquad \text{iff} \qquad  \gamma= \frac{\log(d)}{\pi d} \qquad \text{with} \qquad d = 2j +1 
\end{align*}
This result is obtained by assuming the same spin $j$ for all the punctures. We see already that the fine tuning of $\gamma$ depends on this choice.
We can conclude that at least for the micro-canonical entropy of the quantum isolated horizon, real Loop Quantum Gravity does predict $\gamma$-dependent results.

This can be interpreted as an anomaly of the quantum theory for the following reason. It is a well known fact that the real Ashtekar-Barbero connection on which the quantum theory is based does not transform properly under time-diffeomorphism \cite{ch4-Sam1}. This could a priori generate some anomalies both at the kinematical level and at the dynamical level, such as the presence of $\gamma$ in the kinematical and physical prediction of the quantum theory. However, this problematic behavior of the real connection is cured when one works with the self dual connection, i.e. with $\gamma = \pm i$. It is therefore tempting to postulate that the initial self dual Ashtekar connection is the right variable to deal with the dynamic and perhaps with the semi classical limit of the quantum theory. This point of view is strongly supported by earlier works all pointing in the same direction \cite{ch4-Pranz1, ch4-Noui1}. Moreover, the interpretation of the status of the Immirzi parameter as a regulator allowing to wick rotate the theory was enhanced by a careful study of a toy model of three dimensional gravity that we present in the next chapter. Altought $2+1$ gravity is much more simpler than its $3+1$ counterpart, it provides a laboratory to test new idea in order to quantize gravity.
The idea that $\gamma$ should disappear at the end of the quantization process in Loop Quantum Gravity turns out to be true in this dimensional reduced model \cite{ch4-BA1, ch4-BA2, ch4-Geiller1}. Whether this result extends to the four dimensional theory is still to be understood.
Yet, the canonical analysis of the Holst action without fixing any gauge was developed by Alexandrov in \cite{ch4-Alex2}. Although very complicated, the resulting phase space is $\gamma$-independent. Consequently, the presence of $\gamma$ in the $SU(2)$ Ashtekar-Barbero phase space seems to be a gauge artifact and the physical predictions of the quantum theory based on this phase space should not depend on $\gamma$ \cite{ch4-Alex3, ch4-Alex4, ch4-Alex5}.

We will see that in the context of spherically isolated horizon, one can explicitly build an analytic continuation of the quantum degeneracy (i.e. of the micro canonical entropy) from $\gamma \in \mathbb{R}$ to the self dual case $\gamma = \pm i$. This will defined the so called ``self dual'' black hole in Loop Quantum Gravity. We will show that this new object does not suffer from the same ambiguity than its real counterpart, supporting the idea that we should use the ``self dual'' variables in order to obtain the right semi classical limit in Loop Quantum Gravity. The strategy is similar to the one proposed by Thiemann twenty years ago, with the difference that we are now able to extract some physics from the self dual quantum theory.

The first attempt to define this ``self dual'' black hole was proposed in \cite{ch4-Geiller2}. We first review their derivation and its limits and we present then the rigorous construction of the analytic continuation undertaken in \cite{ch4-BA3}.

\section{First attempt : analytic continuation of the Verlinde formula and its limits}

Motivated by the previous discussion, the first attempt to define the dimension of the black hole Hibert space with a purely imaginary Immirzi parameter was proposed in \cite{ch4-Geiller2}. In this work, the starting point was the Verlinde formula as a discrete sum. As we have seen in the previous chapter, this formula is the key object to derive the entropy of the black hole in the micro canonical ensemble in LQG. In this approach, a black hole can be described by a Chern Simon field living on a punctured $2$-sphere, i.e. the horizon, and the Verlinde formula gives the dimension of its Hilbert space:
\begin{equation}
g_{k}(n , d_{l}) = \frac{2}{2+k} \sum^{k+1}_{d=1} \sin^{2}(\frac{\pi d}{k+2}) \prod^{n}_{l=1} \frac{\sin(\frac{\pi}{k+2}dd_{l})}{\sin(\frac{\pi }{k+2}d)}.
\end{equation}

Mimicing this analytic continuation of the entropy for the BTZ black hole performed in \cite{ch4-Geiller3}, the first idea to analytically continue the Verlinde formula was to send the Chern Simons level to a purely imaginary value: $k = \pm i$. However, such a procedure leads to a number of difficulties, the first one being that the dimension of the Hilbert space remain complex at the end. Moreover, since $k$ enters in the upper bound of the sum of the Verlinde formula, one has to modify the sum and use its module $|k|$ to give a sense to the sum.
This idea was then replaced by another procedure which is based on the spins labels $j_{l}$.
In \cite{ch4-Geiller2}, the analytic continuation prescription is given by:
 \begin{equation}
\gamma = \pm i  \;\;\;\;\;\; \text{and} \;\;\;\;\;\;  j_{l} \rightarrow i s_{l} - \frac{1}{2} \;\;\;\;\;\;\;\;\;\;\; \text{therefore} \;\;\;\;\;\;\;\;\;\;\;\;  d_{l} = 2j_{l} + 1 =  2 i s_{l}
\end{equation}

This prescription maps the discrete $SL(2,\mathbb{C})$-representations, i.e. the spinorial representation of the $SU(2)$ subgroup, into the continuous representation of the $SU(1,1)$-subgroup.
Under this map, the usual area spectrum of LQG becomes:
 \begin{equation}
a_{H} (j_{l}) = 8 \pi l^{2}_{p} \gamma \sum_{l} \sqrt{j_{l}(j_{l} + 1)} =  8 \pi l^{2}_{p}  \sum_{l} \sqrt{s^{2}_{l} + 1/4}
\end{equation}
where we choose that $\sqrt{-1} = \mp i$ for $\gamma = \pm i$. The spectrum becomes continuous but remains real.
If we had used instead the discrete representations of the $SU(1,1)$ subgroup, we would have obtain a complex area spectrum. This is why the continuous principal serie is picked up among all.
In the semi classical limit, this new area spectrum reads:
 \begin{equation}
a_{H} (j_{l}) =  8 \pi l^{2}_{p}  \sum_{l} s_{l}  \;\;\;\;\;\;\;  \text{for} \;\;\;\;\;\;\; s_{l} \gg 1
\end{equation}
Now let us apply the precedent prescription and rederive the result obtain in \cite{ch4-Geiller2}.  
\begin{align*}
g_{k}(n, d_{l} = 2i s_{l}) & = \frac{2}{2+k} \sum^{k+1}_{d=1} \sin^{2}(\frac{\pi d}{k+2}) \prod^{n}_{l=1} \frac{\sin(\frac{\pi}{k+2}d \; 2 i \; s _{l})}{\sin(\frac{\pi }{k+2}d)} \\
& = \frac{2}{2+k} \sum^{k+1}_{d=1} \sin^{2 - n}(\frac{\pi d}{k+2}) \prod^{n}_{l=1} \sin(\frac{\pi}{k+2} \; d \; 2 i \; s _{l}) \\
& = \frac{2 \; (i)^{n}}{2+k} \sum^{k+1}_{d=1} \sin^{2 - n}(\frac{\pi d}{k+2}) \prod^{n}_{l=1} \sinh(\frac{2 \pi }{k+2}d \; s _{l}) 
\end{align*}

Working in the semiclassical limit, we can simplify the last line. Indeed, the area of the hole being related to the Chern Simon level, we have at the semi classical limit: $k \gg 1$ (large area).
Moreover, it was shown in \cite{ch4-Amit2} that the large spins dominate this regim, so we can use safely $s_{l} \gg 1$.
Using those approximations, we observe that the sum will be dominated by the exponential with the largest argument, i.e. $ d = k+1$:
\begin{align*}
g_{k \gg 1}(n , s_{l} ) & = \frac{2 \; (i)^{n}}{2+k} \sum^{k+1}_{d=1} \sin^{2 - n}(\frac{\pi d}{k+2}) \prod^{n}_{l=1} \sinh(\frac{2 \pi }{k+2}d \; s _{l}) \\
 & \simeq  \frac{(i)^{n}}{ 2^{n-1}k} \sin^{2-n}(\frac{\pi (k+1)}{k+2}) \prod^{n}_{l=1} \text{exp} ( \; \frac{2 \pi (k + 1)}{k+2} \; s _{l} ) \\
 & =  \frac{(i)^{n}}{2^{n-1} k} \epsilon^{2-n} \prod^{n}_{l=1} \text{exp} ( \; 2 \pi \; s _{l} ) \\
 & =  \frac{(i)^{n}}{2^{n-1} k} \epsilon^{2-n}  \text{exp} ( \; 2 \pi \; \sum^{n}_{l=1} \; s _{l} ) \\
 & =  \frac{(i)^{n}}{2^{n-1} k} \epsilon^{2-n}  \text{exp} ( \frac{a_{H}}{4 l^{2}_{p}}) \\
\end{align*}

Now, taking the logarithm of $g_{k \gg 1}(n, d_{l} )$, we obtain the result found in \cite{ch4-Geiller1}:
\begin{align*}
\text{log} \big{(} g_{k \gg 1}(n, s_{l} ) \big{)} & = \; \frac{a_{H}}{4 l^{2}_{p}} + ... \\
\end{align*}

This result is striking. Working with $\gamma = \pm i$, which means to work with the (anti)-self dual  Ashtekar variables, permit to recover exactly the Bekenstein Hawking area law at the leading term.
The problem of solving the reality condition is by passed by choosing the continuous representation in the principal serie of the $SU(1,1)$-representation. Indeed those representations are the only one which leads to a positive and real area spectrum in the self dual theory. The fact that we can derive the Hawking result from self dual Ashtekar gravity, without any fine tuning, points towards the peculiar status of the complex Ashtekar variables for quantum gravity. Up to now, no one knows how to quantize directly the complex Ashtekar gravity, because of the highly non trivial reality conditions to imposed at the quantum level.
However, the analytic continuation presented above could in principle open a new road. Indeed, one could first work out the quantization of gravity in term of the real Ashtekar Barbero variables as a first step, and then used the analytic continuation prescription to obtain the physical prediction of the self dual quantum theory as a second step.

However, although very encouraging, the precedent result suffers from at least two points. First, the derivation was not done in a rigorous fashion, and the analytic continuation has to be understood by a rigorous mathematical construction before any conclusion. Secondly, the sub leading term have not been derived and it is of first importance to obtain them.

We will now present the rigorous construction of the analytic continuation introduced above.
This work was undertaken in \cite{ch4-BA3} and clarify the two weak points explained above. 

\section{Second attempt : going to the complex plane}

Our goal is to define the dimension of the Hilbert space of the Chern Simons theory living on a punctured two sphere for $\gamma = \pm i$.
In the real theory, for $\gamma \in \mathbb{R}$, the dimension of the Hilbert space is given by the Verlinde formula:
\begin{equation}
g_{k}(n , d_{l}) = \frac{2}{2+k} \sum^{k+1}_{d=1} \sin^{2}(\frac{\pi d}{k+2}) \prod^{n}_{l=1} \frac{\sin(\frac{\pi}{k+2}dd_{l})}{\sin(\frac{\pi }{k+2}d)}.
\end{equation}
and the area of the hole, the Chern Simons level $k$ and the Immirzi parameter $\gamma$ are related through the following relation:
\begin{equation}
\frac{a_{H}}{l^{2}_{p}} = \frac{2 \pi \gamma k}{(1- \gamma^{2})}
\end{equation}
Other variant of this expression exist in the litterature but their precise form do not change the conclusion.
The very first observation that one can make when trying to define the precedent formula for $\gamma = \pm i$ is that we need to send at the same time the Chern Simons level $k$ to a purely imaginary value, in order to keep the area of the hole real and positive. We give the transformation:
\begin{equation}
\gamma = \pm i  \;\;\;\;\;  \text{imply} \;\;\;\;\;\;  k = \mp i \lambda  \;\;\;\;\;\; \lambda \in \mathbb{R}^{+} \;\;\;\;\;\;\;\;   \rightarrow \;\;\;\;\;\;\;   \frac{a_{H}}{l^{2}_{p}} =  \pi \lambda \in \mathbb{R}^{+}
\end{equation}
Now as we have argue before, to study the consequences of sending the Chern Simons level $k$ to a purely imaginary value, the Verlinde formula written as a discrete sum is not well suited.
Indeed, $k$ enters in the upper bound of the sum and make the analytic continuation ambiguous.
However, one can reexpress the sum in at least two ways. 

First, the Verlinde formula can be understood as the Riemann sum of a function $f(\theta)$ on the interval $[0, \pi]$, the spacing of the sum being $\frac{\pi}{k +2}$.
It reads:
\begin{equation}
g_{k}(n , d_{l}) =  \frac{2}{\pi} \int^{\pi}_{0} d \theta \; f(\theta)   \;\;\;\; \;\;\;\; \text{where} \;\;\;\; \;\;\;\;  f(\theta) = \sin^{2} \theta \prod^{n}_{l=1} \frac{\sin(d_{l} \theta)}{\sin \theta} 
\end{equation}
Unfortunately, the Chern Simons level remains hidden in the spacing of the Rieman sum and we cannot go further with this expression.

The second way is to interpret the Verlinde formula as the sum of the residues of an holomorphic function $f(z)$ in the complex plane. This integral reads:
\begin{equation}
g_{k}(n, d_{l}) =  \frac{i}{\pi} \oint_{\mathcal{C}}  dz \; \sinh^{2} z \prod^{n}_{l=1} \frac{\sinh (d_{l} z)}{\sinh z} \coth((k+2) z) 
\end{equation}

For the rest of the discussion, we use the following notation : $F(z) = \coth((k+2) z)$ while the rest of the integrand is called $G(z)$.
The poles of the integrand are located on the imaginary axis at $z_{p} = \frac{i \pi p}{k +2}$ and come only from the term $F(z)$. The contour $\mathcal{C}$ goes around the imaginary axis, going through $(0, i\pi)$.
We choose to work with this expression because the Chern Simons level $k$ appears explicitly and we have therefore a better control on it. We will see that sending $k$ to a purely imaginary value modify the poles of the integrand and we will derive the unique analytical continuation prescription leading to a consistent result. Before diving in the technicalities, we mention the very pedagogical review of Witten on the analytic continuation of Chern Simons theory which turns out to be very helpful for our purposes \cite{ch4-Witten1}.

 \begin{figure}
\begin{center}
	{\includegraphics[width = 0.3\textwidth]{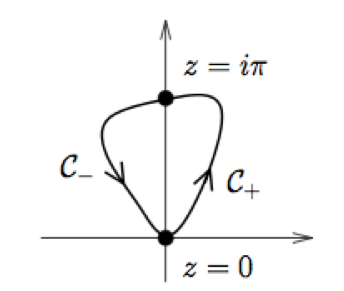}}
	\caption{Contour $\mathcal{C} =\mathcal{C}_{+} + \mathcal{C}_{-}$.}
\end{center}
\end{figure}

First we need to clarify a point concerning the countour we have choosen. We are interested in the semi classical limit, where the area $A$ is large, i.e. where $k \gg 1$. However, the function $F(z)$ is not continuous when taking the large $k$ limit: 
\begin{align}
 \;  \lim\limits_{k \to \infty}  \; \coth((k+2) z) = \frac{1 + e^{- 2 (k +2) R(z)} e^{- 2 (k+2) Im(z)}}{1 - e^{- 2 (k +2) R(z)} e^{- 2 (k+2) Im(z)}}  \rightarrow  \left \{ \begin{array}{l}
1\;  \text{if} \;  R(z) > 0 \\
- 1 \; \text{if} \; R(z) < 0 
\end{array} \right .
\end{align}
This limit is valid for $z \in \mathbb{C} / i \mathbb{R}$. When $z$ is purely imaginary, the precedent expression diverges.
Therefore, the function $F(z)$ being ill defined on the imaginary axis, we need to divide the countour (see Figure $2$) into two pieces: $\mathcal{C} = \mathcal{C}_{+} + \mathcal{C}_{-}$ where $\mathcal{C}_{+}$ goes from $z = 0$ to $z= i \pi + \epsilon$ with $R(z) > 0$ and $\mathcal{C}_{-}$ goes from $z = i \pi - \epsilon$ to $z= 0$ with $R(z) < 0$. Then, we have to take the limit: $\epsilon \rightarrow 0$.
The integral becomes:
\begin{align*}
g_{\infty}(n, d_{l})  & = \lim\limits_{k \to \infty} \frac{i}{\pi} \int_{\mathcal{C_{+}}}  dz \sinh^{2} z \prod^{n}_{l=1} \frac{\sinh (d_{l} z)}{\sinh z} \coth((k+2) z)  + \lim\limits_{k \to \infty}  \frac{i}{\pi} \int_{\mathcal{C_{-}}}  dz \sinh^{2} z \prod^{n}_{l=1} \frac{\sinh (d_{l} z)}{\sinh z} \coth((k+2) z) \\
& =  \; \frac{i}{\pi} \int_{\mathcal{C_{+}}}  dz \; \sinh^{2} z \prod^{n}_{l=1} \frac{\sinh (d_{l} z)}{\sinh z}  -  \frac{i}{\pi} \int_{\mathcal{C_{-}}}  dz \; \sinh^{2} z \prod^{n}_{l=1} \frac{\sinh (d_{l} z)}{\sinh z} \\
& =  \; \frac{2 i}{\pi} \int_{\mathcal{C_{+}}}  dz \; \sinh^{2} z \prod^{n}_{l=1} \frac{\sinh (d_{l} z)}{\sinh z} \\
\end{align*}

 Now that this point has been make clear, we can come back to the previous expression and study the poles of the integrand depending on the nature of the two free parameters $(k, d_{l})$.
Let us study the different cases. We start from the following expression:
\begin{equation}
g_{k}(n , d_{l}) =  \frac{i}{\pi} \oint_{\mathcal{C}}  dz \; \sinh^{2} z \prod^{n}_{l=1} \frac{\sinh (d_{l} z)}{\sinh z} \coth(k \; z) 
\end{equation}
where we have redefine the Chern Simons level $k' = k+2$ and that we note $k$ again.

\begin{itemize}
\item The first case corresponds to work with $\gamma \in \mathbb{R}$, i.e. to work with the real Ashtekar-Barbero variables.
The Chern Simons level $k$ and the colors of the spin $d_{l}$ are integers, i.e. $(k,d_{l}) \in \mathbb{N}$.
$F(z)$ admits poles on the imaginary axis at:
\begin{equation}
z_{p} = i \;  \frac{\pi p}{k} \;\;\;\;\;\;\;\;  p \in \mathbb{N} \;\;\;\;\;\;\;\;  \text{with} \;\;\;\;  (k,d_{l}) \in \mathbb{N}
\end{equation}
while $G(z)$ do not admit any poles. 
This is the usual case leading to the area law provided we fix the real Immirzi parameter $\gamma$ to a fix valued. This case was work out in details in the precedent chapter.

\item Now what happen if we ask for $\gamma = \pm i$ ? In this case, we need to send the Chern Simons level to a purely imaginary value.
Therefore, the second case corresponds to $k = i \lambda$  with $\lambda \in \mathbb{R}^{+}$ while $d_{l}$ remains an integer, i.e. $d_{l} \in \mathbb{N}$.
In this situation, the poles of the functions $F(z)$ are rotated and belong now to the real axis at:
\begin{equation}
z_{p} = -  \;  \frac{\pi p}{\lambda} \;\;\;\;\;\;\;\;  p \in \mathbb{N} \;\;\;\;\;\;\;\;  \;\;\;\;  \text{with} \;\;\;\;  k \in i\mathbb{R} \;\;\; \text{and} \;\;\;\; d_{l} \in \mathbb{N}
\end{equation}
The function $G(z)$ still do not admit any poles.
Since all the poles of the integrand are on the real axis, the countour $\mathcal{C}$ doesn't enclose any poles and the integral vanish by mean of the residues theorem.
Therefore, if we restrict the analytic continuation to this prescription, i.e. $k \in i \mathbb{R}$ , the result is inconsistent and do not lead to a well defined dimension of the Hilbert space for $\gamma = \pm i$.

\item We can now generalize and study a third case where the Chern Simons level $k  \in \mathbb{C}$ is complex and not just purely imaginary, i.e. $k= a + i b$.
In this case, the poles of $F(z)$ belong to a straight line in the complex plane which goes through:
\begin{equation}
z_{p} = -  \;  \frac{\pi p}{a^{2} - b^{2}} ( b + i a) \;\;\;\;\;\;\;\;  p \in \mathbb{N} \;\;\;\;\;\;  \;\;\;\;  \text{with} \;\;\;\;  k \in \mathbb{C} \;\;\; \text{and} \;\;\;\; d_{l} \in \mathbb{N}
\end{equation}
while the rest of the integrand $G(z)$ is still free of poles.
In this case, the area of the hole becomes complex and its expression reads:
\begin{equation}
\frac{a_{H}}{l^{2}_{p}} = \frac{2 \pi \gamma k}{(1- \gamma^{2})}   \;\;\;\;\;\;  \rightarrow \;\;\;\;\;\;  \frac{a_{H}}{l^{2}_{p}} =  \pi ( - b + i a)
\end{equation}
Therefore, we disregard this case which lead to an unphysical notion of area.
For the moment, the attempt to define the dimension of the Hilbert space for $\gamma = \pm i$ lead to a non sense, i.e. either a vanishing result for the dimension of the Hilbert space, either a complex area for the black hole.
However, there is an elegant way to obtain a well defined result. This is the fourth and last case.

\item  In addition to sending the Chern Simons level to a purely imaginary value, i.e. $k = i \lambda$ with $ \lambda \in \mathbb{R}^{+}$, we also require that the colors of the spins become purely imaginary, i.e. $d_{l}= i s_{l}$ with $s_{l} \in i \mathbb{R}^{+}$.
In this case, both $F(z)$ and $G(z)$ admit poles. There are respectively located at:
 \begin{equation}
z_{p} = -  \;  \frac{\pi p}{\lambda}  \;\;\;\;\;\;\;\;\;  p \in \mathbb{N} \;\;\;\;\;\;\;\;  \text{and} \;\;\;\;\;\;\;\;\;  z_{m} = i \pi m  \;\;\;\;\;\;\;  m \in \mathbb{N}^{*}  \;\;\;\;\;\;\;  \text{with} \;\;\;\;  (k,d_{l}) \in i \mathbb{R}
\end{equation}
Since the integrand depends only on the two parameter $(k, d_{l})$, once we have send $k$ to a purely imaginary value, $d_{l}$ is the only free parameter with which we can modify the poles of the integrand.
Asking for $d_{l}$ purely imaginary is the unique way to obtain unambiguously poles on the imaginary axis, and therefore to have a non vanishing integral in the complex plane regarding the countour $\mathcal{C}$.
More, we will see that this analytical continuation is the unique one which leads to the Bekenstein Hawking area law.
\end{itemize}

A remark is worthwhile at this stage. Since the function $F(z)$ admit among its poles $z_{0}=0$, it could seem at first that the integral is ill defined. However, this divergence is cured by the term $\sinh^{2}(z)$ in the function $G(z)$.
There is therefore no problem with this pole. 

In light of the precedent discussion, we are led to define the analytic continuation prescription as follow:
\begin{equation}
\gamma = \pm i  \;\;\;\;\;  \text{imply} \;\;\;\;\;\;  k = \mp i \lambda  \;\;\; \text{with}\;\;\; \lambda \in \mathbb{R}^{+} \;\;\;\;\;\;\;\;  \text{and} \;\;\;\;\;\;  j_{l} = \frac{1}{2} ( i s_{l} - 1) \;\;\; \text{with}\;\;\;  s_{l} \in \mathbb{R}^{+} 
\end{equation}

The form of $j_{l}$ is not unique and the more general form compatible with our needs is given by : $j_{l} = \alpha i s_{l} - 1/2$ where $\alpha$ is not restricted.
However, the $1/2$ part is the only possibility which lead the colors $d_{l}$ to be purely imaginary, i.e. $d_{l} = 2 \alpha i s_{l}$.
To simplify the notation and work with $d_{l} = i s_{l}$, we choose to work with $\alpha = 1/2$.
With this prescription, the integral in the complex plane takes the following form:
\begin{equation}
g_{\lambda}(n, s_{l}) =  \frac{i}{\pi} \oint_{\mathcal{C}}  dz \; \sinh^{2} z \prod^{n}_{l=1} \frac{\sinh ( i s_{l}z)}{\sinh z} \coth(i \lambda \; z) 
\end{equation}

Since we are interested in the semi classical limit, we need to study the limit of $F(i \lambda z)$ for large $\lambda$. The precedent integral can be recast as:
\begin{align*}
g_{\lambda}(n, s_{l}) & =  \frac{i}{\pi} \oint_{\mathcal{C}}  dz \; \sinh^{2} z \prod^{n}_{l=1} \frac{\sinh (i s_{l}z)}{\sinh z} (-1 + \nu_{\lambda}) \\
& = g_{\infty}(n, s_{l}) + g_{cor}(s_{l})
\end{align*}
since:
\begin{align*}
\coth(i \lambda \; z) & = - 1 + \nu_{\lambda} \;\;\;\;\;\;\;  \text{where} \;\;\;\;\;\;  \nu_{\lambda} (i\pi)=  \frac{2}{1 - \exp ( 2  \lambda \pi )} \simeq e^{- 2 \lambda \pi}  \;\;\; \text{for} \;\;\; \lambda \gg 1
\end{align*}
Now we need to study $g_{cor}(s_{l})$ in the vicinity of $z_{m=1}$ and $z_{m = 0}$.
The behaviour of $g_{cor}(s_{l})$ in $z_{m=1} = i \pi$ is direct and converge to $0$ for $\lambda \rightarrow \infty$.
Its behavior in $z_{m=0} = 0$ do not bring any divergence because of the term $\sinh^{2}(z)$ in the integrand which take care of the apparent divergence.
Therefore, in the semi classical limit, i.e. $\lambda \gg 1$, the integral reduces to:
 \begin{align*}
g_{\infty}(n, s_{l}) & = -  \frac{i}{\pi} \oint_{\mathcal{C}}  dz \; \sinh^{2} z \prod^{n}_{l=1} \frac{\sinh (i s_{l}z)}{\sinh z}  \\
\end{align*}

At this stage, we have obtained a precise definition of the dimension of the Hilbert space for the quantum black hole when $\gamma = \pm i$.
This expression is valid for semi classical black holes, where it is assumed that $a_{H} \gg l^{2}_{p}$.
Now we can start from this expression to study the thermodynamical properties of the so defined ``complex'' black hole.


The very first step in order to compute those quantities is to evaluate the integral at the semi classical limit.
In order to do so, we note that the precedent integral, which gives the degeneracy of the black hole, can be recast into the form:
 \begin{align*}
g_{\infty}(n, s_{l}) & =  \frac{1}{i\pi} \oint_{\mathcal{C}}  dz \; \sinh^{2} (z) \; e^{n S(z)}   \;\;\;\;\;\;  \text{with} \;\;\;\;\;  S(z) = \log \big{(}  \frac{\sinh (i s z)}{\sinh z}  \big{)} \\
\end{align*}

In the precedent expression, we have assumed that all the punctures carry the same color $s$ and we denote this model the one color model. This simplifies the expression but we can easily relax this assumption and treat the model where there are several different colors. 
The factor in front of the integral has been rewritten in a more convenient form for the following.
The general form of the integral is very appealing to proceed to a semi classical study of the black hole. Indeed, since $n$ is large in the semi classical limit, we can use the stationary phase method to evaluate the integral in this regim. \\

\textit{Studying the critical points}\\

The first task is then to identify the critical points of the action $S(z)$. Their equation is given by:
 \begin{align*}
S'(z) = 0 \;\;\;\;\;\;\;\;\;  is \tan (z) = \tanh ( is z) \\
\end{align*}

To solve this equation, we decompose $z = x + i y$ and we obtain:
 \begin{align*}
\frac{\sin (2sx) + i \sinh(2sy)}{\sinh (2x) + i \sin(2y)} = \; s \;  \frac{\cos (2sy) + \cosh(2sx)}{\cosh (2x) + \cos(2y)}  \in \mathbb{R}
\end{align*}
Therefore, the imaginary part of the term on the left has to vanish, yielding the following condition:
 \begin{align*}
f(x) = g(y)  \;\;\;\;\;\;\;  \text{with} \;\;\;\;\;\;   f(x) = \frac{\sinh (2sy)}{\sin(2y)} \;\;\;\;\;\;\;\;  \text{and} \;\;\;\;\;\;\;  g(y)= \frac{\sin (2sx)}{\sinh (2x)} 
\end{align*}

Studying the behaviour of those two functions, we can show that: $|f| \leqslant f(0) = s$ while $|g| \geqslant g(0) = s$.
Therefore, the precedent equation is solved either for $x=0$ either for $y=0$.
The solutions are both on imaginary axis and on the real axis.

For $z=x \in \mathbb{R}$, the equation for the critical points becomes:
 \begin{align*}
\tan ( s x) = s \tanh (x)\\
\end{align*}
Graphically, we observe that the solutions are located close to the points were $\tan(sx) \rightarrow \infty$, i.e. at $sx_{n} \simeq \frac{\pi}{2} (2n +1)$. 
Consequently, we can write the critical points as $x_{n} = \frac{\pi}{2s} (2n +1) + \epsilon$.
To determine $\epsilon$, we note that $\tanh(x_{n}) \simeq 1$, which gives:
 \begin{align*}
\tan ( s x_{n}) = \tan ( (2n +1) \frac{\pi}{2} + \epsilon) = \frac{\sin( n\pi + \pi /2 + \epsilon)}{\cos(n\pi + \pi /2 + \epsilon)} = - \frac{\cos \epsilon}{\sin \epsilon} = \frac{1}{\epsilon} +\mathcal{O} (1)\simeq s  \;\;\;\;\;  \epsilon \simeq - \frac{1}{s}
\end{align*}
Therefore, the real critical points, i.e. $z= x$, are located at:
 \begin{align*}
x_{n} \simeq \frac{\pi}{2s} (2n +1) - \frac{1}{s}  \;\;\;\;\;\;\;  \text{and at} \;\;\;\;\  x=0
\end{align*}

For $z= iy \in i  \mathbb{R}$, we proceed to the same study. For those imaginary critical points, the equations becomes:
 \begin{align*}
\tanh ( s y) = s \tan (y)\\
\end{align*}

 \begin{figure}
\begin{center}
	\fbox{\includegraphics[width = 0.5\textwidth]{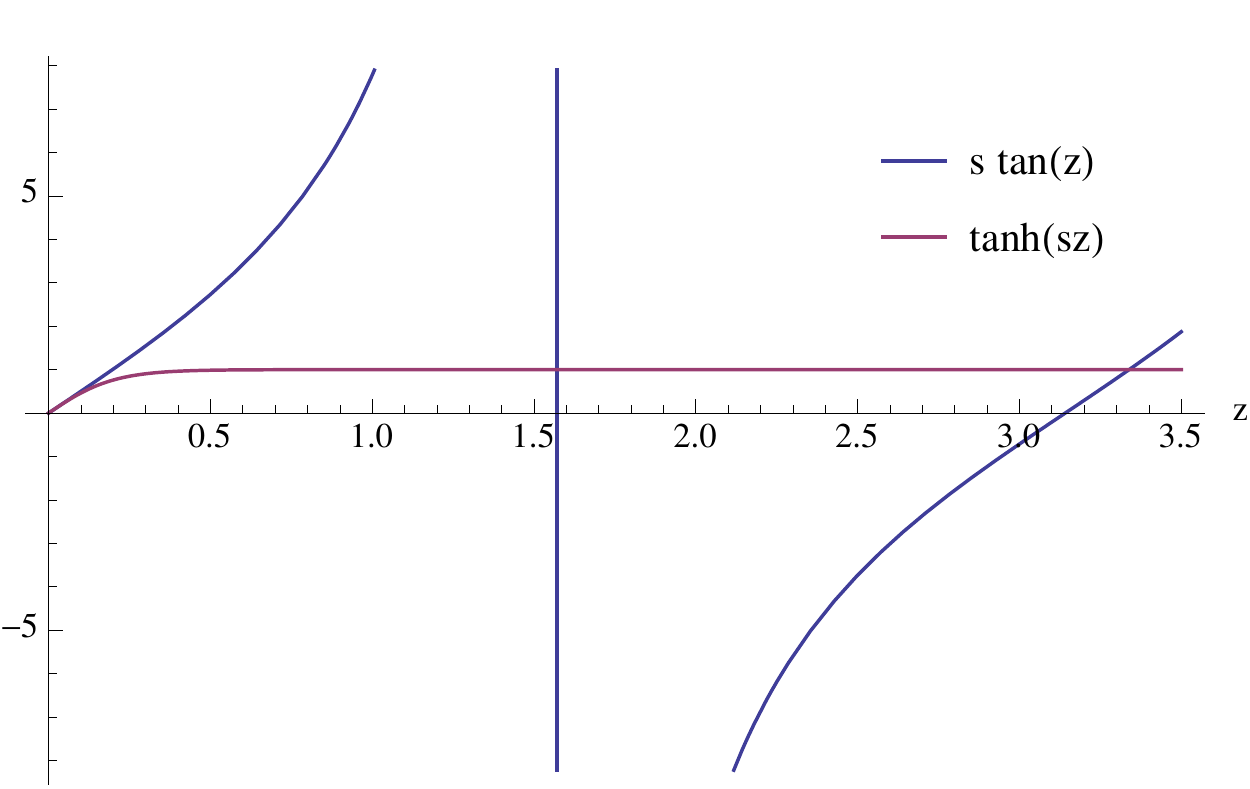}}
	\caption{Graphical representation of the first solutions to the equation for the imaginary critical points. While the first one is located at $y=0$, the second one is located at $y = \pi + \epsilon$ where $\epsilon \ll 1$. There are the two imaginary critical points which will be relevant for our choice of contour $\mathcal{C}$.}
	\label{fig:complex}
\end{center}
\end{figure}
  
Graphically, we observe that on the positif axis, the critical points are located at $y_{m} = m\pi + \epsilon$ with $m \in \mathbb{N}^{+}$. Because of our choice of contour $\mathcal{C}$, we will be concerned only by the interval $[0, 3 \pi /2]$. On this interval, there are two critical points at $y_{0} = 0$ and at a second point close to $y = \pi$ that we note $y_{1} = \pi + \epsilon$. We need to find the expression of $\epsilon$.
For this, using the equation for the critical points evaluated at $y_{1} = \pi + \epsilon$, we write:
 \begin{align*}
s \tan (\pi + \epsilon)  = s \frac{sin( \pi + \epsilon) }{cos ( \pi + \epsilon)} \simeq  s \epsilon + o(\epsilon)   \;\;\; \text{and} \;\;\; \tanh(sy_{1}) = \frac{1}{1+ e^{-2 sy_{1}}} - \frac{e^{-2sy_{1}}}{1+e^{-2sy_{1}}} \simeq 1  \;\;\; \text{for} \;\;\; s \gg 1
\end{align*}
therefore, we can conclude that at first order:
 \begin{align*}
 \epsilon = \frac{1}{s}  \;\;\;\;\; \text{whence} \;\;\;\; y_{1} = \pi + \frac{1}{s}
\end{align*}
Consequently, there are two imaginary critical points of interest for our choice of contour and there are located at:
 \begin{align*}
 y_{1} = i ( \pi + \frac{1}{s})  \;\;\;\;\;\;\;  \text{and} \;\;\;\;\;\; y_{0} = 0
\end{align*}

Now that we have identified all the critical points, we need to study how they will affect the large expansion of the integral.
For this, we evaluate the value of the real part of the action at those points in order to compare the dominant contribution.

The real critical points $x_{n}$ contribute as follow:
 \begin{align*}
 S( x_{n}) = \frac{i \pi}{2} + \log \big{(} \frac{\sin( s x_{n})}{\sinh x_{n}}  \big{)}  
\end{align*}
The real part of the action admits a maximum at $x_{0}=0$ and then decrease exponentially. Therefore, the point $x_{0}$ contributes the most at the semi classical limit and the other contributions are suppressed.
The real part of the action at $x_{0}$ reads $R(S(x_{0})) = \log(s)$.

The are two imaginary critical points $(iy_{0}, iy_{1})$. While $y_{0}$ is the same as $x_{0}$, we already know its contribution. The point $y_{1}$ contributes as follow:
 \begin{align*}
 S( y_{1}) & =  \log \big{(} \frac{\sinh( i s iy_{1})}{\sinh i y_{1}}  \big{)}  = \log \big{(} - \frac{\sinh( s y_{1})}{ i \; \sin y_{1}}  \big{)} = \log \big{(} - \frac{\sinh( s \pi + 1)}{ i \; \sin ( \pi + 1/s)}  \big{)} \simeq \log \frac{e^{s \pi +1}}{- i \frac{2}{s}} \\
 & \simeq - \frac{i \pi}{2} +  s \pi + \log \big{(}  \frac{e s}{2}\big{)} 
\end{align*}

Therefore when $s \gg 1$,  $S(iy_{1}) \gg S(x_{0}) $ and the large $n$ expansion of the integral is totally dominated by what happen at the point $z = i ( \pi + 1/s)$.
We can neglect the contribution of all the other critical points. This close the study of the critical points. We can now compute the large $n$ behaviour of the integral and obtain the dimension of the Hilbert space of the complex black hole in the ski classical limit. \\

\textit{Large n expansion of the integral}\\

Working in the one color model, the dimension of the black hole for $\gamma = \pm i$ in the semi classical limit, i.e. $k \gg 1$, is now defined as:
 \begin{align*}
g_{\infty} ( n, s ) = \oint_{\mathcal{C}} \mu(z) e^{n S(z)} \;\;\;\;\;\;  \text{with} \;\;\;\;\;  \mu(z) = \frac{1}{i \pi} sinh^{2} (z) \;\;\;\;  \text{and} \;\;\;\;\; S(z) = \log \big{(}  \frac{\sinh (i s z)}{\sinh z}  \big{)}
\end{align*}

where $n$ is the number of punctures and $s$ is the color carried by each punctures. From the previous discussion, we know that the large $n$ expansion is dominated by the contribution of the critical point $z_{c} = i( \pi + 1/s)$.
In the first part, we have choosen a countour $\mathcal{C}$ which went through the point $z = i \pi$. We will now shift this countour a bit above this point and let it go through the critical point $z_{c} =  i( \pi + 1/s)$.
This modification of the countour is the only one which lead to the Bekenstein Hawking result at the end of the computation.

 \begin{align*}
g_{\infty} ( n, s ) & = \oint_{\mathcal{C}} \mu(z) e^{n S(z)}  = \mu(z_{c}) e^{n S(z_{c})} \int^{\infty}_{-\infty} dx \; e^{n \frac{S''(z_{c})}{2} x} \\
& = \mu(z_{c}) e^{n S(z_{c})}  \sqrt{ \frac{2\pi}{- n S''(z_{c})}}
\end{align*}

We already know the value of $S(z_{c})$. The value of $\mu(z_{c})$ and of $S''(z_{c})$ are given by:
 \begin{align*}
\mu(z_{c}) & = \frac{1}{i \pi} \sinh^{2}(z_{c}) = \frac{1}{i \pi} \sinh^{2} ( i ( \pi + 1/s)) = \frac{- 1}{i \pi} \sin^{2} ( 1/s) \simeq \frac{i}{\pi s^{2}} \\
S''(z_{c}) & = - ( s^{2} + 1) + \frac{1}{\tanh^{2} (z_{c})} - \frac{s^{2}}{tan^{2} (sz_{c})} \simeq - s^{2}
\end{align*}

Plugging those results in the precent expression for $g_{\infty}(n,s)$, we obtain finally:
 \begin{align*}
g_{\infty} ( n, s ) & = \sqrt{\frac{2}{\pi}} \frac{1}{s^{3} \sqrt{n}} \big{(}  \frac{s e}{2} \big{)}^{n}  e^{-(1-n) i \pi /2} e^{ns\pi} 
\end{align*}

The final formula gives the dimension of the Hilbert space of the black hole for $\gamma = \pm i$, which corresponds to its quantum degeneracy.
We observe that the degeneracy has a phase that we can rewrite as:
 \begin{align*}
g_{\infty} ( n, s ) & = \sqrt{\frac{2}{\pi}} \frac{1}{s^{3} \sqrt{n}} \big{(}  \frac{s e}{2} \big{)}^{n}  (i)^{n-1}  e^{ns\pi}  
\end{align*}

Therefore, for $g_{\infty} ( n, s )$ to be real and positive, $n$ has to satisfy the following condition:
 \begin{align*}
n = 4 p + 1  \;\;\;\;\;\;  \text{with} \;\;\;\;\;\  p \in \mathbb{N}^{+}
\end{align*}
Since the formula for the degeneracy has been computed for $n \gg 1$, the only requirement is that $n$ belongs to $\{ 1, 5, 9, ... \; \}$.
Finally, we observe that the quantum degeneracy of the black hole has an holographic behaviour which is given by the term $e^{ns\pi} = e^{a_{H} / 4 l^{2}_{p}}$ supplemented with some power law corrections.
It is interesting to note that in this model ($\gamma = \pm i$), the holographic behaviour of the degeneracy is a result of a computation from first principles, while it has to postulated in the ``gas of punctures'' model for real black hole  ($\gamma \in  \mathbb{R}$) presented in the precedent chapter.

Having the explicit quantum degeneracy $g_{\infty} ( n, s )$ for the black hole, we are ready to compute the different thermodynamical quantities of interest.

\section{The thermodynamical study of the complex black hole}

In this section, we will derive the different thermodynamical quantities of interest for the semi classical black hole, i.e. its mean energy $\bar{E}$ ( or equivalently its mean area $\bar{A}$), the mean color for the punctures $\bar{s}$, the mean numbers of punctures $\bar{n}$ and finally its entropy $S$.

For the one color model, we note that the area of the block hole for $\gamma = \pm i$ is given by:
 \begin{align*}
a_{H} = 4 \pi l^{2}_{p} \sum^{n}_{l=1} \sqrt{s^{2}_{l} + 1}  \simeq  4 \pi l^{2}_{p} n s   \;\;\;\;\;\;  \text{for} \;\;\;\;\;\  s \gg 1
\end{align*}

Moreover, it was shown in [] , (and rederive in the precedent chapter) that the mean color $\bar{s}$ and the meannumber of punctures  $\bar{n}$ scale as $\sqrt{a_{H}}$. Therefore, we used the following expression for $n$ and $s$:
 \begin{align*}
 n = \nu \frac{\sqrt{a_{H}}}{l_{p}} \;\;\;\;\;\;\;  \text{and} \;\;\;\;\;\;\;  s = \sigma  \frac{\sqrt{a_{H}}}{l_{p}}
\end{align*}
where $(\nu, \sigma)$ are real constants that we will determine afterwards. Finally, for $n \in \{1, 5, 9 , ...\}$ but $n \gg 1$ and omitting the overall factor $\sqrt{\frac{2}{\pi}}$, the degeneracy reads:
 \begin{align*}
g_{\infty} ( n, s ) & = \frac{1}{s^{3} \sqrt{n}} \big{(}  \frac{s e}{2} \big{)}^{n}  e^{ns\pi}  
\end{align*}
\\

\textit{The micro canonical ensemble}\\

We first compute the entropy of the system in the micro canonical ensemble.
Its entropy is given by the logarithm of its quantum degeneracy that we have just computed.
 \begin{align*}
S_{MC} & = \log Z_{MC} = \log  \big{(} g_{\infty} (n,s) \big{)} \\
& = n s \pi + n \log \big{(}  \frac{s e}{2} \big{)} - 3 \log (s) - \frac{1}{2} \log (n) + \mathcal{O}(1) \\
& = \frac{a_{H}}{4l^{2}_{p}} + \frac{\nu \sqrt{a_{H}}}{l_{p}} \log \big{(}  \frac{\sigma e\sqrt{a_{H}}}{2l_{p}} \big{)} - 3 \log  \big{(}  \frac{\sigma \sqrt{a_{H}}}{l_{p}} \big{)} - \frac{1}{2} \big{(}  \frac{\nu \sqrt{a_{H}}}{l_{p}} \big{)} + \mathcal{O}(1) \\
& = \frac{a_{H}}{4l^{2}_{p}} + \frac{\nu \sqrt{a_{H}}}{l_{p}} \log \big{(}  \frac{\sqrt{a_{H}}}{l_{p}} \big{)}  + \frac{\nu \sqrt{a_{H}}}{l_{p}} \log \big{(}  \frac{\sigma e}{2} \big{)}- \frac{7}{2} \log  \big{(}  \frac{ \sqrt{a_{H}}}{l_{p}} \big{)} + \mathcal{O}(1) \\
& = \frac{a_{H}}{4l^{2}_{p}} + \frac{\nu \sqrt{a_{H}}}{2 l_{p}} \log \big{(}  \frac{a_{H}}{l^{2}_{p}} \big{)}  + \frac{\nu \sqrt{a_{H}}}{l_{p}} \log \big{(}  \frac{\sigma e}{2} \big{)}- \frac{7}{4} \log  \big{(}  \frac{ a_{H}}{l^{2}_{p}} \big{)} + \mathcal{O}(1) \\
\end{align*}

We recover at the leading term the Bekenstein Hawking area law which is in agreement with the first analytic continuation described in the precedent section. However, our approach gives us also the subleading term, i.e. the quantum corrections to the entropy. Those quantum corrections are usually expected to be logarithmic, mainly because very different approach recover this logarithmic behaviour. We observe that we obtain some logarithmic correction, i.e. the last term, plus some other contributions which are proportional to $\frac{\sqrt{a_{H}}}{ l_{p}} \log \big{(}  \frac{a_{H}}{l^{2}_{p}} \big{)}$ and to $\sqrt{a_{H}}$. Those supplementary contributions are too large w.r.t the expected logarithmic behaviour. However, as we shall see, we can add some input in the precedent computation to remove the second term.

As we have argue in the precedent chapter, one lesson of quantum mechanic and quantum field theory is that particles are indistinguishable. The usual way to implement this aspect of particles in statistical mechanics is to introduce a Gibbs factor in the partition function. The new partition function reads:
  \begin{align*}
Z_{MC} & =  \frac{g_{\infty} (n,s)}{n!}  \;\;\;\;\;\;\;\;  \text{where} \;\;\;\;\;\;  n ! \simeq \sqrt{n} \big{(} \frac{n}{e} \big{)}  \;\;\;\;\;  \text{for}  \;\;\;\;  n \gg 1\\
\end{align*}
where we have used the Stirling formula since $n$ is large at the semi classical limit.
With this new input, the micro canonical entropy reads:
 \begin{align*}
S_{MC} & = \log  \big{(} \frac{g_{\infty} (n,s)}{n!} \big{)} \\
& =  \log  \big{(} g_{\infty} (n,s) \big{)} - \log ( \sqrt{n} n^{n} ) + n \log (e) \\
& = \log  \big{(} g_{\infty} (n,s) \big{)} - \frac{1}{2} \log \big{(}  \frac{\sqrt{a_{H}}}{l_{p}} \big{)}  - \frac{\nu \sqrt{a_{H}}}{l_{p}} \log \big{(}  \frac{\sqrt{a_{H}}}{l_{p}} \big{)}   -  \frac{\nu \sqrt{a_{H}}}{l_{p}} \log \big{(}  \nu \big{)} +   \frac{\nu \sqrt{a_{H}}}{l_{p}} \log \big{(}  e \big{)} \\
& =  \frac{a_{H}}{4l^{2}_{p}}  + \frac{\nu \sqrt{a_{H}}}{l_{p}} \log \big{(}  \frac{\sigma e^{2}}{2 \nu} \big{)}- 2 \log  \big{(}  \frac{ a_{H}}{l^{2}_{p}} \big{)} + \mathcal{O}(1) \\
\end{align*}

The quantum corrections proportional to $\sqrt{a_{H}} \log \big{(} a_{H} \big{)} $ have been removed and we are left with some logarithmic quantum correction supplemented with quantum corrections proportional to $\sqrt{a_{H}}$.
The leading term is obviously not affected. This close the computation of the entropy of the black hole defined for $\gamma = \pm i$ in the micro canonical ensemble. \\

\textit{The canonical ensemble}\\

We compute now the entropy of the black hole, its mean energy, its mean number of punctures and the mean color of the punctures in the canonical ensemble.
To do so, we used the same strategy as in the precedent chapter and we adopt the ``gas of punctures '' picture for the quantum black hole. That is to say, we will use the same inputs we have introduced in the precedent chapter to write the partition function $Z_{C}(\beta)$.
Indeed, to do so, we need a local notion of energy that a local observer will assign to the horizon. This notion of energy is given by the Frodden-Gosh-Perez notion of energy and reads:
  \begin{align*}
E(n,s) = \frac{a_{H}(n,s)}{8 \pi L} 
\end{align*}
Its expression has been rederive and justify in the precedent chapter. We recall here that importing this classical notion of energy at the quantum level is the main hypothesis of the model.
From the beginning, we will consider the punctures as indistinguishable and introduce a Gibbs factor as above. With those assumptions, the partition function in the canonical ensemble reads:
  \begin{align*}
Z_{C} (\beta) = \int ds \; Z_{s}(\beta) = \int ds \sum_{n} \frac{g_{\infty} (n,s)}{n!} \; e^{- \beta E(n,s)}
\end{align*}
where we use an integral over the color since the area spectrum is now continuous.
In this one color model, where all the punctures carry the same spin $s$, we have:
  \begin{align*}
E(n,s) = \frac{l^{2}_{p}}{2L} ns = \frac{\pi n s}{\beta_{U}}  \;\;\;\;\;  \text{where} \;\;\;\;  \beta_{U} = \frac{2 \pi L}{ l^{2}_{p}}
\end{align*}

$\beta_{U}$ is the inverse Unruh temperature and has been derived in []. $L\ll 1$ is the small distance at which the local observer stands from the horizon.
With those notations, we have:
 \begin{align*}
Z_{s} (\beta) = \frac{1}{s^{3}} \sum^{\infty}_{n=1} \frac{1}{\sqrt{n}}  \frac{q^{n}}{n !} \;\;\;\;\; \text{with} \;\;\;\;\;  q = \frac{se}{2} \; e^{- xs}  \;\;\;\;  \text{and} \;\;\; x = \pi ( \tilde{\beta} - 1) \;\;\;\;  \tilde{\beta} = \frac{\beta}{\beta_{U}}
\end{align*}
We observe that $Z_{s}(\beta)$ is defined only for $ x > 0$, i.e. for $\beta > \beta_{U}$. The system is only defined for a temperature smaller than the Unruh temperature: $T < T_{U}$.
The thermodynamical limit corresponds to $x \rightarrow 0$, or equivalently to $\beta \rightarrow \beta_{U}$. At the semi classical limit, $s$ and $q$ are large.

Let us compute the sum entering in $Z_{s}(\beta)$.
We have:
\begin{align*}
I (q) = \sum^{\infty}_{n=1} \frac{1}{\sqrt{n}}  \frac{q^{n}}{n !} & = \frac{1}{\sqrt{\pi}} \int^{+\infty}_{-\infty} du \; \sum^{\infty}_{n=1} \frac{(q e^{-u^{2}}) ^{n}}{n !}  \;\;\;\;\;\;\;\;  \text{where} \;\;\;\;\;   \frac{1}{\sqrt{n}}  = \frac{1}{\sqrt{\pi}} \int^{+\infty}_{-\infty} du \; e^{- n u^{2}} \\
& = \frac{1}{\sqrt{\pi}} \int^{+\infty}_{-\infty} du \; ( e^{q e^{-u^{2}}} - 1 ) 
\end{align*}
 The last integral  can be computed using the stationary phase method because $q$ is large in the semi classical regim. The integral is dominated by the critical point of $f(u) = e^{-u^{2}}$.
 We have that $f'(u) = 0$ for $u = 0$ and that $f''(u) = - 2$ at the same point. Therefore, the sum becomes:
 \begin{align*}
I (q) & = \frac{1}{\sqrt{\pi}} \int^{+\infty}_{-\infty} du \; ( e^{q e^{-u^{2}}} - 1 )  \simeq \frac{1}{\sqrt{\pi}} \int^{+\infty}_{-\infty} \text{exp} ( q( 1- u^{2})) du \\
& = \frac{e^{q}}{\sqrt{\pi}} \int^{+\infty}_{-\infty} e^{q u^{2}} du = \frac{e^{q}}{\sqrt{\pi}} \sqrt{\frac{\pi}{q}} = \frac{e^{q}}{\sqrt{q}}
\end{align*}

Therefore, the function $Z_{s}(\beta)$ reads:
 \begin{align*}
Z_{s} (\beta) = \frac{1}{s^{3}} \frac{e^{q}}{\sqrt{q}}
\end{align*}

We can now proceed to the study of the thermodynamical limit for the full canonical partition function $Z_{C}(\beta)$. This limit is reached for $x \rightarrow 0$.
 \begin{align*}
Z_{C} (\beta) & = \int^{+\infty}_{s_{0}} \frac{ds}{s^{3}} \frac{e^{q}}{\sqrt{q}} =  \int^{+\infty}_{s_{0}} \frac{ds}{s^{3}}  \big{(} \frac{2}{se} e^{xs} \big{)}^{1/2} \text{exp} \big{(} \frac{se}{2} e^{-xs}\big{)} \\
& = \int^{0}_{e^{-xs_{0}}} \frac{du}{- x u}  \big{(} \frac{x}{ \log(u)}\big{)}  \big{(} \frac{2x}{-  u \log(u) e} \big{)}^{1/2} \text{exp} \big{(} \frac{ \log(u) e}{2x} u\big{)} \;\;\;\;  \text{with} \;\;\; u = e^{-xs} \\
& =  \sqrt{\frac{2}{e}} \int^{e^{-xs_{0}}}_{0} \frac{du}{ u^{3}}  \;  \frac{x^{5/2}}{ (- \log(u))^{7/2}} \; \text{exp} \big{(} - \frac{ u \log(u)}{2x} \big{)}
\end{align*}

Since we are interested in the semi classical limit where $x \rightarrow 0$, we can proceed to a gaussian approximation of the precedent expression. The partition function will be dominated by the critical point of the function $f(u)= u \ln (u)$. It is direct to show that $f'(u)= 0$ for $u = e^{-1}$ and that $f''(u)= \frac{1}{e}$ at this point. We get
 \begin{align*}
Z_{C} (\beta) & =  \sqrt{\frac{2}{e}} e^{3/2} x^{5/2} \exp {\frac{1}{2x}} \; \int^{+ \infty}_{-\infty}  du \; \text{exp} \big{(} - \frac{ e^{2}}{2x} \frac{u^{2}}{2} \big{)} \\
& = \sqrt{\frac{2}{e}} e^{3/2} x^{5/2} \exp {\frac{1}{2x}} \; \sqrt{\frac{4 \pi x}{e^{2}}} = \sqrt{8 \pi} x^{3} \; \exp{\frac{1}{2x}}
\end{align*}

Using the expression of $x$, we obtain $Z_{C}(\beta)$ in term of the inverse temperature explicitly:
 \begin{align*}
Z_{C} (\beta) & = \sqrt{\pi}{8} \frac{l^{6}_{p}}{L^{3}} ( \beta - \beta_{U})^{3} \exp{\frac{L}{l^{2}_{p}(\beta - \beta_{U})}}  \;\;\;\;\; \text{with} \;\;\;\;\;  x = \pi ( \tilde{\beta} - 1) = \frac{l^{2}_{p}}{2L} ( \beta - \beta_{U})
\end{align*}

Now, we are ready to compute the mean energy $\bar{E}$ or equivalently the mean area of the hole $\bar{a}_{H} = 8 \pi L \bar{E}$. Those quantities are given by:
 \begin{align*}
\bar{E} & = - \frac{\partial}{\partial \beta} \; \log {Z_{C} (\beta) } =  - \frac{\pi}{\beta_{U}}\frac{\partial}{\partial x} \; \log {Z_{C} (\beta) }\\
\bar{a}_{H} & = - \frac{8 \pi^{2} L}{\beta_{U}} \frac{\partial}{\partial x} \; \big{\{ } \; 3 \log{x} +  \frac{1}{2x} \; \big{\} } = \frac{8 \pi^{2} L}{\beta_{U}} \; \big{\{ } \; \frac{1}{2x^{2}} - \frac{3}{x} \; \big{\} }=  \frac{2 \pi l^{2}_{p}}{x^{2}} \; \big{\{ } \; 1 - 6x \; \big{\} } \\
& = \frac{8 \pi L^{2}}{l^{2}_{p}(\beta - \beta_{U})^{2}} \; \big{\{ } \; 1 - \frac{3 l^{2}_{p}}{L} (\beta - \beta_{U}) \; \big{\} }  \;\;\;\;\;  \text{with} \;\;\;\;\;\;  x = \frac{\pi}{\beta_{U}} ( \beta - \beta_{U})
\end{align*} 

Therefore the mean area $\bar{a}_{H}$ scales as $x^{-2}$. At the semi classical limit, for $x \rightarrow 0$, the area of the hole becomes larger and larger.

The mean color $\bar{s}$ is computed as follow:
 \begin{align*}
\bar{s} = \frac{1}{Z_{C} (\beta)} \int ds \;  s \;  Z_{s} (\beta) = \frac{1}{Z_{C}(\beta)} \int \frac{ds}{s^{2}} \frac{e^{q}}{\sqrt{q}} = \int \frac{ds}{s^{2}}  \big{(} \frac{2}{se} e^{xs} \big{)}^{1/2} \exp{\frac{se}{2} e^{-xs}}
\end{align*}
Just as before, since we are interested in the regim $x \rightarrow 0$, the form of the integral permits to proceed to a saddle point approximation. Proceeding to the change of variable $u = e^{-xs}$, the integral can be recast in a simpler form. This integral is dominated by the critical point of the function $f(u) = u \log {u}$ which is located at $u = e^{-1}$. Computing the integral with this approximation, we obtain:
 \begin{align*}
\bar{s} & \simeq \frac{1}{Z_{c}(\beta)} \sqrt{\frac{2}{e}} x^{3/2} e^{3/2} \sqrt{\frac{4 \pi x}{e^{2}}} \;  \exp{\frac{1}{2x}} = \frac{\sqrt{8 \pi} x^{2} \;  \exp{\frac{1}{2x}}}{Z_{C}(\beta)} = \frac{1}{x} \;\;\;\;\;  \text{so} \;\;\;\; \bar{s} = \frac{2 L}{l^{2}_{p} ( \beta - \beta_{U})}
\end{align*}

Therefore, in the semi classical regim where $x$ is close to zero, the mean color $\bar{s}$ is large which is consistent with our first assumptions. We recover the result obtain in [].
Moreover, since the area of the hole scales as $x^{-2}$ and the mean color scales as $x^{-1}$, we have that $\bar{s} \propto \sqrt{\bar{a}_{H}}$. We can now determine the coefficient $\sigma$ introduced above.
Indeed, using the expression of the mean area at the leading order, we have:
 \begin{align*}
\bar{s} = \sigma \frac{\sqrt{a_{H}}}{l_{p}} =  \frac{\sigma}{l_{p}} \frac{\sqrt{2\pi} l_{p}}{x} = \frac{1}{x} \;\;\;\;\;\;  \text{whence} \;\;\;\;\;  \sigma = \frac{1}{\sqrt{2 \pi}}
\end{align*}

The mean number of puncture $\bar{n}$ is given by:
 \begin{align*}
\bar{n} = \frac{1}{Z_{C} (\beta)} \int ds \sum_{n} n \; \frac{g_{\infty}(n,s)}{n !} e^{- \beta E(n,s)} = \frac{1}{Z_{C} (\beta)} \int \frac{ds}{s^{3}} \sum_{n} \frac{n}{\sqrt{n}} \frac{q^{n}}{n!} \;\;\;\;\; \text{with} \;\;\;\;\;  q = \frac{se}{2} \; e^{- xs} 
\end{align*}

Using the same trick than previously, we compute the following sum:
 \begin{align*}
I(q) & = \sum_{n} \frac{n}{\sqrt{n}} \frac{q^{n}}{n!} = \sum_{n} \frac{1}{\sqrt{n}} \frac{q^{n}}{(n-1)!} = \sum^{+ \infty}_{n = 0} \frac{1}{\sqrt{n + 1}} \frac{q^{n + 1}}{n!} = q \sum^{+ \infty}_{n = 0} \frac{1}{\sqrt{n + 1}} \frac{q^{n}}{n!} \\
& = \frac{q}{\sqrt{\pi}} \int^{+ \infty}_{- \infty}  du \; \sum^{+ \infty}_{n = 0} \frac{(q e^{- u^{2}})^{n} }{n !} e^{- u^{2}} = \frac{q}{\sqrt{\pi}} \int^{+ \infty}_{- \infty}  du \;  e^{(q e^{- u^{2}}) } e^{- u^{2}} 
\end{align*}

Using again the gaussian approximation for the ``action'' $S(u) = e^{- u^{2}}$ since $q$ is large in the semi classical regim, we obtain:
 \begin{align*}
I(q) & =  \frac{q}{\sqrt{\pi}} \int^{+ \infty}_{- \infty}  du \;  e^{q( 1 -  u^{2}) } e^{- u^{2}} =  \frac{q}{\sqrt{\pi}} e^{q} \; \int^{+ \infty}_{- \infty}  du \;  e^{- (q+1) u^{2}} \\
& = \frac{q}{\sqrt{q+1}} e^{q} \simeq \sqrt{q} \; e^{q}
\end{align*}

The mean number of punctures is therefore given by:
 \begin{align*}
\bar{n} & =  \frac{1}{Z_{C} (\beta)} \int \frac{ds}{s^{3}} \; \sqrt{q} \; e^{q} = \frac{1}{Z_{C} (\beta)} \int \frac{ds}{s^{3}} \big{(} \frac{se}{2} e^{- xs} \big{)}^{1/2} \exp{\frac{se}{2} e^{-xs}} \\
& =  \frac{1}{Z_{C} (\beta)} \int \frac{du}{x \; u} \big{\{} \; \frac{x}{- \log{u}} \; \big{\}}^{3} \sqrt{\frac{e}{2}} \big{\{} \frac{- u \log{u}}{x} \big{\}}^{1/2} \exp{\frac{e u \log{u}}{2x}}  \;\;\;\; \text{with} \;\;\; u = e^{-xs} \\ 
\end{align*}

Applying the saddle point approximation for the ``action'' $S(u) = u \log{u}$ juste as before, we observe that the dominant contribution comes from the only critical point at $u = e^{-1}$.
The integral becomes in this approximation:
 \begin{align*}
\bar{n} & = \frac{1}{Z_{C} (\beta)} \sqrt{\frac{e}{2}}  \; x^{3/2} \int \frac{du}{u^{1/2}}  \big{\{} \; \frac{-1}{ \log{u}} \; \big{\}}^{5/2} \exp{\frac{e u \log{u}}{2x}}  \\ 
& \simeq \frac{1}{Z_{C}(\beta)} \frac{e}{\sqrt{2}} \; x^{3/2} \; \exp{\frac{1}{2x}} \int ^{+ \infty}_{-\infty}du  \; \exp{ \big{\{} - \frac{e^{2}}{2x} \frac{u^{2}}{2} \big{\}}}  \\
& = \frac{1}{Z_{C}(\beta)} \frac{e}{\sqrt{2}} \; x^{3/2} \; \exp{\frac{1}{2x}} \sqrt{4 \pi x}{e^{2}} = \frac{1}{Z_{C}(\beta)} \sqrt{2\pi} \; x^{2} \; \exp{\frac{1}{2x}} = \frac{1}{2x}
\end{align*}

Therefore, $\bar{n}$ becomes large at the semiclassical limit and scales as $x^{-1}$ or equivalently, as $\sqrt{a_{H}}$. As expected, we recover the result of [] which are consistent with our first assumptions.
Just as for $\bar{s}$, we can determined the coefficient $\nu$ introduced earlier:
 \begin{align*}
\bar{n} = \nu \frac{\sqrt{a_{H}}}{l_{p}} =  \frac{\nu}{l_{p}} \frac{\sqrt{2\pi} l_{p}}{x} = \frac{1}{2x} \;\;\;\;\;\;  \text{whence} \;\;\;\;\;  \nu = \frac{1}{\sqrt{8 \pi}}
\end{align*}

The mean color $\bar{s}$ and the mean number of punctures $\bar{n}$ are related through $\bar{s} = 2 \bar{n}$.
Finally, the last quantity we need to compute is the canonical entropy. We have all the ingredient at hand to proceed.
The canonical entropy is given by:
 \begin{align*}
S_{C} (\beta) & = \beta \bar{E} + \log{Z_{C}(\beta)} \\
\end{align*}

Evaluating the two term at the Unruh temperature, i.e. at the semi classical limit, we obtain:
 \begin{align*}
\beta \bar{E} \big{|}_{\beta_{U}} & = \beta_{U} \frac{a_{H}}{8\pi L} = \beta_{U} \frac{2\pi l^{2}_{p}}{8 \pi L x^{2}} ( \; 1 + 6 x )= \frac{\pi}{2x^{2}} (1+6x) \\
\log{Z_{C}(\beta)} & =  3 \log{x} + \frac{1}{2x}\\
\end{align*}

Finally we obtain for the canonical entropy:
 \begin{align*}
S_{C} (\beta_{U}) & =   \frac{\pi}{2x^{2}} (1+6x) + 3 \log{x} + \frac{1}{2x}\\\
& =  \frac{\pi}{2x^{2}} + 3 \log{x} + \frac{1}{2x}( 1 + 6 \pi ) \\
& = \frac{a_{H}}{4 l^{2}_{p}} - \frac{3}{2} \log{\frac{a_{H}}{l^{2}_{p}}} + \sqrt{ \frac{a_{H}}{8 \pi l^{2}_{p}}}( 1 + 6 \pi )
\end{align*}

where we have used in the last line the equation of state at the leading term: $\bar{a}_{H} \simeq \frac{2 \pi l^{2}_{p}}{x^{2}}$
We recover the Bekenstein Hawking area law at the leading order, which is expected since the micro canonical and the canonical ensemble agree at the leading term at the thermodynamical limit.
However, at the subleading order, the two systems are not expected to have the same quantum corrections. Indeed, we observe that even if the general form of the subleasing term is the same for the two systems, the precise pre factor in front of them are different. In the canonical ensemble, we recover the prefactor $ - 3/2$ in front of the logarithmic corrections. 
Having recovered the expected logarithmic quantum corrections with the usual pre factor, we are still facing the problem of the too large quantum corrections proportional to $\sqrt{a_{H}}$.
We will see that such term can be remove in the context of the grand canonical ensemble.\\

\textit{The grand canonical ensemble} \\

We proceed to the very same derivation but in the context of the grand conical ensemble.
The additional ingredient in this context is the introduction of a chemical potential associated to the punctures. We have already discuss this notion in the precedent chapter and we refer the reader to this section.
Obvisously, the grand canonical ensemble share the same thermodynamical limit than the two precedent ensemble and we will recover at the leading term the Bekenstein Hawking area law.
However, the quantum corrections will now depend on the chemical potential and we shall show that a natural choice of the chemical potential cancel the too large quantum corrections encounters precedently.
This mechanism was already used in the precedent chapter when computing the entropy of the black hole for $\gamma \in \mathbb{R}$. It is therefore interesting to note that the behaviour of the quantum corrections proportional to $\sqrt{a_{H}}$ do not depend on the nature of $\gamma$.

Denoting the chemical potential $\mu$, the grand canonical partition function is given by:
  \begin{align*}
Z_{GC} (\beta, \mu)  = \int ds \sum_{n} \frac{g_{\infty} (n,s)}{n!} \; e^{- \beta ( E(n,s) - n \mu)}
\end{align*}

Its expression can be recast such that:
  \begin{align*}
Z_{GC} (\beta, \mu)  = \int \frac{ds}{s^{3}} \sum_{n} \frac{1}{\sqrt{n}}\frac{Q^{n}}{n!} \;\;\;\;\;\;  \text{with} \;\;\;\; Q = \frac{se}{2} e^{-xs} e^{\beta\mu} \;\;\; \text{and} \;\;\; x = \pi ( \tilde{\beta} - 1) \;\;\;\;  \tilde{\beta} = \frac{\beta}{\beta_{U}} 
\end{align*}

Repeating the computation presented previously in the canonical ensemble, we have:
  \begin{align*}
Z_{GC} (\beta, \mu)  & = \int \frac{ds}{s^{3}} \frac{e^{Q}}{\sqrt{Q}} =  \int^{+ \infty}_{s_{0}} \frac{ds}{s^{3}} \big{(} \; \frac{2}{se} e^{xs} e^{- \beta\mu} \; \big{)}^{1/2} \exp{\big{(} \; \frac{se}{2} e^{-xs} e^{\beta\mu} \; \big{)}} \\
& =  \int^{e^{-xs_{0} + \beta\mu}}_{0} \frac{du}{xu}   \big{(} \; \frac{x}{\beta \mu - \log{u}} \; \big{)}^{3} \big{(} \; \frac{2x}{(\beta\mu - \log{u}) ue} \; \big{)}^{1/2} \exp{\big{(} \; \frac{\beta\mu - \log{u}}{2x} eu \; \big{)}}  \;\;\;\; u = e^{-xs} e^{\beta \mu} \\
& =  \int^{e^{-xs_{0} + \beta \mu}}_{0} du  \; \frac{x^{5/2}}{u^{3/2}(\beta\mu - \log{u})^{7/2}} \sqrt{\frac{2}{e}} \exp{\big{(} \; \frac{\beta\mu - \log{u}}{2x} eu \; \big{)}} 
\end{align*}

This integral is dominated by the critical points of $S(u) = \tilde{\beta} \tilde{\mu} u - u \log{u}$. The function $S(u)$ admits one critical point at $u_{c} = \frac{z}{e}$ where we have posed $z = \exp{\beta\mu} $ (usually called the fugacity). Since $S(u_{c}) = z/e$ and $S''(u_{c})= - e /z$, under the gaussian approximation, the partition function reduces to:
  \begin{align*}
Z_{GC} (\beta, \mu)  & \simeq   \sqrt{\frac{2}{e}}  \frac{x^{5/2}}{z^{3/2}} e^{3/2} \exp{\frac{z}{2x}}\int^{+ \infty}_{-\infty} du  \; exp{\big{(} \; - \frac{e^{2}}{4 zx} u^{2} \; \big{)}} \\
& = \sqrt{\frac{2}{e}} \frac{x^{5/2}}{z^{3/2}} e^{3/2} \exp{\frac{z}{2x}} \sqrt{\frac{4\pi z x}{e^{2}}} \\
& = \sqrt{8 \pi} \; \frac{x^{3}}{z} \exp{\frac{z}{2x}}
\end{align*}

The mean area $\bar{a}_{H}$ and the mean number of punctures $\bar{n}$ can be computed directly:
  \begin{align*}
\bar{a}_{H} & = - \frac{8 \pi^{2} L}{\beta_{U}} \frac{\partial}{\partial x} \; \log{Z_{GC} (\beta, \mu)} = - \frac{8 \pi^{2} L}{\beta_{U}} \frac{\partial}{\partial x} \;  \big{\{ } \; 3 \log{x}  +  \frac{z}{2x} \; \big{\} } =  \frac{2 \pi l^{2}_{p} z}{x^{2}} \; \big{\{ } \; 1 - 6\frac{x}{z} \; \big{\} } \\
\bar{n} & = z \frac{\partial}{\partial z} \; \log{Z_{GC} (\beta, \mu)} = z \; \big{\{ } \; - \log{z} +  \frac{z}{2x} \; \big{\} }  = \frac{z}{2x} - 1
\end{align*}

We find again the same scaling for $\bar{a}_{H}$ and $\bar{n}$, i.e. the first one scales as $x^{-2}$ while the second one scales as $x^{-1}$. The equation $\bar{a}_{H} ( x, \mu )$ has to be understood as the equation of states for the semi classical black hole. Now we are ready to compute the grand canonical entropy at the semi classical limit, i.e. for $x \rightarrow 0$. For this, we calculate separately the different term involved. The precedent equation and the definition of $x$ and $z$ give:
\begin{align*}
\beta & = \beta_{U} ( 1 + \frac{x}{\pi} )\\
z & = e^{\mu \beta} = e^{\mu \beta_{U} }e^{\mu \beta_{U} x /\pi} = z_{U} ( 1 + \mu \beta_{U} \frac{x}{\pi}  + \mathcal{O}(\mu \beta_{U} \frac{x}{\pi})) \\
\log{z} & = \mu \beta = \mu \beta_{U} ( 1 + \frac{x}{\pi} ) = \mathcal{O}(1) \\
\frac{z}{2x} & = \frac{z_{U}}{2x}+ \mathcal{O}(1) \\
\beta \mu \bar{n} & = \frac{ \mu \beta_{U} z_{U} }{2x}( 1 + \frac{x}{\pi}) ( 1 + \mu \beta_{U} \frac{x}{\pi}) = \frac{\mu \beta_{U} z_{U}}{2x} + \mathcal{O}(1) \\
\end{align*}

Finally, the grand canonical entropy reads:
\begin{align*}
S_{GC}(\beta_{U}, \mu) & = \beta \frac{\bar{a}_{H}}{8 \pi L} + \log{Z_{GC}(\beta, \mu)} - \mu \beta \bar{n} \\
& = \beta \frac{\pi}{2 \beta_{U} x^{2}} \; ( z + 6 x) + 3 \log{x} - \log{z} + \frac{z}{2x} -  \mu \beta \bar{n}\\
& =  \frac{\pi}{2 x^{2}} ( 1 + \frac{x}{\pi})\; ( z_{U}(1 +  \mu \beta_{U} \frac{x}{\pi})  + 6 x) + 3 \log{x} + \frac{z_{U}}{2x} - \frac{\mu \beta_{U} z_{U}}{2x} + \mathcal{O}(1) \\
& =  \frac{\pi z_{U}}{2 x^{2}} + \frac{\pi}{x} ( \frac{z_{U}}{ \pi} + 3 ) + 3 \log{x} + \mathcal{O}(1) \\
& = \frac{\bar{a}_{H}}{4 l^{2}_{p}} + 3 \log{x}  + \frac{z_{U}}{x} ( 1 - \frac{\mu \beta_{U}}{2} ) + \mathcal{O}(1)  \;\;\;\;\;  \text{since} \;\;\;\; \frac{\pi z_{U}}{2 x^{2}} = \frac{\bar{a}_{H}}{4 l^{2}_{p}}  - \frac{z_{U}\mu \beta_{U}}{2x} - \frac{3 \pi}{x} \\
& = \frac{\bar{a}_{H}}{4 l^{2}_{p}} - \frac{3}{2} \log{\frac{\bar{a}_{H}}{l^{2}_{p}}}  + \frac{z_{U}}{x} ( 1 - \frac{\mu \beta_{U}}{2} ) + \mathcal{O}(1)
\end{align*}

As expected, we recover the Bekenstein Hawking area law supplemented with some logarithmic quantum corrections with the expected prefactor $- 3/2$.
The last term, i.e. the quantum correction proportional to $\sqrt{a_{H}}$, depends on the chemical potential. One can remove it by simply fixing the chemical potential to the Unruh temperature, ie $\mu = 2T_{U}$.
This fine tuning is unique and leads to the final formula for the grand canonical entropy:
\begin{align*}
S_{GC}(\beta_{U}, 2T_{U}) &= \frac{\bar{a}_{H}}{4 l^{2}_{p}} - \frac{3}{2} \log{\frac{\bar{a}_{H}}{l^{2}_{p}}}  + \mathcal{O}(1)
\end{align*}
Provided the fine tuning of the chemical potential to two times the Unruh temperature, we recover in the most general case, i.e. the grand canonical situation, the Bekenstein Hawking area law at the leading order supplemented with the expected logarithmic quantum corrections, the pre factor of those quantum corrections being the ``universal'' $- 3/2$.
We conclude from this result that working with the self dual Ashetkar's variables seems to provide a more satisfactory way to obtain the right semi classical limit in the context of quantum spherically isolated horizons . The fine tuning of the chemical potential, as mentioned in the precedent chapter, is not related to the analytic continuation process since the very same fine tuning is observed in the real case ($\gamma \in \mathbb{R}$) in order to remove the share root quantum correction to the entropy.

Here we computed the entropy for the ``one color model'', i.e. assuming that all the punctures carry the same continuous spin $s$. It is possible to relax this assumptions and work out the general case where different colors are allowed. The case where the quantum isolated horizon is colored by $p$ (where $p \in \mathbb{R}$) different colors have been study at the end of []. We refer the interested reader to the paper. For completeness, we give the partition function obtained after taking the semi classical limit:
\begin{align*}
Z_{p}(\beta, \mu) \simeq \frac{\sqrt{8\pi} p^{p}}{p! (p-1)!} \frac{x^{4-p}}{z p} \text{exp} (\frac{pz}{2x})
\end{align*}

From this partition function, we obtain the very same result than for the ``one colre model'', both for the leading term and for the way to remove the square root quantum correction through a fine tuning of the chemical potential to $\mu = 2 T_{U}$. The difference is that now, the prefactor in front of the logarithmic quantum corrections is $p$-dependent. The entropy for this quantum isolated horizon is given by:
\begin{align*}
S_{p} ( \beta_{U}, 2 T_{U}) = \frac{\bar{a}_{H}}{4l^{2}_{p}} + \frac{p-4}{2} \log{\frac{\bar{a}_{H}}{l^{2}_{p}}} + o (\log{\frac{\bar{a}_{H}}{l^{2}_{p}}} )
\end{align*}
The pre factor $-3/2$ is recovered for $p=1$, which one can easily interpreted as the ``quantum spherical case'' where the whole horizon is colored by the very same punctures.
One could then argue this case is the most probable and that it is therefore natural to obtain the ``universal'' pre factor $-3/2$ only in this particular case. Finally, the mean spin $\bar{s}$, the mean number of punctures $\bar{n}$ and the equation of state for this quantum isolated horizon are given by:
\begin{align*}
\bar{s} = \frac{1}{x} \qquad \bar{n} = \frac{pz}{2x} - 1 \qquad a_{H} = 4 \pi l^{2}_{p} ( \frac{pz}{2x^{2}} + \frac{p-4}{x} )
\end{align*}
The equations of state is now $p$-dependent but the scaling of the different quantities remain the same. This results are consistent with our definition of the semi classical limit.
\\

\textit{Summary} \\

We have presented the detailed study of the thermodynamical properties of the ``self dual black hole'' in the micro canonical, canonical and grand canonical ensembles.
Just as in the real case, the black hole can be understood as a gas of indisintguishable punctures which obey the Mawell Boltzman statistic. Due to the analytic continuation procedure, each puncture is now labelled by a continuous spin denoted $s$. We have worked in a simplified model where we have assumed that all the punctures carry the same spin. The semi classical limit corresponds to taking the limit of large $n$ and large $s$. 

The first important result is that the degeneracy of the ``self dual'' black hole has an holographic behaviour supplemented with some power law corrections.
Therefore, we do not need to postulate an holographic degeneracy in this model since it is derived from first principle. Moreover, the power law corrections to this holographic degeneracy conspired to give at the semi classical limit (in the canonical and the grand canonical ensemble) the expected logarithmic quantum corrections to the entropy.

The second result of this theromydnamical study is that the Bekenstein hawking area law is recovered exactly without any fine tuning (at the leading term). Obviously, the quantum corrections to the entropy depends on the situation considered.
In the micro canonical ensemble, we obtained logarithmic corrections but also correction proportional to the square root of the area.
Going to the canonical ensemble, the situation remains the same but the logarithmic quantum corrections inherit already the ``universal'' pre factor $-3/2$.
Finally, the grand canonical ensemble enables us to cancel the too large square root quantum correction by fixing the chemical potential to the value $\mu = 2 T_{U}$, leaving us only with the Bekanstein Hawking entropy supplemented with the expected logarithmic quantum corrections.

Therefore, this model provides a computation of the entropy of the black hole in (self dual) Loop Quantum Gravity which cure the ambiguity of fixing the Immirzi parameter present in the real computation.
 Our computation answers also to the two goals explained at the end of section $2$, i.e. obtain a rigorous derivation of this analytic continuation procedure and have a precise knowledge of the subleading terms.

Let us now discuss the procedure that we have used to analytically continued the degeneracy of the black form $\gamma \in \mathbb{R}$ to $\gamma = \pm i$.

\section{Discussion on the analytic continuation prescription}

The model presented above relies on an analytic continuation of the dimension of the Hilbert space of the $U_{q}(SU(2))$ Chern Simons theory living on the punctured $2$-sphere. This quantity depends on the Chern Simons level $k$ and on the colors of the spin labeling the punctures, i.e. $d_{l} = 2j_{l} +1$. Working with the self dual variables, i.e. with $\gamma = \pm i$, implies that we have to send the Chern Simons level to a purely imaginary too, i.e. $k \in i \mathbb{R}$.
The question whether one can properly analytically continued Chern Simons theory from a real Chern Simons level $k \in \mathbb{R}$ to a complex or purely imaginary one, i.e. $k \in \mathbb{C}$ or $k \in i \mathbb{R}$, was studied by Witten five years ago \cite{ch4-Witten1}. Using some of the tools presented in his work, we have derived an analytic continuation prescription for the quantum degeneracy of the black hole. This prescription is the unique one which enable us to keep the area positive and real while working with the complex Ashtekar's variables and in the same time have a well defined mathematical formula for the quantum degeneracy. It turns out that this procedure yields to the right semi classical limit supplemented with the expected quantum corrections. The procedure is summarized as follow:
\begin{align*}
\gamma = \pm i  \;\;\;\;\;  \text{imply} \;\;\;\;\;\;  k = \mp i \lambda  \;\;\; \text{with}\;\;\; \lambda \in \mathbb{R}^{+} \;\;\;\;\;\;\;\;  \text{and} \;\;\;\;\;\;  j_{l} = \frac{1}{2} ( i s_{l} - 1) \;\;\; \text{with}\;\;\;  s_{l} \in \mathbb{R}^{+} 
\end{align*}
where $j_{l}$ are the spin labelling the $l^{th}$ puncture.

Equiped with this prescription, we can now propose (at least in principle) a new strategy in order to study the physical predictions of self dual Ashtekar's gravity and deal with the reality conditions.
Starting with the real Ashtekar-Barbero phase space, one can follow the Loop quantization program and obtain the usual kinematical Hilbert space of real Loop Quantum Gravity.
At this step, one analytically continues the quantum theory using the prescription presented above. The reality conditions would be directly solved in the resulting quantum theory.
The last step would be to implement the self dual hamiltonian constraint which turns out to be much simpler than its real counterpart. It is worth mentioning that such strategy is commonly used when dealing with lorentzian three dimensional quantum gravity, for instance when computing the entropy of the Lorentzian BTZ black hole from the Cardy formula []. It is the only way to extract some information about three dimensional quantum gravity when working with a non compact group. We are simply proposing to apply this strategy in the context of four dimensional self dual Loop Quantum Gravity, providing a concrete procedure for this analytic continuation.

However, while this strategy is very appealing on the paper, it encounters different obstacles.

Since it is sometimes easier to analytically continue the equations for an object than the object itself, we do not know how the kinematical quantum states of real LQG are modified after our procedure.
In our work, we analytically continued only the equation for the degeneracy but not directly the quantum states of the black hole. However, if one applies our prescription to the kinematical Hilbert space, he/she needs to control the resulting quantum states to apply the quantum self dual scalar constraint. 
  
Our starting point is a quantum black hole for which the quantum states are given by tensors product of $U_{q}(SU(2))$ irreps. How the analytic continuation procedure modify those representations ?
Put it differently, what is the target Hilbert space of this mapping ? This question remains open up to now. 

Yet, one can use the analytically continued area spectrum to propose a candidate. It is well known that the mapping we used for the spin, ie $j = \frac{1}{2} ( i s- 1)$ where $s \in \mathbb{R}^{+} $ is the mapping which analytically continues the Casimir (as well as the character) of the $SU(2)$ group to the Casimir of the $SU(1,1)$ in the continuous representations. We are therefore trading the discrete $SU(2)$ area spectrum for a continuous $SU(1,1)$ area spectrum
\begin{align*}
a_{H} (j_{l}) = 8 \pi l^{2}_{p} \gamma \sum_{l} \sqrt{j_{l}(j_{l} + 1)}  \qquad \qquad a_{H}(s_{l}) =  4 \pi l^{2}_{p}  \sum_{l} \sqrt{s^{2}_{l} + 1}
\end{align*}
where we have choosen that $\sqrt{-1} = \mp i$ for $\gamma = \pm i$. Note that when $s=0$, there is still an area gap.
It is therefore tempting to interprete our procedure as a mapping from quantum states defined as $U_{q}(SU(2))$ tensor product to another quantum states defined as $U_{q}(SU(1,1))$ tensor product.

However, the structure of the $SU(1,1)$ group is much more complicated than the $SU(2)$ group. Since it is a non compact group, one have access to different irreducible representations which are related either to a discrete Casimir, either to a continuous one. At the level of the area spectrum, working with the self dual variables and satisfying the reality conditions seems to select only the continuous representations. This would imply that our analytically continued gauge invariant quantum states are spin network colored with continuous irrep of $SU(1,1)$. However, a spin network is build from three ingredients, a graph, irreps of a given group on each link and finally a choice of intertwined at each node. 

The difficulty when dealing with the $SU(1,1)$ group (and with any non compact group) is twofold. The two difficulties can be called the problem of the recoupling and the problem of the measure.

The first one arise when one considers the choice of intertwiner since the recoupling theory of the $SU(1,1)$ group is also more involved.
Indeed, two continuous irreps can recouple into a discrete one, which therefore bring us out of the previous selected irreps. 
This question has to be studied at the level of the volume operator. Indeed, the computation of the volume spectrum involves the intertwiner between irreps.
Applying our analytic continuation prescription to the volume spectrum derivation could provide interesting insights about the choice of intertwiner required to keep the spectrum real and positive.

The second one concerns the measure that one can use in order to compute transition amplitudes. The natural measure on the space of cylindrical functions is the Ashetkar-Lewandowski measure. It is a copy of the Haar measure. Since the $SU(1,1)$ group is non compact, its Haar measure blows up and one cannot integrate over $SU(1,1)$ spin networks. The situation seems therefore disastrous. 

Up to now, there is hope that the first problem can be manage but the second remains a large obstacle. While the recoupling theory of the $SU(1,1)$ group could have a good behaviour under our prescription, the problem of the measure is intrinsically related to the non compactness of the group. One could well introduce some weight in the measure in order to make it convergent but it would spoil the $SU(1,1)$ symmetry.
Yet, our analytic continuation was derived in the context of the quantum group $U_{q}(SU(2))$ and not for the classical group. If the target space concerns the quantum version of $SU(1,1)$, it could have some implications one the two precedent problems. Finally, one could argue that if we apply our prescription at the very end of the quantization program, when dynamic have been imposed (if one manages to do so), the problem of the non compact measure is by passed and our procedure becomes simply a way to extract physical predictions from the self dual quantum theory.

Since the development of the Wick rotation by Thiemann twenty years ago, no particular insights have been derived regarding the self dual quantum theory.
Although the status of the target space is still an open question, our analytic continuation prescription derived in this chapter provides the first concrete proposal to deal with the self dual quantum Ashtekar's gravity and extract concrete physical predictions. Either one applies it at the kinematical quantum level or at the end of the quantization program, at the dynamical quantum level.
In the first case, this could provides a way to identify the kinematical quantum states of self dual quantum gravity and then to apply the self dual scalar constraint. In this perspective, due to the non compactness of the group behind self dual Ashetkar's gravity., one encounters the two obstacles explained precedently, i.e. the mathematical problem of the recoupling theory and of the measure.
In the second case, the physical quantum states of real LQG has to be known and well controlled in order to derive physical predictions. Only those predictions would be analytically continued, providing the predictions of self dual quantum Ashtekar gravity and by passing the measure problem.

\clearemptydoublepage

\chapter{Three dimensional gravity as a guide}
\label{ch:3DLQG}
\minitoc

This chapter is devoted to clarifying the interplay between the role played by the Barbero-Immirzi parameter,  the self dual variables and the group $SU(1,1)$.
Indeed, the analytic continuation prescription derived in the precedent chapter in the context of the black hole entropy computation seems to point towards the peculiar role played by the complex Ashtekar's variables to obtain the right semi classical limit in LQG. From this observation, the real Immirzi parameter seems to played the role of a regulator which permit to start the quantization procedure from the real Ashtekar-Barbero phase, and then analyticaly continue the result to the self dual quantum theory. In spirit its presence allows one to perform a Wick rotation form the real quantum theory to the self dual one.
The precise prescription for this Wick rotation has been derived and explained in the precedent chapter

As emphasis in the final discussion, the status of this prescription is quite obscur up to now, since one does not have any control on the target Hilbert space.
For instance, we do not know how the $SU(2)$ spin networks are mapped by this Wick rotation. At the end of the precedent chapter, we proposed to use the analytic continuation of the area spectrum to get some insights about those kinematical self dual quantum states. Applying the prescription to the area spectrum, we noticed that the wick rotated area spectrum can be identified with the Casimir of the $SU(1,1)$ group related to the continuous serie of its irreducible representations.

However, it is only an observation and the appearance of the $SU(1,1)$ non compact group is still to be understood.

We know that starting from an $SL(2, \mathbb{C})$ phase space formulation, General Relativity can be reformulate into an $SU(2)$ gauge theory by fixing the so called time gauge.
Since the $SU(1,1)$ group is also a (non compact) subgroup of the  initial  $SL(2, \mathbb{C})$ group, one can naturally ask if it would be possible to gauge fixed the theory in order to obtain a $SU(1,1)$ phase space formulation of General Relatvity. 

Once equipped with this phase space, a natural question arises, motivated by the appearance of the $SU(1,1)$ continuous Casimir in the analytic continuation of the area spectrum.
Could the quantization of self dual Ashtekar gravity supplemented with its reality conditions be equivalent to the quantization of this $SU(1,1)$ phase space formulation of General Relativity ?
Is our analytic continuation prescription a map from the Hilbert space of real Loop Quantum Gravity to the Hilbert space of the quantum version of this $SU(1,1)$ formulation ?

In order to study those questions, one could tackle directly the hamiltonian analysis of the four dimensional Holst action in a gauge which selects the $SU(1,1)$ group from the initial $SL(2, \mathbb{C})$ group.
However, contrary to the $SU(2)$ group which acts naturally (by rotations) over the spacelike hyper surfaces $\Sigma$ of the foliation, the $SU(1,1)$ group is not well suited with respect to $3+1$ decomposition.
Therefore, performing the computation directly will turn out to be quite involved.

A natural idea is to turn towards a simplified model of General Relativity, where this hamiltonian analysis can be performed easily.
A very appealing candidate can be found in $2+1$ dimensions,  where General Relativity becomes a topological theory, i.e. without any local degrees of freedom (if there is no boundary).
Indeed, three dimensional gravity is a very interesting theoretical laboratory to test ideas about quantum gravity, since while higly simplified, the theory keeps all the interesting and difficult features related to the construction of a quantum theory of gravity.

\section{Three dimensional classical and quantum gravity, a laboratory}

The subject of three dimensional gravity is vast and we refer the interested reader to the book and review of Carlip for a general overview \cite{ch5-Carlip1}.
The interest raised by the $2+1$ version of General Relativity lies in the fact that while very simple compared to its $3+1$ counterpart, it allows a very rich gravitational physics.

The action of $2+1$ General Relativity is given by:
\begin{align*}
S_{3D} = \int_{\mathcal{M}}  \epsilon_{IJK} \; (\; e^{I}\wedge F^{JK} + \Lambda \; e^{I} \wedge e^{J} \wedge e^{K} )
\end{align*}

where $\Lambda$ is the cosmological constant. The group structure behind this theory is given by the three dimensional version of the Lorentz group, i.e. the $SO(2,1)$ group or equivalently, its double cover, i.e. the $SU(1,1)$ group. The internal indices $(I,J,K)$ run over $\{0,1,2\}$ and are raised by the three dimensional Minkowski metric $\eta_{IJ} = \text{diag}(-1, +1, +1)$.

Varying the action according to the tetrad and the spin connection, one obtains respectively the following field equations: 
\begin{align*}
\epsilon_{IJK} \; ( F^{JK} + 3 \Lambda \; e^{J} \wedge e^{K} ) = 0 \;\;\;\;\;\;   T^{I} = 0 
\end{align*}

The second equation implies that the torsion field is vanishing, and therefore that the spin connection reduces to the only metric-compatible and torsionless connection, i.e. the Levi-Civita one.
Then the first equation implies that the geometry and therefore the isometry group of the space-time solutions are entirely determined by the value of the cosmological constant.
If $\Lambda$ is vanishing, the space-time is flat everywhere and the resulting spin-connection is called a flat connection.
If $\Lambda$ is positive, the space-time is De Sitter-like while if $\Lambda$ is negative, one obtains a Anti-De Sitter space-time.
They are the three possible solutions to the Einstein equations in three dimensions.

In order to understand the topological nature of the theory, let us count the local degrees of freedom.
In a metric formulation, those degrees of freedom are all contained in the three dimensional metric tensor $g_{\mu\nu}$.
This field has $9$ components that the symmetry of the tensor reduces to $6$ independent components. The Einstein equations, which are given in this metric formulation by the symmetric Einstein tensor, give rise to 6 constraints, fixing all the metric components. There is therefore no dynamical degrees of freedom encoded in the metric tensor, contrary to the $3+1$ case. 

However, while free from local degrees of freedom, the theory remains non trivial.
Indeed, the interest for this dimensional reduced version of General Relativity was enhanced when some authors discovered a black hole solution to the field equations.
The so called BTZ black hole was derived by simply requiring the ``spherical symmetry'' in $2+1$ dimensions \cite{ch5-Tet1}.
This simple solution turned out to possess the same thermodynamical properties than a four dimensional black hole, such as a temperature and an  entropy. Moreover, it was shown to experience an evaporation through Hawking radiation.
Explaining such thermodynamical properties for a system having no local degrees of freedom became a major task and generate a very intense field of research \cite{ch5-Carlip2}.
This black hole solution exists only in an Anti De Sitter space-time, i.e. for a negative cosmological constant. This space-time admits a closed boundary, which generates the presence of local boundary degrees of freedom.
Those degrees of freedom, located on the boundary, are though to be responsible of the entropy of the black hole. The entropy of the hole was computed via different technics and rely on the so called Cardy formula. The reader can refer to \cite{ch5-Carlip2} for a pedagogical review.
However, from the point of view of the Lorentzian theory,  there are no control on the quantum states that we are counting and the entropy is obtained by mean of an analytic continuation.

From the point of view of LQG, the BTZ entropy was computed in \cite{ch5-Nouio1}, obtaining the right semi classical limit.
This computation was then investigated from the spin foam approach in \cite{ch5-Nouio2}.

While the derivation of the thermodynamical properties of the quantum BTZ black hole from a quantum theory of gravity represents a very exciting challenge, it is truly the possibility of completing the quantization to the full $2+1$ gravitational field that raise so much interest in three dimensional gravity.

One of the most important work in this field is undoubtedly the quantization of $2+1$ gravity by Witten \cite{ch5-Witten 3d}. This quantization was realized by using the Chern Simons formulation of three dimensional gravity.
The  loop quantization of $2+1$ gravity was undertaken in \cite{ch5-Smolinn1} and then investigated further in \cite{ch5-Marolf1, ch5-Ashte1}. It was shown that the length spectrum of a space-like length is given by the continuous Casimir of the $SU(1,1)$ group while the quantum time-like length is given by the discrete $SU(2)$ Casimir \cite{ch5-Frei1}. Moreover, euclidean three dimensional gravity was used to demonstrate the equivalence between the canonical and the spin foam approach to Loop Quantum Gravity \cite{ch5-Noui3}.

In the following, we will use this dimensional reduced version of General Relativity to investigate the fate of the Barbero-Immirzi parameter in Loop Quantum Gravity, and hopefully clarify the role of the $SU(1,1)$ group in our analytic continuation prescription. The first task is to introduce an Immirzi parameter in the action. The difficulty comes from the fact that there is not a Holst term available in three dimensions.
We will therefore present a new action for three dimensional gravity where an Immirzi parameter is introduced by a dimensional reduction of the four dimensional Holst action. The symmetry group will be the one of four dimensional gravity, i.e. $SL(2,\mathbb{C})$. This will allow us to study different gauge fixing and mimic what happen in four dimension from the point of view of the internal symmetry.
We will perform the canonical analysis of this action both in the three dimensional version of the time gauge selecting the compact subgroup $SU(2)$ from $SL(2,\mathbb{C})$, and in a non compact gauge which selects the non compact $SU(1,1)$ subgroup.
Already at this evel we will show that the presence of the Immirzi parameter is related to the choice of working with the compact group $SU(2)$ and is therefore, at least in three dimensional gravity, a pure gauge artifact.
We will then discuss the implications on the quantum theory and focus mainly in the area operator. The comparison of the $SU(2)$ and the $SU(1,1)$ area spectrum will show that at the end of the day, the analytic continuation prescription derived in the precedent chapter map also the two kinematical quantum theory based on the $SU(2)$ and the $SU(1,1)$ phase space.

Let us first describe the dimensional reduction used to introduce the Immirzi parameter in three dimensional gravity.

\section{Introducing the Immirzi parameter in $2+1$ gravity, a new toy model}

\noindent  In this section, we introduce our toy model and the $\gamma$-dependent action for three dimensional gravity. 

In four spacetime dimensions, the first order action for general relativity that serves as a starting point for canonical loop quantum gravity is given by \cite{ch5-holst}
\be\label{holst action}
S_\text{4D}[e,\omega] \equiv \frac{1}{4}\int_\mathcal{M} \left(\frac{1}{2}\eps_{IJKL}e^I\wedge e^J \wedge 
F^{KL}+\frac{1}{\gamma}\delta_{IJKL}e^I\wedge e^J\wedge F^{KL}\right).
\ee
The dynamical variables are the tetrad one-form fields $e^I_\mu$ and the $\sl(2,\mathbb{C})$-valued connection $\omega^{IJ}_\mu$, whose curvature is denoted by $F=\de\omega+(\omega\wedge\omega)/2$. The totally antisymmetric tensor $\eps_{IJKL}$ is the Killing form on $\sl(2,\mathbb{C})$, and $\delta_{IJKL}=(\eta_{IK}\eta_{JL}-\eta_{IL}\eta_{JK})/2$, with $\eta_{IJ}=\text{diag}(-1,1,1,1)$ the flat metric, is the other independent invariant bilinear form on $\sl(2,\mathbb{C})$ (with a suitable normalization). 

Without fixing the time gauge, the canonical analysis of this action is quite involved, and was performed originally in \cite{ch5-alexandrov1,ch5-alexandrov2}. Once the second class constraints are taken into account, it leads to a phase space where the symplectic structure given by the Dirac bracket involves a non-commutative $\sl(2,\mathbb{C})$ connection and its conjugate momentum. Furthermore, the Dirac bracket and the constraints become completely independent of $\gamma$, which drops out of the theory as expected from the Lagrangian analysis. This strongly suggests that in this formulation the Barbero-Immirzi parameter will play no role at the quantum level. Unfortunately, no representation of the associated quantum algebra has ever been found (see however \cite{ch5-alexandrov6} for an attempt), and the quantization of the so-called Lorentz-covariant formulation of loop gravity has never been achieved. It has however been argued by Alexandrov that this quantization could lead to a continuous area spectrum with no dependency on $\gamma$ \cite{ch5-alexandrov5}.

In loop quantum gravity, one chooses to work in the time gauge, which consists in breaking the $\SL(2,\mathbb{C})$ gauge group into an $\SU(2)$ maximal compact subgroup by imposing the conditions $e_a^0=0$. In this case, the canonical analysis simplifies dramatically \cite{ch5-holst,ch5-GN}, and the phase space is parametrized by an $\su(2)$-valued connection known as the Ashtekar-Barbero connection, together with its conjugate densitized triad field. The quantization in the time gauge is much easier to perform than in the Lorentz-covariant case, and leads to a mathematically well-defined kinematical Hilbert space because of the compactness of the gauge group. At the kinematical level, the area and volume operators exhibit discrete spectra, and the Barbero-Immirzi parameter can be interpreted as a measure of the area gap in Planck units.

Evidently, there seems to be a discrepancy between the predictions of the manifestly Lorentz-covariant quantization and the results derived in the time gauge. However, the problem is that up to now none of these derivations are fully understood. Indeed, as we have just mentioned above, the kinematical states are not even defined in the Lorentz-covariant quantization due to the non-commutativity of the connection, and in the quantization in the time gauge we do not have full control over the physical Hilbert space and the geometrical operators are only defined at the kinematical. It is nonetheless honest to say that the quantization of the $\SU(2)$ theory in the time gauge is much more advanced and mathematically well-defined, although it is very interesting and intriguing that the Lorentz-covariant theory points towards important issues concerning the status of the Barbero-Immirzi parameter and the relevance of the $\SU(2)$ Ashtekar-Barbero connection.

We are going to present a formulation of three-dimensional gravity that can help understand the tensions that we have just described. This model was originally introduced in \cite{ch5-GN2} in the context of spin foam models, and further studied in \cite{ch5-GN} in order to illustrate the interplay between the gauge-fixing of the Holst action and the role of the Barbero-Immirzi parameter. It can be obtained by a reduction of the four-dimensional Holst action to three-dimensions. In this section, we will present this model in details (for the sake of completeness) and recall its classical properties.

\subsection{Symmetry reduction from 4 to 3 dimensions}

\noindent Starting with the four-dimensional Holst action (\ref{holst action}), we perform a space-time reduction without a priori reducing the internal gauge group. As a consequence, the resulting three-dimensional model will be Lorentz-invariant. We assume that the four-dimensional space-time has the topology $\mathcal{M}_4=\mathcal{M}_3\times\mathbb{I}$ where $\mathcal{M}_3$ is a three-dimensional space-time, and $\mathbb{I}$ is a space-like segment with coordinates $x^3$. In this way, we single out the third spatial component $\mu=3$. Let us now impose the conditions
\be\label{symmetry conditions}
\partial_3=0,\qquad\omega_3^{IJ}=0.
\ee
The first condition means that the fields do not depend on the third spatial direction $x^3$. The second one means that the parallel transport along  $\mathbb{I}$ is trivial. Therefore, the covariant derivative of the fields along the direction $\mu=3$ vanishes. A direct calculation shows that the four-dimensional Holst action reduces under the conditions (\ref{symmetry conditions}) to
\be\label{action form}
S_\text{red}=-\int_\mathbb{I}\de x^3\int_{\mathcal{M}_3}\de^3x\,\eps^{\mu\nu\rho}\left(\f{1}{2}\eps_{IJKL}e_3^Ie^J_\mu F^{KL}_{\nu\rho}+
\frac{1}{\gamma}\delta_{IJKL}e_3^Ie^J_\mu F^{KL}_{\nu\rho}\right),
\ee
where $\mu=0,1,2$ is now understood as a three-dimensional spacetime index, and $\de^3 x\,\eps^{\mu\nu\rho}$ is the local volume form on the three-dimensional
space-time manifold. Apart from a global multiplicative factor that is not relevant at all, and provided that we set $x^I\equiv e_3^I$, we recover the three-dimensional action with Barbero-Immirzi parameter introduced in \cite{ch5-GN,ch5-GN2}, i.e.
\be\label{3D action}
S[e,x;\omega]=\int_{\mathcal{M}_3}\de^3x\,\eps^{\mu\nu\rho}\left(\frac{1}{2}\eps_{IJKL}x^Ie_\mu^JF_{\nu \rho}^{KL}+\frac{1}{\gamma}\delta_{IJKL}x^Ie_\mu^JF_{\nu \rho}^{KL}\right).
\ee
From now on, we will denote the three-dimensional space-time manifold $\mathcal{M}_3$ simply by $\mathcal{M}$.

\subsection{Lagrangian analysis}

\noindent It is not immediately obvious that the action (\ref{3D action}) is equivalent to that of three-dimensional gravity, simply because its expression is rather different from the standard first order BF action. First of all, it seems that we have introduced an additional degree of freedom represented by the variable $x$, and secondly the internal gauge group is $\SL(2,\mathbb{C})$ instead of the usual gauge group $\SU(1,1)$ of Lorentzian three-dimensional gravity. Furthermore, the action now features a Barbero-Immirzi parameter\footnote{A Barbero-Immirzi-like parameter was previously introduced in \cite{ch5-Bonzon Livine} in the context of three-dimensional gravity, based on the existence of two independent bilinear invariant forms on the symmetry group of the Chern-Simons formulation. Unfortunately, this parameter does not feature the properties of its four-dimensional counterpart appearing in the Holst action. In particular, it does not disappear when one passes from the first to the second order formulation of the theory.}. Despite all these differences, it can be shown that the action (\ref{3D action}) represents a valid formulation of three-dimensional gravity \cite{ch5-GN,ch5-GN2}.

There are many ways to see that this is indeed the case. The easiest one consists in showing that the action (\ref{3D action}) reproduces the standard Einstein-Hilbert action when one goes from the first order to the second order formulation. This method does also show straightforwardly that the parameter $\gamma$ disappears exactly as it does in four dimensions, i.e. when one expressed the theory in the metric form. To make this statement concrete, it is convenient to decompose the connection $\omega$ into its self-dual and anti self-dual components $\omega^\pm$ according to the decomposition of $\sl(2,\mathbb{C})=\su(2)_\mathbb{C}\oplus\su(2)_\mathbb{C}$ into its self-dual and anti-self-dual complex subalgebras. Then, the action (\ref{3D action}) can be expressed as a sum of two related BF action as follows:
\be
S[e,x;\omega]=\left(1+\frac{1}{\gamma}\right)S[B^+,\omega^+]+\left(1-\frac{1}{\gamma}\right)S[B^-,\omega^-],
\ee
where $S[B^\pm,\omega^\pm]$ is the standard $\su(2)_\mathbb{C}$ BF action
\be
S[B^\pm,\omega^\pm]=\frac{1}{2}\int\de^3x\,\eps^{\mu \nu \rho}\,\text{tr}\left(B_\mu^\pm,F_{\nu \rho}^\pm\right),
\ee
and
\be
B_\mu^{\pm i}=\pm\mathrm{i}(x\time e_\mu)^i+x^0e_\mu^i-x^ie^0_\mu
\ee
is calculated from the relations $B_\mu^{\pm i}=B^{IJ}_\mu T^{\pm i}_{IJ}$ and $B^{IJ}_\mu=\eps^{IJ}_{~~KL}x^Ke^L_\mu$. In this BF action, the trace $\tr$ denotes the normalized Killing form on $\su(2)_\mathbb{C}$. 

As usual, going from the first order to the second order formulation of gravity requires to solve for the components of the connection $\omega^\pm$ in terms of the $B$ variables. This can be done by solving the equations of motion obtained by varying the action with respect to the connection $\omega^\pm$, which are nothing but the torsion-free conditions
\be
T(B^\pm,\omega^\pm)=0.
\ee
If $\det(B^\pm)\neq0$, this torsion-free condition can be inverted to find the torsion-free spin connection $\omega(B)$. This latter, when plugged back into the original action, leads to the sum of two second order Einstein-Hilbert actions,
\be\label{einstein-hilbert}
S_\text{EH}[g_{\mu\nu}^+,g_{\mu\nu}^-]=\f{1}{2}\epsilon^+\int_\mathcal{M}\de^3x\sqrt{|g^+|}\,\mathcal{R}[g_{\mu\nu}^+]+\f{1}{2}\epsilon^-\int_\mathcal{M}\de^3x\sqrt{|g^-|}\,\mathcal{R}[g_{\mu\nu}^-],
\ee
each being defined with respect to a two-dimensional Urbantke-like metric \cite{ch5-urbantke} $g_{\mu\nu}^\pm=B_\mu^\pm\cdot B_\nu^\pm$ (in the sense that each metric comes from a $B$ field). In this expression, $\epsilon^\pm$ denotes the sign of $\det(B^\pm)$. It is straightforward to show that the signs $\epsilon^\pm$ are identical \cite{ch5-GN2}. To see that this is indeed the case, one can write the fields $B^\pm$ as follows:
\be\label{BdecompositionG}
B^{\pm i}_\mu=\pm\mathrm{i}\eps^i_{~jk}x^j e_\mu^k+x^0e^i_\mu-x^ie^0_\mu=\big(\pm\mathrm{i}x_0^{-1}\underline{x}+  \mathds{1} \big)L_\mu^i,
\ee
with $L_\mu^i\equiv\eps^i_{~jk}B^{jk}_\mu/2=x^0e^i_\mu-x^ie^0_\mu$, and where we have introduced the three-dimensional matrix
\be
\underline{x}=
 \begin{pmatrix}
0 & -x_3 & x_2\\
x_3 & 0 & -x_1\\
-x_2 & x_1 & 0
 \end{pmatrix}
\ee
associated to $x$ such that $\underline{x}\alpha^i=\eps^i_{~jk}x^j\alpha^k$ for any $\alpha\in\mathbb{R}^3$. With this notation, we can compute the determinant
\be
\det(B^\pm)=\det\big(\pm\mathrm{i}x_0^{-1}\underline{x}+    \mathds{1} \big)\det(L^i_\mu)=\big(1-x_0^{-2}(x_1^2+x_2^2+x_3^2)\big)\det(L^i_\mu),
\ee
where $L_\mu^i$ is considered as a $3\times3$ matrix. Therefore, we conclude that $\epsilon^+=\epsilon^-$, as announced above. Furthermore, a simple calculation shows that the two Urbantke metrics $g_{\mu\nu}^\pm$ are identical and given by
\be\label{urbantke metrics}
g_{\mu\nu}^\pm=g_{\mu\nu}=(e_\mu\cdot e_\nu)\big(x^2-(x^0)^2\big)-x^2e^0_\mu e^0_\nu+x^0x\cdot(e^0_\mu e_\nu+e^0_\nu e_\mu),
\ee
with $x^2\equiv x^i x_i$.

Gathering these results on the Urbantke metrics and the sign factors $\epsilon^\pm$, we can conclude that, once the torsion-free condition is imposed, the action (\ref{3D action}) reduces to the standard Einstein-Hilbert action
\be
S_\text{EH}[g_{\mu\nu}]=\int_\mathcal{M}\de^3x\sqrt{|g|}\,\mathcal{R}[g_{\mu\nu}].
\ee
This shows that the theory that we are dealing with corresponds indeed to three-dimensional gravity, and that the Barbero-Immirzi parameter disappears once the torsion is vanishing. This is exactly what happens with the four-dimensional Holst action, which motivates the use of the three-dimensional model (\ref{3D action}) in order to test the fate of the Barbero-Immirzi parameter at the physical level.

\subsection{Lagrangian symmetries}

\noindent Before presenting the Hamiltonian analysis in details, let us finish the Lagrangian analysis with a study of the symmetries. This will be helpful in what follows. Obviously, the action (\ref{3D action}) is invariant under $\SL(2,\mathbb{C})$, and admits therefore the infinite-dimensional gauge group $\mathcal{G}\equiv C^\infty\big(\mathcal{M},\SL(2,\mathbb{C})\big)$ as a symmetry group. An element $\Lambda\in\mathcal{G}$ is an $\SL(2,\mathbb{C})$-valued function on the spacetime $\mathcal{M}$, which acts on the dynamical variables according to the transformation rules
\be\label{gauge}
e_\mu\longmapsto\Lambda\cdot e_\mu,\qquad
x\longmapsto\Lambda\cdot x,\qquad
\omega_\mu\longmapsto\text{Ad}_\Lambda(\omega_\mu)-\partial_\mu\Lambda\Lambda^{-1},
\ee
where $(\Lambda\cdot v)^I=\Lambda^{IJ}v_J$ denotes the fundamental action of $\Lambda$ on any four-dimensional vector $v$, and $\text{Ad}_\Lambda(\xi)=\Lambda\xi\Lambda^{-1}$ is the adjoint action of $\SL(2,\mathbb{C})$ on any Lie algebra element $\xi\in\sl(2,\mathbb{C})$.  

From the expression (\ref{3D action}) of the action as the integral of a three-form, it is immediate to see that the theory is also invariant under spacetime diffeomorphisms, as it should be for gravity. Infinitesimal diffeomorphisms are generated by vector fields $v=v^\mu\partial_\mu$ on $\mathcal{M}$, and their action on the dynamical variables is simply given by the following Lie derivatives:
\be\label{diffeos}
e\longmapsto\mathcal{L}_ve,\qquad
x\longmapsto\mathcal{L}_vx=v^\mu\partial_\mu x,\qquad
\omega\longmapsto\mathcal{L}_v\omega, 
\ee
where $\mathcal{L}_v\varphi=(v^\nu\partial_\nu\varphi_\mu+\varphi_\nu\partial_\mu v^\nu)\de x^\mu$ for any one-form $\varphi$.

The previous symmetries are expected from a theory of gravity formulated in first order variables. But a theory in three space-time dimensions with only these symmetries would reduce to $\SL(2,\mathbb{C})$ BF theory, which is not what our model is. Thus, our Lagrangian should admit additional symmetries. This is indeed the case, and it is immediate to notice that the action (\ref{3D action}) in invariant under a rescaling symmetry and a translational symmetry. The former is generated by non-vanishing scalar fields $\alpha$ on $\mathcal{M}$ according to the transformation rules
\be\label{rescale}
e^I_\mu\longmapsto\alpha e^I_\mu,\qquad
x^I\longmapsto\frac{1}{\alpha}x^I.
\ee
The translational symmetry is generated by one-forms $\beta=\beta_\mu\de x^\mu$ according to
\be\label{extra trans}
e^I_\mu\longmapsto e^I_\mu+\beta_\mu x^I.
\ee
The presence of these two symmetries follow from the fact that the variables $x$ and $e$ appear in the action (\ref{3D action}) in the form $x^{[I}e^{J]}=(x^I e^J-x^J e^I)/2$. Note that they do not affect the connection $\omega$.

The transformations (\ref{gauge}), (\ref{diffeos}), (\ref{rescale}), and (\ref{extra trans}), encode all the symmetries of the action. We will make use of some of these invariance properties to simplify the canonical analysis in the following section. Furthermore, due to the $\SL(2,\mathbb{C})$ invariance, the sign of $x^2=x^Ix_I=x^I\eta_{IJ}x^J$ is an invariant of the theory, even if its value is not fixed because of the rescaling invariance. Thus, to define the theory, one has to fix this sign, and we choose it to be positive:
\be\label{sign condition}
x^2=x^I \eta_{IJ}x^J>0.
\ee
As we will see in the next section, this choice will make the time gauge accessible.

\section{Hamiltonian analysis in the non compact gauge}

In this subsection, $\mu,\nu,\dots\in\{0,1,2\}$ are three-dimensional spacetime indices, $a,b,\dots \in\{1,2\}$ are spatial indices, $I,J,\dots\in\{0,1,2,3\}$ are internal $SL(2,\mathbb{C})$ indices, and $i,j,\dots\in\{0,1,2\}$ are internal $SU(1,1)$ indices. The indices $i,j,\dots$ are lowered and raised with the flat three-dimensional Minkowski metric $\eta_{ij}=\text{diag}(-1,+1,+1)$. We will use the cross-product notation $v\times w$ to denote the vector $z$ whose components are given by $z^i=\epsilon^{ijk}v_jw_k$, and $v\cdot w$ for the scalar product $v^iw_i=v^i\eta_{ij}w^j$.

The gauge group $SL(2,\mathbb{C})$ is broken into the subgroup $SU(1,1)$ by fixing, in the action (4), the field $x^I = e^{I}_{3}$ to the special value $(0,0,0,1)$\footnote{Note the slight difference with [], where the gauge was chosen to be $x^I=(1,0,0,0)$}. This choice is compatible with the condition (13), and the rescaling symmetry (17) can be used to fix the norm of $x^{3}$ to one for simplicity. The resulting $SU(1,1)$ symmetry corresponds precisely to the isotropy group of $x$. 
The time gauge in our case becomes:
\begin{align}
x^{i} = e^{i}_{3} = 0 \;\;\;\;\;\;\; \text{where} \;\;\;\;\;\; i \in \{0, 1, 2\}
\end{align}

Since this gauge choice singularizes the third internal space-like component, it is natural to decompose the connection $\omega^{IJ}$ into its $su(1,1)$ components, denoted by $\omega^i$, and the complement denoted by $\omega^{(3)i}$.
\begin{align*}
& \omega^{i}_{\mu} = \frac{1}{2} \epsilon^{i}{}_{jk} \omega^{jk}_{\mu} \;\;\;\;\;\;\;\;\;  \text{and} \;\;\;\;\;\;\;\;\; \omega^{(3)i}_{\mu}
\end{align*}
The curvature tensor $F^{IJ}$ also decomposes into its $su(1,1)$ components $F^i=\epsilon^i_{~jk}F^{jk}/2$, and the remaining part $F^{(3)i}$. Denoting by $F$ the vector with components $F^i$, and by $F^{(3)}$ the vector with components $F^{(3)i}$, we have the explicit expressions
\begin{align}
& F^{(3)i}_{\mu\nu}  = \partial_{\mu} \omega^{(3)i}_{\nu} - \partial_{\nu} \omega^{(3)i}_{\mu} -( \omega_{\mu} \times \omega^{(3)}_{\nu} )^{i} + ( \omega_{\nu}\times \omega^{(3)}_{\mu})^{i} \\
& F^{i}_{\mu\nu} = \frac{1}{2} \epsilon^{i}{}_{jk}F^{jk}_{\mu\nu} = \partial_{\mu} \omega^{i}_{\nu} -  \partial_{\nu} \omega^{i}_{\mu} - (\omega_{\mu} \times \omega_{\nu})^{i}  -(\omega^{(3)}_{\mu}\times \omega^{(3)}_{\nu})^{i}
\end{align}

Let us proceed to the decomposition of the action with those notations. In order to impose our new gauge condition, we split the internal indices between the indice $3$ and the indice $i \in \{0, 1, 2\}$. We use the convention $\epsilon_{3ijk} = 1$ and $\epsilon^{ab0} = 1$. This reads:
\begin{align*}
S &= \frac{1}{2} \int_{M_{3}} dx^{3}  \epsilon^{\mu\nu\rho}\; \{ \; \frac{1}{2} \epsilon_{IJKL} x^{I} e^{J}_{\mu} F^{KL}_{\nu\rho} + \frac{1}{\gamma} x^{I} e^{J}_{\mu} F_{\nu\rho \; IJ} \; \} \\
& = \frac{1}{2} \int_{M_{3}} dx^{3}  \epsilon^{\mu\nu\rho}\; \{ \;  \frac{1}{2} \epsilon_{3ijk} x^{3}  e^{i}_{\mu} F^{jk}_{\nu\rho} + \frac{1}{2}  \epsilon_{i3jk} x^{i}  e^{3}_{\mu} F^{jk}_{\nu\rho}  +   \epsilon_{ij3k} x^{i}  e^{j}_{\mu} F^{3k}_{\nu\rho} \\
& \;\;\;\;\;\;\;\;\;\;\;\;\;\;\; \;\;\;\;\;\;\;\;\;\; \;\;\;\; +  \frac{1}{\gamma} ( \; x^{3} e^{i}_{\mu} F_{\nu\rho \; 3i} + x^{i} e^{3}_{\mu} F_{\nu\rho \; i3} + x^{i} e^{j}_{\mu} F_{\nu\rho \; ij} \; ) \} \\
& = \frac{1}{2} \int_{M} dx^{3} \epsilon^{\mu\nu\rho} e_{\mu} . \{ F_{\nu\rho} + \frac{1}{\gamma} F^{(3)}_{\nu\rho}\} \\
& = \frac{1}{2} \int_{M} dx^{3} \; \{ \; - 2 \epsilon^{ab} e_{a} . (F_{0b} + \frac{1}{\gamma} F^{(3)}_{0b} )  + \epsilon^{ab} e_{0} . ( F_{ab} + \frac{1}{\gamma} F^{(3)}_{ab}) \; \} \\
& = \frac{1}{2} \int_{M} dx^{3} \; \{ \; L_{C}  + L_{S} \; \}
\end{align*}

With those decomposition, we can now compute the two terms resulting from the decomposition of the action. We introduce the electric field $E^{b} = \epsilon^{ab}e_{a}$.
The first term $L_{C}$ becomes:
\begin{align*}
L_{C} & = - 2 E^{b} . \{ \; \partial_{0} \omega_{b} - \partial_{b} \omega_{0} - \omega_{0} \times \omega_{b} - \omega^{(3)}_{0} \times \omega^{(3)}_{b} + \frac{1}{\gamma} \; ( \; \partial_{0} \omega^{(3)}_{b} - \partial_{b} \omega^{(3)}_{0} - \omega_{0} \times \omega^{(3)}_{b} + \omega_{b} \times \omega^{(3)}_{0} \; ) \; \} \\
& = - 2 E^{b} . \{ \; \partial_{0} ( \omega_{b} + \frac{1}{\gamma} \omega^{(3)}_{b} ) - \partial_{b} ( \omega_{0} + \frac{1}{\gamma} \omega^{(3)}_{0} ) - ( \omega_{0} + \frac{1}{\gamma} \omega^{(3)}_{0} ) \times ( \omega_{b} + \frac{1}{\gamma} \omega^{(3)}_{b} ) \\
& \;\;\;\;\;\; + ( \omega_{0} + \frac{1}{\gamma} \omega^{(3)}_{0} ) \times ( \omega_{b} + \frac{1}{\gamma} \omega^{(3)}_{b} ) - \omega_{0}\times \omega_{b} - \omega^{(3)}_{0} \times \omega^{(3)}_{b} - \frac{1}{\gamma} \omega_{0} \times \omega^{(3)}_{b} + \frac{1}{\gamma} \omega_{b} \times \omega^{(3)}_{0} \; \} \\
& = - 2 E^{b} . \{ \; \partial_{0} \tilde{A}_{b} - \partial_{b} \tilde{A}_{0} - \tilde{A}_{0} \times \tilde{A}_{b} + ( \frac{1}{\gamma^{2}} - 1) \omega^{(3)}_{0} \times \omega^{(3)}_{b} \; \} \\
& = - 2 \{ \; E^{b} . \partial_{0} \tilde{A}_{b} + \tilde{A}_{0} . ( \partial_{b} E^{b} - \tilde{A}_{b} \times E^{b} ) + ( \frac{1}{\gamma^{2}} - 1) \omega^{(3)}_{0} . ( \omega^{(3)}_{b} \times E^{b} ) \; \} \\
& =  2 \{ \; E^{b} . \partial_{0} A_{b} + A_{0} . ( \partial_{b} E^{b} + A_{b} \times E^{b} ) + ( 1 - \frac{1}{\gamma^{2}}) \omega^{(3)}_{0} . ( \omega^{(3)}_{b} \times E^{b} ) \} \\
& = 2 \{ E^{b} . \partial_{0} A_{b} + A_{0} . G + ( 1 - \frac{1}{\gamma^{2}}) \omega^{(3)}_{0} . \phi \}
\end{align*}

where we have used the trade in the fifth line the connection $\tilde{A}$ for the connection $A = - \tilde{A}$ in order to have a proper Gauss constraint $G$. We note that we could have equivalently express $L_{C}$ as:
\begin{align*}
L_{C} & = - 2 \{ E^{b} . \partial_{0} A_{b} + (\omega_{0} + \gamma \omega^{(3)}_{0}) . G + (\frac{1}{\gamma} - \gamma) \omega^{(3)}_{0} . \chi \} \;\;\;\;\; \text{where} \;\;\;\;\; \chi = \partial_{b}E^{b} - \omega_{b} \times E^{b}
\end{align*}

In this case, the constraint $\phi$ is simpler to solve than the usual constraint $\chi$. The expression of $L_{C}$ implies that the true dynamical variables of the theory are given by:
\begin{align}
A^{i}_{a} = - ( \omega^{i}_{b} + \frac{1}{\gamma} \omega^{(3)i}_{b}) \;\;\;\;\;\;  \text{and} \;\;\;\;\;\; E^{b} = \epsilon^{ab} e_{a}
\end{align}
The first variable is an $SU(1,1)$ real connection, i.e. the analogue of the Ashtekar-Barbero connection with this particular gauge fixing, while the second variable is the real $SU(1,1)$ electric field canonically associated.
The second term can be computed as follow:
\begin{align*}
L_{S} & = \epsilon^{ab} e_{0} . \{ \; \partial_{a} \omega_{b} - \partial_{b} \omega_{a} - \omega_{a} \times \omega_{b} - \omega^{(3)}_{a} \times \omega^{(3)}_{b} + \frac{1}{\gamma} \; ( \; \partial_{a} \omega^{(3)}_{b} - \partial_{b} \omega^{(3)}_{a} - \omega_{a} \times \omega^{(3)}_{b} + \omega_{b} \times \omega^{(3)}_{a} \; ) \; \} \\
& = \epsilon^{ab} e_{0} . \{ \; \partial_{a}( \omega_{b} + \frac{1}{\gamma} \omega^{(3)}_{b} ) - \partial_{b} ( \omega_{a} + \frac{1}{\gamma} \omega^{(3)}_{a} ) - \omega_{a} \times \omega_{b} - \omega^{(3)}_{a} \times \omega^{(3)}_{b} - \frac{1}{\gamma} \omega_{a} \times \omega^{(3)}_{b} + \frac{1}{\gamma} \omega_{b} \times \omega^{(3)}_{a} \; \} \\
& = \epsilon^{ab} e_{0} . \{ \; \partial_{a}\tilde{A}_{b}  - \partial_{b} \tilde{A}_{a} - \tilde{A}_{a} \times \tilde{A}_{b} + ( \frac{1}{\gamma^{2}} - 1 ) \omega^{(3)}_{a} \times \omega^{(3)}_{b} \; \} \\
& = \epsilon^{ab} e_{0} . \{ \;  - \partial_{a} A_{b}  + \partial_{b} A_{a} - A_{a} \times A_{b} + ( \frac{1}{\gamma^{2}} - 1 ) \omega^{(3)}_{a} \times \omega^{(3)}_{b} \; \} \\
& = -  \epsilon^{ab} e_{0} . \{ \; \mathcal{F}_{ab} + ( 1 - \frac{1}{\gamma^{2}})  \omega^{(3)}_{a} \times \omega^{(3)}_{b} \; \} \\
\end{align*}

With those expressions, the action becomes:
\begin{align}
S = -  \int_{\mathbb{R}} dt \int_{M_{2}} dx^{2} & \; \{ \; E^{b} . \partial_{0} A_{b} + A_{0} . G + ( 1 - \frac{1}{\gamma^{2}}) \omega^{(3)}_{0} . \phi  +   e_{0}.  H + \mu_{a} . S^{a}\; \}
\end{align}
The analysis of the action in this non compact gauge shows that the theory can be formulated in terms of the variables $E^{a}$, $A_{a}$, $\omega^{(3)}_{a}$, $\omega^{(3)}_{0}$ and $\omega_{0}$. From the analysis of the canonical term, one can see that only $E$ and $A$ are a priori dynamical while $\omega^{(3)}_{0}$ and $\omega_{0}$ are Lagrange multipliers.
However, from the canonical point of view, one has to consider the $\omega^{(3)}_{a}$ as a dynamical variable and therefore, and introduce its conjugated momentum $\pi^{a}$ together with the constraint:
\begin{align*}
S^{a} = \pi^{a} \simeq 0
\end{align*}
As a consequence, the symplectic structure is defined by the following Poisson bracket:
\begin{align}
\{ E^{a}_{i}(x), A^{j}_{b}(y)\} = \delta^{a}_{b} \delta^{i}_{j} \delta^{2}(x-y)= \{\pi^{a}_{i}(x) , \omega^{(3)j}_{b}(y)\}
\end{align}
The primary constraints between those canonical variables are:
\begin{align*}
G = \partial_{b} E^{b} + A_{b} \times E^{b}  \;\;\;\;\;\;  H = \frac{1}{2}\epsilon^{ab}  ( \; \mathcal{F}_{ab} + ( 1 - \frac{1}{\gamma^{2}})  \omega^{(3)}_{a} \times \omega^{(3)}_{b} \;) \;\;\;\;\; \phi = \omega^{(3)}_{b} \times E^{b} \;\;\;\; S = \pi^{a} 
\end{align*}
Among the four primary constraints $G, H, \phi, S$, only the last one generates secondary constraints when evolving it in time. Let us impose that its evolution be vanishing.
It reads:
\begin{align*}
\dot{\pi}^{c} = \{\pi^{c} , H_{tot} \} & = ( 1 - \frac{1}{\gamma^{2}} )\int_{M} dx^{3} \; \{ \pi^{c} , \omega^{(3)}_{0} . (\omega^{(3)}_{b} \times E^{b}) + \epsilon^{ab} e_{0} .  (\omega^{(3)}_{a} \times \omega^{(3)}_{b}) \} \\
& = ( 1 - \frac{1}{\gamma^{2}} )\int_{M} dx^{3} \; \{  \delta^{c}_{a} \epsilon^{ab} \; (e_{a} \times \omega^{(3)}_{0}) + \frac{1}{2} \epsilon^{ab} \delta^{c}_{a} \; (\omega^{(3)}_{b} \times e_{0}) + \frac{1}{2} \epsilon^{ab} \delta^{c}_{b} \; ( e_{0} \times \omega^{(3)}_{a}) \}\\
& = ( 1 - \frac{1}{\gamma^{2}} )\int_{M} dx^{3} \;  \epsilon^{ac} ( \; e_{a}\times \omega^{(3)}_{0} + \omega^{(3)}_{b} \times e_{0} \; ) \simeq 0
\end{align*}

The 6 equations involve Lagrangian multipliers as well as dynamical variables, and as such, they can be separated into two sets. The first set of equations fixes the values of the Lagrange multipliers $e_{0}$ and $\omega^{(3)}_{0}$, and the second set is formed by secondary constraints. To extract these secondary constraints, it is convenient to combine the precedent 6 equations with the 3 primary constraint $\phi \simeq 0$ derived above.
Indeed, these 9 equations can be written in the compact form:
\begin{align*}
& \epsilon^{0ab} e_{a} \times \omega^{(3)}_{b} = \epsilon^{ab} e_{a} \times \omega^{(3)}_{b} = E^{b} \times \omega^{(3)}_{b} = - \phi \\
& \epsilon^{a0b} e_{0} \times \omega^{(3)}_{b} + \epsilon^{ab0} e_{b} \times \omega^{(3)}_{0} = \epsilon^{ab} ( e_{b}\times \omega^{(3)}_{0} - e_{0} \times \omega^{(3)}_{b}) \simeq 0
\end{align*}
Therefore, a compact form for those 9 equation is:
\begin{align}
\epsilon^{\mu\nu\rho} e_{\nu} \times \omega^{(3)}_{\rho}  \simeq 0 
\end{align}
As a consequence, if $e$ is invertible (which is what we are assuming from the beginning), the original 9 equations are equivalent to the 9 equations $\omega^{(3)}_{\mu} \simeq 0$. It is clear that the vanishing of $\omega^{(3)}_{0}$ is a fixation of Lagragne multipliers, whereas the remaining 6 equations $\omega^{(3)}_{a} \simeq 0$ are secondary constraints. Moereover, these constraints together with the primary constraint $S^{a}$ form a second class system ( i.e. they do not generate symmetries), and can be solved strongly. Setting $\omega^{(3)}_{a}$ to zero in the constraint $H$ shows that the Barbero-Immirzi parameter disappears completely, and we end up with the standard action of Lorentzian three dimensional gravity, (i.e. with a flat connection). This closes the canonical analysis of the action in the non compact gauge.

This result is consistent with the observation that we made earlier at the Lagrangian level concerning the irrelevance of the Barbero-Immirzi parameter in the classical theory. As a consequence, $\gamma$ will play no role in the canonical quantum theory once we work in the non compact gauge. This is already an interesting observation, since it seems to be in conflict with the situation in four dimensions where $\gamma$ plays a crucial role (at least) at the kinematical level. However, to make this conclusion stronger and more meaningful, we have to cast our three-dimensional model in a form that is closer to the four-dimensional Ashtekar-Barbero phase space, and then take this as the starting point for the quantization. This can be done by using the three dimensional time gauge as we will show now.

\section{Hamiltonian analysis in the compact gauge}
We present now the hamiltonian analysis of the toy model defined above in the three dimensional ``time gauge''. 
In this subsection, $\mu,\nu,\dots\in\{0,1,2\}$ are three-dimensional spacetime indices, $a,b,\dots \in\{1,2\}$ are spatial indices, $I,J,\dots\in\{0,1,2,3\}$ are internal $SL(2,\mathbb{C})$ indices, and $i,j,\dots\in\{1,2,3\}$ are internal $SU(2)$ indices. The indices $i,j,\dots$ are lowered and raised with the flat three-dimensional euclidean metric $\delta_{ij}=\text{diag}(+1,+1,+1)$. We will use the cross-product notation $v\times w$ to denote the vector $z$ whose components are given by $z^i=\epsilon^{ijk}v_jw_k$, and $v\cdot w$ for the scalar product $v^iw_i=v^i\delta_{ij}w^j$.

The gauge group $SL(2,\mathbb{C})$ is broken into the subgroup $SU(2)$ by fixing the field $x^I = e^{I}_{3}$ to the special value $(0,1,1,1)$\footnote{Note the slight difference with [], where the gauge was chosen to be $x^I=(1,0,0,0)$} and the tetrad components $e^{0}_{a}$ to zero. This choice is compatible with the condition (13), and the rescaling symmetry (17) can be used to fix the norm of $x^{3}$ to one for simplicity. 

Our starting point is the precedent action. We use the convention $\epsilon_{ijk0} = 1$ and $\epsilon^{ab0} = 1$. 
The time gauge in our case becomes:
\begin{align}
x^{0} =  e^{0}_{3} = 0 \;\;\;\;\;\;\; \text{and} \;\;\;\;\;\; e^{0}_{a} = 0 \;\;\;\;\; \text{where} \;\;\;\;\; a \in \{1,2\}
\end{align}
The splitting in the internal and space-time indices reads:
\begin{align*}
S &= \frac{1}{2} \int_{M_{3}} dx^{3}  \epsilon^{\mu\nu\rho}\; \{ \; \frac{1}{2} \epsilon_{IJKL} x^{I} e^{J}_{\mu} F^{KL}_{\nu\rho} + \frac{1}{\gamma} x^{I} e^{J}_{\mu} F_{\nu\rho \; IJ} \; \} \\
& = \frac{1}{2} \int_{M_{3}} dx^{3}  \epsilon^{\mu\nu\rho}\; \{ \;  \frac{1}{2} \epsilon_{ijk} x^{i}  e^{0}_{\mu} F^{jk}_{\nu\rho} -  \epsilon_{ijk} x^{i}  e^{j}_{\mu} F^{0k}_{\nu\rho}  + \frac{1}{\gamma} x^{i} e^{0}_{\mu} F_{\nu\rho \; i0} + \frac{1}{\gamma} x^{i} e^{j}_{\mu} F_{\nu\rho \; ij} \; \} \\
& = \frac{1}{2} \int_{M_{3}} dx^{3}  \epsilon^{\mu\nu\rho}\{ \;  \frac{1}{2} \epsilon_{ijk} x^{i}  N F^{jk}_{ab}   -  \epsilon_{ijk} x^{i}  N^{j} F^{0k}_{ab}  + 2\epsilon_{ijk} x^{i} e^{j}_{a} F^{0k}_{ab}  + \frac{1}{\gamma} x^{i} N F_{ab \; i0} + \frac{1}{\gamma} x^{i} N^{j} F_{ab \; ij} - \frac{2}{\gamma} x^{i} e^{j}_{a} F_{0b \; ij} \} \\
& = \frac{1}{2}  \int_{M_{3}} dx^{3}  \epsilon^{ab} \; \{ \; 2( x\times e_{a}) \; . \; ( F^{0}_{0b} - \frac{1}{\gamma} F_{0b} ) - N^{a}( x \times e_{a}) \; .\; ( F^{0}_{ab} - \frac{1}{\gamma} F_{ab} ) + N x \; .\; ( F_{ab} + \frac{1}{\gamma} F^{0}_{ab} ) \; \} \\
& =  \frac{1}{2} \int_{M_{3}} dx^{3} \; ( \; L_{C} + L_{V} + L_{S} \; )
\end{align*}
where we have set $N = e^{0}_{0}$ and $N^{i} = N^{a} e^{i}_{a}$. They are the lapse scalar function and the shift vector. Note that in this time gauge, there is one more piece in the decomposition of the action, i.e. denoted $L_{V}$.
In order to obtain the explicit expression of the three terms, we need to decompose the curvature of the spin connection. Just as in the first chapter where the hamiltonian analysis was performed, we have:
\begin{align}
& F^{0i}_{\mu\nu} = \partial_{\mu} \omega^{(0)i}_{\nu} - \partial_{\nu} \omega^{(0)i}_{\mu} +( \omega_{\nu} \times \omega^{(0)}_{\mu} )^{i}- ( \omega_{\mu}\times \omega_{\mu})^{i} \\
& F^{i}_{\mu\nu} = \frac{1}{2} \epsilon^{i}{}_{jk}F^{jk}_{\mu\nu} = \partial_{\mu} \omega^{i}_{\nu} -  \partial_{\nu} \omega^{i}_{\mu} - (\omega_{\mu} \times \omega_{\nu})^{i}  +(\omega^{(0)}_{\mu}\times \omega^{(0)}_{\nu})^{i}
\end{align}

With this decomposition of the curvature, the first term becomes:
\begin{align*}
L_{C} & = \frac{2}{\gamma}  \epsilon^{ab} (x \times e_{a}) \;.\; ( \gamma F^{0}_{0b} -  F_{0b} ) \\
& = \frac{2}{\gamma} E^{b} \; . \; \{ \partial_{0}( \gamma \omega^{(0)}_{b} - \omega_{b} ) - \partial_{b} \gamma \omega^{(0)}_{0} + ( \gamma \omega_{b} + \omega^{(0)}_{b}) \times \omega^{(0)}_{0} + \partial_{b} \omega_{0}  + ( - \omega_{b} + \gamma \omega^{(0)}_{b}) \times \omega_{0} \} \\
& = \frac{2}{\gamma} E^{b} \; . \; \{ \partial_{0} A_{b} +  \omega^{(0)}_{0} \; . \; [ \gamma ( \partial_{b} E^{b} - \omega_{b} \times E^{b} ) - \omega^{(0)}_{b} \times E^{b} ] - \omega_{0} \;.\; ( \partial_{b}E^{b} + A_{b} \times E^{b} ) \; \} \\
& = \frac{2}{\gamma} \; . \; \{ E^{b} . \partial_{0} A_{b} - \omega_{0} . G + \omega^{(0)}_{0} . ( \gamma \phi - \omega^{(0)}_{b} \times E^{b} ) \; \} \\
& =  \frac{2}{\gamma} \; . \; \{ E^{b} . \partial_{0} A_{b} - \alpha . G + \gamma ( 1 + \frac{1}{\gamma^{2}} ) \; \omega^{(0)}_{0} . \phi  \; \} \\
\end{align*}
where $\alpha^{i} =  ( \frac{1}{\gamma} \omega^{(0)i}_{0} + \omega^{i}_{0})$.

We have also used the following notations:
\begin{align}
A_{a} = \gamma \omega^{(0)}_{a} - \omega_{a} \;\;\;\;\;\;\; E^{a} = \epsilon^{ab} x \times e_{a}
\end{align}
They are the canonically conjugated variables of the theory. $A_{a}$ is the three dimensional analogue of the real $su(2)$ Ashtekar Barbero connection while $E^{a}$ is the $su(2)$ electric field. 
The two Lagrange multipliers $\alpha$ and $K_{0}$ enforce the two constraints:
\begin{align}
G_{i} = \partial_{b}E^{b}_{i} + (A_{b} \times E^{b} )_{i} \;\;\;\;\;\;\;\;\;\;  \phi_{i} = \partial_{b} E^{b}_{i} - (\omega_{b} \times E^{b})_{i} \;\;\;\;\; \text{whence} \;\;\;\;\; \omega^{(0)}_{b} \times E^{b} = - \frac{1}{\gamma} ( G - \phi)
\end{align}

In order to write the second term in the action, we remark that the curvature of the three dimensional Ashtekar Barbero connection $A_{a}$ can be decomposed as:
\begin{align*}
\mathcal{F}_{ab}(A) &= \partial_{a} A_{b} - \partial_{b} A_{a} + A_{a} \times A_{b} \\
& = \gamma ( \partial_{a}\omega^{(0)}_{b} - \partial_{b} \omega^{(0)}_{a} - \omega_{a} \times \omega^{(0)}_{b} + \omega^{(0)}_{a} \times \omega_{b})  + ( \partial_{b} \omega_{a} - \partial_{a} \omega_{b} + \omega_{a} \times \omega_{b} - \omega^{(0)}_{a} \times \omega^{(0)}_{b}) \\ 
& \;\;\; + ( 1 + \gamma^{2}) \omega^{(0)}_{a} \times \omega^{(0)}_{b} \\
& = \gamma F^{0}_{ab}(\omega) - F_{ab}(\omega) + (1+ \gamma^{2}) \omega^{(0)}_{a} \times \omega^{(0)}_{b}
\end{align*}

Therefore, we have for the second term:
\begin{align*}
L_{V} & = \frac{1}{\gamma} \epsilon^{ab}N^{c}( x \times e_{c}) \; .\; ( \gamma F^{0}_{ab} - F_{ab} ) \\
& = \frac{1}{\gamma}N^{c} \epsilon^{ab} \epsilon_{dc}\epsilon^{dc}( x \times e_{c}) \; .\; (\mathcal{F}_{ab}(A) -(1+ \gamma^{2}) \omega^{(0)}_{a} \times \omega^{(0)}_{b}) \\
& = \frac{2}{\gamma}N^{a}  \{ \; E^{b} . \mathcal{F}_{ab}(A) - (1+ \gamma^{2}) \omega^{(0)}_{a} . (\omega^{(0)}_{b} \times E_{b}) \; \} \\
& = \frac{2}{\gamma}N^{a} \{ \; E^{b} .\mathcal{F}_{ab}(A) + (\frac{1}{\gamma}+ \gamma) \omega^{(0)}_{a} . (G- \phi) \; \} \\
\end{align*}
The precedent decomposition of the curvature can be modified in order to find the last term of the action.
This modification reads:
\begin{align}
\mathcal{F}_{ab}(A) &=  \gamma F^{0}_{ab}(\omega) - F_{ab}(\omega) + (1+ \gamma^{2}) \omega^{(0)}_{a} \times \omega^{(0)}_{b} \\
& =  \gamma F^{0}_{ab}(\omega) - F_{ab}(\omega) + (1+ \gamma^{2}) ( F_{ab} - R_{ab}) \\
& = \gamma ( F^{0}_{ab} + \gamma F_{ab} )(\omega) - (1+ \gamma^{2}) R_{ab}
\end{align}
where we have used:
\begin{align}
\omega^{(0)}_{a} \times \omega^{(0)}_{b} = F_{ab} (\omega)- R_{ab}(\omega)  \;\;\;\;\;\; \text{where} \;\;\;\;\; R_{ab} = \partial_{a} \omega_{b} - \partial_{b} \omega_{a} - \omega_{a} \times \omega_{b}
\end{align}
Therefore, the very last term in the action is given by:
\begin{align}
L_{S} & = \frac{1}{\gamma^{2}} \epsilon^{ab}N x \; .\; \gamma ( \gamma F_{ab} + F^{0}_{ab} ) \\
& = \frac{1}{\gamma^{2}}\epsilon^{ab}N x \; .\; ( \mathcal{F}_{ab}(A) + (1+ \gamma^{2}) R_{ab})
\end{align}

Therefore, under this decomposition, the three dimensional Holst action becomes:
\begin{align}
S =  \int_{\mathbb{R}} dt \int_{M_{2}} dx^{2} & \; \{ \; \frac{1}{\gamma} \;  E^{b} . \partial_{0} A_{b} -  \frac{1}{\gamma} \alpha . G +  ( 1 + \frac{1}{\gamma^{2}} ) \; \omega^{(0)}_{0} . \phi  + \frac{1}{\gamma} N^{a}H_{a}+ \frac{1}{2\gamma^{2}} N \; H\; \}
\end{align}
We note that the lapse $N$ and the shift vector $N^{a}$ are both Lagragne multipliers which enforce the constraints $H$ and $H_{a}$, just as in four dimensions.
The explicit expressions of the constraints read:
\begin{align}
\phi_{i} & =  \partial_{b} E^{b}_{i} - (\omega_{b} \times E^{b})_{i} \\
G _{i} &= \partial_{b}E^{b}_{i} + (A_{b} \times E^{b} )_{i} \simeq 0 \\
H_{a} & =  E^{b} . \; \mathcal{F}_{ab}(A) + (\frac{1}{\gamma}+ \gamma) \omega^{(0)}_{a} . (G- \phi) \simeq 0\\
H & =  \epsilon^{ab} x \; .\; ( \mathcal{F}_{ab}(A) + (1+ \gamma^{2}) R_{ab}) \simeq 0
\end{align}
Those primary constraints mimic closely their four dimensional counterparts.
By imposing the two primary constraints $G$ and $\phi$, the three dimensional versions of the vectorial and the scalar constraints reduces to:
\begin{align*}
 H_{a} \simeq \epsilon_{ab} E^{b} . \; \mathcal{F}_{12}  \;\;\;\;\;\;\;\;\;\;  H = x . \; ( \mathcal{F}_{12} + (1+\gamma^{2}) R_{12} )
\end{align*}
The analysis of the action in the time gauge shows that the theory can be formulated in terms of the variables $E^{a}$, $A_{a}$, $\omega_{a}$, $\omega_{0}$ and $\omega^{(0)}_{0}$. Therefore, the initial component $\omega^{(0)}_{a}$ can be replaced by the three dimensional version of the Ashtekar-Barbero variables $A_{a}$, and $\omega_{a}$, as it is the case in four dimensions. From the analysis of the canonical term, one can see that only $E$ and $A$ are a priori dynamical, whereas all the other variables have vanishing conjugated momenta. however, this does not mean that all these non dynamical variables can be treated as genuine Lagrange multipliers. $\omega_{0}$ and $\omega^{(0)}_{0}$ can be treated as Lagrange multipliers, but $\omega_{a}$ has to be associated to a momenta $\pi^{a}$, just as in the previous section. Because of this, the theory inherits new primary constraints enforcing the vanishing of $\pi^{a}$:
\begin{align*}
S^{a} =\pi^{a} \simeq 0
\end{align*}
We end up with the following sympletcic structure:
\begin{align*}
\{E^{a}_{i} (x), A^{j}_{b} (y)\} = \gamma \delta^{j}_{i} \delta^{a}_{b} \delta^{2}(x-y) \;\;\;\;\;\;\;\;\; \{\pi^{a}_{i} (x), \omega^{j}_{b} (y)\} = \delta^{j}_{i} \delta^{a}_{b} \delta^{2}(x-y) 
 \end{align*}
Note the presence of the Barbero-Immirzi parameter in the Poisson bracket of the Ashtekar-Barbero connection and its conjugated momenta.
$\gamma$ shows up because we are working with the connection $A = - \gamma \tilde{A}$ instead of $\tilde{A}$. This is required in order to have a well defined $SU(2)$ Gauss constraint.

Finally, the Hamiltonian reads:
\begin{align}
H_{tot} = \int_{\Sigma} dx^{2} \{ \; \Omega_{0} . \; G + \Lambda_{0} . \; \phi + \frac{1}{\gamma}N^{a} H_{a} + \frac{1}{2\gamma^{2}}N \; H + \lambda_{a} \pi^{a} \; \} 
\end{align}
where the condensed Lagrange multipliers $\Omega_{0}$ and $\Lambda_{0}$ are given by:
\begin{align*}
\Omega_{0} = - \alpha + ( 1 + \frac{1}{\gamma^{2}} ) N^{a} \omega^{(0)}_{a} \;\;\;\;\;\;\;\;\;\;  \Lambda_{0} = ( 1 + \frac{1}{\gamma^{2}} ) \; ( \omega^{(0)}_{0} - N^{a} \omega^{(0)}_{a})
\end{align*}

We now study the stability of the primary consrtaints. Just as in the previous section, only the constraint $\pi^{a} \simeq 0$ generates secondary constraints.
Their time evolution is given by the following Poisson bracket:
\begin{align*}
\dot{\pi}^{c} = \{ \; \pi^{c} , H_{tot} \} & = \int_{M} dx^{3} \; \{ \pi^{c} , \Lambda_{0} \phi + \frac{1}{2\gamma^{2}} N \; H \; \}  \\
& = \int dx^{3} \; \{ \; \pi^{c} , \Lambda_{0} ( \partial_{a}E^{a} - \omega_{a} \times E^{a}) + \frac{1+ \gamma^{2}}{2\gamma^{2}} \epsilon^{ab} N \; x . \; ( \partial_{a} \omega_{b} - \partial_{b} \omega_{a}  - \omega_{a} \times \omega_{b} ) \; \}  \\ 
& = \int dx^{3} \; ( \; \delta^{c}_{a}  \; E^{a} \times \Lambda_{0} + \frac{1+ \gamma^{2}}{2\gamma^{2}}  \epsilon^{ab} (- \partial_{a}(Nx) \delta^{c}_{b} + \partial_{b}(Nx) \delta^{c}_{a} -  N ( \omega_{b} \times x \; \delta^{c}_{a} -  \omega_{a} \times x \; \delta^{c}_{b} )) \\
& = \int dx^{3} \; ( \; E^{c} \times \Lambda_{0} + \frac{1+ \gamma^{2}}{\gamma^{2}}  \epsilon^{cb} ( \partial_{b}(Nx)  -  N\;  \omega_{b} \times x ) \; )\\
& =  \int dx^{3} \; ( \; E^{c} \times \Lambda_{0} + \frac{1+ \gamma^{2}}{\gamma^{2}}  \epsilon^{cb} ( x \; \partial_{b}N  -  N\; (\partial_{b} x -  \omega_{b} \times x ) \; ) \; ) \simeq 0\\
\end{align*}
This secondary constraint can be written in a more compact form.
To do so, we first project the equation on the direction $E^{a}$ since we now that $x . E^{a} = x . (\epsilon^{ab} e_{b} \times x)= 0$. This gives:
\begin{align*}
E^{a} . ( E^{c} \times \Lambda_{0} + \frac{1+ \gamma^{2}}{\gamma^{2}}  \epsilon^{cb} ( x \; \partial_{b}N  -  N\; (\partial_{b} x -  \omega_{b} \times x ) \; ) ) &= - \Lambda_{0} \; E^{a}\times E^{c} - \frac{1+ \gamma^{2}}{\gamma^{2}}  E^{a}\epsilon^{cb}  N \; (\partial_{b} x -  \omega_{b} \times x ) \\ &\simeq 0
\end{align*}
We can then symmetrize the indices $(a,c)$ in order to eliminate the term with $\Lambda_{0}$.
\begin{align*}
\Lambda_{0} \; E^{(a}\times E^{c)} - \frac{1+ \gamma^{2}}{\gamma^{2}}  E^{(a} \epsilon^{c)b}  N\; (\partial_{b} x -  \omega_{b} \times x ) = - \frac{1+ \gamma^{2}}{\gamma^{2}}  E^{(a} \epsilon^{c)b}  N\; (\partial_{b} x -  \omega_{b} \times x )  \simeq 0
\end{align*}
Finally, assuming that the lapse $N$ is never vanishing, we end up with the secondary constraint:
\begin{align*}
\Psi^{ca} = E^{(a} \epsilon^{c)b}  \; (\partial_{b} x -  \omega_{b}\times x )  \simeq 0 
\end{align*} 
This secondary constraint does not generate tertiary constraint and the Dirac algorithm ends here. \\

\textit{Solving the second class constraints}\\

We note that the set of constraint $S^{a}$, $\phi$ and $\Psi^{ca}$ form a second class system.
\begin{align*}
S^{a} = \pi^{a} \;\;\;\;\;\; \phi = \partial_{a} E^{a} - \omega_{a} \times E^{a} \;\;\;\;\;\; \Psi^{ca} = \epsilon^{b(c}E^{a)} . ( \partial_{b} x - \omega_{b} \times x)
\end{align*}
As usual in canonical analysis, we need to solve those second class constraints prior quantization.
Let us explicit the constraints $(\phi, \Psi^{ca})$:
\begin{align*}
& E^{1} . ( \partial_{1} x - \omega_{1} \times x) \simeq 0 \;\;\;\;\;\;\; E^{2} . ( \partial_{1} x - \omega_{1} \times x) \simeq 0 \\
& E^{1} . ( \partial_{2} x - \omega_{2} \times x) \simeq 0 \;\;\;\;\;\;\; E^{2} . ( \partial_{2} x - \omega_{2} \times x) \simeq 0 \\
& E^{1} . ( \partial_{a} E^{a} - \omega_{2} \times E^{2}) \simeq 0 \;\;\;\;\;\;\; E^{2} . ( \partial_{a} E^{a} - \omega_{1} \times E^{1}) \simeq 0 \\
\end{align*}
Since the electric field $E^{a}$ decomposes into:
\begin{align*}
E^{a} = \epsilon^{ab} e_{b} \times x \;\;\;\;\;\;\; \text{whence} \;\;\;\;\;\;\; E^{1} = e_{2} \times x  \;\;\;\;\;\;\; E^{2} = - e_{1} \times x 
\end{align*}
the set $\{E^{1}, E^{2}, x\}$ forms a basis of the internal space. On can show the useful properties:
\begin{align*}
(E^{1} \times E^{2})^{m} & = - \epsilon_{ijk} e^{j}_{2} x^{k} \epsilon^{mil} \epsilon_{lpq} e^{p}_{1} x^{q} = \epsilon_{kpq} e^{m}_{2} x^{k} e^{p}_{1}x^{q} - \epsilon_{jpq} e^{j}_{2} x^{m} e^{p}_{1} x^{q} \\
& = e^{m}_{2} \; x . ( e_{1} \times x) - x^{m} \; e_{2} . (e_{1} \times x) = x^{m} \; (e_{1} \times e_{2}) . x \\
& = x^{m} \text{det}(e)
\end{align*}
and we note also: $\text{det} (E) = (E^{1} \times E^{2}) . x = \text{det}(e) x^{2}$.
The resolution of the second class constraint reduces to express the component $\omega_{a}$ as a function $f(E,x)$. Therefore, we decompose $\omega_{a}$ on the internal basis $\{E^{1}, E^{2}, x\}$ as:
\begin{align*}
\omega_{a} = \alpha_{a} E^{1} + \beta_{a} E^{2} + \xi_{a} x
\end{align*}
Injecting this expression in the 4 first equations above, we obtain:
\begin{align*}
& \beta_{1} ( E^{1} \times E^{2}) . x =  E^{1} \partial_{1} x \;\;\;\;\;\;\;\;\;\;\; \beta_{2} ( E^{1} \times E^{2} ) .x = - E^{1} . \partial_{2} x  \;\;\;\; \text{whence} \;\;\;\;\; \beta_{a} = \frac{E^{1}}{\text{det} (E) } \partial_{a} x\\
&  \alpha_{1} ( E^{1}\times E^{2}).x = - E^{2} . \partial_{1} x    \;\;\;\;\;\;\; \alpha_{2} ( E^{1} \times E^{2} ) .x = - E^{2} . \partial_{2} x  \;\;\;\; \text{whence} \;\;\;\;\; \alpha_{a} = \frac{E^{2}}{\text{det} (E) } \partial_{a} x
\end{align*}

To find $\xi_{a}$, we use the two last equations above in which we explicit $\beta_{a}$ and $\alpha_{a}$. This gives:
\begin{align*}
\xi_{1} \; x. (E^{1} \times E^{2}) = E^{2} . \partial_{a} E^{a} \;\;\;\;\;\;\;  \xi_{2} \; x. (E^{1} \times E^{2} ) = - E^{1} . \partial_{a} E^{a}  \;\;\;\; \text{whence} \;\;\;\;\; \xi_{a} = \frac{1}{\text{det} (E) } \epsilon_{ab} E^{b} \partial_{c} E^{c}
\end{align*}

Therefore, we can write down the expression of $\omega_{a}$. To do so, we use the identity between the vectorial product of three vectors: $  v\times ( w \times z) = (v . z ) w - ( v.w) z$ and the expression of the Gauss constraint.
\begin{align*}
\omega_{a} & = \frac{1}{\text{det} (E) } \; \{ \; - ( E^{2} . \partial_{a} x) E^{1} + ( E^{1} . \partial_{a} x) E^{2} + \epsilon_{ab} (E^{b} . \partial_{c} E^{c}) x \; \} \\
& = \frac{1}{\text{det} (E) } \; \{ \; -  \partial_{a} x \times ( E^{1} \times E^{2} ) + \epsilon_{ab} (E^{b} . ( G - A_{c} \times E^{c} )  )x \; \} \\
& = \frac{1}{\text{det} (E) } \; \{ \; -  \partial_{a} x \times ( E^{1} \times E^{2} ) + \epsilon_{ab} A_{c} \; .  \; (E^{b} \times E^{c} ) x + \epsilon_{ab} E^{b} . \; G \; x \; \} \\
& = \frac{1}{\text{det} (E) } \; \{ \; -  \partial_{a} x \times ( E^{1} \times E^{2} ) - A_{a} \; .  \; (E^{1} \times E^{2} ) x + \epsilon_{ab} E^{b} . \; G \; x \; \}
\end{align*}
where we have used that:  $\epsilon_{ab} A_{c} \; .  \; (E^{b} \times E^{c} ) x = - A_{a} \; .  \; (E^{1} \times E^{2} ) x$. Moreover, we can write the last line in a more compact form using the useful properties derived precedently.
The middle term can be written as:
\begin{align*}
\frac{1}{\text{det} (E) } A_{a} \; . \; ( E^{1} \times E^{2} ) x = \frac{1}{\text{det} (E) } (A_{a} \; . \; x ) \frac{\text{det} (E)}{x^{2}} x = (\; A_{a} \; .\;  \frac{x}{\sqrt{x^{2}}} \; ) \;  \frac{x}{\sqrt{x^{2}}} = (\; A_{a} \; .\;  u \; ) \;  u
\end{align*}
where we introduce the condensed notation: $u = \frac{x}{\sqrt{x^{2}}}$. Obviously, we have $u^{2} = 1$.

Finally, since $\sqrt{x^{2}}$ is a scalar, one can write the first term in the expression of $\omega_{a}$ as:
\begin{align*}
\frac{1}{\text{det} (E) } \partial_{a} x \times ( E^{1} \times E^{2} ) = \frac{1}{\text{det} (E) } x^{2} (\partial_{a} \frac{x}{\sqrt{x^{2}}} \times \frac{x}{\sqrt{x^{2}}} ) \; \frac{\text{det} (E) }{x^{2}} = \partial_{a}u \times u
\end{align*}
The final expression for $\omega_{a}$ is given by:
\begin{align*}
\omega_{a} = u \times \partial_{a} u - ( A_{a} \times u ) \; . \; u +  \frac{1}{\text{det} (E) } \epsilon_{ab} E^{b} . \; G \; x 
\end{align*}
This conclude the resolution of the second class constraints of the system.
Let us derive one more property with the notation introduce above. It will be useful in the next part of the analysis. Using the second class constraints $\psi^{ca}$, we have that:
\begin{align*}
E^{1} . ( \partial_{1} x - \omega_{1} \times x) = \sqrt{x^{2}} \; E^{1} . ( \partial_{1} u - \omega_{1} \times u) \simeq 0 \;\;\;\;\;\; \text{whence} \;\;\;\;\;\;\; E^{b} . ( \partial_{a} u - \omega_{a} \times u) \simeq 0 
\end{align*}
since it is true for the four equations listed above. $E^{b}$ being invertible, only the term in parenthesis vanishes.
Now it will be interesting to write our scalar constraint in a similar form than its four dimensional counterpart. For this, we explicit the term $x . R_{12} (\omega)$ appearing in the scalar constraint and we use the expression of $\omega_{a}$ just derived. We list the properties that we will use to obtain our final expression:
\begin{align*}
\partial_{a} u^{2} = 2 u . \partial_{a} u = 0  \;\;\;\;\;\; u \; . \; \omega_{a} = - A_{a} \; . \; u \;\;\;\;\;\; u \times \omega_{a} = \partial_{a} u 
\end{align*}
\begin{align*}
x .  R_{12}(\omega) & = \sqrt{x^{2}} \; u . R_{12}(\omega) =\sqrt{x^{2}}  u . ( \partial_{1} \omega_{2} - \partial_{2} \omega_{1} - \omega_{1} \times \omega_{2} ) \\
& = \sqrt{x^{2}} \{ \; \partial_{1} ( u . \omega_{2}) - (\partial_{1} u ) . \omega_{2} - \partial_{2} ( u . \omega_{1} ) + (\partial_{2} u ) . \omega_{1} - u . (\omega_{1} \times \omega_{2} ) \; \} \\
& = \sqrt{x^{2}}  \{ \; - \partial_{1} ( A_{2} .u) - (\omega_{1} \times u ) . \omega_{2}  + \partial_{2} ( u . A_{1}) + (\omega_{2} \times u) . \omega_{1} - u . ( \omega_{1} \times \omega_{2}) \; \} \\
& = \sqrt{x^{2}}  \{ \; - A_{2} .(\partial_{1} u) - u. ( \partial_{1} A_{2}) + u.(\omega_{1} \times \omega_{2}) + u. (\partial_{2}A_{1}) + A_{1} . (\partial_{2} u) - u.(\omega_{1} \times \omega_{2}) - u.(\omega_{1} \times \omega_{2}) \; \} \\
& =  \sqrt{x^{2}}  \{ \; - u . (  \partial_{1} A_{2} - \partial_{2} A_{1} - \omega_{1} \times A_{2} + \omega_{2} \times A_{1} - \omega_{1} \times \omega_{2})\; \} \\ 
& =    -  x. ( \mathcal{F}_{12} (A) - A_{1} \times A_{2} - \omega_{1} \times A_{2} + \omega_{2} \times A_{1} - \omega_{1} \times \omega_{2})\;  \\ 
& =  -x . ( \mathcal{F}_{12} (A) - (A_{1} + \omega_{1}) \times (A_{2} + \omega_{2}) )\;  \\ 
& = - x . ( \mathcal{F}_{12} (A) - K_{1} \times K_{2} ) \;  \\ 
\end{align*}

where we have introduced the quantity $K_{a} = A_{a} + \omega_{a}$. It is easy to show that:
\begin{align*}
K_{a} & = A_{a} + u \times \partial_{a}u - ( A_{a} \times u) u \\
& = u \times \partial_{a}u + A_{a} (u.u) - ( A_{a} \times u) u = u \times ( \partial_{a} u + A_{a} \times u )
\end{align*}
which shows that $K_{a}$ is orthogonal to $u$ and that $K_{a} \times K_{b}$ is along the direction of $u$.
With all those intermediate results, one can now rewrites the scalar constraint in a similar form than the one of the four dimensional case.
The scalar and vectorial constraint sare now given by:
\begin{align*}
  H & = x . \mathcal{F}_{12} + (1+\gamma^{2}) x . R_{12}   \;\;\;\;\;\;\;\;\;\;\;\;\;\; \;\;\;\;\;\;\;\; \;\;\;\; \;\;\;\; \;\;\;\; \;\;\;\; \;\;\;\; \;\;\;\;  H_{a}  \simeq \epsilon_{ab} E^{b} . \; \mathcal{F}_{12}  \\
 & = x .  \mathcal{F}_{12}  - (1+\gamma^{2}) x . \mathcal{F}_{12} + (1+\gamma^{2}) x . (K_{1} \times K_{2}) \\
 & = - \gamma^{2} \; x . ( \; \mathcal{F}_{12} - (1+\gamma^{-2}) K_{1} \times K_{2} \; )
\end{align*}

We can therefore regroup the precedent vectorial and the scalar constraints under the same expression:
\begin{align*}
\mathcal{F}_{12} - (1+\gamma^{-2}) K_{1} \times K_{2}  \simeq 0
\end{align*} \\

\textit{Summary} \\

We have presented the canonical analysis of the toy model in the time gauge. 
Let us summarize our results. In this gauge, which select the compact subgroup $SU(2)$ from the initial $SL(2, \mathbb{C})$ group, the phase space mimics closely the four dimensional Ashtekar-Barbero phase space.
The symplectic structure is given by:
\begin{align*}
\{E^{a}_{i} (x), A^{j}_{b} (y)\} = \gamma \delta^{j}_{i} \delta^{a}_{b} \delta^{2}(x-y) 
 \end{align*}
This is the three dimensional version of the Ashtekar Barbero connection, together with its conjugated momenta.
Just as in the four dimensional case, the Barbero-Immirzi enters in the expression of the Poisson bracket.
After the simplification performed previously and the resolution of the second class constraints of the theory, only the following first class constraints remain:
\begin{align*}
G  &= \partial_{b}E^{b} + (A_{b} \times E^{b} ) \simeq 0 \;\;\;\;\;\;\;\;\;\;  \mathcal{F}_{12} - (1+\gamma^{-2}) K_{1} \times K_{2}  \simeq 0
\end{align*}\\

\textit{On the role of the Barbero-Immirzi parameter} \\

 Before studying the quantization of the theory, let us conclude this section about the classical analysis with a discussion on the role of the Barbero-Immirzi parameter. As already emphasized in \cite{ch5-alexandrov2, ch5-GN} and reviewed in the previous subsections, the presence (or absence) of $\gamma$ in the description of the classical phase space seems to be closely related to the partial gauge fixing of the internal Lorentz group.

With the two gauge choices that we have just studied, it seems apparent that the three-dimensional Barbero-Immirzi parameter $\gamma$ ``knows something'' about the Lorentzian signature of the gauge group of the action (2.3). Indeed, when using the time gauge and reducing the Lorentz group to $SU(2)$, the Lorentzian signature is lost and, if $\gamma$ was not present, there would be no way of knowing that the original action that we started our analysis with was Lorentzian and not Euclidean\footnote{Just like in the four-dimensional Holst theory, once the gauge group (either $SO(4)$ or $SL(2,\mathbb{C})$) of the action is reduced to $SU(2)$ by using the time gauge, the only remaining information at the level of the phase space about the gauge group of the non-gauge-fixed action is a relative sign in the scalar constraint.}. Therefore, everything happens as if $\gamma$ was keeping track of the fact that we started with a Lorentzian signature. By contrast, when $SL(2,\mathbb{C})$ is reduced to $SU(1,1)$ by using the non compact gauge of the first section, the Lorentzian signature is still encoded in the gauge group after the gauge fixing, and $\gamma$ completely drops out of the theory because it becomes just superfluous.

Another important observation is that the connection variables that appear once we perform the two gauge choices have very different properties. In addition to their structure group being different because of the gauge fixing, their transformation behavior under diffeomorphisms differ. Indeed, the $su(1,1)$ connection transforms as a one-form under space-time diffeomorphisms, whereas the $su(2)$ connection transforms correctly only under spatial diffeomorphisms. Here again, the analogy between our model and the four-dimensional theory holds, and the anomalous transformation behavior of the $su(2)$ connection is exactly analogous to the anomalous transformation behavior of the four-dimensional Ashtekar-Barbero connection. This comes from the fact that the Ashtekar-Barbero connection is not the pullback of a spacetime connection \cite{ch5-samuel}.

What the three-dimensional model that we are studying here strongly suggests, is that $\gamma$ should be irrelevant at the quantum level. Indeed, we have seen that there exists a gauge in which the dynamical variable is an $su(1,1)$ connection (which in addition transforms correctly under space-time diffeomorphisms), and where $\gamma$ plays no role at all since it disappears already in the classical Hamiltonian theory. The quantization of this $su(1,1)$ theory is far from being trivial, but it can be done for example using the combinatorial quantization scheme \cite{ch5-Cath and I2}, and it is clear that $\gamma$ will play no role in this construction and not appear in the spectrum of any observable. This is a strong indication that $\gamma$ should not play any role at the quantum level even in the time gauge. This would otherwise lead to anomalies, i.e. different quantum predictions in two different gauges, which is not physically acceptable. The two different gauge choices have to lead to equivalent physical predictions in the quantum theory and, as a consequence, the Barbero-Immirzi parameter should be eliminated by a suitable mechanism once all the constraints of the time gauge formulation are imposed. In the next section, we are going to argue that this is indeed the case.

\section{Quantum theory}
\label{sec:quantum theory}

\noindent Since three-dimensional gravity admits only topological and no local degrees of freedom, for a long time if was thought to be too simple to be physically or mathematically interesting. The seminal work of Witten \cite{ch5-Witten 3d} based on its formulation as a Chern-Simons theory \cite{Achucarro-Townsend} showed that it was actually an exactly soluble system with incredibly rich underlying mathematical structures, and provided an unforeseen link with topological invariants \cite{ch5-Witten Jones}. This amazing result triggered an intense research activity around three-dimensional quantum gravity, which lead in particular to the introduction by Ponzano and Regge \cite{ch5-ponzano-regge} and later on Turaev and Viro \cite{ch5-turaev-viro}, of the first spin foam models. These models inspired later on, in four-dimensions, the attempts to represent the covariant dynamics of loop quantum gravity \cite{ch5-BC,ch5-EPR,ch5-EPRL,ch5-livine-speziale1,ch5-livine-speziale2,ch5-FK}, and in \cite{ch5-Noui:2004iy, ch5-Noui:2004iz} the link between three-dimensional loop quantum gravity and spin foam models was establish in the case of a vanishing cosmological constant (see \cite{ch5-AGN} for a more general review). This illustrates concretely the relevance of three-dimensional quantum gravity as a way to investigate the unknown aspects of the higher-dimensional theory. We show in this section that three-dimensional quantum gravity can also be used to investigate the role of the Barbero-Immirzi parameter in the physical Hilbert space of canonical loop quantum gravity.

This section is organized as follows. First we discuss the quantization of the three-dimensional model in the non-compact $\SU(1,1)$ gauge. We argue that the combinatorial quantization scheme can give a precise definition of the physical Hilbert space even if the gauge group is non-compact. By contrast, the loop quantization gives a clear definition of the kinematical Hilbert space but a more formal description of the physical Hilbert space. The rest of the section is devoted to the quantization in the $\SU(2)$ time gauge of subsection \ref{sec:compact gauge}. We adapt and apply the loop quantization to construct the physical Hilbert space, by first turning the initial Ashtekar-Barbero connection into a complex (self-dual) connection, and then rewriting the associated reality conditions as a linear simplicity-like constraint. Finally, we show that the resolution of this constraint lead to the elimination of the Barbero-Immirzi parameter at the quantum level.

\subsection{Quantization in the non-compact gauge}

\noindent In this subsection, we recall a few facts about the quantization of the $G=\SU(1,1)$ BF theory that is obtained from the action (\ref{3D action}) in the non-compact gauge.

The total symmetry group $G_\text{tot}$ of a BF theory is bigger than the gauge group $G$, and is totally determined by the signature of the spacetime (or equivalently the ``signature'' of the gauge group $G$) and the sign of the cosmological constant $\Lambda$. In the case that we are interested in, $G=\SU(1,1)$ and $\Lambda=0$, and the total symmetry group  is the three-dimensional Poincar\'e group $G_\text{tot}=\ISU(1,1)$. In other words, $G$ is somehow augmented with the group of translations. The invariance under translations and the action of $G$ is equivalent (when the $B$ field satisfies invertibility properties) to the invariance under spacetime diffeomorphisms. The total symmetry group $G_\text{tot}$ has a clear geometrical interpretation as the isometry group of the three-dimensional Minkowski space $\mathbb{M}^3$, and any solution to the Einstein equations in the Lorentzian regime with vanishing cosmological constant is locally $\mathbb{M}^3$. In fact, such a BF theory is equivalent to a three-dimensional Chern-Simons theory whose gauge group is precisely $G_\text{tot}$. The Chern-Simons connection takes values in the Lie algebra $\su(1,1)\oplus\mathbb{R}^3$, and admits two components, an $\su(1,1)$ one and a translational one. The $\su(1,1)$ component is the original BF connection whereas the translational component is given essentially by the $B$ field (with the correct dimension).

The symmetry group $\ISU(1,1)$ is non-compact, and inherits the non-compactness of both $\SU(1,1)$ and the group of translations $\mathbb{R}^3$. This makes the quantization quite involved, and is the reason for which quantum BF theory was originally studied in the Euclidean case with a positive cosmological constant. Indeed, this is the only case in which the total symmetry group, $\SU(2)\times\SU(2)$, is compact. In this case, the path integral can be given a well-defined meaning, and gives (three-manifolds or knots) topological invariants. In the non-compact case the definition of the path integral is still an open problem. The most recent attempts to address this issue are based on analytic continuation methods to go from the compact case to the non-compact one \cite{ch5-Witten analytic}. To our knowledge, the Hamiltonian quantization offers a more efficient framework to study Chern-Simons theory with a non-compact group.

Among the different canonical quantizations methods for three-dimensional gravity, the loop and the combinatorial quantizations are certainly the most powerful ones. In fact, the two schemes are closely related as it was shown in \cite{ch5-Cath and I}. They are both based on a discretization of the spatial surface $\Sigma$, which is replaced by an oriented  graph $\Gamma$ sufficiently refined to resolve the topology of $\Sigma$. To simplify the discussion, we will assume that $\Sigma$ has no boundaries and does not contain any particles. Then, the graph $\Gamma$ is necessary closed, and contains $L$ links and $V$ vertices.

\subsubsection{The combinatorial quantization scheme}

\noindent The combinatorial approach consists in quantizing the theory in its Chern-Simons formulation. The dynamical variable is the $\isu(1,1)$ Chern-Simons connection, and to each oriented link $\ell$ of the graph $\Gamma$ is associated an element $U_\ell\in\ISU(1,1)$. After introducing a regularization scheme (based on the choice of a linear order at each vertex of $\Gamma$), the set of elements $U_\ell$ forms a quadratic Poisson algebra known as the Fock-Rosly Poisson bracket. The Fock-Rosly bracket involves classical $r$-matrices of $\isu(1,1)$, and its quantization naturally leads to the quantum double $\DSU(1,1)$ which plays a central role for the algebra of quantum operators. The precise definition of $\DSU(1,1)$ can be found for instance in \cite{ch5-Cath and I}, where it is shown that $\DSU(1,1)$ can be interpreted as a quantum deformation of the algebra of functions on the Poincar\'e group $\ISU(1,1)$. As a consequence, the combinatorial quantization clearly shows that, at the Planck scale, classical isometry groups are turned into quantum groups, and classical smooth (homogeneous) manifolds become non-commutative spaces. To make a very long story short, physical states are constructed from the representation theory of $\DSU(1,1)$.  The combinatorial quantization is a very powerful techniques that allows (at least in principle) to construct the physical Hilbert space of three-dimensional gravity for any Riemann surface $\Sigma$, even in the presence of point particles (see \cite{ch5-BNR} or \cite{ch5-Cath and I2} for instance). 

\subsubsection{The loop quantization}

\noindent The loop quantization is based on the BF formulation of three-dimensional gravity. In the continuous theory, the basic variables are the $\su(1,1)$-valued connection $A$ and its conjugate variable $E$. Given a graph $\Gamma$, one introduces the holonomies $U_\ell\in\SU(1,1)$ along the links $\ell$ of $\Gamma$, and the ``fluxes'' $X_\ell$ of the electric field along edges dual to the links of $\Gamma$. These discretized variables $U_\ell$ and $X_\ell$ form the holonomy-flux Poisson algebra. The quantization promotes these classical variables to operators, the set of which forms a non-commutative algebra which can be represented, as usual in loop quantum gravity, on the Hilbert space
\be
\mathcal{H}_0(\Gamma)=\left(\mathcal{C}(\SU(1,1)^{\otimes L}),\de\mu(\Gamma)\right)
\ee
of continuous functions on the tensor product $\SU(1,1)^{\otimes L}$ endowed with the measure $\de\mu(\Gamma)$. At this stage, the measure is defined as the product of $L$ measures $\de\mu_0$ on $\SU(1,1)$. We notice immediately that the situation is more subtle than in the four-dimensional case because of the non-compactness of the group $\SU(1,1)$. Indeed, for the Hilbert space $\mathcal{H}_0(\Gamma)$ to be well-defined, one should restrict the space of continuous functions to the space of square integrable functions with respect to $\de\mu(\Gamma)$, i.e. $L^2\left((\SU(1,1)^{\otimes L},\de\mu(\Gamma)\right)$. However, any solution to the Gauss constraints is, by definition, invariant under the action of $\SU(1,1)$ at the vertices $v$ of $\Gamma$, and therefore cannot belong to the set of square integrable functions due to the infinite volume of $\SU(1,1)$. As a consequence, the construction of the kinematical Hilbert space requires a regularization process, which amounts to dividing out the volume of the gauge group. This has been studied and well-understood in \cite{ch5-Freidel Livine}. For this construction, it is useful to consider the simplest graph $\Gamma$ that resolves the topology of $\Sigma$. When $\Sigma$ is a Riemann surface (with no punctures and no boundaries) of genus $g$, the simplest graph $\Gamma$ consists in only one vertex $v$ and $L=2g$ loops starting and ending at $v$. Such a graph is called for obvious reason a flower graph, and each loop is in one-to-one correspondence with a generator of the fundamental group $\Pi_1(\Sigma)$. Since the problem of defining the kinematical Hilbert space is a consequence of the invariance of kinematical states under the action of $\SU(1,1)$ at each vertex, this difficulty is considerably reduced by choosing $\Gamma$ to be a flower graph, and one can construct rigorously the kinematical Hilbert space
\be\label{kinematicalH}
\mathcal{H}_\text{kin}(\Gamma)=\left(\mathcal{\cal C}^\text{inv}(\SU(1,1)^{\otimes L}),\de\mu^\text{reg}(\Gamma)\right),
\ee
where ``inv'' stands for invariant and ``reg'' for regularized. One has
\be
f\in\mathcal{H}_\text{kin}(\Gamma)\quad\Longrightarrow\quad
f(U_1,\dots,U_L)=f(VU_1V^{-1},\dots,VU_LV^{-1}),\qquad
\int|f|^2\de\mu^\text{reg}(\Gamma)<\infty,
\ee
for $U_1,\dots,U_L$ and $V$ elements in $\SU(1,1)$. We refer the reader to \cite{ch5-Freidel Livine} for explicit details about this construction.

Once the Gauss constraint is imposed at the quantum level, the flatness condition has to be implemented. This was addressed in the Euclidean regime (where the gauge group is compact) in \cite{ch5-Noui:2004iy}. More precisely, it was shown that one can define a ``projector'' from the kinematical state space into the moduli space of flat $\SU(2)$ connections. This allows to construct rigorously the physical scalar product between kinematical states. The idea is very simple, and consists in replacing the measure $\de\mu_\text{kin}(\Gamma)$ in the kinematical Hilbert space associated to the graph $\Gamma$ by 
\be\label{physicalH}
\de\mu_\text{phys}(\Gamma)=\de\mu_\text{kin}(\Gamma)\prod_{f\in\Gamma}\delta\left(\overrightarrow{\prod_{\ell\subset f}}U_\ell\right),
\ee
where the first product runs over the set of faces $f$ in $\Gamma$ that can be represented by an ordered sequence $(U_1,\dots,U_n)$ of $n$ links, $\delta$ is the Dirac distribution on $\SU(2)$ and $U_\ell$ is the group element associated to the oriented link $\ell$. The physical scalar product can be shown (under certain hypothesis) to be well-defined, and to reproduce exactly the spin foam amplitudes of the Ponzano-Regge model. Even if one does not obtain generically (for any Riemann surface $\Sigma$) an explicit basis for the physical Hilbert space, one can concretely compute the physical scalar product between any two kinematical states. In principle, one could adapt this construction in order to define the physical scalar product in the non-compact case of $\SU(1,1)$ BF theory, and replace the measure on the kinematical Hilbert space (\ref{kinematicalH}) by a measure similar to (\ref{physicalH}) but with $\delta$ the Dirac distribution on $\SU(1,1)$ instead. Even if the presentation that we have done here is incomplete and formal, the technical details are not needed for the main purpose of the paper, which is to address the quantization in the $\SU(2)$ time gauge. We tackle this in the next subsection.

\subsection{Quantization in the time gauge}
	
\noindent We now study the quantization of the theory in the time gauge. There are essentially two ways of doing so. The first one is to mimic exactly four-dimensional loop quantum gravity, where one starts with the construction of the kinematical Hilbert space and then finds a regularization of the Hamiltonian constraint \`a la Thiemann in order to find the physical solutions. The second one relies on a reformulation of the classical phase space in a way that looks again like a BF theory. Let us start with a discussion about the first strategy. We use the same notations as in the previous subsection: $\Sigma$ is the spatial manifold, and $\Gamma$ a graph in $\Sigma$ with $L$ links and $V$ vertices.

In $\SU(2)$ loop quantum gravity, the construction of the kinematical Hilbert space leads to
\be\label{SU2 kinematical}
\mathcal{H}_\text{kin}(\Gamma)=\left(\mathcal{C}(\SU(2)^{\otimes L}),\de\mu(\Gamma)\right),
\ee
where $\mathcal{C}(G)$ denotes the space of continuous functions on the group $G$, and $\de\mu(\Gamma)$ is the usual Ashtekar-Lewandowski measure defined as a product of $L$ Haar measures $\de\mu_0$ on $\SU(2)$. Contrary to what happens in four dimensions where it is necessary to consider all possible graphs on $\Sigma$ (and to take a projective limit), here is it sufficient to fix only one graph (appropriately refined to resolve the topology of $\Sigma$) in order to define the kinematical Hilbert space. Since the gauge group is compact, the kinematical Hilbert space is well-defined, and $\mathcal{H}_\text{kin}(\Gamma)$ carries a unitary representation of the three-dimensional holonomy-flux algebra. The action of a flux operator $X_\ell$ on any kinematical state $\psi\in\mathcal{H}_\text{kin}(\Gamma)$ can be deduced immediately from the action on the representation matrices $\mathbf{D}^{(j)}(U_\ell)$ (which are the building blocks of the spin networks), where $\mathbf{D}^{(j)}:\SU(2)\rightarrow\mathbb{V}^{(j)}$ is the $\SU(2)$ spin-$j$ representation on the space $\mathbb{V}^{(j)}$ of dimension $d_j=2j+1$. This action is given by
\be\label{action of E}
X_\ell^i\triangleright\mathbf{D}^{(j)}(U_{\ell'})=\mathrm{i}\gamma\lp\delta_{\ell,\ell'}\mathbf{D}^{(j)}(U_{\ell<c})J_i\mathbf{D}^{(j)}(U_{\ell>c}),
\ee
where $c$ denotes the intersection $\ell\cap\ell'$. The constants $\gamma$ and $\lp=\hbar G_\text{N}$ are the Barbero-Immirzi parameter and the three-dimensional Planck length. As in four dimensions, the spin network states diagonalize the three-dimensional analogue of the area operator, namely $\sqrt{X_\ell^2}$, whose eigenvalues are $\gamma\lp\sqrt{j_\ell(j_\ell+1)}$. Therefore, one arrives at the conclusion that in the time gauge the kinematical length operator has a discrete spectrum given by the Casimir operator of $\SU(2)$, and is furthermore proportional to the Barbero-Immirzi parameter, which can be interpreted as the fundamental length scale in Planck units. Just like in the four-dimensional case, one inherits a $\gamma$-dependency in the quantum theory, which as we will argue later on is completely artificial and an artifact of the gauge choice.

It is however legitimate to ask what happens if we try to push further the derivation of physical results based on this $\SU(2)$ formulation. For example, mimicking once again what is done in the four-dimensional theory, one could try to compute the entropy of a black hole, which in this three-dimensional model would correspond to a BTZ black hole. Using the notion of observables in the Ponzano-Regge spin foam model, one can reproduce the calculation of \cite{ch5-BTZ-SF} and choose the fundamental length elements to be such that the perimeter $L$ of the black hole is given by
\be
L=8\pi\gamma\lp\sum_{\ell=1}^p\sqrt{j_\ell(j_\ell+1)},
\ee
where $p$ is the number of spin network links $\ell$ puncturing the horizon (we have here reintroduced the appropriate numerical factors). Then, the computation of the number of microstates leads at leading order to an entropy formula of the type
\be
S_\text{BH}=\f{L}{4\lp}\f{\gamma_0}{\gamma},
\ee
and one can proceed by fixing the value of the three-dimensional Barbero-Immirzi parameter to be $\gamma_0$, whose value can be computed explicitly. What is remarkable is that this value agrees with that derived in the four-dimensional case. For example, if we choose for simplicity all the spins puncturing the horizon to be equal, i.e. $j_\ell=j$ for all $\ell \in \llbracket1,p\rrbracket$, then on has to fix the value of $\gamma$ to $\gamma_0=\log(2j+1)/\big(2\pi\sqrt{j(j+1)}\big)$. This observation is a further indication that our three-dimensional model does indeed mimic exactly its four-dimensional counterpart, and that the behavior of the Barbero-Immirzi parameter is the same in both cases (once we use the time gauge and the $\SU(2)$ formulation).

Finally, once the kinematical structure is established, one should impose at the quantum level the three remaining constraints $H\simeq0$. These appear as the sum of two terms, $H=H_\text{E}-(1+\gamma^{-2})H_\text{L}$, where $H_\text{E}=\mathcal{F}_{12}$ and $H_\text{L}=K_1\time K_2$ are respectively called the Euclidean and the Lorentzian part of the constraints. In four dimensions, one has to consider separately the vector constraint and the scalar constraint, but we have seen that the peculiarity of three-dimensional gravity is that these can be treated as a single set. The set $H$ of constraints needs to be regularized in order to have a well-defined action on $\mathcal{H}_\text{kin}(\Gamma)$, and it is clear that the regularization of $H_\text{L}$ will lead to the same ambiguities that are present in four-dimensional canonical loop quantum gravity \cite{ch5-ale-amb,ch5-Bonzom:2011jv}. Since we know (from the quantization of three-dimensional gravity in the usual BF or spin foam setting) what the physical states should look like, one could potentially investigate these regularizations ambiguities of the Hamiltonian constraint and maybe try to clarify them. Although this would be a very interesting task that could have important consequences for the construction of the four-dimensional theory, we are going to follow instead the second strategy mentioned above, which consists in rewriting the $\SU(2)$ Ashtekar-Barbero phase space in the form of a BF theory.

\subsubsection{Equivalence with a complex BF theory}

\noindent Our aim is to reformulate the phase space of the three-dimensional theory in the time gauge in a way equivalent to a BF theory. More precisely, we are looking for a pair $(\mathbf{A}_a^i,\mathbf{E}^a_i)$ of canonical variables such that the constraints $G\simeq0$ and $H\simeq0$ are equivalent to the constraints:
\begin{eqnarray}\label{new constraints}
\mathbf{G}=\partial_a\mathbf{E}^a+\mathbf{A}_a\time\mathbf{E}^a\simeq0,\qquad
\mathbf{F}_{12}\simeq0, 
\end{eqnarray}
where $\mathbf{F}_{12}$ is the curvature of $\mathbf{A}$. We use the following ansatz for the expressions of the new variables in terms of the old ones:
\be
\mathbf{A}=A+\alpha L+\beta K,\qquad
\mathbf{E}=\zeta E+\xi u\time E,
\ee
where $K=u\time L$ was introduced, $L=\de u+A\time u$, and $\alpha$, $\beta$, $\zeta$, and $\xi$ are constants that have to be fixed by the 
relations (\ref{new constraints}). In fact, this ansatz gives the most general expression for a connection $\mathbf{A}$ and an electric field $\mathbf{E}$ that transform correctly under $\SU(2)$ gauge transformations. This results from the fact that $u$, $L$ and $K$ transform as vectors under such gauge transformations.

It is useful to derive some properties of the quantities $L$ and $K$. A direct calculation shows that $L$ and $K$ satisfy the equations
\begin{subequations}
\ba
&&\partial_1L_2-\partial_2L_1+A_1\time L_2+L_1\time A_2=\mathcal{F}_{12}\time u,\\
&&\partial_1K_2-\partial_2K_1+A_1\time K_2+K_1\time A_2=u\time(\mathcal{F}_{12}\time u)+2L_1\time L_2.
\ea
\end{subequations}
Furthermore, $L_1\time L_2 = K_1\time K_2$. Thus, the curvature $\mathbf{F}$ of $\mathbf{A}$ can be written in terms of $\mathcal{F}$ and $K$ as follows:
\be
\mathbf{F}_{12}=\mathcal{F}_{12}+(\alpha^2+\beta^2+2\beta)K_1\time K_2+\alpha\mathcal{F}_{12}\time u+
\beta(u\time\mathcal{F}_{12})\time u.
\ee
Due to the fact that $K_1\time K_2$ is in the direction of $u$, the curvature $\mathbf{F}$ takes the form
\be
\mathbf{F}_{12}=(1-\alpha\underline{u}-\beta\underline{u}^2)\big(\mathcal{F}_{12}+(\alpha^2+\beta^2+2\beta)K_1\time K_2\big),
\ee
where, for any vector $a$, $\underline{a}$ denotes the matrix that acts as $\underline{a}b_i=\eps_{i}^{~jk}a_jb_k$ on any vector $b$. As a consequence, since the three-dimensional matrix $1-\alpha\underline{u}-\beta\underline{u}^2$ is invertible, the constraints $H\simeq0$ are equivalent to $\mathbf{F}_{12}\simeq 0$ if and only if
\be
\alpha^2+(1+\beta)^2+\gamma^{-2}=0.
\ee
Before considering the fate of the Gauss constraint, we already see that the connection $\mathbf{A}$ must be complex when the Barbero-Immirzi parameter $\gamma$ is real. Indeed, the general solution of the previous equation is given by
\be
\alpha=z\sin\theta,\qquad
\beta=z\cos\theta-1,
\ee
with $z^2+\gamma^{-2}=0$, and where $\theta$ in an arbitrary angle. We will discuss the complexification in more detail later on.

Now, we compute the new Gauss constraint $\mathbf{G}$ in term of the original variables. A long but straightforward calculation shows that
\be
\mathbf{G}=\zeta G+\xi u\time G+(\zeta\beta-\xi\alpha) (G\cdot u)u+
(\xi+\alpha\zeta+\beta\xi)\big(\partial_au\time E^a+(A_a\cdot E^a)u\big),
\ee
which can be written as follows:
\be\label{capital G}
\mathbf{G}=MG+(\xi+\alpha\zeta+\beta\xi)\big(\partial_au\time E^a+(A_a\cdot E^a)u\big),
\ee
where $M$ is the matrix $M=\zeta(1+\beta)-\xi\alpha+\xi\underline{u}+(\zeta\beta-\xi\alpha)\underline{u}^2$. A necessary condition for $\mathbf{G} =0$ to be equivalent to $G=0$ is that the coefficient $(\xi+\alpha\zeta+\beta\xi)$ in front of the second term in (\ref{capital G}) be vanishing. This implies that $\xi=-\lambda\alpha$ and $\zeta=\lambda(1+\beta)$ with an arbitrary (but non-vanishing) coefficient $\lambda$ which in addition makes the matrix $M$ necessarily invertible. As a consequence, the general solution of the new constraints (\ref{new constraints}) is given by
\be\label{general sol}
\mathbf{A}=A+z\sin\theta L+(z\cos\theta-1)K,\qquad
\lambda^{-1}\mathbf{E}=z\cos\theta E-z\sin\theta(u\times E),
\ee
where $\theta$ is an arbitrary angle, $\lambda\neq0$, and $z^2+\gamma^{-2}=0$. Since $\lambda$ affects only the Poisson bracket between $\mathbf{A}$ and $\mathbf{E}$, we can set it to $\lambda=1$ for simplicity without loss of generality. 

At this point, there is a priori no reason for $\mathbf{A}$ and $\mathbf{E}$ to be canonically conjugated, and even $\mathbf{A}$ itself might be non-commutative. This would prevent the phase space of the theory in the time gauge from being equivalent to that of a BF theory. Fortunately, the previous expressions can be simplified considerably by noticing that all the solutions (\ref{general sol}) are in fact equivalent. More precisely, for any solution (\ref{general sol}), there exists a $\Lambda\in \SU(2)$ that sends this solution to the simple one corresponding to $\theta=0$. As a consequence, one can take $\theta=0$ without loss of generality, and this makes the study of the new variables much simpler. To see that this is indeed the case, let us compute how a solution (\ref{general sol}) transforms under the action of a rotation $\Lambda(u,\alpha)$ of angle $\alpha$ in the plane normal to $u$. Such an element is represented by the matrix
\be
\Lambda(u,\alpha)=\cos\left(\f{\alpha}{2}\right)+2\sin\left(\f{\alpha}{2}\right)J\cdot u
\ee
in the fundamental (two-dimensional) representation of $\SU(2)$, where $J_i$ are the $\su(2)$ generators satisfying the Lie algebra
\be
[J_i,J_j]=\eps_{ij}^{~~k}J_k.
\ee
If we identified any vector $a\in\mathbb{R}^3$ with an elements of $\su(2)$ according to the standard map $a\longmapsto a\cdot J$, the transformation laws for $\mathbf{A}\longmapsto\mathbf{A}^{\!\Lambda}$ and $\mathbf{E}\longmapsto\mathbf{E}^{\Lambda}$ under the action of $\Lambda$ are given by
\be\label{transform A and E}
\mathbf{A}^{\!\Lambda}=\Lambda^{-1}\mathbf{A}\Lambda+\Lambda^{-1}\de\Lambda,\qquad
\mathbf{E}^{\Lambda}=\Lambda^{-1}\mathbf{E}\Lambda.
\ee
To go further, we need to compute the adjoint action of $\SU(2)$ on its Lie algebra, and the differential form in the expression of $\mathbf{A}^\Lambda$:
\begin{subequations}
\ba
&&\text{Ad}_\Lambda J=\Lambda^{-1}J\Lambda=\cos\alpha J+\sin\alpha u\time J+2\sin^2\left(\f{\alpha}{2}\right) (u\cdot J)u,\\
&&\Lambda^{-1}\de\Lambda=\left(1+\sin^2\left(\f{\alpha}{2}\right)J\cdot u\right)\de\alpha+\sin\alpha J\cdot\de u-2\sin^2\left(\f{\alpha}{2}\right)J\cdot u\times\de u,
\ea
\end{subequations}
where we used the relation
\be
J_iJ_j=-\f{1}{4}\delta_{ij}+\f{1}{2}\eps_{ij}^{~~k} J_k
\ee
satisfied  by the 
$\su(2)$ generators in the fundamental representation. We can now compute the transformations (\ref{transform A and E}), and after some long but direct calculations, we obtain
\ba
\mathbf{A}^{\!\Lambda}&=&z\big(\cos(\theta-\alpha)E-\sin(\theta+\alpha)u\time E\big)\cdot J,\\
\mathbf{E}^{\Lambda}&=&A\cdot J+z\sin(\theta+\alpha)L\cdot J+\big(z\cos(\theta+\alpha)-1\big)K\cdot J,
\ea
when $\alpha$ is assumed to be constant, i.e. $\de\alpha=0$. Taking $\alpha=\theta$ simplifies the previous expressions and reduces the variables $\mathbf{A}^{\!\Lambda}$ and $\mathbf{E}^{\Lambda}$ to $\mathbf{A}^{\Lambda}$ and $\mathbf{E}^{\Lambda}$ given in (\ref{general sol}) where $\theta=0$. Finally, as announced, all the solutions of the type (\ref{general sol}) are equivalent. Therefore, we will now fix $\theta=0$, and use again the notation $\mathbf{A}$ and $\mathbf{E}$ to denote
\be\label{complex conn}
\mathbf{A}=A+(z-1)K,\qquad\mathbf{E}=zE.
\ee
As a conclusion, there is only one choice (up to $\SU(2)$ gauge transformations) of canonical variables that reduces the constraints obtained in the time gauge to BF-like constraints. However, it is immediate to notice that $\mathbf{E}$ and $\mathbf{A}$ are not canonically conjugated, and also that the components of the connection do not commute with respect to the Poisson bracket. This is a priori problematic since it makes the symplectic structure of the time gauge theory very different from that of BF theory. Fortunately, there is a simple and very natural explanation for this fact. Instead of the connection $\mathbf{A}$, let us consider the connection
\be\label{asd}
\mathcal{A}_a^{(\gamma z)}=\mathbf{A}_a+\frac{z-1}{|E|}\eps_{ab}(E^b\cdot G)x=zA_a+(z-1)\omega_a,
\ee
which differs from $\mathbf{A}$ only by a term proportional to the Gauss constraint, and where $\omega$ is the solution of the second class constraints written in (\ref{sol omega}). The choice of the notation $\mathcal{A}^{(\gamma z)}$ will become clear in what follows. Clearly, adding a term proportional to the Gauss constraint does not change anything to the previous analysis. Moreover, if we go back to the very first definition of $A=\gamma\omega^{(0)}-\omega$ in terms of the boost $\omega^{(0)}$ and rational $\omega$ components of the initial $\sl(2,\mathbb{C})$ connection, we see immediately that, depending on the sign of $\gamma z\in\{+,-\}$, the object
\be\label{shifted connection}
\mathcal{A}^{(\gamma z)}_a=zA_a+(z-1)\omega_a=\gamma z\omega^{(0)}_a-\omega_a=\pm\mathrm{i}\omega^{(0)}_a-\omega_a
\ee
is the (anti) self-dual component of the initial $\sl(2,\mathbb{C})$ connection. This comes from the fact that $z=\pm\mathrm{i}\gamma^{-1}$. In other words, reducing the phase space of the time gauge theory to that of a BF theory has mapped the initial $\su(2)$ Ashtekar-Barbero connection to the (anti) self-dual connection, as one could have anticipated. This property ensures that $\mathbf{E}$ and $\mathbf{A}$ (up to the Gauss constraint) satisfy the ``good'' canonical relations.

Since $z=\pm\mathrm{i}\gamma^{-1}$ is purely imaginary when the Barbero-Immirzi parameter $\gamma$ is real, $\mathbf{A}$ is complex and can be interpreted as an $\sl(2,\mathbb{C})$-valued connection. If we denote by $P_i \in \sl(2,\mathbb{C})$ the infinitesimal boost generators that satisfy the Lie algebra
\be
[J_i,P_j]=\eps_{ij}^{~~k}P_k,\qquad
[P_i,P_j]=-\eps_{ij}^{~~k}J_k,
\ee
and make explicit the Lie algebra generators that serve as a basis for the components of the connection, then the complex connection (\ref{complex conn}) can be identified with
\be\label{complex connection}
\mathbf{A}=A\cdot J+(z-1)K\cdot J=(A-K)\cdot J\pm\mathrm{i}\gamma^{-1}K\cdot J=(A-K)\cdot J\mp\gamma^{-1}K\cdot P
\ee
since $P=-\mathrm{i}J$ in the fundamental representation. Since the fundamental representation is faithful, the identity (\ref{complex connection}) can be extended to the Lie algebra, which is what we are going to do.

\subsubsection{Emergence of the linear simplicity-like condition}

\noindent To construct the quantum theory, we start with the $\SU(2)$ kinematical Hilbert space (\ref{SU2 kinematical}). Kinematical states are cylindrical functions that can be expanded into spin network states. Any spin network is composed of an assignment of unitary irreducible representations $j_\ell$ of $\SU(2)$ to the links $\ell\in\Gamma$ and of intertwiners $\iota_v$ to the vertices $v$ of the graph $\Gamma$. To simplify the problem and without loss of generality, we take for $\Gamma$ the flower that contains only one vertex. We expect that (\ref{SU2 kinematical}) cannot contain physical states that are annihilated by the quantum Hamiltonian constraint. This space should therefore be extended by considering $\SL(2,\mathbb{C})$ cylindrical functions instead of $\SU(2)$ ones. However, because of the non-compactness of the Lorentz group, it then becomes difficult to treat the $\SL(2,\mathbb{C})$ invariance at the vertex of $\Gamma$ (i.e. the construction of $\SL(2,\mathbb{C})$ intertwiners) and to construct a positive-definite physical scalar product. These two issues are technically very difficult to address in full generality, and in what follows we will only need to study the vector space structure of the set of physical states where intertwiners are formally defined.

The extension $\mathcal{H}_\text{kin}^\text{ext}(\Gamma)$ to $\SL(2,\mathbb{C})$ cylindrical functions can be viewed as a complexification of the initial kinematical Hilbert space since $\SL(2,\mathbb{C})$ itself is a complexification of $\SU(2)$. Elements of $\mathcal{H}_\text{kin}^\text{ext}(\Gamma)$ can be (formally) expanded into $\SL(2,\mathbb{C})$ spin networks, which consist of an assignment of irreducible representations of $\SL(2,\mathbb{C})$ to the links $\ell$, and of an intertwiner to the unique vertex $v$ of $\Gamma$. Any representation of $\SL(2,\mathbb{C})$ is labelled by a couple $(\chi_0,\chi_1)$ of complex numbers such that, for the principal series, $\chi_0=m+\mathrm{i}\rho$ and $\chi_1=-m+\mathrm{i}\rho$, with $\rho\in\mathbb{R}$ and $m\in\mathbb{N}$. So far, we have not fixed the category of representations that color the links of the graph, and, as we said above, the intertwiner is only formally defined. Nonetheless, let us emphasize once again that our construction will still be well-defined at the end of the day. Of course, we have the inclusion $\mathcal{H}_\text{kin}(\Gamma)\subset\mathcal{H}_\text{kin}^\text{ext}(\Gamma)$, and any element of the extended kinematical state space belongs to the original one when the $\SL(2,\mathbb{C})$ representations  are chosen to be the self-dual or anti self-dual ones, which corresponds to $\chi_0=0$ or $\chi_1=0$.

The elements of $\mathcal{H}_\text{kin}^\text{ext}(\Gamma)$ that we are interesting in are the cylindrical functions (or equivalently the spin network states) of the complex connection $\mathbf{A}$ defined in (\ref{complex connection}). These can be viewed as generated by the action of an operator on the original kinematical Hilbert space (\ref{SU2 kinematical}). Indeed, cylindrical functions associated to $\mathbf{A}$ are operators acting non-trivially on the $\SU(2)$ Hilbert space $\mathcal{H}_\text{kin}(\Gamma)$, and their action on the vacuum state (colored by trivial representations) gives elements of $\mathcal{H}_\text{kin}^\text{ext}(\Gamma)$. In this sense, $\mathcal{H}_\text{kin}^\text{ext}(\Gamma)$ is constructed from the kinematical Hilbert space $\mathcal{H}_\text{kin}(\Gamma)$. Contrary to the $\SU(2)$ spin network states, the $\SL(2,\mathbb{C})$ ones contain a priori a non-trivial boost component. To understand more precisely the structure of these extended spin networks, let us decompose the connection $\mathbf{A}$ as follows:
\be\label{decomposition of A}
\mathbf{A}=\left[-u\time\de u\cdot(J\mp\gamma^{-1}P)\right]+
\left[(A\cdot u)(u\cdot J)\pm\gamma^{-1}(A\time u)\cdot(P\time u)\right].
\ee
This expression is a straightforward consequence of (\ref{complex conn}) and (\ref{Kformula}), and shows that $\mathbf{A}$ possesses two different parts, which are the two terms between the square brackets. Let us start by interpreting the second one. For this, we introduce the three $\sl(2,\mathbb{C})$ elements $\widetilde{J}_3=J\cdot u$ and $\widetilde{P}_\alpha=P\time u\cdot v_\alpha$, where the two vectors $v_\alpha^i\in\mathbb{R}^3$ ($\alpha=1,2$) are such that $v_\alpha\cdot v_\beta=\delta_{\alpha\beta}$ and $v_1\time v_2=u$. Therefore, $(v_1,v_2,u)$ forms an orthonormal basis of $\mathbb{R}^3$. The two vectors $v_\alpha$ are defined up to a rotation in the plane orthogonal to $u$, but this is not relevant for what follows. It is immediate to see that $(\widetilde{J}_3,\widetilde{P}_1,\widetilde{P}_2)$ forms the Lie algebra
\be
[\widetilde{P}_1,\widetilde{P}_2]=-\widetilde{J}_3,\qquad
[\widetilde{P}_2,\widetilde{J}_3]=\widetilde{P}_1,\qquad
[\widetilde{J}_3,\widetilde{P}_1]=\widetilde{P}_2,
\ee
and therefore generate an $\su(1,1)$ subalgebra of the initial $\sl(2,\mathbb{C})$. In the literature, the Lie algebra $\su(1,1)$ is usually defined as being generated by the three elements $(F_0,F_1,F_2)$ satisfying
\be
[F_1,F_2]=\mathrm{i}F_0,\qquad
[F_0,F_2]=\mathrm{i}F_1,\qquad
[F_0,F_1]=-\mathrm{i}F_2.
\ee
These generators are related to the previous ones through the map $(\widetilde{J}_3,\widetilde{P}_1,\widetilde{P}_2)\longmapsto\mathrm{i}(F_0,F_1,F_2)$. As a consequence, the second component (\ref{decomposition of A}) defines an $\su(1,1)$-valued one-form which has to be interpreted as an $\su(1,1)$ connection. This condition is necessary in order to avoid anomalies, i.e. different quantum theories in two different gauges. We saw in the previous section that there is a gauge in which the original three-dimensional Holst action takes the form of an $\SU(1,1)$ BF theory, and therefore it is natural to recover an $\su(1,1)$ connection even if we work in the time gauge. However, for this to be true, the first term in the expression (\ref{decomposition of A}) for the $\sl(2,\mathbb{C})$ connection must vanish. Since $u\time\de u$ is a dynamical variable, one must impose the conditions
\be\label{simplicity modified}
J\mp\gamma^{-1}P=0.
\ee
Interestingly, this is the same relation as the linear simplicity constraint used in the new spin foam models. Just like in the construction of spin foam models, this constraint selects representations of $\sl(2,\mathbb{C})$.

To understand how the representations of $\sl(2,\mathbb{C})$ are constrained by (\ref{simplicity modified}), one has to notice that this equation has two families of solutions.
\begin{itemize}
\item[1.] The first one consists in modifying the action of the infinitesimal boosts $P_i$ while keeping the action of the infinitesimal rotations $J_i$ unchanged in such a way that the linear condition is satisfied, i.e. by setting $P=\pm\gamma J$. This is exactly what is done in the construction of spin foam models. By doing this, the action of the boosts is somehow compactified, and if one replaces $P$ by $\pm\gamma J$ in the expression (\ref{decomposition of A}), the $\sl(2,\mathbb{C})$ connection reduces to $\mathbf{A}=A\cdot J$. This solution is the initial $\su(2)$-valued Ashtekar-Barbero connection, and therefore it cannot lead to an $\su(1,1)$ connection\footnote{If one still wants to interpret this object as an $\su(1,1)$ connection by writing it on the basis $(F_0,F_1,F_2)$, one has to go back to the fundamental two-dimensional representation. In this representation, the generators $P_i$ and $J_i$ are related by a global factor of $\mathrm{i}$ only, i.e. $P_i=-\mathrm{i}J_i$. Therefore, the connection takes the form
\be\label{su(2) connection as su(1,1)}
\mathbf{A}=A\cdot J=A_1F_1+A_2F_2+\mathrm{i}A_3F_0,
\ee
where we have identified the $\su(2)$ generators $(J_1,J_2,J_3)$ with $(F_1,F_2,\mathrm{i}F_0)$. As expected $\su(2)$ and $\su(1,1)$ are related by a simple (complex) redefinition of the generators, and the connection (\ref{su(2) connection as su(1,1)}) is $\su(2)$-valued because of the factor of $\mathrm{i}$ in its third internal direction.}. This first sector of solutions to the simplicity-like constraints (\ref{simplicity modified}) selects the maximal compact subalgebra $\su(2)$ of $\sl(2,\mathbb{C})$.
\item[2.] The second solution consists in modifying the action of the infinitesimal rotations $J_i$ by setting $J=\pm\gamma^{-1}P$, while keeping unchanged the action of the infinitesimal (non-compact) boosts $P_i$. In this sense, the rotations are ``decompactified''. With this solution, the complex connection (\ref{decomposition of A}) reduces to the non-compact element $\mathbf{A}=\pm\gamma^{-1}A\cdot P$, which does not a priori define an $\su(1,1)$ connection since the $P$'s do not form a Lie subalgebra of $\sl(2,\mathbb{C})$. However, this is only an apparent problem. Indeed, let us recall that our extension from $\su(2)$ to $\sl(2,\mathbb{C})$ has been done in the two-dimensional representation. In this representation, the generators $P_i$ and $J_i$ are related by a global factor of $\mathrm{i}$, i.e. $P_i=-\mathrm{i}J_i$. As a consequence, in the fundamental representation, the second solution can be written as follows:
\be\label{second solution}
\mathbf{A}=\pm\gamma^{-1}A\cdot P=\mp\mathrm{i}\gamma^{-1}A\cdot J=\mp\mathrm{i}\gamma^{-1}\left(A_1F_1+A_2F_2+\mathrm{i}A_3F_0\right),
\ee
where we have identified the $\su(1,1)$ generators $(F_0,F_1,F_2)$ with $-\mathrm{i}(J_3,P_1,P_2)=(-\mathrm{i}J_3,J_1,J_2)$. It is therefore clear that the second solution selects the non-compact $\su(1,1)$ connection in the initial Lorentz algebra.
\end{itemize}

We have presented here what seems to be the only two consistent ways of interpreting the constraint (\ref{simplicity modified}). Indeed, this constraint should select a three-dimensional subalgebra of $\sl(2,\mathbb{C})$ in order for the resulting connection to be well-defined. The first solution selects the compact one and corresponds to the choice made in spin foam models. In the context of our analysis, this solution is not physically relevant since we expect the resulting connection to be valued in $\su(1,1)$ and not in the Lie algebra of a compact group. By contrast, the second solution looks much more appealing since it leads to an $\su(1,1)$-valued connection and also to the disappearance of the Barbero-Immirzi parameter in the spectrum of the geometrical operators due to the overall factor of $\gamma^{-1}$ in $\mathbf{A}$ (\ref{second solution}). Thus, in the time gauge, the resolution of (part of) the quantum dynamics turns the initial Ashtekar-Barbero connection to an $\su(1,1)$ connection, and the theory becomes strictly equivalent to an $\SU(1,1)$ BF theory. This results is totally consistent with the results obtained in the non-compact gauge, where the theory reduces to an $\SU(1,1)$ BF theory as well. Since the equivalence to a BF theory is established, the full quantization of the theory (i.e. the imposition of the quantum flatness constraint) can in principle be performed and does not pose any conceptual problems even if it can be mathematically involved.

\subsubsection{Action of the flux operator}

\noindent Let us finish this section with a  quick discussion on the disappearance of the Barbero-Immirzi parameter from the spectra of the geometrical operators. Once the second family of solutions to the constraint (\ref{simplicity modified}) is selected, we have the $\su(1,1)$ connection (\ref{second solution}) and the spin network states are colored with unitary irreducible representations of $\SU(1,1)$ that we label by $r$. We know that the action of the flux operator $X_\ell^i$ defined by the triad $E^a_i$ is given by $-\mathrm{i}\hbar\delta/\delta A^i_a$. Moreover, since classically $\lb E^a_i,A^j_b\rb=\gamma\delta^a_b\delta^j_i$, we can compute the action of the flux on the holonomy of the connection (\ref{second solution}) to find
\be\label{action of E}
X_\ell^i\triangleright\mathbf{D}^{(r)}(U_{\ell'})=\mp\lp\delta_{\ell,\ell'}\mathbf{D}^{(r)}(U_{\ell<c})F_i\mathbf{D}^{(r)}(U_{\ell>c}),
\ee
where $c$ denotes the intersection $\ell\cap\ell'$ and $J_3$ is identified with $\mathrm{i}F_0$ according to (\ref{second solution}). An equivalent point of view would be to see the flux operator as acting on the holonomy of the shifted (or self-dual) connection $\mathcal{A}^{(\gamma z)}$ defined in (\ref{shifted connection}), whose Poisson bracket with $E$ is given by $\gamma z=\pm\mathrm{i}$. This action is also independent of $\gamma$. Therefore, we see that it is equivalent to consider the self-dual theory with an imaginary Barbero-Immirzi parameter or the complex theory defined with (\ref{second solution}) and $\gamma\in\mathbb{R}$, since in this case $\gamma$ disappears due to the redefinition of the appropriate variables. Beyond this observation about the role of $\gamma$, it is even more interesting to see that it is possible to obtain a positive-definite length spectrum.

Indeed, the ``gauge invariant'' quadratic operator $X_\ell^2$ is diagonalized by the spin networks and its eigenvalues are given by
\be
X_\ell^2\triangleright\mathbf{D}^{(r)}(U_{\ell'})=\lp^2\delta_{\ell,\ell'}Q^{(r)}\mathbf{D}^{(r)}(U_{\ell}),
\ee
where $Q^{(r)}$ denotes the evaluation of the Casimir operator $Q=F_1^2+F_2^2-F_0^2$ in the representation labelled by $r$. First, it is important to notice that the factor of $\mathrm{i}$ in the third component of the connection (\ref{second solution}) is responsible for the fact that the action of $X_\ell^2$ gives exactly the Casimir of $\su(1,1)$. Second, let us recall that there are two families of unitary irreducible representations of $\su(1,1)$, the continuous series (non-exceptional and exceptional classes) and the discrete series (positive and negative). $Q$ takes negative values for the later and positive values for the former. Therefore, if one requires that $X_\ell^2$ be a positive-definite operator, only the continuous series is admissible. Since $X_\ell^2$ is the building block of the length operator, at the physical level this operator will necessarily have a continuous spectrum. This is to be contrasted with the a priori prediction that could have been made at the kinematical level if we had stopped the analysis of the $\SU(2)$ theory in the time gauge i.e. before recasting the Hamiltonian constraint as a flatness constraint for the complex connection $\mathbf{A}$, and before arriving at the linear constraint that selects the $\su(1,1)$ connection. Indeed, if we had stayed at the superficial level of the $\SU(2)$ kinematics, we would have derived a discrete length spectrum proportional to $\gamma$. The observation that working with the $\SU(1,1)$ representations leads to a continuous spectrum independent of $\gamma$ is completely consistent with the fact that we are describing Lorentzian three-dimensional gravity \cite{ch5-freidel-livine-rovelli}.

\section{Conclusion and perspectives}

\noindent In this chapter, we have studied the role of the Barbero-Immirzi parameter in the construction of a symmetry reduced version of loop quantum gravity. This symmetry reduction consists in imposing invariance along a given spatial direction, which reduces the original four-dimensional Holst action to an action for three-dimensional gravity with a Barbero-Immirzi parameter. This action was originally introduced and analyzed in \cite{ch5-GN2} in its Plebanski form, and further studied in \cite{ch5-GN} in Euclidean signature and for two specific gauge choices. In the Lorentzian theory, these two gauge choices, which we have shown to be consistent with the dynamics of three-dimensional gravity, have drastically different interpretations. The first one reduces the action to that of $\SU(1,1)$ BF theory, and leads to a Hamiltonian formulation without any dependency on the Barbero-Immirzi parameter. The second one, which we refer to as the time gauge, leads just like in the four-dimensional case to an $\SU(2)$ theory written in terms of the Ashtekar-Barbero connection for $\gamma\in\mathbb{R}$ and admitting the same type of first class Gauss, scalar and vector constraints. Since three-dimensional gravity is an exactly soluble (classical and quantum) system, we have argued that this model can serve as a test bed to understand the relevance of the Barbero-Immirzi parameter in the dynamics of quantum gravity.

We have seen in this three-dimensional model that it is possible to rewrite the scalar and vector constraints of the $\SU(2)$ theory in the time gauge in the form of a unique flatness constraint for a complex connection $\mathbf{A}$, and that this latter is closely related to the complex (anti) self-dual Ashtekar-Barbero connection. However, nowhere did we set by hand the Barbero-Immirzi parameter to the value $\gamma=\pm\mathrm{i}$. Then, we have argued that in order for the quantum theory to be consistent with the quantization of Lorentzian three-dimensional gravity (i.e. $\SU(1,1)$ BF theory), the complex connection $\mathbf{A}$ had to be $\su(1,1)$-valued, a requirement that is met only if the generators of $\sl(2,\mathbb{C})$ satisfy the constraint $J\mp\gamma^{-1}P=0$. This constraint is nothing but the linear simplicity constraint used in the construction of four-dimensional spin foam models, and its role is to restrict the representations of the Lorentz group in a way that is compatible with the dynamics of (quantum) general relativity. More specifically, the constraint can a priori have two families of solutions depending on how it is interpreted (i.e. by assigning a more fundamental role to either the rotations or the boosts), and we have seen that only one of them is consistent with the physical content of the theory. Indeed, it is only when replacing the rotation generators $J_i$ by $\pm\gamma^{-1}P_i$ that the original $\su(2)$ Ashtekar-Barbero connection is turned into an $\su(1,1)$ connection, leading in turn to continuous spectra for the (kinematical and physical) geometrical operators and to the disappearance of the Barbero-Immirzi parameter.

These new and surprising observations raise a lot of questions, the most important one certainly being that of their implication for the four-dimensional theory and both its canonical and spin foam quantizations. At first sight, it may seem that we are running into circles. Indeed, it is known that in four-dimensional canonical gravity one can make the choice $\gamma=\pm\mathrm{i}$, which evidently gets rid of the Barbero-Immirzi ambiguity in the theory and simplifies the Hamiltonian constraint, but at the expense of introducing the reality conditions which we do not know how to implement at the quantum level. However, what our three-dimensional model has shown is that the reality conditions can in some sense be traded for the linear simplicity-like constraint $J\mp\gamma^{-1}P=0$. These can indeed be thought of as reality conditions since they constrain the components of the complex connection $\mathbf{A}$ in such a way that the resulting connection is $\su(1,1)$-valued (which, as we have argued, is a physically-consistent requirement since we are describing Lorentzian three-dimensional gravity). Moreover, we have seen that in order to obtain the $\su(1,1)$ connection, the simplicity constraint has to be interpreted in a different way from what is done in spin foam models, i.e. by expressing the rotations in terms of the boosts and not the other way around. Finally, we have pointed out that this construction leads to a positive-definite length spectrum, which translates the reality of the metric.

Without the new ingredient of our construction, which consists in sending the three-dimensional Ashtekar-Barbero phase space back to that of $\SU(1,1)$ BF theory, we would have derived a ``wrong" kinematical structure for three-dimensional gravity. Indeed, if we had worked with the $\SU(2)$ theory we would have obtained a discrete length spectrum proportional to $\gamma$. Alternatively, if we had naively chosen $\gamma=\pm\mathrm{i}$ in order to get rid of the Barbero-Immirzi ambiguity and simplify the Hamiltonian constraint, we would have constructed the kinematical structure with $\sl(2,\mathbb{C})$ spin network states and obtained an incorrect minus sign in the length spectrum (unless the representations entering the Casimir operator are interpreted differently). The key point is therefore the derivation of the simplicity-like constraint, which when solved appropriately selects the $\su(1,1)$ subalgebra as the kinematical arena on top of which to construct the physical Hilbert space.

As far as the full four-dimensional theory is concerned, we now have to think about the implementation of this three-dimensional construction in both the canonical theory and the spin foam models. It is quite likely that in the canonical theory there will be analogous simplicity-like conditions which restrict the type of representations that have to be considered. The situation might be clearer in spin foam models, since their construction relies mainly on properties of the internal symmetry group and not that much on the symmetries of the spacetime, and our three-dimensional model has been constructed without affecting the internal symmetry group. Of course, the internal symmetry group has been affected by our gauge choice, but this is exactly what happens in spin foam models, where the simplicity constraints on the $B$ field induce relations between the $\SL(2,\mathbb{C})$ representations, which in turn define the $\SU(2)$ Ashtekar-Barbero connection starting from the initial Lorentz spin connection. It might very well be that when implementing the linear simplicity constraint with $\gamma=\pm\mathrm{i}$, one has to understand the resulting self-dual $\SL(2,\mathbb{C})$ representations rather as representations in the continuous series of $\SU(1,1)$.

Finally, note that the discrete area spectrum, derived from the quantization of the $SU(2)$ phase space, can be obtained by sending by hand the real Immirzi parameter $\gamma$ to the value $\gamma = \pm i$ and in the same time sending the spin $j$ to the complex value $j= (is - 1)/2$ where $s \in \mathbb{R}^{+}$. This is precisely the analytic continuation derived in the context of black hole entropy and presented in the precedent chapter.
At least in three dimensional gravity, we know that General Relativity can be formulated both as an $SU(2)$ and as an $SU(1,1)$ topological field theory. At the kinematical level, when the scalar constraint has not been taken into account, the two quantum theories are quite different. However, the two kinematical area spectrums can be mapped through our analytic continuation prescription discussed above.

However, one cannot conclude for the four dimensional case. More investigations are needed to understood the role of this analytic continuation and the resulting quantum states.
Up to now, we have tested this prescription in symmetry reduced models, i.e. spherically symmetric space-time (black hole) and three dimensional gravity. We can pursue this task to learn more about this prescription
and see whether it preserves the interesting results obtained in LQG. The natural next step is to apply this analytic continuation to the loop quantization of the homogenous and isotropic space-times.

Indeed, the resolution of the initial singularity in Loop Quantum Cosmology has been proven to be a robust result of the loop quantization of homogenous and isotropic space-time.
Moreover, this quantum cosmology admits the right semi classical limit, i.e. the usual Friedman cosmology. An idea would be to implement our prescription in this context and study how those very appealing results of LQC are modified. This is the subject of the next chapter.

\clearemptydoublepage

\chapter{Analytical continuation of Loop Quantum Cosmology}
\label{ch:LQC}
\minitoc

In this chapter, we describe a new model of Loop Quantum Cosmology, developped in \cite{ch6-BA1}. The basic idea is to apply the analytic continuation prescription introduced in chapter 4 to the simplest Loop Quantum Cosmology model, i.e. the flat universe $k=0$ with no cosmological constant $\Lambda=0$. We have seen that this prescription, first derived in the context of the analytic continuation of black hole entropy, turns out to have very interesting properties. This is the unique prescription (in a precise sense explained in chapter 4) leading to the Hawking area law supplemented with its logarithmic quantum corrections. Moreover, the last chapter convince us that, at least in three dimensional LQG, the presence of the Immirzi parameter in the area spectrum is a gauge artifact. Again, the very same prescription cures this gauge artifact and send the discrete $\gamma$-dependent area spectrum (inherited from the $SU(2)$ gauge choice) into the continuous $\gamma$-independent area spectrum (inherited from the $SU(1,1)$ gauge fixing). In the light of those results, it seems that although very simple, this prescription carries a deep meaning for the full self dual version of Loop Quantum Gravity. The 4D theory being highly non trivial, it is therefore natural to look for simpler models where we can explore the implications and properties of this prescription further. The symmetry reduced models offer in this perspective a good laboratory. The spherically symmetric black hole defined by the isolated horizon was the first one to be studied. However, the system was isolated and there was no dynamics to study. Therefore, we would like now to go a step further and work out our prescription in a reduced symmetry model where the dynamics remains non trivial. In this perspective, symmetry reduced models where the dynamics is available and remains in the same time non trivial correspond to the Loop Quantum Cosmology models. 
\\
Those models are based on the applications of the LQG quantization technics to cosmological settings. We refer the reader to \cite{ch6-Ash1, ch6-Agullo1} for pedagogical reviews. The main success of those models was to provide a resolution of the original singularity of the Universe based on a full quantum theory of cosmology \cite{ch6-Bojowald1, ch6-Ash2, ch6-Ash3}, but also a discussion of the pre-inflationary scenario \cite{ch6-Sloan1, ch6-Barrau1} and a study of the quantum perturbations over a quantum space-time \cite{ch6-Grain1, ch6-Agullo2}. The road to quantum cosmology was inspired from  DeWitt's ideas and initiated in 1969 by Charles Misner. It led to a flurry of activity but the mathematical framework remained formal. This old quantum cosmology is based on the quantization of the Heinsenberg algebra generated by the scale factor $a$ and its conjugated variable $p_{a}$, inherited from the geometrodynamics formulation of General Relativity. One can show that such quantum theory of cosmology do not lead to a singularity resolution \cite{ch6-Ash1}. This statement has to be understood in a precise sense. Asking for the singularity resolution implies to have at our disposal a full quantum theory, its Hilbert space structure containing the physical states and the way to compute transition amplitudes and expectation values of the physical observables. We say that the singularity is resolved when observables which diverge in the classical theory (such as the curvature or the energy density) turns out to have bounded expectation values in the quantum theory. In this precise sense, the Wheeler Dewitt quantum cosmology do not resolve the initial singularity. Instead, one can build the quantization procedure on the algebra commonly used in LQG, i.e. the holonomy-flux algebra. As explained in the second chapter, the representation of this algebra turns out to be unique when we require that it carry a unitary action of the diffeomorphism symmetry. This result is known as the LOST theorem and is the corner stone of Loop Quantum Gravity. The quantum cosmology built on this new algebra is inequivalent to the old one \cite{ch6-Bojowald2}, leading to a rich ensemble of result. As we will show below, the new quantum cosmology do resolve the initial singularity. One can compute the expectations values of the energy density and of the volume operators and they turns out to be bounded.  
As we have explained, we would like now to study, in the simplest setting,  if this non trivial result of real Loop Quantum Cosmology persists when we apply our analytic continuation prescription.
To be more precise, the goal is to answer the three following questions:
\begin{itemize}
\item Is this simple prescription sufficient to preserve the bouncing scenario of real LQCat the effective level ? Put it differently, do the energy density and the curvature of the universe in the effective approach remain bounded once the prescription is applied ?
\item In this reduced symmetry framework, can one developed the full quantum theory of Loop Quantum Cosmology in term of the homogenous and isotropic self dual variables, i.e. with $\gamma = i$ ?
\item If the answer is in the affirmative, what can we learn for the full quantum theory ?
\end{itemize} 

The first question can be answer in the affirmative. After applying our prescription, the effective Hamiltonian remains bounded.
The second question is more tricky. Indeed, we will present the full quantization of the model in the so called exactly solvable framework (states, expectations values ...) but this model remains technicaly more involved than its real version. For instance, a close formula for the maximal energy is not available in the self dual case. Moreover, the quantum representation where the dynamics is given by a difference equation has not been derived yet, also for technical reasons. However, we can still proof the bounded character of the volume operator in our self dual solvable model.
Finally, we will discuss the last question and proposed some ideas which could lead to new insights for the full theory.

This chapter is organized as follow. We first present the classical formulation of the reduced model we are quantizing in Loop Quantum Cosmology, written in term of the isotropic and homogenous Ashtekar' s variables and present its hamiltonian formulation.
This is the very first step before canonical quantization. Then we developed the effective approach, where we implement the holonomy corrections to the hamiltoninan. It turns out that this effective hamiltonian carries already rich informations. For example, one can already observe the singularity resolution at this level.
We embark then in the full quantum theory in the $j=1/2$ representation. Different choices of polarization when quantizing the classical theory lead to two versions of the same quantum theory. In the first one, the dynamics is given by a difference equation, underlying the discrete geometry behind this quantum cosmology. In the second one, i.e.  the exactly solvable model,  the dynamics is given by a differential equation. This version is more suited for practical computations of expectation values. We present in details this two sides of real LQC.
Having presented the complete quantum model, we proceed to its generalization and compute the hamiltonian for any spin $j$ representation.
This is the tool we need to apply our analytic continuation prescription. The last section is devoted to the new model and we present the result in the effective and the exactly solvable approach.

\section{Real Loop Quantum Cosmology and the singularity resolution}
\subsection{The classical setting}

We will only focus on the simplest case, the flat FLRW space-time $k=0$, with a vanishing cosmological constant $\Lambda=0$.  We work with the $\mathbb{R}^{3}$ topology. The metric for such a cosmological space-time reads:

\begin{eqnarray}
ds^2 = -N^2 dt^2 + a^2 \delta_{ij} dx^{i} dx^{j}\;
\end{eqnarray}

The starting point is the action of a scalar field minimally coupled to gravity. In quantum gravity, because of background independence, we are dealing with a ``frozen'' formalism, where no time is available. Indeed, the time evolution turns out to be a symmetry in gravity. To circumvent this problem, it was commonly assumed to used an internal variable, either from the gravitational or matter sector, to play the role of a global time. This is the basic idea of the deparametrization procedure. In the WDW theory, the scale factor $a$ was used for this task. However, to be a ``good'' clock, the choosen variable has to be monotonic. Already in the classical theory, in the context of closed models, the scale factor fails to respect this condition and is therefore not well suited to be used an an internal time. Another idea is to used its conjugated variable $p_{a}$. It was shown in \cite{ch6-Coutant1} that this choice lead to a good behaviour of the time variable. In Loop Quantum Cosmology, we introduced a scalar field for two reasons. First, the scalar field $\phi$ is a good candidate to play the role of an internal clock in the quantum theory and second, it is the simplest non trivial matter content we can think of when trying to build the quantum theory. It is needed to define the energy density of the universe as we will see. Consequently, the action we are interested in is:
\begin{eqnarray}
S = S_{grav} + S_{matter}  \;
     = \frac{1}{\kappa} \int \sqrt{g} ( R - \kappa g^{\mu\nu}\partial_{\mu}\phi \partial_{\nu}\phi ) \;
\end{eqnarray}

Using the real Ashtekar's variables, the gravitational action $S_{grav}$ takes the form: 
\begin{eqnarray}
S_{grav}  = \frac{1}{\kappa} \int \frac{1}{\gamma}E^a_i \dot{A^i_a} + NC + N^a H_a + \beta^iG_i\;
\end{eqnarray}
where $\kappa = 8\pi G$, $N$, $N^a$ and $\beta^i$ are respectively the lapse function, the shift vector and an $su(2)$-vector. As explained in the precedent chapters, they are Lagrange multipliers enforcing repesctively the hamiltoninan constraint $C$, the vectorial constraint $H_a$, and the Gauss constraint $G_{i}$. $\gamma$ is the Immirzi parameter. The electric field $E^a_i$ and the Ashtekar-Barbero connection reads: 
\begin{align*}
E^a_i =  \sqrt{det(e)} e^a_i  \;\;\;\;\;\;  \text{and} \;\;\;\;\;\;\;  A^i_a = \Gamma^i_a +  \gamma K^i_a  \;
\end{align*}
where, as usual, $e^a_i $ is the tetrad, $\Gamma^i_a$ is the spatial $su(2)$ -Levi Civita connection and $K^i_a$ the $su(2)$ - extrinsic curvature.
Finally, the explicit expression of the constraints was derived in the second chapter and read:
\begin{align*}
G & = \partial_{a} E^a + A_a \times E^a \\
 H_a  &= E^a . F_{ab} (A)\\
 C & = - \frac{1}{\gamma^2}  \frac{E^a \times E^b}{2 \; \sqrt{det(E)}}.  (F_{ab}(A) + (1+ \gamma^{-2})R_{ab}(\Gamma)) \\
 \end{align*}
The action for the scalar field can also be decomposed along the same line and we obtain: 
 \begin{align*}
S_{matter}  =  \int g^{00}\partial_{0}\phi \partial_{0}\phi  \; = \int - \frac{a^3}{N} \dot{\phi ^2}  \; = \int p_{\phi} \dot{\phi }  =  \int \frac{1}{2}p_{\phi} \dot{\phi } - \frac{N}{2a^3} p^2_{\phi} \;\;\;\;\; \text{where} \;\;\; p_{\phi}  = - \frac{a^3}{N} \dot{\phi}
\end{align*}

In view of the quantum theory, we are looking for an hamiltonian formulation of the system. This is the royal road to quantization. 
Let us therefore write the full Hamiltonian of the system. 
\begin{align*}
H_{tot}  & =  \int  \frac{1}{\kappa \gamma}E^a_i \dot{A^i_a} + \frac{1}{2}p_{\phi} \dot{\phi } - \mathcal{L}_{grav} - \mathcal{L}_{matter} \\
& = \frac{1}{\kappa} \int  - NC - N^a H_a - \beta^iG_i + \frac{N \kappa}{2a^3} p^2_{\phi} \\
& = \frac{1}{\kappa} \int   N( \frac{ \kappa}{2a^3} p^2_{\phi}  - C) - N^a H_a - \beta^iG_i  
\end{align*} 
Now, we can proceed to the symmetry reduction of the model.
The requirement of homogeneity and isotropy severely restrict the form of the electric field $E^a_i $ and the Ashtekar-Barbero connection $A^i_a $. Because of homogeneity, any spatial partial derivative vanish i.e. $\partial_{a} = 0 $, as well as the spatial part of the Levi Civita connection, i.e. $\Gamma^i_a = 0$. The conjugated variables take the form:
\begin{align*}
E^a_i  = p (t) \delta^a_i \;\;\;\;\;\;  \text{and} \;\;\;\;\;\;\;  A^i_a = \gamma k(t) \delta^i_a= c(t)  \delta^i_a\;
\end{align*}
It turns out that the Gauss constraint $G$ and the vectorial constraint $H_{a}$ are automatically sastified by the homogenous and isotropic version of the conjugated variables. 
Moreover, the new symmetry reduced conjugated variables are related to the scale factor by the following relations: 
\begin{align*}
 E^a_i & =  det(e) e^a_i  = p(t) \delta^a_i = a^2 \delta^a_i \;\;\;\;\;\;\;\;\; p = a^2 \\
 A^i_a & = \gamma K^i_a = \gamma \frac{\dot{a}}{N} \delta^i_a \;\;\;\;\;\;\;\;\;\;\;\;\;\;\;\;\;   k = \frac{c}{\gamma} =  \frac{\dot{a}}{N} \\
\end{align*} 
Finally, a last precaution has to be taken. Indeed, since we are working in a non compact space-time with the $\mathbb{R}^{3}$ topology, i.e. without boundaries, any space-time integral will diverge. To avoid those divergences, we restrict our calculus to a fiducial cell of physical volume : $V_0 = a^3 v_0$. Therefore, the comoving volume $v_0$ will enter in all the formula such as the Poisson bracket. However, we can safely set it to unity without changing the conclusion. This is true only when treating the open universe, $k=0$. 
The Poisson bracket for the homogenous and isotropic variables $(k, p)$ and $(\phi, p_{\phi})$ reads:

\begin{align*}
\{k,p\} =   \frac{\kappa}{3} \;\;\;\;\;\;  \text{and} \;\;\;\;\;\;  \{\phi,p_{\phi}\} = 2 
\end{align*}
 From the three constraints, only the hamiltonian constraint remains. However, plugging the homogenous and isotropic version of the electric field and of the Ashtekar Barbero connection drastically simples its expression: 
 \begin{align*}
  C = & - \frac{1}{\gamma^2}  \frac{E^a \times E^b}{ 2 \; \sqrt{det(E)}}. F_{ab}  = - \frac{1}{2 \; \gamma^2}  \epsilon^{ij}{}_k \delta^a_i \ \delta^b_j \frac{p^2 }{p^{3/2}}. F_{ab} = - \frac{1}{2 \; \gamma^2}  \epsilon^{ij}{}_k \delta^a_i \ \delta^b_j \frac{p^2 }{p^{3/2}}  \epsilon^{k}{}_{mn}  \delta_a^m  \delta_b^n  \gamma^2 k^2 \\
  = &\frac{1}{2}   \sqrt{p} k^2  \epsilon_k{}^{ij} \epsilon^{k}{}_{mn}  \delta^m_i \ \delta^n_j  =  \frac{1}{2}  \sqrt{p} k^2 ( \delta^i_m \ \delta^j_n - \delta^j_m \ \delta^i_n  )\delta^m_i \ \delta^n_j \\
  = & \frac{1}{2} (9-3)\sqrt{p} k^2 = 3  \sqrt{p} k^2
 \end{align*}
 
Finally, including the part due to the scalar field, the classical hamiltonian constraint reduces to:
 \begin{align*}
N [ \frac{ p^2_{\phi} }{2 p^{3/2}}   -  \frac{ 3}{ 8\pi G}  \sqrt{p} k^2 ]= 0 
\end{align*} 

The phase space to quantize is therefore given by the two Poisson brackets above, and the two couples of conjugated variables $(k, p)$ and $(\phi, p_{\phi})$ are constrained by the precedent hamiltonian constraint.

In view of the quantum theory, we can simplify the hamiltonian by different choices of the lapse function $N$.
Indeed, we observe that the denominator of the first term contains the power three of the scale factor $p^{3/2} = a^{3}$, i.e. which is the physical volume of the fiducial cell (we have fixed its comobile volume to unity).
This term would bring difficulties when going to the quantum theory, and we therefore cancel it by multiplying the full hamiltoninan by the same quantity, i.e. asking for: $N = a^{3}$.

 \begin{eqnarray}
N = a^3 = p^{3/2} \;\;\;\;\;\;\;\;\;\;\;\;\;\;\;\;\;\;\;\; \frac{ p^2_{\phi} }{2 }   -  \frac{ 3}{ 8\pi G}  p^{2} k^2 = 0 \;\;\;\;\;\;\;\;\;
\end{eqnarray} 

However, to deal with the effective approach, it will be more interesting to work with $N=a^{-3}$. Indeed, one can directly read the energy density in this form and study the modification of the holonomy corrections at the classical level :
 \begin{eqnarray}
 N = a^{-3} \;\;\;\;\;\;\;\;\;\;\;\;\;\;\;\;\;\;\; \frac{ p^2_{\phi} }{2 p^{3}}   -  \frac{ 3}{ 8\pi G}  \frac{k^2}{p} = 0 \;\;\;\;\;\;\;  \text{with}  \;\;\;\;\;  \rho = \frac{ p^2_{\phi} }{2p^{3} } =  \frac{ p^2_{\phi} }{2V^{2}} 
\end{eqnarray} 

As expected, this last equation turns out to be the first Friedman equation:
 \begin{eqnarray}
 \rho -  \frac{ 3}{ 8\pi G}  \frac{k^2}{p}= 0 \;\;\;\;\;\;\;  \rightarrow \;\;\;\;\;\;\;\;  H^2  = \frac{ 8\pi G}{3}  \rho
\end{eqnarray} 

Having detailed the classical theory and its hamiltonian formulation, we can now study the effective approach of Loop Quantum Cosmology.

\subsection{The effective approach}

We present now the effective approach. The idea is to implement the very first step of the LQG quantization technics, without going until quantization.
As explained in chapter 2, in LQG, we are no longer interested in working with the connection $ A^i_a$ as the fundamental variable, but with its holonomy, i.e. the exponential of the path ordered integral of the connection along an edge embedded in the $3$D manifold:
\begin{align*}
h_{i} (A) & = \text{exp} (\int_{\mu} A^{i}_{a} \tau_{i} dx^{a}) \\
& = \text{exp} (\int_{\mu} \gamma k(t) \delta^{i}_{a} \tau_{i} dx^{a}) \\
& = \text{exp} (\gamma k \mu \tau_{i}) \\
\end{align*}

Now that we have identified the variable we want to work with, we need to express the hamiltoninan, the only constraint remaining, in term of those holonomies. Obviously, different choice exist and we call a given choice a parametrization of the hamiltonian. In the LQC literature, it is commonly assumed to rewrite the hamiltonian using the Baker Campbell Haussdorf formula, which gives the curvature of the connection $A$ in term of the holonomy of the same connection around a closed fundamental square plaquette of an infinitesimal size $\mu$:
 \begin{align*}
h_{\Box_{ij}}= 1 + \mu^2 F^k_{ab} \tau_{k}\delta^a_i  \delta^b_j + \mathcal{O}(\mu^{3}) \;\;\;\;\;\;\;  \text{where} \;\;\;\;\;\;\;\;\;  h_{\Box_{ij}}= h_{i}h_{j}h^{-1}_{i}h^{-1}_{j}
\end{align*}
Isolating the curvature, this formula leads to the so called F-quantization procedure and the curvature of the connection is given by:
 \begin{align*}
 F^k_{ab} & = \frac{1}{Tr_{j}(\tau^k \tau^{k})}\;  \lim\limits_{\mu \to 0}  \;Tr_{j} \{\tau^k  \frac{h_{\Box_{ij}}- 1 }{\mu^2} \}  \delta^i_a   \delta^j_b \\
 & =  - \frac{3}{j(j+1)(2j+1)}\;  \lim\limits_{\mu \to 0}  \;Tr_{j} \{\tau^k  \frac{h_{\Box_{ij}}- 1 }{\mu^2} \}  \delta^i_a   \delta^j_b
\end{align*}

where we have used the following formula (derived from the $su(2)$ algebra) : $Tr(\tau^k \tau^{k}) = - \frac{1}{3}j(j+1)(2j+1)$.
A few comments are worthwhile concerning this pre-quantization procedure. The first point to note is that we are trading a local expression for the curvature for a non local one. This is at the heart of the LQG quantization procedure.
Indeed, the very first step is to forget about the local connection $A^i_a$ and to work instead with the non local variable which is the holonomy of the connection along a given path. 
The second point is to observe that the precedent formula for the non local curvature is a truncation at the second order of an infinite serie.
Therefore, plugging this new truncated formula for the curvature into the hamiltonian lead to a constraint which is no more exactly the one of homogenous and isotropic General Relativity.
However, we will see that the truncated constraint tends to the constraint of homogenous and isotropic General Relativity on large scale, which is precisely what is needed.
As we will see, the modifications brought by the holonomies corrections are important only at high energy, i.e. at the initial singularity. 
For the moment, let us compute explicitly the curvature in order to define the new hamiltonian. 
As we can see, the formula for the non local curvature implies a trace which can be done in any spin-$j$ representation of the $su(2)$ algebra.
However, in the literature, we generally restrict the computation to the fundamental representation which is the simplest one, i.e. the $j=1/2$-representation.
 In this representation, one can compute the holonomy around the fundamental square plaquette of size $\mu$ and one obtain:
\begin{align*}
h_{\Box_{ab}} & = h_{a}h_{b}h^{-1}_{a}h^{-1}_{b} \\
& = \text{exp}(2 \alpha \tau_{a}) \; \text{exp}(2\alpha \tau_{b})\; \text{exp}(- 2\alpha \tau_{a})\; \text{exp}(- 2\alpha \tau_{b})\\
& = ( cos(\alpha) + 2 sin(\alpha) \tau_{a})\; ( cos(\alpha) + 2 sin(\alpha) \tau_{b})\; ( cos(\alpha) - 2 sin(\alpha) \tau_{a})\; ( cos(\alpha) - 2 sin(\alpha) \tau_{b}) \\
& = cos(2\alpha)  + \frac{1}{2} sin^{2} (2 \alpha) + sin^{2} (2 \alpha) \epsilon_{abc} \tau^{c} + sin({2 \alpha} )( 1- cos(2\alpha)) ( \tau_{a}-\tau_{b})
\end{align*}

where we have posed for brevity $2\alpha = \mu \gamma k= \mu c$. Having computed this first term entering in the above trace, we can now obtain the trace involved in the formula of the non local curvature:
\begin{align*}
Tr_{j=1/2} \{h_{\Box_{ab}} \tau^{l}\} = - sin^{2} (\mu c) \epsilon_{ab}{}^{l} - \frac{1}{2}sin({\mu c})( 1- cos(\mu c)) ( \delta^{l}_{a}-\delta^{l}_{b})
\end{align*}
With this result at hand, the computation of the non local curvature is direct. Plugging the precedent result in the general formula, we obtain the non local cuvature in the spin $1/2$-representation: 
\begin{align*}
^{1/2}F_{ab}{}^{l} = \;  \lim\limits_{\mu \to 0}  \; \{ \; - \frac{sin^{2} (\mu c)}{ \mu^{2}} \epsilon_{ab}{}^{l} - \frac{sin(\mu c)( 1- cos(\mu c)) }{\mu^{2}}( \delta^{l}_{a}-\delta^{l}_{b})\; \}
\end{align*}
Now we come at the crucial point. To have an explicit formula for the non local curvature, we need to take the limit where the size of the square plaquette shrink to zero. However, inspired by the area spectrum derived in full kinematical LQG, we know that the null area is not accessible at the quantum level. Taking this point seriously and importing it into this model, we let the physical size of the square plaquette shrink to the minimal value of the area spectrum, i.e. to the area gap $\lambda$. Since we are dealing with the physical size of the plaquette and not with its comobile size, we have:
\begin{align*}
a^{2} \bar{\mu}^{2} = \lambda= 4 \pi \gamma l^{2}_{p} \sqrt{3} \;
\end{align*}
where $a$ is the scale factor.
This the true LQG input of the model.
The non local curvature becomes:
\begin{align*}
^{1/2}F_{ab}{}^{l} & = \;  \lim\limits_{\mu \to \bar{\mu}}  \; \{ \; - \frac{sin^{2} (\mu c)}{ \mu^{2}} \epsilon_{ab}{}^{l} - \frac{sin(\mu c)( 1- cos(\mu c)) }{\mu^{2}}( \delta^{l}_{a}-\delta^{l}_{b})\; \} \\
& = \; \{ \; - \frac{sin^{2} (\bar{\mu }c)}{ \bar{\mu}^{2}} \epsilon_{ab}{}^{l} - \frac{sin(\bar{\mu} c)( 1- cos(\bar{\mu} c)) }{\bar{\mu}^{2}}( \delta^{l}_{a}-\delta^{l}_{b})\; \}
\end{align*}

We can now plug the new expression of the non local curvature into the full hamiltonian constraint, obtaining therefore:
\begin{align*}
C & = N \; \{ \;  \frac{p_{\phi}^{2}}{2 p^{3/2}} - \frac{1}{16 \pi G \gamma^{2}} \epsilon^{ij}{}_{k} \delta^{a}_{i}\delta^{b}_{j} \epsilon_{ab}{}^{k}  p^{1/2} \frac{sin^{2} (\bar{\mu} c)}{\bar{\mu}^{2}} \; \} \\
& = N ( \frac{p_{\phi}^{2}}{2 V} - \frac{3}{8 \pi G \gamma^{2}} V \frac{sin^{2} (\lambda b)}{\lambda^{2}} ) \;\;\;\;\;\;  \text{with} \;\;\;\;\;\;  V = p^{3/2} 
\end{align*}

In the last line, we have used a new set on conjugated variables which make the interpretation of the quantities more transparent.
The couple of variables $(b,V)$ is defined as follow:
\begin{align*}
b = \frac{c}{\sqrt{p}} = \gamma H \;\;\;\;\;\;\;  \text{and} \;\;\;\;\;\;  V = p^{3/2}= a^{3} \;\;\;\;\;\;\;  \{b , V\} = \frac{\kappa}{3}\partial_{c} ( \frac{c}{\sqrt{p}}) \partial_{p} ( p^{3/2}) = \frac{\gamma \kappa}{2}
\end{align*}
The first one correspond to the Hubble parameter while the second one encodes the volume of the universe, the fiducial cell. Finally, the quantity $\lambda$ is defined as : $\lambda = a \bar{\mu}$ and satisfy $\bar{\mu} c = \lambda b$.
It is also useful to introduce already at this level the reduced volume variable $\nu = V / 2 \pi G $, which will be used in the quantum theory. The quantum states will be function of $(\phi, \nu)$. Their interpretation is more easy than function of the variable $p$ precendently used.
With those notations and the choice of the lapse function precedently discussed in the classical settings, i.e. $N = 1/V$, the hamiltonian constraint take the usual form used in the LQC litherature:
\begin{align*}
 \frac{p_{\phi}^{2}}{2 V^{2}} - \frac{3}{8 \pi G \gamma^{2}} \frac{sin^{2} (\lambda b)}{\lambda^{2}} = 0
\end{align*}
One can now defined simply the matter energy density, i.e. the energy density of the scalar field as $\rho =  \frac{\Pi^{2}}{2 V^{2}}$, which leads to:
\begin{align*}
\rho = \frac{3}{8 \pi G \gamma^{2}} \frac{sin^{2} (\lambda b)}{\lambda^{2}} \;\;\;\;\;\;\; \text{hence} \;\;\;\;\;\;  \rho < \rho_{max} = \frac{3}{8 \pi G \gamma^{2}} 
\end{align*}
The energy density of the universe is now bounded and its evolution in the past do not engender a singularity.
The modified Friedman law can be derived from the Poisson bracket of the volume and the precedent hamiltonian which reads:
\begin{align*}
\dot{V} = \{V, C \} = - 8 \pi G \gamma \frac{\partial H_{tot}}{\partial b} =  \frac{3 V}{ \gamma \lambda}sin (\lambda b) cos (\lambda b) 
\end{align*}
In this particular case, the effective Friedman equation, supplemented with the holonomy corrections take a very nice form:
\begin{align*}
H^{2} = \big{(} \frac{\dot{V} }{3V}  \big{)}^{2} =  \frac{1}{ \gamma^{2} \lambda^{2}} sin^{2} (\lambda b) cos^{2} (\lambda b) \;\;\;\; \rightarrow \;\;\;\;\;\;  H^{2} = \frac{8 \pi G}{3} \rho ( 1 - \frac{\rho}{\rho_{max}}) 
\end{align*}

We see that the holonomy modifications dilute the contraction of the universe when $\rho \simeq \rho_{max}$ until stopping it precisely at $\rho = \rho_{max}$.
Therefore, already at the level of the effective equation, we observe that the universe do not experience a singularity, since the density is bounded by a given value.
It follow that the Hubble parameter do not tend to infinity when going backwards, both the curvature and the energy density remains bounded. The initial singularity, corresponding to $(\rho = \infty, H = \infty)$, is resolved and is now replaced by: $(\rho = \rho_{max}, H = 0)$.
However, to conclude properly, we need to develop the full quantum theory and compute the expectation value of those two quantities to see if they remains bounded at the quantum level.

\subsection{The full quantum theory}

We present now the full quantum theory of Loop Quantum Cosmology. The very first step to quantize the theory is to choose the classical algebra we want to quantize. 
The old WDW quantum cosmology was based on the Heinsenberg algebra generated by the classical variables $(c,p)$, i.e. the usual algebra of quantum mechanics. Since this quantum cosmology fails to resolve the initial singularity, we are led to search for another classical algebra. In LQC, we mimic the strategy used in full LQG and we choose to work with the so called homogenous and isotropic holonomy-flux algebra.
This choice is motivated because we are interested in working in a diffeomorphism invariant formalism. It turns out that the holonomy flux algebra has a unique representation which carries a unitary action of the diffeomorphism symmetry. Let us comment on the variables generating the homogenous and isotropic analogue of the  holonomy flux algebra. The electric field being only time dependent and reducing to the function $p(t)$, its flux across a surface $S$ will only be given by $F(p) = p S$. Therefore, we can forget about $S$ and only consider the variable $p$ since $S$ in only a constant.
Finally, we know that in the spin $1/2$-representation, the holonomy of the connection $h_{\mu}(c)$ along an edge $\mu$ can be decomposed as:
 \begin{align*}
h_{\mu}(c) = e^{\mu c / 2 \tau_{i}} =  \text{cos} ( \frac{ \mu c}{2}) \mathbb{I} + 2 \;  \text{sin} ( \frac{ \mu c}{2}) \tau_{i} = \sum_{k} \alpha_{k} e^{i \mu_{k} c}
\end{align*}  
Therefore the building blocks of the homogenous and isotropic holonomy are the elementary functions $N_{\sigma}(c) = e^{i \sigma c}$ where $\sigma$ is real.  We will work with this fundamental functions for the rest of the discussion.
Consequently, the homogenous and isotropic version of the holonomy-flux algebra of LQG reads respectively at the classical and at the quantum level:
 \begin{align*}
\{N_{\sigma}(c) ,  p\} = \frac{\gamma \kappa}{3} i \sigma N_{\sigma} (c)  \;\;\;\;\;  \rightarrow \;\;\;\;  [ \; \hat{N}_{\sigma} (c), \hat{p} \; ] = - \frac{\gamma \kappa}{3} \hbar \sigma N_{\sigma} (c)
\end{align*}

Now that we have identified the quantum algebra $A$ of operators, we can construct the quantum theory \cite{ch6-GNS}. The quantum theory is found by finding a representation $(\mathcal{H}, \pi)$ of the quantum algebra $A$.
The Hilbert space $\mathcal{H}$ is defined by the couple $(V, <;>)$ of a vector space $V$ endowed with a scalar product. The application:
\begin{align*}
\pi : \; & A \rightarrow \text{End(V)} \\
& \; \; a \rightarrow \pi ( a) = \hat{A}
\end{align*} 
 maps an element $a \in A$ into an endomorphism, i.e.  an operator $\hat{A} \in \text{End(V)} $ which acts on $V$.
If some symmetry is present in the classical theory, it appears as an automorphism in the quantum algebra $A$, i.e. a map $\theta : A \rightarrow A$.
In the quantum theory, we say that the representation carries a unitary action of the symmetry $\theta$ if there is a unitary operator $U$ acting on $V$ such that : $\pi(\theta(A)) = U \hat{A} U^{-1}$.	

Let us consider the Heinsenberg algebra, or equivalently the Weyl algebra generated by $U_{\alpha}(c) = e^{i \alpha c}$ and $V_{\beta}(p) = e^{i \beta p}$. From the Von Newman theorem, the representation of the Weyl algebra is unique provided the matrix elements $< \Psi | \pi (U_{\alpha}) | \Psi > $ and $< \Psi | \pi (V_{\beta}) | \Psi > $ for a given quantum state $\Psi \in \mathcal{H}$ are continuous respectively w.r.t $\alpha$ and $\beta$. Therefore, the scalar product $<;>$ has to be continuous. However, this unique representation, i.e. the Heinseberg representation, fails to carry a unitary action of the diffeomorphism symmetry. It is therefore not well suited when dealing with quantum gravity (or quantum cosmology in the present case).

Following the LOST theorem and the results of Fleischak, in turns out that looking for a representation of the Weyl algebra which carries a unitary action of the diffeomorphism symmetry implies to drop out the continuity property of $< \Psi | \pi (U_{\alpha}) | \Psi > $ and $< \Psi | \pi (V_{\beta}) | \Psi > $ w.r.t $\alpha$ and $\beta$. Giving up on the continuity property of the matrix elements of the representation leads to a unique representation, called the Ashtekar Lewandowski representation. This representation is characterized by a discontinuous scalar product $< ; >$.
Therefore, the price to pay to work in a diffeomorphism invariant fashion at the quantum level is to give up on the continuity property of the representation, i.e. of the scalar product. Finally, the requirement of diffeomorphism invariance seems to be much more powerful than any other symmetry requirement, because it select a unique representation. The LOST theorem is the corner stone of LQG and justify a posteriori the choice of the quantum algebra.

Now, as we have explained above, the quantum states $\Psi(c) \in \mathcal{H}$ are build as functional of the homogenous holonomy. Therefore, they are simply given by elementary functions $N_{\sigma}(c) = e^{i \sigma c}$. Since the scalar product is no more continuous w.r.t $\sigma$, we have the following relation between quantum states:
 \begin{align}
< N_{\sigma_{1}}| N_{\sigma_{2}}> ~ = ~  \delta_{\sigma_{1}, \sigma_{2}} =  \left \{ \begin{array}{l}
1\;  \text{if} \;  \sigma_{1} = \sigma_{2} \\
0 \; \text{if} \; \sigma_{1} \neq \sigma_{2}
\end{array} \right .
\end{align}
Therefore, $N_{\sigma}(c) = e^{i\sigma c}$ are quasi periodic function, where $\sigma$ has support only on a countable number of points. Now that we have our quantum theory at hand, we can explicit the actions of the quantum operators on the quantum states. We work in the polarization where the holonomy operator $\hat{N}_{\sigma}(c)$ act as a multiplicative operator while the operator $\hat{p}$ acts by multiplication, i.e.:
\begin{align*}
\hat{N}_{\sigma}(c) \triangleright N_{\mu} (c) & = e^{i \sigma c}  N_{\mu} (c) \;\;\;\;\;\;\;\;\;\;  \hat{p} \triangleright N_{\mu} (c)  = i \hbar \frac{\gamma \kappa}{3} \;  \frac{\partial}{\partial c} N_{\mu} (c) 
\end{align*}
We can also work in the opposite polarization, where the quantum states are defined as the eigenstates of the operator $\hat{p}$, i.e. : $\Psi(p)$. In this polarization, the operator $\hat{p}$ acts by multiplication while the operator $\hat{N}_{\sigma}(c)$ acts via translation on the quantum states :
 \begin{align*}
\hat{N}_{\sigma}(c) \triangleright \Psi (p) & = \sum_{n}\frac{1}{n !} ( \beta \sigma \frac{\partial}{\partial p})^{n} \Psi (p) = \Psi (p + \beta \sigma )   \;\;\;\;\;\;\;\;\;\;  \hat{p} \triangleright \Psi (p)  = p \; \Psi (p)
\end{align*}
with $\beta = 8 \pi \gamma l^{2}_{p} / 3$.

Finally, it is possible to work with the other set of canonical conjugated variables that we called $(b, \nu)$ in the precedent section. We recall that $v = \frac{V}{2 \pi G \gamma}$. In term of those variables, the classical theory is given by: 
 \begin{align*}
\{N_{\sigma}(b) ,  v \} = 2 i \sigma N_{\sigma} (b) \;\;\;\;\;\;  \text{and} \;\;\;\;\;\;  \{\phi,p_{\phi}\} = 2 \;\;\;\;\;\;\;\;\;\;\;\;\;   p^2_{\phi}  -  3 \pi G  v^{2} \frac{sin^{2} (\lambda b)}{\lambda^{2}} \; = 0 
\end{align*}
Applying the same strategy, our quantum algebra is given by:
 \begin{align*}
\{N_{\sigma}(b) ,  v \} = 2 i \sigma N_{\sigma} (b)  \;\;\;\;\;  \rightarrow \;\;\;\;  [ \; \hat{N}_{\sigma} (b), \hat{v} \; ] = - 2 \hbar \sigma N_{\sigma} (b)
\end{align*}
At this point we can follow what is done in standard LQC and introduce the new operator $\hat{\nu}= \hat{v} / \hbar$.  We choose to work in the $\nu$-representation, where the quantum states are of the form: $\Psi( \nu)$.
In this case, the quantum operators acts on the quantum states as follow:
 \begin{align*}
\hat{N}_{\sigma}(b) \triangleright \Psi (\nu) & = \sum_{n}\frac{(2 \sigma)^{n}}{n !} \partial^{n}_{\nu} \Psi (\nu) = \Psi (\nu + 2 \sigma)   \;\;\;\;\;\;\;\;\;\;  \hat{\nu} \triangleright \Psi (\nu)  = \nu \; \Psi (\nu)
\end{align*}

Having the action of the quantum operator at hand, we can explicit the expression of the quantum hamiltonian constraint:
 \begin{align*}
 \partial^{2}_{\phi} \Psi(\phi, \nu) & = \frac{3 \pi G }{4 \lambda^{2}} \hat{\nu} \; ( \; e^{i \lambda b} - e^{-i \lambda b} ) (\; e^{i \lambda b} - e^{-i \lambda b} ) \hat{\nu} \;  \Psi(\phi, \nu) \\
 & =  \frac{3 \pi G }{4 \lambda^{2}} \hat{\nu} \; ( \; e^{i 2\lambda b} - 2 + e^{-i 2 \lambda b} ) \hat{\nu} \;  \Psi(\phi, \nu) \\
& = \frac{3 \pi G }{4 \lambda^{2}} \nu \hat{\nu} \; ( \; \Psi(\phi, \nu + 2\lambda)  - 2\Psi(\phi, \nu) +  \Psi(\phi, \nu - 2\lambda))  \;  \Psi(\phi, \nu) \\
 & = \frac{3 \pi G }{4 \lambda^{2}} \nu  \; ( \; (\nu + 2\lambda)\Psi(\phi, \nu + 2\lambda)  - 2 \nu \Psi(\phi, \nu) + (\nu - 2\lambda) \Psi(\phi, \nu - 2\lambda))  \\
\end{align*}

The dynamical equation for the universe, in this particular representation, i.e. the $\nu$-representation, turns out to be a difference equation. It means that the variable $\nu$ has support on a lattice for which the size is specified by multiple of the quantity $\lambda$ which is the area gap. From this result, we observe that the universe experiences discontinuous evolution from one volume state to another one.
However, it is difficult to get access to the expectation values of Dirac observables in this representation. This is why we will present now the exactly solvable model, which is more suited for those goals.

\subsection{The exactly solvable model}

Now, we turn to the so called exactly solvable model. This model is based on the observation that working instead in the polarization where the quantum states are functions of $b$ and not of $\nu$ permit to solve analytically the model. In this approach, different questions regarding the bounce and the quantum fluctuations can be addressed. Moreover, working in the $b$-representation trade the difference equation of the precedent section into a true differential equation. 
Therefore, the quantum states are now wave functions of the variables $(\phi, b)$: i.e. $\chi(\phi, b)$.
The quantum algebra is generated by the reduced volume operator $\hat{v} = V /2 \pi G \gamma $ and the holonomy operator $\hat{N}_{\lambda} (b) = e^{i \lambda b}$. Their commutations relations are given by: 
 \begin{align*}
\{ \; \hat{N}_{\lambda} (b) , \hat{v} \; \}  & = 2 \lambda \hat{N}_{\lambda} (b)  \;\;\;\;\;\;\;  \rightarrow \;\;\;\;\;\;\;\;\;\; [ \; \hat{N}_{\lambda} (b) , \hat{v} \; ]  = i 2 \hbar \lambda \hat{N}_{\lambda} (b) \\
\end{align*}
Finally, because of the choice of polarization, the holonomy operator acts by multiplication and the reduced volume operator acts as a derivative operator.
Their actions reads:
 \begin{align*}
\hat{N}_{\lambda}(b) \triangleright \chi (\phi, b) & = e^{i \lambda b} \chi (\phi, b) \;\;\;\;\;\;\;\;\;\;  \hat{v} \triangleright \chi (\phi, b) = - 2 i\hbar \;  \frac{\partial}{\partial b} \chi (\phi, b)
\end{align*}
We can rewrite the hamiltonian constraint in term of those operators and we obtain:
 \begin{align*}
p^{2}_{\phi} -  3\pi G v^{2} \frac{sin^{2} (\lambda b)}{\lambda^{2}} \; = 0 \;\;\;\;\;\;\;\;  \rightarrow \;\;\;\;\;\;\;  [ \; \partial^{2}_{\phi} - 12 \pi G \big{(} \frac{sin (\lambda b)}{\lambda} \partial_{b}\big{)}^{2} \; ] \;\chi (\phi, b) = 0
\end{align*}
This new quantum hamiltonian constraint is a differential equation. Since we are working with a periodic function, the range of $b$ can be restricted to: $b \in ( 0, \pi / \lambda)$. We will see that the range of $b$ is modified in the new model we will present in the following sections.
 Now, the quantum hamiltonian constraint can be recast into a Klein Gordon equation using the change of variable:
 \begin{align*}
x = \frac{1}{ \sqrt{12 \pi G}} \text{log} \big{(} \text{tan} ( \frac{\lambda b}{2} ) \big{)}  \;\;\;\;\;\;\;\;\;\;\;\;\;  dx = \frac{1}{\sqrt{12 \pi G}} \frac{\lambda}{ \text{sin} (\lambda b)} db
\end{align*}
where $x \in [- \infty, + \infty]$.
Plugging this change of variable in the precedent equation, we obtain the usual Klein Gordon equation:
 \begin{align*}
( \; \partial^{2}_{\phi} - \partial^{2}_{x} \; ) \;\chi (\phi, b) = 0
\end{align*}
This simple dynamical equation selects the physical quantum states. The physical Hilbert space is given by the positive frequency solutions of the precedent equation, i.e. solutions of:
 \begin{align*}
 -i\; \partial_{\phi} \;\chi (\phi, x)  = \partial_{x} \; \chi (\phi, x) 
\end{align*}
where we can split the solutions into its right moving and left moving parts: $\chi (\phi, x ) = \chi_{L} (x_{+}) + \chi_{R} (x_{-})$ and $x_{\pm} = \phi \pm x$.
Because there is an invariance when we change the triad orientation, the quantum state $\chi (\phi, x)$ satisfies $\chi (\phi, - x) = - \chi (\phi, x)$. This restriction implies that we can rewrite the quantum states as:
$\chi (\phi, x) = ( F(x_{+}) - F(x_{-})/\sqrt{2} $, where $F(x_{\pm})$ are arbitrary positive frequency solutions.
Now, we are interested in computing expectation values of the Dirac observables.
To do so, we choose to work with the natural scalar product, i.e. the Klein Gordon scalar product:
 \begin{align*}
( \chi_{1}, \chi_{2} )_{\text{phys}} = & - \int_{\phi = \phi_{0}} [ \; \bar{\chi_{1}}(\phi, x) \partial_{\phi} \chi_{2}(\phi, x) - \partial_{\phi} \bar{\chi}_{1}(\phi, x) \chi_{2}(\phi, x)] \; dx\\
= & \; - i \int_{\phi = \phi_{0}} [ \; \bar{\chi_{1}}(\phi, x) \partial_{x} \chi_{2}(\phi, x) - \partial_{x} \bar{\chi}_{1}(\phi, x) \chi_{2}(\phi, x)] \; dx\\
= & \;  i \int_{\phi = \phi_{0}} [ \; \partial_{x} \bar{F_{1}}(x_{+}) F_{2}(x_{+}) - \partial_{x} \bar{F}_{1}(x_{-}) F_{2}(x_{-})] \; dx \\
\end{align*}

The volume operator is simply given by $\hat{V} =  2 \pi  \gamma G \hat{v}=  - i 4 \pi \gamma G \hbar \partial_{b}$ where $\hat{v} = - i 2 \hbar \partial_{b}$. Using instead the operator $\hat{\nu} = \hat{v} / \hbar$, we have that:
 \begin{align*}
\hat{\nu} = - i 2 \; \frac{\partial x}{\partial b} \partial_{x} = - i \frac{2 \lambda}{\sqrt{12 \pi G} \sin{\lambda b}} \partial_{x} = - i \frac{2 \lambda}{\sqrt{12 \pi G} } \cosh{(\sqrt{12\pi G} \; x)} \partial_{x}
\end{align*}


Let us compute its expectation value of the self dual version of this operator. For brevity, we note $\alpha = \sqrt{12\pi G}$:
 \begin{align*}
( \chi, \hat{\nu} \chi)_{\text{phys}} = & \;  i \int_{\phi = \phi_{0}} [ \; \partial_{x} \bar{F}(x_{+}) ( \hat{\nu} F(x_{+})) - \partial_{x} \bar{F}(x_{-}) ( \hat{\nu} F(x_{-}))] \; dx \\
= & \;  \frac{2\lambda}{\alpha} \int_{\phi = \phi_{0}} [ \; \partial_{x} \bar{F}(x_{+}) \{ \cosh{(\alpha x)} \; \partial_{x} F(x_{+}) + \partial_{x} \;  \cosh{(\alpha x)}  F(x_{+}) \}  \\
= & \;  \frac{2 \lambda}{\alpha} \int_{\phi = \phi_{0}} [ \; \partial_{\tilde{x}} \bar{F}(\tilde{x}) \partial_{x} F(\tilde{x}) \{ \cosh{(\alpha ( \tilde{x} - \phi))}  + \cosh{(\alpha ( \phi - \tilde{x}))} \}  \\
= & \;  \frac{4\lambda}{\alpha } \int_{\phi = \phi_{0}}  \; |\partial_{\tilde{x}} F(\tilde{x}) |^{2} \; \cosh{(\alpha ( \tilde{x} - \phi))} \\
\end{align*}

Here, we have assumed that the function $F(x)$ falls off rapidly at infinity and that there are square integrable. From the second to the third line, we have applied two change of variables: $x_{+} = \tilde{x} \;\;\;\; dx = d\tilde{x}$ or $x_{-} = \tilde{x}\;\;\;\;\; dx = - d\tilde{x}$. Finally, the last line is obtain making use of the parity of the function $\cosh(x)$. 
Now, using that : $\cosh(x) \geqslant 1$, we conclude that the expectation value of the volume operator is bounded from below. The universe never experiences a zero volume state and do not shrink to a point, resolving therefore the initial singularity of the Big Bang.

 \begin{align*}
< \; V \; > \; = \; 2\pi \gamma l^{2}_{p} ( \chi, \hat{\nu}^{sa} \chi)_{\text{phys}} \geqslant & \;  \frac{8 \pi\gamma l^{2}_{p}\lambda}{\alpha} \int_{\phi = \phi_{0}}  \; |\partial_{\tilde{x}} F(\tilde{x}) |^{2} \; dx \\
\end{align*}

This conclude the presentation of the exactly solvable model.

We have shown in two precedent sections the quantization of the LQC algebra following the two main paths presented in the literature.
We would like now to mention briefly another road, which could be relevant with respect to our program.

\subsection{Beyond the $j=1/2$ representation}

We would like now to go further. As explained above, to obtain the hamiltonian in the effective approach, we need to compute the non local form of the curvature.
This object lives in the $su(2)$ algebra and we can therefore compute it any $j$-representation. In the precedent presentation, the computation was done in the fundamental representation, i.e. in the $j=1/2$ representation.
However, for our purpose, we need the expression of the non local curvature and hence the effective hamiltonian in the $j$ representation. Indeed our prescription requires to analytically continue the spin $j$ to the complex value $j = ( -1 + i s)/2$. from this perspective, we need the explicit dependency of the hamiltonian on $j$.

Therefore, we perform in the following this computation and develop the effective theory in any arbitrary $j$-representation.
As explained in the second section, we choose a regularization of the non local curvature that is expressed in term of the holonomy of the homogenous connection around a fundamental square plaquette of area $\mu^{2}$.
Neglecting the third order term in $\mu$, its expression reads: 
 \begin{align*}
 F^k_{ab} & =  - \frac{3}{j(j+1)(2j+1)}\;  \lim\limits_{\mu \to \bar{\mu}}  \;Tr_{j} \{\tau^k  \frac{h_{\Box_{ij}}- 1 }{\mu^2} \}  \delta^i_a   \delta^j_b
\end{align*}

Therefore the quantity we need to compute is the $Tr_{j} \{ \tau^k h_{\Box_{ij}}\}$ for an arbitrary spin $j$. Let us fix the indices for simplicity, the other combinations being easy to deduce.
The trace can be recast such that:
 \begin{align*}
Tr_{j} \{\tau^3  h_{\Box_{12}}\} = \frac{\partial}{\partial \epsilon } tr_{j} \big{(} h_{\Box_{12}} \; e^{\epsilon \tau^{3}}\big{)} \big{|}_{\epsilon = 0}
\end{align*}

Now let us call $h_{\Box_{12}} \; e^{\epsilon \tau^{3}} = h_{123}$. This object lives in the $SU(2)$ group and its trace is given by the general formula for the $SU(2)$ character:
 \begin{align*}
 Tr_{j} \big{(} h_{\Box_{12}} \; e^{\epsilon \tau^{3}}\big{)} = Tr_{j} \big{(} h_{123} \big{)} = \chi_{j} (\theta (\epsilon)) = \frac{\sin{d_{j} \theta(\epsilon)}}{\sin{\theta(\epsilon)}}
\end{align*}
where $d_{j} = 2j +1$ is the dimension of the representation.
The angle $\theta(\epsilon)$ is the Euler angle where $\theta \in [0, 2 \pi]$. Since the holonomy $h_{\Box_{12}}$ is expressed in term of the variable $b$, we need to evaluate the relation between $\theta$ and $b$.
To do so, we come back to the fundamental representation where we can write:
 \begin{align*}
 \chi_{1/2} (\theta (\epsilon)) & = Tr_{j=1/2} \big{\{} e^{\lambda b \tau^{1}}e^{\lambda b \tau^{2}}e^{-\lambda b \tau^{1}}e^{-\lambda b \tau^{2}}\; e^{\epsilon \tau^{3}} \big{\}} \\
 & = Tr_{1/2} \big{\{} [ \cos{(\lambda b)} + \frac{1}{2} \sin^{2}{(\lambda b)} + \sin^{2}(\lambda b) \tau^{3} \\
 & \;\;\;\;  + \sin{(\lambda b)} ( 1 - \cos{(\lambda b)} ( \tau^{1} -\tau^{2} ) ] [ \cos{\epsilon/2} + 2 \sin{\epsilon/2} \tau^{3} ] \big{\}} 
\end{align*}

Restricting this relation to $\epsilon = 0$, we obtain:
 \begin{align*}
2 \cos{\theta} = 2 \cos{(\lambda b)} + \sin^{2}{(\lambda b)} 
\end{align*}
Here, we define the angle $\theta$ as $\theta = \theta (0)$.
Noting $X =\cos{(\lambda b)} $ and solving the equation of the second degrees for $X$, we obtain: $X = 1 \pm \sqrt{2(1-\cos{\theta})}$.
After some little algebra, one get a condensed form for the relation between $\theta$ and $b$:
 \begin{align*}
\sin{\frac{\theta}{2}} = \sin^{2}{(\frac{\lambda b}{2})} 
\end{align*}

The next step is to evaluate the derivative of $\theta$ w.r.t $\epsilon$ at $\epsilon =0$, i.e. $\frac{\partial \theta}{\partial \epsilon} \big{|}_{\epsilon = 0}$.
For this we use again the fundamental representation where we can write:
 \begin{align*}
\frac{\partial \theta}{\partial \epsilon} \frac{\partial}{\partial \theta} \chi_{1/2}{(\theta)}\big{|}_{\epsilon = 0} = \frac{\partial \theta}{\partial \epsilon} \sin{\theta} \; \big{|}_{\epsilon = 0}  = - \frac{1}{4} \sin^{2} (\lambda b) \;\;\; \rightarrow \;\;\;\;  \frac{\partial \theta}{\partial \epsilon} \; \big{|}_{\epsilon = 0}  = \frac{ \sin^{2} (\lambda b)}{4 \sin{\theta}}
\end{align*}

Putting all those results together, the trace involved in the non local curvature, in the $j$-representation, is given by:
 \begin{align*}
\frac{\partial}{\partial \epsilon } Tr_{j} \big{(} h_{\Box_{12}} \; e^{\epsilon \tau^{3}}\big{)} \big{|}_{\epsilon = 0} = \frac{\partial \theta}{\partial \epsilon} \frac{\partial}{\partial \theta} \chi_{j}{(\theta)}\big{|}_{\epsilon = 0} =  \frac{ \sin^{2} (\lambda b)}{4 \sin{\theta}}\;  \frac{\partial}{\partial \theta} \; \frac{\sin{d_{j} \theta}}{\sin{\theta}}
\end{align*}

Finally the non local curvature in the $j$-representation reads:
 \begin{align*}
 F_{ab}{}^{l}  = \epsilon_{ab}{}^{l} F_{12}{}^{3} & = - \epsilon_{ab}{}^{l} \; \frac{12}{d ( d^{2} -1) \bar{\mu}^{2}} \; \frac{\partial}{\partial \epsilon } Tr_{j} \big{(} h_{\Box_{12}} \; e^{\epsilon \tau^{3}}\big{)} \big{|}_{\epsilon = 0} \\
 & = - \epsilon_{ab}{}^{l} \; \frac{3 |p|}{d ( d^{2} -1) {\lambda}^{2}} \; \frac{ \sin^{2} (\lambda b)}{ \sin{\theta}}\;  \frac{\partial}{\partial \theta} \; \frac{\sin{d_{j} \theta}}{\sin{\theta}}
\end{align*}

where $\lambda$ is the area gap. We can now give the full hamiltonian in the $j$ representation:
 \begin{align*}
 C_{tot} & = \frac{p^{2}_{\phi}}{2V} + \frac{3 V}{8 \pi G \gamma^{2}} \epsilon^{ij}{}_{k} \delta^{a}_{i}\delta^{b}_{j}  F_{ab}{}^{k}  \\
  & = \frac{p^{2}_{\phi}}{2V} + \frac{9 V}{8 \pi G \gamma^{2} \lambda^{2} d (d^{2}-1)}  \frac{ \sin^{2} (\lambda b)}{ \sin{\theta}}\;  \frac{\partial}{\partial \theta} \; \frac{\sin{d_{j} \theta}}{\sin{\theta}}
\end{align*}
where again, $d = 2j +1$. This hamiltonian is the starting point to apply our analytic continuation prescription.
But before presenting the self dual model, a few comments are worthwhile. It is immediate to see that the hamiltonian remains bounded for any $j$. Therefore, the bouncing scenario of real LQC is not affected by the choice of the $SU(2)$ representation we are working with. However, when computing the energy density for the universe, we obtain:
 \begin{align*}
\rho = \frac{p^{2}_{\phi}}{2V^{2}}  = - \frac{9}{8 \pi G \gamma^{2} \lambda^{2} d (d^{2}-1)}  \frac{ \sin^{2} (\lambda b)}{ \sin{\theta}}\;  \frac{\partial}{\partial \theta} \; \frac{\sin{d_{j} \theta}}{\sin{\theta}}
\end{align*}

 \begin{figure}
\begin{center}
	\includegraphics[width = 0.6\textwidth]{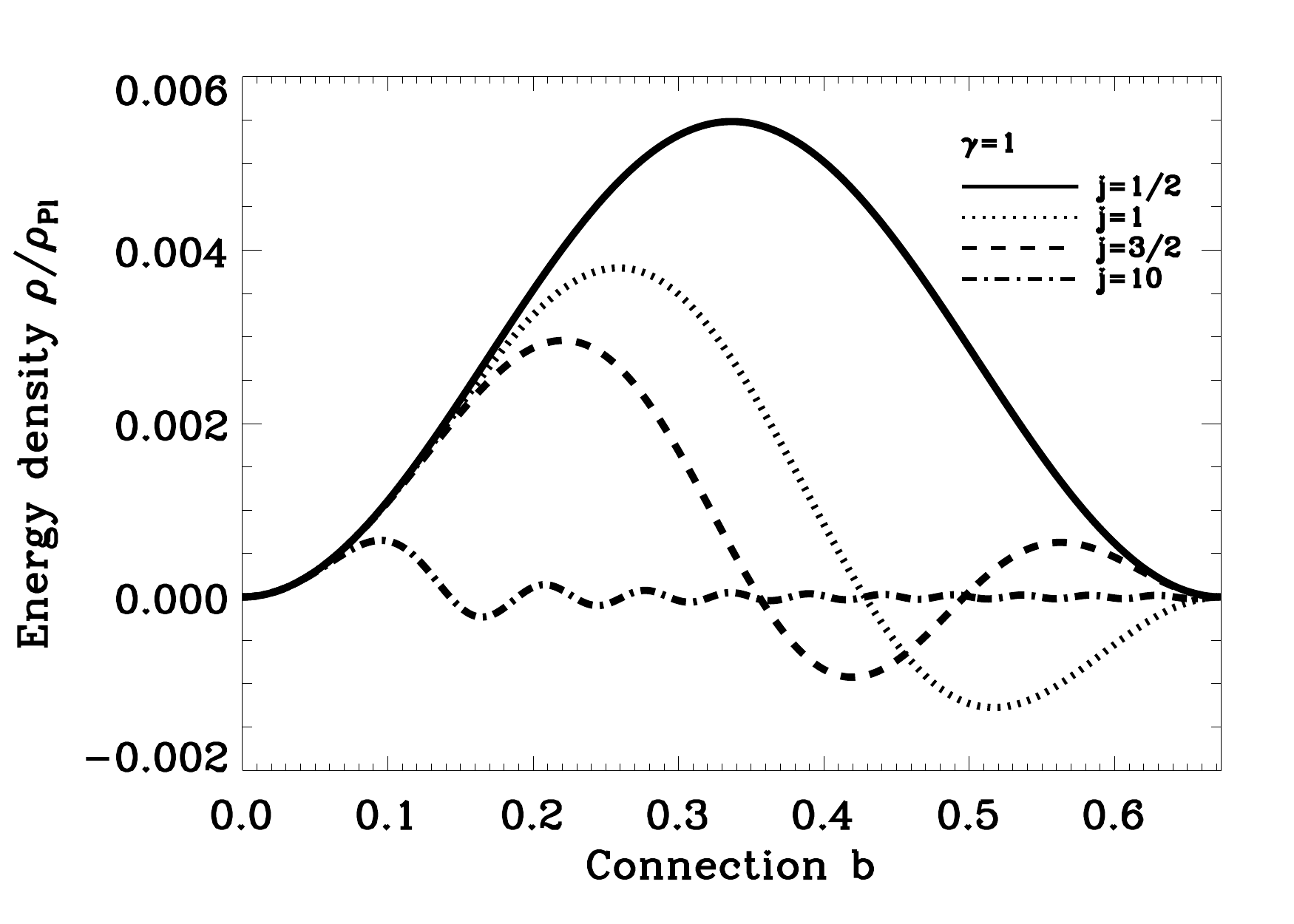}
	\caption{Energy density as a function of $\tilde{b}$ for real loop quantum cosmology for the lowest representation $j=0$ (left panel) and for $s=1/2,~1,~3/2$ and 10.}
	\label{fig:complex}
\end{center}
\end{figure}

While this energy density remains bounded for any $j$, it can experienced some negative excursions. This is clear from the Figure $1$. This observation is in direct conflict with the definition of the energy density $\rho$ which is always positive, i.e. $\rho = \frac{p^{2}_{\phi}}{2V^{2}} > 0$. Even if this point is still not clear up to now, we can argue that because we are dealing with truncated expression for the curvature,  and therefore for the effective hamiltonian, we are no longer describing exact General Relativity and therefore $\rho$ do not describe truly an energy density. Yet, this remark do not cure the tension between the definition and the negative value of $\rho$.

As we can see on the Figure $1$, only the $j=1/2$ representation remains positive on the whole range of $b$. Therefore, this is the unique representation which do not enter in conflict with the definition of $\rho$.
This observation could be interpreted as a kind of selection rule for the $j=1/2$ representation. 

Having developed the effective theory for any arbitrary spin $j$, we are now ready to perform our analytic continuation and define the so called self dual LQC.

\section{The new model : self dual Loop Quantum Cosmology}

This new model was first introduced in \cite{ch6-BA1}. Our goal is to explore the implications of our analytic continuation prescription in the cosmological setting, and more precisely, if the bouncing scenario described by the simplest LQC model is preserved once $\gamma$ become purely imaginary, i.e. when working with the homogenous and isotropic complex Ashtekar variables. We first focus our attention on the effective approach and we devellop in the last section the full quantum theory of self dual LQC, in the framework of the exactly solvable model.

\subsection{The effective approach}

The effective theory of real LQC, in any arbitrary spin $j$ is given by the Poisson bracket of the two conjugated variables , i.e. the gravitational canonical variables $(b,V)$ and the canonical variables describing the scalar field $(\phi, p_{\phi})$.
They reads:
\begin{align*}
\{b , V\} = \frac{\gamma \kappa}{2} \;\;\;\;\;\;\;\;\  \{ \phi, p_{\phi} \} = 1
\end{align*}

The dynamics of those variables is given by the general hamiltonian computed precedently, where the holonomies corrections has been incorporated:
 \begin{align*}
 C_{tot} &  = \frac{p^{2}_{\phi}}{2V} + \frac{9 V}{8 \pi G \gamma^{2} \lambda^{2} d (d^{2}-1)}  \frac{ \sin^{2} (\lambda b)}{ \sin{\theta}}\;  \frac{\partial}{\partial \theta} \; \frac{\sin{d_{j} \theta}}{\sin{\theta}} \;\;\;\;  \text{and} \;\;\;\; \sin{\frac{\theta}{2}} = \sin^{2}{(\frac{\lambda b}{2})} 
\end{align*}

We use this phase space to apply our analytic continuation prescription. Recall that our prescription is given by:
 \begin{align*}
\gamma = \pm i \;\;\;\;\;\;\;\;  \text{and} \;\;\;\;\;\;\;\;  j = \frac{1}{2} (-1 + is) \;\;\;\;  s \in \mathbb{R}^{+}
\end{align*}
It follows that $d = 2j+1 = i s$.

The first task is to identify the presence of $\gamma$ in the variables. Indeed, the variables $b$ that we used from the beginning contains a $\gamma$ and we note it $b = \gamma \tilde{b}$ where $\tilde{b} \in \mathbb{R}$.
Therefore $b$ becomes purely imaginary, i.e. $b \in i\;\mathbb{R}$. This shift from real to purely imaginary for the variable $b$ induces some restrictions on the variable $\theta$.
The relation between $\theta$ and $b$ lead to $ \sinh^{2}( \lambda \tilde{b}/2) = - \sin{\theta /2}$ where $\theta \in \mathbb{R}$. Therefore, we need to restrict the range of $\theta$ where $\sin{\theta} \leqslant 0$. However, this restriction doesn't seem natural.
What happen if we require that $\theta$ become $\theta = i \tilde{\theta}$ ?
In this case, the relation becomes $ \sinh^{2}( \lambda \tilde{b}/2) = - i \sinh{\tilde{\theta} /2}$. In this case, $\tilde{\theta}$ has to be complex. This leads to a energy density which is complexed valued.
Now we can rise the power of this equality until we obtain a consistent relation. Taking the power $4$ of the precedent equality, we obtain $ \sinh^{8}( \lambda \tilde{b}/2) =  \sinh^{4}{\tilde{\theta} /2}$.
In this case, $\tilde{b}$ and $\tilde{\theta}$ remains real. Rising the power of the equality could seems a priori obscure, but the point is that on the real line, for $(b, \theta) \in \mathbb{R}$, all the power of the equality are equivalent. However it is no longer true when $(b, \theta)$ become complex or purely imaginary. In this case, we need to make a choice which is guided by physical consideration, i.e. having an energy density real valued. Therefore, our choice is to work with $b = i \tilde{b}$ and $\theta = i \tilde{\theta}$ where $(\tilde{b}, \tilde{\theta}) \in \mathbb{R}$.

Applying our analytic continuation prescription on the $j$-dependent hamiltonian $C_{tot}$, we obtain the self dual version of the hamiltonian constraint:
 \begin{align*}
 C_{tot} &  = \frac{p^{2}_{\phi}}{2V} + \frac{9 V}{8 \pi G \lambda^{2} s (s^{2}+1)}  \frac{ \sinh^{2} (\lambda \tilde{b})}{ \sinh{\tilde{\theta}}}\;  \frac{\partial}{\partial \tilde{\theta}} \; \frac{\sin{s \tilde{\theta}}}{\sinh{\tilde{\theta}}} \;\;\;\;  \text{and} \;\;\;\; \sinh{\frac{\tilde{\theta}}{2}} = \sinh^{2}{(\frac{\lambda \tilde{b}}{2})} 
\end{align*}

The energy density for the universe is directly readed from the expression of the hamiltonian constraint, i.e. 
 \begin{align*}
\rho_{s} & = \frac{p^{2}_{\phi}}{2V^{2}} = - \frac{9}{8 \pi G \lambda^{2} s (s^{2} +1)}  \frac{ \sinh^{2} (\lambda \tilde{b})}{ \sinh{\tilde{\theta}}}\;  \frac{\partial}{\partial \tilde{\theta}} \; \frac{\sin{s \tilde{\theta}}}{\sinh{\tilde{\theta}}}  
\end{align*}

We can now study the two asymptotics regims, where $(\tilde{\theta}, \tilde{b})$ are negligible, and when $(\tilde{\theta}, \tilde{b})$ are large. We will show below that the variable $\tilde{b}$ increase when the cosmic time descrease, therefore, the two regim corresponds respectively to our present universe and to the far past.
For $(\tilde{\theta}, \tilde{b}) \ll 1$, we have the following behaviour:
  \begin{align*}
\rho_{s} \simeq \frac{3}{8\pi G} \tilde{b}^{2} \;\;\;\;\;\;  \text{with} \;\;\;\;\;  \frac{ \sin^{2} (\lambda \tilde{b})}{ \sinh{\tilde{\theta}}}\;  \frac{\partial}{\partial \tilde{\theta}} \; \frac{\sin{s \tilde{\theta}}}{\sinh{\tilde{\theta}}}  \simeq \frac{s(s^{2} +1)}{3} (\lambda \tilde{b})^{2} + o((\lambda\tilde{b})^{3})\;\;\; \text{and} \;\;\; \tilde{\theta} = \frac{(\lambda \tilde{b})^{2}}{2}
\end{align*}
Only in this regim one recover the usual Friedman expansion law:
  \begin{align*}
H = \frac{\dot{a}}{a} = \frac{\dot{V}}{3V} = - \frac{4\pi G}{3} \frac{\partial \rho_{s}}{\partial \tilde{b}} = - \tilde{b} \;\;\;\; \text{whence} \;\;\;\;\;  H^{2} = \frac{8\pi G}{3} \rho_{s}
\end{align*}
Therefore we recover the right semi classical limit for the present universe, i.e. for $\tilde{b} \ll 1$.
Already at this point, we observe that our analytic continuation preserves the semi classical limit.

For $(\tilde{\theta}, \tilde{b}) \gg 1$, we obtain:
  \begin{align*}
\rho_{s} \simeq \frac{9}{2\pi G s(s^2+ 1) \lambda^{2}} e^{-\tilde{\theta}} ( \; \sin{s\tilde{\theta}} - s\; \cos{\tilde{\theta}} ) \;\;\;\;\;\;\;  \text{where} \;\;\;\;\; e^{2\tilde{\theta}} \simeq \frac{1}{16} e^{4 \lambda \tilde{b}}
\end{align*}
From this two limits,  since $\lim\limits_{\tilde{b} \to 0} \rho_{s} = \lim\limits_{\tilde{b} \to \infty} \rho_{s} = 0$ and since $\rho_{s}$ do not admit any pole, we conclude the function $\rho_{s}$ remains bounded.
Figure $2$  gives the energy density for different value of $s$. As shown analytically, the function $\rho_{s}(\tilde{b})$ remains bounded for any value of $s$. This is the first result of this model.
Already at the effective level, we observe that the universe will not experienced a singularity. Our analytic continuation prescription do preserve the bouncing scenario, which a non trivial result.

 \begin{figure}
\begin{center}
	\includegraphics[width = 0.6\textwidth]{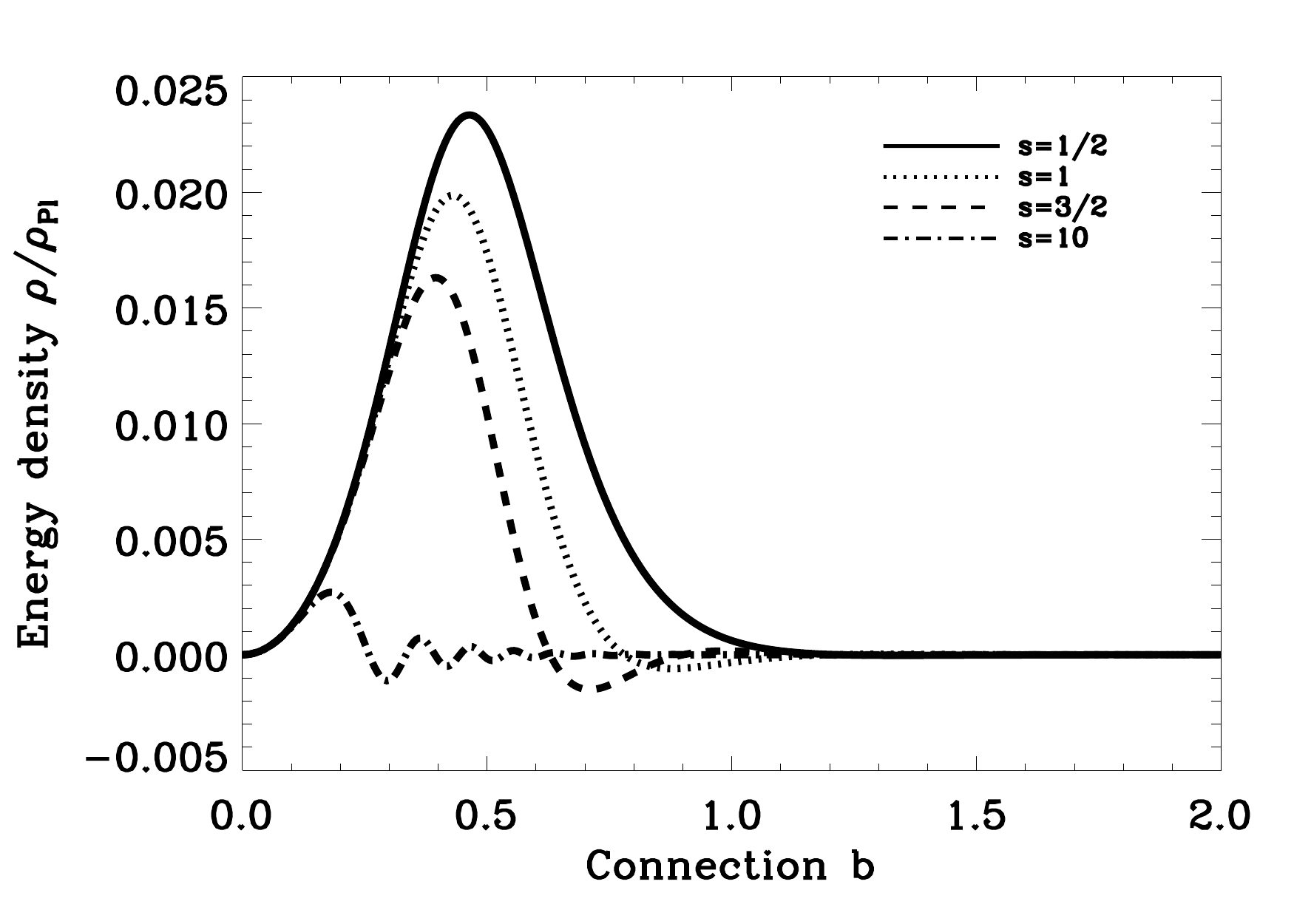}
	\caption{Energy density as a function of $\tilde{b}$ for complex loop quantum cosmology for the lowest representation $s=0$ and for $s=1/2,~1,~3/2$ and 10.}
	\label{fig:complex}
\end{center}
\end{figure}

Moreover, from the Figure $2$, we observe that the energy density can be negative valued for any value of $s > 0$. This is the very same behaviour than the one described for the real case.
The resulting energy density function is in direct conflict with the definition we used for $\rho_{s}$, i.e. $\rho_{s}  = \frac{p^{2}_{\phi}}{2V^{2}} > 0$. Just as in the real case, we remark that only the case $s=0$ leads to a well defined positive energy density on the whole range of $\tilde{b}$. We will therefore privilege this particular model for our computations.

To understand more precisely the plot, let us study how the variable $b$ vary w.r.t the cosmic time $t$.
A straightforward computation gives:
 \begin{align*}
\frac{\partial \tilde{b}}{\partial t} & = \{ \tilde{b}, C_{tot} \}= - 4 \pi G \rho_{s}
\end{align*}
For $s=0$, since $\rho_{0} > 0$, the variable $\tilde{b}$ decreases when the cosmic time increases. Therefore, the regime $b \ll 1$ corresponds to our present universe while the case $b\gg1$ corresponds to the far past, before the bounce. 

Let us compute the general Friedman equation for this universe:
 \begin{align*}
H = \frac{\dot{a}}{a} = \frac{\dot{V}}{3V} = \frac{1}{3V} \{ V , C_{tot} \} = \frac{1}{3V} \{ V , \rho_{s} \} =  - \frac{4\pi G}{3} \frac{\partial \rho_{s}}{\partial \tilde{b}} = -  \frac{4\pi G \lambda}{3} \frac{ \sinh{(\lambda \tilde{b})}}{\cosh{\tilde{\theta}}}\frac{\partial \rho_{s}}{\partial \tilde{\theta}}
\end{align*}

Let us focus on the case $s=0$. In the limit where $b \gg 1$, the energy density and the Hubble constant become:
 \begin{align*}
\rho_{0} & = \frac{9}{8\pi G} \frac{\sinh^{2}{\lambda \tilde{b}}}{\lambda^{2}} \frac{\tilde{\theta} \coth{\tilde{\theta}}- 1}{\sinh^{2}{\tilde{\theta}}} \simeq \frac{9}{8\pi G} \frac{e^{2\lambda \tilde{b} - 2 \tilde{\theta}}}{\lambda^{2}} \tilde{\theta} \simeq \frac{36}{\pi G \lambda} \; \tilde{b}  \; e^{-2\lambda \tilde{b}} \\
H & = - \frac{48}{\lambda}  e^{- 2\lambda \tilde{b}} ( 1 - 2 \lambda \tilde{b}) \simeq  96 \; \tilde{b} \; e^{- 2\lambda \tilde{b}}
\end{align*}

where for $(\tilde{\theta} , \tilde{b})$ large, we have: $ e^{2\tilde{\theta}} = e^{4 \lambda \tilde{\theta}} /4$.
Consequently, before the bounce, the Friedman equation is linear, i.e. the Hubble constant is proportional to the energy density: $H \propto \rho_{0}$.
This peculiar linear expansion law is also found in some models of branes cosmology. However, we don't know if one can extrapolate such result and if so, the interpretation remains unclear.
Its explicit expression reads:
 \begin{align*}
H & \simeq \frac{8\pi G }{3} \; \lambda \; \rho_{0} 
\end{align*}

Finally, we note that after applying our analytic continuation, the new expression for the energy density is no more periodic contrary to the real case.
To summarize the results obtained for the effective self dual LQC, we have shown that our analytic continuation prescription describes a bouncing universe which admit the usual Friedman law at the semi classical limit, and satisfy a linear expansion law before the bounce. The evolution before and after the bounce is no more symmetric in this new cosmology.

To go further, we need to develop the full quantum theory and compute the expectation values of Dirac observables, i.e. the volume operator for instance, in order to conclude on the robustness of our bouncing model.
This is precisely the goal of the next section.

\subsection{The exactly solvable model}

In this section, we proceed to the canonical quantization of the precedent effective classical theory, i.e. we build the full quantum self dual LQC.
We will work in the so called exactly solvable model which is well suited to compute expectation values such as the expectation value of the volume operator.
The starting point is the effective theory defined by the following phase space:
\begin{align*}
& \{\tilde{b} , V\} = \frac{\kappa}{2} \;\;\;\;\;\;\;\;\  \{ \phi, p_{\phi} \} = 1 \\
&  N \; C_{tot}  = N \; (\;  \frac{p^{2}_{\phi}}{2V} + \frac{9 V}{8 \pi G \lambda^{2} s (s^{2}+1)}  \frac{ \sin^{2} (\lambda \tilde{b})}{ \sinh{\tilde{\theta}}}\;  \frac{\partial}{\partial \tilde{\theta}} \; \frac{\sin{s \tilde{\theta}}}{\sinh{\tilde{\theta}}} \; )\;\;\;\;  \text{and} \;\;\;\; \sinh{\frac{\tilde{\theta}}{2}} = \sinh^{2}{(\frac{\lambda \tilde{b}}{2})}  
\end{align*}

We will restrict our model to the case $s=0$ since this is the only one which do not generate a negative energy density: $\rho_{s} \geqslant 0$. Moreover, we choose to work in the harmonic time, i.e. with $N = V$.
Those restriction leads to the following phase space for $s=0$:
\begin{align*}
& \{\tilde{b} , \nu \} = 1 \;\;\;\;\;\;\;\;\  \{ \phi, p_{\phi} \} = 1 \;\;\;\;\;\;\;   p^{2}_{\phi} - ( v \; X(\tilde{b}))^{2} \;  = 0 
\end{align*}
where we have used the reduced volume $v = V / 4\pi G$ and the function $X(\tilde{b})$ is given by:
\begin{align*}
 X(\tilde{b}) = 4 \pi G \sqrt{2 \rho_{0}} \;\;\;\;\; \;\; \text{with} \;\;\;\;\;\;\; \rho_{0} = \frac{9}{8 \pi G \lambda^{2}} \frac{\sinh^{2}{\lambda \tilde{b}}}{\sinh^{2}{ \tilde{\theta}}} \; ( \; \tilde{\theta} \; \coth{\tilde{\theta}} -1 \; )
\end{align*}

In the following, we simplify the notation and forget about $\tilde{b}$, keeping just $b$. Indeed, we have now to distinguish between the classical variable $b$ and the quantum operators with an $\hat{b}$.
The usual quantization procedure leads to the quantum algebra:
\begin{align*}
& [ \; \hat{b} , \hat{v} \; ]= i \hbar \;\;\;\;\;\;\;\; [ \; \hat{\phi}, \hat{p}_{\phi} \; ] = i \hbar
\end{align*}

As usual in the exactly solvable model, the polarization is choosen such that the quantum states are function of the variables $(\phi, b)$ and we note them: $\Psi (\phi, b)$. The quantum operators $\hat{\phi}$ and $\hat{b}$ act simply by multiplication int his polarization while the quantum operators $\hat{\nu}$ and $\hat{p}_{\phi}$ act by derivation as:
\begin{align*}
\hat{v} \; \Psi (\phi, b) = - i \hbar \frac{\partial \Psi (\phi, b)}{\partial b}  \;\;\;\;\;\;\;\;    \hat{p}_{\phi} \; \Psi (\phi, b) = -  i \hbar \frac{\partial \Psi (\phi, b)}{\partial \phi} 
\end{align*}

Therefore, in this polarization, we obtain the dynamical equation, i.e. the Schrodinger equation, as a differential equation which reads:
\begin{align*}
\frac{\partial^{2} \Psi (\phi, b)}{\partial \phi^{2}}  = \; ( \; X(b) \frac{\partial}{\partial b}  \; )^{2} \;  \Psi (\phi, b) 
\end{align*}

We note that in the real case, for $j = 1/2$, the function $X(b)$ was given by a cardinal sinus, i.e. $\sin{(\lambda b)} / \lambda$. Now, the function $X(b)$ is much more involved.
However, it is still possible to proceed to a change of variable defining the following variable:
\begin{align*}
x(b) = \int^{b}_{b_{0}} \frac{du}{X(u)} \;\;\;\;\;\;\; \text{and}  \;\;\;\;\;\;\;  \frac{\partial}{\partial x} = X(b)  \frac{\partial}{\partial b}
\end{align*}
Let us study the asymptotics of the function $x(b)$. This will give us the range of $x$.
\begin{align*}
&\text{when} \;\;\;  b \ll  \;\;\;  \text{then} \;\;\;  x(b) \simeq \frac{1}{\sqrt{12\pi G}} \int^{b}_{b_{0}} \frac{du}{u} \simeq \; \frac{1}{\sqrt{12\pi G}} \log{b} \\
& \text{when} \;\;\;  b \gg  \;\;\;  \text{then} \;\;\;  x(b) \simeq \frac{1}{24} \sqrt{\frac{\lambda}{ \sqrt{2\pi G}}} \int^{b}_{b_{0}}  \; du \; \frac{e^{\lambda u}}{\sqrt{u}} \simeq \;\frac{1}{24} \sqrt{\frac{\lambda}{ \sqrt{2\pi G}}}  \; \frac{e^{\lambda b}}{\sqrt{b}}  \\
\end{align*}
Therefore, $x(b)$ is an increasing function which runs from $- \infty$ to $+ \infty$ for $b \in [- \infty, + \infty]$. 
The dynamical equation for the universe is recast into a Klein Gordon equation:
\begin{align*}
\frac{\partial^{2} \Psi (\phi, x)}{\partial \phi^{2}}  = \;  \frac{\partial^{2} \Psi (\phi, x)}{\partial x^{2}} 
\end{align*}
This equation selects the physical quantum states of the universe.
Just as before, in the real case, the solutions to this equation can be decomposed into right and left moving sectors and we can write: $\Psi (\phi, b) = \Psi_{R} (\phi, b) + \Psi_{L} (\phi, b)$.
For the very same reason than precedently, i.e. the invariance under change of the tetrad orientation, the quantum states becomes: $\Psi (\phi, b) = ( F(x_{+}) + F(x_{-})/\sqrt{2}$ where we note $x_{\pm}= \phi \pm x$.

The Klein Gordon scalar product reads:
 \begin{align*}
( \chi_{1}, \chi_{2} )_{\text{phys}} = & \;  i \int_{\phi = \phi_{0}} [ \; \partial_{x} \bar{F_{1}}(x_{+}) F_{2}(x_{+}) - \partial_{x} \bar{F}_{1}(x_{-}) F_{2}(x_{-})] \; dx \\
\end{align*}

We are now ready to compute the expectation value of the volume operator. We will use the operator $\hat{\nu}= \hat{v} / \hbar$, which is given by:
 \begin{align*}
\hat{\nu} = - i \; \frac{\partial x}{\partial b} \; \partial_{x} = - i \frac{1}{X(x)} \partial_{x} 
\end{align*}

The expectation value of the self dual version of this operator can be computed and leads to:
 \begin{align*}
( \Psi, \hat{\nu} \; \Psi)_{\text{phys}} = & \;  i \int_{\phi = \phi_{0}} [ \; \partial_{x} \bar{F}(x_{+}) ( \hat{\nu} F(x_{+})) - \partial_{x} \bar{F}(x_{-}) ( \hat{\nu} F(x_{-}))] \; dx \\
= & \; \;  \int_{\phi = \phi_{0}} [ \; \partial_{x} \bar{F}(x_{+}) \;  \frac{1}{X(x)} \; \partial_{x} F(x_{+}) + \partial_{x} \bar{F}(x_{-}) \;  \frac{1}{X(x)} \; \partial_{x} F(x_{-}) \; ]  \\
= & \; \;  \int_{\phi = \phi_{0}}  \; | \partial_{x} F(\tilde{x})|^{2} \{ \frac{1}{X( \tilde{x} - \phi)}  + \frac{1}{X(\phi - \tilde{x})} \}  \\
\end{align*}

Here, we have assumed that the function $F(x)$ falls off rapidly at infinity and that there are square integrable. From the second to the third line, we have applied two change of variables: $x_{+} = \tilde{x} \;\;\;\; dx = d\tilde{x}$ or $x_{-} = \tilde{x}\;\;\;\;\; dx = - d\tilde{x}$. The plot of the function $X(x)$ is given in figure $2$. We observe that it falls off rapidly to zero at $\pm \infty$ and it admit a maximal value at $x_{0}$. Therefore, we conclude that the expectation value of the volume operator is bounded from below. The universe never experiences a zero volume state and do not shrink to a point, resolving therefore the initial singularity of the Big Bang.
Its minimal volume is given by:
 \begin{align*}
< \; V \; > \; = 4 \pi l^{2}_{p} ( \Psi, \hat{\nu} \;  \Psi)_{\text{phys}} \geqslant & \;  \frac{8 \pi l^{2}_{p}}{X_{max}} \int_{\phi = \phi_{0}}  \; |\partial_{\tilde{x}} F(\tilde{x}) |^{2} \; dx \\
\end{align*}

where $X_{max}$ has the dimension of $[L^{-1}]$.
This conclude the presentation of the exactly solvable model of self dual LQC.

\section{Discussion}

We have presented in details the new model of self dual LQC \cite{ch6-BA1}. We first would like to emphasis on the different results obtained and then compare with recent works dealing with the complex Ashtekar variables in Loop Quantum Cosmology.

As we have shown, the analytic continuation prescription applied to the simplest LQC model leads to a bouncing cosmology.
This bouncing universe admits the right semi classical limit, i.e. the usual Friedman cosmology at large scale after the bounce.
Those two results are highly non trivial since there is a priori no reasons for our ad hoc prescription to generate a coherent dynamic both at the semi classical level and at the quantum level.
Yet, this is precisely what happen. It is striking that such simple prescription defined in the context of the black hole entropy derivation in order to obtain the right semi classical limit lead in the same time to a coherent picture of the quantum cosmology based on the LQC technics. 

Then we have seen that, for real LQC just as for self dual LQC, the quantum universe experience a new kind of cosmology before the bounce, i.e. the dynamics before and after the bounce is no more symmetric for a general $j$ or a general $s$. Moreover, the energy density which depends on the spin $j$ representation (resp $s$) can experience some negative values which is in contradiction with its definition: $\rho = p^{2}_{\phi} / 2 V^{2} > 0$.
This particular behaviour has to be interpretated with due care. Indeed, the hamiltonian responsible for those obscure behaviour is computed only in the spin $j$ representation (resp $s$).

What happen if one build an hamiltonian as a sum of all the $j$ (resp $s$) contributions ? 

In the real case, one can show that the energy density computed with a general combinations of $j$ remains always positive and that the behavior before and after the bounce stays symmetric.
Therefore, those issues seems to be related to the way we build our hamiltonian, at least in the real case. Consequently, the asymmetric evolution of the quantum universe and the negative excursion of the energy density  should not be interpreted as a a relevant physical prediction of the model but as an artefact of the regularization procedure we have used.

The situation in the self dual case is not clear. Indeed, the analytic continuation prescription trades periodic function for aperiodic functions.
From this observation, it is obvious that the hamiltonian of self dual LQC will loose the periodicity of its real counterpart. The quantum universe will inevitably be asymmetric with regard to the bounce.
However, we have not study what are the modifications brought by an hamiltonian built as a sum of contributions of different spins $s$.

It is worth mentioning that the regularization of the hamiltonian constraint that we have used can be generalized. Indeed, one can always introduce a function $f : \mathbb{R} \rightarrow \mathbb{R}$ and used the modified regularization prescription for the non local curvature:
 \begin{align*}
 F_{ab}{}^{l}  & = - \epsilon_{ab}{}^{l} \; \frac{12}{d ( d^{2} -1) \bar{\mu}^{2}} \; f ( \; \frac{\partial}{\partial \epsilon } Tr_{j} \big{(} h_{\Box_{12}} \; e^{\epsilon \tau^{3}}\big{)} \big{|}_{\epsilon = 0} ) \\
 & = - \epsilon_{ab}{}^{l} \; \frac{3 |p|}{d ( d^{2} -1) {\lambda}^{2}} \; f( \; \frac{ \sin^{2} (\lambda b)}{ \sin{\theta}}\;  \frac{\partial}{\partial \theta} \; \frac{\sin{d_{j} \theta}}{\sin{\theta}})
\end{align*}
provided that the function $f$ satisfies:
\begin{align*}
 f \; ( \; \frac{ \sin^{2} (\lambda b)}{ \sin{\theta}}\;  \frac{\partial}{\partial \theta} \; \frac{\sin{d_{j} \theta}}{\sin{\theta}}) ) \rightarrow b^{2} \;\;\;\;\;\;\;\;\; \text{when}  \;\;\;\;\;\;\; b, \theta \ll 1
\end{align*}
This ensures that one recovers the right semi classical limit, i.e. the classical Friedman equation. Obviously, there is a large class of functions satisfying this condition, and there is therefore a very large choice of possible regularizations. Either one takes this ambiguity as a new freedom in the model to obtain new phenomenology, either one can argue that such ambiguity in the regularization is quite disastrous since one can obtain pretty much what he wants. It could be interesting to study if some other constraints can be derived for the function $f$, selecting therefore particular dynamics for the quantum universe.
This point underlines the fact that the choice of regularization is of first importance in this context. Consequently, only the bouncing property and the right semi classical limit are the true generic properties of the real and self dual models while the ``phenomenology'' derived in this model is highly regularization dependent. 

Finally, it is interesting to compare our work to the recent paper of Wilson-Ewing \cite{ch6-EW1}. This work deals with the explicit imposition of the reality conditions in the closed FLRW quantum cosmology.
In this case, the reality conditions can be expressed in terms of the homogenous and isotropic conjugated variables $(c,p)$ as:
\begin{align*}
c + \bar{c} = l_{0} \;\;\;\;\;\;\; \text{and} \;\;\;\;\;\;\; \bar{p} = p
\end{align*}
where $V_{0} = l^{3}_{0}$ is the volume of the fiducial cell.
In order to implement those quantum conditions, two steps are needed. First, the author introduces the so called ``generalized holonomies'' in order to have a well defined loop variables in the self dual version of LQC.
Then, the measure of the Hilbert space has to be modified through distributional constraints on the quantum states. With this two ingredients, a self dual version of LQC is worked out and the bouncing scenario is recovered. In this model, the dynamics turns out to be very similar to the real case.
For the flat universe, taking the limit $l_{0} = 0$, one recovers exactly the results of standard real LQC.
The two models being based on very different strategy, the comparison is quite difficult.
While both models recover a bouncing scenario when working with $\gamma = \pm i$, the construction of \cite{ch6-EW1} leads to very different conclusions than the self dual model presented in this chapter. 
The first main difference is that while in \cite{ch6-EW1}, one work in the spin $1/2$ representation, our model is worked out for a general $j$ representation. Then we apply the analytic continuation prescription and restrict the study to the smallest spin $s=0$ of the continuous representation of $SU(1,1)$. Therefore, contrary to the strategy used in \cite{ch6-EW1}, our procedure consisting in mapping the usual $SU(2)$ representation into the $SU(1,1)$ continuous representations  brings important modifications to the quantum dynamics. However, as explained above, the procedure we have used has the advantage to provide also a well defined procedure in very different contexts, such as for the black hole entropy derivation \cite{ch6-BA2, ch6-BA3} and in three dimensional quantum gravity \cite{ch6-BA4, ch6-BA5}. Quite surprisingly, it also preserves the bouncing scenario of standard Loop Quantum Cosmology.

\clearemptydoublepage


\chapter*{Conclusion\markboth{Conclusion}{Conclusion}}
\addstarredchapter{Conclusion}
\minitoc

We conclude by giving some general comments on the results presented in the thesis, and we proposed some perspectives.
For a details conclusion on each of the results, the reader can refers to the end of each chapter.

In the context of black hole thermodynamics, the model of the ``gas of punctures'' seems to provides an efficient tool to describe the quantum black hole.
This theoretical laboratory allows one to test the impact of different physical inputs on the semi classical predictions of the theory, such as the quantum statistic of the punctures, the presence of a chemical potential or the holographic degeneracy.

Indeed, we have seen that the introduction of the chemical potential associated to the punctures modifies the semi classical results. As explained, the usual loop computation of the entropy results in a $\gamma$-dependent entropy, which requires to be fine tuned in order to match the Bekenstein-Hawking entropy. It turns out that the introduction of a chemical potential $\mu$ associated to the punctures (plus the hypothesis of an holographic degeneracy) shifts this $\gamma$-dependency from the leading term to the subleading terms.
More, the $\gamma$-dependency of the entropy drops out when one assumes a Bose-Einstein statistic for the punctures for some particular regimes \footnote{However, $\gamma$ does not totally disappear, since it enters in the expression of other quantities.}. It turns out that for those regimes, a Bose-Einstein condensation of the punctures in the smallest spin $j=1/2$ occurs and the subleading quantum corrections to entropy are purely logarithmic. While interesting, the link observed between the Bose-Einstein condensation and the presence of purely logarithmic quantum corrections remains obscure up to now. Finally, although the interpretation of the role of the chemical potential $\mu$ in this context is not direct, it should be somehow related to the conversion process of the punctures into matter degrees of freedom (and the reverse). From this perspective, it is a very appealing candidate to describe a potential evaporation process. This direction remains to be investigated.

Finally, in the context of this model, the hypothesis of a holographic degeneracy seems to be crucial to recover the right semi classical limit. Yet, this holographic behavior of the degeneracy is not predicted in the 
loop quantization of spherically isolated horizon within the real $SU(2)$ framework. One needs therefore to postulate it without much justification \footnote{The holographic degeneracy hypothesis could well be motivated by some extrapolation of the loop quantum dynamics of the horizon coupled to matter , but this can not be trusted until one solves explicitly the quantum dynamics of LQG for the system horizon coupled to matter.}. This could be a first hint that the quantization based on the real Ashtekar-Barbero connection has to be modified at some point, in order to deal with the semi classical limit of LQG. 

The idea developed in this PhD was to study to what extend the use of the self dual variables could cure this problem. 

Quite surprisingly, it turns out that working with the self dual variables, i.e. requiring that $\gamma = \pm i$, leads to the right semi classical limit.
In this context, the holographic character of the quantum degeneracy of the horizon is derived from the ``self dual quantum theory''\footnote{Note the we have not at our disposal the self dual quantum theory, since we have no access to its Hilbert space. The only thing that we obtained is the dimension of this self dual Hilbert space.}. This prediction of the self dual quantum theory of the spherically isolated horizon is obtained by mean of an analytic continuation procedure. This procedure turns out to be unique, and provides for the first time a mapping between the predictions of the real $SU(2)$ quantum theory and the predictions of the self dual quantum theory (at least in the context of black hole thermodynamics). In this context, the holographic degeneracy becomes a prediction of the loop quantization based on the self dual variables.
More, the holographic degeneracy is supplemented with some power law corrections which conspired, at the semi classical limit, to give the right logarithmic quantum corrections.
Working out the ``gas of punctures'' model based on the self dual loop quantization, one ends up with a stronger construction with respect to its initial assumptions. Indeed, one can remove the holographic hypothesis, since one obtained it as a prediction of the quantum theory. At the end of the day, the fine tuning of $\gamma$ has disappeared and the semi classical limit is reached in a much more satisfying way.

This result underlines the peculiar status of the self dual variables for the loop quantization program which was already noticed by different authors during those last three years. Those complex variables seems to be the right variables to use in order to deal with the semi classical limit, at least in the context of black hole physics.

It is therefore crucial to study further the analytic continuation procedure introduced above. It could provides the prescription to wick rotate the real $SU(2)$ quantum theory to the self dual quantum theory.
Obviously, this prescription should first transform the real variables for the complex ones, and in the same time, solves somehow the reality conditions inherent to the complex variables.
The prescription derived in the context of black hole thermodynamics fulfills those two goals. It reads:
\begin{align*}
\gamma= \pm i \;\;\;\;\;\;\;\;\;\;\;\;\;\; \text{and} \;\;\;\;\;\;\;\;\;\;\;\;\;\; j = \frac{1}{2}(is - 1)  \;\;\;\;\;\;\;\; \text{where} \;\;\;\;\;\; s \in \mathbb{R}^{+}
\end{align*}
The first part maps trivially the real Ashtekar-Barbero connection to the self dual Ashtekar's complex connection. The second part takes care of the reality conditions, at least for the area spectrum.
This mapping of the half integer spin $j$ into a complex continuous spin is a well known prescription, which maps the Casimir (and the character) of the $SU(2)$ group into the Casimir (and character) of the $SU(1,1)$ group associated to its continuous representation serie. The important point to stress is that this prescription was derived and not simply introduced for our purposes. Moreover, it is the unique prescription leading to the right semi classical result, i.e. the Bekenstein-Hawking area law.

Concerning the area spectrum, as explained above, the analytic continuation maps the discrete $SU(2)$ spectrum into a continuous one, given by the Casimir of $SU(1,1)$ associated to its continuous representation serie.
The resulting continuous area spectrum is positive, real and admits an area gap for $s=0$ even if we are now working with $\gamma = \pm i$.

Consequently, this analytic continuation prescription is a very appealing candidate to ``solve'' the old problem of the quantum reality conditions present in the self dual quantum theory. However, although very encouraging, we do not have any control on the resulting self dual quantum states up to now.
Indeed, we do not know how the $SU(2)$ spin networks are transformed under this prescription. The reason is that the analytic continuation was performed on the dimension of the Hilbert space (of the quantum black hole) and not on the quantum states.
This is a fairly common fact that it is often easier to analytically continue the equations for an objet than the object itself and this is precisely what happens here.
 
 However, at the level of the area spectrum, it seems that the group $SU(1,1)$ (in its continuous representation) replaces the $SU(2)$ group. Consequently, there is a nice interplay between the self dual theory, the disappearance of the Immirzi parameter in the quantum theory, and the appearance of the non compact $SU(1,1)$ group. The precise link between those different facets of our construction remains obscure up to now. Yet, it is tempting to identify the self dual kinematical quantum states as $SU(1,1)$ spin networks, each edges being colored by a continuous representation labelled by a continuous spin $s$. This is required to have a physical real and positive area associated to each edge when working with $\gamma = \pm i$. However, considering such spin network, we run directly to some difficulties.
 
 As explain at the end of chapter 4, the problems of dealing with non compact $SU(1,1)$ spin networks are twofold.
 
 On one hand, the recoupling theory of this group is much more involved. Indeed, a spin network is built up from a graph, irreps associated to each edges and intertwiners associated to each nodes. Those intertwiners are the ingredients which make the quantum states gauge invariant by recoupling the ingoing and outgoing representations. Since two continuous representations of $SU(1,1)$ can recouple into a discrete one, it brings us out from the representations selected to solve the reality conditions. In order to understand deeper this question of the recoupling, one has to investigate the fate of the $SU(2)$ volume spectrum under our analytic continuation. Indeed, the computation of the volume spectrum (let say of a tetrahedron) involves the (four valent) intertwiner. Applying our prescription to this computation and studying under which conditions one obtains a real and positive volume spectrum would select the intertwiner we need in order to solve the reality conditions at the level of the volume operator. This point needs to be investigated.
 
On the other hand, the $SU(1,1)$ group being non compact, its Haar measure blows up and the corresponding Ashtekar-Lewandowski measure cannot be used to computed scalar product.
This is the well known drawback of working with non compact groups. While this point can be overcome in the quantization of three dimensional gravity, where one can define the notion of non compact spin networks, the situation in four dimensions is disastrous. However, the problem of the measure arises only when one computes some transition amplitudes. Therefore, in principle, if one succeeds to implement the dynamics in the $SU(2)$ real quantum theory and compute some predictions, one could then apply the analytic continuation at this point and never encounters the problem of the unbounded measure. Therefore, the analytic continuation would provides a map between physical predictions of the $SU(2)$ quantum theory and the one predicted by the self dual quantum theory.

At which level one has to apply the analytic continuation, before or after implementing the dynamics, is still to be understood.

Concerning the appearance of the $SU(1,1)$ group, a possible interpretation could be that when quantizing the four dimensional self dual theory which relies on an complex $SL(2,\mathbb{C})$ connection, the imposition of the reality conditions selects a real $SU(1,1)$ connection, leading to a real $SU(1,1)$ phase space. This interpretation relies on the results obtained in the context of the toy model of three dimensional gravity presented in chapter 5.
In this framework, the $\gamma$-dependent $SU(2)$ real phase space has been shown to be equivalent to a phase space based on complex variables supplemented with some reality conditions.
Interestingly, those reality conditions take the form of simplicity-like constraints encountered in spin foam models. Once solved, the complex variables reduce to real $SU(1,1)$ variables and the resulting $SU(1,1)$ phase space turns out to be $\gamma$-independent. It turns out that the predictions of the $SU(2)$ kinematical quantum theory are related to the one of the $SU(1,1)$ kinematical quantum theory precisely via our analytic continuation.

Obviously, one cannot extend this result to the four dimensional case yet. Indeed, it could be that this mechanism only occurs due the topological nature of the toy model. However, it offers an interesting example where the presence of the Immirzi parameter in the classical phase space is a pure gauge artifact and therefore, its appearance in the kinematical quantum theory is recognized as an anomaly due to the initial gauge fixing.
Whether this fact extends to the four dimensional case has to be investigated. The first step would be to reproduce the canonical analysis in the non compact gauge used in this toy model.
If one succeeds to derive an $SU(1,1)$ phase space for four dimensional General Relativity where the Immirzi parameter is absent, one could argue that the presence of $\gamma$ in the real Ashtekar-Barbero phase space is a pure gauge artifact, just as in three dimensions. If this is the case, the presence of $\gamma$ in the kinematical and physical predictions of real LQG would be understood as an anomaly of the quantization. 

Equipped with an $SU(1,1)$ phase space for General Relativity, one could still argue that there is no hope to even quantize this theory since it relies on a non compact group.
However, one could simply proceed to the quantization of the $SU(2)$ phase space, and eventually complete it and compute some predictions. Then one would be able to extract the corresponding predictions of the $SU(1,1)$ quantum theory by applying the analytic continuation prescription presented in this PhD. Obvisously, we are still far from this program, but it represents an interesting perspective that need to be investigated further.

Finally, we have shown that our prescription allows to reproduces to expected semi classical limit in the context of black hole thermodynamics  and that it admits a clear interpretation in three dimensional gravity, as being the map from the predictions of the $SU(2)$ kinematical quantum theory to the ones of the $SU(1,1)$ kinematical quantum theory. In the last chapter, we have shown additionally that this prescription preserves the bouncing scenario of Loop Quantum Cosmology as well as the semi classical limit of the quantum universe. This is a non trivial results since there is a priori no reasons for this prescription to lead to a coherent bouncing quantum universe.
The new model resulting from the analytic continuation of real LQC is quite involved but represent an interesting model of self dual LQC. In particular, one should compare it to the recents results obtained in solving directly the reality conditions associated to the homogenous and isotropic complex Ashtekar variables. 

The results presented in this thesis open a new road to deal with the old problem of solving the reality conditions inherent to the self dual quantum theory.
We have already mention different perspectives and possible future investigations. Let us mention some others.

First, in the context of symmetry reduced models, the case of spherically symmetric models has been investigated intensively both in the context of the real and the self dual variables.
It has been shown that it is possible to solve the reality conditions completely in this framework. One could therefore consider the version based on the real variables and apply the analytic continuation on this model.
The comparison with the self dual version could provides interesting insights.

From the point of view of the full theory, different directions of research are possible.
The first one would be to understand the link (if it exists) between our prescription and the wick rotation proposed by Thiemann in order to solve the reality conditions.
For instance, one could implement it on our toy model of three dimensional gravity and compare the results. The same exercise could be implemented on the real LQC model.

Finally, it would be interesting to see how one can apply our prescription to the construction of a spin foam model.
Those tasks represent a very exciting program.  

To conclude, it seems that the quantum theory based on the self dual variables reproduces in a much more satisfying way the semi-classical limit in the context of black hole thermodynamics.
The self dual quantum theory can be investigated by mean of the analytic continuation prescription presented in this PhD, which find a clear interpretation in the case of three dimensional gravity as extracting the predictions of the self dual theory from the one derived from the real quantum theory. Finally, the prescription additionally preserves the important results obtained in loop quantum cosmology, both from the point of view of the bounce and of the semi classical limit. Much more investigations are needed to fully understand the status of the prescription in four dimensions but it seems to provides an very exciting candidate to study the predictions of self dual loop quantum gravity and by-passed the old problem of the reality conditions associated to the complex Ashtelar's variables.

\clearemptydoublepage



\listoffigures
\clearemptydoublepage


\bibliographystyle{unsrt}
\bibliography{../Bibliography/References}

\begin{thebibliography}{99}

\bibitem{I-HiggsD}
CMS collaboration, Observation of a new boson at a mass of 125 GeV with the CMS experiment at the LHC, Phys. Lett. B 716 (2012) 30, arXiv:1207.7235 [hep-ex]

\bibitem{I-BE}
R. Brout and F. Englert, Broken Symmetry and the Mass of Gauge Vector Mesons, 10.1103/PhysRevLett.13.321, (1964)

\bibitem{I-H}
P. Higgs, Broken Symmetries and the Masses of Gauge Bosons, 10.1103/PhysRevLett.13.508, (1964)

\bibitem{I-Planck}
Planck Collaboration, Planck 2013 results. XVI. Cosmological parameters, Astron.Astrophys. 571 (2013) A16, arXiv:1303.5076 [astro-ph.CO] 

\bibitem{I-PlanckInf}
Planck Collaboration, Planck 2015 results. XX. Constraints on inflation, (2015) arXiv:1502.02114 [astro-ph.CO]

\bibitem{I-QM1}
W. Heisenberg, Über den anschaulichen inhalt der quantentheorischen Kinematik und Mechanik. Zeitschrift für Physik, 43, 172-198 (1927)

\bibitem{I-QM2}
M. Born and P. Jordan, Zur Quantenmechanik I, Z. Phys. 34, 858-888, (1925)

\bibitem{I-QM3}
M. Born, W. Heisenberg, and P. Jordan,  Zur Quantenmechanic II, Z. Phys. 35, 557-615, (1926)

\bibitem{I-QM4}
P.  A.  M.  Dirac, The fundamental equations of quantum mechanics, Proc. R. Soc. London, Ser. A, 109, 642–653, (1925)

\bibitem{I-QM4}
E. Schrodinger, An Undulatory Theory of the Mechanics of Atoms and Molecules, 10.1103- PhysRev.28.1049, (1926)

\bibitem{I-QFT}
T. Aoyama, M. Hayakawa, T. Kinoshita and M. Nio, Revised value of the eight-order QED contribution to the anomalous magnetic moment of the electron, Phys.Rev.D77:053012,2008, arXiv:0712.2607 [hep-th] (2008)


\bibitem{I-Al1}
Albert Einstein, Feldgleichungen der Gravitation (The Field Equations of Gravitation), Preussische Akademie der Wissenschaften, Sitzungsberichte, part 2, p(844-847) (1915).

\bibitem{I-BKI}
L. Smolin, The case for background independence, Rickles, D. (ed.) et al.: The structural foundations of quantum gravity 196-239, hep-th/0507235

\bibitem{I-Ed1}
Edmund  Bertschinger, Symmetry transformations, the Einstein-Hilbert action, and gauge invariance , Massachussets Institute of Technology, (2002).

\bibitem{I-Rov1}
C. Rovelli and M. Gaul ,  Loop quantum gravity and the meaning of diffeomorphism invariance , Towards Quantum Gravity , p (277--324), (2000) , Springer 

\bibitem{I-STh}
T. Thiemann, The LQG string: Loop quantum gravity quantization of string theory I: Flat target space, Class.Quant.Grav. 23 (2006) 1923-1970, hep-th/0401172 

\bibitem{I-SKa}
W. J. Fairbairn and K. Noui, Canonical Analysis of the Algebraic String Actions, Sep 2009. JHEP 1001 (2010) 045, arXiv:0908.0953 [hep-th] 

 \bibitem{I-Reuter}
M. Reuter and F. Saueressig, Functional Renormalization Group Equations, Asymptotic Safety, and Quantum Einstein Gravity, Lectures given at Conference: C07-04-23.2, arXiv:0708.1317 [hep-th] 

\bibitem{I-ADM}
 Arnowitt, Deser and Misner , The dynamics of general relativity , Gravitation:  An Introduction to Current Research, pp (227-265) Wiley New York, (1962) , arXiv:gr-qc/0405109 

 \bibitem{I-Ash1}
 A. Ashtekar, New variables for classical and quantum gravity, Phys. Rev. Lett. 57,
2244 (1986).
 
 \bibitem{I-Ash2}
A. Ashtekar, New Hamiltonian formulation of general relativity, Phys. Rev. D 36, 1587 (1987).

 \bibitem{I-JacSmo1}
T. Jacobson and L. Smolin, Nonperturbative Quantum Geometries, Jul 1987. Nucl.Phys. B299 (1988) 295

 \bibitem{I-JacSmo2}
 A. Ashtekar, T. Jacobson and L. Smolin, A New Characterization of Half Flat Solutions to Einstein's Equation, Commun.Math.Phys. 115 (1988) 631 
 
  \bibitem{I-RovSmo1}
C. Rovelli and L. Smolin, Loop Space Representation of Quantum General Relativity,  Apr 18, 1989. Nucl.Phys. B331 (1990) 80-152

  \bibitem{I-Gam1}
R. Gambini and A. Trias, Gauge Dynamics in the C Representation, Apr 1986. Nucl.Phys. B278 (1986) 436

  \bibitem{I-PolSca}	
A. Ashtekar, J. Lewandowski and H. Sahlmann, Polymer and Fock representations for a scalar field, (2002). Class.Quant.Grav. 20 (2003) L11-1, gr-qc/0211012 

  \bibitem{I-PolMax}	
A. Ashtekar, C. Rovelli and L. Smolin, A Loop representation for the quantum Maxwell field, (1991). Class.Quant.Grav. 9 (1992) 1121-1150, hep-th/9202063

  \bibitem{I-PolGraviton}
A. Ashtekar, C. Rovelli and L. Smolin, Gravitons and loops, (1991). Phys.Rev. D44 (1991) 1740-1755: hep-th/9202054 | PDF

 \bibitem{I-Bar1}
 J. F. Barbero, Real Ashtekar variables for Lorentzian signature space-times, Phys. Rev. D 51 5507 (1995), arXiv:gr-qc/9410014
 
 \bibitem{I-As1}
 A. Ashtekar and C. J. Isham, Representations of the holonomy algebras of gravity and nonAbelian gauge theories, Class.Quant.Grav. 9 (1992) 1433-1468, hep-th/9202053
 
\bibitem{I-As2}
 A. Ashtekar and J. Lewandowski, Representations theory of analytic holonomy C* alge- bras, 1993, CGPG-93-8-1, gr-qc/9311010
  
\bibitem{I-As3}
A. Ashtekar and J. Lewandowski, Completeness of Wilson loop functionals on the moduli space of SL(2,C) and SU(1,1) connections, Class.Quant.Grav. 10 (1993) L69-L74, gr- qc/9304044

\bibitem{I-As4}
A. Ashtekar and J. Lewandowski, Projective techniques and functional integration for gauge theories, J.Math.Phys. 36 (1995) 2170-2191, gr-qc/9411046
  
 \bibitem{I-As5}
A. Ashtekar and J. Lewandowski, Differential geometry on the space of connections via graphs and projective limits, J.Geom.Phys. 17 (1995) 191-230, hep-th/9412073
    
\bibitem{I-As6}
A. Ashtekar, J. Lewandowski, D. Marolf, J. Mourao and T. Thiemann, Quantiza- tion of diffeomorphism invariant theories of connections with local degrees of freedom, J.Math.Phys. 36 (1995) 6456-6493, gr-qc/9504018

  \bibitem{I-RovSmo2}
C. Rovelli and L. Smolin, Spin networks and quantum gravity, Phys.Rev. D52 (1995) 5743-5759, gr-qc/9505006 

  \bibitem{I-RovSmo3}
C. Rovelli and L. Smolin, Discreteness of area and volume in quantum gravity, Nov 1994, Nucl.Phys. B442 (1995) 593-622, Nucl.Phys. B456 (1995) 753, gr-qc/9411005 


\bibitem{I-ThRC}
T. Thiemann, Reality conditions inducing transforms for quantum gauge field theory and quantum gravity, Class. Quant. Grav. Class.Quant.Grav. 13 (1996) 1383-1404, arXiv:gr-qc/9511057v1.

\bibitem{I-RovBH}
C. Rovelli, Black hole entropy from loop quantum gravity, Phys.Rev.Lett. 77 (1996) 3288-3291, gr-qc/9603063 

\bibitem{I-ThHC1}
T. Thiemann, Anomaly - free formulation of nonperturbative, four-dimensional Lorentzian quantum gravity, Jun 1996. Phys.Lett. B380 (1996) 257-264, gr-qc/9606088

\bibitem{I-ThHC2}
T. Thiemann, Quantum spin dynamics (QSD), Jun 1996. Class.Quant.Grav. 15 (1998) 839-873, gr-qc/9606089

\bibitem{I-ThHC3}
T. Thiemann, Quantum spin dynamics (qsd). 2. Jun 1996. 27 pp. Class.Quant.Grav. 15 (1998) 875-905, gr-qc/9606090 

\bibitem{I-Ooguri}
H. Ooguri, Topological lattice models in four-dimensions, 1992. Mod.Phys.Lett. A7 (1992) 2799-2810, hep-th/9205090 

\bibitem{I-RovReis1}
M. Reisenberger and C. Rovelli, Sum over surfaces form of loop quantum gravity,1996. Phys.Rev. D56 (1997) 3490-3508, gr-qc/9612035

\bibitem{I-RovReis2}
M. Reisenberger and C. Rovelli, Space-time as a Feynman diagram: The Connection formulation, 2000. Class.Quant.Grav. 18 (2001) 121-140, gr-qc/0002095 

\bibitem{I-RovReis3}
M. Reisenberger and C. Rovelli, Spin foams as Feynman diagrams, 2000. Conference: C01-09-03.3, p.431-448 Proceedings, gr-qc/0002083

\bibitem{I-BarCrane1}
J. W. Barrett and L. Crane, A Lorentzian signature model for quantum general relativity,1999. Class.Quant.Grav. 17 (2000) 3101-3118, gr-qc/9904025

\bibitem{I-BarCrane2}
J. W. Barrett and L. Crane, Relativistic spin networks and quantum gravity, 1997. J.Math.Phys. 39 (1998) 3296-3302, gr-qc/9709028 

\bibitem{I-SSThKas1}
T. Thiemann, H.A. Kastrup, Canonical quantization of spherically symmetric gravity in Ashtekar's selfdual representation, 1992. Nucl.Phys. B399 (1993) 211-258, gr-qc/9310012 

\bibitem{I-SSThKas2}
T. Thiemann, H.A. Kastrup, Spherically symmetric gravity as a completely integrable system, 1993. Nucl.Phys. B425 (1994) 665-686, gr-qc/9401032

\bibitem{I-SSBoAs}
A. Ashtekar and M. Bojowald, Quantum geometry and the Schwarzschild singularity, 2005. Class.Quant.Grav. 23 (2006) 391-411, gr-qc/0509075 

\bibitem{I-LQCBo1}
M. Bojowald, Absence of singularity in loop quantum cosmology, 2001. Phys.Rev.Lett. 86 (2001) 5227-5230, gr-qc/0102069 

\bibitem{I-LQCBo2}
M. Bojowald, Loop quantum cosmology. I. Kinematics, 1999. Class.Quant.Grav. 17 (2000) 1489-1508, gr-qc/9910103 

\bibitem{I-Var}
M. Varadarajan, Fock representations from U(1) holonomy algebras, Phys. Rev.D61,104001 (2000)

\bibitem{I-AsLew1}
A. Ashtekar and J. Lewandowski, Relation between polymer and Fock excitations, 2001. Class.Quant.Grav. 18 (2001) L117-L128, gr-qc/0107043

\bibitem{I-PolFP}
A. Ashtekar and J. L. Willis, Quantum Gravity, shadow states and quantum mechanics, 2002. Class.Quant.Grav. 20 (2003) 1031-1062, gr-qc/0207106v3

\bibitem{I-PolCorichi1}
A. Corichi , T. Vukasinac and J. A. Zapata, Hamiltonian and physical Hilbert space in polymer quantum mechanics, 2007. Class.Quant.Grav. 24 (2007) 1495-1512, gr-qc/0610072

\bibitem{I-PolCorichi2}
A. Corichi , T. Vukasinac and J. A. Zapata, Polymer quantum mechanics and its continuum limit, 2007. Phys.Rev. D76. (2007) 044016, gr-qc/0704.0007

\bibitem{I-LOST}
J. Lewandowski, A. Okolow, H. Sahlmann and T. Thiemann, Uniqueness of diffeomorphism invariant states on holonomy-flux algebras, (2005). Commun.Math.Phys. 267 (2006) 703-733, gr-qc/0504147

\bibitem{I-Flei}
C. Fleishchack, Representations of the Weyl algebra in quantum geometry, Commun. Math.Phys. 285, 67-140 (2009)

\bibitem{I-NouiSF} 
K. Noui and A. Perez, Three-dimensional loop quantum gravity: Physical scalar product and spin foam models, (2004). Class.Quant.Grav. 22 (2005) 1739-1762, gr-qc/0402110

\bibitem{I-ThCS1}
T. Thiemann, Gauge field theory coherent states (GCS): 1. General properties, (2000). Class.Quant.Grav. 18 (2001) 2025-2064, hep-th/0005233

\bibitem{I-ThCS2}
T. Thiemann, Complexifier coherent states for quantum general relativity, (2002).  Class.Quant.Grav. 23 (2006) 2063-2118, gr-qc/0206037 

\bibitem{I-FreiCS}
F. Conrady and L. Freidel, On the semiclassical limit of 4d spin foam models, (2008). Phys.Rev. D78 (2008) 104023, arXiv:0809.2280

\bibitem{I-LivCS}
E.R. Livine and S. Speziale, A new spinfoam vertex for quantum gravity,  Phys. Rev, D76:084028, (2007), arXiv:0705.0674.

\bibitem{I-BianchiCS}
Bianchi, Eugenio, Magliaro, Elena, and Perini, Claudio. 2010a. Coherent spin-networks. Phys. Rev., D82, 24012, arXiv:0912.4054.

\bibitem{I-FK} 
L. Freidel and K. Krasnov, A New Spin Foam Model for 4d Gravity, (2007). Class.Quant.Grav. 25 (2008) 125018, arXiv:0708.1595 [gr-qc]

\bibitem{I-EPRL} 
J. Engle, E. Livine, R. Pereira, and C. Rovelli, LQG vertex with finite Immirzi parameter, Nucl. Phys , B799:136–149, 2008, arXiv:0711.0146

\bibitem{I-Barrett1}
J. W. Barrett, R.J. Dowdall, W. J. Fairbairn, F. Hellmann, R. Pereira, Lorentzian spin foam amplitudes: Graphical calculus and asymptotics, (2009). Class.Quant.Grav. 27 (2010) 165009, arXiv:0907.2440 [gr-qc].

\bibitem{I-Barrett2}
J. W. Barrett, R.J. Dowdall, W. J. Fairbairn, H. Gomes, F. Hellmann, A Summary of the asymptotic analysis for the EPRL amplitude, (2009). AIP Conf.Proc. 1196 (2009) 36, arXiv:0909.1882 [gr-qc]

\bibitem{I-OritiGFT}
D. Oriti, The microscopic dynamics of quantum space as a group field theory, (2011), Conference: C09-08-10.2, p.257-320 Proceedings, arXiv:1110.5606 [hep-th] 

\bibitem{I-CarrozzaPhD}
S. Carrozza, Tensorial methods and renormalization in Group Field Theories, (2013). arXiv:1310.3736 [hep-th] 

\bibitem{I-BianchaCoarse1}
B. Dittrich, F. C. Eckert, M. Martin-Benito, Coarse graining methods for spin net and spin foam models, (2011) ,New J.Phys. 14 (2012) 035008, arXiv:1109.4927 [gr-qc]

\bibitem{I-BianchaCoarse2}
B. Dittrich and M. Martin-Benito and E. Schnetter, Coarse graining of spin net models: dynamics of intertwiners, (2013) , New J.Phys. 15 (2013) 103004, arXiv:1306.2987 [gr-qc]

\bibitem{I-Bon1}
V. Bonzom, Spin foam models and the Wheeler-DeWitt equation for the quantum 4-simplex, (2011). Phys.Rev. D84 (2011) 024009, arXiv:1101.1615 [gr-qc] 

\bibitem{I-Bon2}
V. Bonzom and E. Livine, Generating Functions for Coherent Intertwiners, (2012). Class.Quant.Grav. 30 (2013) 055018, arXiv:1205.5677 [gr-qc]

\bibitem{I-Bon3}
V. Bonzom and L. Freidel, The Hamiltonian constraint in 3d Riemannian loop quantum gravity, (2011). Class.Quant.Grav. 28 (2011) 195006, arXiv:1101.3524 [gr-qc]

\bibitem{I-Bon4}
V. Bonzom, E. Livine and V. Bonzom, Recurrence relations for spin foam vertices, Class.Quant.Grav. 27 (2010) 125002, arXiv:0911.2204 [gr-qc] 

\bibitem{I-BchGei1}
B. Bahr, B. Dittrich and M. Geiller, A new realization of quantum geometry
Benjamin Bahr (2015). arXiv:1506.08571 [gr-qc] 

\bibitem{I-BchGei2}
B. Dittrich and M. Geiller, Flux formulation of loop quantum gravity: Classical framework, (2014). Class.Quant.Grav. 32 (2015) 13, 135016, arXiv:1412.3752 [gr-qc]

\bibitem{I-TGFrei1}
L. Freidel and S. Speziale,Twisted geometries: A geometric parametrisation of SU(2) phase space, (2010). Phys.Rev. D82 (2010) 084040, arXiv:1001.2748 [gr-qc]

\bibitem{I-TGFrei2}
L. Freidel and S. Speziale, From twistors to twisted geometries, (2010). Phys.Rev. D82 (2010) 084041, arXiv:1006.0199 [gr-qc] 

\bibitem{I-SGFrei}
L. Freidel and J. Ziprick, Spinning geometry = Twisted geometry, (2013). Class.Quant.Grav. 31 (2014) 4, 045007, arXiv:1308.0040 [gr-qc] 

\bibitem{I-Twistor1}
S. Speziale and W. M. Wieland, The twistorial structure of loop-gravity transition amplitudes, (2012). Phys.Rev. D86 (2012) 124023, arXiv:1207.6348 [gr-qc]

\bibitem{I-TwistorAction}
W. M. Wieland, A new action for simplicial gravity in four dimensions, (2014). Class.Quant.Grav. 32 (2015) 1, 015016, arXiv:1407.0025 [gr-qc] 

 \bibitem{I-AshBH1}	
A. Ashtekar, C. Beetle and S.Fairhurst, Isolated horizons: A Generalization of black hole mechanics, (1998). Class.Quant.Grav. 16 (1999) L1-L7, gr-qc/9812065

 \bibitem{I-AshBH2}
A. Ashtekar, C. Beetle and B. Krishnan, Isolated horizons: Hamiltonian evolution and the first law, (2000). Phys.Rev. D62 (2000) 104025, gr-qc/0005083

 \bibitem{I-AshBH3} 	
A. Ashtekar, C. Beetle, O. Dreyer, S. Fairhurst and B. Krishnan, Isolated horizons and their applications, (2000).  Phys.Rev.Lett. 85 (2000) 3564-3567, gr-qc/0006006 

 \bibitem{I-AshBH4}
A. Ashtekar, J. C. Baez and K. Krasnov, Quantum geometry of isolated horizons and black hole entropy, (2000). Adv.Theor.Math.Phys. 4 (2000) 1-94, gr-qc/0005126

 \bibitem{I-AlejBH1}	
Jonathan Engle, Karim Noui, Alejandro Perez, Daniele Pranzetti, Black hole entropy from an SU(2)-invariant formulation of Type I isolated horizons, (2010). Phys.Rev. D82 (2010) 044050, arXiv:1006.0634 [gr-qc] 

 \bibitem{I-AlejBH2}
Jonathan Engle, Alejandro Perez and Karim Noui, Black hole entropy and SU(2) Chern-Simons theory, (2009). Phys.Rev.Lett. 105 (2010) 031302, arXiv:0905.3168 [gr-qc]

 \bibitem{I-AlejBH3}
J. Engle, K. Noui, A. Perez, D. Pranzetti,The SU(2) Black Hole entropy revisited, (2011). JHEP 1105 (2011) 016, arXiv:1103.2723 [gr-qc] 

 \bibitem{I-Fro1}
E. Frodden, A. Ghosh and A. Perez, Quasilocal first law for black hole thermodynamics, (2011). Phys.Rev. D87 (2013) 12, 121503, arXiv:1110.4055 [gr-qc] 

 \bibitem{I-Amit1}
A. Ghosh and A . Perez, Black hole entropy and isolated horizons thermodynamics, (2011). Phys.Rev.Lett. 108 (2012) 16990, arXiv:1107.1320 [gr-qc] 

 \bibitem{I-Amit2}
A. Ghosh and A . Perez, The scaling of black hole entropy in loop quantum gravity, (2012). arXiv:1210.2252 [gr-qc]

 \bibitem{I-Amit3}
A. Ghosh, K. Noui and A. Perez, Statistics, holography, and black hole entropy in loop quantum gravity, (2013). Phys.Rev. D89 (2014) 8, 084069, arXiv:1309.4563 [gr-qc]

 \bibitem{I-BHOriti}
D. Oriti, D. Pranzetti and L. Sindoni, Entropy of isolated horizons from quantum gravity condensates, (2015), arxiv:1510.06991[gr-qc]

 \bibitem{I-Asin1}
O. Asin, J. Ben Achour, M. Geiller, K. Noui and A. Perez, Black holes as gases of punctures with a chemical potential: Bose-Einstein condensation and logarithmic corrections to the entropy, (2014). Phys.Rev. D91 (2015) 084005, arXiv:1412.5851 [gr-qc] 

 \bibitem{I-Fro2}
E. Frodden, M. Geiller, K. Noui and A. PerezStatistical Entropy of a BTZ Black Hole from Loop Quantum Gravity, (2012). JHEP 1305 (2013) 139, arXiv:1212.4473 [gr-qc]

 \bibitem{I-Fro3}
E. Frodden, M. Geiller, K. Noui and A. Perez, Black Hole Entropy from complex Ashtekar variables, (2012). Europhys.Lett. 107 (2014) 10005, arXiv:1212.4060 [gr-qc]


 \bibitem{I-BA1}
J. Ben Achour, A. Mouchet and K. Noui, Analytic Continuation of Black Hole Entropy in Loop Quantum Gravity, (2014). JHEP 1506 (2015) 145, arXiv:1406.6021 [gr-qc]

 \bibitem{I-Muxin} 
 M. Han, Black Hole Entropy in Loop Quantum Gravity, Analytic Continuation, and Dual Holography, (2014). arXiv:1402.2084 [gr-qc] 

 \bibitem{I-Pranz1}
D. Pranzetti, Geometric temperature and entropy of quantum isolated horizons, (2013). Phys.Rev. D89 (2014) 10, 104046, arXiv:1305.6714 [gr-qc]

 \bibitem{I-Pranz2}
D. Pranzetti and H. Sahlmann, Horizon entropy with loop quantum gravity methods, (2014). Phys.Lett. B746 (2015) 209-216, arXiv:1412.7435 [gr-qc]

 \bibitem{I-Pranz3}
A. Ghosh and D. Pranzetti, CFT/Gravity Correspondence on the Isolated Horizon, (2014). Nucl.Phys. B889 (2014) 1-24, arXiv:1405.7056 [gr-qc]

 \bibitem{I-Neiman1}
Y. Neiman, Imaginary part of the gravitational action at asymptotic boundaries and horizons, (2013). Phys.Rev. D88 (2013) 2, 024037, arXiv:1305.2207 [gr-qc]

 \bibitem{I-Neiman2}
Y. Neiman, The imaginary part of the gravity action and black hole entropy, (2013). JHEP 1304 (2013) 071, arXiv:1301.7041 [gr-qc]

 \bibitem{I-BA2}
J. Ben Achour and K. Noui, Analytic continuation of real Loop Quantum Gravity: Lessons from black hole thermodynamics (2015).
Conference: C14-07-15.1, arXiv:1501.05523 [gr-qc] 

 \bibitem{I-BA3}
J. Ben Achour, M. Geiller, K. Noui and C. Yu, Spectra of geometric operators in three-dimensional loop quantum gravity: From discrete to continuous, (2013). Phys.Rev. D89 (2014) 064064, arXiv:1306.3246 [gr-qc] | PDF

 \bibitem{I-BA4}
J. Ben Achour, M. Geiller, K. Noui and C. Yu, Testing the role of the Barbero-Immirzi parameter and the choice of connection in Loop Quantum Gravity, (2013). Phys.Rev. D91 (2015) 104016, arXiv:1306.3241 [gr-qc] 

 \bibitem{I-AsLQC1}
A. Ashtekar and P. Singh, Loop Quantum Cosmology: A Status Report, (2011).Class.Quant.Grav. 28 (2011) 213001, arXiv:1108.0893 [gr-qc]

 \bibitem{I-IvLQC1}
I. Agullo and A. Corichi, Loop Quantum Cosmology, (2013). arXiv:1302.3833 [gr-qc]

 \bibitem{I-IDLQC}
A. Ashtekar, T. Pawlowski and P. Singh, Quantum Nature of the Big Bang: Improved dynamics, (2006). Phys.Rev. D74 (2006) 084003, gr-qc/0607039 

 \bibitem{I-BianchiLQC1}
A. Ashtekar and E. Wilson-Ewing, Loop quantum cosmology of Bianchi I models, (2009). Phys.Rev. D79 (2009) 083535, arXiv:0903.3397 [gr-qc]

 \bibitem{I-BianchiLQC2} 	
A. Ashtekar and E. Wilson-Ewing, Loop quantum cosmology of Bianchi type II models, (2009). Phys.Rev. D80 (2009) 123532, arXiv:0910.1278 [gr-qc] 

 \bibitem{I-BianchiLQC3}
E. Wilson-Ewing, Loop quantum cosmology of Bianchi type IX models, (2010). Phys.Rev. D82 (2010) 043508, arXiv:1005.5565 [gr-qc]

 \bibitem{I-GowdyLQC1}
M. Martin-Benito, G. A. Mena Marugan and E. Wilson-Ewing, Hybrid Quantization: From Bianchi I to the Gowdy Model, (2010). Phys.Rev. D82 (2010) 084012, arXiv:1006.2369 [gr-qc]

 \bibitem{I-LQCUC}
A. Ashtekar, T. Pawlowski, P. Singh and K. Vandersloot, Loop quantum cosmology of k=1 FRW models, (2006). Phys.Rev. D75 (2007) 024035, gr-qc/0612104

 \bibitem{I-LQCOU}
K. Vandersloot, Loop quantum cosmology and the k = - 1 RW model, (2006). Phys.Rev. D75 (2007) 023523, gr-qc/0612070

 \bibitem{I-LQCRad}
T. Pawlowski, R. Pierini and E. Wilson-Ewing, Loop quantum cosmology of a radiation-dominated flat FLRW universe, (2014). Phys.Rev. D90 (2014) 12, 123538, arXiv:1404.4036 [gr-qc]

 \bibitem{I-LQCDust}
E. Wilson-Ewing, The Matter Bounce Scenario in Loop Quantum Cosmology, (2012). JCAP 1303 (2013) 026, arXiv:1211.6269 [gr-qc]

 \bibitem{I-LQCRob}
A. Ashtekar, A. Corichi and P. Singh, Robustness of key features of loop quantum cosmology, (2007). Phys.Rev. D77 (2008) 024046, arXiv:0710.3565 [gr-qc] 

 \bibitem{I-EteraLQC}
E. R. Livine and M. Martin-Benito, Group theoritical Quantization of Isotropic Loop Cosmology, (2012). Phys.Rev. D85 (2012) 124052, arXiv:1204.0539 [gr-qc]


 \bibitem{I-Dressedmetric}
I. Agullo, A. Ashtekar and W. Nelson, The pre-inflationary dynamics of loop quantum cosmology: Confronting quantum gravity with observations  Class.Quant.Grav. 30 (2013) 085014, arXiv:1302.0254 [gr-qc] 

 \bibitem{I-LQCDA1}
T. Cailleteau, J. Mielczarek, A. Barrau and J. Grain, Anomaly-free scalar perturbations with holonomy corrections in loop quantum cosmology, Class.Quant.Grav. 29 (2012) 095010, arXiv:1111.3535 [gr-qc] 

 \bibitem{I-LQCDA2}
A. Barrau, M. Bojowald, G. Calcagni, J. Grain and M. Kagan, Anomaly-free cosmological perturbations in effective canonical quantum gravity,  JCAP 1505 (2015) 05, 051, arXiv:1404.1018 [gr-qc]

 \bibitem{I-LQCProbInf1}
A. Ashtekar and D. Sloan, Probability of Inflation in Loop Quantum Cosmology, Gen.Rel.Grav. 43 (2011) 3619-3655, arXiv:1103.2475 [gr-qc]

 \bibitem{I-LQCProbInf2}
A. Ashtekar and D. Sloan, Loop quantum cosmology and slow roll inflation, Phys.Lett. B694 (2010) 108-112, arXiv:0912.4093 [gr-qc]

 \bibitem{I-SFLQC1}
A. Ashtekar, M. Campiglia and A. Henderson, Casting Loop Quantum Cosmology in the Spin Foam Paradigm, Class.Quant.Grav. 27 (2010) 135020, arXiv:1001.5147 [gr-qc] 

 \bibitem{I-DSLQCKev}
K. Vandersloot, On the Hamiltonian constraint of loop quantum cosmology, Phys.Rev. D71 (2005) 103506, gr-qc/0502082

 \bibitem{I-DSLQCNoui}
J. Ben Achour, J. Grain and K. Noui, Loop Quantum Cosmology with Complex Ashtekar Variables, Class.Quant.Grav. 32 (2015) 025011, arXiv:1407.3768 [gr-qc]

 \bibitem{I-SDLQC}
E. Wilson-Ewing, Loop quantum cosmology with self-dual variables, (2015). arXiv:1503.07855 [gr-qc]

 \bibitem{I-SFLQCAs}
A. Ashtekar, M. Campiglia and A. Henderson, Casting Loop Quantum Cosmology in the Spin Foam Paradigm, Class.Quant.Grav. 27 (2010) 135020, arXiv:1001.5147 [gr-qc] 

 \bibitem{I-SFLQCRov}
E. Bianchi, C. Rovelli and F. Vidotto, Towards Spinfoam Cosmology, Phys.Rev. D82 (2010) 084035, arXiv:1003.3483 [gr-qc]

 \bibitem{I-LQCGF1}
 S. Gielen, D. Oriti and L. Sindoni, Cosmology from Group Field theory formalism for Quantum Gravity, Phys. Rev. Lett 111 (2013) 3, 031301, [gr-qc]: 1303.3576

 \bibitem{I-LQCGF2}
 G. Calcagni, S. Gielen and D. Oriti, Group field cosmology: a cosmological field theory of quantum geometry, Class.Quant.Grav. 29 (2012) 105005, ArXiv: 1201.4151 [gr-qc]
 
  \bibitem{I-LQCGF3}
S. Gielen, D. Oriti and L. Sindoni, Homogenous cosmologies as group field theory condensates, JHEP 1406 (2014) 013, ArXiv:1311.1238 [gr-qc]

  \bibitem{I-LQCGF4}
D. Oriti, D. Pranzetti, J. P. Ryan and L. Sindoni, Generalized quantum gravity condensates for homogenous geometries and cosmology, (2015), ArXiv:1501.00936 [gr-qc]

 \bibitem{I-ObsLQC1}
A. Barrau and J. Grain, Loop quantum gravity and observations, (2014).  arXiv:1410.1714 [gr-qc] 

 \bibitem{I-ObsLQC2}
A. Ashtekar and A. Barrau, Loop quantum cosmology: From pre-inflationary dynamics to observations, (2015), arXiv:1504.07559 [gr-qc] 

 \bibitem{I-Conne}
 A. Connes and M. Marcolli, A Walk in the noncommutative garden, (2006).  C05-09-11.3, p.1-128 Proceedings, math/0601054 
 
   \bibitem{I-Gam}
R. Gambini, O. Obregon and J. Pullin, Yang-Mills analogs of the Immirzi ambiguity, Phys.Rev. D59 (1999) 047505 gr-qc/9801055
 
  \bibitem{I-RovTh}
C. Rovelli and T. Thiemann, The Immirzi parameter in quantum general relativity, Phys.Rev. D57 (1998) 1009-1014, gr-qc/9705059

 \bibitem{I-Merc1}
S. Mercuri, A Possible topological interpretation of the Barbero--Immirzi parameter, (2009). arXiv:0903.2270 [gr-qc]

 \bibitem{I-Merc2}
S. Mercuri and A. Randono, The Immirzi Parameter as an Instanton Angle, Class.Quant.Grav. 28 (2011) 025001, arXiv:1005.1291 [hep-th]

 \bibitem{I-Merc3}
S. Mercuri and V. Taveras, Interaction of the Barbero-Immirzi Field with Matter and Pseudo-Scalar Perturbations, Phys.Rev. D80 (2009) 104007, arXiv:0903.4407 [gr-qc]

 \bibitem{I-Merc4}
G. Calcagni and S. Mercuri, The Barbero-Immirzi field in canonical formalism of pure gravity, Phys.Rev. D79 (2009) 084004, arXiv:0902.0957 [gr-qc]

 \bibitem{I-Krasnov}
A. Torres-Gomez and K. Krasnov, Remarks on Barbero-Immirzi parameter as a field, Phys.Rev. D79 (2009) 104014, arXiv:0811.1998 [gr-qc]


 \bibitem{I-Rand1}
A. Randono, A Generalization of the Kodama state for arbitrary values of the Immirzi parameter, (2005). gr-qc/0504010 

 \bibitem{I-Rand2}
A. Randono, Generalizing the Kodama state. II. Properties and physical interpretation, (2006).  gr-qc/0611074 

 \bibitem{I-AlejRov}
A. Perez and C. Rovelli, Physical effects of the Immirzi parameter, (2005). Phys.Rev. D73 (2006) 044013, gr-qc/0505081

 \bibitem{I-Merc5}
S. Mercuri, Fermions in Ashtekar-Barbero connections formalism for arbitrary values of the Immirzi parameter, Phys.Rev. D73 (2006) 084016, gr-qc/0601013

 \bibitem{I-FreiImm}
G. de Berredo-Peixoto, L. Freidel, I.L. Shapiro and C.A. de Souza, Dirac fields, torsion and Barbero-Immirzi parameter in Cosmology, JCAP 1206 (2012) 017, arXiv:1201.5423 [gr-qc]

 \bibitem{I-Liv}
C. Charles and E. Livine, Ashtekar-Barbero holonomy on the hyperboloid: Immirzi parameter as a Cut-off for Quantum Gravity, (2015). arXiv:1507.00851 [gr-qc] 

\bibitem{I-Alex1}
S. Alexandrov, SO(4,C)-covariant Ashtekar-Barbero gravity and the Immirzi param- eter, Class. Quant. Grav. 17 4255 (2000), arXiv:gr-qc/0005085.

\bibitem{I-Alex2}
S. Alexandrov and D. V. Vassilevich, Area spectrum in Lorentz-covariant loop grav- ity, Phys. Rev. D 64 044023 (2001), arXiv:gr-qc/0103105.

\bibitem{I-AlexLiv}
S. Alexandrov and E. Livine, SU(2) Loop Quantum Gravity seen form the Covariant Theory, Phys.Rev.D67:044009, (2003), arXiv:gr-qc/0209105


 \end{thebibliography}

\begin{thebibliography}{99}

\bibitem{ch1-Al1}
Albert Einstein, Feldgleichungen der Gravitation (The Field Equations of Gravitation), Preussische Akademie der Wissenschaften, Sitzungsberichte, part 2, p(844-847) (1915).

\bibitem{ch1-Ed1}
Edmund  Bertschinger, Symmetry transformations, the Einstein-Hilbert action, and gauge invariance , Massachussets Institute of Technology, (2002).

\bibitem{ch1-Rov1}
C. Rovelli and M. Gaul ,  Loop quantum gravity and the meaning of diffeomorphism invariance , Towards Quantum Gravity , p (277--324), (2000) , Springer 

\bibitem{ch1-Gour}
E. Gourgoulhon, 3+1 formalism and bases of numerical relativity, arXiv:gr-qc: 0703035v1 (2007)

\bibitem{ch1-ADM}
 Arnowitt, Deser and Misner , The dynamics of general relativity , Gravitation:  An Introduction to Current Research, pp (227-265) Wiley New York, (1962) , arXiv:gr-qc/0405109

\bibitem{ch1-Simone1}
 P. Dona and S. Speziale , Introduction to Loop Quantum Gravity, pre-print (2010) , arXiv:1007.0402

\bibitem{ch1-Th1}
T. Thiemann , Modern Canonical Quantum General Relativity, (2007) , Cambridge University press 

\bibitem{ch1-Zan1}
J. Zanelli , Lecture notes on Chern-Simons (super-) gravities. (2005) , arXiv preprint hep-th/0502193

\bibitem{ch1-Perez1}
D. J. Rezende and A. Perez , Four-dimensional Lorentzian Holst action with topological terms , (2009),  Phys RevD, 79, 6, 064026 

\bibitem{ch1-Peldan1}
P. Peldan , Actions for gravity, with generalizations: A Review , Published in Class.Quant.Grav. 11 (1994) 1087-1132 , DOI: 10.1088/0264-9381/11/5/003 , e-Print: gr-qc/9305011 

 \bibitem{ch1-Sen1} 
 A. Sen, Gravity as a spin system, Phys. Lett. B 119, 89 (1982).
 
 \bibitem{ch1-Ash1}
 A. Ashtekar, New variables for classical and quantum gravity, Phys. Rev. Lett. 57,
2244 (1986).
 
  \bibitem{ch1-Ash2}
A. Ashtekar, New Hamiltonian formulation of general relativity, Phys. Rev. D 36, 1587 (1987).

 \bibitem{ch1-Bar1}
 J. F. Barbero, Real Ashtekar variables for Lorentzian signature space-times, Phys. Rev. D 51 5507 (1995), arXiv:gr-qc/9410014
 
\bibitem{ch1-Imm1}
G. Immirzi, Real and complex connections for canonical gravity, Class. Quant. Grav. 14 L177 (1997), arXiv:gr-qc/9612030.

\bibitem{ch1-RCTH}
T. Thiemann, Reality conditions inducing transforms for quantum gauge field theory and quantum gravity, Class. Quant. Grav. Class.Quant.Grav. 13 (1996) 1383-1404, arXiv:gr-qc/9511057v1.

\bibitem{ch1-ThieRov}
C. Rovelli and T. Thiemann, The Immirzi parameter in Quantum General Relativity, Phys.Rev. D57 (1998) 1009-1014, arXiv:gr-qc/9705059

\bibitem{ch1-Sam1}
J. Samuel, A Lagrangian basis for Ashtekar's reformulation of canonical gravity, Pra- mana J. Phys. 28, L429 (1987).

\bibitem{ch1-Lee1}
 T. Jacobson and L. Smolin, The left-handed connection as a variable for quantum gravity, Phys. Lett. B 196, 39 (1987).
 
 \bibitem{ch1-Lee2}
 T. Jacobson and L. Smolin, Covariant action for Ashtekar's form of canonical gravity, Class. Quant. Grav. 5 583 (1988).

\bibitem{ch1-Holst1}
S. Holst, Barbero's Hamiltonian derived from a generalized Hilbert-Palatini action, Phys. Rev. D 53 5966 (1996), arXiv:gr-qc/9511026.

\bibitem{ch1-Noui1}
M. Geiller and K. Noui, A note on the Holst action, the time gauge and the Barbero-Immirzi parameter General Relativity and Gravitation. Gen.Rel.Grav. 45 (2013) 1733-1760,  arXiv:1212.5064 [gr-qc] 

\bibitem{ch1-Sam2}
J. Samuel, Is Barbero's Hamiltonian formulation a gauge theory of Lorentzian gravity ?, Class. Quant. Grav. 17 L141 (2000), arXiv:gr-qc/0005095.

\bibitem{ch1-Alex1}
S. Alexandrov, SO(4,C)-covariant Ashtekar-Barbero gravity and the Immirzi param- eter, Class. Quant. Grav. 17 4255 (2000), arXiv:gr-qc/0005085.

\bibitem{ch1-Alex2}
S. Alexandrov and D. V. Vassilevich, Area spectrum in Lorentz-covariant loop grav- ity, Phys. Rev. D 64 044023 (2001), arXiv:gr-qc/0103105.

\bibitem{ch1-Alex3}
S. Alexandrov, On choice of connection in loop quantum gravity, Phys. Rev. D 65 024011 (2001), arXiv:gr-qc/0107071.

\bibitem{ch1-Alex4}
S. Alexandrov, Hilbert space structure of covariant loop quantum gravity, Phys. Rev. D 66 024028 (2002), arXiv:gr-qc/0201087.

\bibitem{ch1-Alex6}
S. Alexandrov and E. Livine, SU(2) Loop Quantum Gravity seen form the Covariant Theory, Phys.Rev.D67:044009, (2003), arXiv:gr-qc/0209105

\bibitem{ch1-Alex5}
S. Alexandrov, Reality conditions for Ashtekar gravity from Lorentz-covariant formulation, Class.Quant.Grav. 23 (2006) 1837-1850 e-Print: gr-qc/0510050

\bibitem{ch1-WT2}
Guillermo A. Mena Marugan, Geometric interpretation of Thiemann's generalized Wick transform, Apr 1997. 14 pp. Grav.Cosmol. 4 (1998) 257, e-Print: gr-qc/9705031 

\bibitem{ch1-Ash3}
A. Ashtekar, A Generalized Wick Transform for Gravity, Phys.Rev. D53 (1996) 2865-2869 , e-Print: gr-qc/9511083

\bibitem{ch1-R1}
H A. Morales-Tecotl , L F. Urrutia, J. David Vergara,  Reality conditions for Ashtekar variables as Dirac constraints,  May 1996. 16 pp. Class.Quant.Grav. 13 (1996) 2933-2940, e-Print: gr-qc/9607044

\bibitem{ch1-R2} 
J.M. Pons, D.C. Salisbury, L.C. Shepley, Gauge group and reality conditions in Ashtekar's complex formulation of canonical gravity, Dec 1999. 17 pp. Phys.Rev. D62 (2000) 064026, e-Print: gr-qc/9912085 

\bibitem{ch1-R3}
G. Yoneda and H. Shinkai, Constraints and Reality Conditions in the Ashtekar Formulation of General Relativity, Class.Quant.Grav. 13 (1996) 783-790, arXiv:gr-qc/9602026

\bibitem{ch1-R4}
G. Yoneda and H. Shinkai, Lorentzian dynamics in the Ashtekar gravity, (1997) arXiv:gr-qc/9710074

\bibitem{ch1-R5}
W M. Wieland, Complex Ashtekar Variables and Reality Conditions for Holst's Action, (2012) arXiv:gr-qc/1012.1738v2

\bibitem{ch1-Th3}
T. Thiemann, H.A.  Kastrup, Canonical quantization of spherically symmetric gravity in Ashtekar's selfdual representation, (1992) Nucl.Phys. B399 (1993) 211-258, e-Print: gr-qc/9310012 

\bibitem{ch1-Th4}
T. Thiemann,  Canonical quantization of a minisuperspace model for gravity using selfdual variables , (1994) Proceedings of Conference: C93-12-12, e-Print: gr-qc/9910012 

\bibitem{ch1-BA1}
J. Ben Achour and K. Noui, Analytic continuation of real Loop Quantum Gravity: lessons from black hole thermodynamics, (2015) Proceeding of Conference: C14-07-15.1, e-Print: arXiv:1501.05523 [gr-qc] 

\bibitem{ch1-BA2}
J. Ben Achour, A. Mouchet and K. Noui,  Analytic Continuation of Black Hole Entropy in Loop Quantum Gravity  (2014) JHEP, e-Print: arXiv:1406.6021 [gr-qc] 

\bibitem{ch1-BA3}
J. Ben Achour, J. Grain and K. Noui,  Loop Quantum Cosmology with complex Ashtekar variables (2014) Class.Quant.Grav. 32 (2015) 2, 025011, e-Print: arXiv:1407.3768 [gr-qc]

\bibitem{ch1-BA4}
J. Ben Achour, M. Geiller, K. Noui and C. Yu,  Testing the role of the Barbero-Immirzi parameter and the choice of connection in Loop Quantum Gravity, Phys.Rev. D91 (2013) 10, 104016, e-Print: arXiv:1306.3241 [gr-qc] 

\bibitem{ch1-BA5}
J. Ben Achour, M. Geiller, K. Noui and C. Yu,  Spectra of geometric operators in three-dimensional loop quantum gravity: From discrete to continuous, Phys.Rev. D89 (2013) 6, 064064, e-Print: arXiv:1306.3246 [gr-qc] 

\bibitem{ch1-Witten1}
E. Witten, Analytic Continuation Of Chern-Simons Theory, (2010) arXiv:1001.2933 [hep-th]




 \end{thebibliography}

\begin{thebibliography}{99}

\bibitem{ch2-Th1} T. Thiemann, Modern Canonical Quantum General Relativity, (2007), Cambridge University Press 

\bibitem{ch2-Rove1} C. Rovelli, Quantum Gravity, (2004), Cambridge University Press

\bibitem{ch2-Rove2} C. Rovelli and F. Vidotto, Covariant Loop Quantum Gravity: an elementary introduction to Spinfoam Theory, (2013) Cambridge University Press

\bibitem{ch2-Alejandro1} A. Perez, Introduction to loop quantum gravity and spin foams. (2004), arXiv: gr-qc/0409061.

\bibitem{ch2-Hanno1} K. Giesel and H. Sahlmann, From classical to quantum gravity: Introduction to loop quantum gravity. (2012), arXiv:1203.2733.

\bibitem{ch2-Jerzy1} A. Ashtekar and J. Lewandowski, Background independent quantum gravity: a status report, Class. Quant. Grav. 21 R53 (2004), arXiv:gr-qc/0404018.

\bibitem{ch2-Mat1} H. J. Matschull, Dirac's canonical quantization programme. (1996). arXiv: quant-ph/9606031.

\bibitem{ch2-GNS} A. Jaffe, The GNS construction (2014), harvard.edu/fs/docs/icb.topic1466080.files/GNS.pdf

\bibitem{ch2-Br1} M P. Bronstein, Kvantovanie gravitatsionnykh voln (Quantization of Gravitational Waves). Zh. Eksp. Tear. Fiz., 6, 195. (1936)

\bibitem{ch2-Br2} M P. Bronstein, Quantentheorie schwacher Gravitationsfelder. Phys. Z. Sowjetunion, 9, 140-157 (1936)

\bibitem{ch2-BCKI} L. Smolin, The case for background independence, Rickles, D. (ed.) et al.: The structural foundations of quantum gravity* 196-239, hep-th/0507235

\bibitem{ch2-Rove3} M. Gaul and C. Rovelli, Loop Quantum Gravity and the meaning of the diffeomorphism invariance, Lect.Notes Phys. 541 (2000) 277-324, C99-02-02.2, p.277-324 Proceedings, gr-qc/9910079 

\bibitem{ch2-Ed1} Edmund Bertschinger, Symmetry transformations, the Einstein-Hilbert action, and gauge invariance , Massachussets Institute of Technology, (2002).

\bibitem{ch2-Frei1} L. Freidel, M. Geiller and J. Ziprick, Continuous formulation of Loop Quantum Gravity phase space, Class.Quant.Grav. 30 (2013) 085013, arXiv:1110.4833 [gr-qc] 

\bibitem{ch2-LOST} J. Lewandowski, A. Okolow, H. Sahlmann, T. Thiemann, Uniqueness of diffeomorphism invariant states on holonomy-flux algebras, Commun.Math.Phys.267:703-733, (2006), arXiv:gr-qc/0504147

\bibitem{ch2-Fl1} C. Fleichhack, Representations of the Weyl algebra in quantum geometry, Commun. Math. Phys. 285 67 (2009), arXiv:math-ph/0407006.

\bibitem{ch2-KosRep} 
T. A. Koslowski, Dynamical Quantum Geometry (DQG Programme), arXiv:0709.3465 [gr-qc]

\bibitem{ch2-HannoRep} 
H. Sahlmann, On loop quantum gravity kinematics with non-degenerate spatial background, (2010). Class.Quant.Grav. 27 (2010) 225007, arXiv:1006.0388 [gr-qc] 

\bibitem{ch2-VarRep} 
M. Varadarajan, Towards new background independent representations for Loop Quan-tum Gravity, Class. Quant. Grav.25 (2008) 105011 [arXiv:0709.1680 [gr-qc]]

\bibitem{ch2-Ashtekar} A. Ashtekar, Some surprising implications of background independence in canonical quantum gravity, Gen.Rel.Grav. 41 (2009) 1927-1943, arXiv:0904.0184 [gr-qc]

\bibitem{ch2-Geiller1} B. Dittrich and M. Geiller, A new vacuum for Loop Quantum Gravity, Class.Quant.Grav. 32 (2015) 11, 11200, arXiv:1401.6441 [gr-qc] 

\bibitem{ch2-Geiller2} B. Dittrich and M. Geiller, Flux formulation of loop quantum gravity: Classical framework , Class.Quant.Grav. 32 (2015) 13, 135016, arXiv:1412.3752 [gr-qc] 

\bibitem{ch2-Geiller3} B. Bahr, B. Dittrich and M. Geiller, A new realization of quantum geometry, arXiv:1506.08571 [gr-qc]

\bibitem{ch2-Rove6} Carlo Rovelli  and Lee Smolin, Loop Space Representation of Quantum General Relativity, (1989), Nucl.Phys. B331 (1990) 80-152

\bibitem{ch2-Ashte1} A. Ashtekar and C. J. Isham, Representations of the holonomy algebras of gravity and nonAbelian gauge theories, Class.Quant.Grav. 9 (1992) 1433-1468,  hep-th/9202053

\bibitem{ch2-Ashte2} A. Ashtekar and J. Lewandowski, Representations theory of analytic holonomy C* algebras, 1993, CGPG-93-8-1, gr-qc/9311010 

\bibitem{ch2-Ashte3} A. Ashtekar and J. Lewandowski, Completeness of Wilson loop functionals on the moduli space of SL(2,C) and SU(1,1) connections, Class.Quant.Grav. 10 (1993) L69-L74, gr-qc/9304044

\bibitem{ch2-Ashte4} A. Ashtekar and J. Lewandowski, Projective techniques and functional integration for gauge theories, J.Math.Phys. 36 (1995) 2170-2191, gr-qc/9411046 

\bibitem{ch2-Ashte5} A. Ashtekar and J. Lewandowski, Differential geometry on the space of connections via graphs and projective limits, J.Geom.Phys. 17 (1995) 191-230, hep-th/9412073

\bibitem{ch2-Ashte6} A. Ashtekar, J. Lewandowski, D. Marolf, J. Mourao and T. Thiemann, Quantization of diffeomorphism invariant theories of connections with local degrees of freedom, J.Math.Phys. 36 (1995) 6456-6493, gr-qc/9504018

\bibitem{ch2-Ashte7} A. Ashtekar and C. Rovelli, A Loop representation for the quantum Maxwell field, Class.Quant.Grav. 9 (1992) 1121-1150,hep-th/9202063

\bibitem{ch2-Ashte8} A. Ashtekar, Carlo Rovelli  and Lee Smolin, Gravitons and loops,  Phys.Rev. D44 (1991) 1740-1755, hep-th/9202054 

\bibitem{ch2-Ashte9} A. Ashtekar, J. Lewandowski and H. Sahlmann, Polymer and Fock representations for a scalar field, Class.Quant.Grav. 20 (2003) L11-1, gr-qc/0211012 

\bibitem{ch2-Ashte10} A. Ashtekar and J. Lewandowski, Projective techniques and functional integration for gauge theories, J.Math.Phys. 36 (1995) 2170-2191, CGPG-94-10-6, gr-qc/9411046

\bibitem{ch2-Edmond1} Edmunds

\bibitem{ch2-Smolin2} Carlo Rovelli  and Lee Smolin, Spin networks and quantum gravity, (1995), Phys.Rev. D52 (1995) 5743-5759, CGPG-95-4-4, IASSNS-HEP-95-27, gr-qc/9505006 

\bibitem{ch2-Ga1} D. Giulini and D. Marolf, On the generality of refined algebraic quantization, Class. Quant. Grav. 16 2479 (1999), arXiv:gr-qc/9812024.

\bibitem{ch2-Rovelli} W. Fairbairn and C. Rovelli, Separable Hilbert space in loop quantum gravity, J. Math. Phys. 45 2802 (2004), arXiv:gr-qc/0403047.

\bibitem{ch2-HCTh1} 
 T. Thiemann, Anomaly - free formulation of nonperturbative, four-dimensional Lorentzian quantum gravity, Jun 1996. Phys.Lett. B380 (1996) 257-264, gr-qc/9606088
 
 \bibitem{ch2-HCTh2} 
  T. Thiemann, Quantum spin dynamics (QSD), Jun 1996. Class.Quant.Grav. 15 (1998) 839-873, gr-qc/9606089
 
 \bibitem{ch2-HCTh3} 
 T. Thiemann, Quantum spin dynamics (qsd). 2. Jun 1996. 27 pp. Class.Quant.Grav. 15 (1998) 875-905, gr-qc/9606090
 
 \bibitem{ch2-MCTh1}
T. Thiemann, The Phoenix project: Master constraint program for loop quantum gravity, Class.Quant.Grav. 23 (2006) 2211-2248, gr-qc/0305080 
 
  \bibitem{ch2-MCTh2}  
B. Dittrich and T. Thiemann, Testing the master constraint programme for loop quantum gravity. I. General framework, Class.Quant.Grav. 23 (2006) 1025-1066, gr-qc/0411138
  
\bibitem{ch2-MCTh3}
B. Dittrich and T. Thiemann, Testing the master constraint programme for loop quantum gravity. II. Finite dimensional systems, Class.Quant.Grav. 23 (2006) 1067-1088, gr-qc/0411139   
   
 \bibitem{ch2-MCTh4}
T. Thiemann, Quantum spin dynamics. VIII. The Master constraint, Class.Quant.Grav. 23 (2006) 2249-2266, gr-qc/0510011

 \bibitem{ch2-Alex1}   
S. Alexandrov, On the choice of connection in Loop Quantum Gravity, Phys.Rev. D65 (2002) 024011, gr-qc/0107071 

\bibitem{ch2-Samuel} J. Samuel, Is Barbero's Hamiltonian formulation a gauge theory of Lorentzian gravity?, Class. Quant. Grav. 17 L141 (2000), arXiv:gr-qc/0005095.

\bibitem{ch2-Liv}
C. Charles and E. Livine, Ashtekar-Barbero holonomy on the hyperboloid: Immirzi parameter as a Cut-off for Quantum Gravity, (2015), arXiv:1507.00851 [gr-qc]

\bibitem{ch2-BC}  J. W. Barrett and L. Crane, Relativistic spin networks and quantum gravity, J. Math. Phys. 39 3296 (1998), arXiv:gr-qc/9709028.

\bibitem{ch2-FK} L. Freidel and K. Krasnov, A new spin foam model for 4d gravity, Class. Quant. Grav. 25 125018 (2008), arXiv:gr-qc/0708.1595 [gr-qc].

\bibitem{ch2-EPRL} J. Engle, E. R. Livine, R. Pereira and C. Rovelli, LQG vertex with finite Immirzi parameter, Nucl. Phys. B 799 136 (2008), arXiv:0711.0146 [gr-qc]

\bibitem{ch2-Jerzy3} E. Alesci, M. Assanioussi and J. Lewandowski, Curvature operator for loop quantum gravity, Phys.Rev. D89 (2014) 12, 124017, arXiv:1403.3190 [gr-qc] 

\bibitem{ch2-Jerzy4} E. Alesci, M. Assanioussi and J. Lewandowski, New scalar constraint operator for loop quantum gravity, arXiv:1506.00299 [gr-qc] 

\bibitem{ch2-Rove4} Carlo Rovelli  and Lee Smolin, Discreteness of area and volume in quantum gravity, Nucl.Phys. B442 (1995) 593-622, CGPG-94-11-1, gr-qc/9411005 

\bibitem{ch2-Ashte11} A. Ashtekar, J. Lewandowski, Quantum theory of geometry. 1: Area operators, Class.Quant.Grav. 14 (1997) A55-A82, gr-qc/9602046 

\bibitem{ch2-BT1} B. Dittrich and T. Thiemann, Are the spectra of geometrical operators in Loop Quantum Gravity really discrete ? ,  J.Math.Phys. 50 (2009) 012503, arXiv:0708.1721 [gr-qc]

\bibitem{ch2-Ro1} C. Rovelli, Comment on `Are the spectra of geometrical operators in Loop Quantum Gravity really discrete?' by B. Dittrich and T. Thiemann, arXiv:0708.2481 [gr-qc] 

\bibitem{ch2-Rovelli1} C. Rovelli, Partial observables, Phys. Rev. D 65 124013 (2002), arXiv:gr-qc/0110035.

\bibitem{ch2-Tam1} J. Tambornino, Relational observables in gravity: A review, SIGMA 8 017 (2012), arXiv:1109.0740 [gr-qc]

\bibitem{ch2-Bonzom1} V. Bonzom, Quantum geometry in spin foams: From the topological BF theory to General Relativity, (2010). arXiv:1009.5100 [gr-qc]

\bibitem{ch2-Ashte12} A. Ashtekar, J. Lewandowski, Quantum theory of geometry. 2. Volume operators, Adv.Theor.Math.Phys. 1 (1998) 388-429, CGPG-97-11-1, gr-qc/9711031

\bibitem{ch2-J1} J. Brunnemann and D. Rideout, Properties of the volume operator in loop quantum gravity I: Results, Class. Quant. Grav. 25 065001 (2008), arXiv:0706.0469 [gr-qc].

\bibitem{ch2-J2}  J. Brunnemann and D. Rideout, Properties of the volume operator in loop quan- tum gravity II: Detailed presentation, Class. Quant. Grav. 25 065002 (2008), arXiv:0706.0382 [gr-qc].

\bibitem{ch2-Th2}  Johannes Brunnemann and T. Thiemann, Simplification of the spectral analysis of the volume operator in loop quantum gravity, (2004), Class.Quant.Grav. 23 (2006) 1289-1346, gr-qc/0405060 

\bibitem{ch2-L1} T. Thiemann, A length operator for canonical quantum gravity, J. Math. Phys. 39 3372 (1998), arXiv:gr-qc/9606092.

\bibitem{ch2-L2} E. Bianchi, The length operator in loop quantum gravity, Nucl. Phys. B 807 591 (2009), arXiv:0806.4710 [gr-qc].

\bibitem{ch2-L3} Y. Ma, C. Soo and J. Yang, New length operator for loop quantum gravity, Phys. Rev. D 81 124026 (2010), arXiv:1004.1063 [gr-qc].

\bibitem{ch2-TW1} C. Rovelli and S. Speziale, Geometry of loop quantum gravity on a graph, Phys. Rev. D 82 044018 (2010), arXiv:gr-qc/1005.2927.

\bibitem{ch2-TW2} L. Freidel and S. Speziale, Twisted geometries: A geometric parametrisation of SU(2) phase space, Phys. Rev. D 82 084040 (2010), arXiv:1001.2748 [gr-qc].

\bibitem{ch2-TW3}  L. Freidel and S. Speziale, From twistors to twisted geometries, Phys. Rev. D 82 084041 (2010), arXiv:1006.0199 [gr-qc].

\bibitem{ch2-SP1} L. Freidel and J. Ziprick, Spinning geometry = Twisted geometry, Class.Quant.Grav. 31 (2014) 4, 045007, arXiv:1308.0040 [gr-qc]

   \bibitem{ch2-Gamb}
R. Gambini, O. Obregon and J. Pullin, Yang-Mills analogs of the Immirzi ambiguity, Phys.Rev. D59 (1999) 047505 gr-qc/9801055
 
  \bibitem{ch2-RovTh}
C. Rovelli and T. Thiemann, The Immirzi parameter in quantum general relativity, Phys.Rev. D57 (1998) 1009-1014, gr-qc/9705059

 \bibitem{ch2-Merc1}
S. Mercuri, A Possible topological interpretation of the Barbero--Immirzi parameter, (2009). arXiv:0903.2270 [gr-qc]

 \bibitem{ch2-Merc2}
S. Mercuri and A. Randono, The Immirzi Parameter as an Instanton Angle, Class.Quant.Grav. 28 (2011) 025001, arXiv:1005.1291 [hep-th]

 \bibitem{ch2-Merc3}
S. Mercuri and V. Taveras, Interaction of the Barbero-Immirzi Field with Matter and Pseudo-Scalar Perturbations, Phys.Rev. D80 (2009) 104007, arXiv:0903.4407 [gr-qc]

 \bibitem{ch2-Merc4}
G. Calcagni and S. Mercuri, The Barbero-Immirzi field in canonical formalism of pure gravity, Phys.Rev. D79 (2009) 084004, arXiv:0902.0957 [gr-qc]

 \bibitem{ch2-Krasnov}
A. Torres-Gomez and K. Krasnov, Remarks on Barbero-Immirzi parameter as a field, Phys.Rev. D79 (2009) 104014, arXiv:0811.1998 [gr-qc]

 \bibitem{ch2-Rand1}
A. Randono, A Generalization of the Kodama state for arbitrary values of the Immirzi parameter, (2005). gr-qc/0504010 

 \bibitem{ch2-Rand2}
A. Randono, Generalizing the Kodama state. II. Properties and physical interpretation, (2006).  gr-qc/0611074 

 \bibitem{ch2-AlejRov}
A. Perez and C. Rovelli, Physical effects of the Immirzi parameter, (2005). Phys.Rev. D73 (2006) 044013, gr-qc/0505081

 \bibitem{ch2-Merc5}
S. Mercuri, Fermions in Ashtekar-Barbero connections formalism for arbitrary values of the Immirzi parameter, Phys.Rev. D73 (2006) 084016, gr-qc/0601013

 \bibitem{ch2-FreiImm}
G. de Berredo-Peixoto, L. Freidel, I.L. Shapiro and C.A. de Souza, Dirac fields, torsion and Barbero-Immirzi parameter in Cosmology, JCAP 1206 (2012) 017, arXiv:1201.5423 [gr-qc]

 \bibitem{ch2-BA1}
J. Ben Achour, A. Mouchet and K. Noui, Analytic Continuation of Black Hole Entropy in Loop Quantum Gravity, (2014). JHEP 1506 (2015) 145, arXiv:1406.6021 [gr-qc]

 \bibitem{ch2-BA2} 
  J. Ben Achour and K. Noui, Analytic continuation of real Loop Quantum Grav- ity: Lessons from black hole thermodynamics (2015). Conference: C14-07-15.1, arXiv:1501.05523 [gr-qc]
  
   \bibitem{ch2-BA3}
   J. Ben Achour, M. Geiller, K. Noui and C. Yu, Spectra of geometric operators in three- dimensional loop quantum gravity: From discrete to continuous, (2013). Phys.Rev. D89 (2014) 064064, arXiv:1306.3246 [gr-qc] | PDF
   
    \bibitem{ch2-BA4}
    J. Ben Achour, M. Geiller, K. Noui and C. Yu, Testing the role of the Barbero-Immirzi parameter and the choice of connection in Loop Quantum Gravity, (2013). Phys.Rev. D91 (2015) 104016, arXiv:1306.3241 [gr-qc]
    
     \bibitem{ch2-Fro3}
E. Frodden, M. Geiller, K. Noui and A. Perez, Black Hole Entropy from complex Ashtekar variables, (2012). Europhys.Lett. 107 (2014) 10005, arXiv:1212.4060 [gr-qc]
    
     \bibitem{ch2-Muxin} 
 M. Han, Black Hole Entropy in Loop Quantum Gravity, Analytic Continuation, and Dual Holography, (2014). arXiv:1402.2084 [gr-qc] 

 \bibitem{ch2-Pranz1}
D. Pranzetti, Geometric temperature and entropy of quantum isolated horizons, (2013). Phys.Rev. D89 (2014) 10, 104046, arXiv:1305.6714 [gr-qc]

 \bibitem{ch2-Pranz2}
D. Pranzetti and H. Sahlmann, Horizon entropy with loop quantum gravity methods, (2014). Phys.Lett. B746 (2015) 209-216, arXiv:1412.7435 [gr-qc]

 \bibitem{ch2-Pranz3}
A. Ghosh and D. Pranzetti, CFT/Gravity Correspondence on the Isolated Horizon, (2014). Nucl.Phys. B889 (2014) 1-24, arXiv:1405.7056 [gr-qc]

 \bibitem{ch2-Neiman1}
Y. Neiman, Imaginary part of the gravitational action at asymptotic boundaries and horizons, (2013). Phys.Rev. D88 (2013) 2, 024037, arXiv:1305.2207 [gr-qc]

 \bibitem{ch2-Neiman2}
Y. Neiman, The imaginary part of the gravity action and black hole entropy, (2013). JHEP 1304 (2013) 071, arXiv:1301.7041 [gr-qc]

 \bibitem{ch2-Alex2}
S. Alexandrov, SO(4,C) covariant Ashtekar-Barbero gravity and the Immirzi parameter, Class.Quant.Grav. 17 (2000) 4255-4268, gr-qc/0005085

 \bibitem{ch2-Alex3}
S. Alexandrov and D. Vassilevich, Area spectrum in Lorentz covariant loop gravity, Phys.Rev. D64 (2001) 044023, gr-qc/0103105

 \bibitem{ch2-Alex4}
S. Alexandrov and E. Livine, SU(2) loop quantum gravity seen from covariant theory, Phys.Rev. D67 (2003) 044009, gr-qc/0209105 

\end{thebibliography}

\begin{thebibliography}{99}

\bibitem{ch3-Boj1} A. Ashtekar and M. Bojowald, Quantum geometry and the Schwarzschild singularity, (2005). Class.Quant.Grav. 23 (2006) 391-411, IGPG-05-09-01, AEI-2005-132, e-Print: gr-qc/0509075 

\bibitem{ch3-Gam1}  R. Gambini, J. Pullin, An introduction to spherically symmetric loop quantum gravity black holes, (2013). AIP Conf.Proc. 1647 (2015) 19-22, LSU-REL-121913, C13-05-27.4 Proceedings, arXiv:1312.5512 [gr-qc]

\bibitem{ch3-Gam2} 
R. Gambini, J. Pullin, Loop quantization of the Schwarzschild black hole, 2013. Phys.Rev.Lett. 110 (2013) 21, 211301, LSU-REL-022113, arXiv:1302.5265 [gr-qc]

\bibitem{ch3-Gam3}
M. Campiglia, R. Gambini and J. Pullin, Loop quantization of spherically symmetric midi-superspaces : The Interior problem. (2007). AIP Conf.Proc. 977 (2008) 52-63, LSU-REL-120507, arXiv:0712.0817 [gr-qc] 

\bibitem{ch3-Asht1} A. Ashtekar, Interface of general relativity, quantum physics and statistical mechanics: Some recent developments
(1999). Annalen Phys. 9 178-198, CGPG-99-10-3, CGPG-99-8-4,  Conference: C99-09-12.2 Proceedings,  gr-qc/9910101

\bibitem{ch3-Asht2} A. Ashtekar, J. C. Baez and K. Krasnov, Quantum geometry of isolated horizons and black hole entropy, (2000). Adv.Theor.Math.Phys. 4 NSF-ITP-99-153, gr-qc/0005126

\bibitem{ch3-Asht3} A. Ashtekar, C. Beetle and S. Fairhurst, Isolated horizons: A Generalization of black hole mechanics, (1998). Class.Quant.Grav. 16 (1999) L1-L7,  gr-qc/9812065 

\bibitem{ch3-Alej1} A. Ghosh and A. Perez, Black hole entropy and isolated horizons thermodynamics, (2011). Phys.Rev.Lett. 108 169901, arXiv:1107.1320 [gr-qc] 

\bibitem{ch3-Asin1} O. Asin, J. Ben Achour, M. Geiller, K. Noui and A. Perez, Black holes as gases of punctures with a chemical potential: Bose-Einstein condensation and logarithmic corrections to the entropy, (2014), Phys.Rev. D91 (2015) 8, 084005, arXiv:1412.5851 [gr-qc]

\bibitem{ch3-Ashtekar4} A. Ashtekar, S. Fairhurst, B. Krishnan, Isolated horizons: Hamiltonian evolution and the first law, 2000. Phys.Rev. D62 (2000) 104025, gr-qc/0005083 

\bibitem{ch3-Nouiii1} J. Engle, K. Noui, A. Perez and D. Pranzetti, Black hole entropy from an SU(2)-invariant formulation of type I isolated horizons, Phys. Rev. D 82 044050 (2010), arXiv:1006.0634 [gr-qc].

\bibitem{ch3-Frodden1} E. Frodden, On the thermodynamic and Quantum properties of Black Holes, PhD Thesis, University of Aix-Marseille, CPT (2012)

\bibitem{ch3-Hay1} S. Hayward, General laws of black hole dynamics, Phys. Rev. D49, 6467-6474 (1994)

\bibitem{ch3-Asht5} A. Ashtekar and B. Krishnan, Dynamical horizons and their properties, (2003). Phys.Rev. D68 (2003) 104030, CGPG-03-07-3, NSF-KITP-03-57, gr-qc/0308033

\bibitem{ch3-Frei1} L. Freidel, Y. Yokokura, Non-equilibrium thermodynamics of gravitational screens, (2014). arXiv:1405.4881 [gr-qc]

\bibitem{ch3-Frei2} L. Freidel, Gravitational Energy, Local Holography and Non-equilibrium Thermodynamics, (2013). Class.Quant.Grav. 32 (2015) 5, 055005, arXiv:1312.1538 [gr-qc]

\bibitem{ch3-Hawking1} S. Hawking, Particle creation by black holes, Comm. Math. Phys. \textbf{43} 199 (1975).

\bibitem{ch3-Asht6} A. Ashtekar, C. Beetle, Jerzy Lewandowski, Mechanics of rotating isolated horizons, (2001). Phys.Rev. D64 (2001) 044016, gr-qc/0103026 

\bibitem{ch3-Asht7} A. Ashtekar,Christopher Beetle and Jerzy Lewandowski Geometry of generic isolated horizons, (2001). Class.Quant.Grav. 19 (2002) 1195-1225, CGPG-01-11-3, gr-qc/0111067

\bibitem{ch3-Asht8} A. Ashtekar, A. Corichi and K. Krasnov  Isolated horizons: The Classical phase space, (1999). Adv.Theor.Math.Phys. 3 (1999) 419-478, NSF-ITP-99-33, NSF-ITP-9-33, CGPG-99-5-5, gr-qc/9905089 

\bibitem{ch3-Noui2} J. Engle, A. Perez, K. Noui, and D. Pranzetti, Black hole entropy and SU(2) Chern-Simons theory, Phys. Rev. Lett. 105 (2010) 031302 arXiv:gr-qc/0905.3168.

\bibitem{ch3-Rovelli1} C. Rovelli and F. Vidotto, Covariant Loop Quantum Gravity : an elemantery introduction to Quantum Gravity and Spinfoam Theory, Cambridge University Press, (2013)

\bibitem{ch3-CSCPP} G. Grignani  and G. Nardelli, Gravity in 2 + 1 dimensions coupled to point-like sources: a flat Chern-Simons gauge theory equivalent to Einstein's, MIT, Cambridge,
Physics Letters B (1991) 10.1016/0370-2693(91)90701-Q

\bibitem{ch3-PP} M. Wellin, Some approaches to $2+1$ dimensional gravity coupled to point-like particles, arXiv: 9511211, hep-th (1995)

\bibitem{ch3-Noui3} K. Noui, A. Perez, Three-dimensional loop quantum gravity: Coupling to point particles, (2004). Class.Quant.Grav. 22 (2005) 4489-4514, gr-qc/0402111

\bibitem{ch3-Hannno} D. Pranzetti and Hanno Sahlmann, Horizon entropy with loop quantum gravity methods, Phys.Lett. B746 (2015) 209-216, e-Print: arXiv:1412.7435 [gr-qc] , arXiv:1305.6714 [gr- qc].

\bibitem{ch3-Livine1} L. Freidel and E. Livine, Ponzano-Regge model revisited III: Feynman diagrams and effective field theory, Class. Quant. Grav. 23, 2021-2062 (2006)
\bibitem{ch3-Livine2}  L. Freidel and E. Livine, Effective 3D quantum gravity and non-commutative quantum field theory, Phys. Rev. Lett. 96: 221301 (2006)

\bibitem{ch3-Noui4} E. Joung, J. Mourad  and K. Noui Three Dimensional Quantum Geometry and Deformed Poincare Symmetry, (2008). J.Math.Phys. 50 (2009) 052503, arXiv:0806.4121 [hep-th]

\bibitem{ch3-Witten1}
E. Witten,
(2+1)-Dimensional Gravity as an Exactly Soluble System,
Nucl. Phys. \textbf{B 311} 46 (1988).

\bibitem{ch3-Witten2}
E. Witten,
Quantum Field Theory and the Jones polynomial,
Comm. Math. Phys. \textbf{121} 351 (1989).

\bibitem{ch3-QG} V. Chari and A. Pressley, A guide to quantum groups,  Cambridge, UK: Univ. Pr. (1994) 651 p.

\bibitem{ch3-Noui5} J. Engle, K. Noui, A. Perez and D. Pranzetti,
The $\SU(2)$ black hole entropy revisited,
JHEP 1105 (2011), \texttt{arXiv:1103.2723 [gr-qc]}.

\bibitem{ch3-RB} R. Bousso, The holographic principle

\bibitem{ch3-Frodden2} E. Frodden, A. Ghosh, and A. Perez, Quasilocal first law for black hole thermodynamics, Phys.Rev. D87 (2013) 12, 121503, e-Print: arXiv:1110.4055 [gr-qc] 

\bibitem{ch3-Alej2} Amit Ghosh, Alejandro Perez, The scaling of black hole entropy in loop quantum gravity, (2012), arXiv:1210.2252 [gr-qc] 

\bibitem{ch3-Alej3} Amit Ghosh, Karim Noui and Alejandro Perez, Statistics, holography, and black hole entropy in loop quantum gravity, (2013). Phys.Rev. D89 (2014) 8, 084069, arXiv:1309.4563 [gr-qc] 

\bibitem{ch3-Stat} R. K. Pathria, Statistical Mechanics, ISO 160, (1972)

\bibitem{ch3-Anyo1} A G. A. Pithis and H-C. R. Euler, Anyonic statistics and large horizon diffeomorphisms for Loop Quantum Gravity black holes, Phys. Rev. D 91, 064053 (2015), arXiv: 1402.2274v2 [qr-qc]


\bibitem{ch3-Carlip1}  S.Carlip, Logarithmic corrections to black hole entropy from the Cardy formula, Class.Quant.Grav.
17 4175-4186 (2000), arXiv:gr-qc/0005017.

\bibitem{ch3-Das1} S. Das, P. Majumdar and R. K. Bhaduri, General logarithmic corrections to black hole entropy,
Class. Quant. Grav. 19 2355-2368 (2002), arXiv:hep-th/0111001.

\bibitem{ch3-Sen1}  A. Sen, Logarithmic corrections to Schwarzschild and other non-extremal black hole entropy in differ-
ent dimensions, JHEP 4 156 (2013), arXiv:1205.0971 [hep-th].

 \bibitem{ch3-Bousso1} R. Bousso, The Holographic principle, 2002. Rev.Mod.Phys. 74 (2002) 825-874, NSF-ITP-02-17, hep-th/0203101

\bibitem{ch3-Ryu1}  T. Nishioka, S. Ryu and T. Takayanagi, Holographic entanglement entropy: An overview, J. Phys.
A 42, 504008 (2009), arXiv:0905.0932 [hep-th].


\bibitem{ch3-Geiller1}  E. Frodden, M. Geiller, K. Noui and A. Perez, Black hole entropy from complex Ashtekar variables,
EuroPhys. Lett. 107 10005 (2014), arXiv:1212.4060 [gr-qc].

\bibitem{ch3-BA1}
J. Ben Achour, A. Mouchet and K. Noui,  Analytic Continuation of Black Hole Entropy in Loop Quantum Gravity  (2014) JHEP, e-Print: arXiv:1406.6021 [gr-qc] 

\bibitem{ch3-BA2}
J. Ben Achour and K. Noui, Analytic continuation of real Loop Quantum Gravity: lessons from black hole thermodynamics, (2015) Proceeding of Conference: C14-07-15.1, e-Print: arXiv:1501.05523 [gr-qc] 

\end{thebibliography}

\begin{thebibliography}{99}

\bibitem{ch4-Amit1}
E. Frodden, A. Ghosh, and A. Perez, Quasilocal first law for black hole thermodynamics, Phys.Rev. D87 (2013) 12, 121503, e-Print: arXiv:1110.4055 [gr-qc] .

\bibitem{ch4-Ash1}
A. Ashtekar, New variables for classical and quantum gravity, Phys. Rev. Lett. 57, 2244 (1986)

\bibitem{ch4-Alex1}
S. Alexandrov, Reality conditions for Ashtekar gravity from Lorentz-covariant formulation, Class.Quant.Grav. 23 (2006) 1837-1850 e-Print: gr-qc/0510050 

\bibitem{ch4-Th1}
T. Thiemann, Reality conditions inducing transforms for quantum gauge field theory and quantum gravity, Class.Quant.Grav. 13 (1996) 1383-1404, e-Print: gr-qc/9511057 

\bibitem{ch4-WT1}
Guillermo A. Mena Marugan, Geometric interpretation of Thiemann's generalized Wick transform, Apr 1997. 14 pp. Grav.Cosmol. 4 (1998) 257, e-Print: gr-qc/9705031 

\bibitem{ch4-Ash2}
A. Ashtekar, A Generalized Wick Transform for Gravity, Phys.Rev. D53 (1996) 2865-2869 , e-Print: gr-qc/9511083

\bibitem{ch4-R1}
H A. Morales-Tecotl , L F. Urrutia, J. David Vergara,  Reality conditions for Ashtekar variables as Dirac constraints,  May 1996. 16 pp. Class.Quant.Grav. 13 (1996) 2933-2940, e-Print: gr-qc/9607044

\bibitem{ch4-R2} 
J.M. Pons, D.C. Salisbury, L.C. Shepley, Gauge group and reality conditions in Ashtekar's complex formulation of canonical gravity, Dec 1999. 17 pp. Phys.Rev. D62 (2000) 064026, e-Print: gr-qc/9912085 

\bibitem{ch4-R3}
G. Yoneda and H. Shinkai, Constraints and Reality Conditions in the Ashtekar Formulation of General Relativity, Class.Quant.Grav. 13 (1996) 783-790, arXiv:gr-qc/9602026

\bibitem{ch4-R4}
G. Yoneda and H. Shinkai, Lorentzian dynamics in the Ashtekar gravity, (1997) arXiv:gr-qc/9710074

\bibitem{ch4-Bar1}
J. F. Barbero, Real Ashtekar variables for Lorentzian signature space-times, Phys. Rev. D 51 5507 (1995), arXiv:gr-qc/9410014

\bibitem{ch4-Th2}
T. Thiemann , Modern Canonical Quantum General Relativity, (2007) , Cambridge University press

\bibitem{ch4-Sam1}
J. Samuel, Is Barbero's Hamiltonian formulation a gauge theory of Lorentzian gravity ?, Class. Quant. Grav. 17 L141 (2000), arXiv:gr-qc/0005095.

\bibitem{ch4-Pranz1}
D. Pranzetti,
Black hole entropy from KMS-states of quantum isolated horizons,
\texttt{arXiv:1305.6714 [gr-qc]}.

\bibitem{ch4-Pranz2}
D. Pranzetti and Hanno Sahlmann, Horizon entropy with loop quantum gravity methods, Phys.Lett. B746 (2015) 209-216, e-Print: arXiv:1412.7435 [gr-qc] , arXiv:1305.6714 [gr-qc].

\bibitem{ch4-Pranz3}
D. Pranzetti, Geometric temperature and entropy of quantum isolated horizons, Phys.Rev. D89 (2014) 10, 104046, e-Print: arXiv:1305.6714 [gr-qc]

\bibitem{ch4-Pranz4}
A. Ghosh and D. Pranzetti, CFT/Gravity Correspondence on the Isolated Horizon, Nucl.Phys. B889 (2014) 1-24, e-Print: arXiv:1405.7056 [gr-qc] 

\bibitem{ch4-Y0}
N. Bodendorfer and Y. Neiman,
``Imaginary action, spinfoam asymptotics and the `transplanckian' regime of loop quantum gravity'',
Class. Quant. Grav. \textbf{30} 195018 (2013), \texttt{arXiv:1303.4752 [gr-qc]}.

\bibitem{ch4-Y1}
Y. Neiman,
The imaginary part of the gravity action and black hole entropy,
JHEP \textbf{} 71 (2013), \texttt{arXiv:1212.2922 [gr-qc]}.

\bibitem{ch4-Y2}
Y. Neiman,
The imaginary part of the gravity action at asymptotic boundaries and horizons,
Phys. Rev. \textbf{D88} 024037 (2013), \texttt{arXiv:1305.2207 [gr-qc]}.

\bibitem{ch4-Y3}
Y. Neiman,
``The imaginary part of the gravity action and black hole entropy'',
JHEP  \textbf{04} (2013) 71, \texttt{arXiv:1301.7041 [gr-qc]}.

\bibitem{ch4-Mu1}
M. Han,
``Black hole entropy in loop quantum gravity, analytic continuation, and dual holography'',
(2014), \texttt{arXiv:1402.2084 [gr-qc]}.

\bibitem{ch4-Noui1}
M. Geiller and K. Noui,
Near-Horizon Radiation and Self-Dual Loop Quantum Gravity,
 Europhys. Lett. \textbf{105} (2014) 60001, \texttt{arXiv:1402.4138 [gr-qc]}.



\bibitem{ch4-BA1} 
J. Ben Achour, M. Geiller, K. Noui and C. Yu,
Testing the role of the Barbero--Immirzi parameter and the choice of connection in loop quantum gravity,
(2013), \texttt{arXiv:1306.3241 [gr-qc]}.

\bibitem{ch4-BA2}
J. Ben Achour, M. Geiller, K. Noui and C. Yu,
Spectra of geometric operators in three-dimensional LQG: From discrete to continuous,
accpeted to Phys. Rev. \textbf{D} (2013), \text{arXiv:1306.3246 [gr-qc]}.

\bibitem{ch4-Geiller1}
M. Geiller and K. Noui,
A note on the Holst action, the time gauge, and the Barbero-Immirzi parameter,
Gen. Rel. Grav. \textbf{45} 1733 (2013), \texttt{arXiv:1212.5064 [gr-qc]}.

\bibitem{ch4-Alex2} S. Alexandrov, SO(4,C)-covariant Ashtekar-Barbero gravity and the Immirzi param- eter, Class. Quant. Grav. 17 4255 (2000), arXiv:gr-qc/0005085.
\bibitem{ch4-Alex3}  S. Alexandrov and D. V. Vassilevich, Area spectrum in Lorentz-covariant loop grav- ity, Phys. Rev. D 64 044023 (2001), arXiv:gr-qc/0103105.
\bibitem{ch4-Alex4}  S. Alexandrov, On choice of connection in loop quantum gravity, Phys. Rev. D 65 024011 (2001), arXiv:gr-qc/0107071.
\bibitem{ch4-Alex5}  S. Alexandrov, Hilbert space structure of covariant loop quantum gravity, Phys. Rev. D 66 024028 (2002), arXiv:gr-qc/0201087.


\bibitem{ch4-Geiller2} E. Frodden, M. Geiller, K. Noui and A. Perez,
Black hole entropy from complex Ashtekar variables,
accepted to Eur. Phys. Lett. (2014), 
\texttt{arXiv:1212.4060 [gr-qc]}.

\bibitem{ch4-Geiller3} E. Frodden, M. Geiller, K. Noui and A. Perez,
Statistical entropy of a BTZ black hole from loop quantum gravity,
JHEP \textbf{5} 139 (2013), \texttt{arXiv:1212.4473 [gr-qc]}.

\bibitem{ch4-Amit2} 
A. Ghosh, K. Noui and A. Perez,
Statistics, holography, and black hole entropy in loop quantum gravity, 
Phys. Rev. \textbf{D} (2014),
 \texttt{arXiv:1309.4563 [gr-qc]}.
 
 \bibitem{ch4-BA3} 
J. Ben Achour, A. Mouchet, K. Noui,
Analytic continuation of black holes entropy in Loop Quantum Gravity, JHEP 1506 (2015) 145, e-Print: arXiv:1406.6021 [gr-qc] 

\bibitem{ch4-Witten1}
E. Witten,
``Analytic continuation of Chern--Simons theory'',
(2010), \texttt{arXiv:1001.2933 [hep-th]}.










\end{thebibliography}

\begin{thebibliography}{99}

\bibitem{ch5-Carlip1} S. Carlip, Quantum gravity in 2+ 1 dimensions (Vol. 50). (2003). Cambridge University Press.

\bibitem{ch5-Tet1} M. Banados, C. Teitelboim, and J. Zanelli, Phys. Rev. Lett.69, 1849 (1992)

\bibitem{ch5-Carlip2} S. Carlip, What we don't know about BTZ black hole entropy. Classical and Quantum Gravity, 15(11), 3609. (1998).

\bibitem{ch5-Nouio1} E. Frodden, M. Geiller, K. Noui and A. Perez, Statistical entropy of a BTZ black hole from loop quantum gravity, JHEP 5 139 (2013), arXiv:1212.4473 [gr-qc].

\bibitem{ch5-Nouio2} M. Geiller and K. Noui, BTZ Black Hole Entropy and the Turaev-Viro model, Dec 5, (2013), Annales Henri Poincare 16 (2015) 2, 609-640, e-Print: arXiv:1312.1696 [gr-qc]

\bibitem{ch5-Witten 3d} E. Witten,
$2+1$-dimensional gravity as an exactly soluble system,
Nucl. Phys. B \textbf{311} 46 (1988).

\bibitem{ch5-Smolinn1} L. Smolin, Loop representation for quantum gravity in 2+1 dimensions, in: Knots, topology and quantum field theory, Proceedings of the 12th Johns Hopkins Workshop

\bibitem{ch5-Marolf1} D.M. Marolf, Loop representations for 2+1 gravity on a torus, Class. Quant. Grav.10, (1993) 2625-2647

\bibitem{ch5-Ashte1} A. Ashtekar and R. Loll, New loop representations for (2+1) gravity, Class.Quant.Grav. 11 (1994) 2417-2434, CGPG-94-5-1, e-Print: gr-qc/9405031 

\bibitem{ch5-Frei1} L. Freidel, E. R. Livine, C. Rovelli, Spectra of length and area in (2+1) Lorentzian loop quantum gravity, Dec 2002, Class.Quant.Grav. 20 (2003) 1463-1478, e-Print: gr-qc/0212077

\bibitem{ch5-Noui3} K. Noui and A. Perez, Three-dimensional loop quantum gravity: Physical scalar product and spin foam models, Feb 2004, Class.Quant.Grav. 22 (2005) 1739-1762, e-Print: gr-qc/0402110

\bibitem{ch5-holst} S. Holst,
Barbero's Hamiltonian derived from a generalized Hilbert-Palatini action,
Phys. Rev. \textbf{D 53} 5966 (1996), \texttt{arXiv:gr-qc/9511026}.

\bibitem{ch5-alexandrov1} S. Alexandrov and D. V. Vassilevich,
Path integral for the Hilbert-Palatini and Ashtekar gravity,
Phys. Rev. \textbf{D 58} 124029 (1998), \texttt{arXiv:gr-qc/9806001}.

\bibitem{ch5-alexandrov2} S. Alexandrov,
SO(4,C)-covariant Ashtekar-Barbero gravity and the Immirzi parameter,
Class. Quant. Grav. \textbf{17} 4255 (2000), \texttt{arXiv:gr-qc/0005085}.

\bibitem{ch5-alexandrov6} S. Alexandrov,
Hilbert space structure of covariant loop quantum gravity,
Phys. Rev. \textbf{D 66} 024028 (2002), \texttt{arXiv:gr-qc/0201087}.

\bibitem{ch5-alexandrov5} S. Alexandrov,
On choice of connection in loop quantum gravity,
Phys. Rev. \textbf{D 65} 024011 (2001), \texttt{arXiv:gr-qc/0107071}.

\bibitem{ch5-GN2} M. Geiller and K. Noui,
Testing the imposition of the spin foam simplicity constraints,
Class. Quant. Grav. \textbf{29} 135008 (2012), \texttt{arXiv:1112.1965 [gr-qc]}.

\bibitem{ch5-GN} M. Geiller and K. Noui,
A note on the Holst action, the time gauge, and the Barbero-Immirzi parameter,
(2012), \texttt{arXiv:1212.5064 [gr-qc]}.


\bibitem{ch5-samuel} J. Samuel,
Is Barbero's Hamiltonian formulation a gauge theory of Lorentzian gravity?,
Class. Quant. Grav. \textbf{17} L141 (2000), \texttt{arXiv:gr-qc/0005095}.

\bibitem{ch5-Cath and I2} C. Meusburger and K. Noui,
Combinatorial quantization of the Euclidean torus universe,
Nuclear Physics B \textbf{841} 463 (2010), \texttt{arXiv:1007.4615 [gr-qc]}.


\bibitem{ch5-barbero} J. F. Barbero,
Real Ashtekar variables for Lorentzian signature spacetimes,
Phys. Rev. \textbf{D 51} 5507 (1995).



\bibitem{ch5-immirzi} G. Immirzi,
Real and complex connections for canonical gravity,
Class. Quant. Grav. \textbf{14} L177 (1997), \texttt{arXiv:gr-qc/9612030}.

\bibitem{ch5-complexashtekar} A. Ashtekar,
New variables for classical and quantum gravity,
Phys. Rev. Lett. \textbf{57} 2244 (1986).

\bibitem{ch5-rovelli-thiemann} C. Rovelli and T. Thiemann,
The Immirzi parameter in quantum general relativity,
Phys. Rev. \textbf{D 57} 1009 (1998), \texttt{arXiv:gr-qc/9705059}.

\bibitem{ch5-menamarugan} G. A. Mena Marugan,
Extent of the Immirzi ambiguity in quantum general relativity,
Class. Quant. Grav. \textbf{19} L63 (2002), \texttt{arXiv:gr-qc/0203027}.

\bibitem{ch5-fermions1} L. Freidel, D. Minic and T. Takeuchi,
Quantum gravity, torsion, parity violation and all that,
Phys. Rev. \textbf{D 72} 104002 (2005), \texttt{arXiv:hep-th/0507253}.

\bibitem{ch5-fermions2} S. Mercuri,
Fermions in Ashtekar-Barbero connections formalism for arbitrary values of the Immirzi parameter,
Phys. Rev. \textbf{D 73} 084016 (2006), \texttt{arXiv:gr-qc/0601013}.

\bibitem{ch5-fermions3} A. Perez and C. Rovelli,
Physical effects of the Immirzi parameter,
Phys. Rev. \textbf{D 73} 044013 (2006), \texttt{arXiv:gr-qc/0505081}.

\bibitem{ch5-taveras-yunes} V. Taveras and N. Yunes,
The Barbero-Immirzi parameter as a scalar field: K-inflation from loop quantum gravity?,
Phys. Rev. \textbf{D 78} 064070 (2008), \texttt{arXiv:0807.2652 [gr-qc]}.

\bibitem{ch5-fermions4} M. Bojowald and R. Das,
Canonical gravity with fermions,
Phys. Rev. \textbf{D 78} 064009 (2008), \texttt{arXiv:0710.5722 [gr-qc]}.

\bibitem{ch5-fermions5} S. Alexandrov,
Immirzi parameter and fermions with non-minimal coupling,
Class. Quant. Grav. \textbf{25} 145012 (2008), \texttt{arXiv:0802.1221 [gr-qc]}.

\bibitem{ch5-mercuri-taveras} S. Mercuri and V. Taveras,
Interaction of the Barbero--Immirzi field with matter and pseudo-scalar perturbations,
Phys. Rev. \textbf{D 80} 104007 (2009), \texttt{arXiv:0903.4407 [gr-qc]}.

\bibitem{ch5-mercuri-gamma} S. Mercuri,
Peccei--Quinn mechanism in gravity and the nature of the Barbero--Immirzi parameter,
Phys. Rev. Lett. \textbf{103} 081302 (2009), \texttt{arXiv:0902.2764 [gr-qc]}.

\bibitem{ch5-mercuri-gamma2} S. Mercuri,
A possible topological interpretation of the Barbero-Immirzi parameter,
(2009), \texttt{arXiv:0903.2270 [gr-qc]}.

\bibitem{ch5-mercuri-randono} S. Mercuri and A. Randono,
The Immirzi parameter as an instanton angle,
Class. Quant. Grav. \textbf{28} 025001 (2011), \texttt{arXiv:1005.1291 [hep-th]}.

\bibitem{ch5-dittrich-ryan2} B. Dittrich and J. P. Ryan,
On the role of the Barbero-Immirzi parameter in discrete quantum gravity,
(2012), \texttt{arXiv:1209.4892 [gr-qc]}.


\bibitem{ch5-aleSF} A. Perez,
The spin foam approach to quantum gravity,
Living Rev. Relativity \textbf{16} 3 (2013), \texttt{arXiv:1205.2019 [gr-qc]}.

\bibitem{ch5-FGNP} E. Frodden, M. Geiller, K. Noui and A. Perez,
Black hole entropy from complex Ashtekar variables,
(2012), \texttt{arXiv:1212.4060 [gr-qc]}.

\bibitem{ch5-BST} N. Bodendorfer, A. Stottmeister and A. Thurn,
Loop quantum gravity without the Hamiltonian constraint
Class. Quant. Grav. \textbf{30} 082001 (2013), \texttt{arXiv:1203.6525 [gr-qc]}.

\bibitem{ch5-BN} N. Bodendorfer and Y. Neiman,
Imaginary action, spinfoam asymptotics and the 'transplanckian' regime of loop quantum gravity,
(2013), \texttt{arXiv:1303.4752 [gr-qc]}.


\bibitem{ch5-Alekseev:1994pa} A. Y. Alekseev, H. Grosse and V. Schomerus,
Combinatorial quantization of the Hamiltonian Chern-Simons theory,
Commun. Math. Phys. \textbf{172} 317 (1995), \texttt{arXiv:hep-th/9403066}.

\bibitem{ch5-Alekseev:1994au} A. Y. Alekseev, H. Grosse and V. Schomerus,
Combinatorial quantization of the Hamiltonian Chern-Simons theory. 2.,
Commun. Math. Phys. \textbf{174} 561 (1995), \texttt{arXiv:hep-th/9408097}.

\bibitem{ch5-BNR} E. Buffenoir, K. Noui and P. Roche,
Hamiltonian quantization of Chern-Simons theory with $\SL(2,\mathbb{C})$ group,
Class. Quant. Grav. \textbf{19} 4953 (2002), \texttt{arXiv:hep-th/0202121}.

\bibitem{ch5-Catherine} C. Meusburger and B. Schroers,

\bibitem{ch5-wieland} W. Wieland,
Complex Ashtekar variables and reality conditions for Holst's action,
Annales H. Poincar\'e 1 (2011), \texttt{arXiv:1012.1738 [gr-qc]}.


\bibitem{ch5-Bonzon Livine} V. Bonzom and E. Livine,
A Immirzi-like parameter for 3d quantum gravity,
Class. Quant. Grav. \textbf{25} 195024 (2008), \texttt{arXiv:0801.4241 [gr-qc]}.

\bibitem{ch5-urbantke} H. Urbantke,
On integrability properties of SU(2) Yang-Mills fields. I. Infinitesimal part,
J. Math. Phys. \textbf{25} 2321 (1984).

\bibitem{ch5-Sergey unpublished} S. Alexandrov,
(2012), unpublished.

\bibitem{ch5-mitra-rajaraman} P. Mitra and R. Rajaraman,
Gauge-invariant reformulation of an anomalous gauge theory,
Phys. Lett. B \textbf{225} 267 (1989).


\bibitem{ch5-Achucarro-Townsend} A. Ach\'ucarro and P. K. Townsend,
A Chern-Simons action for three-dimensional anti-de Sitter supergravity theories,
Phys. Lett. B \textbf{180} 89 (1986).

\bibitem{ch5-Witten Jones} E. Witten,
Quantum field theory and the Jones polynomial,
Commun. Math. Phys. \textbf{121} 351 (1989).

\bibitem{ch5-ponzano-regge} G. Ponzano and T. Regge,
in \textit{Spectroscopy and group theoretical methods in physics},
ed. by F. Block (North Holland, 1968).

\bibitem{ch5-turaev-viro} V. G. Turaev and O. Y. Viro,
State sum invariants of 3-manifolds and quantum 6j-symbols,
Topology \textbf{31} 865 (1992), \texttt{arXiv:gr-qc/0402110}.

\bibitem{ch5-BC} J. W. Barrett and L. Crane,
Relativistic spin networks and quantum gravity,
J. Math. Phys. \textbf{39} 3296 (1998), \texttt{arXiv:gr-qc/9709028}.

\bibitem{ch5-EPR} J. Engle, R. Pereira and C. Rovelli,
Flipped spinfoam vertex and loop gravity,
Nucl. Phys. \textbf{B 798} 251 (2008), \texttt{arXiv:0708.1236 [gr-qc]}.

\bibitem{ch5-EPRL} J. Engle, E. R. Livine, R. Pereira and C. Rovelli,
LQG vertex with finite Immirzi parameter,
Nucl. Phys. \textbf{B 799} 136 (2008), \texttt{arXiv:0711.0146 [gr-qc]}.

\bibitem{ch5-livine-speziale1} E. R. Livine and S. Speziale,
Consistently solving the simplicity constraints for spinfoam quantum gravity,
Europhys. Lett. \textbf{81} 50004 (2008), \texttt{arXiv:0708.1915 [gr-qc]}.

\bibitem{ch5-livine-speziale2} E. R. Livine and S. Speziale,
A new spinfoam vertex for quantum gravity,
Phys. Rev. \textbf{D 76} 084028 (2007), \texttt{arXiv:gr-qc/0705.0674}.

\bibitem{ch5-FK} L. Freidel and K. Krasnov,
A new spin foam model for 4d gravity,
Class. Quant. Grav. \textbf{25} 125018 (2008), \texttt{arXiv:gr-qc/0708.1595 [gr-qc]}.

\bibitem{ch5-Noui:2004iy} K. Noui and A. Perez,
Three dimensional loop quantum gravity: Physical scalar product and spin foam models,
Class. Quant. Grav. \textbf{22} 1739 (2005), \texttt{arXiv:gr-qc/0402110}.

\bibitem{ch5-Noui:2004iz} K. Noui and A. Perez,
Three dimensional loop quantum gravity: Coupling to point particles,
Class. Quant. Grav. \textbf{22} 4489 (2005), \texttt{arXiv:gr-qc/0402111}.

\bibitem{ch5-AGN} S. Alexandrov, M. Geiller and K. Noui,
Spin foams and canonical quantization,
Sigma \textbf{8} 055 (2012), \texttt{arXiv:1112.1961 [gr-qc]}.

\bibitem{ch5-Witten analytic} E. Witten,
Analytic continuation of Chern-Simons theory,
(2010), \texttt{arXiv:1001.2933 [hep-th]}.

\bibitem{ch5-Cath and I} C. Meusburger and K. Noui,
The Hilbert space of 3d gravity: quantum group symmetries and observables,
Adv. Theor. Math. Phys. \textbf{14} 1651 (2010), \texttt{arXiv:0809.2875 [gr-qc]}.


\bibitem{ch5-Freidel Livine} L. Freidel and E. Livine,
Spin networks for non-compact groups,
Math. Phys. \textbf{44} 1322 (2003), \texttt{arXiv:hep-th/0205268}.

\bibitem{ch5-BTZ-SF} J. M. Garcia-Islas,
BTZ black hole entropy: A spin foam model description,
Class. Quant. Grav. \textbf{25} 245001 (2008), \texttt{arXiv:0804.2082 [gr-qc]}.

\bibitem{ch5-ale-amb} A. Perez,
On the regularization ambiguities in loop quantum gravity,
Phys. Rev. \textbf{D 73} 044007 (2006), \texttt{arXiv:gr-qc/0509118}.

\bibitem{ch5-Bonzom:2011jv} V. Bonzom and A. Laddha,
Lessons from toy-models for the dynamics of loop quantum gravity,
Sigma \textbf{8} 055 (2012), \texttt{arXiv:1110.2157 [gr-qc]}.

\bibitem{ch5-freidel-livine-rovelli} L. Freidel, E. R. Livine and C. Rovelli,
Spectra of length and area in 2+1 Lorentzian loop quantum gravity,
Class. Quant. Grav. \textbf{20} 1463 (2003), \texttt{arXiv:gr-qc/0212077}.

\bibitem{ch5-EPN} J. Engle, A. Perez and K. Noui,
Black hole entropy and SU(2) Chern-Simons theory,
Phys. Rev. Lett. \textbf{105} 031302 (2010), \texttt{arXiv:0905.3168 [gr-qc]}.

\bibitem{ch5-ENPP} J. Engle, K. Noui, A. Perez and D. Pranzetti,
Black hole entropy from an SU(2)-invariant formulation of type I isolated horizons,
Phys. Rev. \textbf{D 82} 044050 (2010), \texttt{arXiv:1006.0634 [gr-qc]}.

\end{thebibliography}

\begin{thebibliography}{99}

\bibitem{ch6-BA1}  J. Ben Achour, J. Grain, K. Noui, Loop Quantum Cosmology with complex Ashtekar variables, (2014), Class.Quant.Grav. 32 (2015) 2, 025011, e-Print: arXiv:1407.3768 [gr-qc]

\bibitem{ch6-Ash1} A. Ashtekar and and  P. Singh, Loop Quantum Cosmology: A Status Report, Class.Quant.Grav. 28 (2011) 213001, e-Print: arXiv:1108.0893 [gr-qc]

\bibitem{ch6-Agullo1} I. Agullo, A. Corichi, Loop Quantum Cosmology, (2013). e-Print: arXiv:1302.3833 [gr-qc] 

\bibitem{ch6-Bojowald1} M. Bojowald, Absence of singularity in loop quantum cosmology, (2001), Phys.Rev.Lett. 86 (2001) 5227-5230, CGPG-01-2-1, e-Print: gr-qc/0102069 

\bibitem{ch6-Ash2} A. Ashtekar, The Big Bang and the Quantum, (2010). AIP Conf.Proc. 1241 (2010) 109-121, Plenary talk at Conference: C09-06-29.1 Proceedings, e-Print: arXiv:1005.5491 [gr-qc]

\bibitem{ch6-Ash3} A. Ashtekar, A. Corichi and P. Singh, Robustness of key features of loop quantum cosmology, Phys.Rev. D77 (2008) 024046, IGC-07-10-01, PI-QG-61, e-Print: arXiv:0710.3565 [gr-qc]

\bibitem{ch6-Sloan1} Abhay Ashtekar and David Sloan, Probability of Inflation in Loop Quantum Cosmology, (2011). Gen.Rel.Grav. 43 (2011) 3619-3655, IGC-11-03-02, e-Print: arXiv:1103.2475 [gr-qc] 

\bibitem{ch6-Barrau1} A. Ashtekar and A. Barrau, Loop quantum cosmology: From pre-inflationary dynamics to observations, (2015), e-Print: arXiv:1504.07559 [gr-qc]

\bibitem{ch6-Grain1} A. Barrau, M. Bojowald, G. Calcagni, J. Grain, M. Kagan, Anomaly-free cosmological perturbations in effective canonical quantum gravity, (2014). JCAP 1505 (2015) 05, 051, e-Print: arXiv:1404.1018 [gr-qc]

\bibitem{ch6-Agullo2} I Agullo, A. Ashtekar and W. Nelson, Extension of the quantum theory of cosmological perturbations to the Planck era, (2012). Phys.Rev. D87 (2013) 4, 043507, IGC-12-11-3, e-Print: arXiv:1211.1354 [gr-qc] 

\bibitem{ch6-Bojowald2} A. Ashtekar,  M. Bojowald  and J. Lewandowski, Mathematical structure of loop quantum cosmology, (2003). Adv.Theor.Math.Phys. 7 (2003) 233-268, CGPG-03-4-4, e-Print: gr-qc/0304074 

\bibitem{ch6-Coutant1} A. Coutant, Unitary and nonunitary transitions around a cosmological bounce, Phys.Rev. D89 (2014) 12, 123524, e-Print: arXiv:1404.5634 [gr-qc] 

\bibitem{ch6-Ash4}  A. Ashtekar, T. Pawlowski and  P. Singh,  Quantum Nature of the Big Bang: Improved dynamics, (2006), Phys.Rev. D74 (2006) 084003, IGPG-06-7-2, e-Print: gr-qc/0607039 

\bibitem{ch6-EW1} E. Wilson-Ewing, Loop quantum cosmology with self-dual variables, (2015), e-Print: arXiv:1503.07855 [gr-qc] 

 	
\bibitem{ch6-Nelson1} Ivan Agullo, Abhay Ashtekar and William Nelson, The pre-inflationary dynamics of loop quantum cosmology: Confronting quantum gravity with observations, (2013). Class.Quant.Grav. 30 (2013) 085014, e-Print: arXiv:1302.0254 [gr-qc] 

\bibitem{ch6-BA2} J. Ben Achour and K. Noui, Analytic continuation of real Loop Quantum Gravity: lessons from black hole thermodynamics, (2015) Proceeding of Conference: C14-07-15.1, e-Print: arXiv:1501.05523 [gr-qc]
\bibitem{ch6-BA3} J. Ben Achour, A. Mouchet and K. Noui, Analytic Continuation of Black Hole Entropy in Loop Quantum Gravity (2014) JHEP, e-Print: arXiv:1406.6021 [gr-qc]

\bibitem{ch6-BA4} J. Ben Achour, M. Geiller, K. Noui and C. Yu, Testing the role of the Barbero-Immirzi parameter and the choice of connection in Loop Quantum Gravity, Phys.Rev. D91 (2013) 10, 104016, e-Print: arXiv:1306.3241 [gr-qc]
\bibitem{ch6-BA5}  J. Ben Achour, M. Geiller, K. Noui and C. Yu, Spectra of geometric operators in three- dimensional loop quantum gravity: From discrete to continuous, Phys.Rev. D89 (2013) 6, 064064, e-Print: arXiv:1306.3246 [gr-qc]


\end{thebibliography}

\end{document}